\documentclass[11pt]{article}
\pdfoutput=1

\usepackage{jcappub}
\usepackage{lineno}
\usepackage{color, xcolor, appendix,latexsym,amsmath,amssymb,graphicx,booktabs,epsfig,hyperref,url,nicefrac,diagbox,multirow,relsize}
\usepackage{tablefootnote}
\usepackage{bm} 
\usepackage[acronym]{glossaries} 
\makeglossaries
\newacronym{et}{ET}{Einstein Telescope}
\newacronym{lvc}{LVC}{LIGO-Virgo Collaboration}
\newacronym{gw}{GW}{Gravitational Wave}
\newacronym{bbh}{BBH}{Binary Black Hole}
\newacronym{bh}{BH}{Black Hole}
\newacronym{bns}{BNS}{Binary Neutron Star}
\newacronym{nsbh}{NSBH}{Neutron Star-Black Hole}
\newacronym{grb}{GRB}{Gamma Ray Burst}
\newacronym{3g}{3G}{Third Generation}
\newacronym{ce}{CE}{Cosmic Explorer}
\newacronym{hf}{HF}{High-Frequency}
\newacronym{lf}{LF}{Low-Frequency}
\newacronym{osb}{OSB}{Observational Science Board}
\newacronym{pls}{PLS}{Power Law Sensitivity}
\newacronym{isb}{ISB}{Instrument Science Board}
\newacronym{cbc}{CBC}{Compact Binary Coalescence}
\newacronym{lvki}{LVKI}{LIGO Hanford, LIGO Livingston, Virgo, KAGRA, LIGO India}
\newacronym{asd}{ASD}{Amplitude Spectral Density}
\newacronym{snr}{SNR}{Signal-to-Noise Ratio}
\newacronym{imr}{IMR}{Inspiral Merger Ringdown}
\newacronym{hm}{HM}{Higher Modes}
\newacronym{pn}{PN}{Post Newtonian}
\newacronym{gr}{GR}{General Relativity}
\newacronym{imbh}{IMBH}{Intermediate Mass Black Hole}
\newacronym{em}{EM}{Electromagnetic}
\newacronym{kn}{KN}{Kilonova}
\newacronym{eos}{EoS}{Equation of State}
\newacronym{fov}{FoV}{Field of View}
\newacronym{uv}{UV}{Ultra Violet}
\newacronym{ir}{IR}{Infrared}
\newacronym{sxi}{SXI}{Soft X-ray Imager}
\newacronym{xgis}{XGIS}{X/Gamma-ray Imaging Spectrometer}
\newacronym{vro}{VRO}{Vera Rubin Observatory}
\newacronym{sgwb}{SGWB}{Stochastic Gravitational Wave Background}
\newacronym{agwb}{AGWB}{Astrophysical Gravitational Wave Background}
\newacronym{cgwb}{CGWB}{Cosmological Gravitational Wave Background}
\newacronym{cib}{CIB}{Cosmic Infrared Background}
\newacronym{cmb}{CMB}{Cosmic Microwave Background}
\newacronym{pli}{PLI}{Power-Law Integrated}
\newacronym{sfh}{SFH}{Star Formation History}
\newacronym{csd}{CSD}{Cross-power Spectral Density}
\newacronym{nn}{NN}{Newtonian Noise}
\newacronym{qnm}{QNM}{Quasinormal Mode}
\newacronym{eco}{ECO}{Exotic Compact Object}
\newacronym{isco}{ISCO}{Innermost Stable Circular Orbit}
\newacronym{tov}{TOV}{Tolman–Oppenheimer–Volkoff}
\newacronym{hmns}{HMNS}{Hypermassive Neutron Star}
\newacronym{smns}{SMNS}{Supramassive Neutron Star}
\newacronym{nr}{NR}{Numerical Relativity}
\newacronym{pbh}{PBH}{Primordial Black Hole}
\newacronym{dm}{DM}{Dark Matter}
\newacronym{lsst}{LSST}{Large Synoptic Survey Telescope}
\newacronym{bao}{BAO}{Barionic Acoustic Oscillation}
\newacronym{sn}{SN}{Supernova}
\newacronym{frw}{FRW}{Friedmann-Robertson-Walker}
\newacronym{gwtc}{GWTC}{Gravitational Wave Transient Catalog}
\newacronym{fopt}{FOPT}{First-Order Phase Transition}
\newacronym{cw}{CW}{Continuous Wave}
\newacronym{tcw}{tCW}{Transient Continuous Wave}
\newacronym{cr}{CR}{Critical Ratio}
\newacronym{cdo}{CDO}{Compact Dark Object}
\newacronym{far}{FAR}{False Alarm Rate}
\newacronym{orf}{ORF}{Overlap Reduction Function}
\newacronym{de}{DE}{Dark Energy}
\newacronym{psd}{PSD}{Power Spectral Density}

\numberwithin{equation}{section}
\usepackage[english]{babel}
\usepackage{subcaption}
\newcommand{\PBH}{\text{\tiny PBH}}
\newcommand{\nv}{\hat{\bf n}}
\newcommand{\resp}{{\cal A}}

\newcommand{\Kappa}[0]{\scalebox{1.5}{$\kappa$}}

\newcommand{\rosso}{} 

\definecolor{MyBlue}{rgb}{0.15,0.15,0.70}
\hypersetup{
colorlinks=true,
citecolor=MyBlue,
linkcolor=MyBlue,
urlcolor=MyBlue
}
\definecolor{lightgray}{gray}{0.9}

\newcommand{\dgw}{d_L^{\,\rm gw}}
\newcommand{\dem}{d_L^{\,\rm em}}

\newcommand{\nn}{\nonumber}

\renewcommand\({\left(}
\renewcommand\){\right)}

\newcommand{\ra}{\rightarrow}

\def\lsim{\raise 0.4ex\hbox{$<$}\kern -0.8em\lower 0.62
ex\hbox{$\sim$}}

\def\gsim{\raise 0.4ex\hbox{$>$}\kern -0.7em\lower 0.62
ex\hbox{$\sim$}}

\def\lbar{{\hbox{$\lambda$}\kern -0.7em\raise 0.6ex
\hbox{$-$}}}

\newcommand\eq[1]{eq.~(\ref{#1})}
\newcommand\eqs[2]{eqs.~(\ref{#1}) and (\ref{#2})}

\newcommand\eqst[2]{eqs.~(\ref{#1})--(\ref{#2})}

\newcommand\p{\partial}

\newcommand\ee{\end{equation}}
\newcommand\be{\begin{equation}}
\def\bea{\begin{array}}
\def\eea{\end{array}}\def\ea{\end{array}}
\newcommand\ees{\end{eqnarray}}
\newcommand\bees{\begin{eqnarray}}
\def\nn{\nonumber}

\def\d{\delta}

\def\dslash{\hspace{-1mm}\not{\hbox{\kern-2pt $\partial$}}}
\def\Dslash{\not{\hbox{\kern-2pt $D$}}}
\def\pslash{\not{\hbox{\kern-2.1pt $p$}}}
\def\kslash{\not{\hbox{\kern-2.3pt $k$}}}
\def\qslash{\not{\hbox{\kern-2.3pt $q$}}}

\newcommand{\vk}{{\bf k}}

\def\p1{{\bf p}_1}
\def\p2{{\bf p}_2}
\def\k1{{\bf k}_1}
\def\k2{{\bf k}_2}

\newcommand{\dddM}{\kern 0.2em \raise 1.9ex\hbox{$...$}\kern -1.0em \hbox{$M$}}
\newcommand{\dddQ}{\kern 0.2em \raise 1.9ex\hbox{$...$}\kern -1.0em \hbox{$Q$}}
\newcommand{\dddI}{\kern 0.2em \raise 1.9ex\hbox{$...$}\kern -1.0em\hbox{$I$}}
\newcommand{\dddJ}{\kern 0.2em \raise 1.9ex\hbox{$...$}\kern-1.0em
\hbox{$J$}}
\newcommand{\dddcalJ}{\kern 0.2em \raise 1.9ex\hbox{$...$}\kern-1.0em
\hbox{${\cal J}$}}

\newcommand{\dddO}{\kern 0.2em \raise 1.9ex\hbox{$...$}\kern -1.0em
\hbox{${\cal O}$}}
\def\dddz{\raise 1.5ex\hbox{$...$}\kern -0.8em \hbox{$z$}}
\def\dddd{\raise 1.8ex\hbox{$...$}\kern -0.8em \hbox{$d$}}
\def\dddbd{\raise 1.8ex\hbox{$...$}\kern -0.8em \hbox{${\bf d}$}}
\def\ddbd{\raise 1.8ex\hbox{$..$}\kern -0.8em \hbox{${\bf d}$}}
\def\dddx{\raise 1.6ex\hbox{$...$}\kern -0.8em \hbox{$x$}}

\newcommand{\msun}{M_{\odot}}

\newcommand{\Ogw}{\Omega_{\rm gw}}

\newcommand{\oma}{\Omega_{M}}

\newcommand{\ola}{\Omega_{\Lambda}}

\newcommand{\rde}{\rho_{\rm DE}}
\newcommand{\wde}{w_{\rm DE}}

\definecolor{orange}{rgb}{0.98, 0.6, 0.01}
\definecolor{darkolivegreen}{rgb}{0.33, 0.42, 0.18}
\definecolor{tealblue}{rgb}{0.21, 0.46, 0.53}

\graphicspath{figures/} 

\title{Science with the Einstein Telescope: \\ a comparison of different designs}

\author[1,2]{Marica Branchesi,}
\author[3,4]{Michele Maggiore,}
\author[5]{David Alonso,}
\author[6]{Charles Badger,}
\author[1,2]{Biswajit Banerjee,}
\author[7]{Freija Beirnaert,}
\author[3,4]{Enis Belgacem,}
\author[8,9]{Swetha Bhagwat,}
\author[10 ,11]{Guillaume Boileau,}
\author[12]{Ssohrab Borhanian,}
\author[13]{Daniel David Brown,}
\author[14]{Man Leong Chan,}
\author[15,3,4]{Giulia Cusin,}
\author[16,17]{Stefan L. Danilishin,}
\author[18]{Jerome Degallaix,}
\author[19]{Valerio De Luca,}
\author[20]{Arnab Dhani,}
\author[21,22]{Tim Dietrich,}
\author[1,2]{Ulyana Dupletsa,}
\author[3,4]{Stefano Foffa,}
\author[8]{Gabriele Franciolini,}
\author[23,16]{Andreas Freise,}
\author[24]{Gianluca Gemme,}
\author[1,2]{Boris Goncharov,}
\author[7]{Archisman Ghosh,}
\author[25]{Francesca Gulminelli,}
\author[20]{Ish Gupta,}
\author[16,26]{Pawan Kumar Gupta,}
\author[1,2]{Jan Harms,}
\author[1,2,27]{Nandini Hazra,}
\author[16,17]{Stefan Hild,}
\author[28]{Tanja  Hinderer,}
\author[29]{Ik Siong Heng,} 
\author[3,4]{Francesco Iacovelli,}
\author[16,26]{Justin Janquart,}
\author[10,11]{Kamiel Janssens,}
\author[30]{Alexander C. Jenkins,}
\author[16,26,31]{Chinmay  Kalaghatgi,}
\author[32,33]{Xhesika Koroveshi,}
\author[34,35]{Tjonnie~G.~F.~Li,}
\author[36]{Yufeng Li,}
\author[1,2]{Eleonora Loffredo,}
\author[22]{Elisa Maggio,}
\author[3,4,37,38]{Michele Mancarella,}
\author[39,40,41]{Michela Mapelli,}
\author[6]{Katarina Martinovic,}
\author[1,2]{Andrea Maselli,}
\author[42]{Patrick Meyers,}
\author[43,16,26]{Andrew L. Miller,}
\author[25]{Chiranjib Mondal,}
\author[3,4]{Niccol\`o Muttoni,}
\author[16,26]{Harsh Narola,} 
\author[44]{Micaela Oertel,}
\author[1,2]{Gor Oganesyan,}
\author[8,37,38]{Costantino Pacilio,}
\author[45]{Cristiano Palomba,}
\author[8]{Paolo Pani,}
\author[46]{Antonio Pasqualetti,}
\author[47,48]{Albino Perego,}
\author[39,40,41]{Carole P\'{e}rigois,}
\author[49, 50]{Mauro Pieroni,}
\author[51]{Ornella Juliana Piccinni,}
\author[16,26]{Anna Puecher,}
\author[45]{Paola Puppo,}
\author[52,39,40]{Angelo Ricciardone,}
\author[3,4]{Antonio Riotto,}
\author[1,2]{Samuele Ronchini,}
\author[6]{Mairi Sakellariadou,}
\author[21]{Anuradha Samajdar,}
\author[39,40,41]{Filippo Santoliquido,}
\author[20,53,54]{B.S. Sathyaprakash,}
\author[16,17]{Jessica Steinlechner,}
\author[16,17]{Sebastian Steinlechner,}
\author[16, 17]{Andrei Utina,}
\author[16,26]{Chris Van Den Broeck,}
\author[9,17]{and Teng Zhang}

\affiliation[1]{Gran Sasso Science Institute (GSSI), I-67100 L'Aquila, Italy}
\affiliation[2]{INFN, Laboratori Nazionali del Gran Sasso, I-67100 Assergi, Italy}
\affiliation[3]{D\'epartement de Physique Th\'eorique,
Universit\'e de Gen\`eve, 24 quai Ansermet, CH-1211 Gen\`eve 4, Switzerland}
\affiliation[4]{Gravitational Wave Science Center (GWSC), Universit\'e de Gen\`eve, CH-1211 Geneva, Switzerland}
\affiliation[5]{Department of Physics, University of Oxford, Denys Wilkinson Building, Keble Road, Oxford OX1 3RH, United Kingdom}
\affiliation[6]{Theoretical Particle Physics and Cosmology Group,  Physics Department, King's College London, University of London, Strand, London WC2R 2LS, United Kingdom}
\affiliation[7]{Department of Physics and Astronomy, Ghent University, 9000 Ghent, Belgium}
\affiliation[8]{Dipartimento di Fisica, ``Sapienza'' Universit\`a di Roma \& Sezione INFN Roma1,\\ Piazzale Aldo Moro 5, 00185, Roma, Italy}
\affiliation[9]{Institute for Gravitational Wave Astronomy \& School of Physics and Astronomy,
University of Birmingham, Edgbaston, Birmingham B15 2TT, UK}
\affiliation[10]{Universiteit Antwerpen, Prinsstraat 13, 2000 Antwerpen, Belgium}
\affiliation[11]{Universit\'e C\^{o}te d'Azur, Observatoire de la C\^{o}te d'Azur, CNRS, ARTEMIS, 06304 Nice, France}
\affiliation[12]{Theoretisch-Physikalisches Institut, Friedrich-Schiller-Universit{\"a}t Jena, 07743, Jena, Germany}
\affiliation[13]{ARC Centre of Excellence for Gravitational Wave Discovery,
School of Physical Sciences, University of Adelaide, 5005
Australia}
\affiliation[14]{Department of Physics and Astronomy, The University of British Columbia, Vancouver, BC V6T 1Z4, Canada}
\affiliation[15]{Sorbonne Université, CNRS, UMR 7095, Institut d'Astrophysique de Paris, 75014 Paris, France}
\affiliation[16]{Nikhef -- National Institute for Subatomic Physics, 
Science Park 105, 1098 XG Amsterdam, The Netherlands}
\affiliation[17]{Department of Gravitational Waves and Fundamental Physics, Maastricht University, 6200 MD Maastricht, The Netherlands}
\affiliation[18]{Universit\'e Lyon, Universit\'e Claude Bernard Lyon 1, CNRS, Laboratoire des Mat\'eriaux Avanc\'es (LMA), IP2I Lyon / IN2P3,F-69622 Villeurbanne, France}
\affiliation[19]{Center for Particle Cosmology, Department of Physics and Astronomy,
University of Pennsylvania 209 S. 33rd St., Philadelphia, PA 19104, USA}
\affiliation[20]{Institute for Gravitation and the Cosmos, Department of Physics, The Pennsylvania State University, University Park, PA 16802, USA}
\affiliation[21]{Institut f\"{u}r Physik und Astronomie, Universit\"{a}t Potsdam, Haus 28, Karl-Liebknecht-Str. 24/25, 14476, Potsdam, Germany}
\affiliation[22]{Max Planck Institute for Gravitational Physics (Albert Einstein Institute), D-14476 Potsdam, Germany}
\affiliation[23]{Department of Physics and Astronomy, VU Amsterdam, De Boelelaan 1081, 1081, HV, Amsterdam, The Netherlands}
\affiliation[24]{INFN, Sezione di Genova, I-16146 Genova, Italy}
\affiliation[25]{Normandie Univ, ENSICAEN, UNICAEN, CNRS/IN2P3, LPC Caen, 14000 Caen, France} 
\affiliation[26]{Institute for Gravitational and Subatomic Physics (GRASP), 
Utrecht University, Princetonplein 1, 3584 CC Utrecht, The Netherlands}
\affiliation[27]{INAF - Osservatorio Astronomico d'Abruzzo, Teramo, Italy}
\affiliation[28]{Institute for Theoretical Physics, Utrecht University, Princetonplein 5, 3584 CC Utrecht, The Netherlands}
\affiliation[29]{SUPA, School of Physics and Astronomy, University of Glasgow, Glasgow G12 8QQ, United Kingdom}
\affiliation[30]{Department of Physics and Astronomy, University College London, London WC1E 6BT, United Kingdom}
\affiliation[31]{Institute for High-Energy Physics, University of Amsterdam, Science Park 904, 1098 XH Amsterdam,  The Netherlands}
\affiliation[32]{ Institute of Technical Thermodynamics and Refrigeration: Refrigeration and Cryogenics, Karlsruhe Institute of Technology (KIT), 76131 Karlsruhe, Germany}
\affiliation[33]{ Institute of Beam Physics and Technology, Karlsruhe Institute of Technology (KIT), 76344 Eggenstein-Leopoldshafen, Germany}
\affiliation[34]{Institute for Theoretical Physics, Department of Physics and Astronomy, KU Leuven, Celestijnenlaan 200D, B-3001 Leuven, Belgium}
\affiliation[35]{STADIUS, Department of Electrical Engineering (ESAT), KU Leuven, Kasteelpark Arenberg 10, B-3001 Leuven, Belgium}
\affiliation[36]{Department of Astronomy, Beijing Normal University, Beijing 100875, China}
\affiliation[37]{Dipartimento di Fisica ``G. Occhialini'', Universit\'a degli Studi di Milano-Bicocca, Piazza della Scienza 3, 20126 Milano, Italy}
\affiliation[38]{INFN, Sezione di Milano-Bicocca, Piazza della Scienza 3, 20126 Milano, Italy}
\affiliation[39]{Physics and Astronomy Department Galileo Galilei, University of Padova, Vicolo dell'Osservatorio 3, I--35122, Padova, Italy}
\affiliation[40]{INFN -- Padova, Via Marzolo 8, I--35131 Padova, Italy}
\affiliation[41]{INAF -- Osservatorio Astronomico di Padova, Vicolo dell'Osservatorio 5, I--35122 Padova, Italy}
\affiliation[42]{Theoretical Astrophysics Group, California Institute of Technology, Pasadena, CA 91125, USA}
\affiliation[43]{Université catholique de Louvain, B-1348 Louvain-la-Neuve, Belgium}
\affiliation[44]{Laboratoire Univers et Th\'eories, CNRS, Observatoire de Paris, Universit\'e PSL, Universit\'e Paris Cit\'e, 5 place Jules Janssen, 92195 Meudon, France}
\affiliation[45]{INFN, Sezione di Roma,  I-00185 Roma, Italy}
\affiliation[46]{ European Gravitational Observatory (EGO), I-56021 Cascina, Pisa, Italy}
\affiliation[47]{Dipartimento di Fisica, Università di Trento, Via Sommarive 14, 38123 Trento, Italy}
\affiliation[48]{INFN-TIFPA, Trento Institute for Fundamental Physics and Applications, Via Sommarive 14, I-38123 Trento, Italy}
\affiliation[49]{Theoretical Physics Department, CERN, 1211 Geneva 23, Switzerland}
\affiliation[50]{Blackett Laboratory, Imperial College London, South Kensington Campus, London, SW7 2AZ, UK}
\affiliation[51]{Institut de Fisica d’Altes Energies (IFAE), The Barcelona Institute of Science and Technology, Campus UAB, 08193 Bellaterra (Barcelona) Spain}
\affiliation[52]{Dipartimento di Fisica ``E. Fermi'', Universit\'a di Pisa, I-56127 Pisa, Italy}
\affiliation[53]{Department of Astronomy and Astrophysics, The Pennsylvania State University, University Park, PA 16802, USA}
\affiliation[54]{School of Physics and Astronomy, Cardiff University, Cardiff, CF24 3AA, UK}

\emailAdd{marica.branchesi@gssi.it}
\emailAdd{michele.maggiore@unige.ch}

\date{}

\abstract{The Einstein Telescope (\acrshort{et}), the European project for a third-generation gravitational-wave detector, has a reference configuration based on a triangular shape consisting of three nested detectors with 10 km arms, where each detector has a `xylophone' configuration made of an interferometer tuned toward high frequencies, and an interferometer tuned toward low frequencies and working at cryogenic temperature. Here, we examine the scientific perspectives under possible variations of this reference design. We perform a detailed evaluation of the science case for  a single triangular geometry observatory, and we compare it with the results obtained for a network of two L-shaped detectors (either parallel or misaligned) located in Europe, considering different choices of arm-length for both the triangle and the 2L geometries. We also study how the science output changes in the  absence of the low-frequency instrument, both for the triangle and the 2L configurations. We examine a broad class of  simple `metrics' that quantify the science output, related to compact binary coalescences, multi-messenger astronomy and stochastic backgrounds, and we then examine the impact of different detector designs on a more specific set of scientific objectives.}

\begin{document}
\maketitle
\flushbottom

\section{Introduction}

Thanks to the extraordinary discoveries of the LIGO-Virgo Collaboration (\acrshort{lvc}) in the last few years \cite{LIGOScientific:2016aoc,TheLIGOScientific:2017qsa,Monitor:2017mdv,LIGOScientific:2017ync,LIGOScientific:2020ibl,LIGOScientific:2021djp,LIGOScientific:2021psn}, the field of gravitational waves (\acrshort{gw}s) is blossoming. After three observing runs the current catalog of GW detections contains about 90 binary black hole (\acrshort{bbh}) coalescences, which in the last run have been detected  on a weekly basis, as well as  two binary neutron stars (\acrshort{bns}s)  and  two neutron star--black hole (\acrshort{nsbh}) binaries \citep{LIGOScientific:2020ibl,LIGOScientific:2021djp}. These discoveries are already having a significant impact on our understanding of the population properties (see, e.g. ref.~\cite{LIGOScientific:2021psn}) and provide  first results on fundamental physics and cosmology (see, e.g. ref.~\cite{LIGOScientific:2021sio,LIGOScientific:2021aug}). The tremendous potential of combining multi-messenger observations including gravitational waves has been shown by GW170817, the first gravitational-wave observation from the merger of a binary neutron-star system which was detected in all the electromagnetic bands from gamma rays (GRB 170817A), X-ray, ultraviolet-optical-near infrared (AT2017gfo), to radio \cite{LIGOScientific:2017ync}. This event had important implications in many fields of astrophysics, from relativistic astrophysics, nuclear physics, nucleosynthesis in the Universe, and cosmology.

However, the current  infrastructures have intrinsic limitations, and for more than a decade the European community has been preparing the jump toward a `third--generation' (\acrshort{3g}) European GW detector. The Einstein Telescope (ET)~\cite{Hild:2008ng,Punturo:2010zz,Hild:2010id} is the European observatory designed to detect gravitational-wave sources along the cosmic history up to the early Universe. 
The US community is participating in the efforts toward 3G detectors, with the  Cosmic Explorer (\acrshort{ce}) project~\citep{Reitze:2019iox,Evans:2021gyd}.
Third-generation detectors such as ET and CE will provide an improvement in sensitivity by one order of magnitude and a significant enlargement of the  bandwidth,  both toward   low and high frequencies, and  will have extraordinary potential for discoveries in astrophysics, cosmology, and fundamental physics. 
Building on many previous works, the science case for ET  has been  summarized in \cite{Maggiore:2019uih, Sathyaprakash:2012jk}, while a recent comprehensive study  of the ET capabilities can be found in \cite{Iacovelli:2022bbs} and, for multi-messenger observations, in 
\cite{Ronchini:2022gwk}. For CE and for a more general discussion of 3G detectors see also \cite{Borhanian:2022czq,Kalogera:2021bya}. 

The current design of ET is based on several innovative concepts, with respect to the (second-generation) LIGO-Virgo  detectors. In particular, ET is currently planned to be a single observatory, located  200-300 meters underground, in order to significantly reduce seismic noise; to have  a triangular shape, consisting of three nested detectors, providing redundancy, the possibility of resolving the GW polarizations, and  a null stream, i.e. a combination of outputs where the GW signal cancels, that can act as a  veto against disturbances; and it will feature a `xylophone' configuration, in which each of the three detectors actually consists of two interferometers, one tuned toward high frequencies, and therefore using high laser power, and one tuned toward low-frequency, working at cryogenic temperatures and low laser power.
The first ET Conceptual Design Report dates back to 2011, and laid down the basic structure mentioned above. It has been updated in 2020, in the context of the successful proposal for including ET in the ESFRI Roadmap, the roadmap of large scientific European infrastructures.\footnote{See \url{http://www.et-gw.eu/index.php/relevant-et-documents} for a collection of documents on the ET design study and science case.} However, a detailed comparison of different alternatives in the design and on their impact on the science output has never been performed. 
The concept of ET was proposed well before the first observations of GWs. Unlike then, all the knowledge about GW sources that has accumulated in recent years makes such a study with astrophysically motivated GW source populations possible, and mandatory, today.

The aim of this work is twofold. First, we want to examine the effect of variations in the geometry, comparing a single triangle to a configuration made of two well-separated L-shaped detectors, which we will refer to as the `2L' configuration, while maintaining all other innovative concepts of the ET design (underground, cryogenic, and in a xylophone configuration). Second, we want to study what the impact on the science output would be if some aspects of the design should turn out to be difficult to achieve, at least in the first stage of operations (both for a triangle and for a 2L configuration). In particular, we will study the situation in which
the low-frequency (\acrshort{lf}) instrument is not operative, and we only have the  high-frequency (\acrshort{hf}) instrument in each detector. This will allow us to quantify the impact of the LF instrument   on the expected science output.

This study has been performed in the context of the activities of the Observational Science Board (\acrshort{osb}) of ET, which is in charge of developing the Science Cases and the analysis tools relevant for ET.\footnote{See 
\url{https://www.et-gw.eu/index.php/observational-science-board} for a public repository of papers produced within the ET Observational Science Board.} The study is structured as follows. In Section~\ref{sect:geometry} we will introduce and motivate the different detector geometries considered and we will give the corresponding strain sensitivities. 
We will then discuss the impact of different choices of geometry and design on the science that can be made with ET.
The coalescence of compact binaries, such as  binary black holes (BBHs) and  binary neutron stars (BNSs), is a primary target of  3G detectors. The crucial properties here are the detection rates, the range and distribution in redshift of the detected events, and the accuracy in the reconstruction of the source parameters. 
Properties such as detection rate, range and accuracy in parameter estimation have the advantage of being very general and requiring minimal model-dependent assumptions (which are reported in the paper). Therefore, they  already provide a first solid understanding of the relative advantages/disadvantages of  different configurations. 
In  Section~\ref{sect:CBC} we will then compare, from this point of view, the performance of the different ET configurations introduced in Section~\ref{sect:geometry}.
We will first consider ET observing as single detector (whether in its triangle or 2L configuration), discussing BBHs in Section~\ref{sect:PEBBH} and BNSs in Section~\ref{sect:PEBNS}, and we will then perform in Section~\ref{sect:3Gnetwork} the same study when ET is part of a 3G detector network including  one or two Cosmic Explorer  detectors. 
The effect of the different choices of ET as single observatory on multi-messenger observations of BNSs is studied in Section~\ref{sect:MMO}. In particular, Section~\ref{sect:MMOskylocpremerger} discusses the sky-localization and pre-merger alerts, while Sections~\ref{sect:MMOGRB} and \ref{sect:MMOkilonova} evaluate, respectively, the prospects of detecting short $\gamma$-ray bursts (\acrshort{grb}) and kilonovae associated with BNS mergers. 

Stochastic backgrounds of GWs are another primary target of ET. The effect of different configurations on stochastic searches is studied in Section~\ref{sect:stochastic}. In particular, in Section~\ref{BackgroundSensitivities} we  discuss the sensitivity to stochastic backgrounds (as expressed by the power-law integrated sensitivity, \acrshort{pls}) of the different configurations, while the angular resolution obtained from a multipole decomposition is discussed in  Section~\ref{sect:angularresstochastic}. A comparison of the different sensitivities with predictions for astrophysical backgrounds is performed in  Section~\ref{sect:astroback}, while the impact of correlated noise is discussed in Section~\ref{sect:corrnoise}.

Beside the study of these rather general metrics, it is also important to investigate how these differences in detection rates, parameter estimation, or sensitivity to stochastic backgrounds, reflect on  specific scientific targets of ET around which the ET Science Case is structured~\cite{Maggiore:2019uih}, even if this, unavoidably, may introduce some extra model-dependent assumptions. In Section~\ref{sect:ImpactSpecific} we will then discuss the impact of the different configurations on a broad set of specific scientific targets of ET in the domains of astrophysics, fundamental physics and cosmology. Finally, in Section.~\ref{sect:nullstream} we will discuss the properties of the null stream for the triangular configuration and its potential benefits.
A summary of the main results of this study is presented  in Section~\ref{sect:summary} and, in particular, our conclusions are discussed in Section~\ref{sect:Conclusions}. More technical material is collected in four appendices; in particular, several Tables with useful figures of merits for BBHs and BNSs are collected in App.~\ref{app:TablesCBC}. We also provide, at the end of the paper, a list of acronyms used in the text.

\section{Detector geometries and sensitivity curves}\label{sect:geometry}

The ET initial design included high laser  power  at the interferometer input, leading to high optical power in the cavity of the order of  several MW, optimisation of signal recycling, frequency-dependent light squeezing, increase of the beam size, and heavier mirrors compared to 2G detectors. These specifications led to the so-called ET-B sensitivity curve~\cite{Hild:2008ng}. This design, however,  neglected the fact that high circulating power is difficult to reconcile with cryogenic test masses. This has  led to a `xylophone' concept, in which a detector  is actually composed of two different interferometers, one optimized for low frequencies  (LF) and one for high frequencies (HF); the LF instrument has low power (since laser power is only needed to beat down the shot noise in the high-frequency range) and cryogenic mirrors, while the HF instrument has high power and mirrors at room temperature. This has led to the ET-C sensitivity curve and, after  further refinement of some noise models, to the ET-D sensitivity curve \cite{Hild:2010id}.  The ET-B and ET-C sensitivities must be considered as obsolete, and should not be used. The ET-D sensitivity has been the basis for all recent ET studies.
Actually, we will use a further refinement  of the ET-D sensitivity curve, recently developed in the context of the ET Instrument Science Board (\acrshort{isb}), see below. 

\begin{figure}[t]
    \centering
     \includegraphics[width=0.45\textwidth]{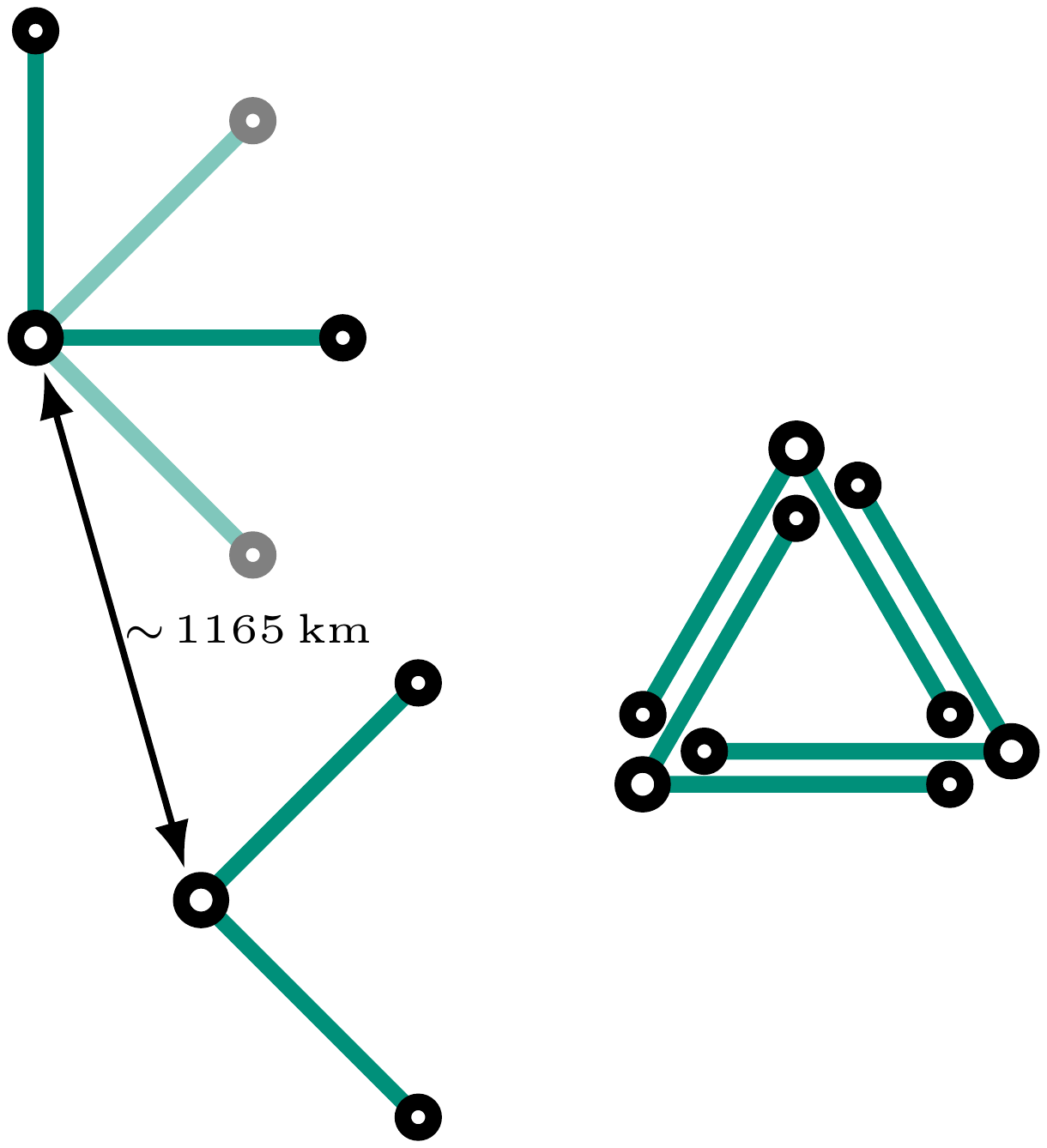}
    \caption{\small A schematic picture of the different geometries considered:  two widely separated L-shaped detectors (either parallel or at $45^{\circ}$), or a triangle made of three nested detectors.}
    \label{fig:CoBA_scheme1}
\end{figure}

As far as the geometry is concerned, we will study triangular configurations, \rosso{ in which three detectors (each one made of a LF and a HF interferometer)} are nested into a triangular shape, as well as a network of two well-separated L-shaped detectors (that we will denote as the `2L' configuration), and will we examine different options for the arm length and, for the 2L configuration, different relative orientations between their arms.  \rosso{ A schematic picture of the different geometries considered is shown in Fig.~\ref{fig:CoBA_scheme1}.}
For a single triangular configuration, currently
the candidate sites  are  the Sos Enattos site in Sardinia, and the Meuse-Rhine  three-border region across  Belgium, Germany and the Netherlands.  When considering a single triangle, we will locate it for definiteness in Sardinia, but there will be no significant difference in the results considering ET in the Meuse-Rhine region. When considering the 2L configuration, we will locate one detector in Sardinia and one in the Meuse-Rhine region.
However, at the current level of precision,  the analysis will be valid, with minor  changes, for any pair of interferometers located at a comparable distance.

For two L-shaped detectors, in the  coplanar limit the setting with parallel arms maximizes the  sensitivity to stochastic backgrounds and the range to compact binaries, while it is the least favorable in terms of the accuracy on angular localization and  reconstruction of the distance of compact binary coalescences (or other individual sources). Conversely, the setting with arms at $45^{\circ}$ is basically blind to stochastic backgrounds, while it maximizes the performances in terms of angular localization and distance of compact binaries, see Section~\ref{BackgroundSensitivities}. When taking into account the Earth's curvature, the relative orientation between the two detectors is usually defined with reference to the great circle that connects the two detectors~\cite{Flanagan:1993ix,Christensen:1996da}. Denoting by $\beta$ the angle describing the relative orientation of the two detectors, defined with reference to this great circle, $\beta=0^{\circ}$ corresponds to the case where the arms of the two interferometers make the same angle with respect to the great circle, while $\beta=45^{\circ}$ is when one of the two interferometers is rotated by $45^{\circ}$ from the $\beta=0^{\circ}$ orientation.
For $\beta=45^{\circ}$, the sensitivity to stochastic backgrounds is minimized, to the extent that it becomes exactly zero in the limit $f\Delta x\ra 0$, where $f$ is the GW frequency and $\Delta x$ the distance between the two detectors. 
In this work, we aim at studying the performance of the 2L configurations close to these limiting cases; however, setting exactly $\beta=45^{\circ}$ is not a convenient choice because it would send practically to zero the sensitivity to stochastic backgrounds; instead, as we will see, even a small deviation from  $\beta=45^{\circ}$ allows us to reach an interesting sensitivity to stochastic backgrounds, without essentially affecting the performances for angular localization and parameter estimation of compact binary coalescences.

We decide to focus on two examples for the relative orientation of the interferometers arms in the 2L configurations, representatives of choices favoring, respectively, either the angular localization and distance determination of compact binaries, or the detection of stochastic backgrounds, but avoiding the limit in which the sensitivity to compact binaries is optimized at the expenses of sending to zero the sensitivity to stochastic backgrounds. We thus consider a simpler definition of the angles that uses  the local North at the two detector sites as reference for the orientations. Denoting by $\alpha$ the relative angle defined in this way, and by $\beta$ the relative angle defined with reference to the great circle connecting the detectors in Sardinia and in the Netherlands, we have $\alpha\simeq \beta+2.51^{\circ}$.\footnote{For the Sardinia site we take (latitude $40^{\circ}$~31', longitude $9^{\circ}$~25'), while for the 
Meuse-Rhine site we take (latitude $50^{\circ}$~43'~23'', longitude $5^{\circ}$~55'~14'').}  
We will study the cases 
$\alpha=0^{\circ}$ and $\alpha=45^{\circ}$, that corresponds to introducing a small offset with respect  
to the `perfectly' parallel ($\beta=0^{\circ}$) and maximally misaligned ($\beta=45^{\circ}$) configurations.
The results for compact binary coalescence (\acrshort{cbc}) parameter estimation depends minimally on the precise value of this offset, as long as this misalignment angle is small. We have indeed checked that the results shown in Section~\ref{sect:CBC} are unaffected, for all practical purposes,  by this small offset; e.g. the differences between the cases $\alpha= 45^{\circ}$ and the truly optimal orientation
$\alpha= 47.51^{\circ}$ would be invisible on the scale of the plots shown in Section~\ref{sect:CBC}, and are below the  variability due to the specific sample realization of the population.
In contrast, the sensitivity to stochastic backgrounds strongly depends on it, as we will discuss in App.~\ref{app:StocasticMisalignement}. 
In the following, we will refer to these choices as parallel (by which we mean $\alpha=0^{\circ}$) and misaligned 
($\alpha=45^{\circ}$).\footnote{If a 2L option should be retained, it will be important to perform a more detailed study of the trade-offs between accuracy in the localization of compact binaries and sensitivity to stochastic backgrounds, as a function of the misalignment angle; such a study, however, cannot be performed abstractly but, especially when discussing misalignment at the $(1-2)^{\circ}$ level,  will have to integrate constraints from the geography and the geology of the candidate sites. For instance, underground tunnels have to follow optimized paths avoiding as much as possible fractures or water springs.\label{foot:angles}} 

In particular, we will consider (with the definition of parallel and at $45^{\circ}$ specified above, i.e. with respect to the local Norths at the two detector sites): 

\begin{enumerate}

\item a triangle with 10~km arms (the current baseline ET geometry)
\item a triangle with 15~km arms 
\item a 2L with 15~km arms, with parallel arms
\item a 2L with 15~km arms, with a relative orientation of $45^{\circ}$
\item a 2L with 20~km arms, with parallel arms
\item a 2L with 20~km arms, with a relative orientation of $45^{\circ}$

\end{enumerate}

We will also occasionally compare with the results that can be obtained with the most advanced network of 2G detectors (LIGO Hanford, LIGO Livingston, Virgo, KAGRA and LIGO India), in order to better appreciate the improvement due to 3G detectors. We use the publicly available best sensitivities that are planned to be achieved by LIGO, Virgo and KAGRA by the end of the O5 run \cite{KAGRA:2020rdx}. We will denote this network \acrshort{lvki}~O5. LIGO India is expected to join the 2G network late in this decade. Both the Virgo and LIGO collaborations are currently discussing plans for further upgrades of the detectors to be implemented after the O5 run  by developing and testing key technologies critical for the 3G detectors \citep{Bailes:2021tot}.
We will also  show some results for a single L-shaped detector of 20~km (denoted as `1L'), that we will locate for definiteness in the Meuse-Rhine region.

It is clear that the configurations considered have different financial costs, so some configurations might be obviously better than others. The aim of this work is to provide scientific input to a broader cost-benefit analysis.\footnote{An  analysis of the costs of the different configurations is well beyond the scope of this study. We note, however, that in the comparison between  geometries, the total linear arm length is not the only relevant parameter for estimating the relative costs. For instance, another significant parameter is the diameter of the tunnels. In an L-shaped configuration there will be only one detector per tunnel, compared to two in the triangle configuration with nested interferometers. As a result,  in an L-shaped detector  the diameter  of the tunnels will be smaller,   approximately 6.5~m,  to be compared with 8~m for the triangle. 
Already just for  what concerns the relative excavation costs, several other factors must be included for a realistic  comparison,  
such as the fact that,   in the triangle configuration, extra tunnel length is needed to displace the end towers of a detector with respect to the input towers of the other, as well as  differences in  the number and size of auxiliary caverns, and in the tunnels serving them. All these aspects can only be evaluated with a detailed dedicated study. \label{foot:tunnels}}

\begin{figure}[t]
    \centering
     \includegraphics[width=0.45\textwidth]{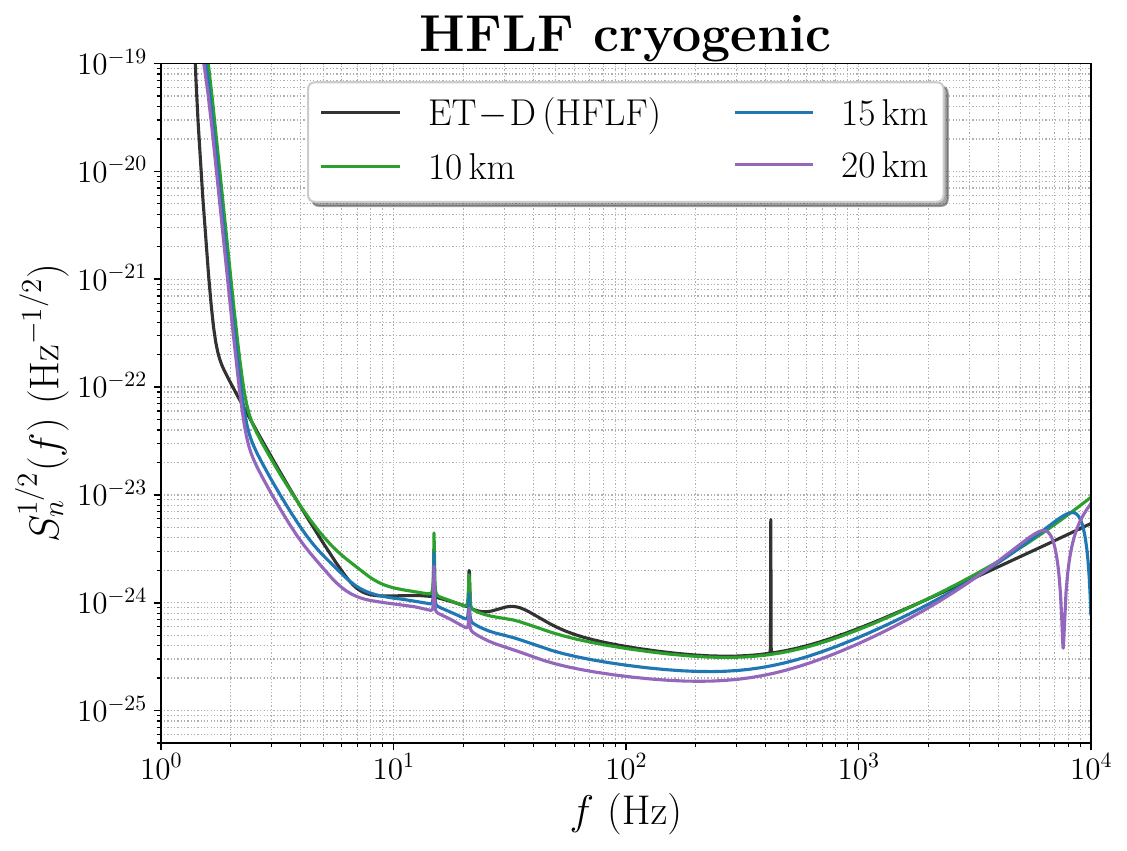}
    \hfill
   \includegraphics[width=0.45\textwidth]{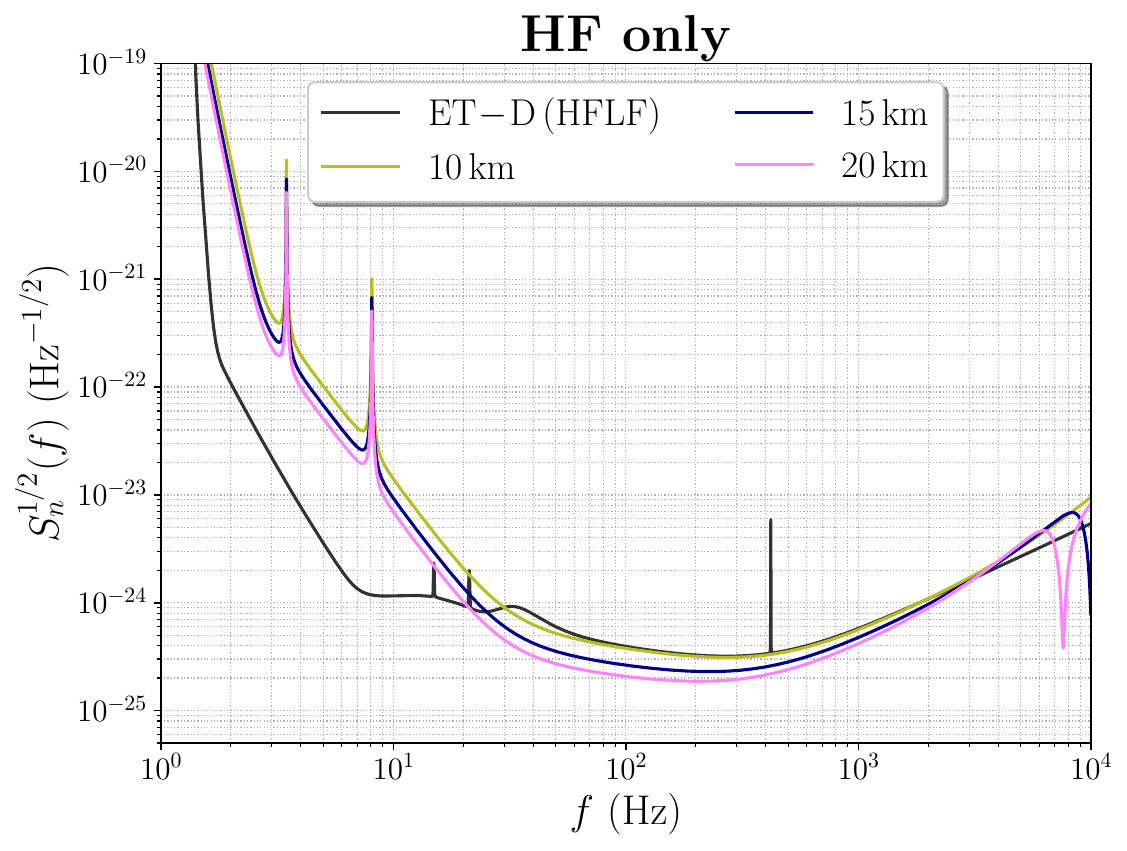}
    \caption{\small Amplitude spectral densities (ASDs, single-sided), used in the present work for different scenarios. In both panels we show the ASD for an interferometer with 10, 15 and 20~km arms. The left panel shows  the full sensitivity curves (obtained from the HF instrument and  the cryogenic LF instrument,  in a xylophone configuration); the right panel is the sensitivity obtained  with the HF instrument only. For comparison, in both panels we also show the  ASD of the 10~km ET-D interferometer (which, by definition, includes the full HFLF sensitivity). 
    Note that the ASD are plotted as a single nested detector (each one made of a LF and a HF interferometer left plot and each one made of HF only right plot). See the text on how to take into consideration the angle between the arms and the number of interferometers for the triangular shape.}
    \label{fig:ASDs_allConf}
\end{figure}

In Fig.~\ref{fig:ASDs_allConf} we show the different noise Amplitude Spectral Densities (\acrshort{asd}s) used in this work to characterise the various detector configurations considered. In particular, in the left panel we show, for various arm lengths, the noise curve attainable in the full design, in which there is a xylophone configuration with an instrument tuned toward high-frequencies (HF) and one tuned toward low frequencies (LF), and the LF instrument works at cryogenic temperatures.  In the  right panel we show the ASDs resulting from exploiting only the HF instruments. These curves  have been computed in the context of  the activities of the ET Instrument Science Board (ISB).\footnote{These curves are available at \url{ https://apps.et-gw.eu/tds/ql/?c=16492}, see in particular the Annex files. They obtained with the help of PyGWINC package 
\cite{2020ascl.soft07020R}.} In particular, the line corresponding to the 10~km length in the panel labelled ``HFLF cryogenic" updates the sensitivity curve known as ET-D, available at \url{https://apps.et-gw.eu/tds/?content=3&r=14065}, and shown as a reference as the black line in both panels of Fig.~\ref{fig:ASDs_allConf},  and reflects  more detailed data on the technology used in the ET-D design. It should be stressed that, in this phase where the design of ET is being more deeply analyzed, the sensitivity curves necessarily evolve, and will continue to evolve. The two curves shown, for the full sensitivity and for the HF instrument only, must be taken as representative curves, within other  possibilities currently under study. Further intermediate sensitivities could also be studied (for instance, corresponding to a possible  intermediate stage where the LF instrument is present, but operates at room temperature), and would give results in between those obtained with the HF-only and the HFLF-cryo sensitivities.

When comparing a triangle to an L-shaped interferometer with the same instrument-noise ASD, one must take into account that the triangle is made of three nested detectors (a detector being an LF and HF interferometer pair), with an opening angle of $60^{\circ}$. 
For the triangle configuration,  one must then  project the GW tensor of the incoming wave onto each of these three  components [see e.g. eqs.~(9)--(11) of \cite{Jaranowski:1998qm} for explicit expressions], and then
combine the results at the level of the signal-to-noise ratio (\acrshort{snr}) and parameter estimation to obtain the ET capabilities.
The accurate evaluations of detection horizons and parameter-estimation capabilities require the above calculation of GW-tensor projections onto the three ET components, as is done for all the results of the present paper, and in all Fisher matrix codes mentioned below. However, to have a simple and approximate rule-of-thumb for comparing  the SNR of a triangle to that obtained, for the same signal, by an L-shaped detector with the same ASD, one must first of all take into account that, for each given component, 
the opening angle  gives an extra  factor  $\sin(60^{\circ})=\sqrt{3}/2$, see again eqs.~(9)--(10) of \cite{Jaranowski:1998qm}. In a first approximation, one can also assume  that the three components of the triangle see the signal with the same SNR (which, however, is only approximately true). Then, since the 
SNR of the three components of the triangle add up in quadrature,  the ${\rm SNR}^2$ in a triangle is approximately larger than the ${\rm SNR}^2$ in a L-shaped detector with the same ASD, by a factor
$3\times (\sqrt{3}/2)^2=9/4$. For this reason, when comparing graphically the ASD of a triangle to that of an L-shaped detector, a more fair comparison is obtained dividing the ASD of the triangle by a factor $3/2$. Since some  source of noise, such as Newtonian noise, are such that the SNR scales as the length of the arms, in a first approximation one could further be tempted to estimate that,  for a triangle,
${\rm SNR}\propto L\, \sqrt{3}\sin(60^{\circ})=1.5 L$, while for a 2L network with the same arm-length ${\rm SNR}\propto L\, \sqrt{2}\sin(90^{\circ})\simeq 1.41 L$.\footnote{In fact,  not all noise relevant for 3G detectors are such that the SNR scales linearly with $L$. Examples of noise whose SNR do not scale with $L$ include coating thermal noise (since it also depends on beam size, which increases with $L$), shot noise (since one needs to adjust the arm-cavity finesse depending on  $L$), controls noise (since $L$ can change stability conditions of the opto-mechanical system, which requires adjustment of control filters) and suspension thermal noise (since one needs larger mirrors in longer detectors, which means higher suspended mass), see \cite{Evans:2016mbw} for discussions. Therefore, the correct comparison is between detectors at fixed ASD, as discussed above, rather than at fixed $L$, and the scaling of ${\rm SNR}\propto L$ is a further approximation, to be used only for a first orientation.}  
However, one must be careful not to infer from this that a 2L and a triangle configurations with similar arm lengths necessarily perform comparably. The scientific performance is determined basically by the quality of parameter estimation, which does not depend just on the SNR. As an obvious counter-example, for an ensemble of  compact binary coalescence signals,   and fixed ASD, 2L parallel have a higher SNR than 2L at $45^{\circ}$; however,  as far as parameter estimation is concerned, 2L at $45^{\circ}$ perform significantly better (as we will see  explicitly for ET in Section~\ref{sect:CBC}), exactly the contrary of what would be inferred using an SNR-only argument. Similarly, for angular localization, the baseline between the detectors is another crucial factor. A comparison between the performances of different geometries can only be obtained with a parameter-estimation study, as we will perform below, and cannot be obtained from considerations based only on the SNR.

\begin{figure}[t]
    \centering
     \includegraphics[width=0.45\textwidth]{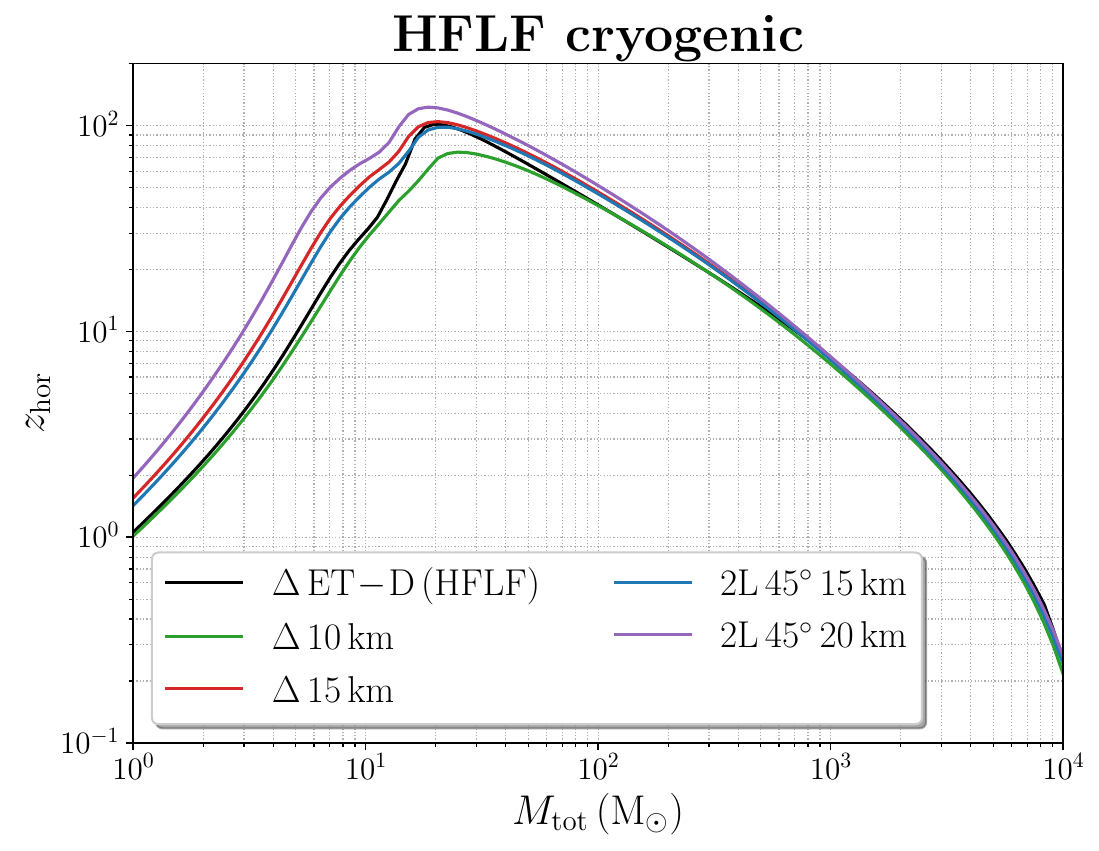}
     \hfill
    \includegraphics[width=0.45\textwidth]{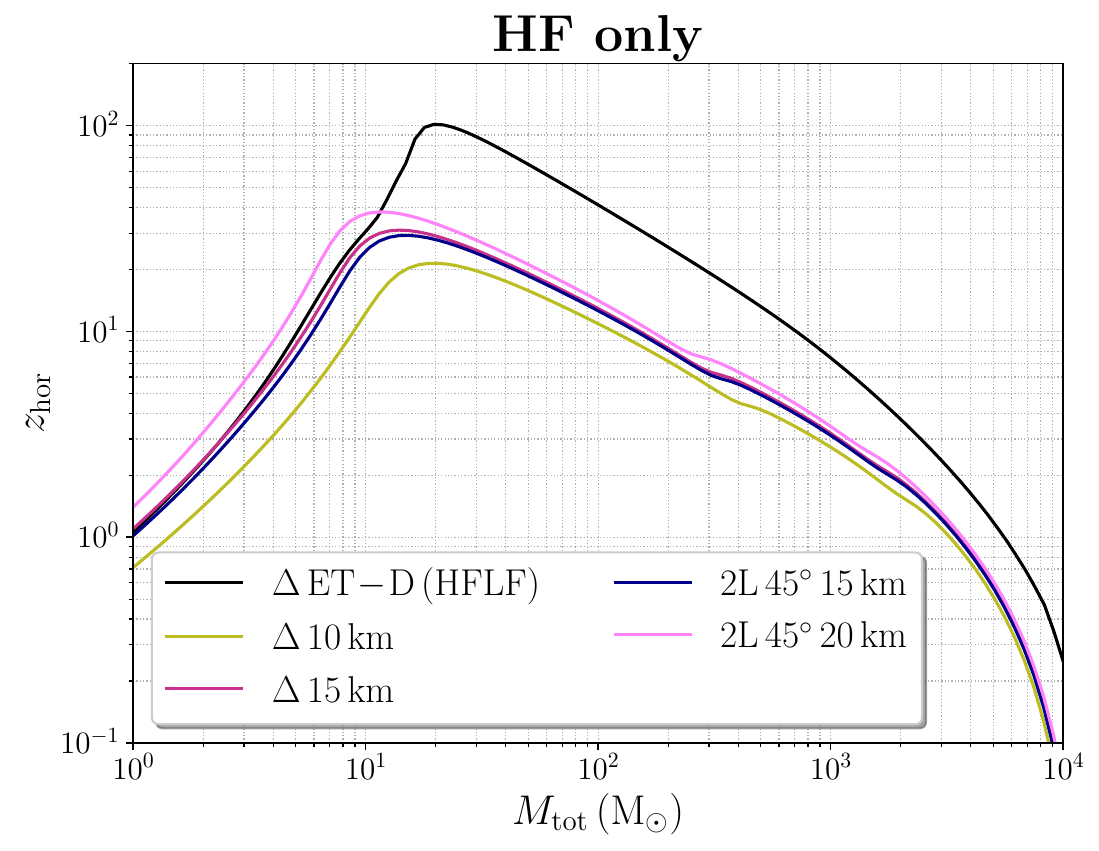}
     \caption{\small Detection horizons for monochromatic populations of equal-mass non-spinning binaries, for the various detector configurations considered. For comparison, we also show in each panel the curve obtained using a triangular detector with the 10~km ET-D ASD (which, by definition, includes the full HFLF sensitivity).}
    \label{fig:Detector_Horizons_BBH_AllConf}
\end{figure}
 
\begin{figure}[t]
    \centering
     \includegraphics[width=0.45\textwidth]
     {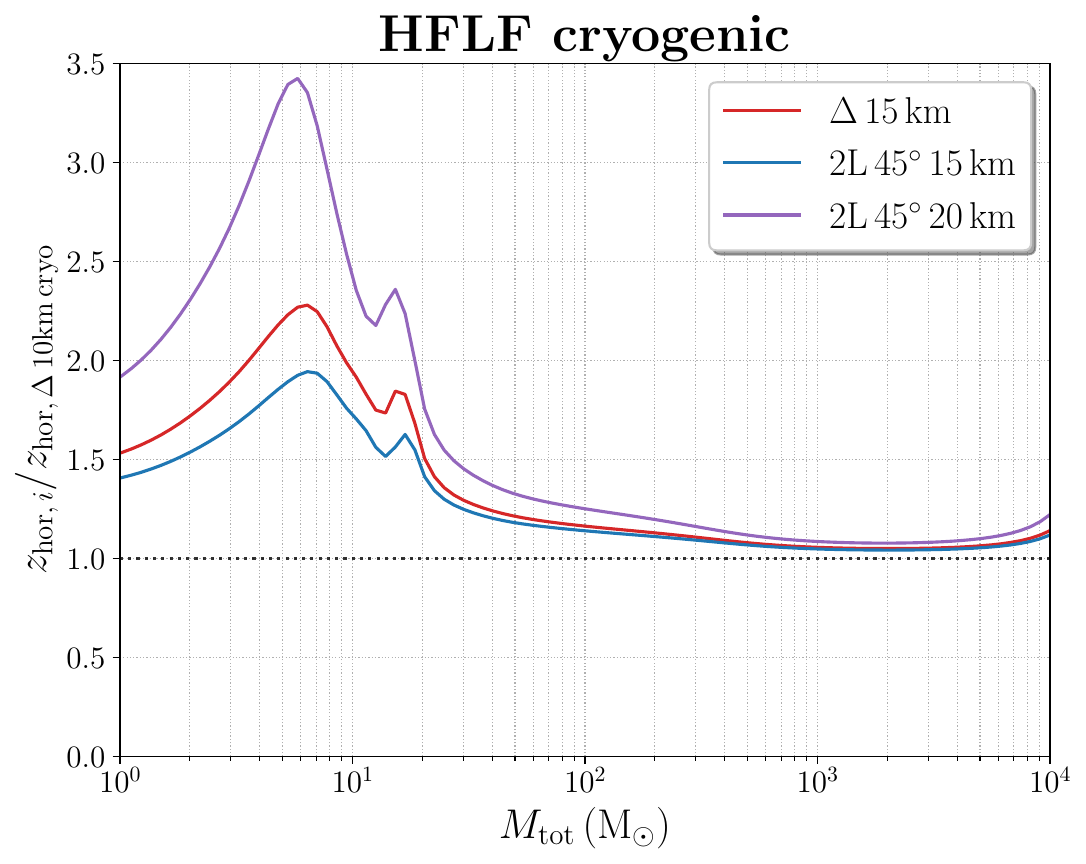}
     \hfill
    \includegraphics[width=0.45\textwidth]
    {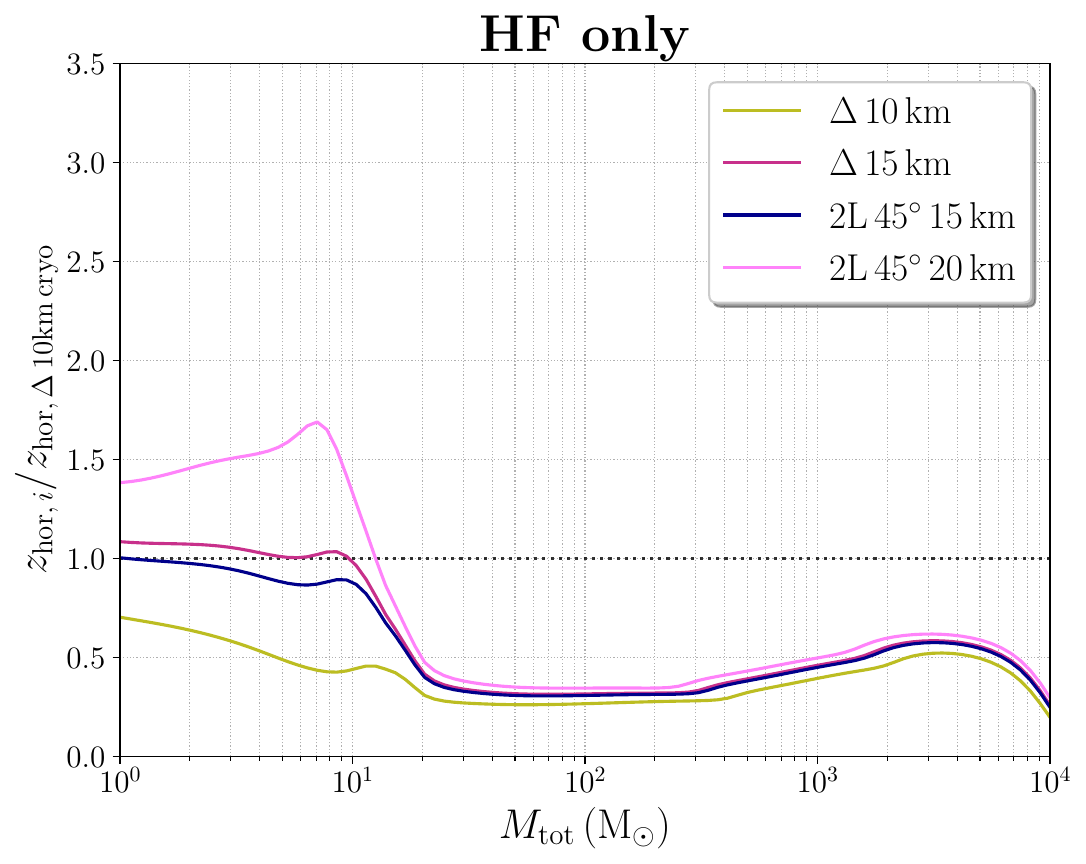}
     \caption{\small Ratios of the detector horizons for the different geometries. Left panel: full  HFLF-cryo sensitivities, normalized to the horizon for the HFLF-cryo sensitivity of the 10~km triangle (i.e. to the green line in the left panel of Fig.~\ref{fig:Detector_Horizons_BBH_AllConf}). Right: HF sensitivities (still normalized to the full HFLF-cryo sensitivity of the 10~km triangle).}
    \label{fig:Detector_Horizons_Difference}
\end{figure}

In Fig.~\ref{fig:Detector_Horizons_BBH_AllConf} we  show the corresponding detection horizons for equal-mass non-spinning binaries, as a function of their source frame total mass, for the various detector configurations considered. These are defined as the maximum redshift out to which a binary with the chosen characteristics, with optimal location and inclination, can be observed by a given detector configuration with a network ${\rm SNR} \geq 8 $. This computation is performed using the IMRPhenomHM waveform model~\cite{London:2017bcn}, and accounting for the effect of Earth rotation, which can give a contribution in the range of masses typical of BNS systems.
For comparison, we also show the results obtained for a triangular-shaped detector  using the `older' ET-D ASD, which reproduces the result  in Fig. 1 of \cite{Hall:2019xmm}.

In Fig.~\ref{fig:Detector_Horizons_Difference}, left panel, we show the ratios of the redshift horizons for three different  geometries (triangle 15~km, 2L-15km-$45^{\circ}$ and  2L-20km-$45^{\circ}$), all in their HFLF-cryo configurations, with respect to  the 10~km triangle, also taken in its HFLF-cryo configuration. In the right panel we show the same ratios for
triangle 15~km, triangle 10~km, 2L-15km-$45^{\circ}$ and  2L-20km-$45^{\circ}$, now all in their HF-only configuration, again with respect  to the horizon of the 10~km HFLF-cryo triangle.  We can appreciate that, especially for total masses below $\sim 10\msun$, arm-length has a very important effect on the detection horizon: for these masses, the 15~km triangle has a range that, in terms of redshift, is larger   by a factor $1.5-2.2$ than that of the 10~km triangle,  and the  2L-15km-$45^{\circ}$ is larger by a factor $1.4-1.8$; as we see from the  right panel, in this range of masses this increase in arm-length would basically compensate the loss of the LF instrument (especially for the 15~km triangle). Once again, however, it is important to recall that the detection horizon is determined just by the SNR. More important, for the science case, is the quality of parameter estimation. For instance, the 2L-15km-$45^{\circ}$, compared to the 2L-15km-$0^{\circ}$, has a smaller horizon reach, but better capacity to determine the parameters of the sources, as we will quantify in the next section. 

\section{Coalescence of compact binaries}\label{sect:CBC}

In this section we study the impact of the different detector geometries and different ASDs on the detection and parameter estimation of BBHs and BNSs. Several parameter estimation codes, tuned toward 3G detectors, have been developed recently, in particular \texttt{GWBENCH}~\cite{Borhanian:2020ypi,Borhanian:2022czq},   \texttt{GWFISH} \cite{Dupletsa:2022wke},  \texttt{GWFAST}~\cite{Iacovelli:2022bbs,Iacovelli:2022mbg}, \texttt{TiDoFM}
\cite{Chan:2018csa,Li:2021mbo}
and the code used in  \cite{Pieroni:2022bbh}.
In the context of the activities of the Observational Science Board (OSB) of ET,  we have performed extended   cross--checks between these codes,
finding very good  consistency. The results presented in this section have been obtained with \texttt{GWFAST} which, thanks to its vectorization properties, is quite convenient for dealing with the large number of runs required,  
while those presented in Section~\ref{sect:MMO} have been obtained with \texttt{GWFISH}, which is also tuned toward multi-band and multi-messenger observations~\cite{Dupletsa:2022wke,Ronchini:2022gwk}. These codes are based on the Fisher matrix formalism. The Fisher matrix formalism has well-known limitations (see e.g.  \cite{Vallisneri:2007ev,Rodriguez:2013mla} for extensive discussion) but is currently the only computationally  practical way
of dealing with parameter estimation for large populations. We refer to the original papers ~\cite{Borhanian:2020ypi,Borhanian:2022czq,Dupletsa:2022wke,Iacovelli:2022bbs,Iacovelli:2022mbg} for technical details on the  application of the  Fisher matrix formalism  to parameter estimation for 3G detectors. It should also be observed that 3G detectors will observe a large number of events with very large SNR. As we will see for instance in Fig.~\ref{fig:AllGeoms_CumulBBH_NdetScale} below, ET, in any of its configurations, will detect ${\cal O}(10^3-10^4)$ BBH/yr with ${\rm SNR} > 100$. These events will largely dominate the  performance of a detector configuration for most aspects of the science case, and for large ${\rm SNR}$ the Fisher matrix approach becomes more and more reliable.

We use state-of-the-art waveforms and population models. For the waveforms, we use IMRPhenomXPHM for BBHs, which includes precessing spins and higher-order modes~\cite{Pratten:2020ceb},   and IMRPhenomD\_NRTidalv2~\cite{Khan:2015jqa,Dietrich:2019kaq} for BNSs, which includes tidal effects. 
The parameters of the waveform are $ \{{\cal M}_c, \eta, d_L, \theta, \phi, \iota, \psi, t_c, \Phi_c, \chi_{1,x}, \chi_{2,x}, \chi_{1,y}, \chi_{2,y}, \chi_{1,z}, \chi_{2,z}, \Lambda_1, \Lambda_2\}$, where ${\cal M}_c$ denotes the detector--frame chirp mass, $\eta$ the symmetric mass ratio, $d_L$ the luminosity distance to the source, $\theta$ and $\phi$ are the sky position coordinates (defined as $\theta=\pi/2-\delta$ and $\phi=\alpha$, with $\alpha$ and $\delta$ right ascension and declination, respectively), $\iota$ the  angle between the orbital angular momentum of the binary and the line of sight, $\psi$ the polarisation angle, $t_c$ the time of coalescence, $\Phi_c$ the phase at coalescence, $\chi_{i,c}$ the dimensionless spin of the object $i=\{1,2\}$ along the axis $c = \{x,y,z\}$ of the coordinate system, chosen with the $z$ axis pointing along the orbital angular momentum (we use the same conventions as in Section~IID of~\cite{Buonanno:2002fy}), and $\Lambda_i$ the dimensionless tidal deformability of the object $i$ (which is present only for systems containing a NS). 
Instead of $\Lambda_1, \Lambda_2$, we will actually use the two combinations $\tilde{\Lambda}$ and $\delta\tilde{\Lambda}$ defined in \cite{Wade:2014vqa},
which have the advantage that $\tilde{\Lambda}$ is the combination that enters the inspiral waveform at 5\,\acrshort{pn}, while $\delta\tilde{\Lambda}$ first enters at 6\,PN. In particular,
\be\label{deftildeLambda}
\tilde{\Lambda} = \dfrac{8}{13} \left[(1+7\eta-31\eta^2)(\Lambda_1 + \Lambda_2) + \sqrt{1-4\eta}\, (1+9\eta-11\eta^2)(\Lambda_1 - \Lambda_2)\right]\, .
\ee
For BBHs we will perform the inference on all parameters (except, of course, $\Lambda_1, \Lambda_2$, that vanish for BHs) and, as in LVK parameter estimations, rather than $\iota$ and $\chi_{i,c}$, we will use $\theta_{JN}$ (i.e. the angle between the \emph{total} angular momentum and the line of sight; note that this is the same as $\iota$ only in the absence of precession), the spin magnitudes and angles, namely $\chi_1,\ \chi_2,\ \theta_{s,1},\ \theta_{s,2},\ \phi_{JL},\ \phi_{1,2}$.\footnote{This choice  has the advantage of using quantities which are less degenerate among each other, leading to better conditioning of the Fisher matrices, thus improving the reliability of this approach.}
For BNSs, instead, we include tidal deformability but, given the small expected values of their spin magnitudes, we only consider the aligned spin components in the analysis, thus performing estimation on $\chi_{1,z}$ and $\chi_{2,z}$.  The labels `1' and `2' always refer, respectively, to the heaviest and lightest component of the binary system. 
After the inversion of the Fisher matrix, we compute the sky localisation area for the events at $90\%$ c.l., $\Delta\Omega_{90\%}$. 
As in \cite{Ronchini:2022gwk,Iacovelli:2022bbs}, we assume an uncorrelated 85\% duty cycle in each L-shaped detector, and in each of the three instruments composing the triangle.
All technical details of the inference process for the results of this section are as described in \cite{Iacovelli:2022bbs}.

We generated the catalogs of binary neutron stars (BNSs) \label{page:catalog}
with the population synthesis code {\sc mobse} \cite{Mapelli:2017hqk,Giacobbo:2017qhh}. {\sc mobse} evolves isolated binary stars until they possibly become binary compact objects. For the models used here, we have evolved $1.8\times{}10^8$ binary systems with 12 different metallicities, ranging from $Z=10^{-4}$ to $0.02$. Primary masses are drawn from a Kroupa initial mass function \cite{Kroupa:2001jy} between 5 and 150 M$_\odot$, while mass ratios, initial orbital periods and eccentricities follow the distributions presented in \cite{Sana:2012px}. {\sc mobse} adopts up-to-date models for stellar winds \cite{Giacobbo:2017qhh}. We model electron-capture, and (pulsational) pair-instability supernovae according to \cite{Giacobbo:2018hze} and \cite{Mapelli:2019ipt}, respectively. For core-collapse supernovae, we use the rapid model by \cite{Fryer:2011cx}, which enforces a mass gap between neutron stars and black holes. The main difference with  respect to \cite{Fryer:2011cx} is that we model neutron star masses according to an uniform distribution between $m_{\rm min}=1.1$ M$_\odot$ and $m_{\rm max}=2.5$ M$_\odot$, which is a better match to current LIGO--Virgo results \cite{LIGOScientific:2021psn}. We draw natal kicks from a Maxwellian distribution with one-dimensional root-mean-square $\sigma=265$ km s$^{-1}$, rescaled by the mass of the compact remnant and by the mass of the ejecta, see \cite{Giacobbo:2019fmo}
for more details. During binary evolution we treat mass transfer, tidal evolution, gravitational-wave decay, and common envelope as described in  \cite{Hurley:2002rf}. For the common envelope, we assume an efficiency parameter $\alpha=3$.
We describe the evolution with redshift by accounting for the metallicity-dependent star formation rate across cosmic time, as described in \cite{Santoliquido:2020axb}. The value for the local merger rate for this model  turns out to be $R_0\simeq 250\, {\rm Gpc}^{-3}\, {\rm yr}^{-1}$, which  is consistent with  the range  
$[10, 1700]\,  \rm Gpc^{-3} yr^{-1}$ obtained from the
LVK results~\cite{LIGOScientific:2021psn}.

We generated the catalogs of binary black holes (BBHs) by mixing the isolated evolution channel (obtained as described above for BNSs) 
with the dynamical formation channel. We used the code {\sc fastcluster} \cite{Mapelli:2021syv,Mapelli:2021gyv}, which evolves binary black holes in young, globular and nuclear star clusters, by accounting for their dynamical pair-up and for hierarchical mergers. In our catalogs, 55\% of all BBHs come from isolated binary evolution, while  28, 15 and 2\% come from dynamical formation in young, globular, and nuclear star clusters. We chose these mixing fractions because they ensure the best agreement with LIGO--Virgo black hole masses in the local Universe, see \cite{Mapelli:2021gyv} for more details on the methodology. 

The BBH and BNS catalogs used for this study are publicly available at  \url{https://apps.et-gw.eu/tds/?content=3&r=18321}.

\subsection{Binary Black Holes}\label{sect:PEBBH}

In this section we present the results obtained for BBHs. The data correspond to one year of data taking. With the assumed BBH population, the total number of BBHs coalescing in one year is about $1.2\times 10^5$.
We first study the effect of varying the geometry, for detectors all with their full HFLF-cryo sensitivity, and we then discuss the effect of varying the ASD.

\subsubsection{Comparison between geometries}
In the upper-left panel of Fig.~\ref{fig:AllGeoms_CumulBBH_NdetScale} 
we show the cumulative distribution of the events with a network SNR larger than a given value, while in all other panels, involving parameter estimation, we restrict to events that pass a detection threshold ${\rm SNR}\geq 12$,  and we show the corresponding cumulative distribution of events with an error smaller than a given value.\footnote{\label{footnote:Fisher_inversion_issues}More precisely, when performing parameter estimation, we restrict to events that, besides passing the threshold on the SNR, also have a well-defined inversion of the Fisher matrix. In some cases, the Fisher matrix is ill-conditioned, so its inversion can lead to large errors and we then discard the corresponding event, see  \cite{Iacovelli:2022bbs} for details. For BBHs, the fraction of ill-conditioned Fisher matrices is of order $1\%$, so has very little impact at this level, and is anyhow about the same percentage for different configurations, so this has essentially no impact on the comparison between configurations. For BNS, between $4\%$ and $10\%$ of the events are discarded, with the higher percentages appearing in the analyses using the ASD including only the HF instrument. There is also a slight dependence on the geometry especially in the HF-only case, with the 2L configurations resulting in worse conditioned matrices as compared to the triangular ones. However, the events discarded in these cases correspond preferentially to the situation when there is only one detector operating due to the duty cycle, in which case the Fisher matrix provides a poorer approximation. 
The ill-conditioned Fisher matrices often correspond at the population level to the events that would have anyhow large errors in parameter reconstruction (while the science output of the detectors depends mostly on the events with good reconstruction), but could also be related to events having a strong degeneracy between some parameters. In particular for BNS systems, where a higher percentage of events has ill-conditioned matrices as compared to BBH, the solution of discarding events, despite being conservative, might then leave a bias in the observed population, especially for specific science cases (as an example, events close to face-on/off are preferentially discarded due to the strong degeneracy present in this case between the inclination angle and the luminosity distance, but they are relevant e.g. for multi-messenger studies with GRBs).
Another solution, adopted for the results presented in Section~\ref{sect:MMO}, is to regularize the close-to-singular matrices according to the procedure described in \cite{Dupletsa:2022wke}. This allows us to analyze the non-degenerate parameters of a larger number of the events, at the price of introducing a dependence on the adopted regularisation threshold. For a recent discussion and comparison of these intrinsic limitations of the Fisher approach related to the inversion procedure see e.g. Appendices B and C of \cite{Iacovelli:2022bbs}. Each method has its own benefits, and it is reassuring that both have been used in the present analysis producing consistent results.} In each panel of this figure we show the results for the six geometries considered, all taken with their best ASD, i.e. in the xylophone configuration with the HF instrument and  the cryogenic  LF instrument (we  use the label `HFLF-cryo' for this ASD). 

We see from the figure that even if, by itself, the performance of the 10~km triangle is extraordinary, the other geometries considered in general provide a further improvement (except for angular localization, where the 2L configurations with parallel arms have the worst performance). This is particularly evident in the reconstruction of the luminosity distance. In our population of $1.2\times 10^5$ events (and in the specific sample drawn from it),  the 10~km triangle can reach an accuracy on $d_L$ equal or better than $1\%$ for  28 events; this raises  to  202 events for 2L with 15~km arms at $45^{\circ}$, 
and 365 events for
2L with 20~km  arms at $45^{\circ}$, which is the configuration that provides the best results (a set of tables with the explicit number of events with SNR above given thresholds, or errors on various parameters below given values, is collected in App.~\ref{app:TablesCBC}).
In general
(not surprisingly), the 2L configuration with the longest arm length, 20~km, and relative  arm orientation of $45^{\circ}$, is the one that performs better for all the parameters.  

We can also appreciate from  Fig.~\ref{fig:AllGeoms_CumulBBH_NdetScale}  that  the configurations 
of 2L with  parallel arms performs quite poorly (comparatively) for  angular localization. While this was generally expected because of the parallel arms, it is quite interesting to see that, in terms of angular resolution, even the 2L with  20~km parallel arms  performs somewhat less well  than a single 10~km triangle, despite the longer arm length and the possibility of performing triangulation (even if only partial, since just two detectors are involved). Again, the best results are obtained with the 2L at $45^{\circ}$. For example, with our choices for the population and in our sample,
the 10~km triangle can detect  35 BBH/yr with angular resolution below $1\, {\rm deg}^2$, to be compared with 92 BBH/yr for the 15~km 2L at $45^{\circ}$ and
155 BBH/yr for the 20~km 2L at $45^{\circ}$ (see Table~\ref{tab:BBHAllConfDeldLDelOm} in App.~\ref{app:TablesCBC} for a more extended set of examples).  
For the chirp mass reconstruction, the arm length sets the hierarchy among these configurations (see also Table~\ref{tab:BBHAllConfDelMcDelchi} in App.~\ref{app:TablesCBC}) so, again, among the configurations considered,  the 10~km triangle is the least accurate.
The results for the reconstruction of the  symmetric mass ratio and of the spins show, instead, less pronounced differences among these configurations (note, however, the logarithmic scale, that visually flattens the differences; see again Table~\ref{tab:BBHAllConfDelMcDelchi} in App.~\ref{app:TablesCBC} for numerical values).

\vspace*{1mm}
From  Fig.~\ref{fig:AllGeoms_CumulBBH_NdetScale} and  Tables~\ref{tab:BBHAllConfDeldLDelOm}, \ref{tab:BBHAllConfDelMcDelchi}  we also see that the 15~km triangle has performances very similar to the 15~km 2L at $45^{\circ}$, except for the accuracy on the luminosity distance, where the 15~km 2L at $45^{\circ}$ is clearly superior; for instance, the 15~km 2L at $45^{\circ}$ would detect 202 BBH/yr with error on the luminosity distance better than $1\%$, to be compared with 77 for the 15~km triangle. 
In App.~\ref{app:corrBBHs} we examine the correlations between the various parameters for typical well-measured events and we find that, for the triangle,
the error on $d_L$ is more sensitive to the error on the localization, polarization angle and orbit inclination, so that errors on these parameters  have a larger effect on the marginalized error on $d_L$, compared to the 2L configuration. \rosso{We also note that this large improvement is specific to the 2L configurations with arms at $45^{\circ}$. For instance, we see from Table~\ref{tab:BBHAllConfDeldLDelOm} that the number of events/yr with $d_L$ measured to better than $1\%$ is 28 for the 10-km triangle, raising to 77 for the 15~km triangle; for the 15~km 2L with arms at $0^{\circ}$ this number is 79, and jumps to 202 for the 15~km 2L at $45^{\circ}$. Therefore, the raise from 28 to 77  (or 79) events/yr can be traced  to the increase in arm length, from 10~km to 15~km; the fact that the 15~km triangle and the  15~km 2L with parallel arms  have a very similar performance (on this metric) means that, for a given arm length, the three nested detectors of the triangle configuration approximately compensate the large baseline of the 2L configuration with parallel arms;  the further increase to order 200 events/yr is due to the better capability of  the 2L configuration at $45^{\circ}$ to measure the source location, polarization and inclination angles, and therefore disentangle them from the luminosity distance.}

\rosso{We can also appreciate the overall similar performance of the different geometries in the reconstruction of the intrinsic parameters (masses and spins). This is due to the fact that their reconstruction is mostly influenced  by the loudness of the signal in the detector, rather than by the localization capability, and is therefore largely determined by the SNR. As we see from the SNR panel in Fig.~\ref{fig:AllGeoms_CumulBBH_NdetScale}, all configurations have similar distributions of the SNR, except for the 10~km triangle, which is disfavored, and this is reflected in the estimation of masses and spins.}

\vspace*{1mm}
Finally, it is interesting to show the joint distribution of the events with respect to  $\Delta d_L/d_L$ and $\Delta\Omega_{90\%}$, since this determines the overall localization volume. To avoid a proliferation of plots, here we limit ourselves to a comparison between 
the 10~km triangle and 15~km 2L at $45^{\circ}$, which is shown in the left panel of  Fig.~\ref{fig:scatter_dLOm_2L4515kmvsT10and15km_BBH}, and between 
the 15~km triangle and 15~km 2L at $45^{\circ}$, shown in the right panel. We see that the 15~km 2L at $45^{\circ}$ has many more events than the triangle in the region of the 
$(\Delta d_L/d_L,\Delta\Omega_{90\%})$ plane corresponding to an interesting three-dimensional  localization, e.g., $\Delta d_L/d_L \leq 1\%$ and 

\vspace*{1mm}

{\em As far as these metrics are concerned, one can conclude that, while the performance of the 10~km triangle is by itself remarkable,  the {\rm 2L} configuration with 15~km arms, oriented at $45^{\circ}$, provides a further improvement (as  the {\rm 2L} configuration with 20~km arms or the 15~km triangle). In contrast, the {\rm 2L} with parallel arms looks quite disfavoured, because of a comparatively poor angular localization capability. The 15~km 2L at $45^{\circ}$ and the 15~km triangle provide results quite comparable, except for the accuracy on luminosity distance,  for which the 15~km 2L at $45^{\circ}$ is clearly superior.}

\vspace*{1mm}
Finally, in Fig.~\ref{fig:ETMR_1L_20km_CumulBBH_NdetScale} we study the performance of a single L-shaped detector with 20~km arms in the full HFLF cryo configuration, and we compare it with the 10~km triangle and with the 15~km 2L at $45^{\circ}$, also at their full sensitivity, as well as with LVKI~O5.   The most apparent feature is that, {\em in terms of angular resolution and reconstruction of luminosity distance,  a single {\rm L} is clearly performing  very poorly, and is even way below the LVKI~O5 forecast. This shows that  a single {\rm L}-shaped detector, not inserted in a global network, is basically useless for
those aspects of the Science Case, such as multi-messenger astronomy or cosmology, that require good estimates of sky-localization and distance of the sources. For the other parameters, however, a single {\rm L} of 20~km, although it provides worse estimates, can be comparable to the 
3G networks that we have considered.}\footnote{This, of course, assumes that spurious transient noise and glitches could be vetoed and eliminated reliably since, for a single detector, it would not be possible to reduce them by using the coincidence between independent detectors, or (as for the triangle) the null stream. In particular, the detection of unmodeled burst signals could not be reliably claimed with a single detector.}

\subsubsection{Effects of a change in the ASD}

We next investigate how these results are affected by changes in ASD. In Fig.~\ref{fig:ETS_T_10km_CumulBBH_NdetScale} we show, for the 10~km triangle, the cumulative distributions of the SNR and of parameter estimation (where again, for the SNR we show the cumulative distribution of events with SNR {\em larger} than a given value, while for parameter estimation we restrict to events with ${\rm SNR}\geq 12$ and we  show the cumulative distribution of events with error {\em smaller} than a given value)
for the full ASD including the HF instrument and a cryogenic LF instruments (denoted `HFLF~cryo'),  and for the ASD in which the LF instrument is completely missing and only the HF instrument is operative (denoted `HF~only'). We also show, for comparison, our forecasts for LVKI~O5. In order to be consistent with the ET results, the analysis for LVKI~O5 is obtained following exactly the same Fisher-Matrix analysis used for ET. For a more precise and realistic forecast, we refer to \citep{KAGRA:2020rdx} which uses a Bayesian framework close to real analysis applied to the current gravitational-wave observations.

{\em An important message that emerges from Fig.~\ref{fig:ETS_T_10km_CumulBBH_NdetScale} is that even if,  at least in the first stage of operations,  the LF instrument were not operational, still ET in its 10~km triangle configuration and HF-only sensitivity would provide a significant jump from 2G detectors, in terms of a number of detections, distribution of SNR and parameter reconstruction of BBHs, with the exception of angular localization (as expected, for a single infrastructure that cannot rely on triangulation).
It is also worth noting that a single observatory, such as the 10-km ET triangle, at its best sensitivity, can localize a comparable number of sources with an accuracy as good as that  of a network comprising five 2G detectors at full O5 sensitivity.}

\vspace{1mm}
Note that, on a single event basis, the events observed by LVKI are almost always better localized than by ET, showing that localization benefits more from the network triangulation than from ET's higher SNR (with fluctuations due to the assumed duty cycle). Nonetheless, the number of detections by ET is more than 10 times higher, and comprises high SNR events with low masses and/or high mass ratios. The high-SNR low-mass BBH mergers can be optimally localized accessing low frequencies and using the imprint of Earth's rotation in the longer observed signal. The localization of high-SNR  events with large mass ratio benefits from the detection of higher-order modes. This results in a similar number of optimally localized sources in the cumulative distribution   up to a sky-localization of about 6 square degrees, see Fig.~\ref{fig:ETS_T_10km_CumulBBH_NdetScale}. The larger amount of BBHs localized with sky-localization uncertainty larger than this threshold by ET compared to LVKI network depends on the detection efficiency; ET has much higher detection efficiency, reaches higher $z$ and has many more distant events where LVKI is no more able to detect sources. Going to larger and larger redshifts, on average the SNR decreases, and the ET localization worsen.

\vspace*{1mm}
Fig.~\ref{fig:ETSMR_2L4515_CumulBBH_NdetScale} shows the corresponding results for the 2L configuration with 15~km arms at $45^{\circ}$, with the two different ASDs. For comparison, in this figure we  also show both the results for the  10~km triangle in the full HFLF-cryo configuration, and the results for LVKI~O5. It is interesting to observe that, even in the configuration in which the LF instrument is completely missing, for BBHs  the
2L with 15~km arms at $45^{\circ}$ is superior to the 10~km triangle with its full (HFLF cryo) sensitivity for the accuracy on the luminosity distance (particularly for the number of events with very good accuracy, say $\Delta d_L/d_L\leq 1\%$, with 56 events against 28, see Table~\ref{tab:BBHAllConfDeldLDelOm} in App.~\ref{app:TablesCBC}) and gives basically equivalent results  for  angular localization, spins and orbit inclination. In contrast, for the chirp mass and symmetric mass reconstruction, 
the full HFLF-cryo triangle is better than the HF-only 2L-15km-$45^{\circ}$.
In the comparison with LVKI~O5 we see that, even in the HF-only configuration, the
2L with 15~km arms at $45^{\circ}$ provides a remarkable jump from 2G detectors on BBH detection number, SNR distribution and parameter reconstruction, except that the angular localization of the best-localized events would no longer be better than that of the five-detector LVKI network at O5 sensitivities (although the number of events with $\Delta\Omega_{90\%}<10^2\, {\rm deg}^2$, relevant for multi-messenger studies, is a factor $\sim 5$  higher for the 2L with 15~km arms at $45^{\circ}$, compared to LVKI~O5). See again Tables~\ref{tab:BBHAllConfDeldLDelOm} and \ref{tab:BBHAllConfDelMcDelchi} in App.~\ref{app:TablesCBC} for a compilation of number events for the different geometries and ASDs, and different cuts on the parameters.

{\em A main message that emerges from  Fig.~\ref{fig:ETSMR_2L4515_CumulBBH_NdetScale}  is that, as far as BBH parameter reconstruction is concerned, even if in a first stage of operations the whole LF instrument should not be operative, a {\rm 2L} configuration 
with 15~km arms at $45^{\circ}$ would still be a very competitive 3G instrument, in fact quite comparable to a 10~km triangle at full sensitivity  (with better performance of luminosity distance, less good performance on mass reconstruction, and equivalent performances on all other parameters and in SNR distribution).\footnote{It should also be observed that, in most studies where the accuracy on both volume localization  and masses is relevant, as for instance in joint studies of population and cosmological parameters, the bottle-neck is in the accuracy on sky localization and luminosity distance, which are always much worse than the relative accuracy on the (detector-frame) masses. As we see e.g. from Fig.~\ref{fig:ETSMR_2L4515_CumulBBH_NdetScale}, the relative error on the (detector-frame) chirp mass and on the symmetric mass ratio is typically well below  $10^{-4}$ and can be as low as  $10^{-6}$ for the chirp mass and $10^{-8}$ for $\eta$, while, for $d_L$, it is difficult to go below the $1\%$ accuracy. 
Therefore, having a better performance  on $d_L$ has a much stronger impact than having a better performance on the detector-frame masses, which in any case are very accurately measured. Note, furthermore, that at the large redshifts explored by 3G detectors, the accuracy of the reconstruction of the actual (source-frame) masses from the detector-frame masses will be determined by the accuracy of the reconstruction of the redshift.}}

\vspace{1mm}
Finally, another important target of 3G detectors are intermediate massive black holes (\acrshort{imbh}s) binaries, with masses in the range $10^2-10^4 \msun$, In this case, there are currently no observational constraints, and thus we do not use any astrophysically motivated IMBH population. Going back to Fig.~\ref {fig:Detector_Horizons_BBH_AllConf}, which showed the horizons for monochromatic non-spinning populations of equal-mass binaries (or Fig.~\ref{fig:Detector_Horizons_Difference}, that directly gives the relative differences with respect to the baseline 10~km triangle), it can be deduced that,  with the best sensitivity curve (HFLF-cryo configuration), for total masses above
about $(300-400)\msun$, the difference between the different ET geometries is at about  $10\%$ level. For masses around $100\msun$ the difference is larger, and for instance for $100\msun$ the detection horizon ranges between $z\simeq 42$ and $z\simeq 50$ for the different geometries; however, in this  redshift range we do not even expect astrophysical IMBH. 
Much more significant is the dependence on ASD. The right panel of Fig.~\ref {fig:Detector_Horizons_BBH_AllConf} show a significant reduction in the IMBH horizon  for the HF-only configuration. {\em This shows the importance of the low-frequency sensitivity for IMBHs.}

\subsubsection{Golden events}
Another important metric to assess the performance of a detector, or of a detector network, 
is provided  by the number and redshift distribution of `golden events', i.e. events with especially good properties in terms of SNR (e.g. BBHs with ${\rm SNR} \geq 100$), or error on luminosity distance (e.g. BBHs with $\Delta d_L/d_L\leq 0.05$) or angular localization (e.g. BBHs localized to better than  $10\, {\rm deg}^2$). Indeed, while for some aspects of the Science Case, such as population studies, the completeness of the sample is a key element, for other aspects, such as precision tests of General Relativity (\acrshort{gr}) or cosmological studies, the result will be largely, if not uniquely, determined by these `golden events'
(see the discussion in Section~5.4 of \cite{Iacovelli:2022bbs} and Section~VI of \cite{Borhanian:2022czq}).
In Fig.~\ref{fig:ET_allgeom_ASD_BBH_distr_vs_z} we show the redshift distribution of the BBHs detected 
with ${\rm SNR} \geq 100$, the distribution of events with  $\Delta d_L/d_L\leq 5\%$, and the distribution of events with  $\Delta \Omega_{90\%}\leq 10~{\rm deg}^2$.\footnote{Of course, these distributions are correlated. For instance, events with large ${\rm SNR}$ typically have small errors on $d_L$. 
The correlation between luminosity distance and angular resolution is shown, for BBHs, in Fig.~\ref{fig:scatter_dLOm_2L4515kmvsT10and15km_BBH}, and the analogous result for BNS will be shown in  Fig.~\ref{fig:scatter_dLOm_2L4515kmvsT10and15km_BNS}.
See Figs.~12 and 17 of \cite{Iacovelli:2022bbs}  and Figs.~5-9 of 
\cite{Ronchini:2022gwk}
for further scatter plots of the correlations among different parameters, for the ET triangle configuration.} In the upper row we show the results for the six detector geometries that we are considering, all taken at full (HFLF cryo) sensitivity. The middle row focuses on the 10~km triangle and shows how the result changes with the ASD, while the lower row shows the same result for the 2L with 15km arms  at $45^{\circ}$ (comparing also with the reference triangle configuration). 
In all plots, we show for comparison the redshift distribution of the BBH population that we have used and, for each detector configuration, the distribution of the events detected with ${\rm SNR}\geq 8$ (which, on this scale, is almost indistinguishable from the whole population, at least for $z\,\lsim\, 2$).

\begin{figure}[th]
    \centering
    \includegraphics[width=1.\textwidth]{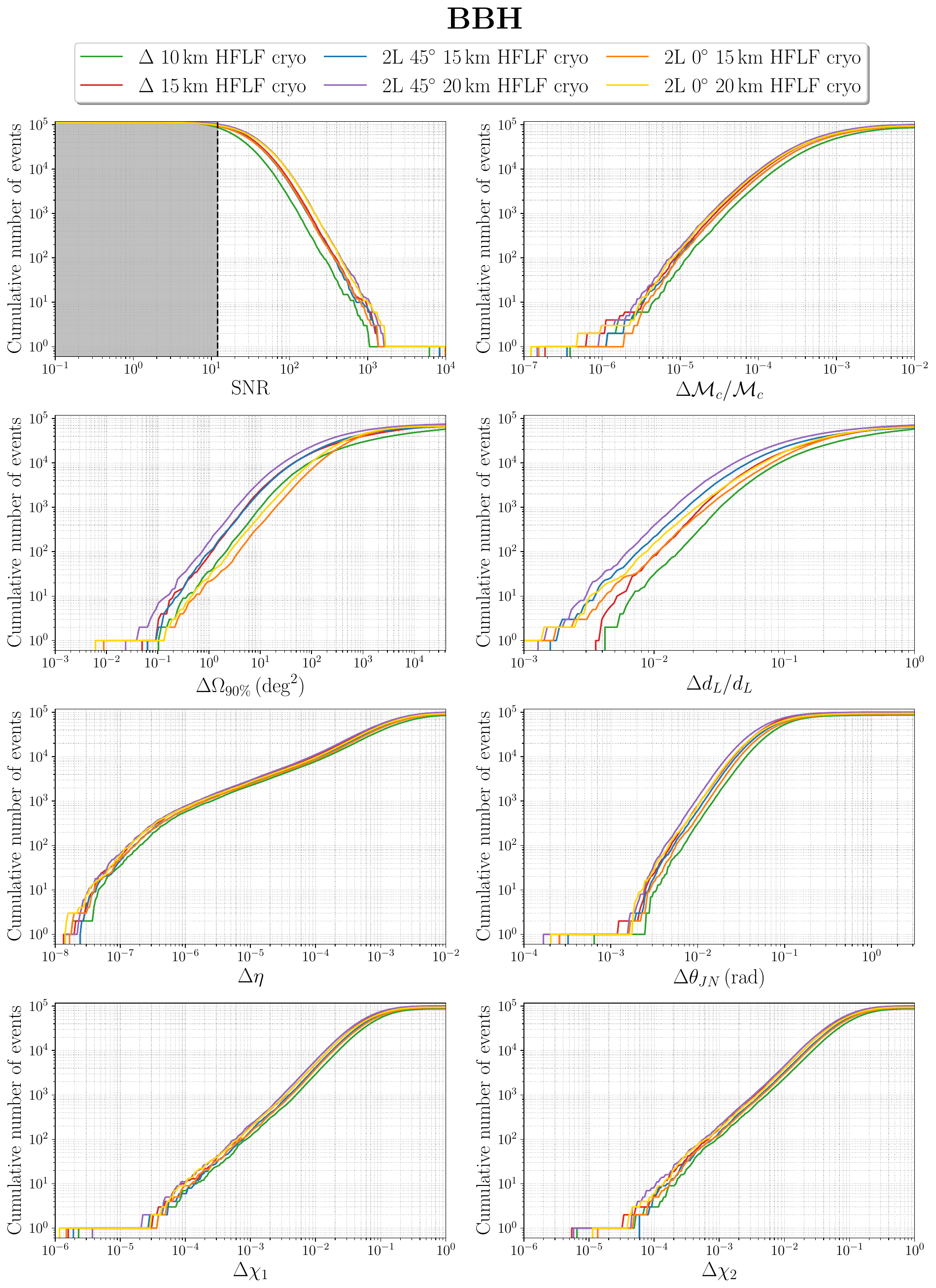}
    \caption{\small Cumulative distributions of the number of detections per year, for the SNRs and for the error on the parameters, for BBH signals, for the six considered  geometries, all with their best ASD, including xylophone configuration and cryogenic LF instrument.}
    \label{fig:AllGeoms_CumulBBH_NdetScale}
\end{figure}

\begin{figure}[t]
    \centering
    \includegraphics[width=0.47\textwidth]{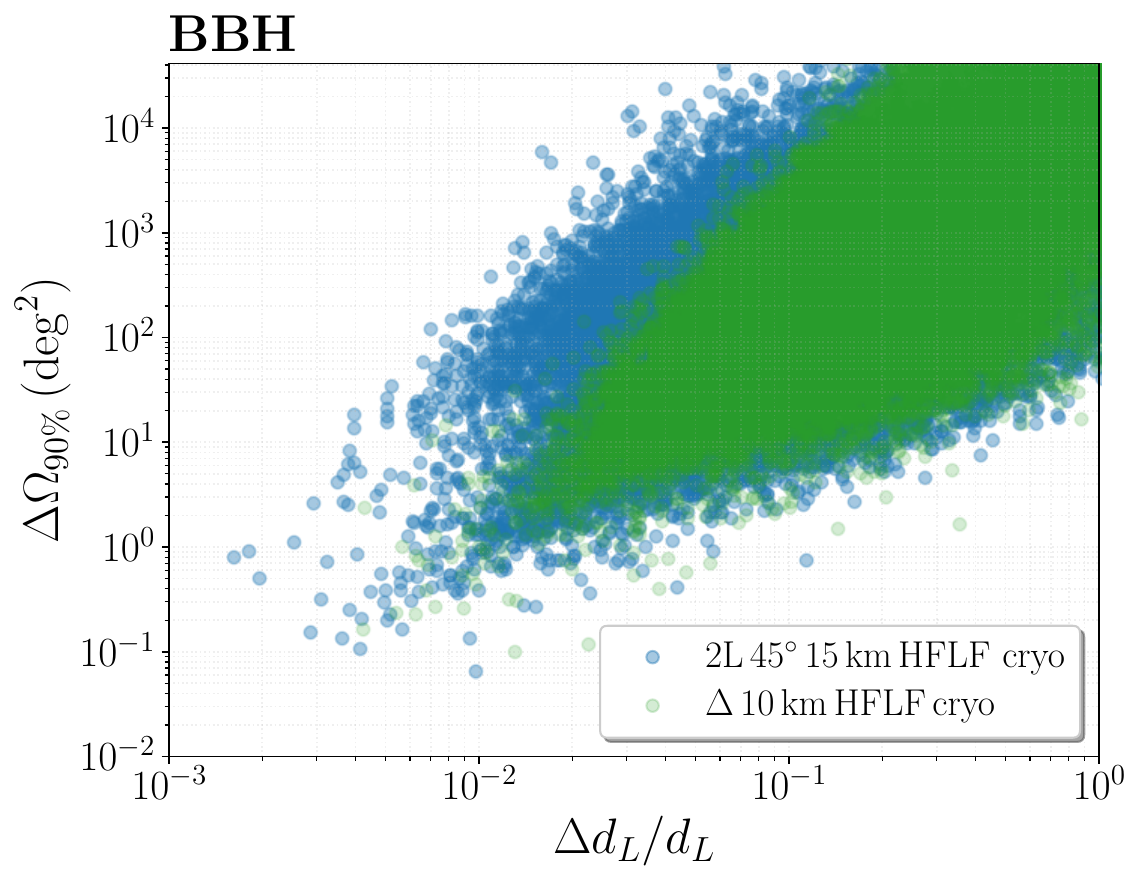}
    \includegraphics[width=0.47\textwidth]
    {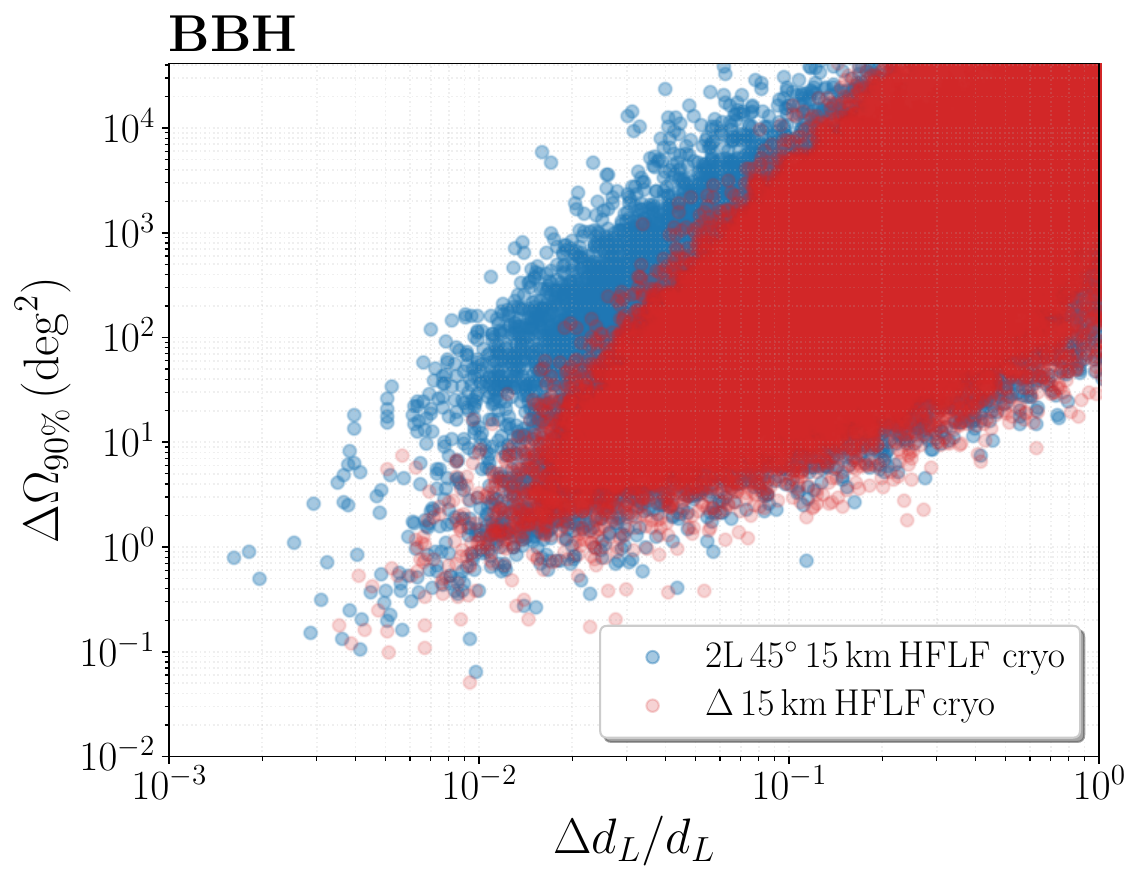}
    \caption{\small The joint accuracy on luminosity distance and angular resolution for  BBHs. Left panel:  the 15~km 2L at $45^{\circ}$ (blue) compared to the 10~km triangle (green). Right panel:
    the 15~km 2L at $45^{\circ}$ (blue) compared to the 15~km triangle  (red).}
    \label{fig:scatter_dLOm_2L4515kmvsT10and15km_BBH}
\end{figure}

From the left panel in the upper row, and Table~\ref{tab:BBHAllConfSNR} in App.~\ref{app:TablesCBC},  we see that, in terms of detections of events with large SNR, the 10~km triangle by itself has remarkable performances, with (in our sample realization) 2298 BBH/yr detected with ${\rm SNR}\geq 100$, and a redshift distribution extending up to $z\sim 5$. However, this is further improved in the other configurations; in particular, in the 2L-15km configurations, the number of BBH/yr with ${\rm SNR}\geq 100$ is  4933 for the misaligned configuration (and 5143 for the aligned-arms configuration), more than twice as large, and extends up to $z\simeq 7$; this further raises to more than 8000 for the 2L-20km configurations.
On the other hand, we already saw from Fig.~\ref{fig:AllGeoms_CumulBBH_NdetScale} that, over the whole ensemble of detected events, the setting with parallel  arms has comparatively poor performance for  angular localization and accuracy on the luminosity distance; this remains true also when we restrict to golden events, as we see from the middle and  the right panel in the first row of Fig.~\ref{fig:ET_allgeom_ASD_BBH_distr_vs_z}. In particular,  the 2L configurations with parallel arms  (whether with 15 or 20~km arm-length) are by far the less performing in terms of events with $\Delta \Omega\leq 10~{\rm deg}^2$ (followed  by the 10~km triangle)
while, in terms of the events with errors smaller than $5\%$ on $d_L$, the 10~km triangle is, comparatively,  the less performing, followed by the 2L configurations with parallel arms,
while the 2L configurations with arms at $45^{\circ}$ give the best results.

{\em These results indicate that, for BBH golden events, the {\rm 2L} configuration with arms at $45^{\circ}$, or the 15~km triangle,  provide the best compromise between detecting
many of them, and out to large redshift, and localizing them, further improving on the already remarkable performances of the  10~km triangle.}

\vspace{1mm}
The middle row of Fig.~\ref{fig:ET_allgeom_ASD_BBH_distr_vs_z} shows how the results depend on the ASD, for the 10~km triangle. From these plots we infer that, {\em also for  golden BBH events, a 10~km triangle
would  provide an extraordinary jump, compared to 2G detectors, even in the HF-only configuration.} Indeed, by comparison, at LVKI~O5, even having assumed the best-planned sensitivities for all five detectors, with our BBH population and the specific sample that we have drawn from it, we find only 4 events with ${\rm SNR} \geq 100$, to be compared with about 3000 for  the 10~km triangle with the full HFLF-cryo ASD, and almost 800 for the  10~km triangle with the HF-only ASD,
see Table~\ref{tab:BBHAllConfDeldLDelOm} in App.~\ref{app:TablesCBC}.

\vspace{1mm}
\noindent
The lower row of Fig.~\ref{fig:ET_allgeom_ASD_BBH_distr_vs_z}  shows the dependence on the ASD for the  2L geometry with 15~km arms at $45^{\circ}$ (showing, for comparison, also the result for the 10~km triangle at full sensitivity). Once again, {\em a significant conclusion is that the  {\rm 2L} geometry with 15~km arms at $45^{\circ}$, even with the HF-only instrument, is comparable to the 
10~km triangle at full sensitivity.}

\begin{figure}[t]
    \centering
    \includegraphics[width=1.\textwidth]{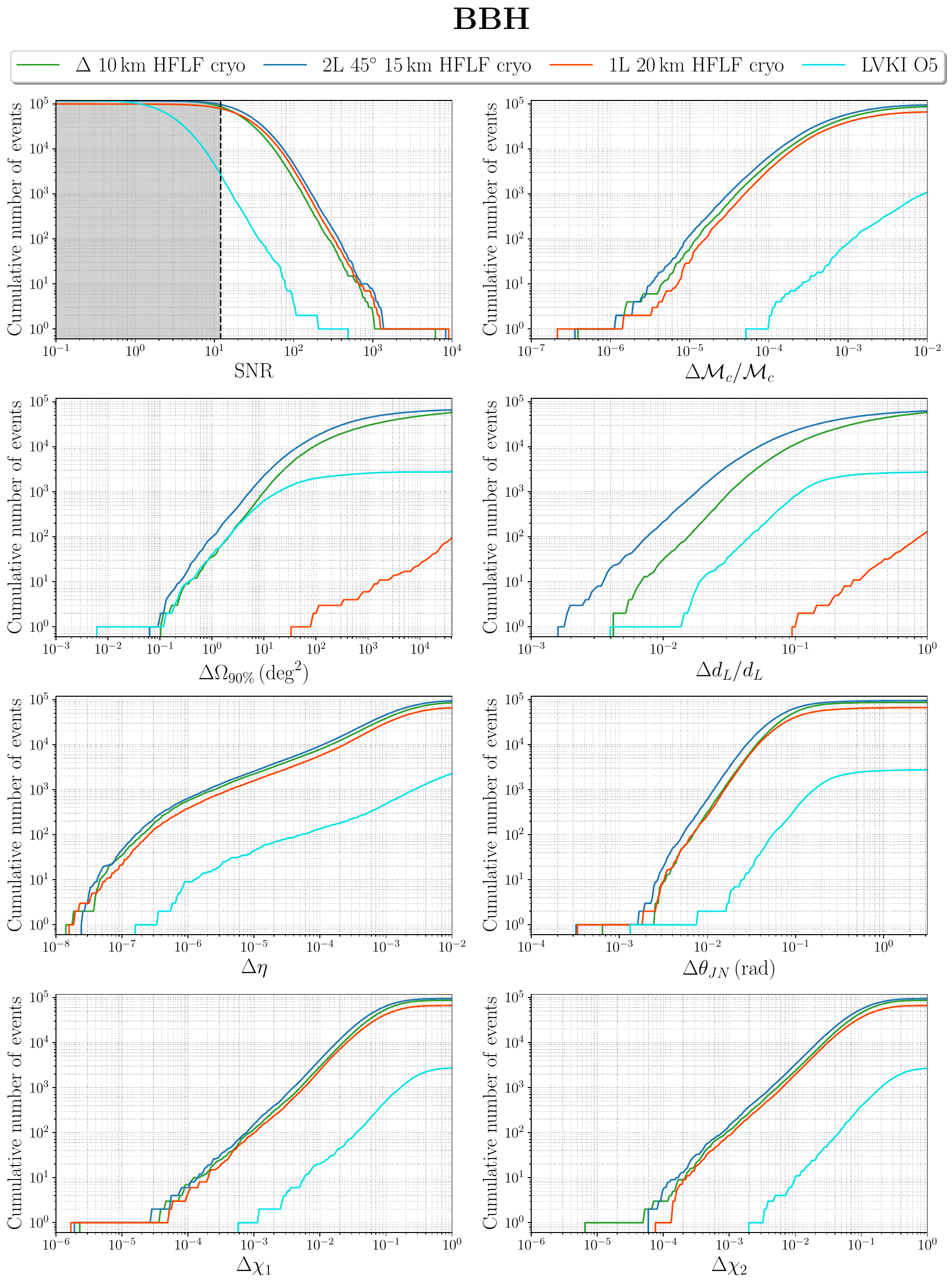}
    \caption{\small Comparison of SNR and parameter estimation error for the 10 km triangle, 15 km 2L misaligned and a single 20 km L  with the HFLF-cryo ASD. By comparison, we also show the forecast for LVKI~O5.}
    \label{fig:ETMR_1L_20km_CumulBBH_NdetScale}
\end{figure}

\begin{figure}[t]
    \centering
    \includegraphics[width=1.\textwidth]{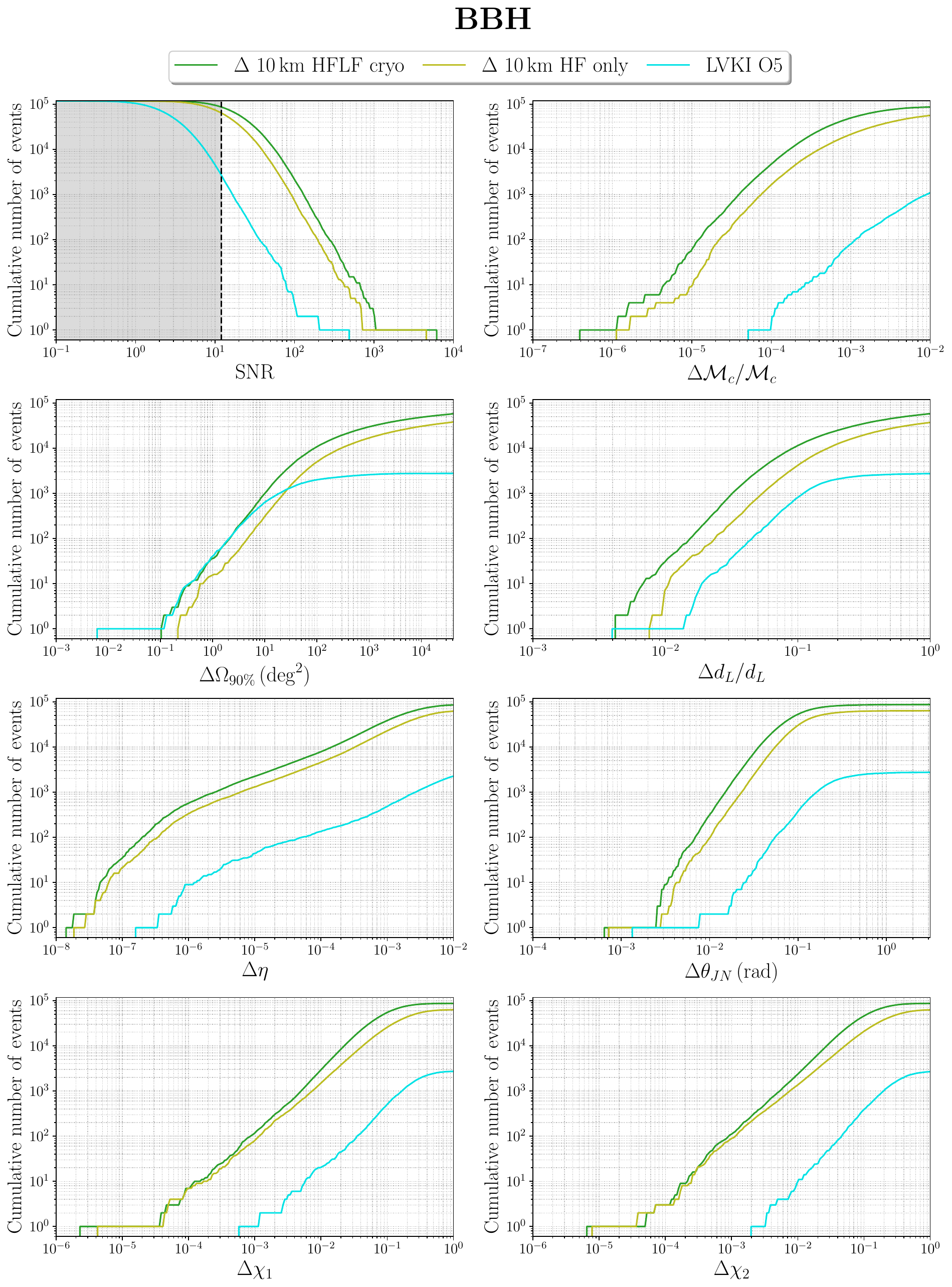}
    \caption{\small Comparison of SNR and parameter estimation error for the 10~km triangle with the HFLF-cryo and the HF-only ASDs. By comparison, we also show the forecast for LVKI~O5.}
    \label{fig:ETS_T_10km_CumulBBH_NdetScale}
\end{figure}

\begin{figure}[t]
    \centering
    \includegraphics[width=1.\textwidth]{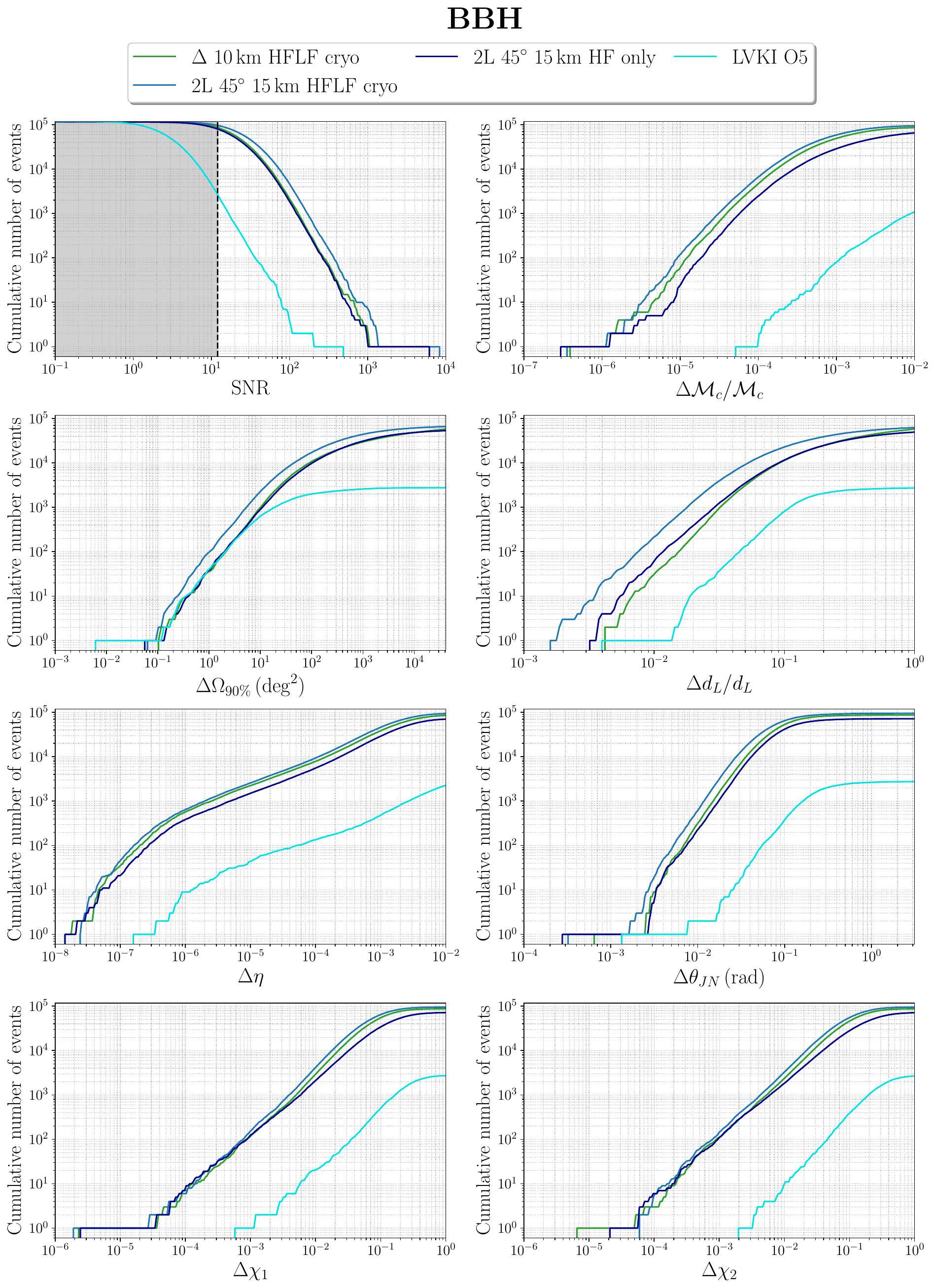}
    \caption{\small Comparison of SNR and parameter estimation error for the 15~km 2L misaligned  with the HFLF-cryo and the HF-only ASDs. For comparison, we also show the result for the 10~km triangle in the HFLF-cryo configuration, and for LVKI~O5.}
    \label{fig:ETSMR_2L4515_CumulBBH_NdetScale}
\end{figure}

\begin{figure}[t]
\hspace{-1.3cm}
\begin{tabular}{l@{\hskip -.02cm}l@{\hskip -.02cm}l}
     \includegraphics[width=5.7cm]{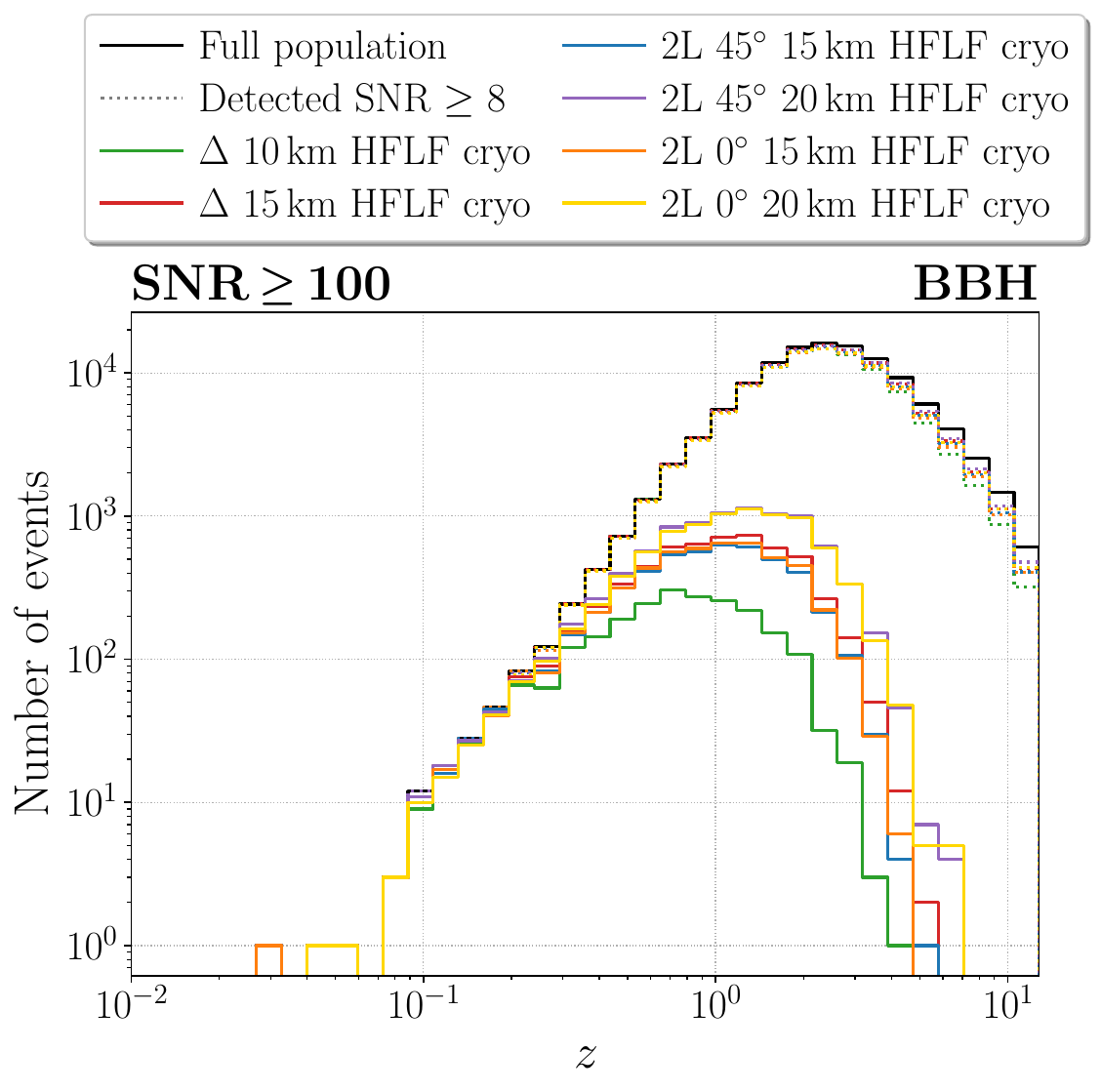} & 
     \includegraphics[width=5.7cm]{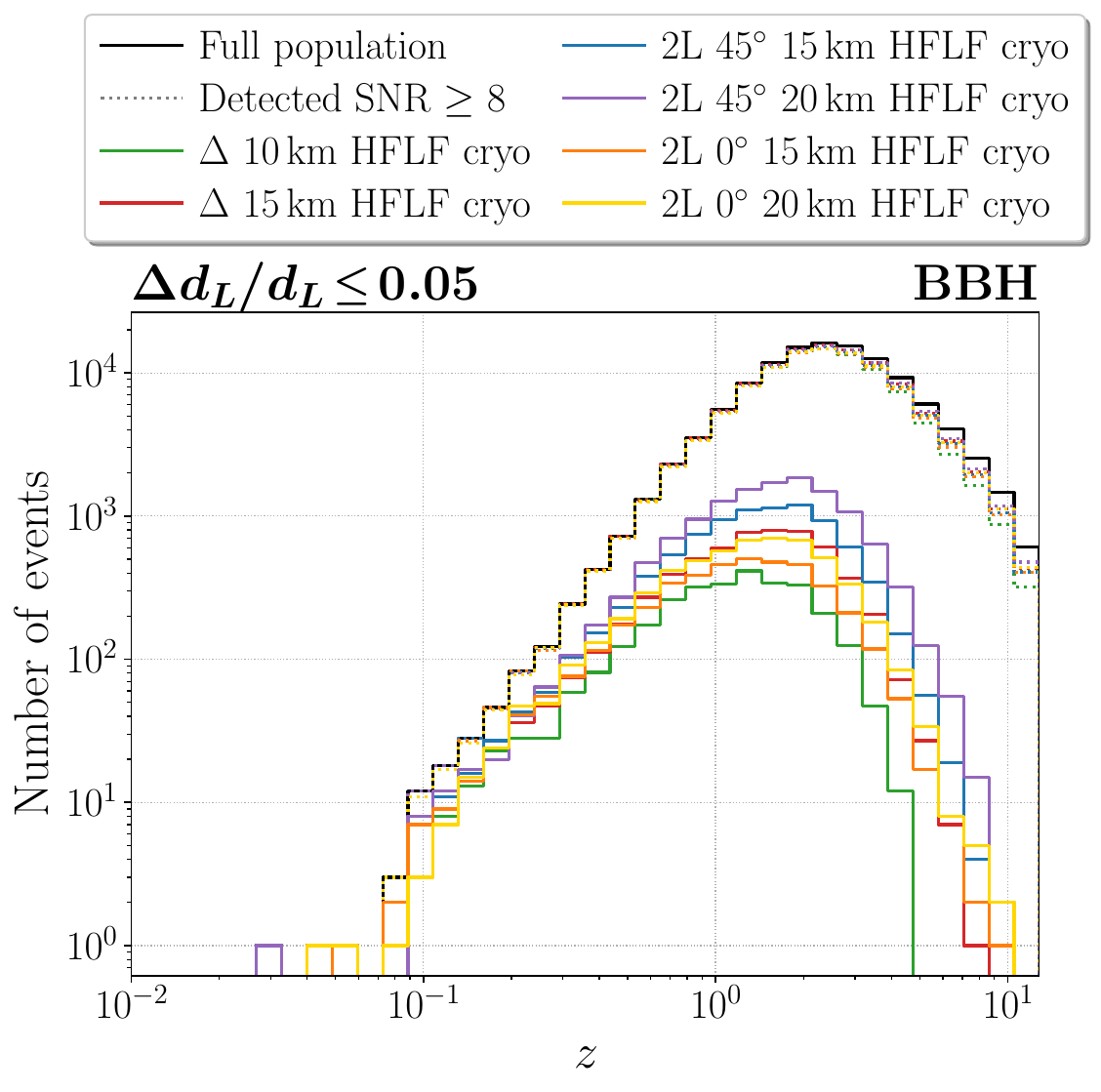} & \includegraphics[width=5.7cm]{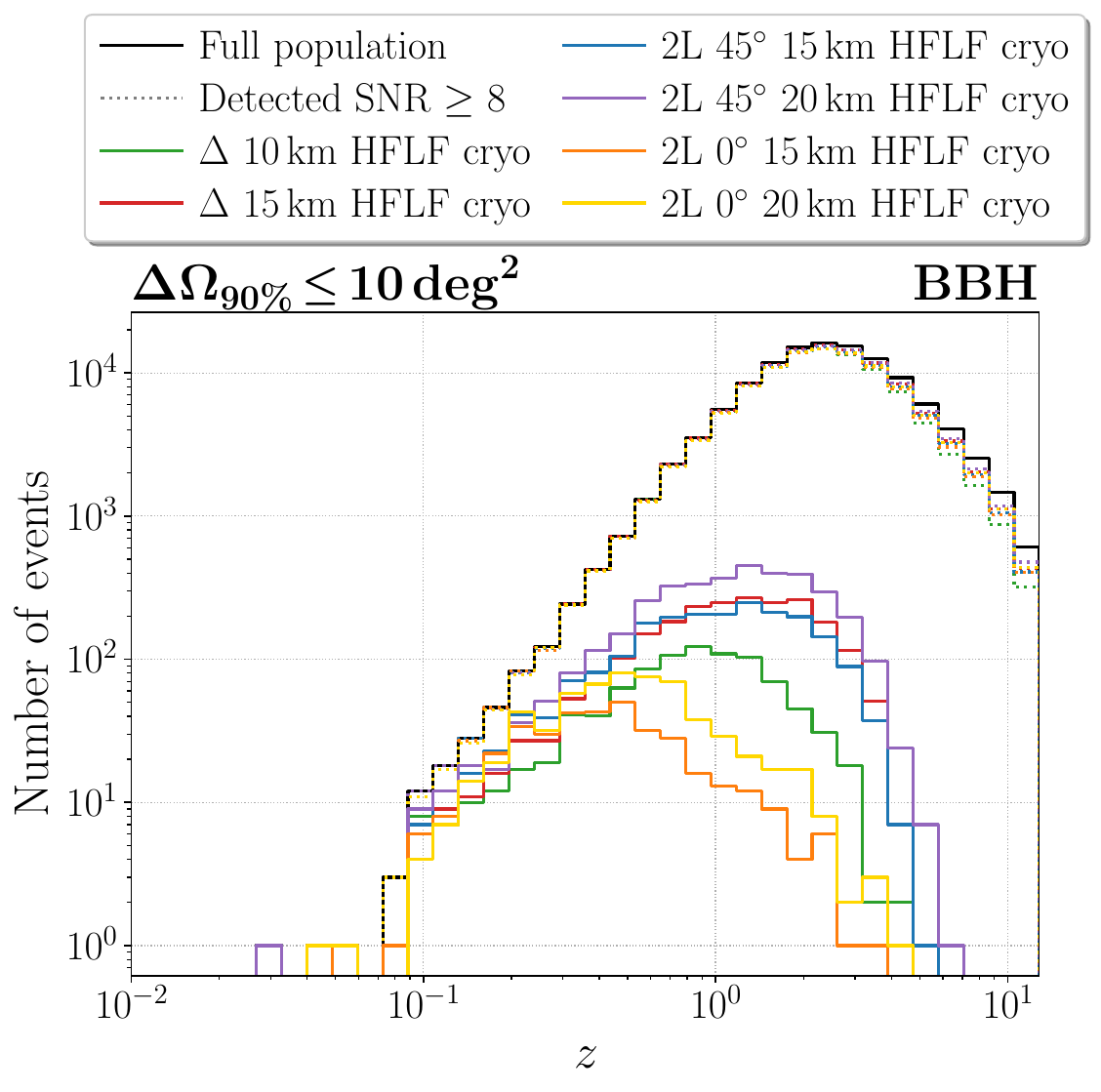} \\
     \includegraphics[width=5.44cm]{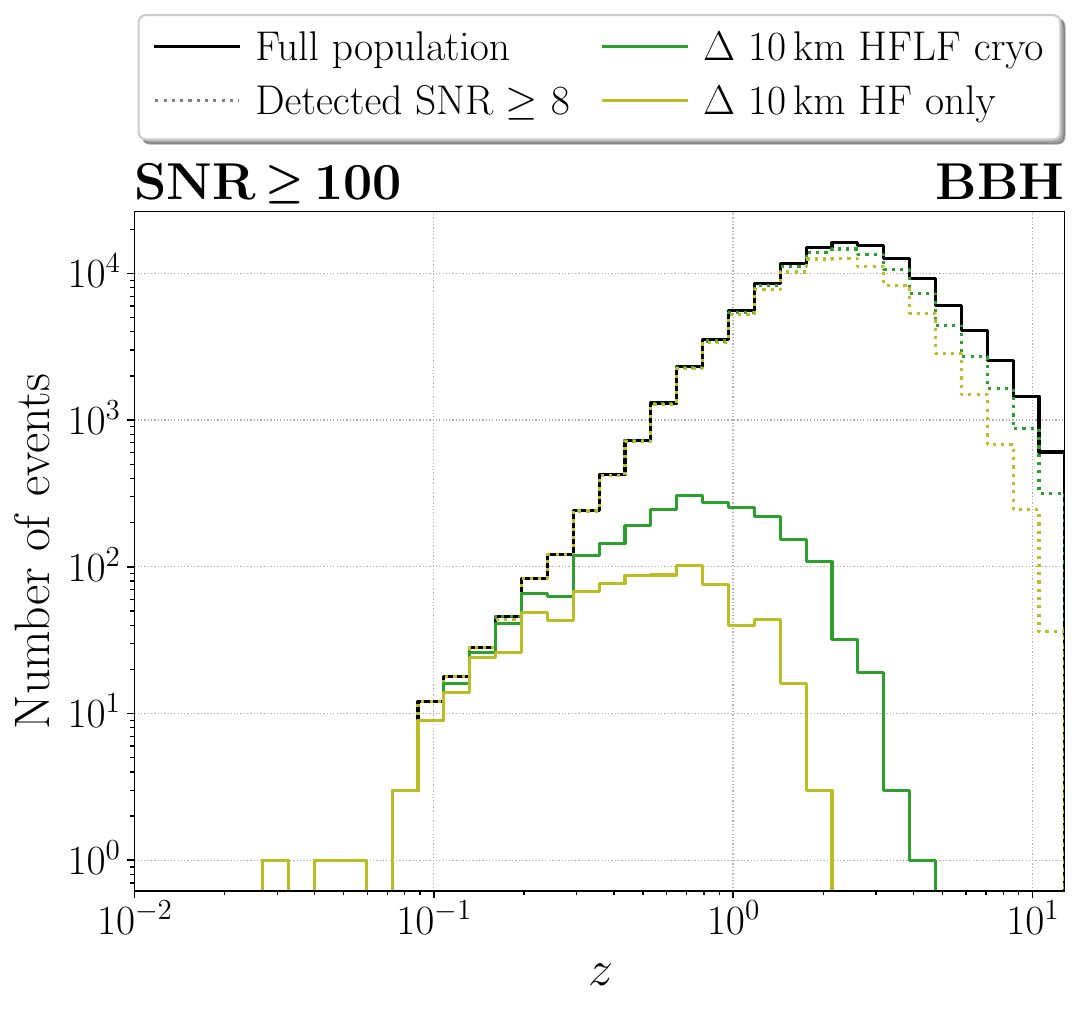} & 
     \includegraphics[width=5.44cm]{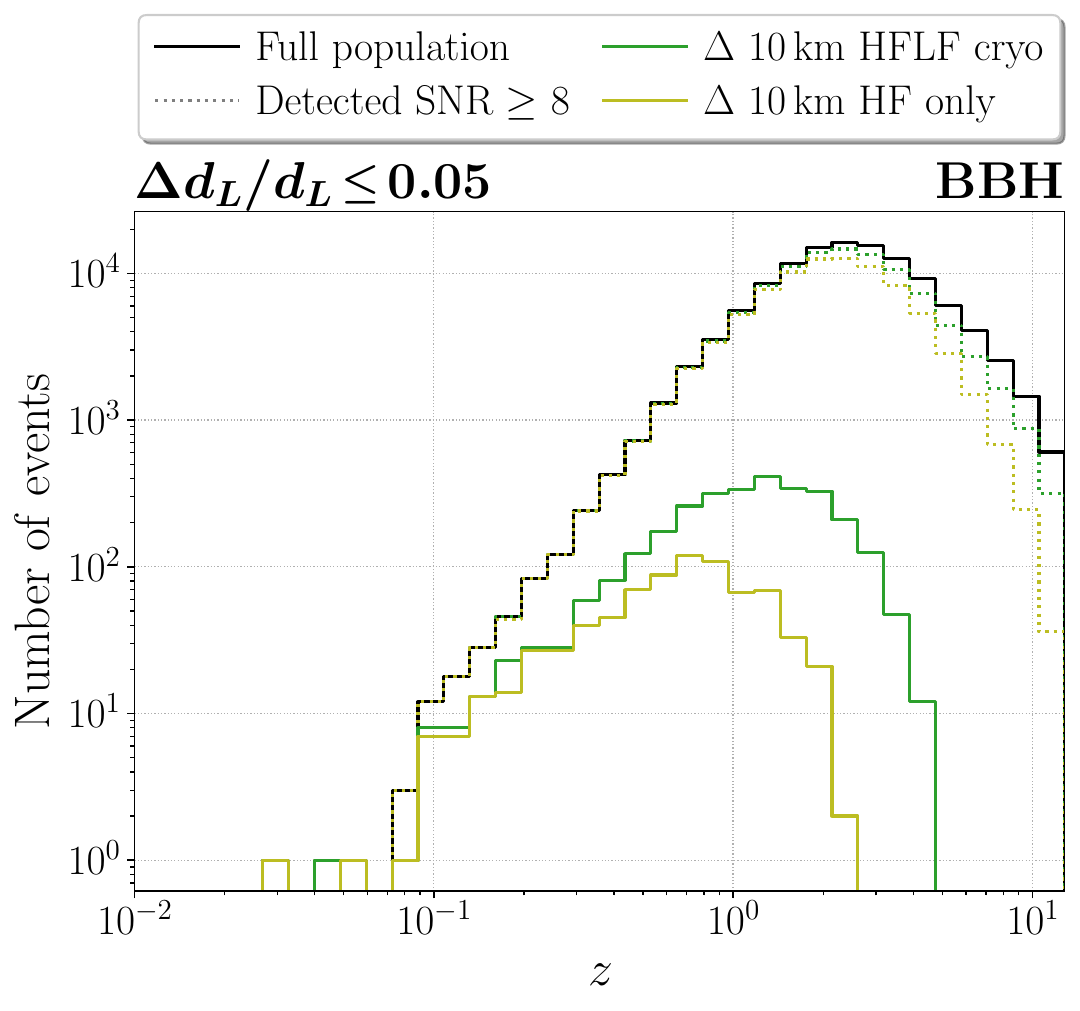} & \includegraphics[width=5.44cm]{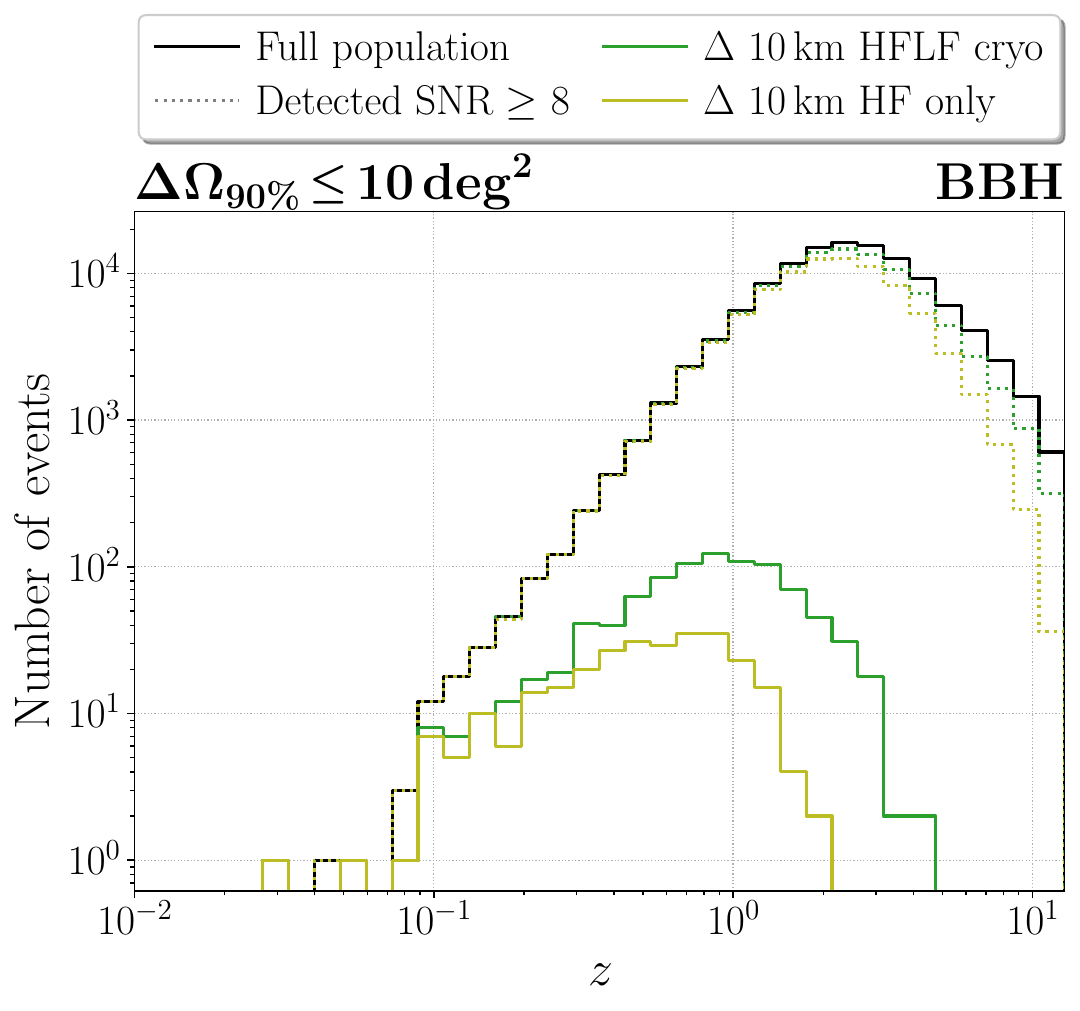} \\
     \includegraphics[width=5.65cm]{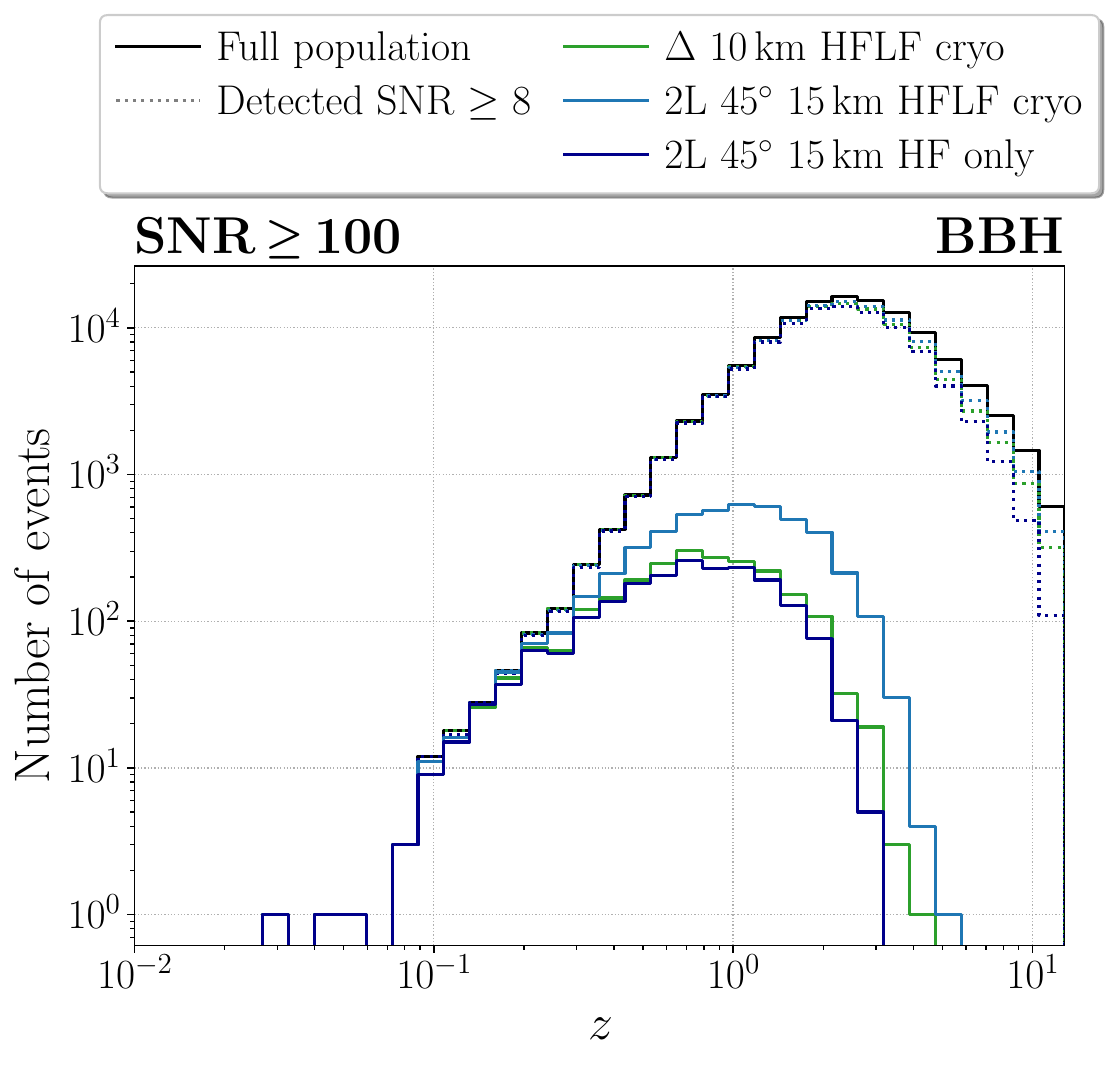} & 
     \includegraphics[width=5.65cm]{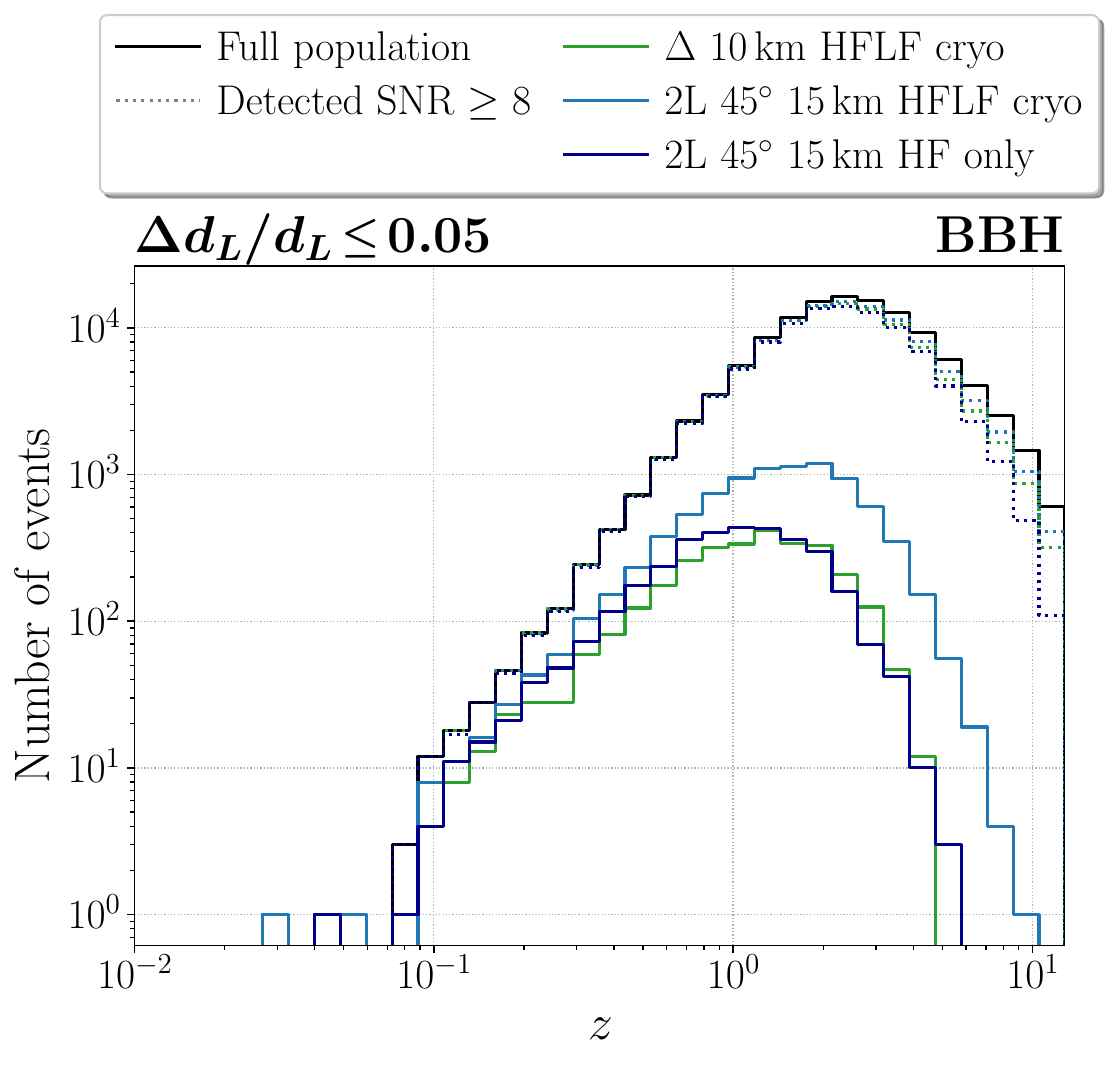} & \includegraphics[width=5.65cm]{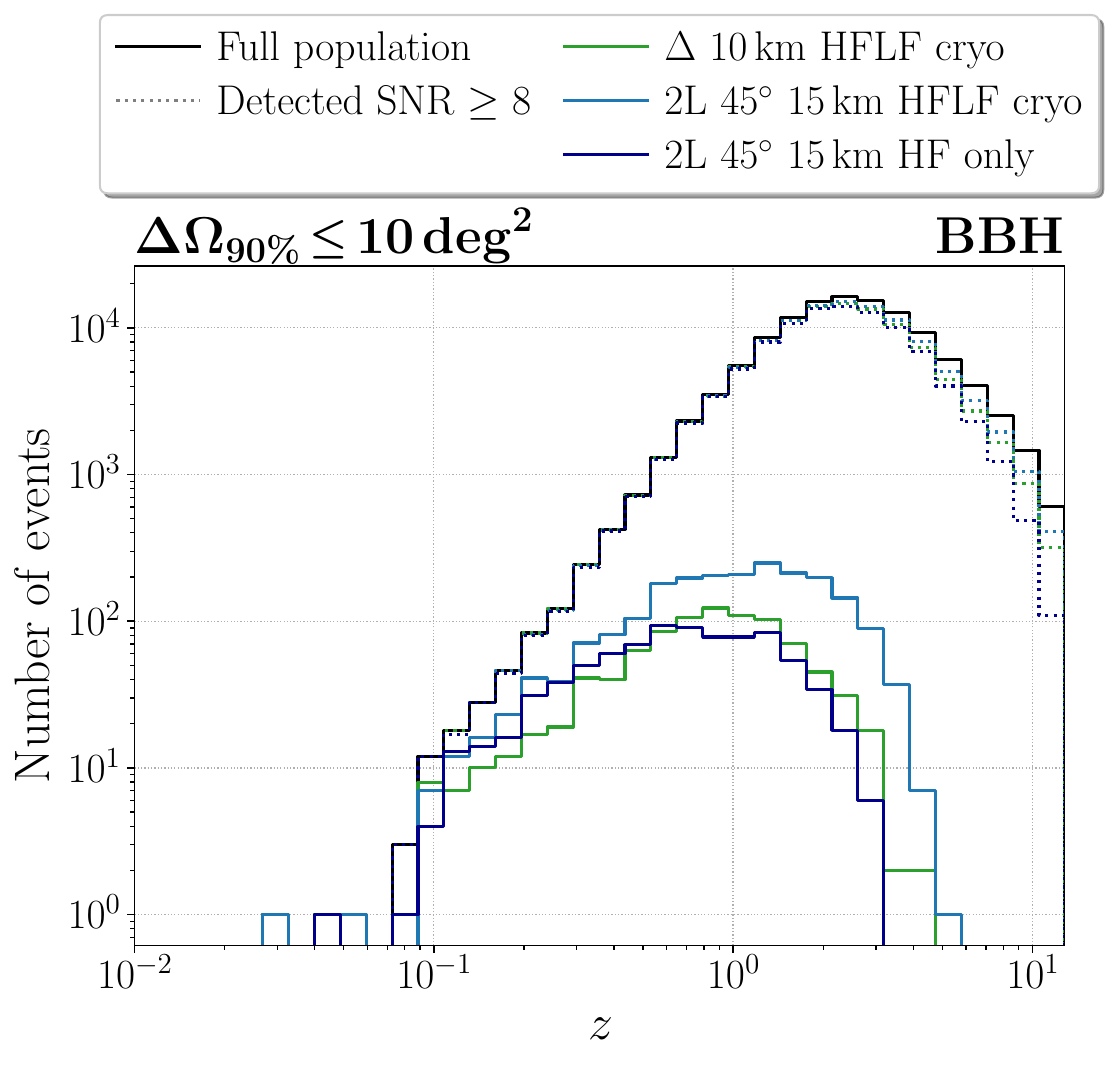}
\end{tabular}
    \caption{\small Redshift distribution of BBHs detected  with ${\rm SNR}\,\geq\,100$ (left column), or with relative error on the luminosity distance $\Delta d_L/d_L\,\leq\,0.05$ (central column), or with sky location $\Delta\Omega\,\leq\,10~{\rm deg}^2$ (right column) for various detector geometries and sensitivity curves. The upper  row shows the results for the six considered geometries, all with their best sensitivity, the central row for the 10~km triangle with the two considered ASDs, and the bottom row for the 2L with 15~km arms at $45^{\circ}$ and the two considered ASDs (together with the 10~km triangle with the full ASD). In each panel we also show, for reference, the redshift distribution of the  BBH population  used (black solid line), and of the events detected in the various configurations with ${\rm SNR}\,\geq\,8$  (dotted lines).}
    \label{fig:ET_allgeom_ASD_BBH_distr_vs_z}
\end{figure}

\clearpage

\subsection{Binary Neutron Stars}\label{sect:PEBNS}

We now perform the same analysis for BNSs. The corresponding results are shown in Figs.~\ref{fig:AllGeoms_CumulBNS_NdetScale}--\ref{fig:ET_allgeom_ASD_BNS_distr_vs_z}. We discuss again separately the dependence on geometry, on ASD, and the `golden events'. 

\subsubsection{Comparison between geometries}
Fig.~\ref{fig:AllGeoms_CumulBNS_NdetScale} confirms the basic message obtained from BBHs in Fig.~\ref{fig:AllGeoms_CumulBBH_NdetScale}: the 10~km triangle has, in itself, remarkable performances and produces  order-of-magnitudes improvements with respect to 2G detectors. However, a further
improvement can be obtained from
the other geometries that we have considered, on almost all parameters.  The only exceptions are the inclination angle $\iota$, for which the 2L configuration with parallel arms  performs, comparatively, quite poorly, both for 15~km and 20~km arm-length,\footnote{The difference with respect to the angle $\theta_{JN}$ shown in the BBH case is that, for BBHs, higher-order modes and precessing spin significantly help to determine the inclination angle.}
and the luminosity distance, for which the 10~km triangle performs similarly to the 2L with 15~km parallel arms. Similarly to what we found for BBHs, also for BNSs {\em the   {\rm 2L}  15~km with arms at $45^{\circ}$ improves on the accuracy estimate of all the parameters with respect to the 10~km triangle}. 
Comparing instead 
the 2L 15~km at $45^{\circ}$ to the 15~km triangle, we see that the two perform similarly on most parameters  except for luminosity distance, for which the 2L 15~km at $45^{\circ}$ is significantly better. The 15~km triangle, however, performs better than   a 15~km 2L with arms at $45^{\circ}$ for the inclination angle $\iota$, while the  2L 15~km at $45^{\circ}$  performs better  for the polarization angle $\psi$. \rosso{To understand better the causes of some differences between geometries, it is interesting to compare the 2L-15km with parallel arms to the triangle with the same arm length. The  2L-15km with parallel arms  is significantly worse than the 15-km triangle for the reconstruction of the inclination angle $\iota$. This is due to the fact that, for the 2L-15km with parallel arms, only a single combination of the two polarization amplitudes is accessible. On the other hand,  the 2L-15km with parallel arms  is better than the 15-km triangle on the angular localization because, thanks to its long baseline, it can partially triangulate. Both the reconstruction of $\iota$ and the angular localization influence the reconstruction of the luminosity distance. However,  we see from the panel on the luminosity distance in Fig.~\ref{fig:AllGeoms_CumulBNS_NdetScale} that the 2L-15km with parallel arms  is still significantly worse than the 15-km triangle in the reconstruction of $d_L$, so the dominant effect is due to the (comparatively) poor reconstruction of $\iota$.
}

For the parameter $\tilde{\Lambda}$  related to the tidal deformability of the two NSs, again  the 10~km triangle  already  obtains remarkable performances, with  $1040$ events/yr with $\tilde{\Lambda}$ measured better than $10\%$, and the best events providing a measurement at a few percent level; however, all other geometries that we considered improve further on this; in particular, the 2L-15km-$45^{\circ}$ detects 2463 events/yr with $\tilde{\Lambda}$ measured better than $10\%$, so an improvement by a factor of 2.5, and the 15~km triangle 2783,
see  Tables~\ref{tab:BNSAllConfSNR},
\ref{tab:BNSAllConfDeldLDelOm} and \ref{tab:BNSAllConfDelMcDelLam} in App.~\ref{app:TablesCBC}. 

Fig.~\ref{fig:scatter_dLOm_2L4515kmvsT10and15km_BNS} shows the
joint accuracy on luminosity distance and angular resolution for  BNSs, comparing    the 15~km 2L at $45^{\circ}$  to the 10~km triangle (upper panel) and to the 15~km triangle (lower panel). Considering, for instance, the events with $\Delta d_L/d_L <0.1$ and
$\Delta\Omega_{90\%} \leq 10^2\, {\rm deg}^2$, we see that the

\vspace{2mm}
Fig.~\ref{fig:ETMR_1L_20km_CumulBNS_NdetScale} shows the  performance of a single L-shaped detector of 20~km for BNSs, compared to other geometries, similarly to Fig.~\ref{fig:ETMR_1L_20km_CumulBBH_NdetScale} for BBH. We observe that, with a single L-shaped detector, no BNS will have an angular resolution $\Delta\Omega_{90\%}$ below $10\, {\rm deg}^2$, and (in our sample realization) only 6 BNS/yr would have $\Delta\Omega_{90\%}\leq 10^2\, {\rm deg}^2$ (to be compared  with  184 BNS/yr for the 10~km triangle and 559 BNS/yr for the {\rm 2L} configuration with 15~km arms at $45^{\circ}$, see Table~\ref{tab:BNSAllConfDeldLDelOm} in  App.~\ref{app:TablesCBC}.).
The reconstruction of distances would also be comparatively poor, with just a handful of events measured to better than $10\%$; so, without other 3G detectors, a single L-shaped detector, even with 20~km arms, would miss all the aspects of the Science Case, such as multi-messenger astronomy  or cosmology, that require good source localization.

\subsubsection{Effects of a change in the ASD}
Fig.~\ref{fig:ETS_T_10km_CumulBNS_NdetScale} shows the result for the two ASDs of the 10~km triangle and compares them with LVKI~O5. Quite  interesting results emerge from the panel on the angular resolution. A five-detector network, such as 
LVKI~O5, would of course have excellent localization capabilities for some very close events, so that it can reach a localization below $1\, {\rm deg}^2$ on a few events per year, which a single 10~km triangle, even with the full ASD, cannot reach. However, this will be limited to a handful of particularly close events.  
What is also relevant, in the panel for $\Delta\Omega_{90\%}$ of Fig.~\ref{fig:ETS_T_10km_CumulBNS_NdetScale}, is the number of BNS with $\Delta\Omega_{90\%}$ below about $10^2 \, {\rm deg}^2$. 
In fact, such an angular localization accuracy can already be sufficient to perform multi-messenger studies with wide field of view (\acrshort{fov}) electromagnetic observatories, as has been demonstrated by the current follow-up of gravitational-wave signals (see e.g. \cite{LIGOScientific:2016qac,LIGOScientific:2017ync}). In this case, the single triangle is quite superior to the LVKI~O5 network, by detecting a number of well-localized sources higher by a factor $\sim 3.6$ (see Table~\ref{tab:BNSAllConfDeldLDelOm} in App.~\ref{app:TablesCBC}  for a compilation of numerical values relative to these figures,  and Section~\ref{sect:MMO} for a detailed discussion and implications for multi-messenger studies). However, this is true only for the triangle at full sensitivity, i.e. in the HFLF-cryo configuration. The ET full sensitivity localization significantly benefits from the use of the effect of Earth's rotation on the longer signal detected, thanks to the access to lower frequencies. The benefit is much larger for closer events where the BNS signal can be followed for a longer time. In terms of the number of BNSs with angular resolution below $10^2 \, {\rm deg}^2$,  the HF-only configuration is therefore much worse than LVKI~O5. For most other parameters, in contrast,  a 10~km triangle provides a significant jump with respect to LVKI~O5 even in the HF-only configuration (except for luminosity distance and polarization, where it is worse; this is because these quantities are more  correlated with angular localization). See again Table~\ref{tab:BNSAllConfDeldLDelOm} in App.~\ref{app:TablesCBC} for a compilation of numerical values.

Fig.~\ref{fig:ETSMR_2L4515_CumulBNS_NdetScale} show the analogous results for the 2L configurations with 15~km arms at $45^{\circ}$; for comparison, we also show both LVKI~O5 and the 10~km triangle at full sensitivity. First of all, we see again that, at the level of full sensitivities, the 
configurations with 15~km arms at $45^{\circ}$ is superior to the 10~km triangle for all parameters.
We also see that,  for angular localization, the HF-only configuration is  much worse than LVKI~O5, at least for events with
$\Delta\Omega_{90\%}<10^2\, {\rm deg}^2$, which are the relevant ones for multi-messenger studies,
while, for all other parameters, even the HF-only configurations provides a remarkable jump compared to LVKI~O5.
However, contrary to the BBH case, the 
HF-only configuration is now very much inferior to the  10~km triangle with the full ASD, particularly for sky localization.   This is due to the fact that the improvement of the low-frequency sensitivity  allows a BNS to remain in the detector bandwidth for hours and even up to a day, as we already mentioned above (see also Fig.~2 of \cite{Iacovelli:2022bbs}), and then the rotation of the Earth improves the BNS localization. This is less relevant for BBHs since most of them have a  total mass   greater than about ${\cal O}(20\, \msun)$, and stay in the detector bandwidth for a time too  short to appreciate the effect of Earth's rotation. Therefore, the low-frequency sensitivity, and therefore the LF instrument, is more important for the localization of BNSs compared to BBHs. {\em This shows the crucial importance of the LF instrument for BNSs}.

\subsubsection{Golden events}
Fig.~\ref{fig:ET_allgeom_ASD_BNS_distr_vs_z} shows the redshift distribution of `golden events' for BNSs, defined here  either by the condition   ${\rm SNR}\geq 30$,  or by $\Delta d_L/d_L\leq 0.2$, or by $\Delta\Omega_{90\%}\leq 10^2\, {\rm deg}^2$.
These conditions are less stringent than those used for BBHs in Fig.~\ref{fig:ET_allgeom_ASD_BBH_distr_vs_z}, reflecting the fact that BNSs are intrinsically less loud than BBHs.  The upper row compares the different geometries, all taken with their best ASD, and basically confirms the picture obtained for BBH golden events, compare with the upper row of  Fig.~\ref{fig:ET_allgeom_ASD_BBH_distr_vs_z}: for the SNR, the 10~km triangle  gives the less good results, while the other configurations are comparable, with the 2L with 20~km  parallel arms being the best; for luminosity distance and angular localization, 
the 2L configurations with arms at $45^{\circ}$ give the best results and, overall, appears to provide the best compromise between having a large number of events with large ${\rm SNR}$ out to large distances, and localizing them well in volume.  

The middle row quantifies the loss of performance of the 10~km triangle when the LF instrument is absent. The loss is particularly dramatic for the accuracy on luminosity distance and sky localization.
{\em This confirms the crucial importance of the LF instrument for BNS}, as we already discussed when commenting Fig.~\ref{fig:ETSMR_2L4515_CumulBNS_NdetScale}.

From the bottom row we  see that, also on this metric, the 2L configurations with arms at $45^{\circ}$ and the HF-only ASD gives results basically identical to  the 10~km triangle with full HFLF-cryo sensitivity for the SNR distribution; in contrast, we already remarked on the full ensemble of events that the LF instrument is crucial for angular localization and accuracy on luminosity distance for BNSs, and we see the same information emerging from golden events, where the HF-only configuration of the 2L configurations with arms at $45^{\circ}$ is very much inferior to  the full 10~km triangle, in contrast to what happens for BBHs, compare with  Fig.~\ref{fig:ET_allgeom_ASD_BBH_distr_vs_z}. As already discussed, this is due to the fact that, with a good low-frequency sensitivity, BNSs can stay in the detector bandwidth for hours to about 1~day, with the corresponding benefit in angular localization and in all parameters correlated with it.

\subsubsection{Dependence on the population model}
\begin{figure}[t]
    \centering
    \includegraphics[width=1.\textwidth]{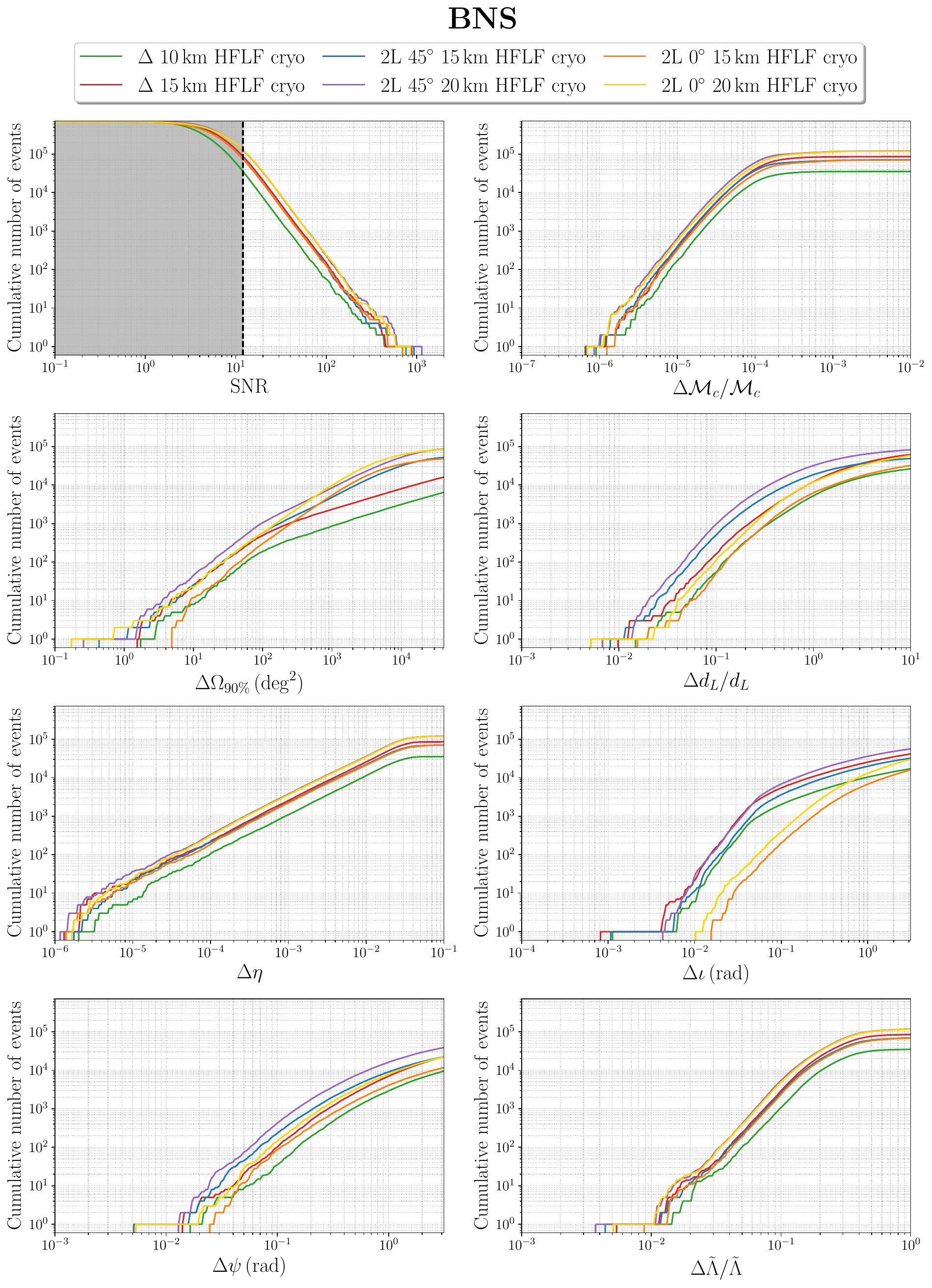}
    \caption{\small Cumulative distributions of the number of detections per year, for the SNRs and for the error on the parameters, for BNS signals, for the six considered  geometries, all with their best ASD, including xylophone configuration and cryogenic LF instrument.}
    \label{fig:AllGeoms_CumulBNS_NdetScale}
\end{figure}

Finally, it is interesting to discuss how these results depend on the model that we have used for the BNS population (and on the specific random sample extracted from it). First of all, we remark that the differences between designs that emerge from these plots, and the conclusions that we draw from them, are not driven by the tails of the distributions (that, even for a given population model, would also depend on the sample realization), but rather by their overall behavior,  which in general involves hundreds of events; even for the `golden events’, there are several tens to several hundreds  events per bin, see Fig.~\ref{fig:ET_allgeom_ASD_BNS_distr_vs_z}.
To understand in full generality how the results depend on the population model used is quite difficult; the full set of runs, on all different detector configurations studied, is  computationally expensive, and  it is not feasible to explore a large parameter space of population models in a reasonable timeframe. Our strategy has rather been to use a state-of-the art population model \cite{Santoliquido:2020axb} for BNS and NSBH
(and \cite{Mapelli:2021gyv} for BBHs),
which are able to match the LVK rates.
However, a useful check, for BNS, can be performed as follows. For the redshift distribution of the BNS rate, current models are quite sophisticated and we do not expect great variations (furthermore,  we are not much interested in the absolute numbers of detections in a given configuration, but rather in the relative performances of the detectors). There is  large uncertainty in the overall normalization, given by the local rate, which for BNS can be up a factor  10 higher or lower than the one that we use. However, while this will influence significantly the absolute number of detections per year, again  it will not affect much the comparison between detectors, which is our main aim. For instance, if the local rate should turn out to be a factor of 2 smaller than the one that we use, the plots that we show will have to be understood as relative to 2 years of data, rather than 1 yr; however, the conclusions for the relative performances of the different design would be basically unaffected (of course, the tails of rare events will be affected for a given fixed observation time but, again, the conclusions are not really dependent on these tails). So,  for our purposes the parameter that is most significant to vary, for BNS, is the mass distribution. As 
 mentioned on page~\pageref{page:catalog}, the results for BNS presented above were  obtained with a flat mass distribution between 1.1 and 2.5 solar masses. We have  repeated the analysis using a Gaussian mass distribution, with mean $1.33\msun$ and standard deviation $0.09\msun$, sampled independently for both masses (we then impose that the heaviest one corresponds to $m_1$). The results, for the six considered geometries in their full HFLF-cryo configuration,  are shown in Fig.~\ref{fig:AllGeoms_CumulBNS_NdetScale_massGauss}. Comparing with Fig.~\ref{fig:AllGeoms_CumulBNS_NdetScale}, we see that the results are very consistent. There is a somewhat smaller number of overall detections (because in the gaussian distribution events with large masses are suppressed), but the parameter estimation is quite similar (with differences in the tails, expected because of  sample variance), and the conclusions for the comparison between detector configurations are unchanged.

\begin{figure}[t]
    \centering
    \includegraphics[width=0.47\textwidth]{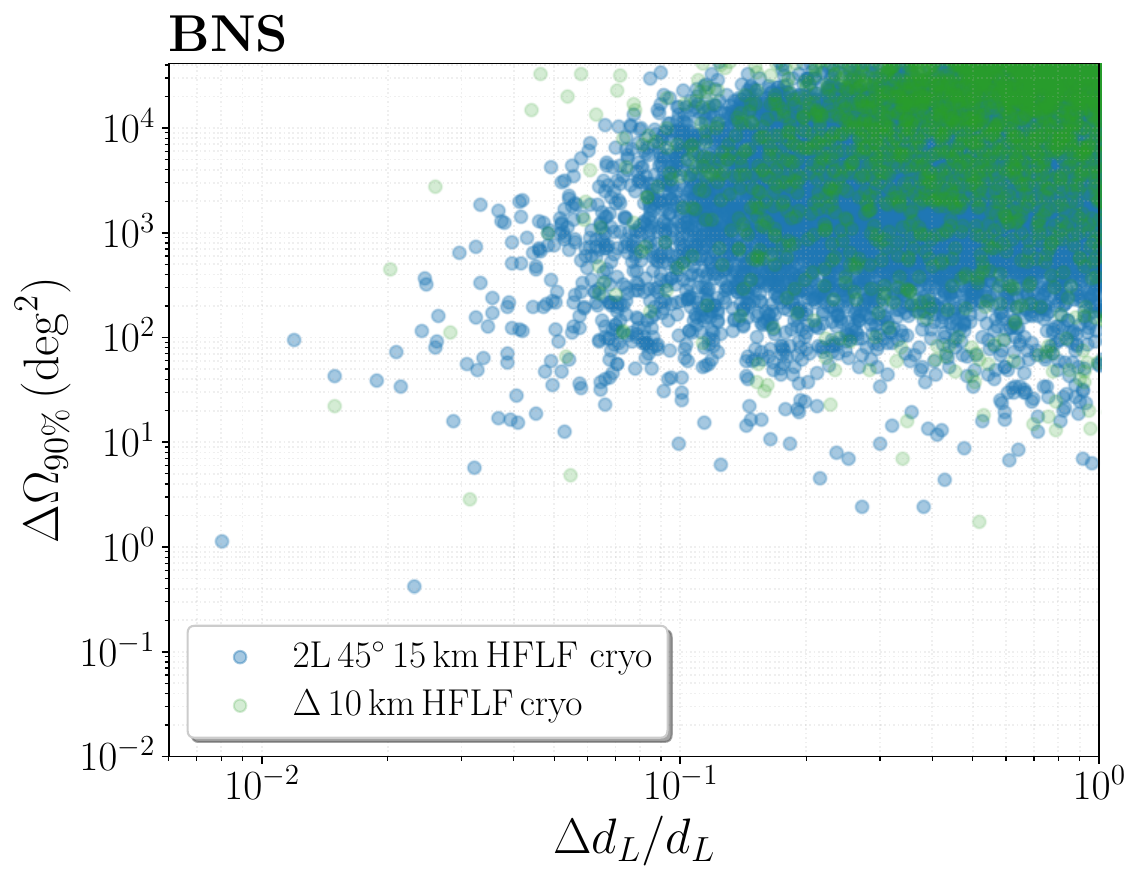}
    \includegraphics[width=0.47\textwidth]
    {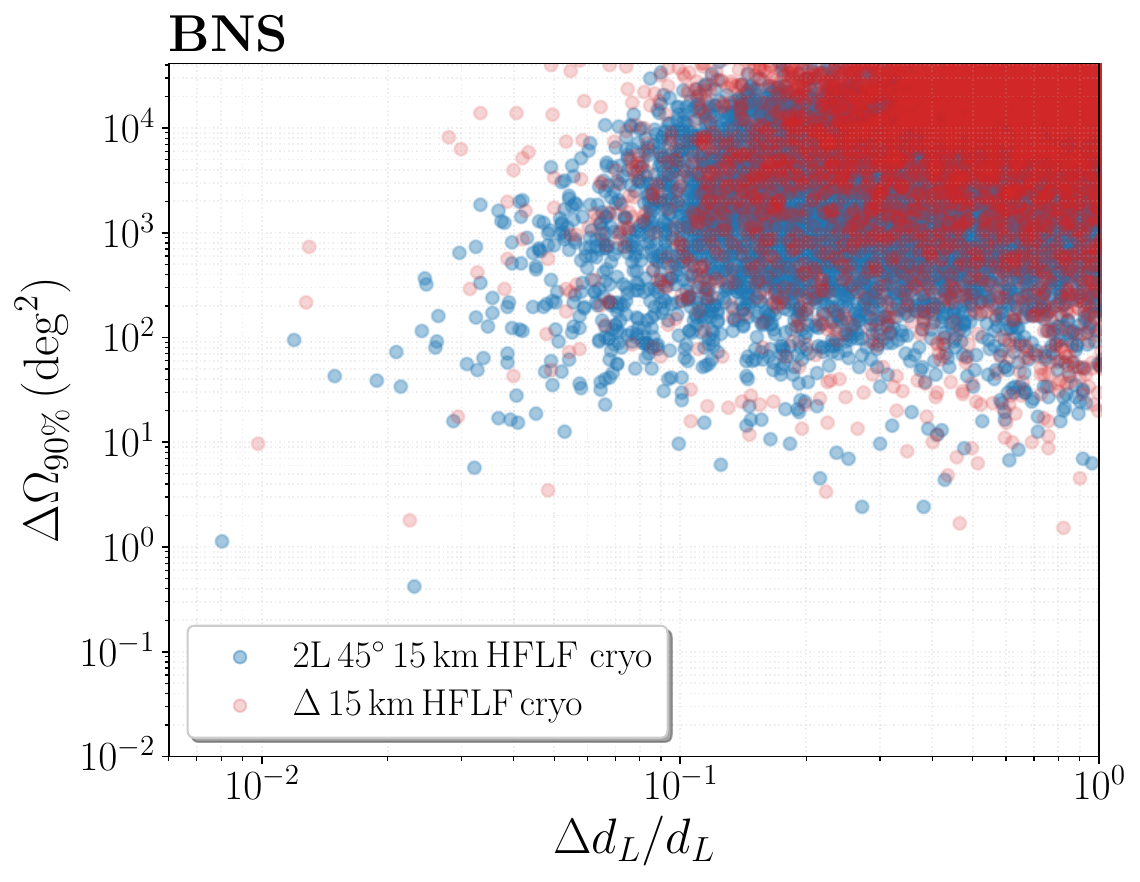}
    \caption{\small The joint accuracy on luminosity distance and angular resolution for  BNSs. Left panel:  the 15~km 2L at $45^{\circ}$ (blue) compared to the 10~km triangle (green). Right panel:
    the 15~km 2L at $45^{\circ}$ (blue) compared to the 15~km triangle  (red).}
    \label{fig:scatter_dLOm_2L4515kmvsT10and15km_BNS}
\end{figure}


\begin{figure}[t]
    \centering
    \includegraphics[width=1.\textwidth]{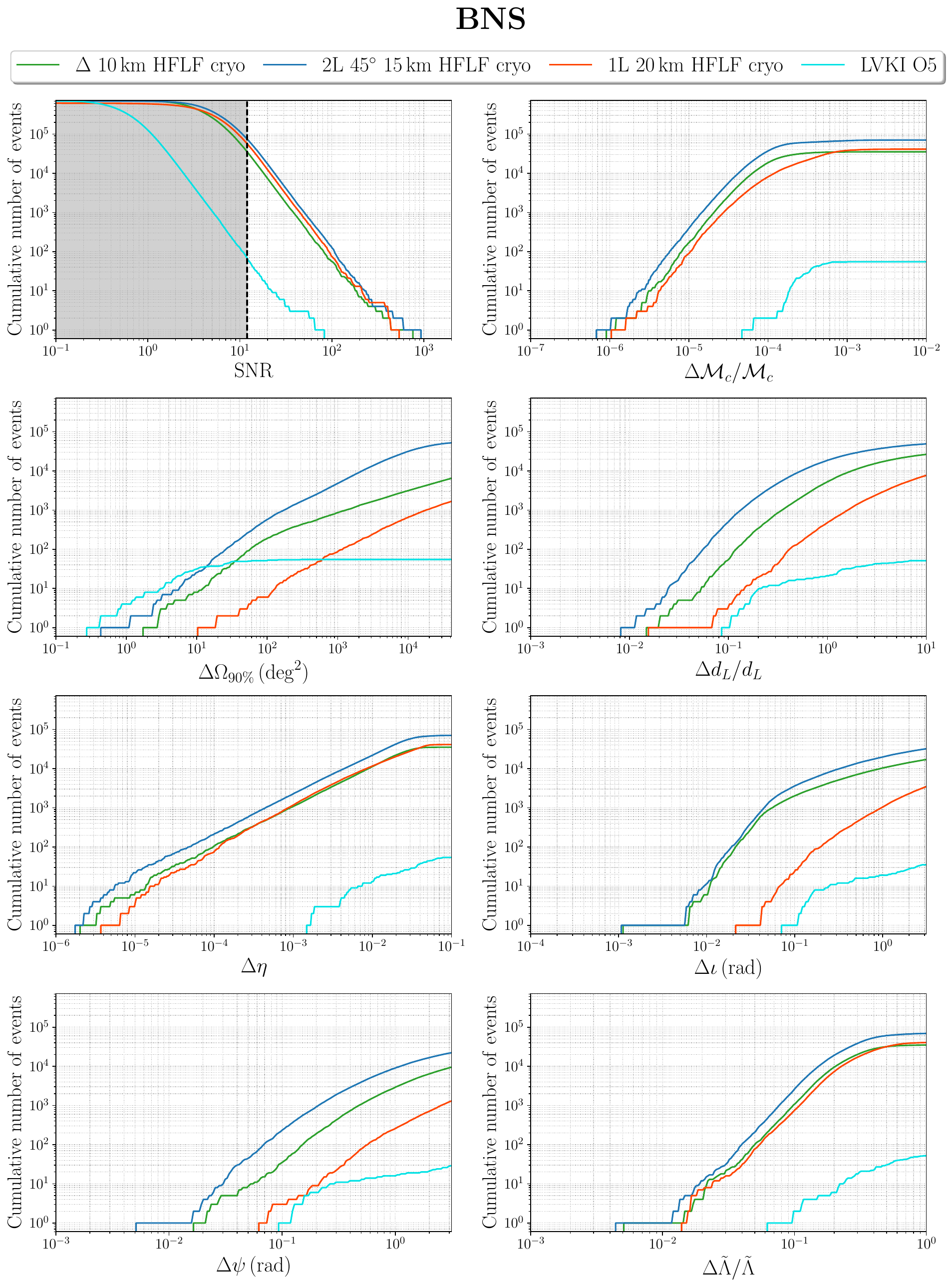}
    \caption{\small Comparison of SNR and parameter estimation error for
     a single 20~km L-shaped detector with the HFLF-cryo ASD, compared to the 10~km triangle, a  2L with 15~km arms at $45^{\circ}$, and the forecast for LVKI~O5.}
    \label{fig:ETMR_1L_20km_CumulBNS_NdetScale}
\end{figure}

\begin{figure}[t]
    \centering
    \includegraphics[width=1.\textwidth]{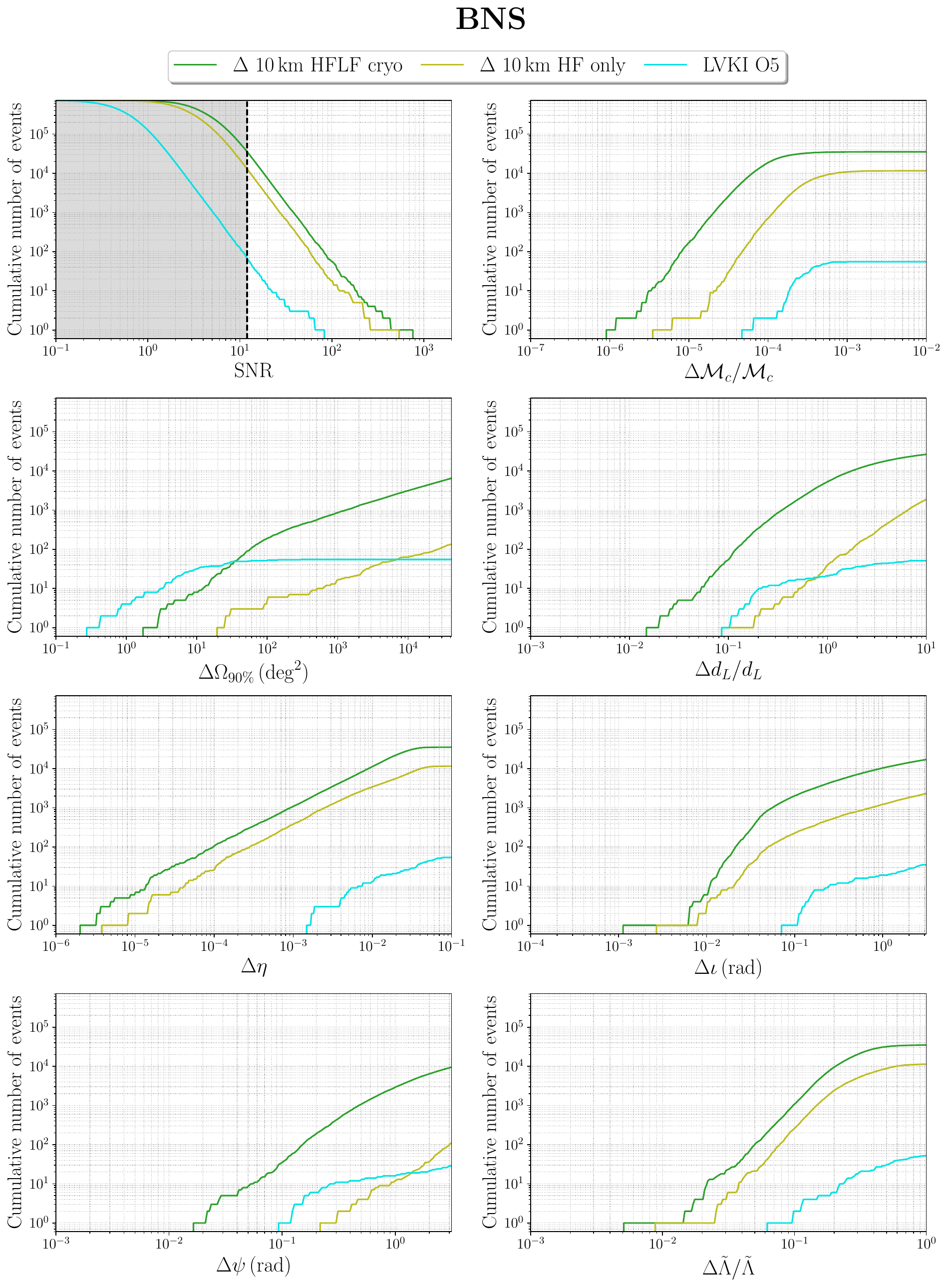}
    \caption{\small Comparison of SNR and parameter estimation error for the 10~km triangle with different ASDs. By comparison, we also show the forecast for LVKI~O5.}
    \label{fig:ETS_T_10km_CumulBNS_NdetScale}
\end{figure}

\begin{figure}[t]
    \centering
    \includegraphics[width=1.\textwidth]{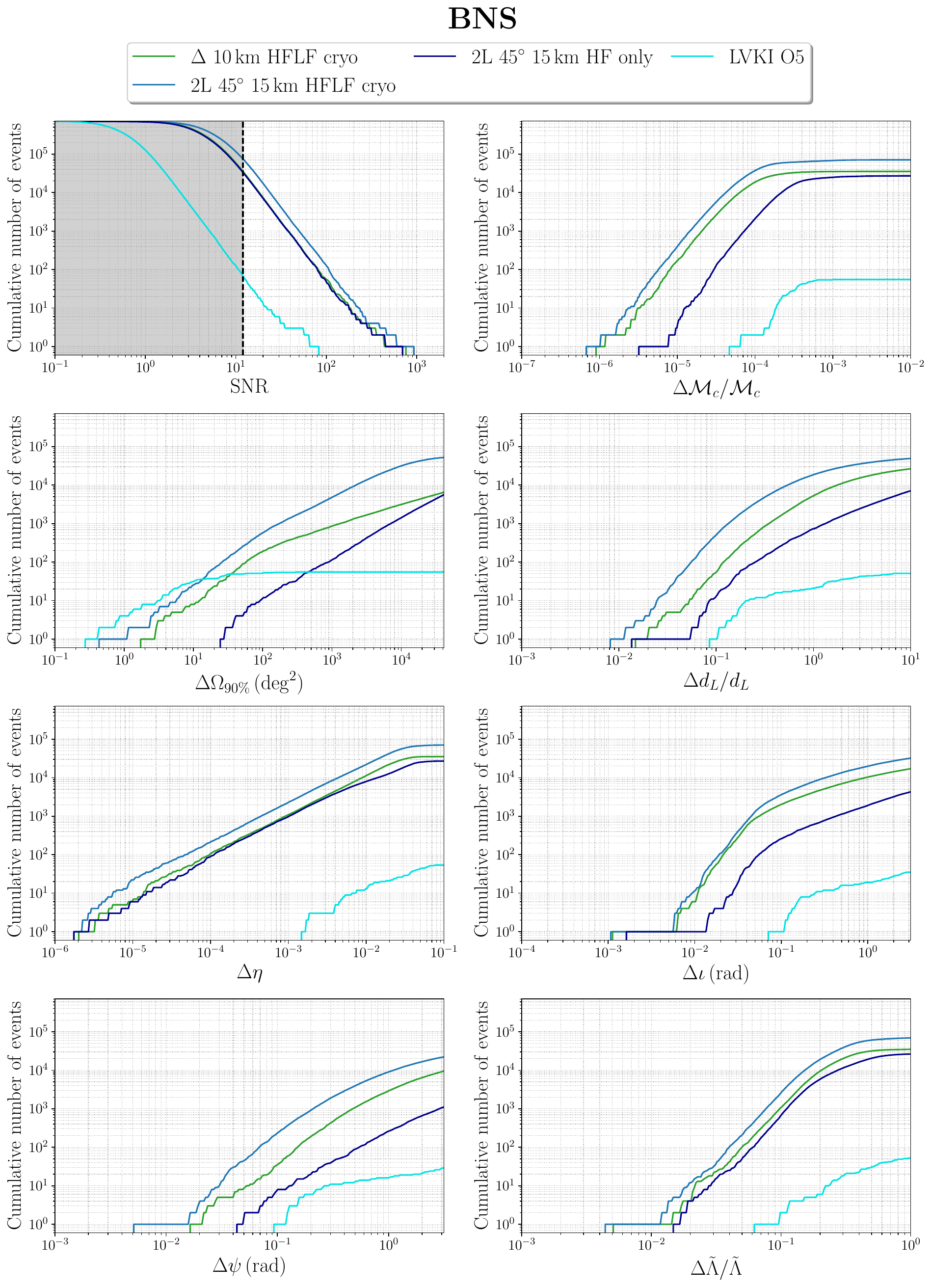}
    \caption{\small Comparison of SNR and parameter estimation error for the  2L with 15~km arms at $45^{\circ}$, with the two different ASDs. For comparison, we also show the 10~km triangle at full sensitivity, and the forecast for LVKI~O5.}
    \label{fig:ETSMR_2L4515_CumulBNS_NdetScale}
\end{figure}

\begin{figure}[t]
\hspace{-1.3cm}
\begin{tabular}{l@{\hskip -.02cm}l@{\hskip -.02cm}l}
     \includegraphics[width=5.7cm]{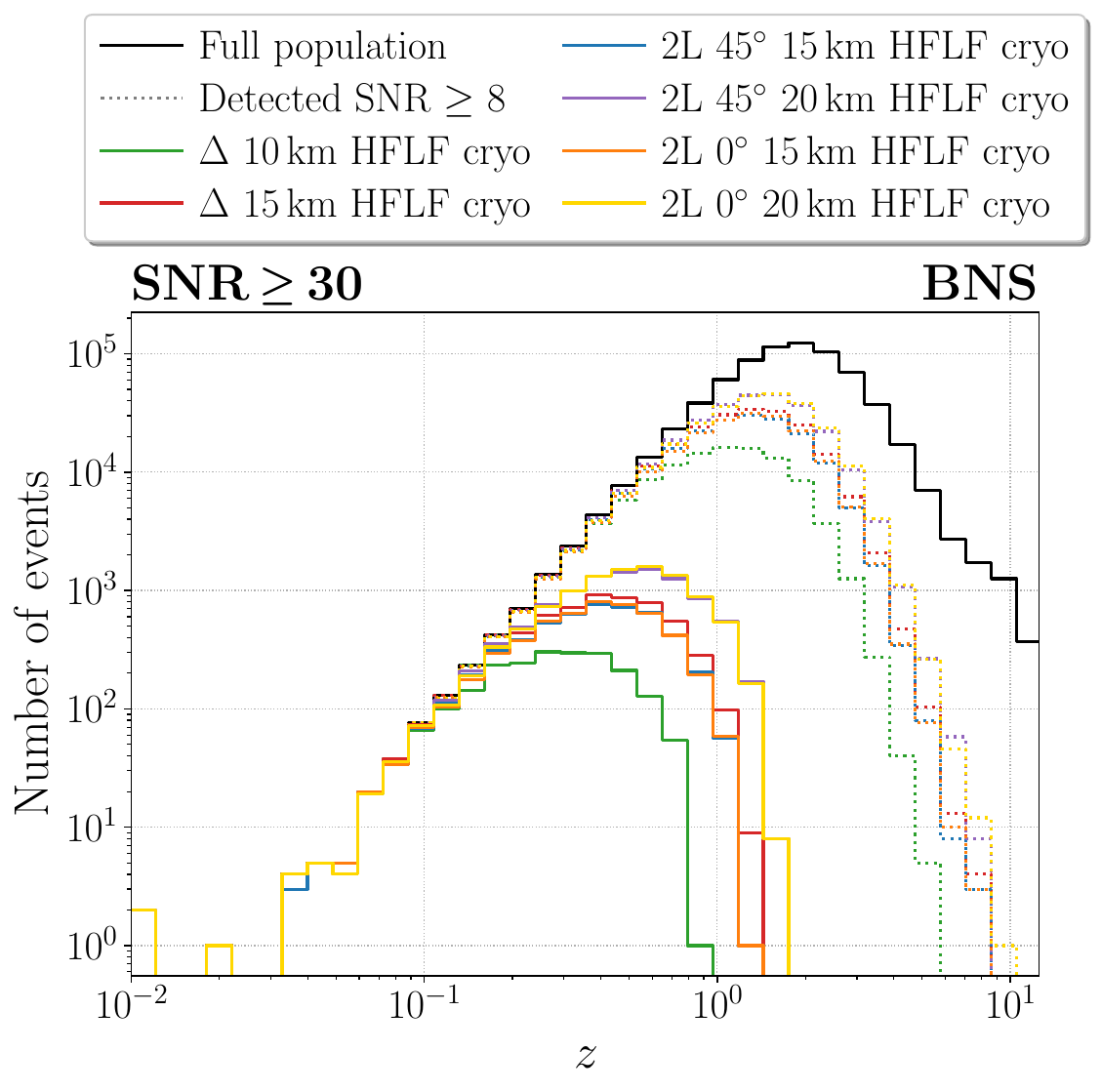} & 
     \includegraphics[width=5.7cm]{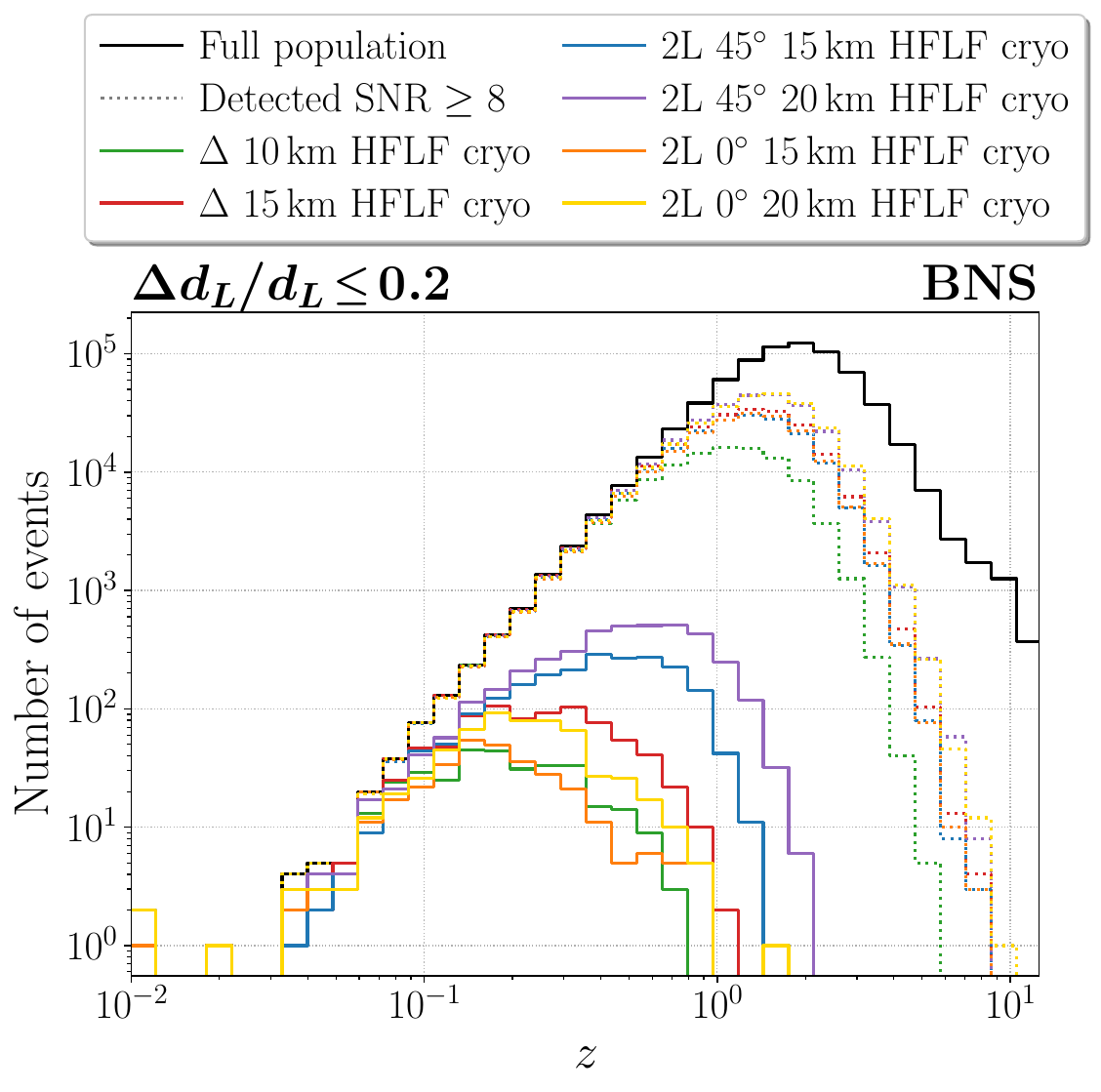} & \includegraphics[width=5.7cm]{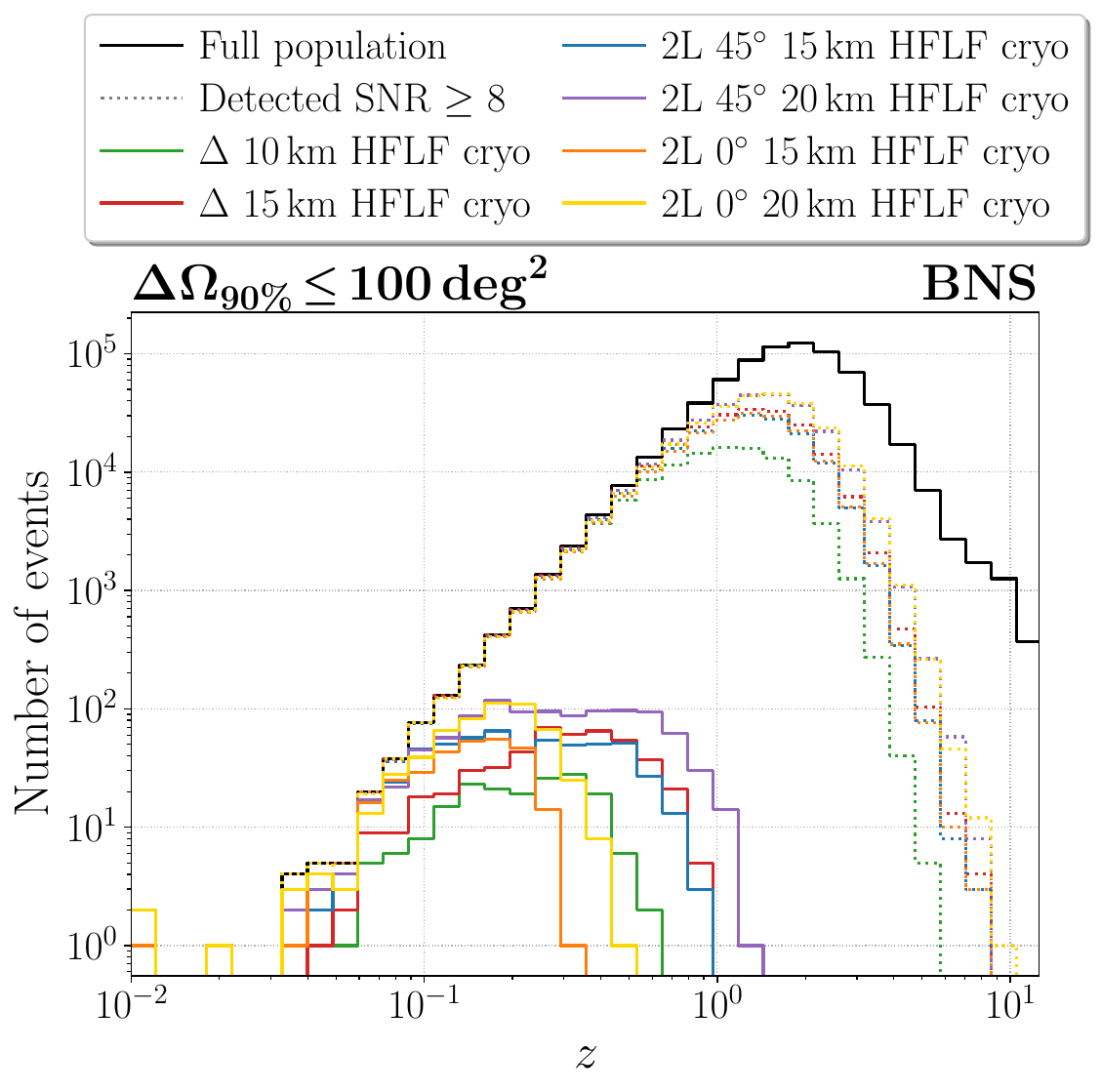} \\
     \includegraphics[width=5.44cm]{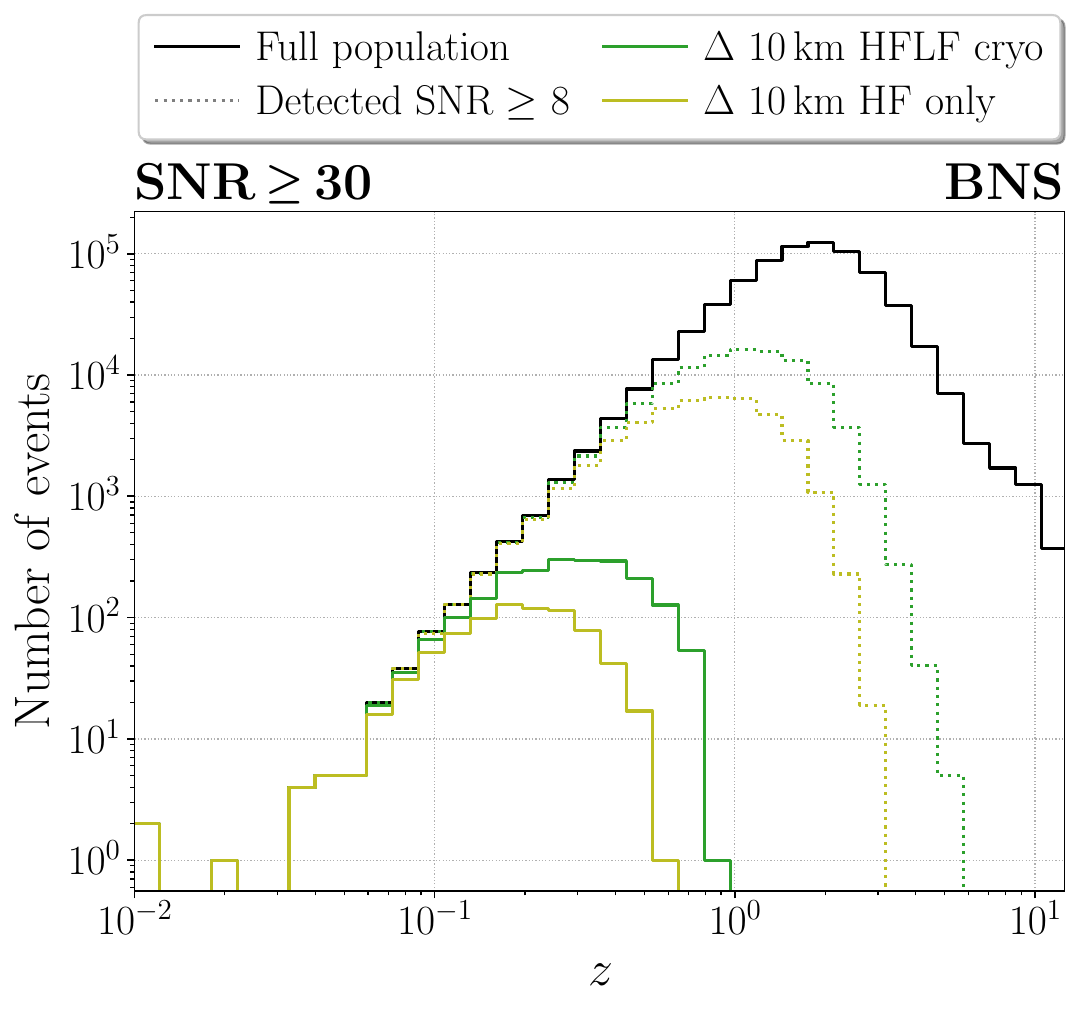} & 
     \includegraphics[width=5.44cm]{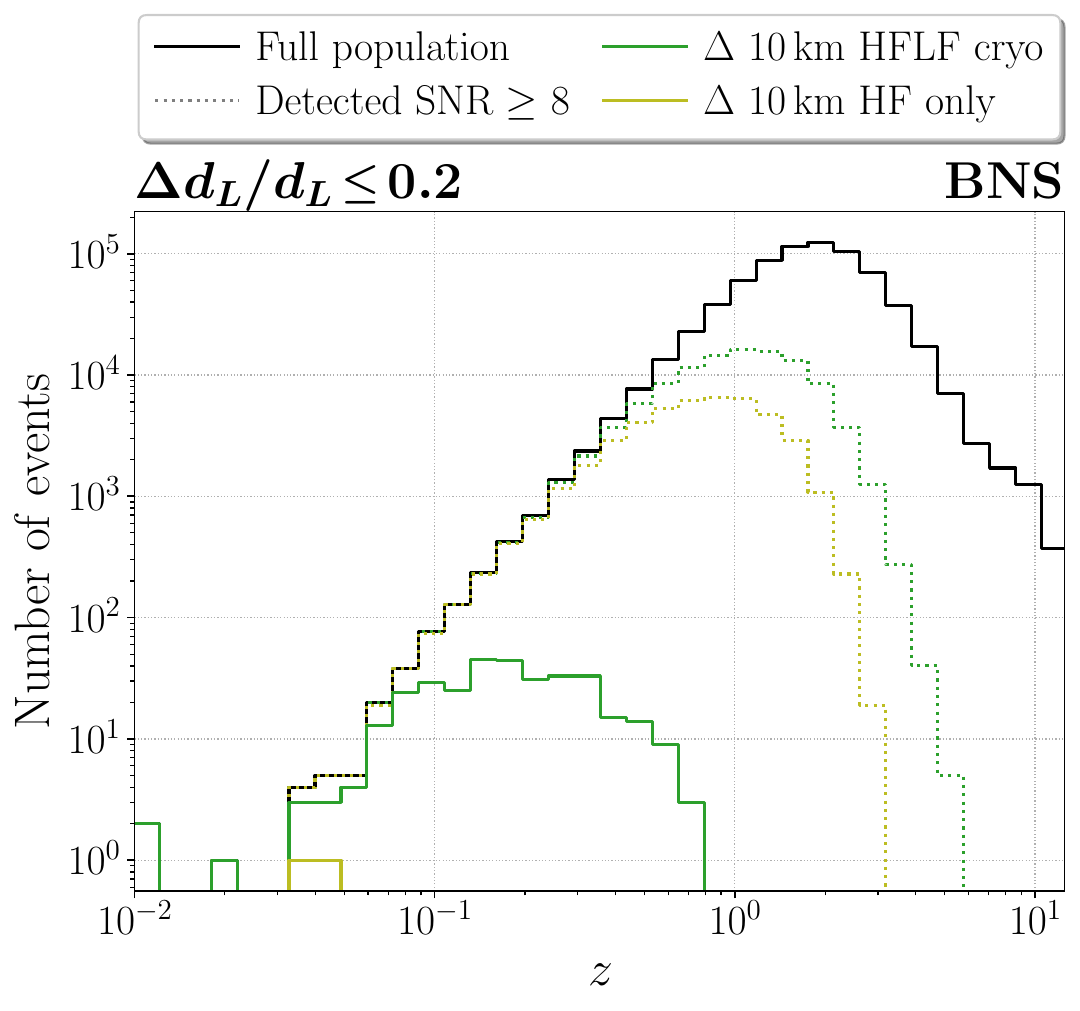} & \includegraphics[width=5.44cm]{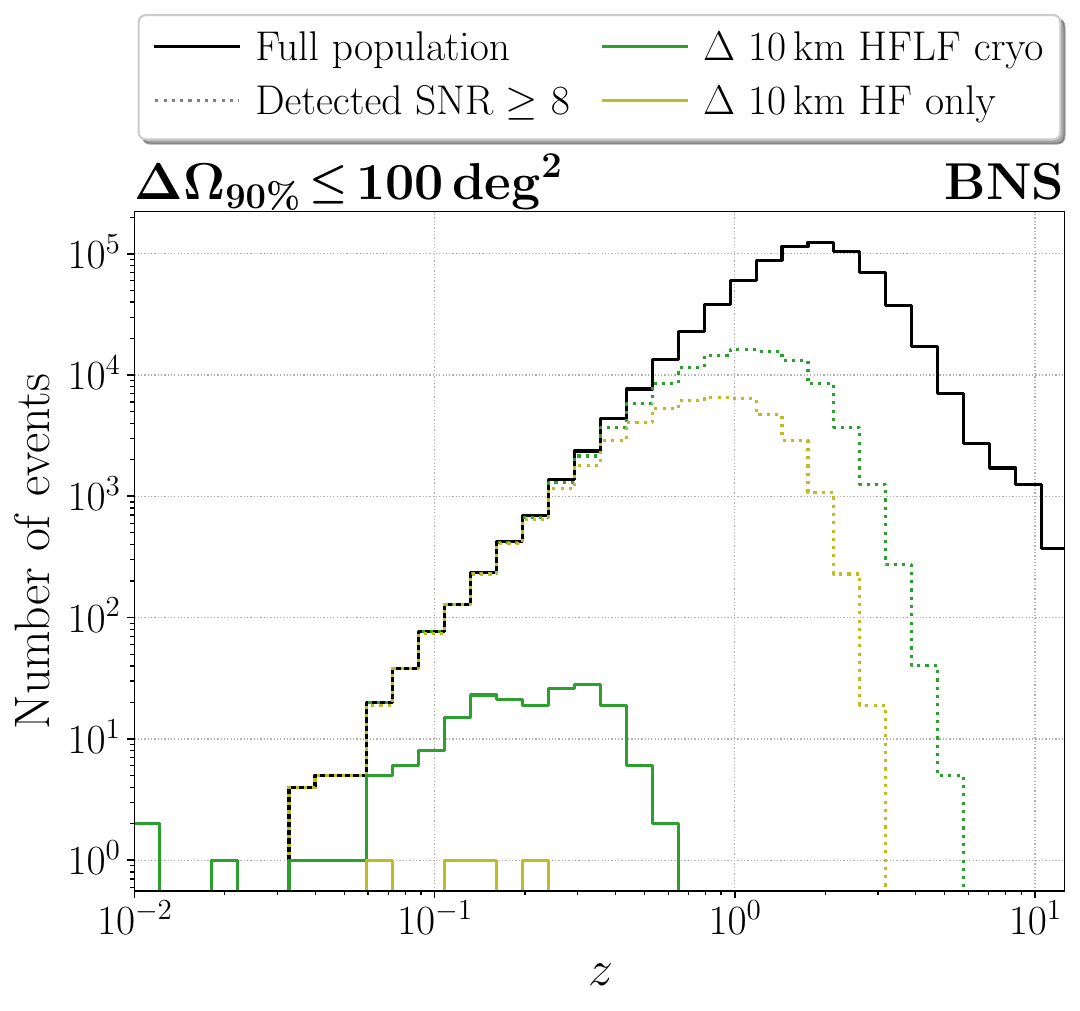} \\
     \includegraphics[width=5.65cm]{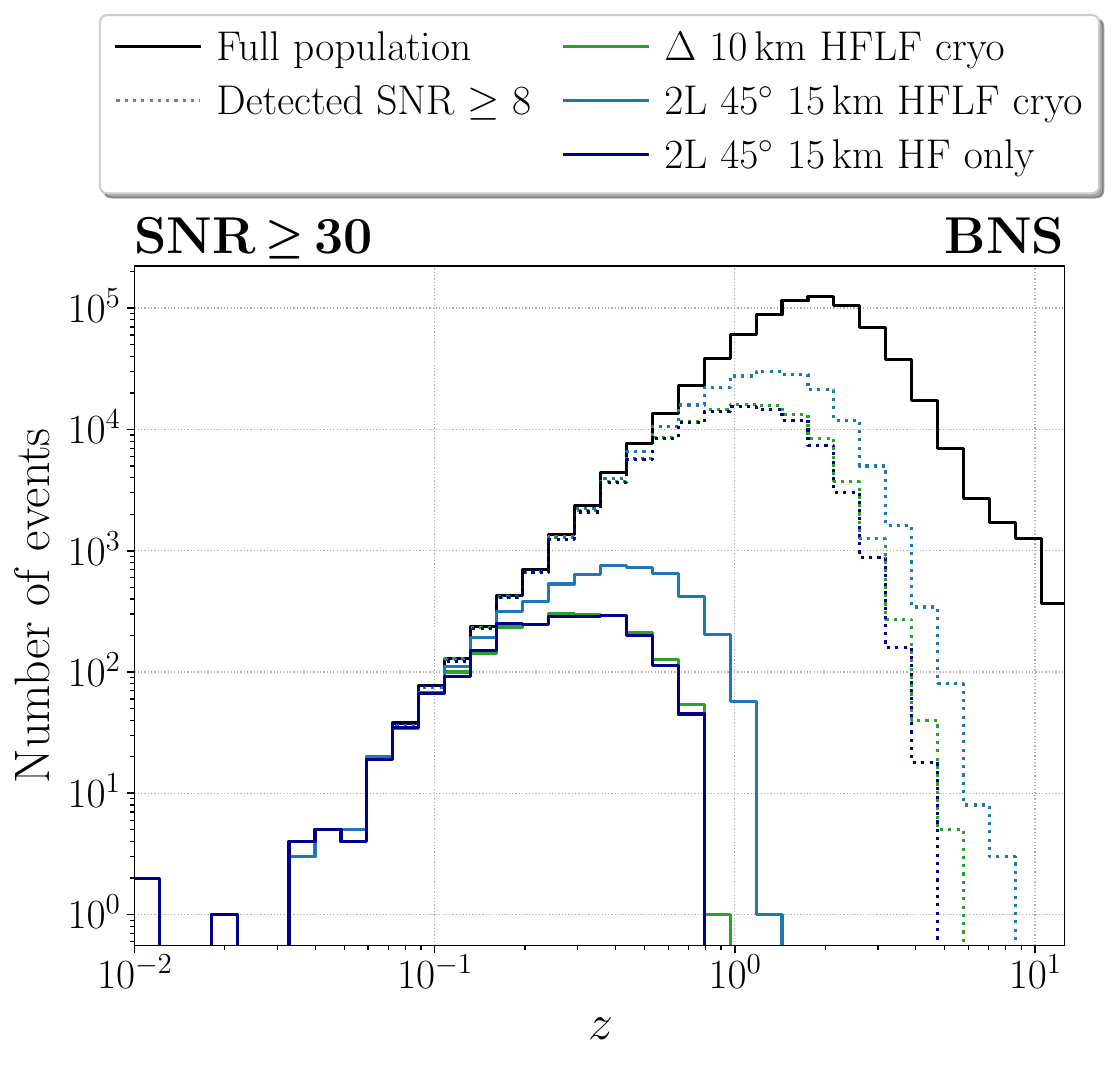} & 
     \includegraphics[width=5.65cm]{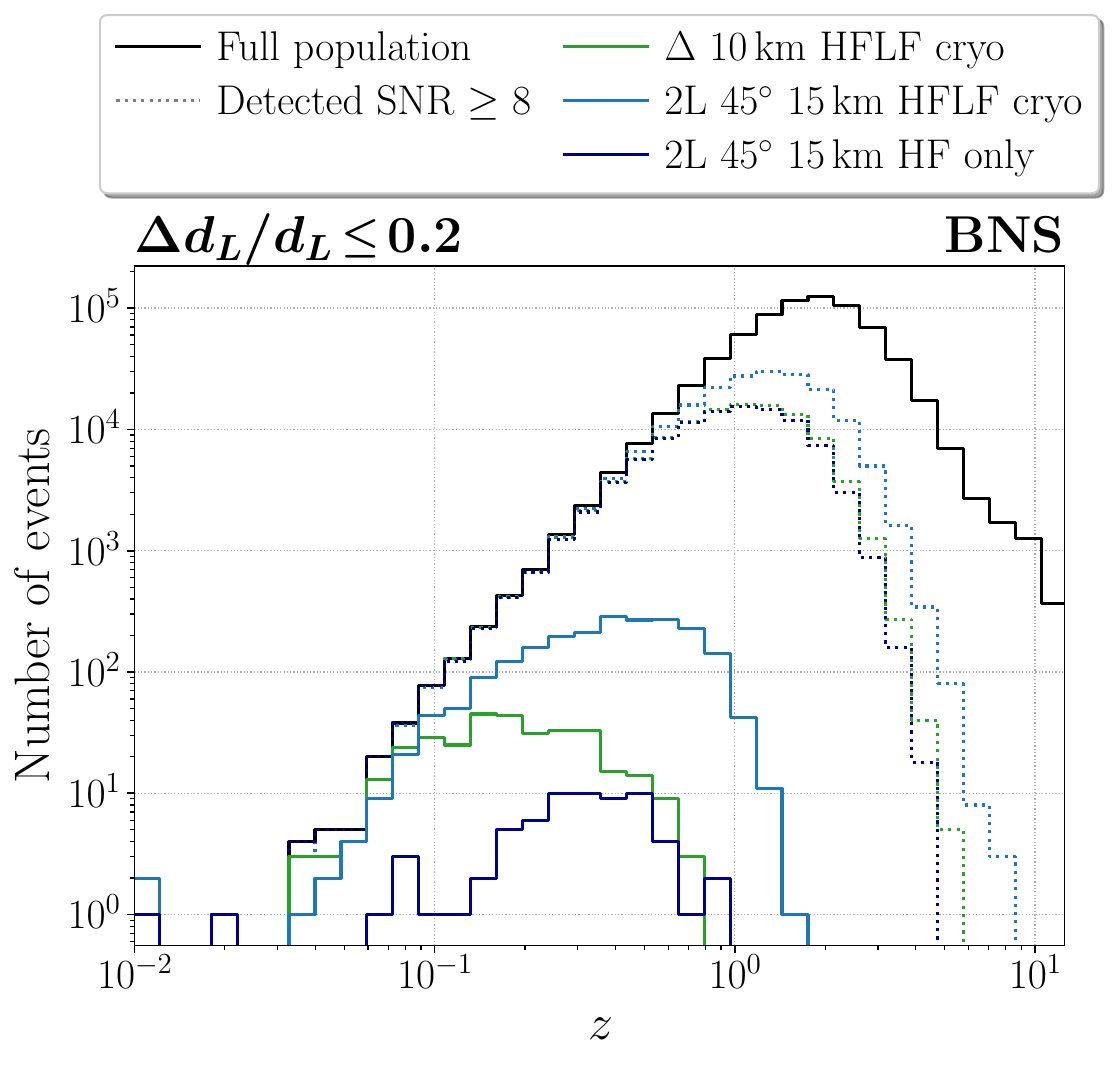} & \includegraphics[width=5.65cm]{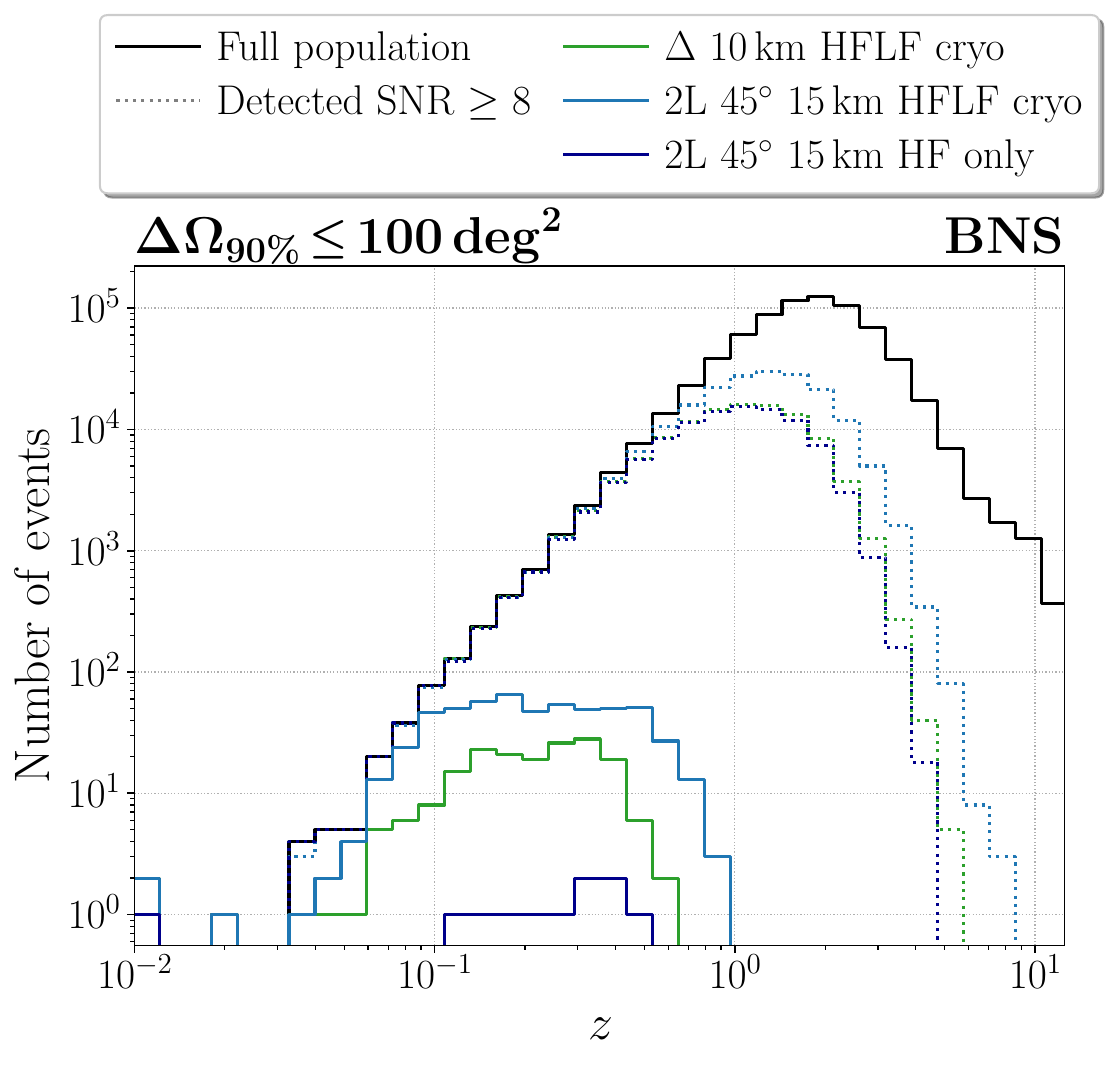}
\end{tabular}
    \caption{\small Redshift distribution of BNS detected  with ${\rm SNR}\,\geq\,30$ (left column), or relative error on the luminosity distance $\Delta d_L/d_L\,\leq\,0.2$ (central column), or with sky location $\Delta\Omega\,\leq\,100~{\rm deg}^2$ (right column) for various detector configurations and sensitivity curves.  The upper  row shows the results for the six considered geometries, all with their best sensitivity, the central row for the 10~km triangle with the two considered ASDs, and the bottom row for the 2L with 15~km arms at $45^{\circ}$ and the two considered ASDs. In each panel, we also show, for reference, the redshift distribution of the  BNS population  used (black solid line) and the events detected in the various configurations with ${\rm SNR}\,\geq\,8$  (dotted lines).}
    \label{fig:ET_allgeom_ASD_BNS_distr_vs_z}
\end{figure}

\begin{figure}[t]
    \centering
    \includegraphics[width=1.\textwidth]{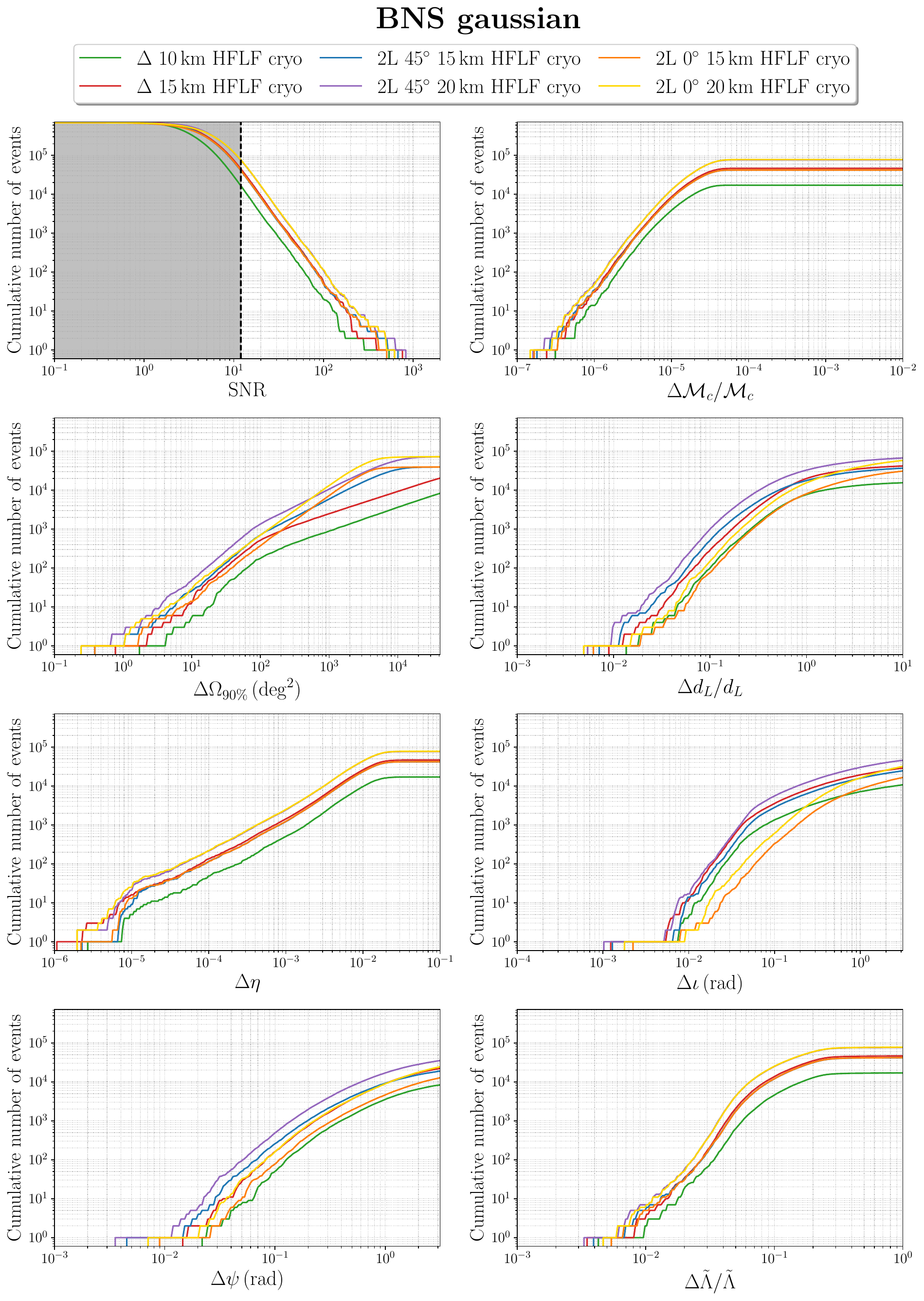}
    \caption{\small As in Fig.~\ref{fig:AllGeoms_CumulBNS_NdetScale}, for a gaussian mass distribution of the two component neutron stars.}
    \label{fig:AllGeoms_CumulBNS_NdetScale_massGauss}
\end{figure}

\clearpage\newpage

\subsection{ET in a network of 3G detectors}\label{sect:3Gnetwork}

The above analysis studied the capabilities of ET alone (whether consisting of a single-site triangle or of two L-shaped detectors on widely separated sites), comparing the results obtained with different possible geometries and sensitivity curves.
We now study how the different geometries for ET (triangle or 2L) perform when they are included in a network of 3G detectors which includes Cosmic Explorer. The reference CE configuration currently 
consists of two L-shaped detectors (on the Earth's surface rather than underground), one with 40~km arms and one with 20~km arms \cite{Evans:2021gyd}. We will refer to this configuration as `2CE' and we will study how  different choices for the ET geometry affect the performance of a network made by  ET and  2CE. Given the current uncertainties, it is also instructive to investigate what happens when ET (in a triangular or 2L geometry)  is part of a network with a single CE detector, which we take to be the 40~km one (`1CE'). In order to avoid a proliferation of plots, we limit here to three options for ET: the 10~km triangle, and the 15~km 2L configurations, either with parallel arms or with arms at $45^{\circ}$. Therefore, we consider the following networks:

\begin{enumerate}

\item ET as a triangle with 10~km arms, together with 2CE (20+40 km)
\item ET as a 2L of 15~km  and parallel arms,
together with 2CE (20+40 km)
\item ET as a 2L of 15~km and arms at $45^{\circ}$, 
together with 2CE (20+40 km)
\item ET as a triangle with 10~km arms,  together with 1CE (40 km)
\item ET as a 2L of 15~km  and parallel arms,
together with 1CE (40 km)
\item  ET as a 2L of 15~km and arms at $45^{\circ}$, 
together with 1CE (40 km)

\end{enumerate}

In particular, the last two options correspond to a global network of three L-shaped detectors, two of 15~km and with the ET  ASD, and one of 40~km and the CE ASD.
In all these configurations, for ET we will consider only the full sensitivity, corresponding to the ASD that we have labelled as `HFLF-cryo' and, for CE, we  use the latest publicly available official sensitivity curves.\footnote{The most recent PSDs of CE
can be found at \url{https://dcc.cosmicexplorer.org/CE-T2000017/public}.
We used the locations and orientations reported in rows C and N of Table III of \cite{Borhanian:2020ypi}. In particular, the 40~km detector is placed in Idaho and the 20~km one in New Mexico. These orientations are such that the relative angle between the two CE detectors, measured with respect to the great circle that joins them, is about $56.0^{\circ}$. When ET is taken to be in the 2L configuration, with our choice of orientations the relative angle between the Sardinia detector and the CE-Idaho is about $191.2^{\circ}$ and that between the Sardinia detector and the CE-New Mexico one is about $237.6^{\circ}$. As in the discussion in footnote~\ref{foot:angles}, if a 2L configuration should be retained for ET, a careful study of the optimization of the relative orientations of the ET+2CE detector network will become necessary.}

\begin{figure}[t]
    \centering
    \includegraphics[width=1.\textwidth]{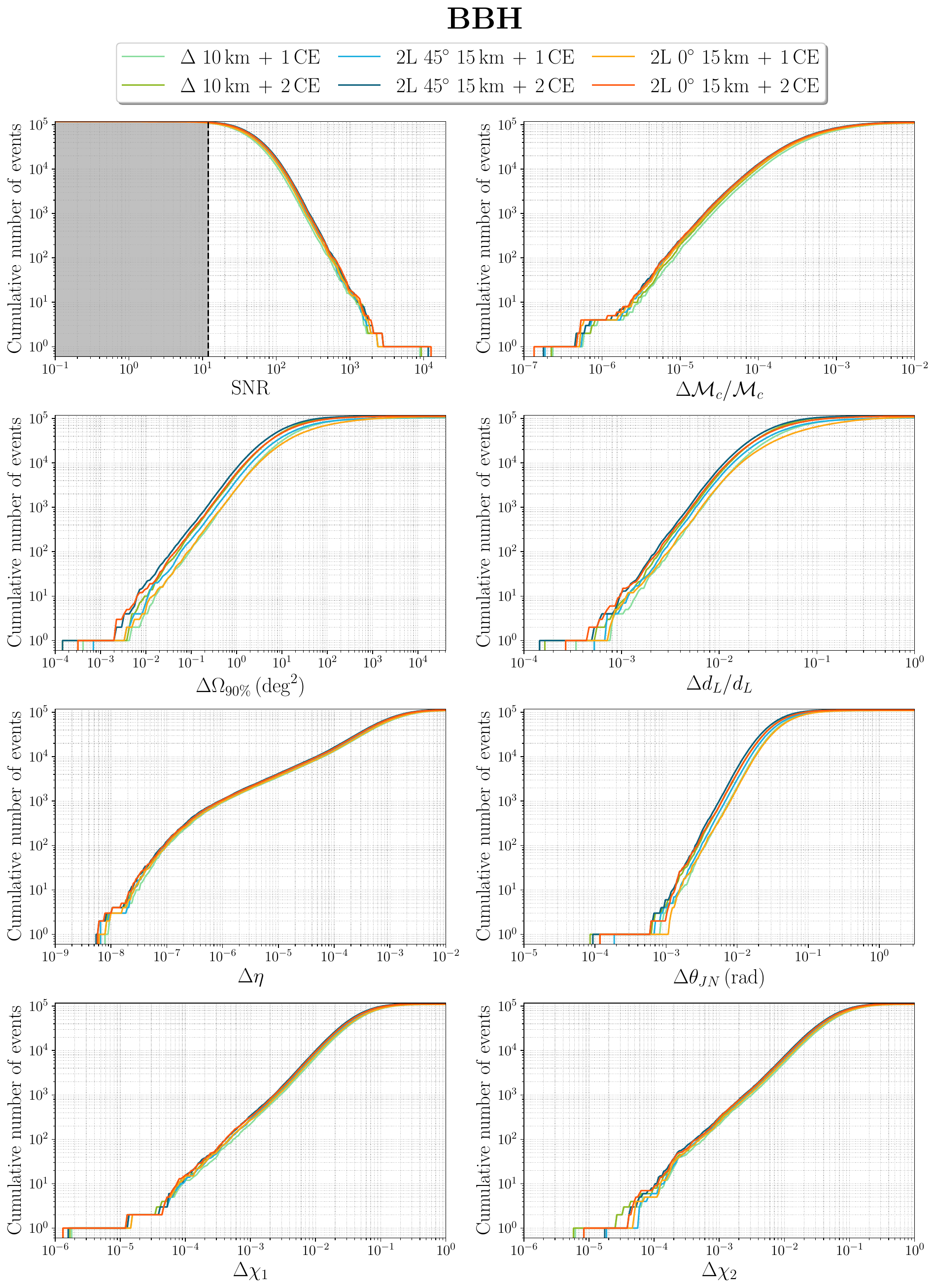}
    \caption{\small Comparison of SNR and parameter estimation error for BBHs, for the six 3G  detector networks considered.}
    \label{fig:ETCE_CumulBBH_NdetScale}
\end{figure}

\begin{figure}[t]
    \centering
    \includegraphics[width=1.\textwidth]{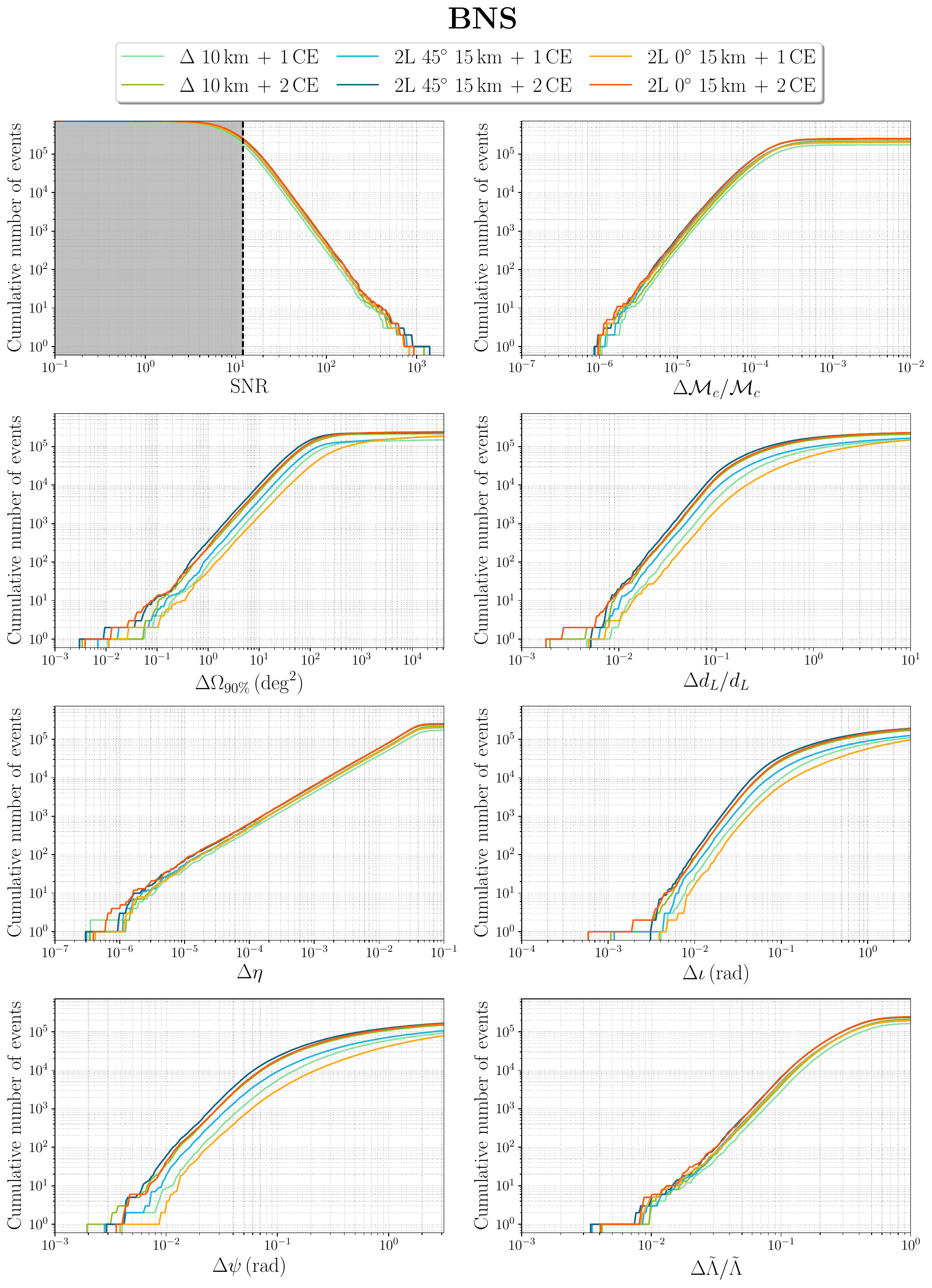}
    \caption{\small Comparison of SNR and parameter estimation error for BNSs, for the six 3G  detector networks considered.}
    \label{fig:ETCE_CumulBNS_NdetScale}
\end{figure}

Fig.~\ref{fig:ETCE_CumulBBH_NdetScale}  
shows the SNR distribution and the errors on parameter estimation for BBHs for the six networks considered, while Fig.~\ref{fig:ETCE_CumulBNS_NdetScale} shows the corresponding results for BNSs. 
From Fig.~\ref{fig:ETCE_CumulBBH_NdetScale}, to be compared with the green, blue and orange lines in
Fig.~\ref{fig:AllGeoms_CumulBBH_NdetScale} referring to ET alone, we see that, as expected, the overall performance of the network is significantly better than that of ET alone for angular localization and distance determination. As an example, the BBH/yr localized to better than $1\, {\rm deg}^2$ raise from a few tens (depending on the ET geometry) to a few times $10^3$.
On the other parameters, the gain from adding one or two CE detectors results typically in an increase by a factor of order 2 of the number of events with accuracy better than a given value of the chosen parameter; e.g., in our specific realization, for the symmetric mass ratio $\eta$, the number of events per year with $\Delta\eta\leq 10^{-6}$ raises from 559 for ET alone in the 10~km triangle configuration, to about 922 in a network with 1CE and 1018 with 2CE; similar results hold for the other parameters.

As we see from Tables~\ref{tab:BBHAllConfSNR}, \ref{tab:BBHAllConfDeldLDelOm}  and \ref{tab:BBHAllConfDelMcDelchi} in App.~\ref{app:TablesCBC}
(factors of 2 are difficult to appreciate on the logarithmic scale of Fig.~\ref{fig:ETCE_CumulBBH_NdetScale}),
in a network with a 40km CE, for BBHs the different geometries chosen for ET  induce overall differences in the results at the level of factors $\sim 1.5$ or smaller. As an example, the number of BBH/yr with $d_L$ measured better than $1\%$ raises from 2901 for the 10~km triangle+1CE, to 4301 for 2L-15km-$45^{\circ}$+1CE, while the number of BBH/yr localized to better than $10\, {\rm deg}^2$ raises from $3.0\times 10^4$ to $3.6\times 10^4$.

\vspace{2mm}
\noindent
For BNS, the effect of the ET geometry in a network with 1 or 2CE  can be seen  from Fig.~\ref{fig:ETCE_CumulBNS_NdetScale}, and especially from Tables~\ref{tab:BNSAllConfSNR}, \ref{tab:BNSAllConfDeldLDelOm}
and \ref{tab:BNSAllConfDelMcDelLam}. For instance, the number of BNS/yr with ${\rm SNR} \geq 150$ raises from 87 for the 10~km triangle+1CE, to 116 for  2L-15km-$45^{\circ}$+1CE, while those with $d_L$ measured better than $10\%$ increases by about a factor of 2, from 
4100 for the 10~km triangle+1CE to 7949 for 2L-15km-$45^{\circ}$+1CE.
Similar differences appear 
for  the chirp mass, symmetric mass ratio and tidal deformability. For instance, the number of BNS/yr with tidal deformability measured to better than $10\%$ raises from 2669  for  the 10~km triangle+1CE to 4753 for  2L-15km-$45^{\circ}$+1CE, and from 3847  for  the 10~km triangle+2CE to 6319 for  2L-15km-$45^{\circ}$+2CE.

So,  among the three geometries considered here, the one that  gives consistently the best results 
 is, again, the 2L with 15~km arms at $45^{\circ}$, 
with typical improvement with respect to the 10~km triangle of about a factor of $\sim (1.5-2)$. The configuration 2L-15km-$0^{\circ}$, when in a network with 1CE, gives similar performances to 2L-15km-$45^{\circ}$ for masses or tidal deformability, but is  worse by a factor $\sim 2.5$ for angular localization and $\sim 4$ for luminosity distance. For instance, the number of BNS/yr with $d_L$ measured better than $10\%$ is 2079 for
2L-15km-$0^{\circ}$+1CE and 7949 for
2L-15km-$45^{\circ}$+1CE. In a network with 2CE the differences become smaller.

{\em The overall conclusion is that the difference  between the performances of the ET geometries remains visible, at the level of factors $\sim  (1.5-2)$ for BNSs, and somewhat smaller for BBHs, also when ET is in a network with 1 or 2~CE.}

\begin{figure}[t]
\hspace{-1.3cm}
\begin{tabular}{l@{\hskip .07cm}l@{\hskip .07cm}l}
     \includegraphics[width=5.6cm]{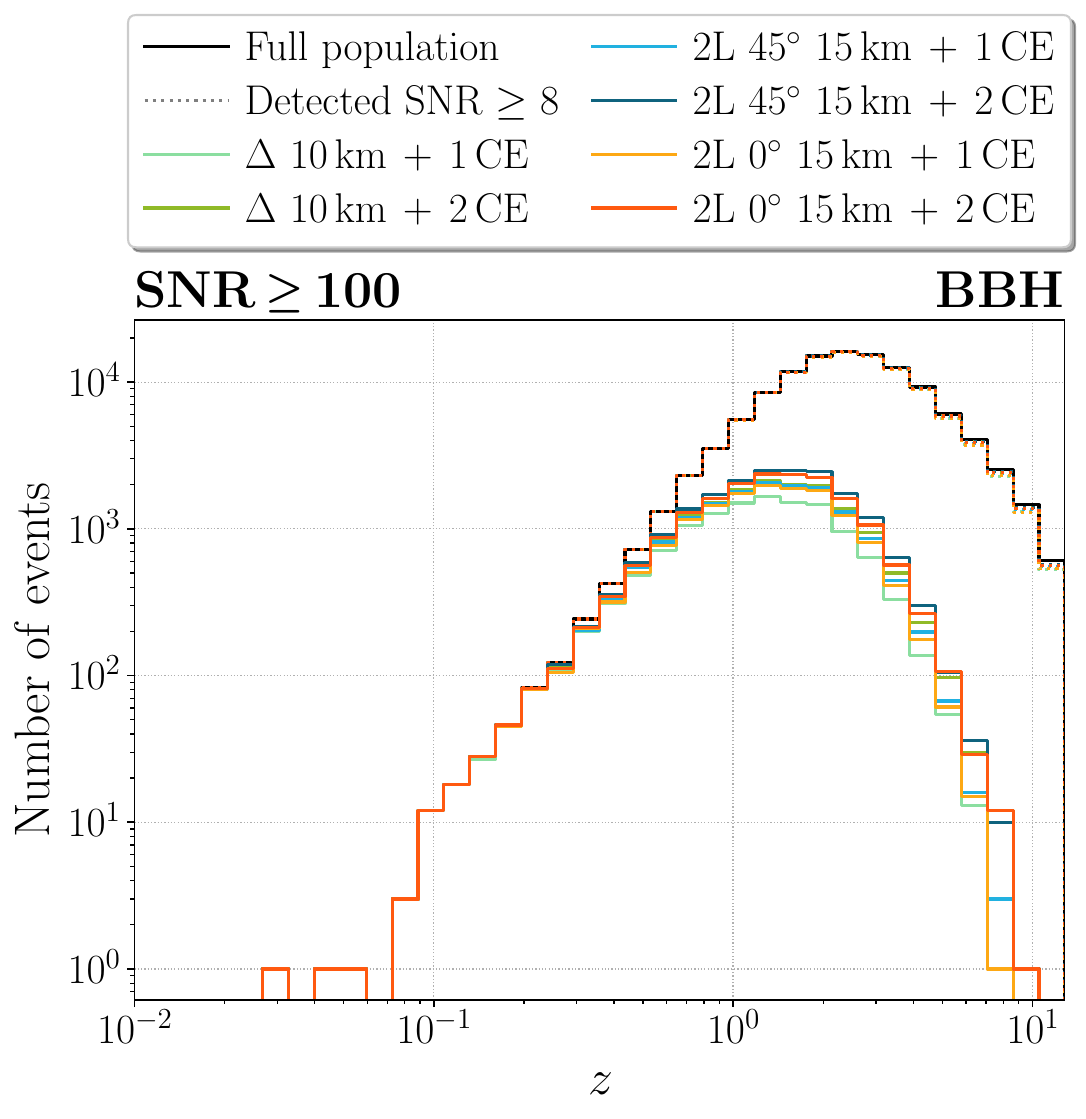} & 
     \includegraphics[width=5.6cm]{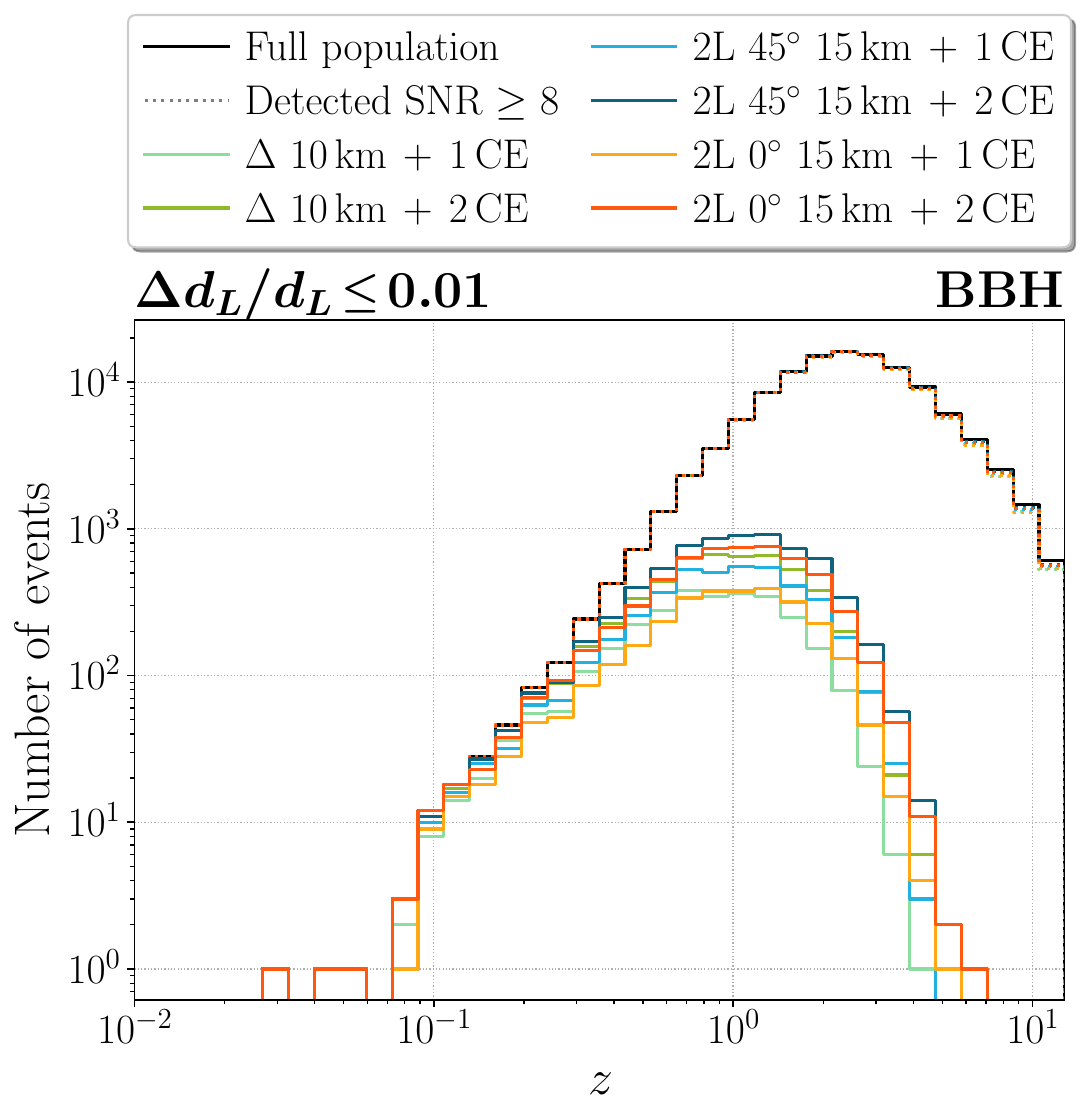} & \includegraphics[width=5.6cm]{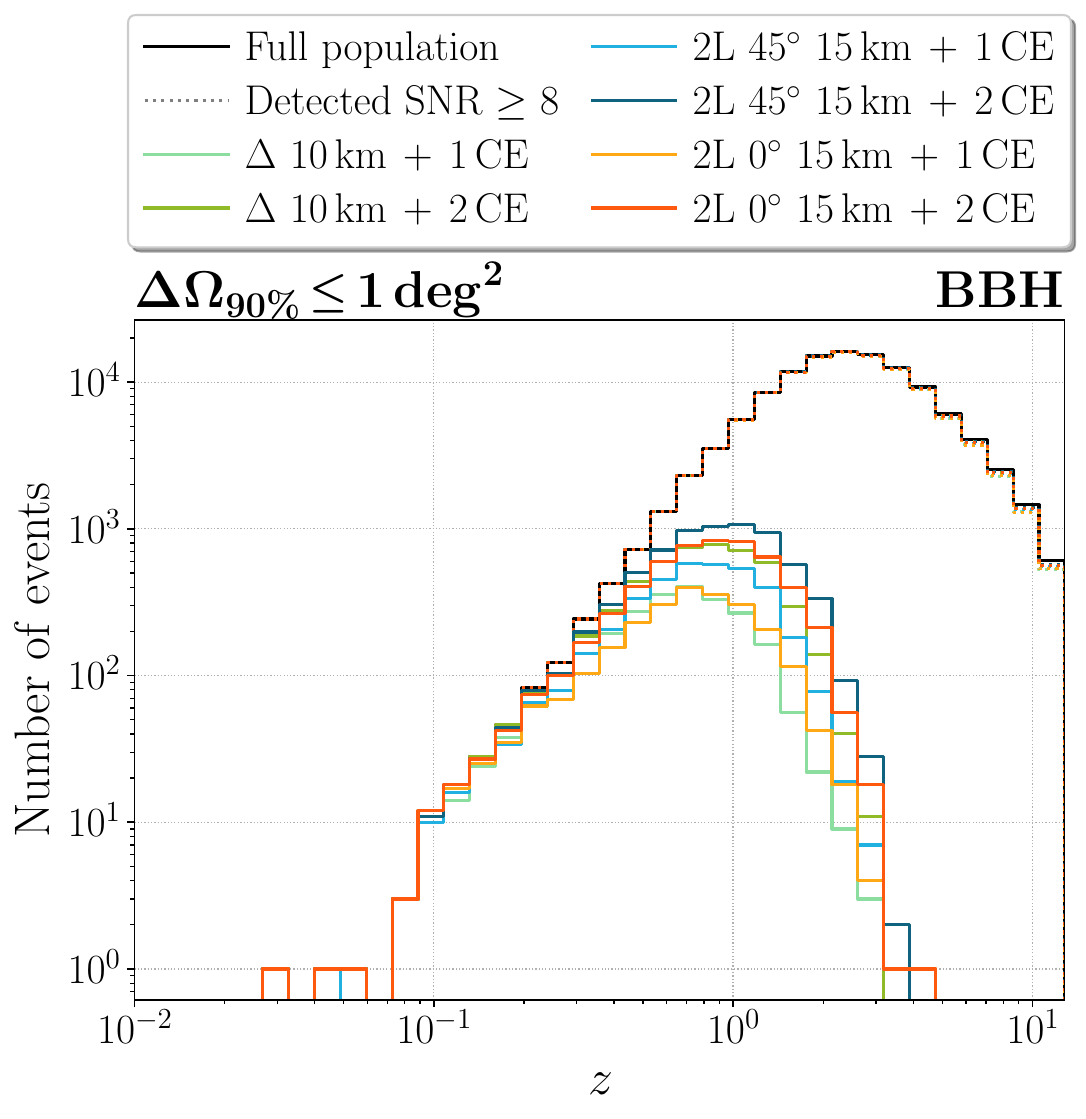}
\end{tabular}
    \caption{\small Redshift distribution of detected BBHs with ${\rm SNR}\,\geq\,100$ (left column), or relative error on the luminosity distance $\Delta d_L/d_L\,\leq\,0.01$ (central column), or sky location $\Delta\Omega\,\leq\,1~{\rm deg}^2$ (right column), for the six 3G the detector networks considered. In each panel, we also show, for reference, the redshift distribution of the  BBH population used (black solid line), and of the events detected in the various configurations with ${\rm SNR}\,\geq\,8$  (dotted lines, graphically almost indistinguishable from the full population on this scale).}
    \label{fig:ETCE_allgeom_BBH_distr_vs_z}
\end{figure}

\begin{figure}[t]
\hspace{-1.3cm}
\begin{tabular}{l@{\hskip .07cm}l@{\hskip .07cm}l}
     \includegraphics[width=5.6cm]{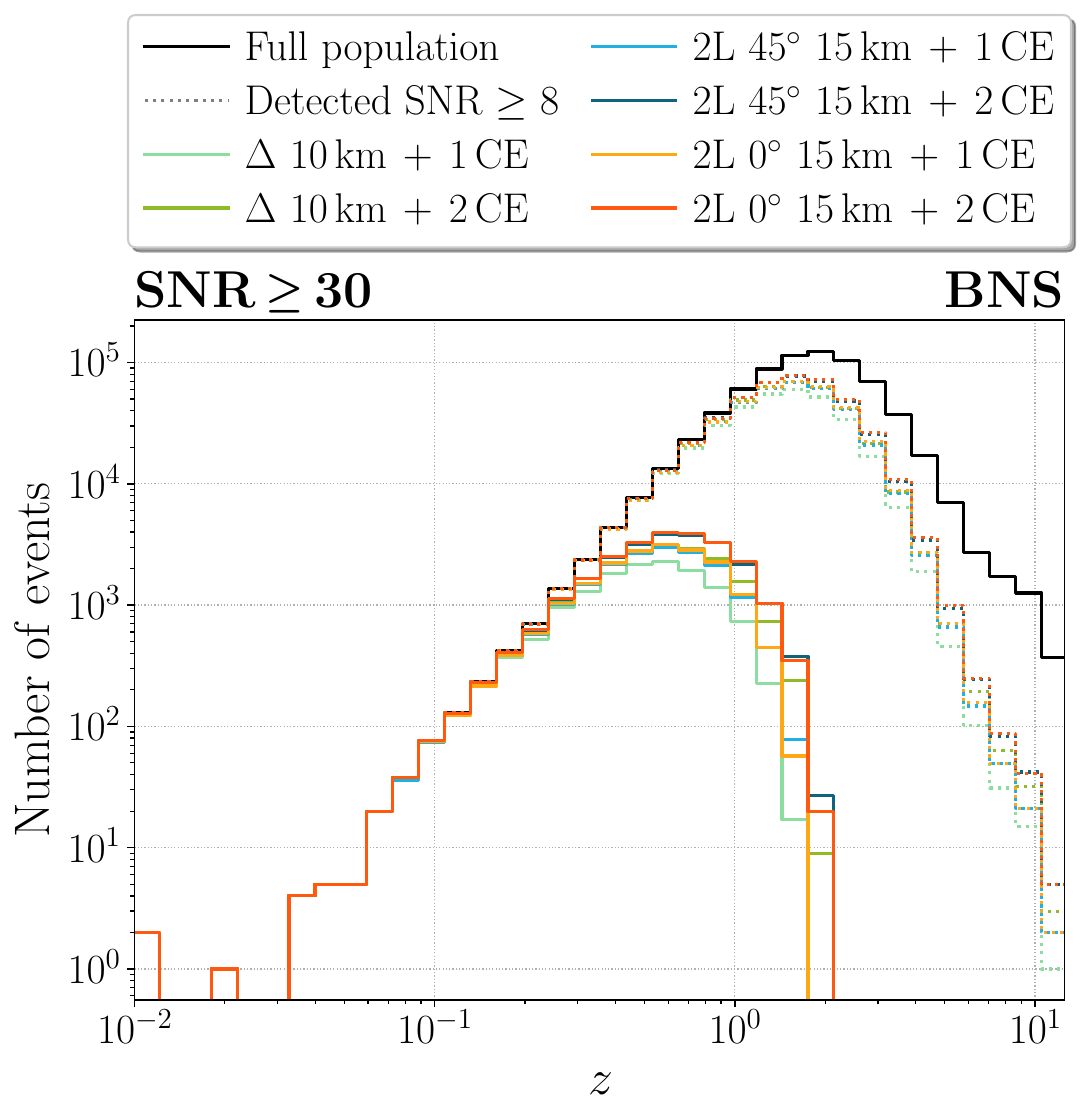} & 
     \includegraphics[width=5.6cm]{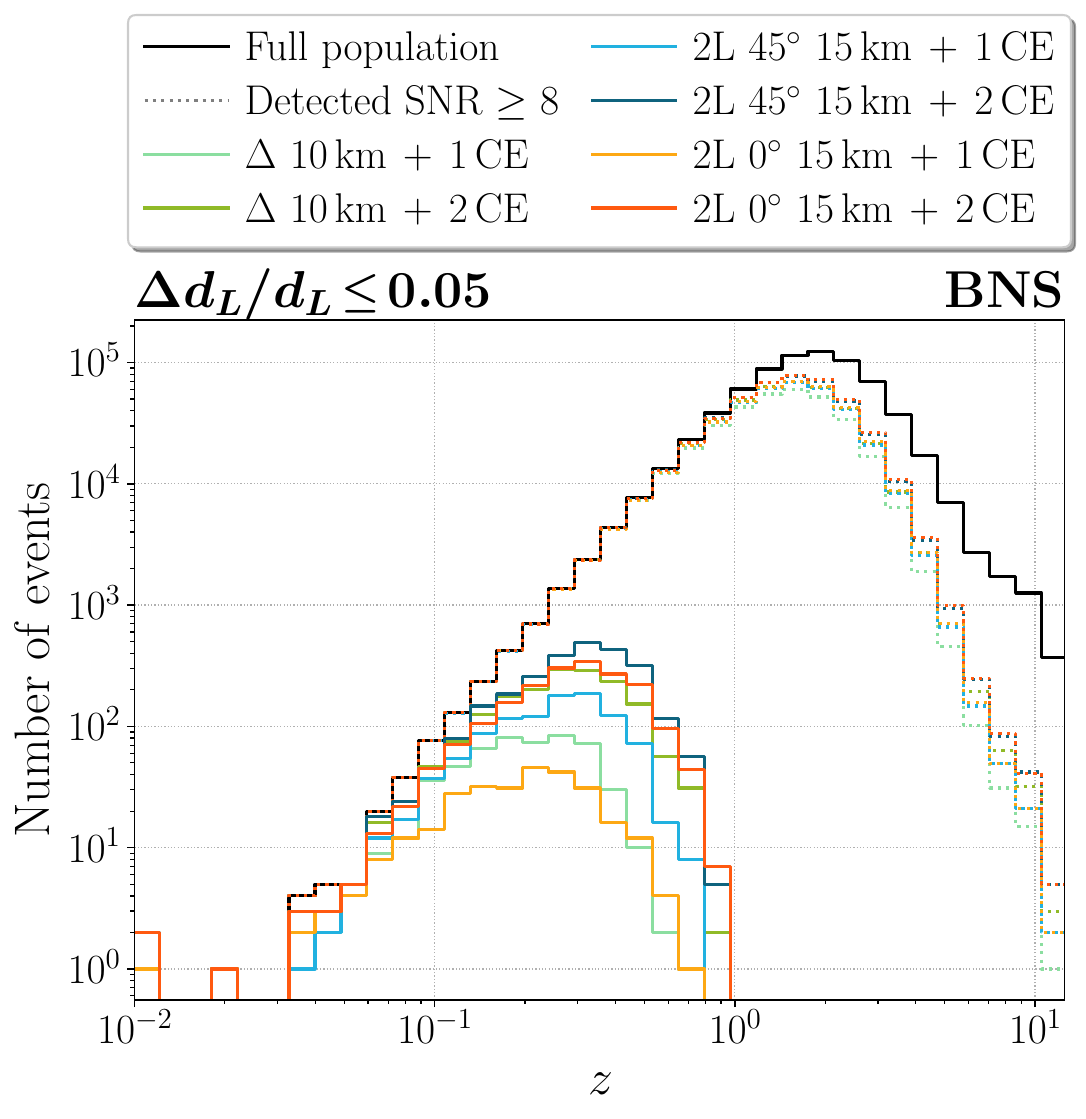} & \includegraphics[width=5.6cm]{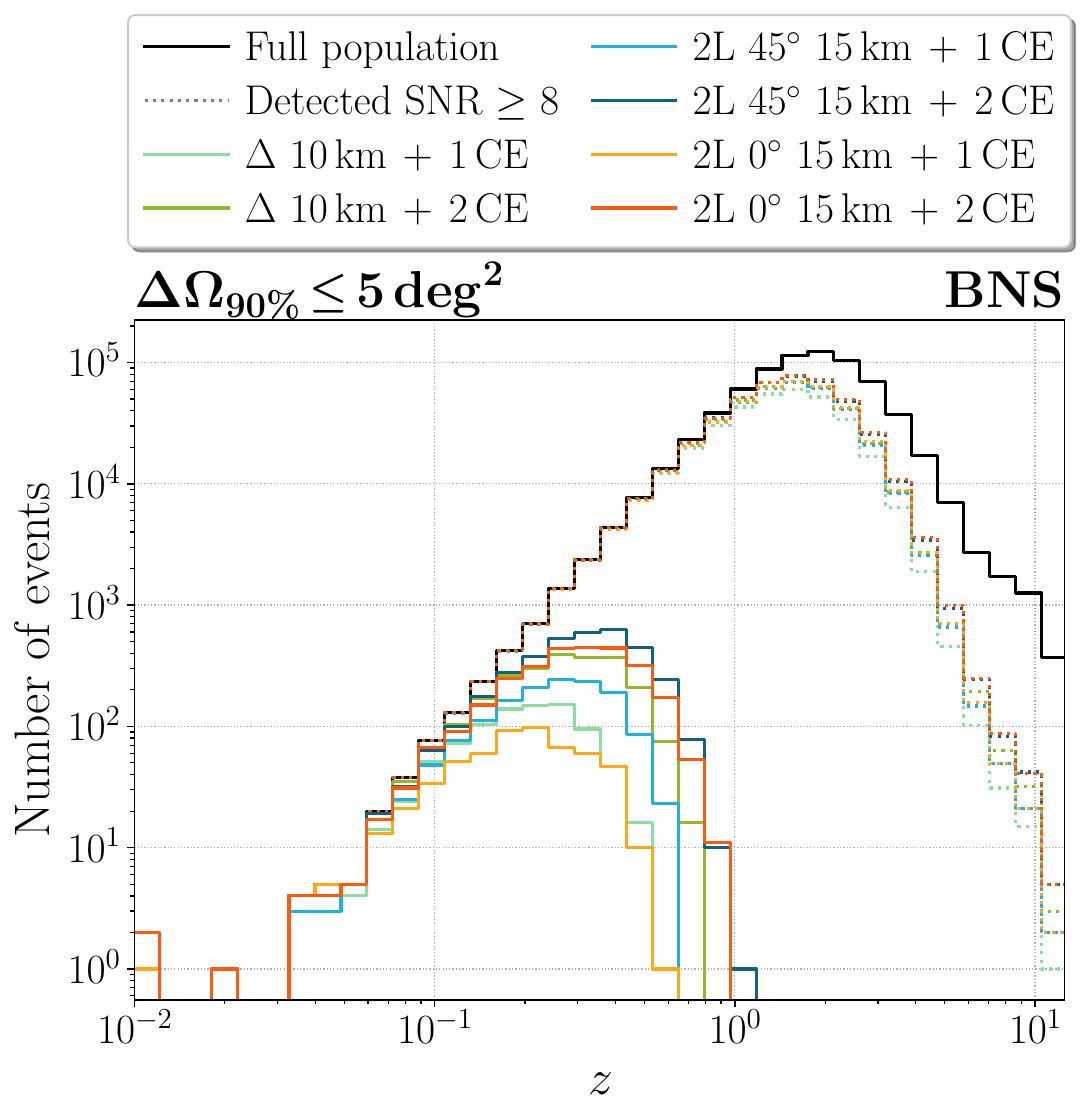}
\end{tabular}
    \caption{\small Redshift distribution of detected BNS with ${\rm SNR}\,\geq\,30$ (left column), or relative error on the luminosity distance $\Delta d_L/d_L\,\leq\,0.05$ (central column), or sky location $\Delta\Omega\,\leq\,5~{\rm deg}^2$ (right column) for the six 3G the detector networks considered. In each panel, we also show, for reference, the redshift distribution of the  BNS population used (black solid line), and of the events detected in the various configurations with ${\rm SNR}\,\geq\,8$  (dotted lines).
    }
    \label{fig:ETCE_allgeom_BNS_distr_vs_z}
\end{figure}

\clearpage\newpage
\section{Multi-messenger astrophysics}
\label{sect:MMO}
In this section we study the impact of the different detector geometries and different ASDs on the multi-messenger capabilities of ET operating in synergy with electromagnetic (\acrshort{em}) observatories. ET will observe together with a new generation of innovative EM observatories, such as CTA~\cite{CTAConsortium:2013}, Athena~\cite{Nandra:2013jka}, the Vera Rubin Observatory~\cite{VeraRubin}, JWST~\cite{Gardner:2006ky}, ELT~\cite{ELT:2008}, SKA~\cite{SKA} and with the possible mission concepts UVEX~\cite{2021arXiv211115608K}, THESEUS~\cite{theseus:2021}, HERMES~\citep{HERMES-SP:2021hvq}, GRINTA, ASTROGAM~\cite{DeAngelis:2021} and AMEGO~\cite{AMEGO:2019gny}. These observatories will probe the formation, evolution, and physics of BNS systems in connection with kilonovae (\acrshort{kn}e) and short gamma-ray bursts (GRBs) along the star formation and chemical evolution of the Universe. The multi-messenger studies will tremendously benefit from the larger sample of joint GW/EM detections and, in the closer Universe, from the greatly improved parameter estimates of ET with respect to the current detectors.

Taking into account the sky-localization capability of ET and the fact that the KN emission is intrinsically fainter than that of GRBs, future multi-messenger astronomy will differ at closer redshifts ($< 0.3-0.4$), where it will be possible to detect KN emission using the wide-field telescopes, such as the Vera Rubin Observatory, from the distant Universe where the detectable counterparts at large redshift will be almost uniquely observed in the high-energy band through the beamed emission of GRBs.  While the joint GW/KN detections will enable unprecedented studies of the enrichment of the Universe with heavy elements, of nuclear physics to constrain the equation of state (\acrshort{eos}) of NSs and cosmology estimating the expansion rate of the Universe, the joint GW/GRBs detections  will unveil the structure and properties of relativistic jets, the emission mechanism of short GRBs, the role of merger remnants such as magnetars, and will make it possible to estimate cosmological parameters and test modified gravity at the cosmological scale.

The main parameters determining the joint GW/EM detections, and thus affecting the multi-messenger performance of ET, are the ability to localize the source and the volume of the Universe up to which ET will be able to detect the source in terms of achieved redshifts. Another important aspect of third-generation GW detectors is the ability to detect BNSs before the merger and send pre-merger alerts with a good estimate of the sky localization of the source. This is critical for detecting prompt and early emissions in EM bands where all-sky monitors are not operating.

Here, we examine the sky-localization capabilities as a function of redshift and the pre-merger alert scenarios, and we evaluate the joint detections for ET observation together with examples of high-energy satellites and optical wide-FoV telescopes. Differently from the  other sections, in this section, we use  an SNR detection threshold of 8 taking into account that the presence of a counterpart increases the significance of the source detection. 

\subsection{BNS sky-localization and pre-merger alerts}
\label{sect:MMOskylocpremerger}
\begin{figure}
\begin{subfigure}{.3\textwidth}
  \centering
  \includegraphics[width=1.0\linewidth]{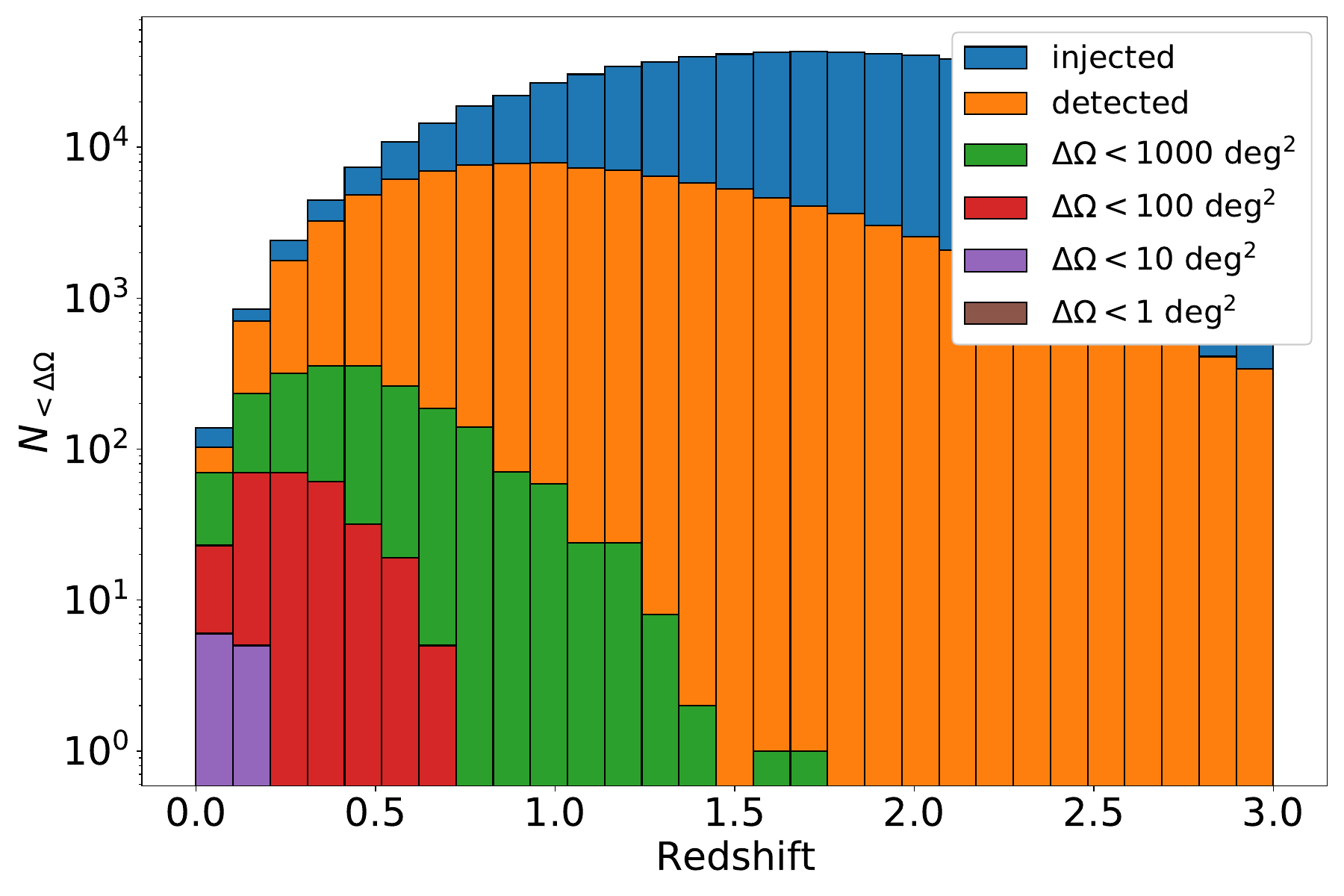} 
   \caption{\small $\Delta$ 10 km HFLF cryo}

\end{subfigure}
\hfill
\begin{subfigure}{.3\textwidth}
  \centering
  \includegraphics[width=1.0\linewidth]{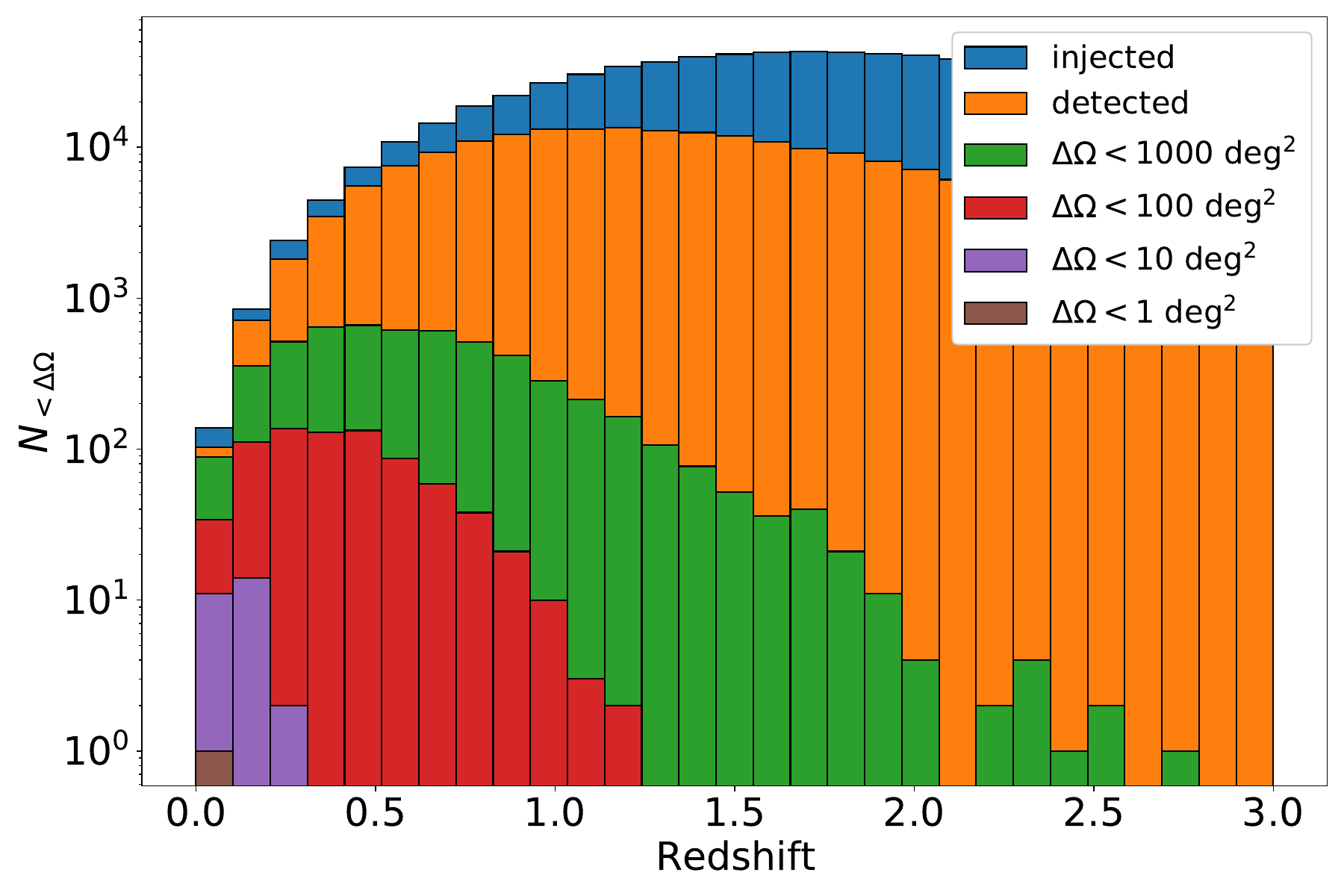}
   \caption{\small $\Delta$ 15 km HFLF cryo}

\end{subfigure}
\hfill
\begin{subfigure}{.3\textwidth}
  \centering
  \includegraphics[width=1.0\linewidth]{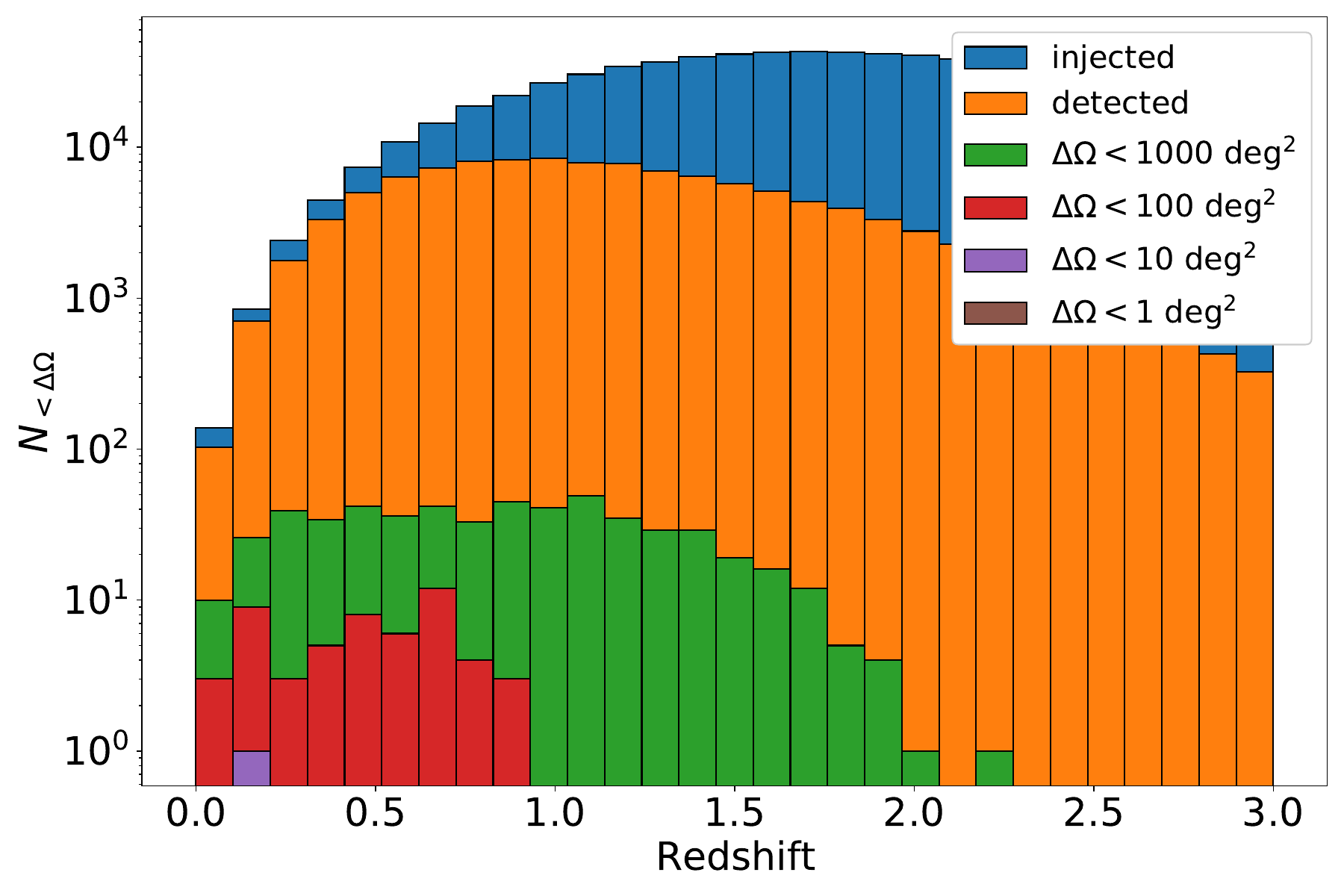} 
 \caption{\small $\Delta$ 15 km HF}
 
\end{subfigure}
\hfill
\begin{subfigure}{.3\textwidth}
  \centering
  \includegraphics[width=1.0\linewidth]{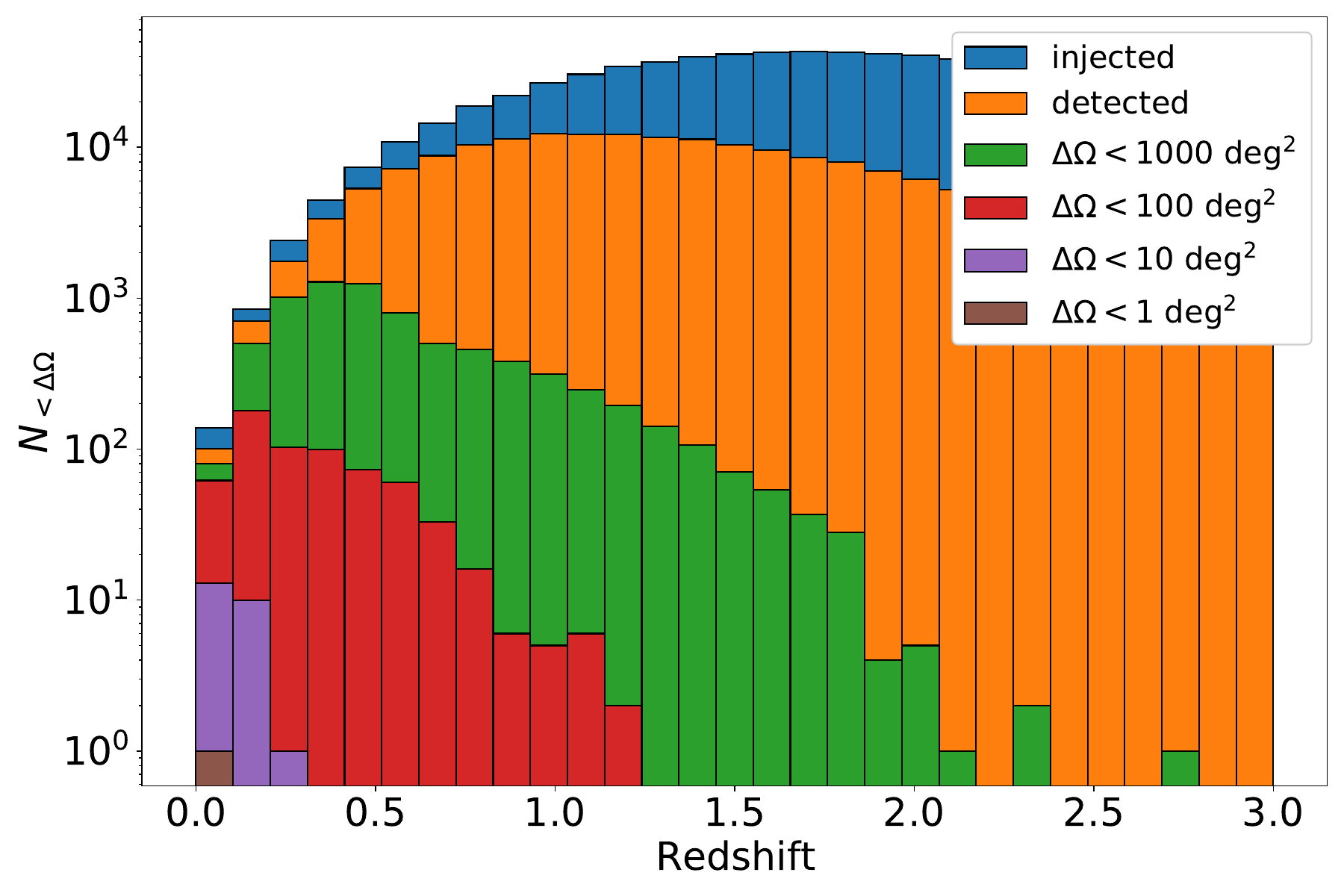} 
   \caption{\small 2L 15 km HFLF cryo}

  \label{fig:sub-third}
\end{subfigure}
\hfill
\begin{subfigure}{.3\textwidth}
  \centering
  \includegraphics[width=1.0\linewidth]{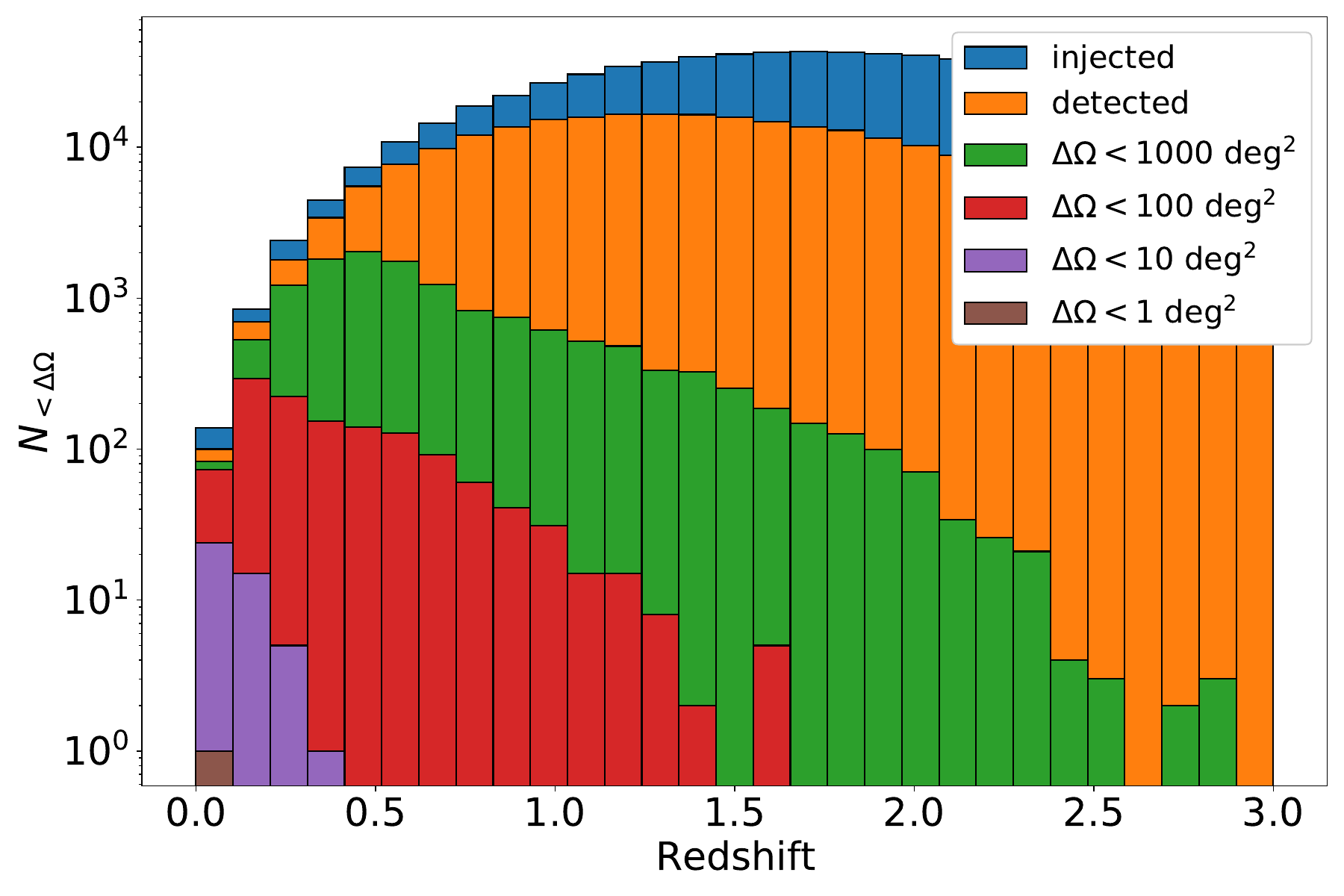} 
\caption{\small 2L 20 km HFLF cryo}

\end{subfigure}
\hfill
\begin{subfigure}{.3\textwidth}
  \centering
  \includegraphics[width=1.0\linewidth]{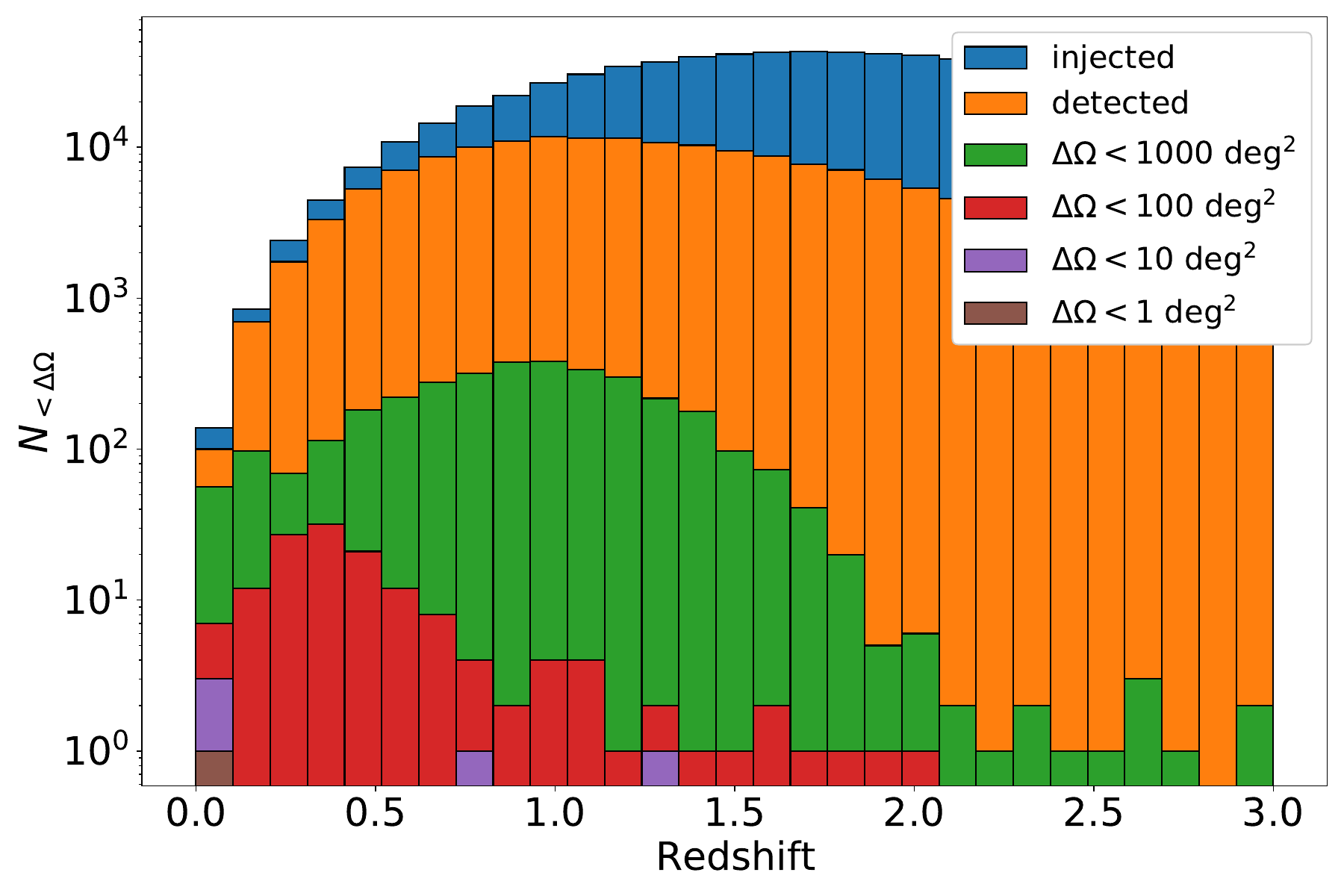}  
   \caption{\small 2L 20 km HF}

\end{subfigure}
\caption{\small Redshift distribution of the sky-localization uncertainty (given as 90$\%$ credible region) of randomly oriented BNS systems as a function of redshift for four detector geometries: the 10~km triangle ($\rm \Delta\, 10\, km$), the 15~km triangle ($\rm \Delta\, 15\, km$), the 2L with 15~km misaligned arms ($\rm 2L\, 15\, km$), and the 2L with 20~km misaligned arms ($\rm 2L\, 20\, km$). The plots on left and central columns show the full (HFLF cryo) sensitivity detectors. In comparison, the plots on the right columns show the performance of the detectors operating only with the HF interferometers. The absolute numbers are relative to one year of observation and assuming a duty cycle of 0.85 as described in the text. \label{fig:MMskyloc}}
\end{figure}

The sky-localization capability of GW detectors has a major impact in the efficiency of the search for a counterpart determining the number of pointings of the EM instruments to cover the GW signal location, the corresponding amount of observational time to be used and the  search sensitivity, and the efficiency to select and remove contaminating transients. Starting from the BNS population described in Section~\ref{sect:CBC}, \texttt{GWFISH} \cite{Dupletsa:2022wke} is used to evaluate the localization capability and build Table~\ref{tab:skyloccryoMM}, which gives the expected number of detections (${\rm SNR} \geq 8$) per year with sky-localization uncertainty (given as 90$\%$ c.l.) smaller than a threshold of 10, 40, 100, 1000 ${\rm deg}^2$ for the different ET configurations (we do not consider the 2L aligned configurations because as shown in Section~\ref{sect:PEBNS} they yield worse localization capabilities with respect to the misaligned configurations). Figure~\ref{fig:MMskyloc} shows the redshift distribution of the sky-localization uncertainty for all the detected BNS events. The distribution is given as a function of redshift for four detector geometries; 10~km triangle, 15~km triangle, 2L with 15~km misaligned arms (i.e. arms at a relative orientation of $45^{\circ}$), and 2L with 20~km misaligned arms. The different colours show the detections localized within a certain threshold on the sky-localization uncertainty. Comparing the plots on the left and central columns of the figure, which show the full (HFLF cryo) sensitivity detectors, it emerges that {\em the 2L with 15~km misaligned arms performs better than the 10~km triangle and is comparable to 15~km triangle} (in agreement with the results shown in the panel for $\Delta\Omega_{90\%}$ in  Fig.~\ref{fig:AllGeoms_CumulBNS_NdetScale}). Looking at the events localized with $\Delta \Omega_{90\%} < 10\,  {\rm deg}^2$ (100 ${\rm deg}^2$), the number of detection for the 2L with 15~km misaligned arms increases by factor 2 reaching $z$ of $0.3\, (1.2)$ with respect to the 10~km triangle, whose well-localized detections reach $z=0.2\,  (0.7)$. The 2L with 20~km misaligned arms increases the number of well-localized BNSs by a factor 4 with respect to the 10~km triangle reaching $z=0.4\, (1.6)$ with events localized within $10\,{\rm deg}^2 (100\, {\rm deg}^2)$. Table~\ref{tab:skyloccryoMM} shows also the expected number of detections per year within certain sky-localization uncertainty selecting events with a viewing angle, $\Theta_v$,\footnote{The viewing angle is defined as $\Theta_v = \iota$ for $0 \leq \iota \leq 90^{\circ} $ and $\Theta_v = |\iota-180^{\circ}|$ for $90^{\circ} < \iota < 180^{\circ} $} smaller than $ 15^{\circ}$. A fraction of these events is expected to produce detectable high-energy emission powered by the GRB relativistic jet, assumed to be perpendicular to the orbital plane. {\em Also for on-axis events, the 2L with 15~km misaligned arms performs better than the 10~km triangle  and is comparable to the 15~km triangle.} 

\begin{table}
\centering
Full (HFLF cryo) sensitivity detectors
\begin{tabular}{|l|c|c|c|c|c|c|c|c|}
\hline
$\Delta\Omega_{90\%} \rm (deg^2)$ & \multicolumn{4}{|c|}{All orientation BNSs} & \multicolumn{4}{c|}{BNSs with viewing angle $ \Theta_{v} < 15^{\circ}$}\\
\hline
	& $\Delta 10$ 							& $\Delta 15$ 						  & 2L 15 & 2L 20 & $\Delta 10 $							& $\Delta 15$ 						  & 2L 15 & 2L 20 \\ \hline
10		&	11	&	27	&	24	&	45 	&	0	&	1	&	2	&	5\\ \hline
40		& 78	&	215	& 162	& 350 &	8	&	22	&	20	&	33	\\ \hline
100		&	280	&	764	&	644	&	1282 &	26	&	74	&	68	&	133	\\ \hline
1000	& 2112	&	5441	&	7478	&	13482 & 272	&	632	&	1045	& 1725 \\ \hline
\end{tabular}
\caption{\small Expected number of detection (${\rm SNR} \geq 8$) per year with sky-localization uncertainty $\Delta \Omega_{90\%} ({\rm deg}^2)$ smaller than the threshold  indicated in the first column. While the columns 2-5 give the detections for BNS systems randomly oriented, the columns 5-9 give the detection of on-axis events, whose viewing angle is smaller than $15^{\circ}$. The numbers are relative to one year of observation assuming a duty cycle of 0.85 as described in the text.}
\label{tab:skyloccryoMM}
\end{table}

\begin{table}
\centering
HF sensitivity detectors
\begin{tabular}{|l|c|c|c|c|c|c|c|c|}
\hline
$\Delta\Omega_{90\%} \rm (deg^2)$ & \multicolumn{4}{|c|}{All orientation BNSs} & \multicolumn{4}{c|}{BNSs with viewing angle $ \Theta_{v} < 15^{\circ}$}\\
\hline
	& $\Delta 10$ 							& $\Delta 15$						  & 2L 15 & 2L 20 & $\Delta 10$ 							& $\Delta 15$ 						  & 2L 15 & 2L 20 \\ \hline
10		&	0	&	1	&	5	&	5 &	0	&	0	&	2	&	2	\\ \hline
40		&	4	&	10	&	20	&	47 &	0	&	5	&	6	&	17	\\ \hline
100		&	14	&	53	&	76	&	144	&	7	&	33	&	35	&	64\\ \hline
1000	&	145	&	548	&	1662	&	3378 &	80	&	336	&	672	&	1302	\\ \hline
\end{tabular}
\caption{\small Same as Table~\ref{tab:skyloccryoMM} but considering the detectors operating with only the HF interferometers.}
\label{tab:skylocHFMM}
\end{table}

\begin{table*}
\centering
\renewcommand{\arraystretch}{1.5}
Full (HFLF cryo) sensitivity detectors\\
\begin{tabular}{| c | l | r | r | r | r | r | r |}
\hline
\multirow{2}{*}{Configuration} & {\hspace{0.3cm}$\Delta\Omega_{90\%}$}      & \multicolumn{3}{c|}{All orientation BNSs} & \multicolumn{3}{c|}{BNSs with $\Theta_{v} < 15^{\circ}$ } \\ \cline{2-8}
& [deg$^{2}$]    & 30 min  &    10 min                   & 1 min  & 30 min  &    10 min                   & 1 min  \\ \hline    
\multirow{4}{*}{$\rm \Delta 10 km$} 
&	 10 &	 0 &	 1 & 5 &	 0 &	 0 & 0\\ \cline{2-8}
 &	 100 &	 10 &	 39 & 113 &	 2 &	 8 & 20\\ \cline{2-8}
 &	 1000 &	 85 & 293 & 819 &	 10 &	 34 & 132 \\ \cline{2-8}
&	 All detected        &	 905 &	 4343  & 23597	 &	 81 & 393 & 2312\\\hline
\multirow{4}{*}{ $\rm \Delta 15 km$} 
&	 10 &	 1 &	 5 & 11 &	 0 & 1 & 1\\ \cline{2-8}
&	100 &	 41 &	 109 & 	 281  &	 6 &	 14 & 36\\ \cline{2-8}
& 1000 & 279 & 806 & 2007  &	 33 &	 102 & 295\\ \cline{2-8}
&	 All detected        &	 2489 &	 11303  & 48127 &	 221 &	 1009 & 4024 \\\hline

\multirow{4}{*}{ 2L  15 km misaligned}
&	 10 &	 0 & 1 & 8 &	 0 &	 0 & 0\\ \cline{2-8}
&	 100 &	 20 & 54 & 169 &	 2 &	 7 & 26 \\ \cline{2-8}
&	 1000 &	 194 & 565 & 1399 &	 23 &	 73 & 199\\ \cline{2-8}
&	 All detected        &	 2172 &	 9598 & 39499 &	 198 &	863 & 3432 \\\hline

\multirow{3}{*}{ 2L  20 km misaligned}
&	 10 &	 2 &	 4 & 15 &	 1 &	 1 & 2\\ \cline{2-8}
&	 100 &	 39 &	 118 & 288  &	 7 &	 19 & 47\\ \cline{2-8}
&	 1000 &	 403 & 1040 & 2427 & 47 &	 128 & 346\\ \cline{2-8}
&	 All detected        &	 4125 &	 17294 & 56611 &	 363 &	 1588 & 4377 \\\hline
\end{tabular}
\caption{\small Number of BNS mergers per year detected  
(${\rm SNR} \geq 8$) before the merger  within $z=1.5$ for the different full (HFLF cryo) sensitivity ET configurations. Three pre-merger scenarios (30, 10, and 1 minute(s) before the merger) are shown in different columns.  For each detector configuration, the rows give the number of detections with sky-localization ($90\%$ c.l.) within 10, 100, 1000 deg$^{2}$, and all the detected sources. The numbers of GW detections per year take into account the ET duty cycle as described in the text.} 
\label{table:premergerFull}
\end{table*}

\begin{table*}
\centering
\renewcommand{\arraystretch}{1.5}
HF sensitivity detectors\\
\begin{tabular}{| c | l | r | r | r | r | r | r |}
\hline
\multirow{2}{*}{Configuration} & {\hspace{0.3cm}$\Delta\Omega_{90\%}$}      & \multicolumn{3}{c|}{All orientation BNSs} & \multicolumn{3}{c|}{BNSs with $\Theta_{v} < 15^{\circ}$ } \\ \cline{2-8}
& [deg$^{2}$]    & 30 min  &    10 min                   & 1 min  & 30 min  &    10 min                   & 1 min  \\ \hline    
\multirow{3}{*}{$\rm \Delta 10 km$}       
 &	 100 &	 0 &	 0 & 0 &	 0 &	 0 & 0\\ \cline{2-8}
 &	 1000 &	 0 &    0 & 4 &	 0 &	 0 & 1 \\ \cline{2-8}
&	 All detected   &    0 & 3 & 317 & 0 &	 0 &	 26  \\\hline
\multirow{3}{*}{ $\rm \Delta 15 km$}       &	100 & 0 &	 0 & 2  &	 0 &	 0 & 0\\ \cline{2-8}
& 1000 &	 0 &     0 &  10  &	 0 &	 0 & 4\\ \cline{2-8}
&	 All detected        &	 2 &	 8  & 891 &	 0 &	 1 & 84 \\\hline

\multirow{3}{*}{ 2L  15 km misaligned}       &	 100 &	 0 & 0 & 0 &	 0 &	 0 & 0 \\ \cline{2-8}
&	 1000 &	 0 &  0 & 7 &	 0 &	 0 & 3\\ \cline{2-8}
&	 All detected        &	 0 & 7 & 743 &	 0 &	 1 & 69 \\\hline

\multirow{3}{*}{ 2L  20 km misaligned}       &	 100 &	 0 & 0 & 3  &	 0 &	 0 & 0\\ \cline{2-8}
&	 1000 &	 0 & 0 & 13 &	 0 &	 0 & 6\\ \cline{2-8}
&	 All detected       & 2 & 11 & 1535 &	 0 &	 1 & 146 \\\hline
\end{tabular}
\caption{\small Same as Table~\ref{table:premergerFull} but for the HF sensitivity ET.} 
\label{table:premergerHF}
\end{table*}

{\em Losing the low-frequency interferometers, the HF GW detectors detect a significantly smaller number of well-localized events. However, as shown comparing Table~\ref{tab:skyloccryoMM} and Table~\ref{tab:skylocHFMM},\footnote{The different numbers, comparing Tables~\ref{tab:skyloccryoMM} and \ref{tab:skylocHFMM} with Table~\ref{tab:BNSAllConfDeldLDelOm} in App.~\ref{app:TablesCBC},  reside in the different  SNR threshold, assumed to be 8 for multi-messenger studies and 12 for the metrics of Section~\ref{sect:CBC} and  App.~\ref{app:TablesCBC}. The larger difference on the HF-only estimates is also related to the different approach of \texttt{GWFISH} (used for the Tables~\ref{tab:skyloccryoMM} and \ref{tab:skylocHFMM}) which regularizes the close-to-singular matrices \citep{Dupletsa:2022wke} instead of discarding the ill-conditioned Fisher matrix as \texttt{GWFAST} (used for the Table \ref{tab:BNSAllConfDeldLDelOm}) \cite{Iacovelli:2022bbs,Iacovelli:2022mbg}. As described in the footnote \ref{footnote:Fisher_inversion_issues}, for the HF-only case the percentage of discarded events approaches 10\%.} the decrease of detections  of well-localized events is more severe for the triangle configurations than for 2L misaligned.} The right column of Fig.~\ref{fig:MMskyloc} shows  the redshift distribution of the sky-localization uncertainty for the HF 15~km triangle and 2L with 20~km misaligned arms. The comparison with the corresponding configurations HFLF cryo shows that a large fraction of well-localized events is already missed at small redshifts. Focusing on the on-axis events, the use of only HF interferometers decreases the number of well-localized events but in a percentage of a smaller fraction with respect to the events randomly oriented. While the HF 2L with 15~km and 20~km misaligned arms
localize worse than the HFLF-cryo 10~km triangle for randomly oriented systems, {\em for on-axis events, the HF 2L with 15~km misaligned arms localization capability is comparable to the HFLF-cryo 10~km triangle and the HF 2L with 20~km better than the HFLF-cryo 10~km triangle}.

The detection of prompt/early multi-wavelength emission from BNS mergers is critical to probe the central engine of GRBs, particularly to understand the jet composition, the particle acceleration mechanism, the radiation and energy dissipation mechanisms \citep[see e.g.][]{2014IJMPD..2330002Z}. Regarding the kilonova emission, it has been shown that the early phase emission is particularly sensitive to the structure of the outer sub-relativistic ejecta (see, e.g., \cite{Banerjee:2020myd}) and can give rise to early \acrshort{uv} emission detectable by wide-field satellites such as ULTRASAT \cite{2021SPIE11821E..0UA}, and mission concepts as UVEX \cite{2021arXiv211115608K} and Dorado~\cite{2022arXiv220609696D} which can benefit of pre-merger alerts. Detecting a BNS signal before the merger makes it possible to eventually detect electromagnetic emission precursors. It makes it possible also to image the sky-localization before the merger that, especially for optical observations, is extremely useful to remove contaminating transients present before the merger. Considering all the benefits of pre-merger alerts for multi-messenger astronomy, here we explore the ET capabilities to detect BNSs and give a sky-localization before the merger.

Table~\ref{table:premergerFull} shows the number of BNS mergers  per year detected (${\rm SNR} \geq 8$) before the merger  within $z=1.5$, for the different full (HFLF cryo) sensitivity ET configurations. We consider three pre-merger scenarios; 30, 10 and 1 minute(s) before the merger. For each ET configuration, we give the number of detection with sky-localization ($90\%$ c.l.) within 10, 100, 1000 ${\rm deg}^{2}$, and all the detected sources. We show the results for the BNSs randomly oriented of our population (columns 3-5) and select among them the BNSs with viewing angles smaller than $15^{\circ}$ (columns 6-8). The on-axis detections are particularly important to detect the prompt/early beamed emission associated with relativistic jets, for example the very high energy prompt emission never detected so far from GRBs (see \cite{Banerjee2022}  for the perspectives with the Cherenkov Telescope Array \cite{CTAConsortium:2013ofs}).   
{\em Focusing on well-localized events detected pre-merger with a sky-localization smaller than 100 ${\rm deg}^2$, we highlight that the 15~km triangle turns out to perform better than the 10~km triangle and the 2L with 15~km misaligned arms, and its performance is similar to the 2L with 20~km misaligned arms. The 2L with 15~km misaligned arms performs better than the 10~km triangle. The results are similar for on-axis events.}

Table~\ref{table:premergerHF} shows the dramatic decrease of pre-merger alerts without low-frequency. {\em For all the ET configurations, there are no localized pre-merger detections, except a few events with sky localization smaller than $\rm 1000\, deg^2$ per year one minute before the merger (a few $ < \rm 100 \,deg^2$ for $\rm 2L 20 km$).}

\subsection{Gamma-ray bursts: joint GW and high-energy detections}
\label{sect:MMOGRB}
Following the approach described in \cite{Ronchini:2022gwk}, we evaluate the expected detections of prompt and afterglow emission of short GRBs associated with BNS mergers. We explore the joint detection capabilities of the different configurations of ET operating with gamma-ray and X-ray satellites. As reference instruments, we use some examples of current and upcoming satellites and mission concepts. We analyze different observational strategies: survey mode and target pointing.  

\subsubsection{Prompt emission}
Starting from our astrophysically-motivated population of BNS mergers we model the high-energy signals associated with the mergers making the BNS population able to reproduce the rate and the properties of  all the short GRBs observed so far (see \cite{Ronchini:2022gwk} for details). 
For the fraction of BNS mergers in our population that are expected to produce a jet (about 20\% of the BNS see \cite{Ronchini:2022gwk}), we inject both the GW and EM signals and recover them based on the detection efficiency of the GW detector and EM satellites, respectively.
 We define a joint detection if the signal is above the detection threshold simultaneously in both the GW and EM observatories. To properly compare the capabilities of different ET configurations and avoid differences due to the uncertainties in the EM emission, for the whole set of GW simulations we associate to each BNS population the same realization of the EM signal derived from the MCMC.

For the detection of the prompt emission, we consider the following $\gamma$-ray instruments: the Gamma-ray Burst Monitor (GBM) on board of Fermi \citep{Meegan:2009qu}, the Gravitational wave high-energy Electromagnetic Counterpart All-sky Monitor \citep[GECAM;][]{gecam:2020},  the High Energy Rapid Modular Ensemble of Satellites \citep[HERMES;][]{HERMES-SP:2021hvq}, the Transient Event Detector (TED) on board of the Gamma-Ray INternational Transient Array Observatory (GRINTA), ASTROGAM \citep{DeAngelis:2021}, and the X/Gamma-ray Imaging Spectrometer (\acrshort{xgis}) on board of the Transient High-Energy Sky and Early Universe Surveyor \citep[THESEUS;][]{theseus:2021}. The properties of these instruments are summarized in Table~\ref{instruments}.

\begin{table*}[t]
\centering
\begin{tabular}{|l|c|c|c|c|c|c|}
\hline
\multirow{2}{*}{INSTRUMENT}&band &$F_{\rm lim}$&\multirow{2}{*}{FoV/$4\pi$}&\multirow{2}{*}{loc. acc.}&\multirow{2}{*}{Status} \\ 
&MeV& ph cm$^{-2}$ s$^{-1}$ && &\\ \hline
\emph{Fermi}-GBM  & 0.01 - 25 &0.5 & 0.75 & 5 deg & Operating mission \\ \hline
GECAM      & 0.006 - 5 &$2\times 10^{-8}$$(^*)$& 1.0 & 1 deg   & Operating mission   \\
\hline
HERMES     & 0.05 - 0.3 &0.2 & 1.0 & 1 deg  & Mission concept \\
& & & & & Pathfinder next few yrs \\
\hline
GRINTA-TED & 0.02-10 & 0.45  &  0.64 & $ 5\, \rm deg(^{**})$  & Mission concept  \\
\hline
ASTROGAM &  0.03-200 & 0.29 &  0.27  & $< 1.8$ deg & Mission concept \\
\hline
THESEUS-XGIS       & 0.002 - 10 &$3\times 10^{-8}$$(^*)$& 0.16 & $<15$ arcmin & Mission concept\\ 
& & & & & Possible launch 2037 \\
\hline

\end{tabular}
\flushleft
\footnotesize{
$(^*)$ expressed in ph erg cm$^{-2}$ s$^{-1}$}\\
\footnotesize{$(^{**}) \, 65\% (42\%)$ of the TED/GW joint detections are detectable by the Hard X-ray Imager HXI on board of GRINTA by repointing HXI on the TED detection in 60s (5 min). HXI is expected to localize at order of 30 arcSection}
    \caption{\small Instruments characteristics.}
    \label{instruments}
\end{table*}

\begin{figure}[t]
    \centering
    \includegraphics[width=0.5\textwidth]{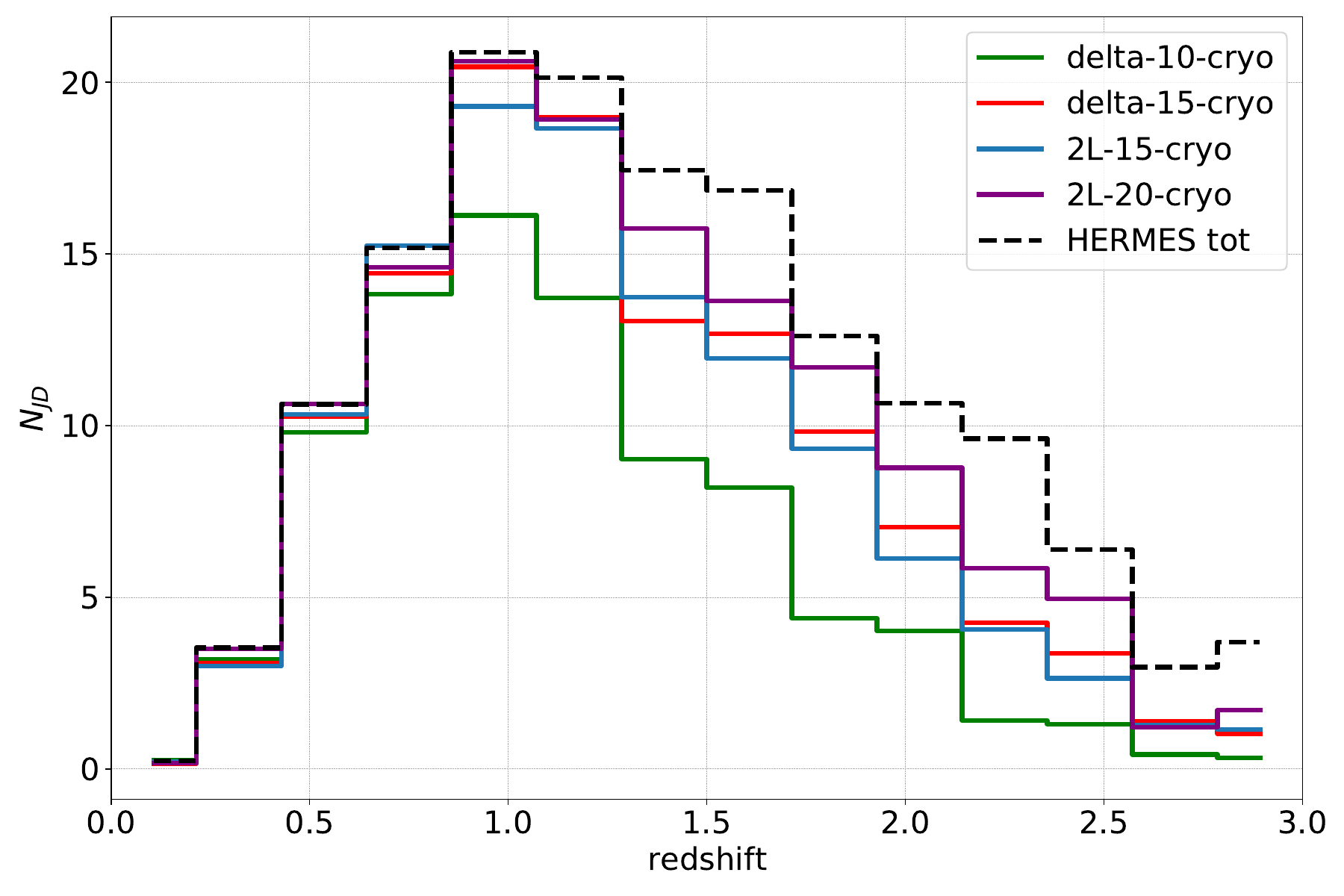}
    \caption{\small Histogram of the joint detection of ET and HERMES observing in survey mode during one year. The solid lines of different colors represent the different configurations of ET full sensitivity. The dashed line show the prompt emission of short $\gamma$-ray bursts expected to be detected by HERMES during one year of observations.}
    \label{fig:HERMESETfulljoint}
\end{figure}

Table~\ref{tab_jointprompt_cryo} shows the numbers of joint GW+$\gamma$-ray detections during one year of observation for the different $\gamma$-ray satellites operating in survey mode together with different ET full sensitivity configurations. These numbers go from ten to one hundred depending on the satellite and ET configurations. We highlight that, while some satellites maximize the number of detections (which is useful for population studies) sacrificing the localization capability, others with a smaller number of detections give the sky-localization necessary to drive the ground-based and space-borne follow-up. This is critical for identifying the host galaxy, evaluating a redshift, and completely characterising the source and its environment. An important piece of information is given in columns 6-9, where we show what is the fraction of $\gamma$-ray bursts detected by each satellite which will have an associated GW signal. Here, we find that these fractions go from $51\%-61\%$ for the 10~km triangle, $74\% - 83\%$ for the 15~km triangle, $70\% - 79\%$ for the 2L with 15~km misaligned arms, $80\% - 89\%$ for the 2L with 20~km misaligned arms. {\em The ET 10~km triangle will make possible remarkable results for GRB science by detecting GW signals for a large fraction of short GRBs observed by $\gamma$-ray satellites, and it represents an impressive step forward from the LIGO, Virgo and KAGRA network, whose BNS horizon strongly limits the ability to detect events with the relativistic jet aligned with the line-of-sight (see e.g. \citep{Patricelli:2022hhr, Colombo:2022zzp}). Then, the percentage of GRBs with a GW counterpart significantly increases going from the 10~km triangle to 15 km and 20 km configurations.} The  15~km triangle is marginally better than the 2L with 15~km misaligned arms. 
Figure~\ref{fig:HERMESETfulljoint} takes as an example HERMES (full constellation) and shows the joint detections during one year. The plot shows that the most significant improvement is from the 10~km triangle to a 15 km configuration. {\em The 15 and 20 km configurations are able to increase the number of joint detection at redshift larger than 0.9 with respect to the 10~km triangle.} This is particularly important to estimate cosmological parameters or to test modified gravity at cosmological scales.

\begin{figure}[t]
\begin{subfigure}{.5\textwidth}
  \centering
    \includegraphics[width=1.0\textwidth]{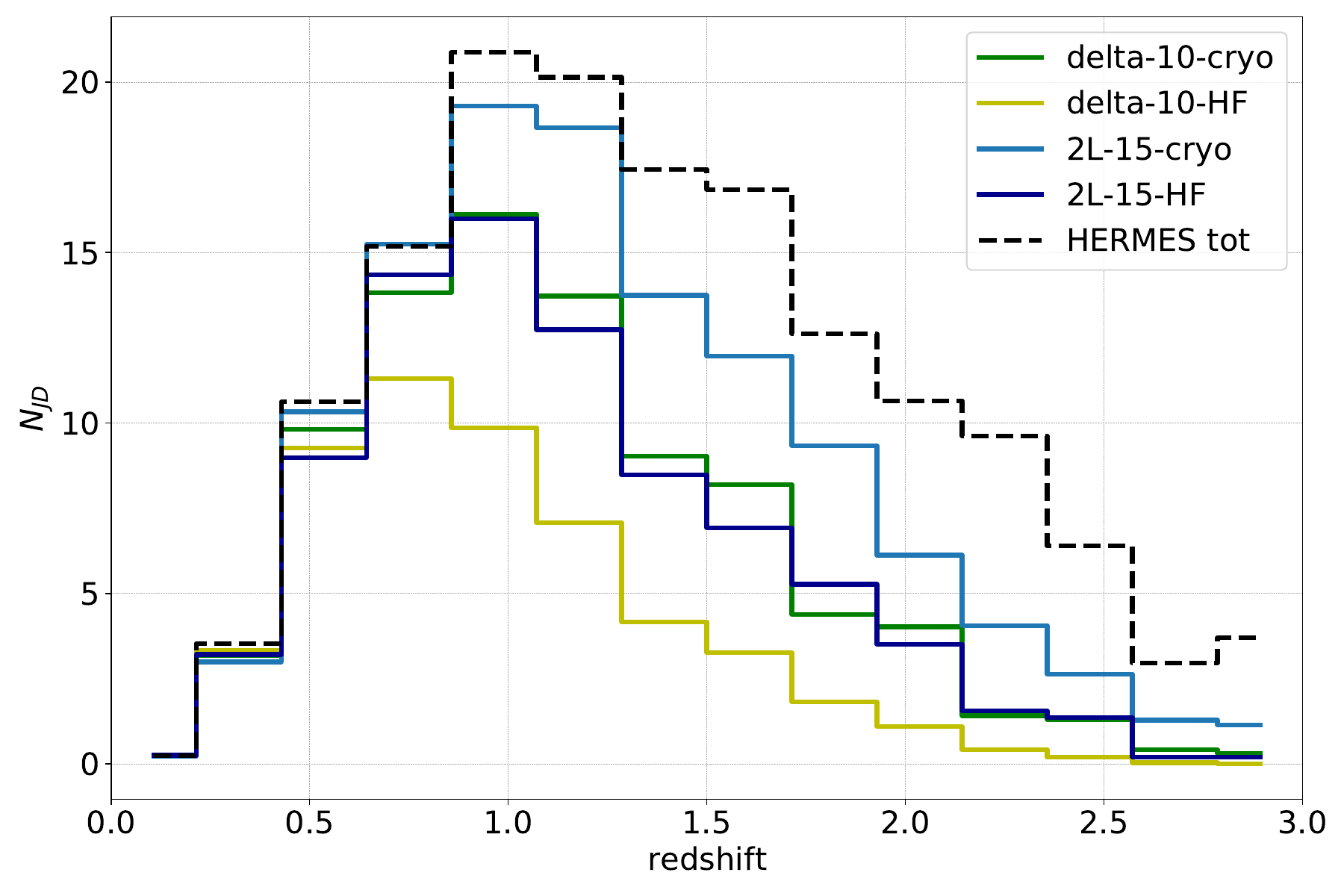}
\end{subfigure}
\begin{subfigure}{.5\textwidth}
  \centering
    \includegraphics[width=1.0\textwidth]{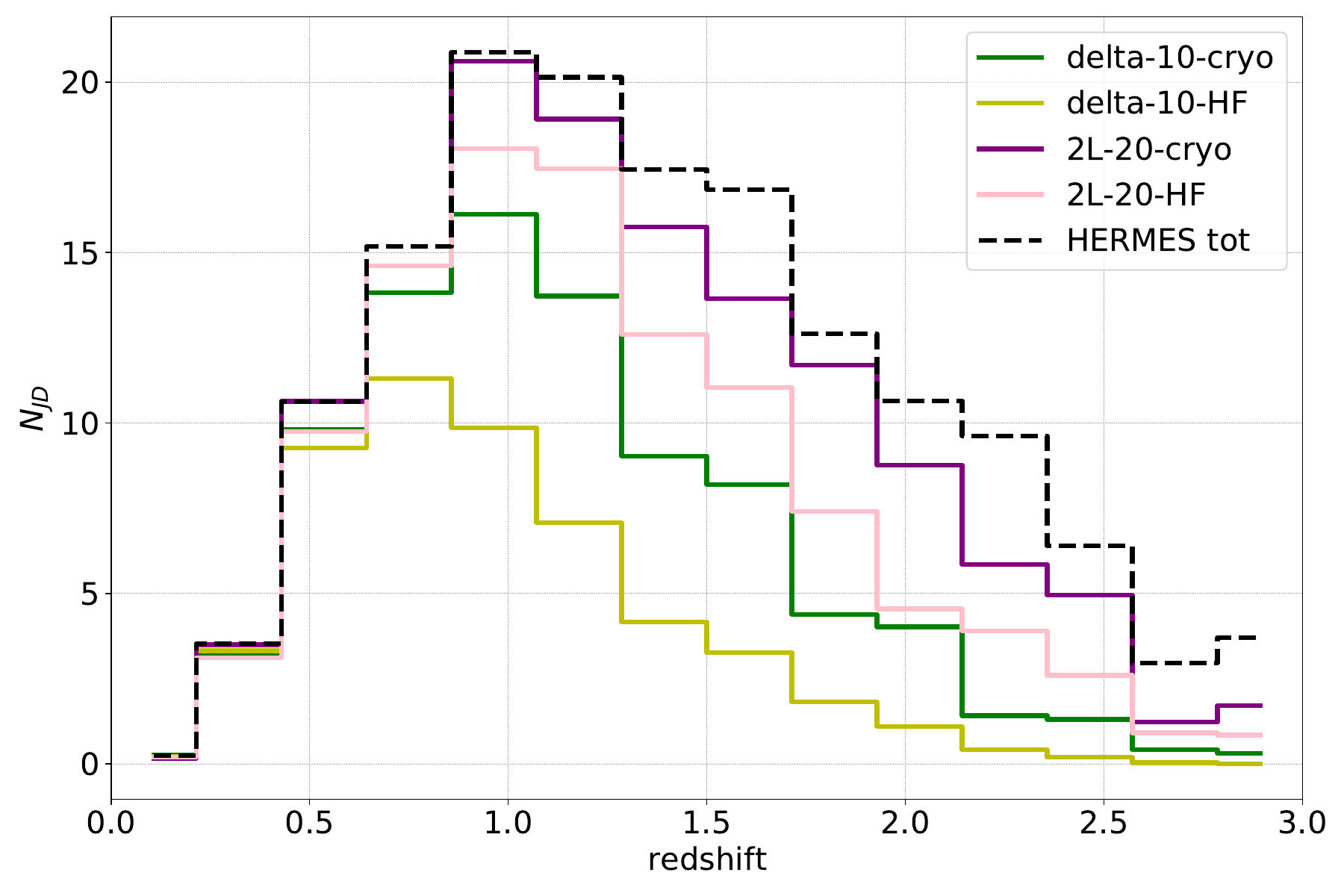}
\end{subfigure}
  \caption{\small Histogram of the joint detection of ET and HERMES observing in survey mode during one year. The plot shows the comparison among ET full sensitivity and ET HF for $\rm \Delta 10 km$ and $\rm 2L 15 km$ (left plot), and  $\rm 2L 20 km$ (right plot). The dashed line show the prompt emission of short $\gamma$-ray bursts expected to be detected by HERMES during one year of observations. }
     \label{fig:HERMESETHFjoint}
\end{figure}

Considering the GW detectors operating with only the high-frequency instrument (see Table~\ref{tab_jointprompt_HF}), we find that the percentage of detected short GRBs with an associated GW signal significantly decreases for each configuration. The percentages become
$29\%-39\%$ for the 10~km triangle, $54\% - 64\%$ for the 15~km triangle, $49\% - 60\%$ for the 2L with 15~km misaligned arms, $66\% - 76\%$ for the 2L with 20~km misaligned arms. {\em In terms of percentage of short GRB with a GW signal, the performance of the 15~km triangle and $\rm 2L 15 km$ HF-only detectors is comparable to the 10~km triangle full sensitivity.} This is also shown in the left plot of Fig.~\ref{fig:HERMESETHFjoint}. The right plot of Fig.~\ref{fig:HERMESETHFjoint} shows that $\rm 2L 20 km$ HF results to be better than the 10~km triangle full sensitivity, but again going to lower frequency significantly increase the $\rm 2L 20 km$ performance making possible to detect the GW signals associated with a large fraction (85$\%$) of short GRBs detected by HERMES. 

\begin{table}[t]
\centering
Full (HFLF cryo) sensitivity detectors
\begin{tabular}{|l|c|c|c|c| c|c|c|c|}
\hline
Instrument 	& $\Delta 10$ 							& $\Delta 15$ 						  & 2L 15 							& 2L 20  & $\Delta 10$ 							& $\Delta 15$ 						  & 2L 15 							& 2L 20\\ \hline
Fermi-GBM	&	$31	_{-9}^{+9}$	&	$42_{-13}^{+11}$	&	$39_{-9}^{+11}$	&	$44_{-11}^{+13}$ & $61_{-11}^{+12}	\%$	&	$83_{-10}^{+9} \%$ &	$79_{-11}^{+8} \%$ &	$89_{-8}^{+4} \%$\\ \hline
GECAM 	 &	$61_{-25}^{+39}$	&	$89_{-34}	^{+54}$	&	$81_{-32}^{+51}$	&	$96_{-36}^{+52}$ & $51_{-6}^{+5}	\%$	&	$74_{-5}^{+5} \%$ &	$70_{-6}^{+3} \%$ &	$80_{-4}^{+4} \%$\\ \hline
HERMES 		&	$86_{-28}^{+31}$	&	$120	_{-31}^{+40}$	&	$117_{-34}^{+37}$	&	$132_{-34}^{+34}$ &	$55_{-7}^{+9} \%$	&	$78_{-7}^{+8} \%$ &	$74_{-9}^{+9} \%$ &	$85_{-6}^{+5} \%$\\ \hline
GRINTA-TED		&	$77_{-25}^{+31}$	&	$107_{-28}^{+31}$	&	$98_{-25}^{+31}$	&	$114_{-28}^{+34}$ &	$57_{-9}^{+10} \%$	&	$79_{-8}^{+8} \%$ &	$74_{-9}	^{+9} \%$ &	$ 85_{-5}^{+5} \%$ \\ \hline
ASTROGAM  &     $18_{-5}^{+8}$	        &	$24_{-7}^{+9}$	        &	$24_{-6}^{+9}$	        &	$27_{-7}^{+8}$	   &  $59_{-9}^{+11} \% $	&	$80_{-8}^{+8} \% $ & $77_{-9}^{+8} \% $   & $86_{-9}^{+6} \% $	\\ 
\hline
THESEUS-XGIS 		&	$10_{-3}^{+3}$	&	$13_{-3}^{+3}$	&	$13_{-3}^{+3}$	&	$15_{-4}^{+3}$ & $57_{-10}^{+9}	\%$	&	$79_{-9}^{+8}	\%$ &	$73	_{-7}^{+11}	\%$ &	$85_{-5}^{+7} \%$ \\ \hline
\end{tabular}
\caption{\small Columns 2-5 give the numbers of joint GW+$\gamma$-ray detections during one year of observation for different combinations of $\gamma$-ray instruments operating in survey mode together with different ET full sensitivity configurations.
The absolute numbers do not assume duty cycle for the satellites. Columns 6-9 give the fraction of detected short GRBs which will have a GW counterparts.
}
\label{tab_jointprompt_cryo}
\end{table}

\begin{table}[t]
\centering
HF sensitivity detectors
\begin{tabular}{|l|c|c|c|c|c|c|c|c|}
\hline
Instrument 	& $\Delta 10$ 							& $\Delta 15$ 						  & 2L 15 							& 2L 20  & $\Delta 10$ 							& $\Delta 15$ 						  & 2L 15 							& 2L 20\\ \hline
Fermi-GBM	&	$20_{-7}^{+8}$	&	$33_{-9}^{+9}$	&	$29_{-9}^{+11}$	&	$38_{-10}^{+12}$ &	$39_{-8}^{+11} \%$&	$64_{-11}^{+12} \%$& $60_{-11}^{+12}	\%$&	$76_{-9}^{+9} \%$ \\ \hline
GECAM 		&	$35_{-15}^{+21}$	&	$62_{-22}^{+38}$	&	$58_{-22}^{+38}$	&	$77_{-30}^{+47}$ &	$29	_{-5}	^{+4}	\%$&	$54_{-5}^{+4} \%$&	$49_{-7}^{+4} \%$&	$66	_{-6}^{+4} \%$\\ \hline
HERMES 		&	$52_{-18}^{+21}$	&	$91_{-29}^{+30}$	&	$83_{-28}^{+28}$	&	$107_{-31}^{+40}$ &	$33_{-8}^{+7} \%$&	$58_{-8}^{+10} \%$&	$53_{-8}^{+10}	\%$&	$71_{-8}^{+8} \%$ \\ \hline
GRINTA-TED		&	$46_{-16}^{+22}$	&	$80_{-25}^{+31}$	&	$74	_{-25}^{+28}$	&	$94_{-23}^{+33}$ & $34_{-9}^{+9} \%$&	$61_{-11}^{+9} \%$&	$55_{-10}^{+9}	\%$&	$72_{-9}^{+9} \%$ \\ \hline
ASTROGAM  & $12_{-5}^{+6}$	& $19_{-5}^{+7}$ & $18_{-6}^{+6}$ & $23_{-7}^{+8}$	& $37_{-10}^{+11} \%$	&	$62_{-11}^{+9} \%$	&	$57_{-9}	^{+10} \%$	& $74_{-10}^{+9}\%$ \\ \hline	
THESEUS-XGIS 		&	$6_{-2}^{+2}$	&	$10_{-3}^{+3}$	&	$9_{-3}^{+3}$	&	$12_{-3}^{+3}$ &	$34_{-9}^{+8} \%$&	$59_{-8}^{+10} \%$&	$54_{-9}^{+10}	\%$ &	$71_{-9}^{+9} \%$ \\ \hline
\end{tabular}
\caption{\small Same as \ref{tab_jointprompt_cryo} but considering the GW detectors operating with only the HF interferometers.}
\label{tab_jointprompt_HF}
\end{table}

\subsubsection{Afterglow: survey and pointing modes}
This section explores the capability of detecting X-ray short GRB afterglows associated with GWs. Using the approach described in \cite{Ronchini:2022gwk}, we model the afterglow taking into account both the forward shock and the high-latitude emission \cite{Oganesyan:2019jij,Ascenzi:2020kxz} and we evaluate the joint GW/X-ray detections by considering two observational strategies; the satellite observing in survey mode or the satellite slewing to point well-localized events. As an example of performance of X-ray wide FoV satellites, we consider the Soft X-ray Imager (\acrshort{sxi}) planned to be on board the THESEUS mission \cite{theseus:2021}. SXI is a wide field of view instrument able to cover 0.5 steradians in one observation. It has a localization accuracy of 1-2 arcmin. Since also XGIS has good sensitivity down to 2 keV in a larger FoV of 2 steradians, We also consider the combination of SXI and XGIS.

\begin{table}[t]
\centering
Full (HFLF cryo) sensitivity detectors\\
\begin{tabular}{|l|c|c|c|c|}
\hline
Instrument 	& $\Delta 10$ 							& $\Delta 15$ 						  & 2L 15 & 2L 20 \\ \hline
THESEUS-SXI	survey	&	$10_{-2}^{+3}$	&	$13_{-4}^{+3}$&	$12_{-3}^{+3}$	&	$12_{-3}^{+3}$  \\ \hline
THESEUS-(SXI+XGIS)	survey	&	$21_{-7}^{+6}$	&	$21_{-6}^{+8}$&	$20_{-5}^{+7}$	&	$21_{-7}^{+7}$  \\ \hline
\end{tabular}\\
\vspace{0.2cm}
HF sensitivity detectors\\
\begin{tabular}{|l|c|c|c|c|}
\hline
Instrument 	& $\Delta 10$ 							& $\Delta 15$ 						  & 2L 15 & 2L 20 \\ \hline
THESEUS-SXI	survey	&	$8_{-3}^{+2}$	&	$11_{-4}^{+2}$ & $10_{-3}^{+2}$	&	$11	_{-2}^{+2}$  \\ \hline
THESEUS-(SXI+XGIS)	survey	&	$16_{-5}^{+6}$	&	$19_{-5}^{+8}$ & $19_{-5}^{+4}$	&	$21	_{-6}^{+8}$  \\ \hline
\end{tabular}
\caption{\small Numbers of joint GW+X-ray detections during one year of observation for THESEUS operating in survey mode considering the instrument SXI and the combination of SXI and XGIS.}    
\label{tab_jointafterglowsurvey}
\end{table}

\begin{figure}[t]
    \centering
    \includegraphics[width=0.8\textwidth]{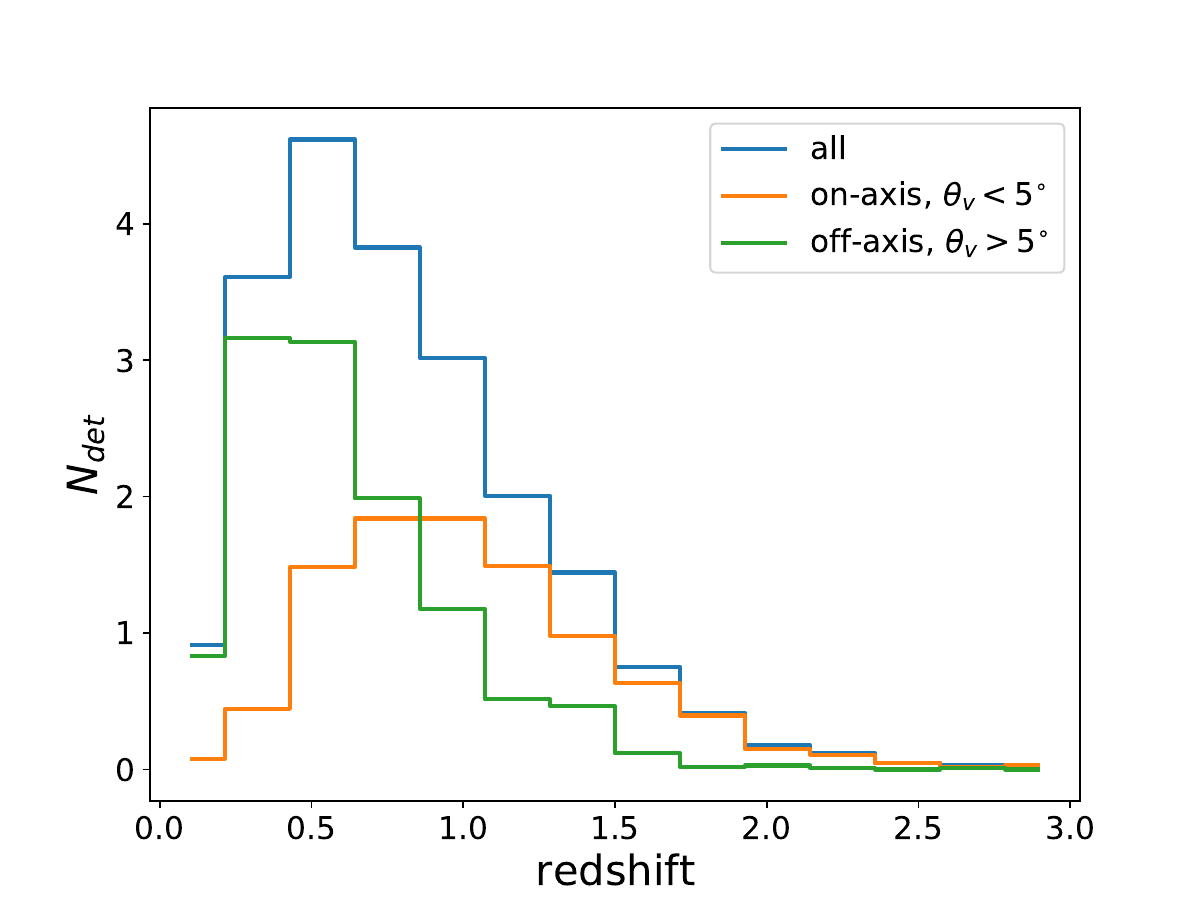}
    \caption{\small Redshift distribution of joint GW+EM detections with THESEUS-(SXI+XGIS) in survey mode, where we distinguish between on-axis and off-axis cases. The ET configuration considered here is 2L 20 km HFLF cryo.}
    \label{fig:onaxisoffaxis}
\end{figure}

Tables~\ref{tab_jointafterglowsurvey} show the numbers of X-ray afterglow detected in survey mode by SXI and the combination of SXI and XGIS on board of THESEUS which will have an associated GW signal. While the top Table shows the joint detection with the ET full sensitivity configurations, the bottom Table shows the ones with ET HF. The number of joint detections is around 10 for SXI (20 for SXI+XGIS) per year independently of the arms lengths and the geometries, and the number remains almost the same also without accessing low-frequencies. Figure~\ref{fig:onaxisoffaxis} shows a histogram of the joint GW+afterglow detections by THESEUS-(SXI+XGIS) in survey mode operating with full sensitivity 2L with 20~km misaligned arms as a function of redshift. The figure shows that the majority of the joint detections are within $z=1$. This explains the reason why {\em both arm-length  and accessible frequency do not change the joint GW/X-ray afterglow detection numbers}; the majority of the afterglows are detectable up to a redshift where there is no significant difference among the GW detection efficiency. The figure also shows that detecting the X-ray afterglow enables to observe events with a jet not aligned to the line of sight. 

For the pointing strategy, we select events with a sky-localization uncertainty smaller than $100\, {\rm deg}^2$, which is well-contained inside the SXI FoV, and we assume that SXI is able to be on target 100 s after the merger. This is a very optimistic scenario because 100 s should include the time to transmit the alert, respond to the trigger and re-point the instrument to the source position. Table~\ref{tab:jointafterglowpoiting} shows that a few detections are possible and going from the 10~km triangle to the full sensitivity 2L 20 km  increases of a factor of about 2 the detections.
As described in \cite{Ronchini:2022gwk}, there are two possible issues for the pointing strategy; being extremely fast to point to the source, and prioritising the events to be followed. No X-ray detection is possible being on source 1 hr after the merger with respect to 100 s because the X-ray afterglow emission is decaying fast. To detect a few events (see Table~\ref{tab:jointafterglowpoiting}) one needs to optimize the event selection among hundreds of events (see Table~\ref{tab:skyloccryoMM}) to be followed-up, to avoid to loose observational time. Selecting the sources localized better than 1000 $\rm deg^2$, the number of joint GW+X-ray detections increases by a factor of 10 but the selection of the event to be followed becomes more complicated. As shown in Table~\ref{tab:skyloccryoMM} the number of sources detected with sky-localization 1000 $\rm deg^2$ are thousands. Also prioritizing the events on the basis of the viewing angle by excluding all the sources edge-on (for which the viewing angle estimate is more precise), the number of events to be followed up remains to be around several hundreds (for the 10~km triangle) to thausands (for the 2L 20~km). While the 10~km triangle gives fewer joint detections with respect to 2L 20 km full sensitivity, the trigger selection is easier because it is among a smaller number of events. The HF-only 2L with 20~km misaligned arms gives a number of joint detection comparable to the 10~km triangle.  

{\em In summary, for the detection of the X-ray afterglow emission associated with the GW BNS signals there are no significant differences for the different configurations of ET.}

\begin{table}[t]
\centering

\begin{tabular}{|l|c|c|c|c|}
\hline
Instrument & $\Delta 10$ cryo & 2L 20 km cryo & $\Delta 10$ HF & 2L 20 km HF 	\\ \hline 
THESEUS-SXI pointing 	& $6_{-4}^{+4}$ &	$11_{-5}^{+6}$ & $1_{-1}^{+2}$ 	& $7_{-4}^{+4}$  \\ \hline
\end{tabular}
\caption{\small Numbers of joint GW+X-ray detections during one year of observation for THESEUS-SXI operating in pointing mode. Only BNS localised better than 100 deg$^2$ are followed-up.}
\label{tab:jointafterglowpoiting}
\end{table}

\subsection{Kilonovae: joint GW and optical detections}
\label{sect:MMOkilonova}
\begin{figure}
\begin{subfigure}{.5\textwidth}
  \centering
  \includegraphics[width=0.9\linewidth]{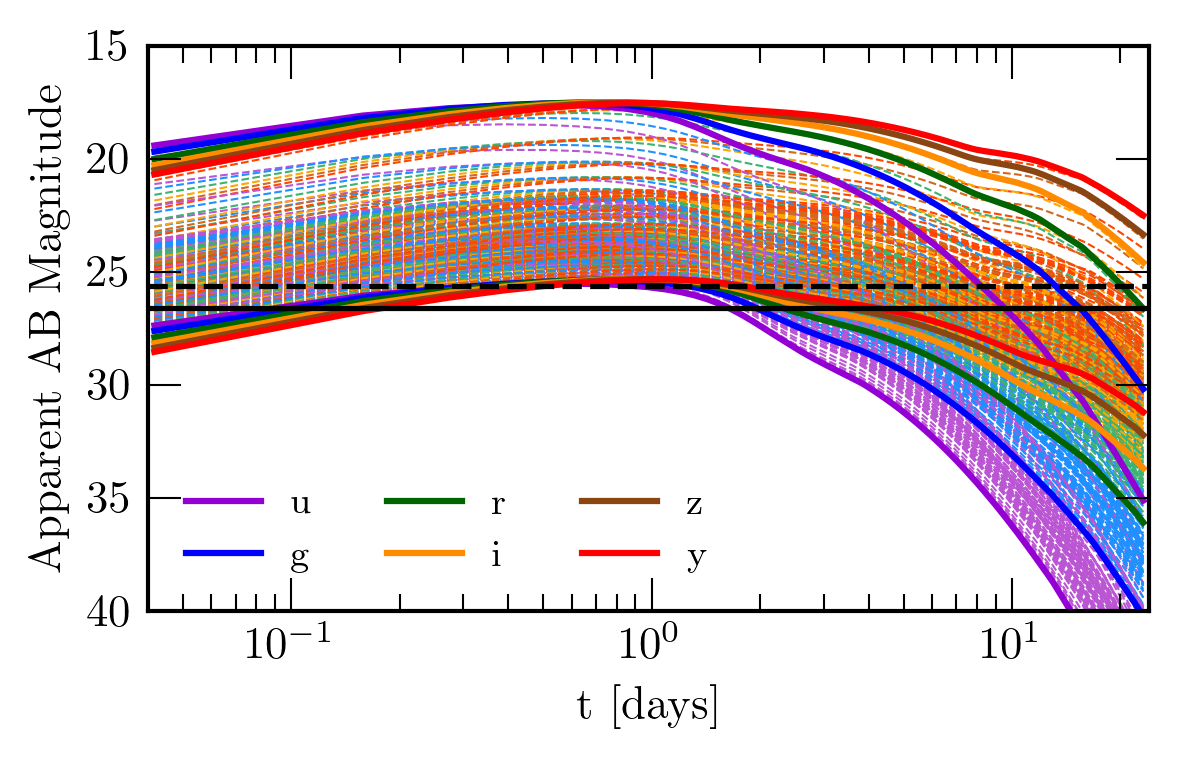} 
   \caption{\small $\Delta$ 10 km HFLF cryo}
\end{subfigure}
\begin{subfigure}{.5\textwidth}
  \centering
  \includegraphics[width=0.9\linewidth]{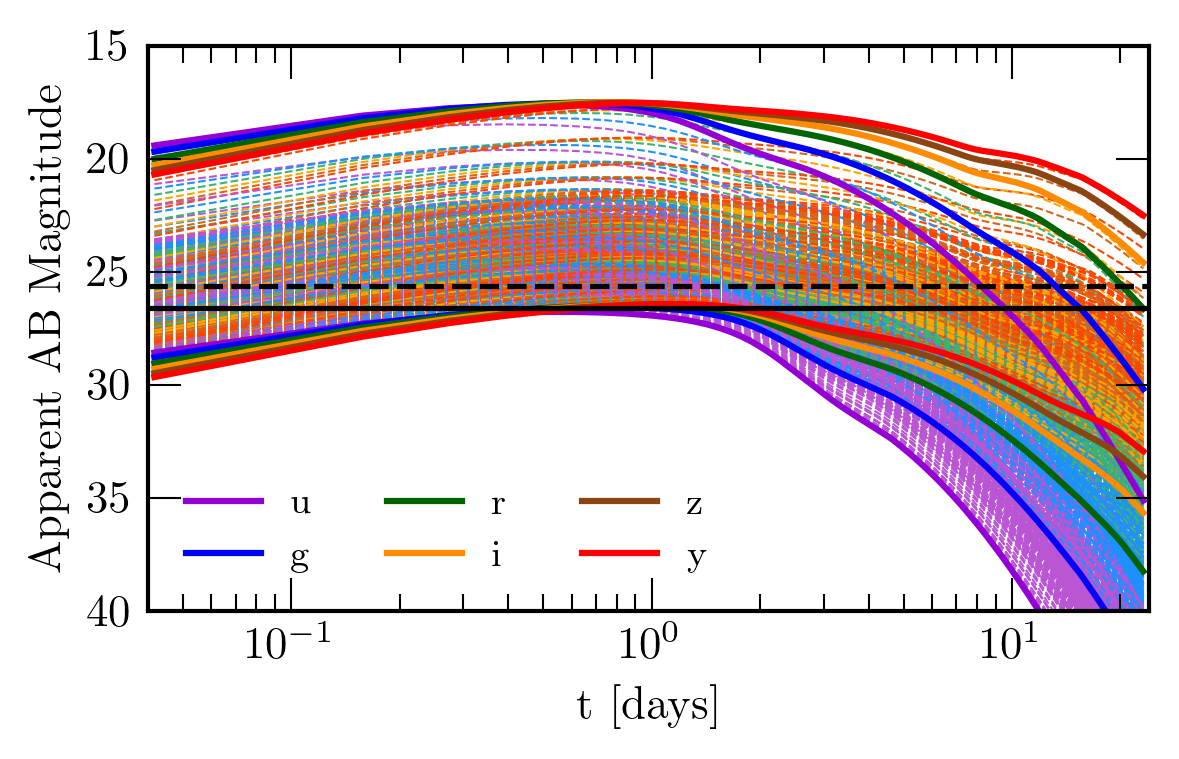}
   \caption{\small $\Delta$ 15 km HFLF cryo}
\end{subfigure}
\begin{subfigure}{.5\textwidth}
  \centering
  \includegraphics[width=0.9\linewidth]{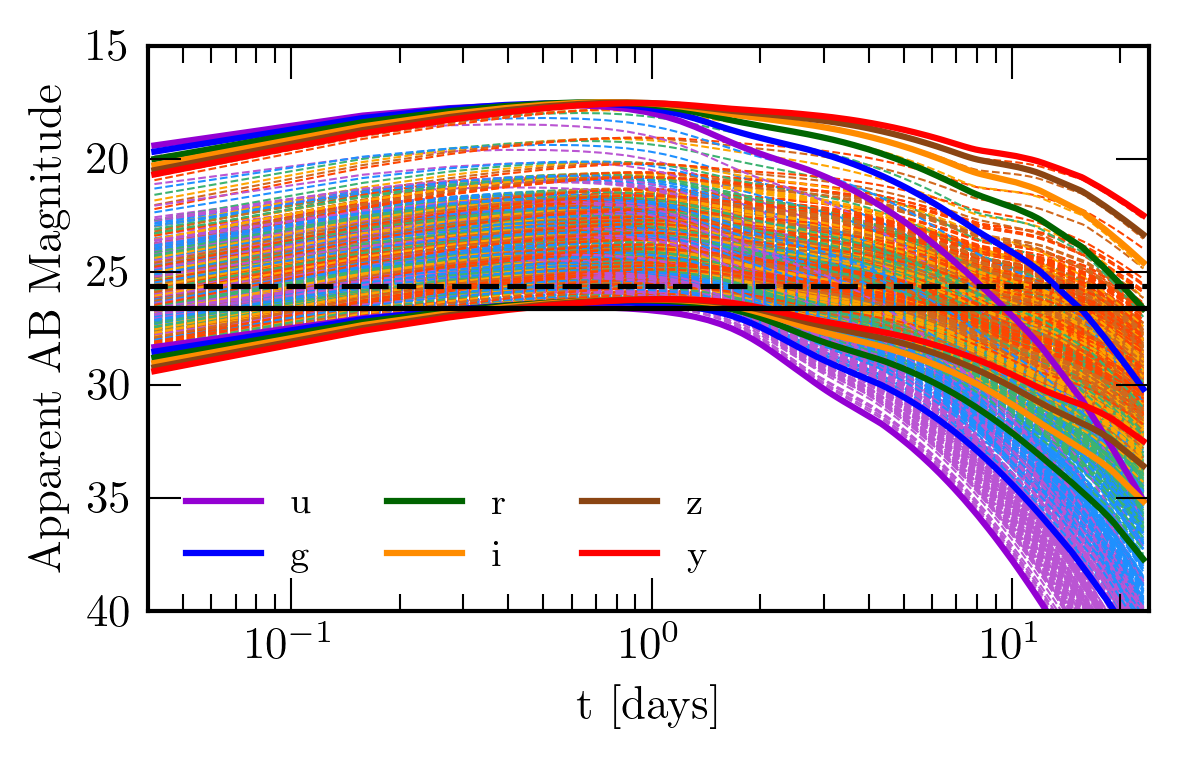} 
   \caption{\small 2L 15 km HFLF cryo}

\end{subfigure}
\begin{subfigure}{0.5\textwidth}
  \centering
  \includegraphics[width=0.9\linewidth]{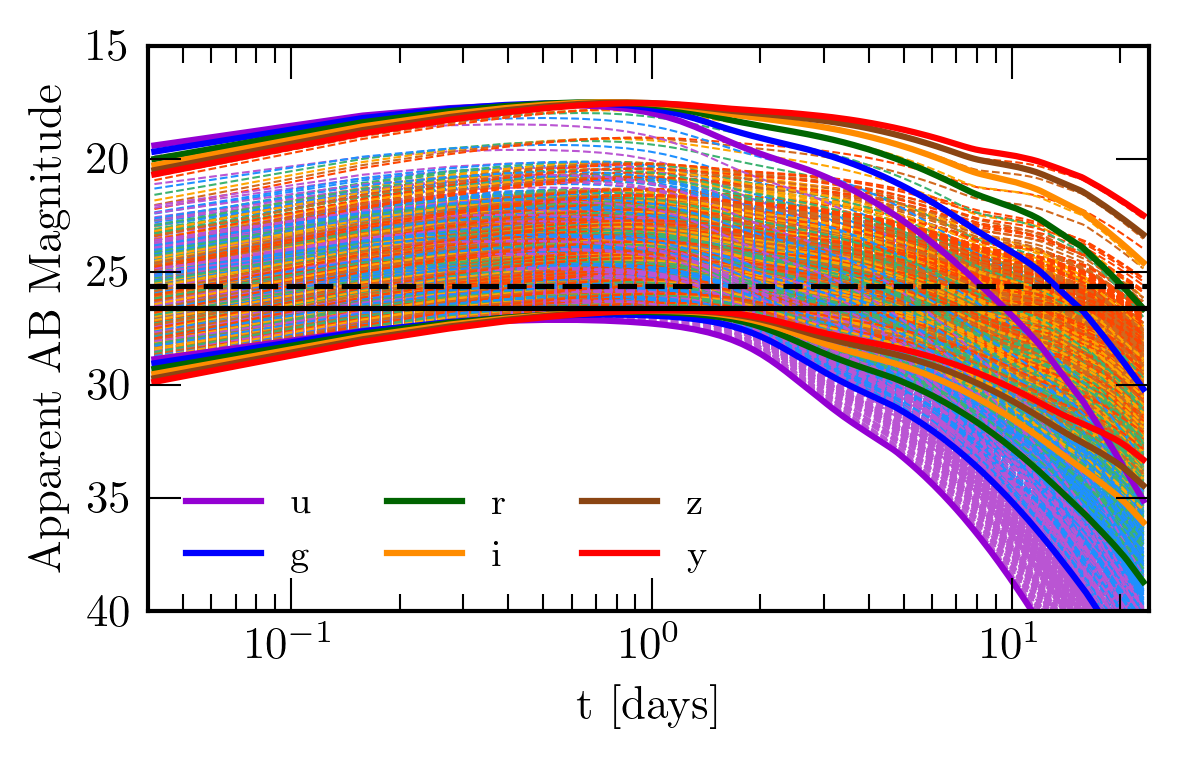} 
\caption{\small 2L 20 km HFLF cryo}
\end{subfigure}
\begin{subfigure}{.5\textwidth}
  \centering
  \includegraphics[width=0.9\linewidth]{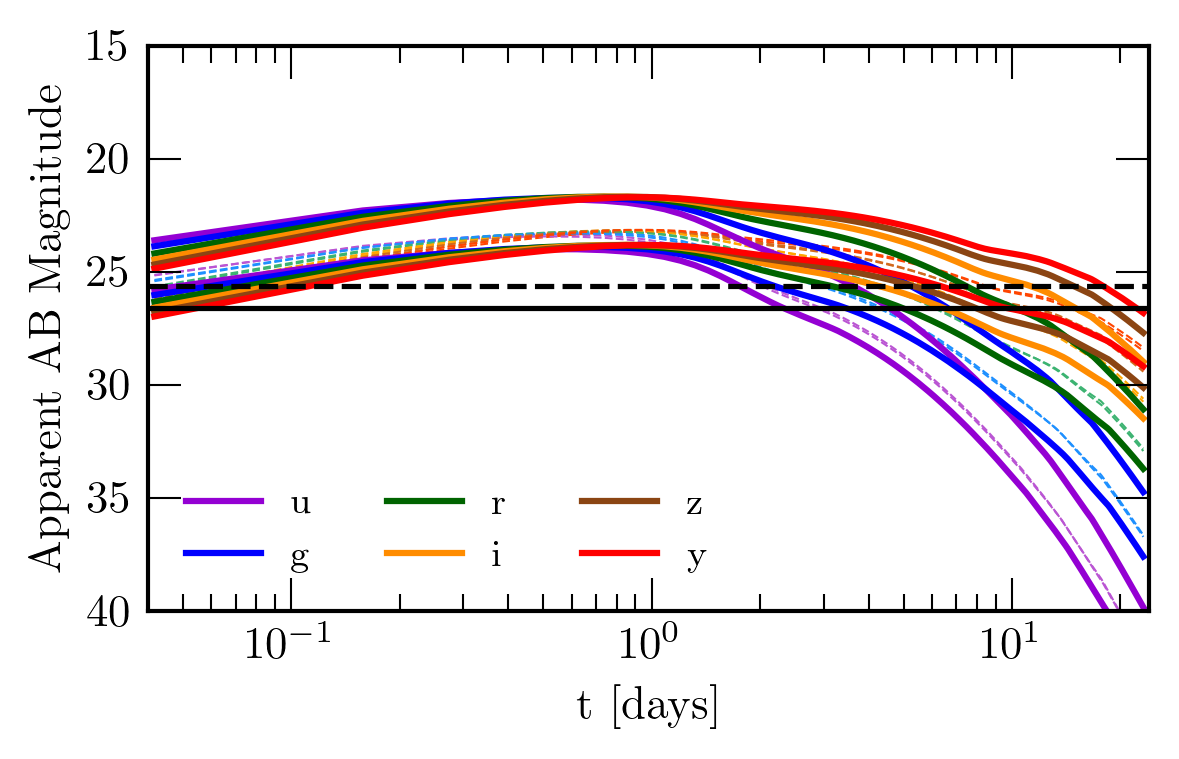}  
   \caption{\small $\Delta$ 10 km HF}
\end{subfigure}
\begin{subfigure}{.5\textwidth}
 \centering
\includegraphics[width=0.9\linewidth]{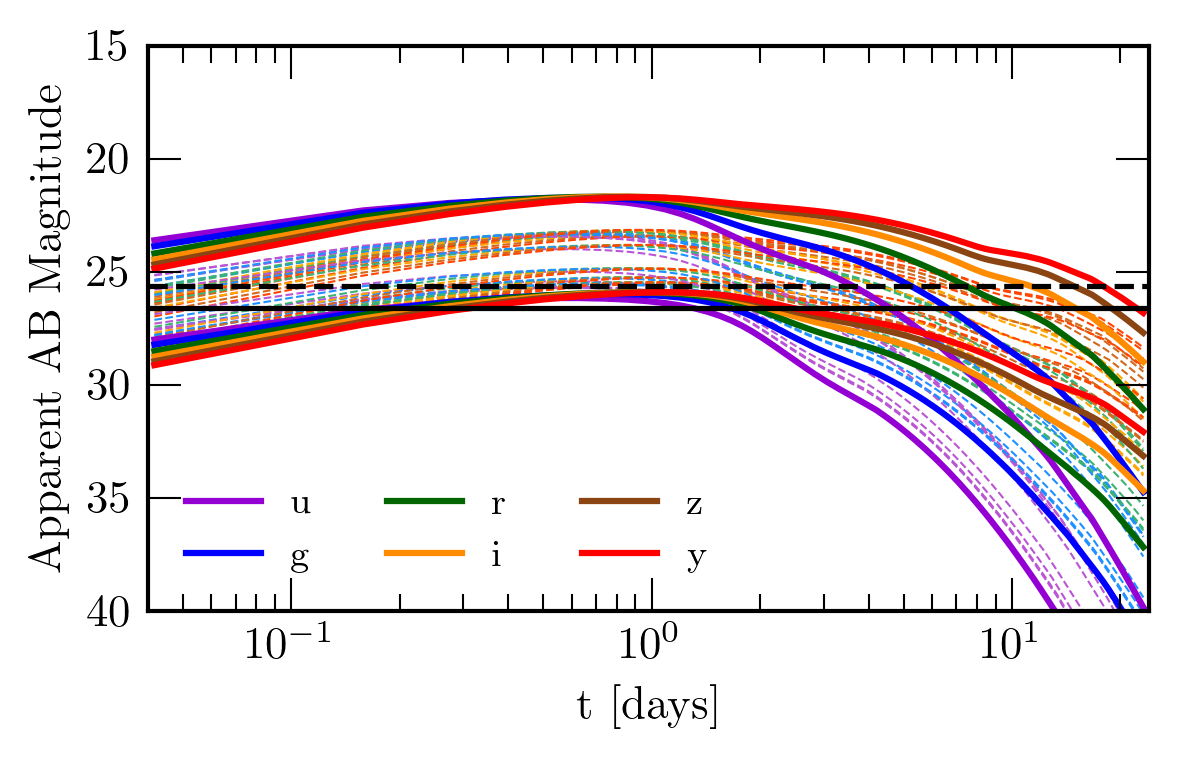} 
 \caption{\small $\Delta$ 15 km HF}
\end{subfigure}
\begin{subfigure}{.5\textwidth}
  \centering
  \includegraphics[width=0.9\linewidth]{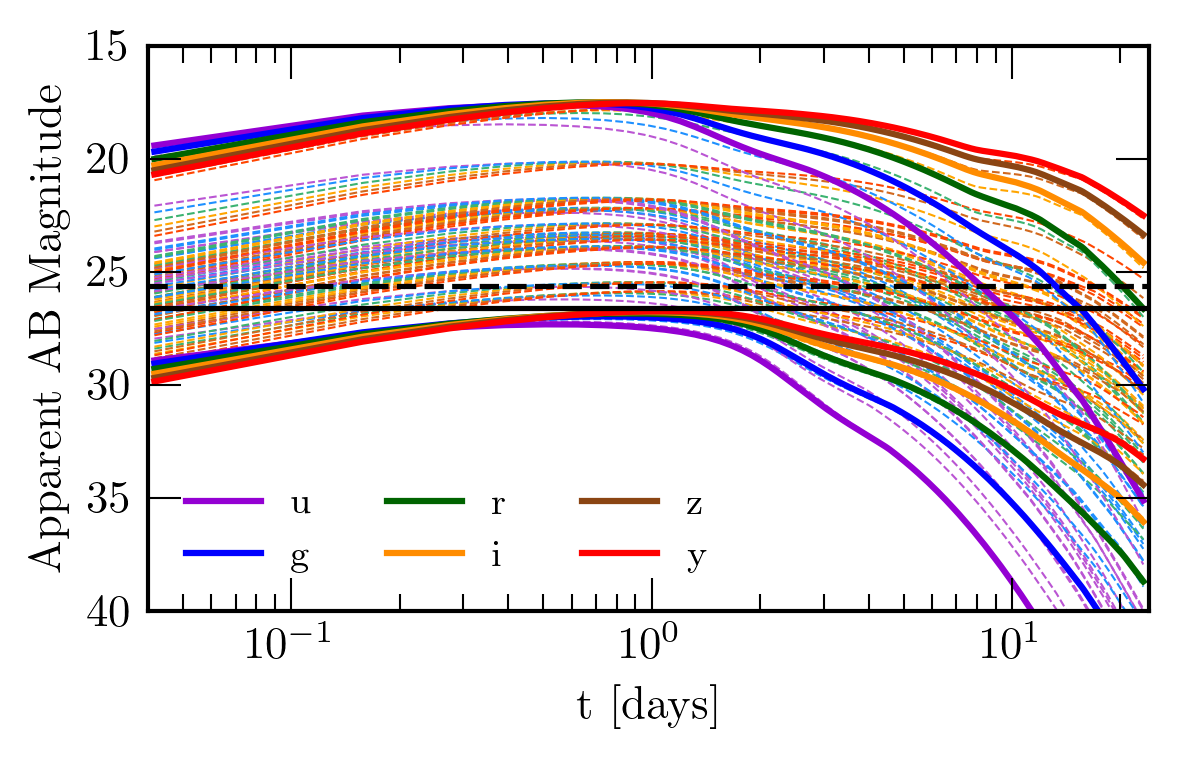}  
   \caption{\small 2L 15 km HF}
\end{subfigure}
\hfill
\begin{subfigure}{.5\textwidth}
  \centering
\includegraphics[width=0.9\linewidth]{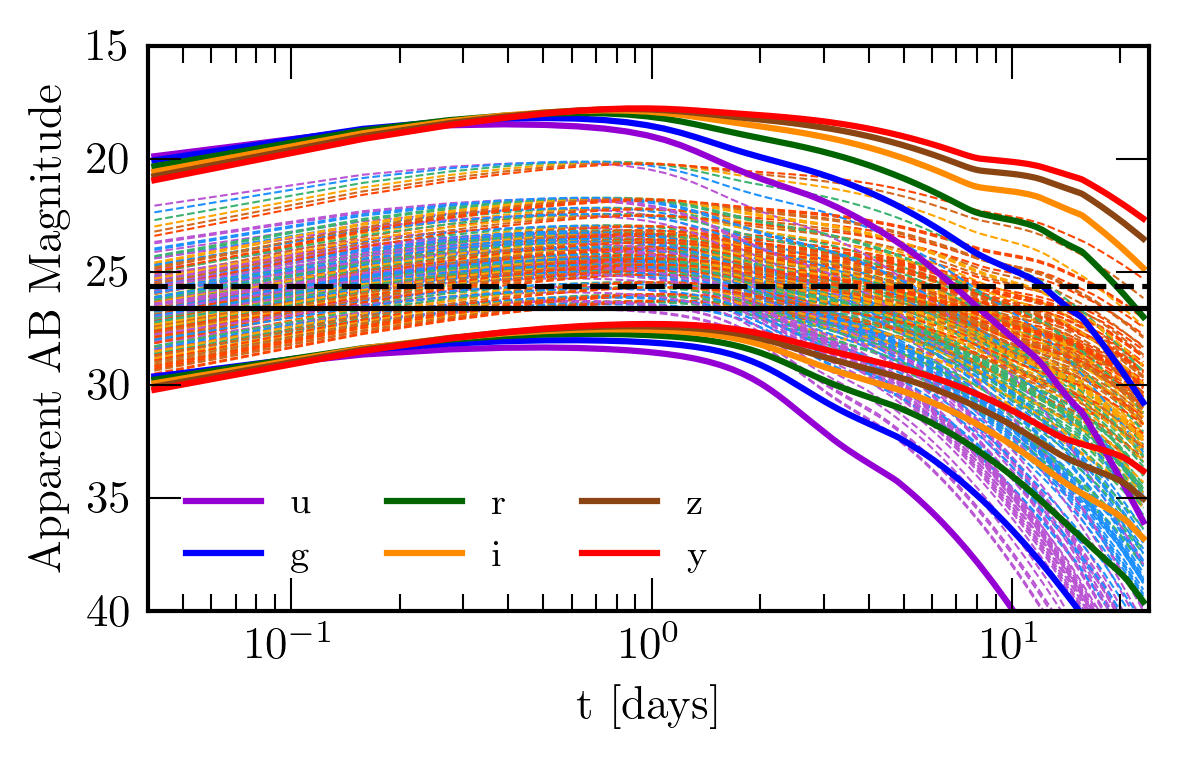} 
 \caption{\small 2L 20 km HF}
\end{subfigure}
\caption{\small Light-curves for GW170817-like signals associated with BNS detected with a sky-localization smaller than 40~$\rm deg^2$ by each of the full (HFLF cryo) sensitivity configurations (four top plot) and by each of the HF sensitivity configurations (four bottom plot). The black lines mark the (5$\sigma$ point-source depth) limiting magnitudes corresponding to 600 s of exposure for each pointing in the g band (solid black line) and the i band (dashed black line).}
\label{fig:GW170817lightcurve}
\end{figure}

\begin{table}[b]
\centering
Full (HFLF cryo) sensitivity detectors
\begin{tabular}{|l|c|c|c|c|c|c|c|}
\hline
Configuration&  $\rm N_{GW,VRO}$ & VRO  &  $\rm N_{GW,VRO}$ & VRO &  $\rm N_{GW,VRO}$ & VRO  \\
 &  {$\rm \Omega < 20\, deg^2$} & time & {$\rm \Omega < 40\, deg^2$} & time & {$\rm \Omega < 100\, deg^2$} & time\\
\hline
$\Delta 10$    &  14 (14)    &  1.1\% (3.3\%) &  36 (39)  &  5.1\%  (15\%)  &  96   &  40\%\\
\hline
$\Delta 15$     &  38 (42) & 3.3\%  (9.8\%)  &  84 (101)   & 14.2\% (42\%)  & 163   &  $>100\%$ \\
\hline
2L 15                  & 28 (28)  &  2.2\% (6.5\%) &  62 (77)  & 10.6\% (31\%) &  189  & 93\%\\
\hline
2L 20                 & 55 (64)   &  5\% (14.9\%)   &  115 (152)  &  23.1\% (68\%) &  324  & $>100$\% \\
\hline
\end{tabular}
\caption{\small Numbers of expected kilonovae detected by the VRO selecting the sources detected by the different configurations of ET with sky localization smaller than 
20~$\rm deg^2$ (column 2), 40~$\rm deg^2$ (column 4) and 
100~$\rm deg^2$ (column 6). Columns 3, 5, and 7 give the percentage of the VRO time necessary to follow up all the sources with sky-localization smaller than the above thresholds and within the VRO visibility sky area. The observational time for each event is obtained considering 600 s for each pointing and a number of the pointing corresponding to the ratio between the sky-localization uncertainty and the VRO FoV plus one pointing to be more conservative. We consider observations in two filter ({\em g} and {\em i}) the first night after the merger and the two filter observations repeated during the second night after the merger. We add 60~s to take into account  filter change, slew time and overheads each night. The percentage of VRO time is evaluated assuming 2600 hours as the total amount of observational time of the VRO in one year. The expected number of detected kilonovae and percentage of VRO time considering 1800~s for each pointing (instead of 600~s) are given in parentheses. For the 100~${\rm deg}^2$ threshold, assuming for each pointing of 1800~s entails that the number of hours requested to cover the entire sky-localization is larger than the ones available in one observing night. This makes our observational strategy unfeasible, and thus we exclude this case. It is also shown that to follow all events with sky-localization $< 100~{\rm deg}^2$, the required observational time becomes prohibitively large and for $\Delta 15$ and 2L 20 larger than the entire VRO observational time.}
\label{tab:jointKNFull}
\end{table}

\begin{table}
\centering
HF sensitivity detectors
\begin{tabular}{|l|c|c|c|c|c|c|c|}
\hline
Configuration &  $\rm N_{GW,VRO}$ & VRO  &  $\rm N_{GW,VRO}$ & VRO  &  $\rm N_{GW,VRO}$ & VRO \\
 &  {$\rm \Omega < 20\, deg^2$} & time & {$\rm \Omega < 40\, deg^2$} & time & {$\rm \Omega < 100\, deg^2$}  & time\\
\hline
$\Delta 10$     &   0 (0) &  0\% (0\%)  & 2 (2)   &  0.3\% (0.8\%) &  4  & 2\%\\
  \hline
$\Delta 15$     &  2 (2) &  0.2\% (0.5\%)  & 3 (4)   &  0.7\% (1.9\%) &  8  & 7.5\%\\
\hline
2L 15                 & 3 (4)  & 0.4\% (1.2\%)   &   7 (7) & 1.3\% (3.9\%)  &  26  & 11\%\\
\hline
2L 20                 & 5 (4)  &  0.6\% (1.6\%)  &   15 (18) &  3.1\% (9.3\%)  &  32  &  20.8\%\\
\hline
\end{tabular}
\caption{\small Same as \ref{tab:jointKNFull} but for ET detectors without the low-frequency.}
\label{tab:jointKNHF}
\end{table}

This section explores the perspectives for the detection of KNe. The KN emission is not beamed as GRBs and is expected from all viewing angles. We consider as reference observatory, the Vera Rubin Observatory (\acrshort{vro}). VRO is the innovative 8.4-meter telescope characterized by a wide field (9.6 ${\rm deg}^2$) camera containing over three billion pixels of solid-state detectors \cite{VeraRubin}. Its compact shape enables to point and slew through large sky-region extremely fast. It represents the ideal instrument to observe the GW sky localizations obtaining deep observations of extremely high quality. In order to evaluate the number of possible joint optical/GW detections, we model the kilonova emission as in \cite{Perego2017:aiu} taking the best-fit parameters obtained to reproduce the multi-filter optical observations of the kilonova AT2017gfo associated with GW170817. Our model includes cosmological and k-correction. We select all the BNS mergers detected with a sky-localization smaller than a threshold and we associate to each of them a GW17087-like kilonova taking into account the source distance and the viewing angle (our modeling starting from GW170817 reproduces the angular dependence of the expected emission). We assume three threshold on the sky-localization by selecting all the events with sky-localization $\rm < 20 deg^2, < 40 deg^2, and < 100 deg^2$. We consider a Target of Opportunity strategy similar to the ones described in \cite{Andreoni2022:oij,Cowperthwaite2019,Margutti2018}; we use observations in two filters ({\em g} and {\em i}) the first night after the merger and we repeat the two filter observations during the second night after the merger. We consider a detection when the kilonova emission is detected at $5\sigma$ the first and second night in both  filters. Having observations in at least two filters is important to reduce the  number of contaminating transients and identify the kilonova associated with the GW signal on the basis of the color evolution. The problem of contaminating transients is solved in real observations also using spectroscopy able to characterize the source, but here for the configurations comparison, this is not taken into account. Our counts take into account the visibility of the VRO.

Figure~\ref{fig:GW170817lightcurve} shows the light-curves for GW170817-like signals associated with our population of BNSs detected with a sky-localization smaller than 40$\rm deg^2$ by the full (HFLF cryo) sensitivity configurations and by the HF sensitivity configurations. The plots show the different numbers of events selected to be followed up for the different configurations. It is immediately clear the significant increase in the number of targets to be followed going from configurations with only high-frequency to the full sensitivity configurations. 

Table~\ref{tab:jointKNFull} gives the number of expected kilonovae by following up all the events within the VRO visibility sky-area and with a sky-localization uncertainty smaller than 20 $\rm deg^2$, 40 $\rm deg^2$ and 100 $\rm deg^2$, and the corresponding percentage of VRO time necessary for the follow-up. Enlarging the threshold on the sky-localization significantly increases the number of events to be followed-up and thus the number of joint detections but at the expense of more observational time. Also increasing the exposure from 600~s to 1800~s for each pointing tends to increase the number of detections but the percentage of time to be used becomes prohibitive. Looking at the full sensitivity detectors, {\em the 2L with 20~km misaligned arms is the best performing in detecting kilonovae. It enables to detect between several tens and a few hundred kilonova counterparts per year. The 15~km triangle is slightly better than the 2L with 15~km misaligned arms giving a number of detection about 30\% larger\footnote{The scenario of following up all the events with sky-localization $< 100~\rm deg^2$ case is not considered because the required observational time of VRO is prohibitive.}. The 15~km triangle is significantly better than the 10~km triangle giving about a factor 2 larger number of detections.}

{\em The presence of low-frequency is critical for ET operating as single observatory to detect a large number of kilonovae counterparts.} Table~\ref{tab:jointKNHF} shows the small number (a few) of detections per year expected with the triangle--HF configurations. This number increases to a few tens for the 2L--HF configurations. 

We highlight that the absolute number per year of VRO detections is affected by the error on the BNS local rate normalization. The astrophysical rates inferred from the observations of the first, second and third run of observations of LIGO and Virgo give a BNS merger rate in the range 10 to 1700 $\rm Gpc^{-3} yr^{-1}$ \cite{LIGOScientific:2021psn}. Since our population corresponds to a local rate $R_0\simeq 250\, {\rm Gpc}^{-3}\, {\rm yr}^{-1}$, the actual numbers could  be up to one order of magnitude smaller or larger than the ones given in the Tables.

\section{Stochastic backgrounds}\label{sect:stochastic}

The stochastic GW background  (\acrshort{sgwb}) 
is formed by
\textbf{}the incoherent superposition of signals emitted by different GW sources in our Universe, primordial or astrophysical, at different redshifts, that we collect at our detector.  Traditionally, we distinguish  between an astrophysical GW  background  (\acrshort{agwb}) and a background of cosmological origin (\acrshort{cgwb}). The latter are tensor modes produced by different processes in the early universe such as  inflation, reheating, phase transitions, cosmic strings, or primordial black holes (see \cite{Maggiore:1999vm,Caprini:2018mtu,Maggiore:2018sht} for reviews). If detected, it would  provide us with direct information about the very first instants of time of the evolution of our Universe. On the other hand, the AGWB is made of the superposition of all GWs emitted by different populations of astrophysical sources, from the onset of stellar activity until the present epoch.  The CGWB and the AGWB are the gravitational counterparts of the Cosmic Microwave Background (\acrshort{cmb}) \cite{durrer_2008} and of the Cosmic Infrared background (\acrshort{cib}) \cite{Hauser_2001} in the electromagnetic context: in the same way as the CIB is a foreground for the CMB, which needs to be properly characterised and subtracted to access the CMB and its cosmological content, astrophysical backgrounds of gravitational waves represent foregrounds for primordial backgrounds of gravitational radiation.  Typically it is expected that different GW sources are characterized by a different frequency profile, that can be used as a tool to disentangle them~\cite{Caprini:2019pxz, Flauger:2020qyi}.
As recalled in App.~\ref{app:PLS}, a stochastic background is characterized by the quantity
\begin{equation} \label{eq:Abkg:def}
\Omega_{\rm gw}(\nv, f) \equiv  \frac{1}{\rho_c} \frac{d\rho_{\rm gw}(\nv,f)}{d\ln f d^2\nv}\, ,
\end{equation}
where $d\rho_{\rm gw}(\nv,f)/d\ln f d^2\nv$ is the energy density per unit logarithmic frequency and unit solid angle that reaches the observer from the direction $\nv$ and with frequency $f$, see \eqst{tildehnv}{eq:Abkg:def:App}.
The isotropic (normalized) energy density is defined  by 
\be
\Omega_{\rm gw}(f)\equiv \int d^2\nv\,\Omega_{\rm gw}(\nv, f)\,,
\ee
and gives the (normalized) energy density of the stochastic background reaching the observer from all directions. It is commonly assumed that it is possible to factorize the dependence on the frequency and that on the direction, i.e. 
\be
\Omega_{\rm gw}(\nv, f)={\cal E} (f) \mathcal{P}(\nv)\,.
\ee
Furthermore, for many cosmological backgrounds (but not, for instance, for the backgrounds generated by cosmological phase transitions)  the scale of variation with frequency is such that, over the bandwidth of a ground-based or a  space-borne GW detector, the frequency dependence  can be approximated by a simple power-law  ${\cal E}(f)=A_{\beta}\,(f/f_{\text{ref}})^{\beta}$, where $A_{\beta}$ is the amplitude at an arbitrarily chosen  reference frequency $f_{\text{ref}}$.

In the presence of statistical isotropy, the statistical properties of the background energy density can be characterised in terms of a frequency-dependent angular power spectrum as 
\be\label{Cl}
\langle \Omega_{\rm gw}(\nv, f) \Omega_{\rm gw}(\nv', f)\rangle=\sum_{\ell} C_{\ell}(f) P_{\ell}(\nv\cdot\nv')\,,
\ee
where $P_{\ell}(\nv\cdot \nv')$ are Legendre polynomials, functions of the angular separation between the two directions $\nv$ and $\nv'$. The angular power spectrum and the frequency spectrum ${\cal E}(f)$ encode most of the statistical properties of a given background component. Recently, it has been shown that, similarly to CMB photons, primordial GWs are also affected by large-scale anisotropies (i.e., Sachs-Wolfe and Integrated Sachs-Wolfe effects)~\cite{Contaldi:2016koz, Geller:2018mwu, Bartolo:2019oiq, Bartolo:2019yeu}. 
Typical CGWB components are expected to have the same level of anisotropy as the CMB: a scale-invariant spectrum $\ell(\ell+1) C_{\ell}\propto {\rm  constant}$, and a level of anisotropy of the order of $\sim10^{-5}$ with respect to the monopole \cite{Geller:2018mwu}. 
In contrast, extra-galactic AGWBs have a scaling given by clustering, resulting in  $(\ell+1) C_{\ell}\propto {\rm constant}$, and a higher level of anisotropy, at the level of $\sim10^{-2}$ with respect to the monopole \cite{Cusin:2017mjm, Cusin:2017fwz, Cusin:2018rsq,  Cusin:2019jpv, Yang:2020usq, Jenkins:2018uac, Jenkins:2018kxc, Jenkins:2018lvb,  Cusin:2019jhg, Jenkins:2019uzp, Jenkins:2019nks,  Cusin:2018avf, Bertacca:2019fnt, Pitrou:2019rjz, Alonso:2020mva} (see ~\cite{Bellomo:2021mer} for a recent public code to compute the anisotropies of the AGWB).
Interestingly, both the CGWB and the AGWB anisotropies show a correlation with CMB anisotropies which can be used both as a source discriminator and for testing systematics~\cite{Adshead:2020bji,Ricciardone:2021kel, Galloni:2022rgg}.

While the production of tensor modes from the early universe can typically be described in terms of a continuous emission, varying on a time scale extremely small with respect to typical observation time-scales,\footnote{For this reason most of the CGWB components represent continuous and stationary background components and they are often referred to as \emph{irreducible} backgrounds.} the AGWB in the frequency band of ET is expected to have a so-called \emph{popcorn-like nature}, due to the discreteness of emissions in time. 
Indeed, in the Hz band, we expect the AGWB to be dominated by mergers of compact objects and, considering the short duration of the merger phase, events are expected to be separated in time and with a limited time overlap (some overlap is expected for BNS mergers). As a consequence, in the frequency band of ET, the angular power spectrum of the AGWB from mergers of compact binaries has an important popcorn component (due to the discrete emission in time),  which adds to the clustering one (due to the discreteness of the spatial galaxy distributions) \cite{Cusin:2019jpv, Jenkins:2019uzp, Jenkins:2019nks, Alonso:2020mva}.  Formally, the total angular power spectrum is given by $
C_{\ell}^{\rm AGWB, tot}=C_{\ell}+N^{\rm pop}$,
where the first term on the right-hand side is the contribution from clustering while the second component represents popcorn noise. The latter is flat in $\ell$-space (it is just an offset) and it is expected to cover the clustering contribution, see \cite{  Cusin:2019jpv, Jenkins:2019uzp, Jenkins:2019nks, Alonso:2020mva}. Even if popcorn noise contains astrophysical information, it does not provide any information about the spatial distribution of sources. A possible way to overcome this problem, i.e. to extract the clustering part out of the popcorn noise threshold, is to consider cross-correlations between an SGWB map and galaxy distributions, see e.g. \cite{Alonso:2020mva}. 
The study of cross-correlations, besides being an independent observable interesting on its own, provides one with a powerful detection tool, as it typically has a higher SNR than the auto-correlation, see e.g. \cite{Cusin:2018rsq, Alonso:2020mva, Cusin:2019jpv,Yang:2020usq, Capurri:2021zli} in the context of the extra-galactic astrophysical background.

Finally, kinematic anisotropies induced by a peculiar motion of the observer with respect to the rest frame of emission are expected to be quite important and potentially detectable if the corresponding monopole has a sufficiently high SNR, see e.g. \cite{Cusin:2022cbb}. Recent studies to extract the cosmic dipole using next-generation detectors have been developed both for the AGWB~\cite{DallArmi:2022wnq, Cusin:2022cbb} and for resolved sources~\cite{Mastrogiovanni:2022nya}.

We conclude this section with a remark. Because of the discrete emission in time, the AGWB in the Hz band is not irreducible: with a perfect instrument with infinite sensitivity (and with infinite computational power), all events would be detectable individually. As the sensitivity of next-generation ground-based detectors will improve more and more events will be detectable individually: in this context, it is interesting to study the so-called \emph{residual background}, i.e. the background contribution that remains after having filtered out resolvable objects. This observable can indeed provide us with a unique way of extracting information on a faint and distant population, that cannot be detected with a catalogue approach. We will get back to this point later in this section.



\subsection{Sensitivity to isotropic stochastic backgrounds }\label{BackgroundSensitivities}

We now study the sensitivity, in different configurations, to a SGWB. To this goal, we compute the 
Power-Law integrated  Sensitivity curve (PLS), whose definition is recalled in App.~\ref{app:PLS},
for the triangular and L-shaped detectors, in the configurations discussed in Section~\ref{sect:geometry}, so for the triangular configuration we consider 10 and 15-km arm, while for the L-shaped we consider 4 configurations: 15 km and 20 km arms, aligned and misaligned (by 45 deg) detectors located one in Sardinia and one in the Netherlands. We then turn to the angular sensitivity for the same ET configurations, i.e. we present how well  the various configurations can reconstruct the $C_{\ell}(f)$ (i.e., the angular power spectrum), defined in \eq{Cl}.

\begin{figure}
    \centering
    \includegraphics[width=12cm]{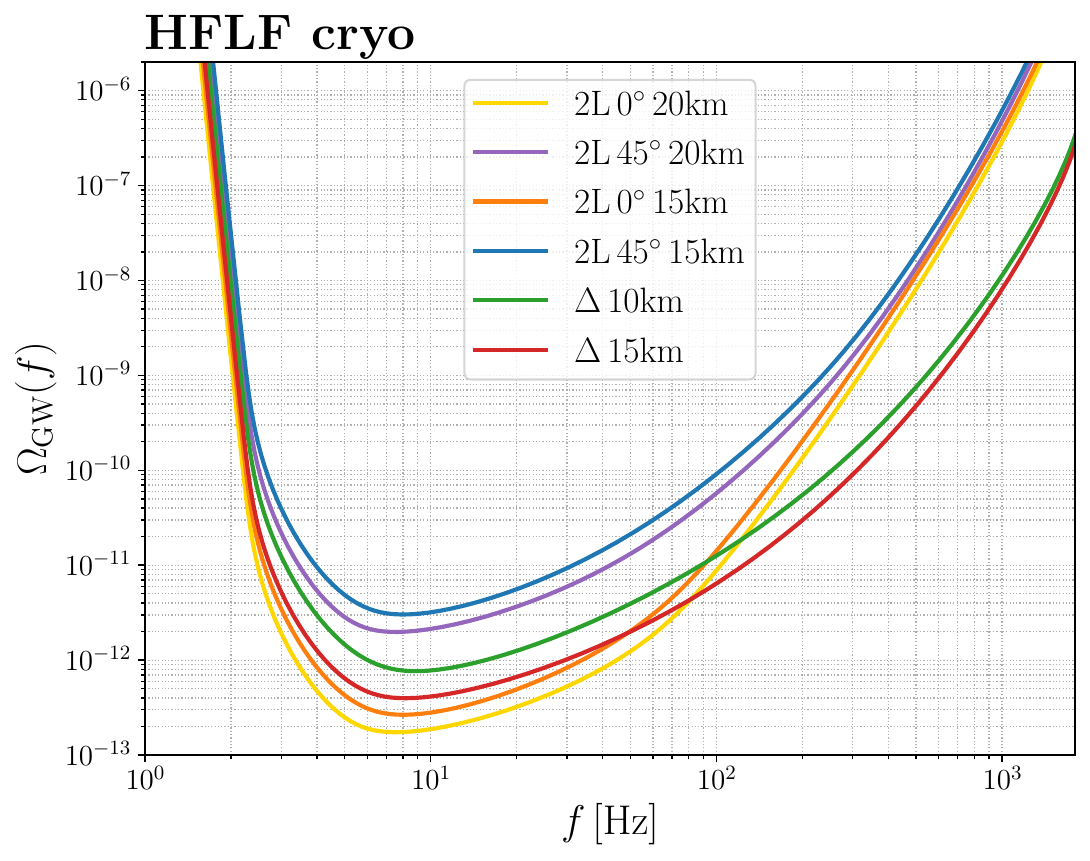}
    \caption{\small Power-law integrated sensitivity (PLS) to $\Omega_{\rm GW}(f)$ for  different ET geometries, in the full HFLF-cryo configuration. We set   ${\rm SNR}_{th} = 1$ and $T_{\rm obs} = 1 {\rm yr}$ and we include only cross-correlations between detectors, excluding auto-correlations.}
    \label{fig:ET_PLS_auto_cross_SNR_1a}
\end{figure}

\begin{figure}
    \centering
    \includegraphics[width = 12cm]{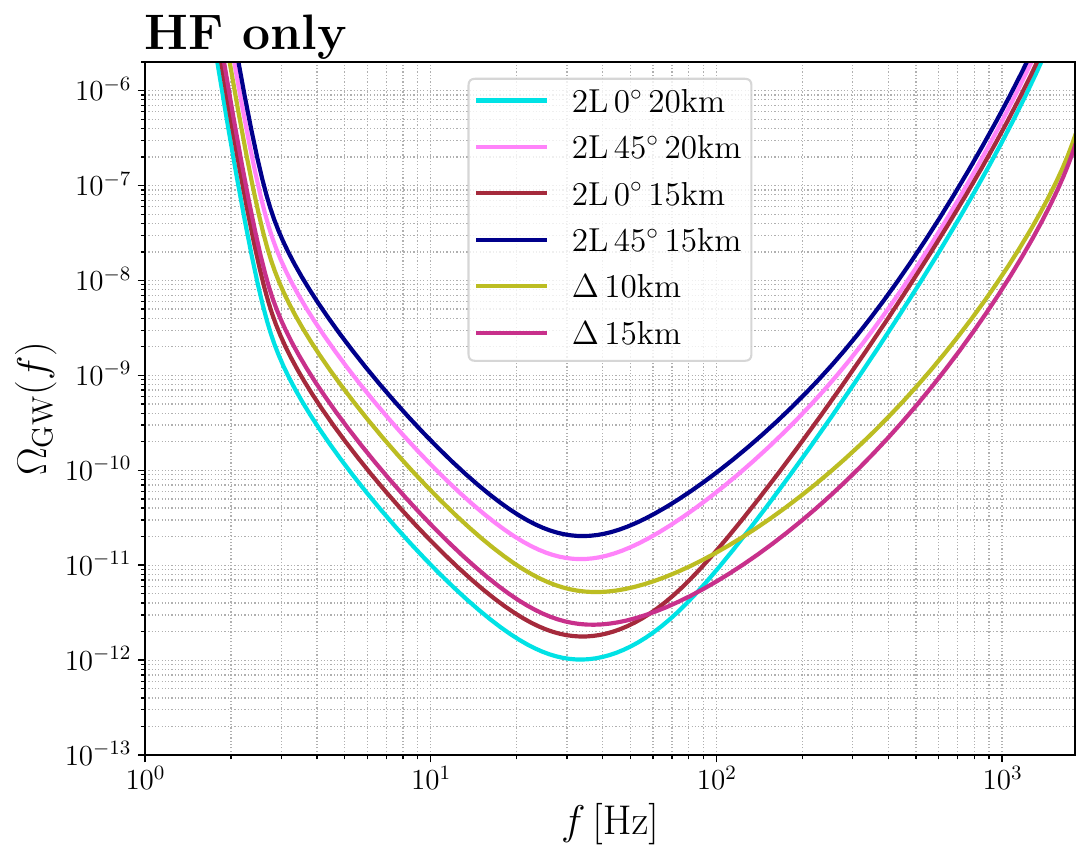}
    \caption{\small As in Fig.~\ref{fig:ET_PLS_auto_cross_SNR_1a}, for the HF-only configurations.}
    \label{fig:ET_PLS_auto_cross_SNR_1b}
\end{figure}

\begin{figure}
    \centering
    \includegraphics[width = 7.cm]{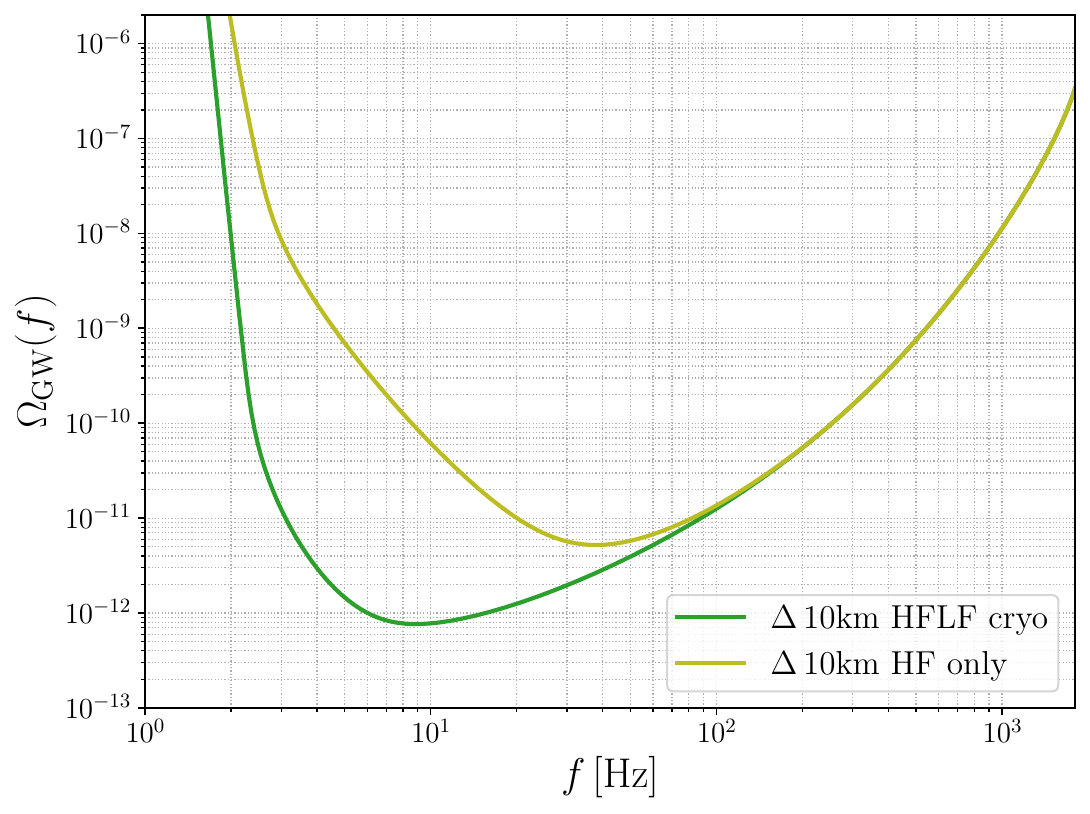}
    \includegraphics[width = 7.cm]{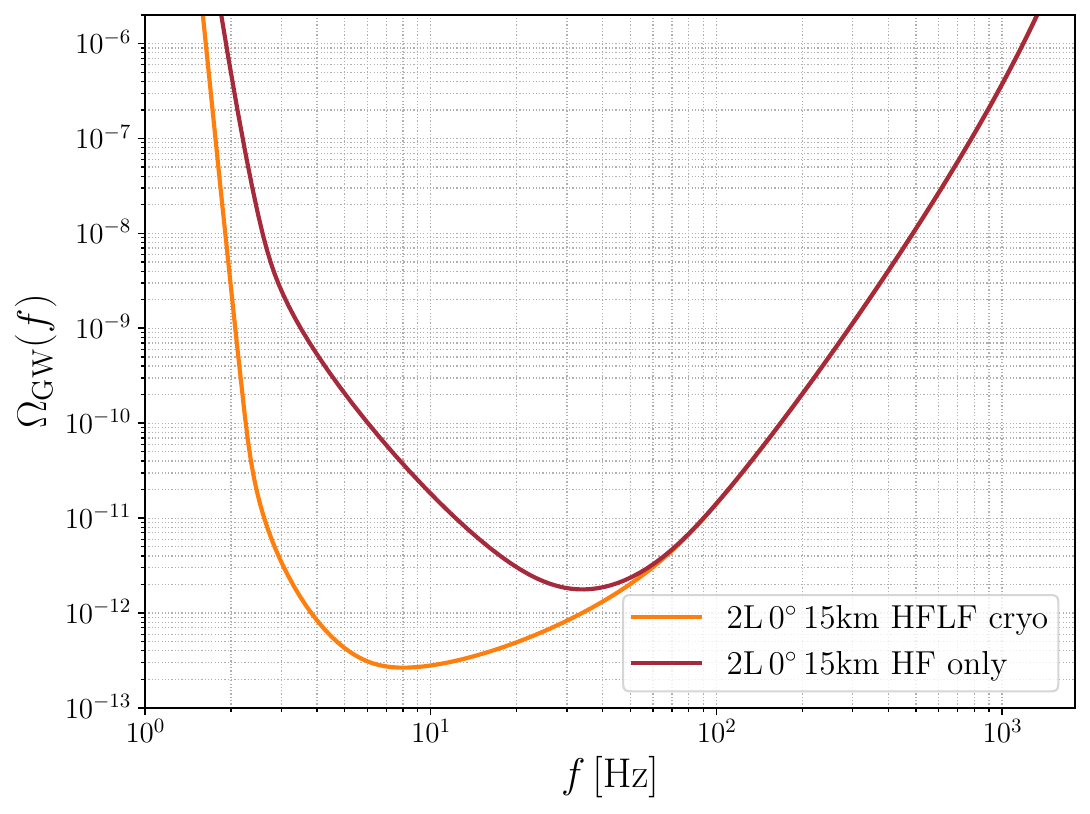}
    \caption{\small Left: ET power-law integrated sensitivity for the triangle configuration for the full HFLF-cryo configuration and the HF-only configuration, assuming ${\rm SNR}_{th} = 1$ and $T_{\rm obs} = 1 {\rm yr}$ and including only cross-correlations between detectors. Right: The same for the 2L-15km-$0^{\circ}$ configuration. }
    \label{fig:ET_PLS_auto_cross_SNR_1_room}
\end{figure}

The result is shown in Fig.~\ref{fig:ET_PLS_auto_cross_SNR_1a} for the different geometries in the full HFLF-cryo configuration, and in Fig.~\ref{fig:ET_PLS_auto_cross_SNR_1b} for the  HF-only configuration.\footnote{Recall from Section~\ref{sect:geometry} that, for the 2L cases,  the alignment between the two detectors has been  defined using the relative angle $\alpha$  measured with respect to the  north of the Sardinia site for both detectors, rather than with respect to the great circle passing through  the  two sites. Therefore the $\alpha=0^{\circ}$  and the $\alpha=45^{\circ}$ cases shown in the figures below do not correspond to the ideal and, respectively, the worst configurations for stochastic searches, but have slight offsets from it. In terms of the angle $\alpha$, the ideally parallel and the perfectly misaligned for the configurations correspond to $\alpha=2.51^{\circ}$ and $\alpha=47.51^{\circ}$, respectively. This is particularly important
for the misaligned case. Indeed, $\alpha=47.51^{\circ}$ would correspond to essentially zero sensitivity to stochastic backgrounds, while we see from Fig.~\ref{fig:ET_PLS_auto_cross_SNR_1a} that, for $\alpha=45^{\circ}$, we already have an interesting sensitivity to stochastic backgrounds, so a small offset from the perfectly misaligned configuration is already very useful for recovering a part of the sensitivity to stochastic searches (without basically affecting the searches for compact binary coalescences, as discussed in
Section~\ref{sect:geometry}).
See App.~\ref{app:StocasticMisalignement} for more details.}

{\em We observe that the triangular configurations perform better at high frequency since the overlap reduction function  remains constant at high frequencies for co-located detectors rather than falling to zero. The 2L configurations do better at low frequencies, up to several tens of Hz, but only for co-aligned detectors (since this maximises the coherence for the stochastic search). The case of 45 deg 2L configurations are universally the worst.}

Fig.~\ref{fig:ET_PLS_auto_cross_SNR_1_room} compares the power-law sensitivities obtained from the HFLF-cryo and the  HF-only ASDs for the 
2L~15km with parallel arms (left panel) and for the
10~km triangle (right). Since  $\Ogw(f)$ is quadratic in the strain, the loss of the low-frequency sensitivity has a huge impact on the sensitivity to stochastic backgrounds at low frequencies. For instance, for
the  2L~15km with parallel arms,
the HF-only configuration is worse than the HFLF-cryo by a factor $\sim 10^2$ at 10~Hz, and  $\sim 10^3$ at 3~Hz.
In contrast, the sensitivity above about 60~Hz is unaffected.

\subsection{Angular sensitivity}\label{sect:angularresstochastic}

In this section we study the angular resolution of different ET configurations in the context of an SGWB. We follow the formalism in \cite{Alonso:2020rar}, whom we refer the reader for details and derivations (see also \cite{Mentasti:2020yyd,Mentasti:2023gmg} for recent forecasts about the ET sensitivity to angular anisotropies).
The sensitivity with which we can measure a given multipole of the background intensity is characterized by a quantity, $N_\ell$, such that the signal-to-noise ratio  of an angular power spectrum is  given by
\be
{\rm SNR}^2\equiv \sum_{\ell} (2\ell+1)\frac{C_{\ell}}{N_{\ell}}\,,
\ee
where the $C_{\ell}$ have been defined in \eq{Cl}.
The $N_\ell$
can be computed from  
      \begin{align}\label{eq:Inoise}
        N_\ell^{-1}\equiv\frac{1}{2}\sum_{ABCD}\int df\,\int dt\,G_\ell^{AB,CD}(t,f)\,,
      \end{align}
where
      \begin{align}\label{Gl}
        &G_\ell^{AB,CD}(t,f)\equiv\left(\frac{2 {\cal E}(f)}{5}\right)^2\left({\sf N}^{-1}_f\right)^{AB}\left({\sf N}^{-1}_f\right)^{CD}\frac{\sum_m{\rm Re}\left(\resp_{BC,\ell m}(t,f)\resp^{*}_{DA,\ell m}(t,f)\right)}{2\ell+1}\,,
        \end{align}
and
\be
\resp_{AB,\ell m}(t,f)\equiv\int d^2\nv\, Y^*_{\ell m}(\nv)\,\resp_{AB}(t,f,\nv)\,,
\ee
is the angular multipole of the antenna pattern;  we explicitly indicated its time dependence. These expressions allow one to predict the angular power spectrum of a given experiment and scan strategy. 
A key assumption made in obtaining this result is that the time-integrated response of the instrument can be compressed into an angular power spectrum. In practice, this will not be the case, since the scan strategy cannot be designed to average the noise over the $m$-modes and the resulting estimator will be sub-optimal. This is a necessary assumption for any analytical estimate; however, in practice, optimal estimators will have to account for inhomogeneous noise in $m$-modes. 
      
Eq.\,(\ref{eq:Inoise}) involves a two-dimensional integral over $t$ and $f$, where the integrand involves at least one computationally expensive spherical harmonic transform of the antenna pattern $\resp$. This can be simplified further for experiments with a constant configuration (e.g. constant arm lengths, angular separations between arms, relative detector positions). 
As explained in \cite{Alonso:2020rar}, in this limit the network changes as a function of time as a rotating solid. Hence the antenna patterns at different times are related to each other through simple three-dimensional rotations (i.e. $\mathcal{A}(t,f,\nv)=\mathcal{A}(f,{\sf R}_t\nv)$, where ${\sf R}_t$ is a time-dependent rotation matrix). 
Recalling that a rotation only mixes the $m$-modes for each fixed $\ell$ in $\mathcal{A}_{\ell m}$, the function $G_\ell$ in (\ref{Gl}) does not depend on $t$ (assuming detector noise to be stationary). In this case one gets the compact expression 
\be\label{eq:Inoise_stationary}
N_{\ell}^{-1}=\frac{T_{\text{obs}}}{2}\sum_{ABCD}\int df\, G_{\ell}^{AB, CD}(f)\,,
\ee
where $T_{\rm obs}$ is the total observing time. 
     
\begin{figure}
    \centering
        \includegraphics[width = 7.1cm]{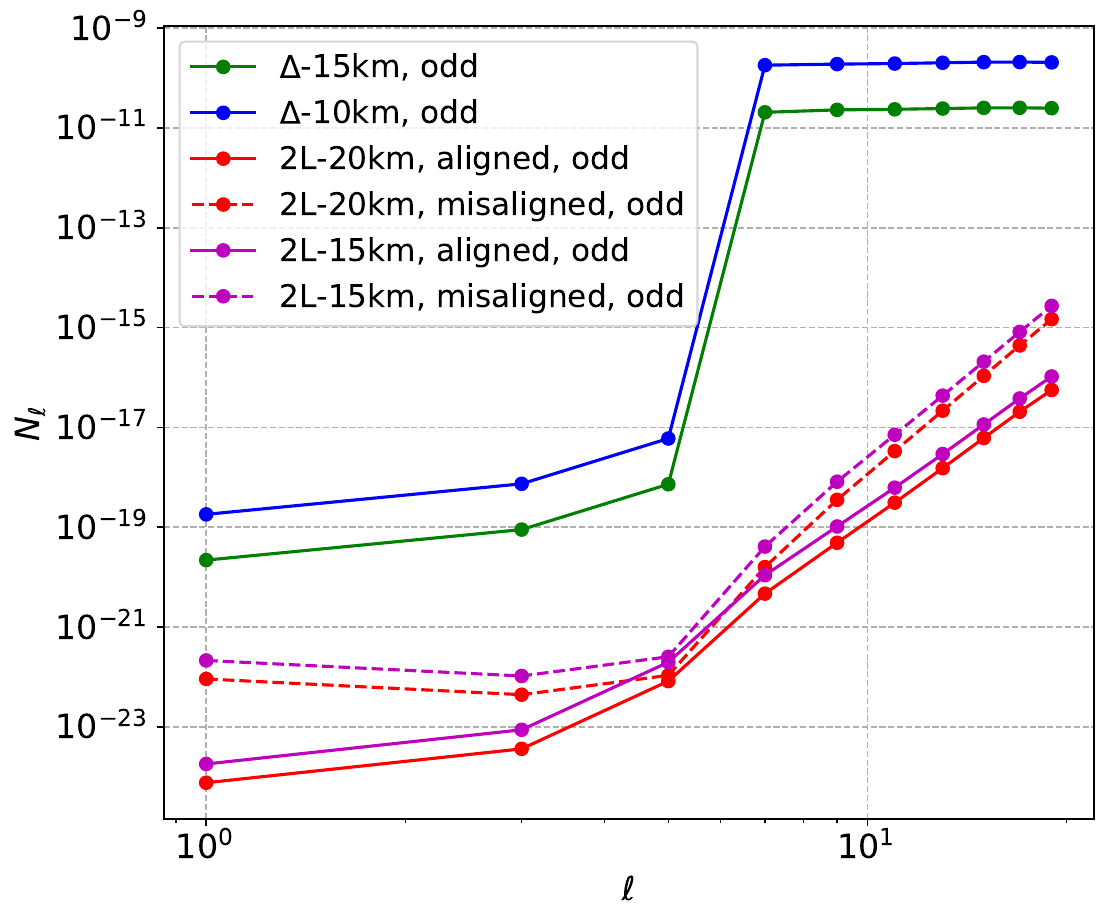}
    \qquad 
     \includegraphics[width = 7.1cm]{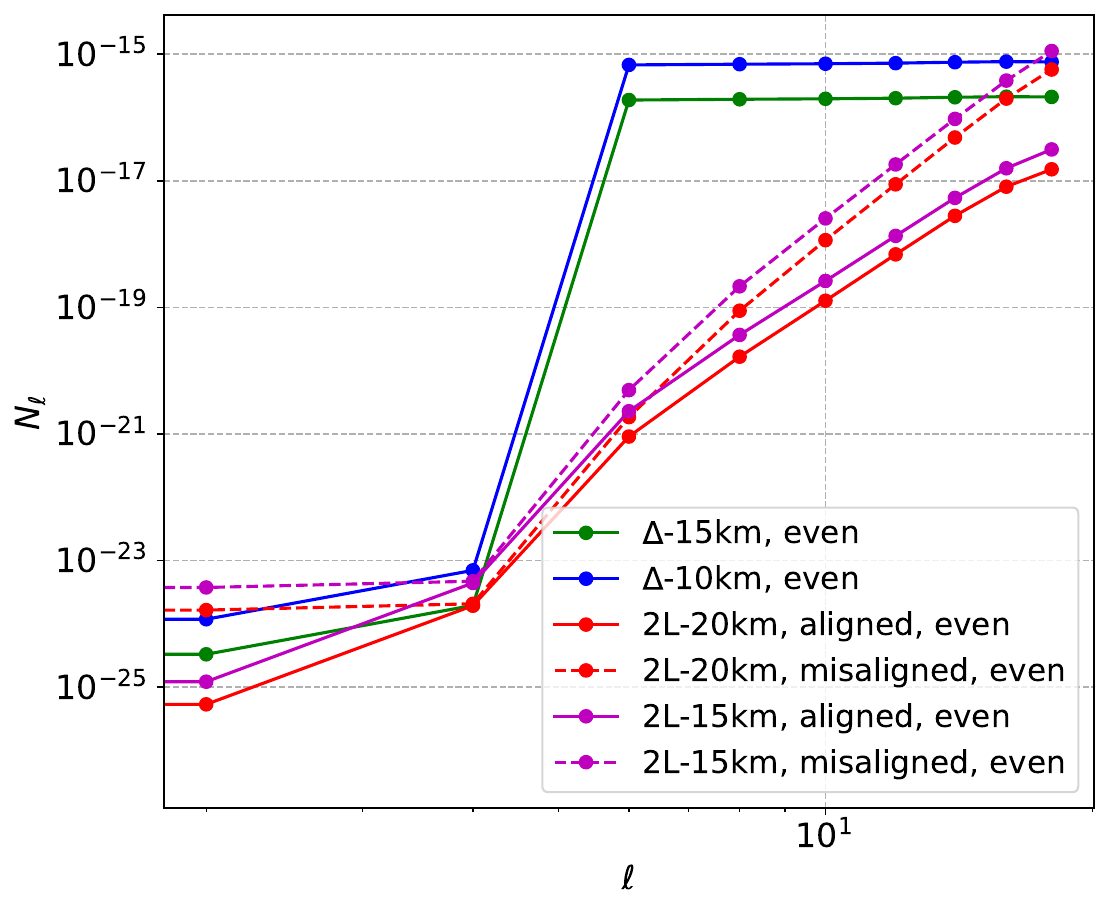}
    \caption{\small \label{fig:Nell} Noise per angular multipole at a reference frequency of $100$ Hz for the four configurations under study. In the left and right panel we show the sensitivity to odd and even multipoles, respectively. All figures are obtained using the ${\tt schNell}$ package \url{https://github.com/damonge/schNell}.}

\end{figure}
        
In Fig.\,\ref{fig:Nell} we plot the noise per multipole $N_{\ell}$ at a reference frequency $f_{\text{ref}}=100$ Hz for different detector configurations. {\em When looking at odd multipoles, we see that the 2L configurations are much more sensitive than triangular configurations.} This is due to the fact that the three nested interferometers composing a triangular detector are separated by a baseline whose size is one arm length, while the two L-shaped detectors of the network have a much larger baseline. As a consequence, the antenna pattern function of triangular configuration is almost completely even under parity, hence the instrument has a very bad sensitivity to odd multipoles, see  \cite{Alonso:2020rar}. Furthermore, as expected, aligned detectors have better sensitivity than misaligned detectors. {\em We observe that, with  the most favourable configurations, i.e. {\rm 2L} aligned with 15km  or 20~km length, we are potentially able to detect a dipole anisotropy at the level of $\sim 10^{-12}$ (hence $N_{\ell=1}\sim 10^{-24}$), which is the size of the kinematic dipole expected from a background component on the edge of being detected by LVC at O5 sensitivity}, see also \cite{Cusin:2022cbb}.\footnote{We observe that the plateau at high-$\ell$ values is symptomatic of the fact that we have lost sensitivity to those multipoles, see \cite{Alonso:2020rar}.}

\subsection{Astrophysical backgrounds}\label{sect:astroback}

Binary mergers nowadays detected by LIGO-Virgo-KAGRA are only the tip  of the iceberg. In fact, a large fraction of the events  at larger distances are not resolved by the detectors and are expected to constitute the dominating contribution of the astrophysical background.
For compact binary coalescences (CBCs), \eq{eq:Abkg:def} can be rewritten in term of energy density flux as \cite{Regimbau:2012ir}
\begin{equation}
    \Omega_{\rm gw}(f) = \frac{1}{c\rho_c}f\mathcal{F}(f),
    \label{omg}
\end{equation}
where $\mathcal{F}(f)$ is defined as 
\begin{equation}
    \mathcal{F}(f) = T^{-1}\sum_{k=1}^{N}\frac{1}{4\pi r^2}\frac{dE_{\rm gw}^k}{df}(f)\, , 
    \label{flux}
\end{equation}
and the sum is over the $N$ individual merger contributions arriving on the detector during an observation time $T$. Rewriting the energy density, one can express the spectrum $\Omega_{\rm gw}(f)$ in terms of frequency domain waveforms: 

\begin{equation}
    \Omega_{\rm gw}(f) = \frac{\pi\,c^2}{2\,G\,\rho_c}f\,T^{-1}\sum^N_{k=1}f^2\,\tilde{h}_k^2(f),
\end{equation}
with $\tilde{h}_k^2(f)$ the sum of the squared Fourier domain GW amplitudes for the two polarization +/$\times$:
\begin{equation}
    \tilde{h}^2_{k}(f) = \tilde{h}^2_{+,k}(f)+\tilde{h}^2_{\times,k}(f).
\end{equation}

\begin{figure}
    \centering
    \includegraphics[width = 12cm]{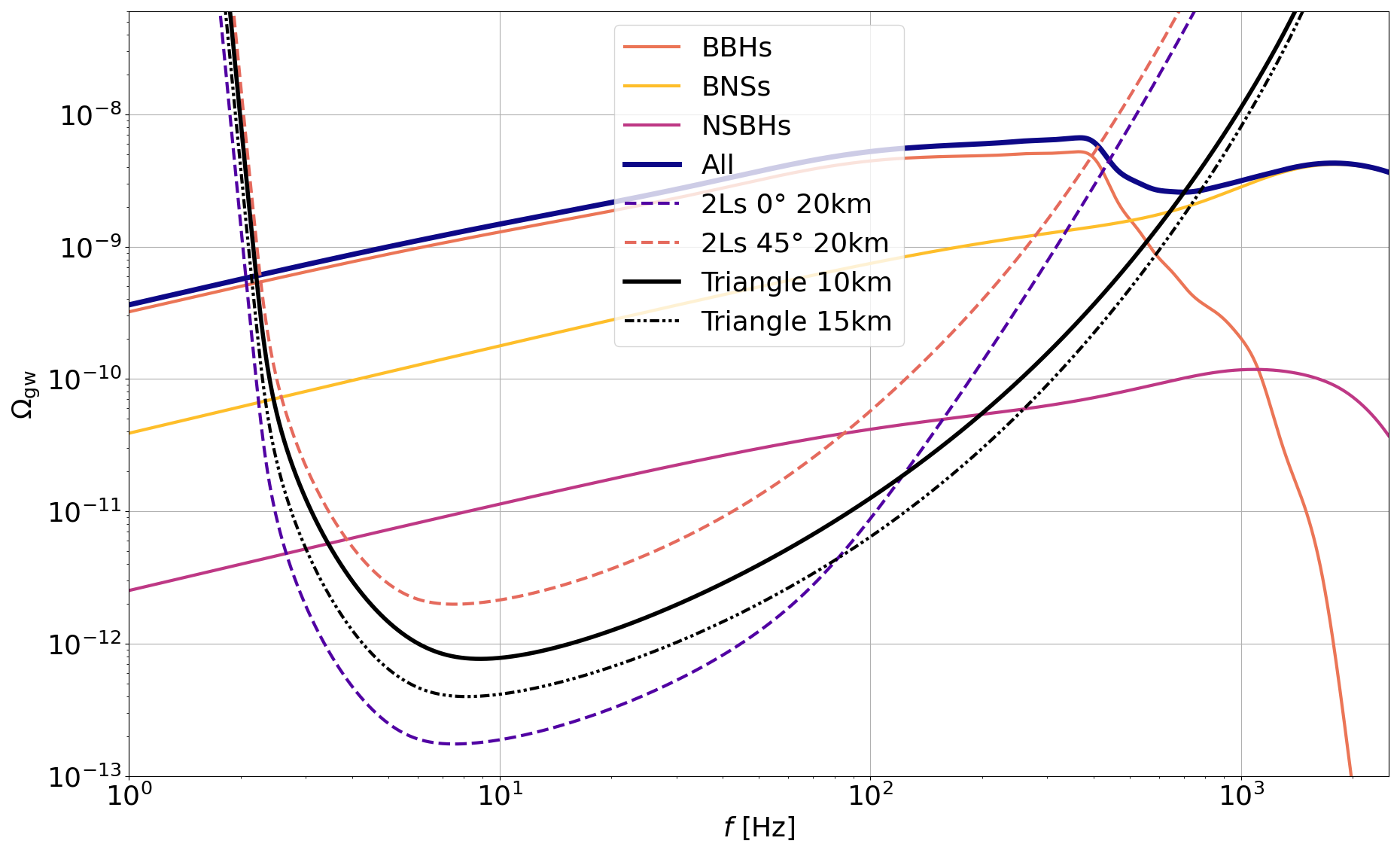}
    \caption{\small Background prediction for the current state-of-the-art astrophysical models, assuming a metallicity spread $\sigma_Z = 0.3$. }
    \label{fig:Total_bkg}
\end{figure}

\begin{figure}
    \centering
    \includegraphics[width = 12cm]{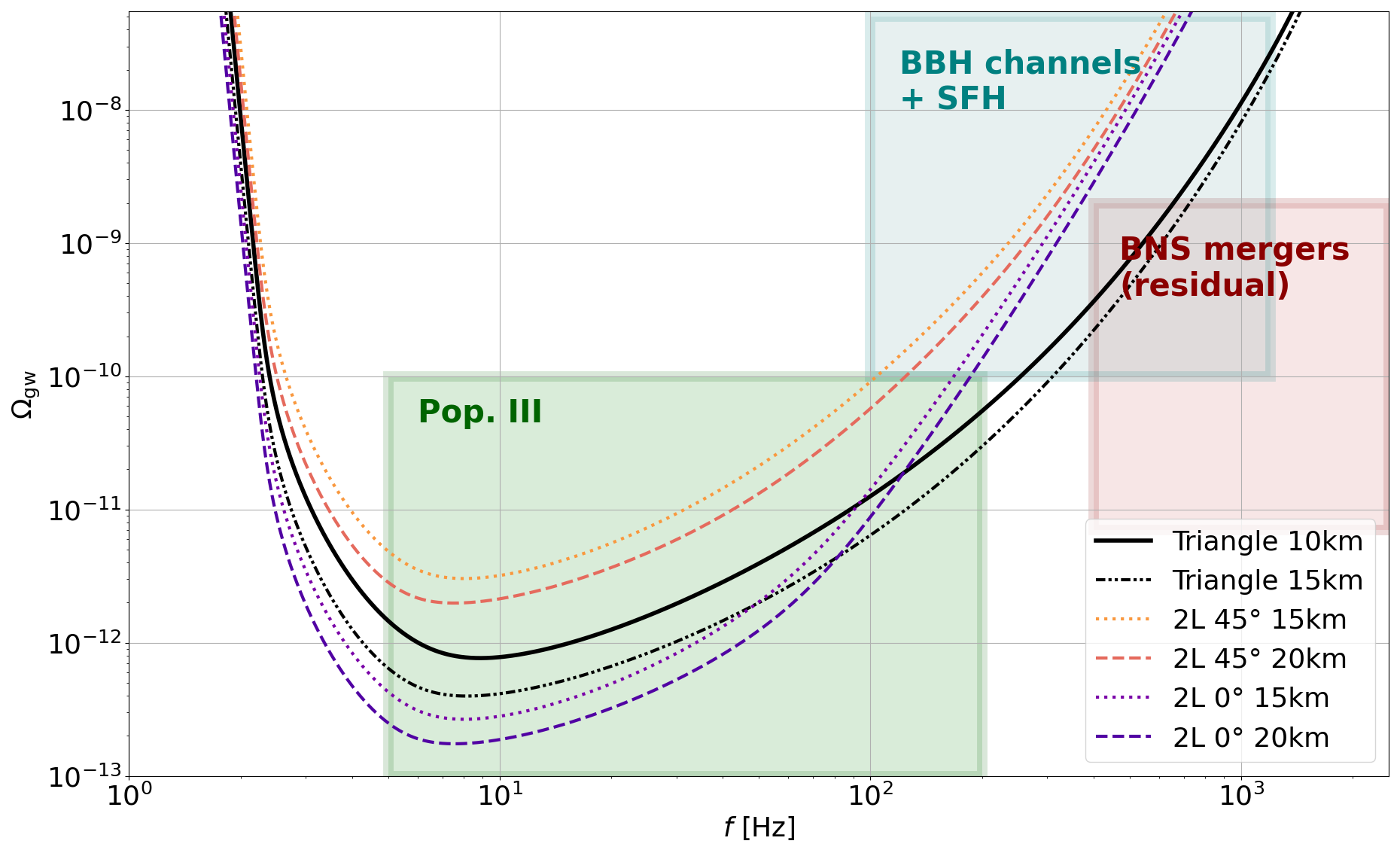}
    \caption{\small Location of the predicted signatures for the three main challenges for the CBC background.}
    \label{fig:Pop_study}
\end{figure}

\noindent
Figure \ref{fig:Total_bkg} shows predictions for the background, derived from the current state-of-the-art of population synthesis models~\cite{Mapelli:2017hqk, Giacobbo:2017qhh, Giacobbo:2018hze, Giacobbo:2019fmo, Mapelli:2019ipt, Santoliquido:2020axb, Mapelli:2021gyv} with a metallicity spread $\sigma_Z = 0.3$ for the three types of binary: binary neutron stars (BNSs), binary black holes (BBHs), and neutrons star-black hole systems (NSBHs). The BNS and BBH population are the ones described in Section~\ref{sect:CBC} and used through this paper to evaluate general metrics and several specific science cases.  On these plots are also shown the PLS for the main configurations of this study, already shown in Fig.~\ref{fig:ET_PLS_auto_cross_SNR_1a}.\footnote{Here, and in Fig.~\ref{fig:Pop_study}, the PLS are computed including only cross-correlation between detectors, and setting ${\rm SNR}_{th}=1$ and $T_{\rm obs} = 1$~yr, as in the plots of Section~\ref{BackgroundSensitivities}.} The low-frequency part of the  spectrum is predicted to have an evolution proportional to $f^{2/3}$, typical for the inspiral phase of compact objects binaries. This shape is common to all CBCs, therefore a sign of a specific population on the background is given by  deviations from the  $f^{2/3}$ behavior. For example, in Fig.~\ref{fig:Total_bkg}  the deviation of the total background (`All', dark blue curve) from the $f^{2/3}$ shape between 90-500Hz is due to the inclusion of BBHs from the cluster, while the deviation above 1000Hz is coming from the BNSs.
Consequently, the background will be a new observation channel for  population studies of compact objects. 

Stellar CBC background challenges in ET can be divided into three main categories: the search of Population III stars, the study of BBH formation channels combined with the star formation history (\acrshort{sfh}), and the residual BNSs background. All these categories are represented on Fig.~\ref{fig:Pop_study} at the location of predicted background signatures,  together with the ET PLS for the different geometries considered (all taken with their best ASD, see Section~\ref{sect:geometry}).

\paragraph*{BBHs channels + SFH: }This part of the spectrum is where (depending on their masses and spins) individual BBHs  stop emitting gravitational waves, so that the overall spectrum drops. Two main factors can imprint some signature in the range [100-1200]~Hz: (1) the SFH, which is a combination of metallicity/redshift, and  star formation rate/redshift relations; and (2) the formation and evolution channel of the binary (i.e. the environment of the binary formation/evolution). The detection of these signatures will be a new channel of observation for the stellar formation and evolution \cite{Perigois:2021ovr, Bavera:2021wmw}. {\em For this specific signature the triangle configuration is preferable to any of  the four {\rm 2L} configurations that we have considered (15 or 20 km arm-length, aligned or at $45^{\circ}$), because its sensitivity to stochastic backgrounds will be better for frequencies above 200 Hz.}

\paragraph*{Population III: } 

Population III stars are the first, metal-free stars that formed in the Universe \cite{Bromm:2003vv}. The peak of the merger rate density of compact remnants left by Population III stars is expected to be at high redshift, $z\sim 8-16$ \cite{Hartwig:2016nde, Kinugawa:2015nla, Belczynski:2016ieo, Liu:2020ufc, Tanikawa:2021qqi}. All of the unresolved population III merger events then combine to an SGWB in the $[5-200]$ Hz frequency range that may deviate from the usual $f^{2/3}$ power-law behaviour \cite{Perigois:2020ymr, Martinovic:2021fzj,Inayoshi:2016hco,Kowalska:2012ba, Suwa:2007du}.
This is due to the fact  that, while population I/II stars
produce  an SGWB in the LIGO/Virgo frequency band during their inspiral phase, population III stars would instead show up at the merger and ringdown phases. {\em In the frequency range where population III stars could leave their imprint, the {\rm 2L} configurations with parallel arms are the best ones.}

\begin{figure}
    \centering
    \includegraphics[width = 12cm]{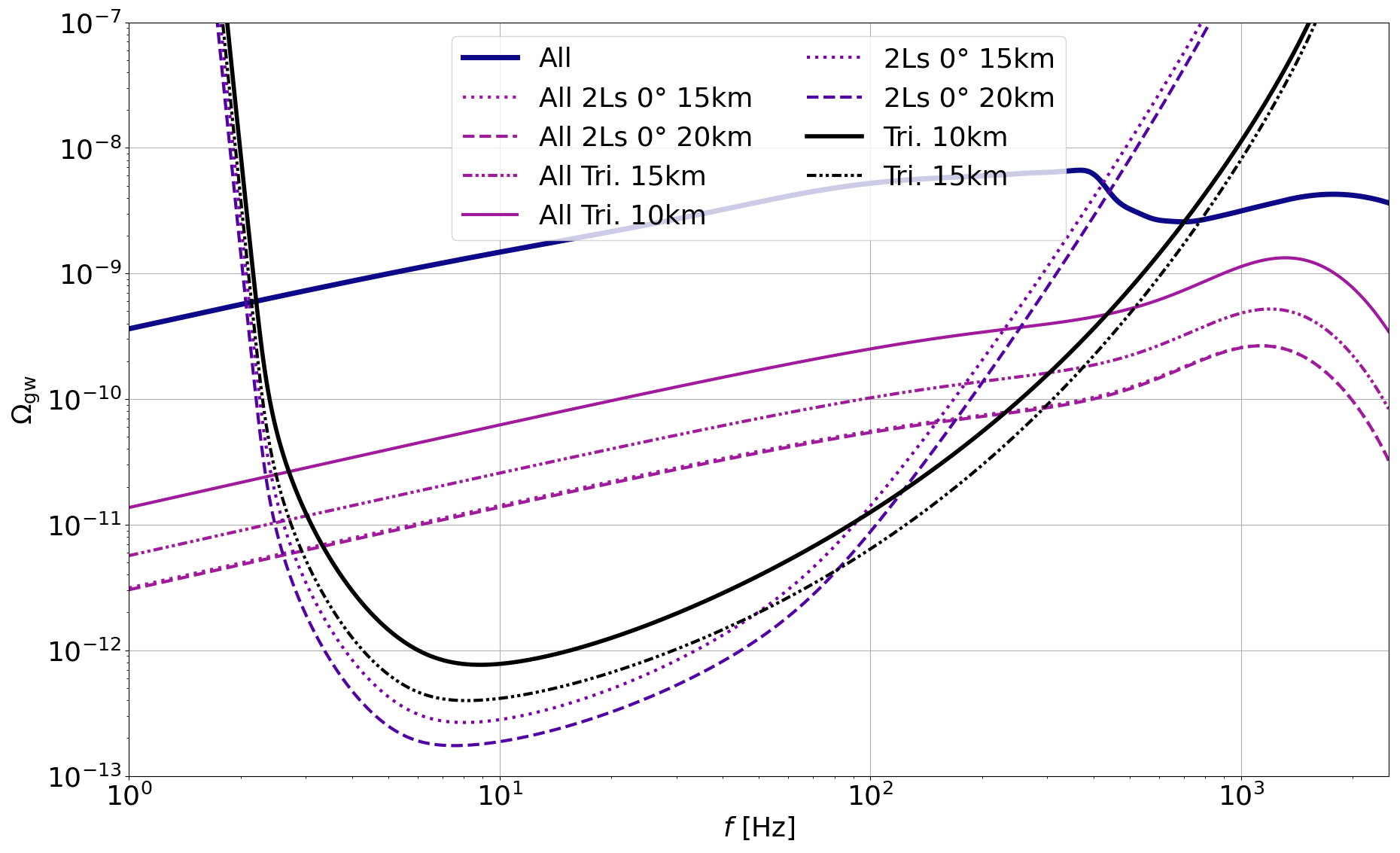}
    \caption{\small Backgrounds from all types of binaries (BBHs+BNSs+NSBHs), compared to the sensitivity of various detector configurations. The blue thick line corresponds to the total background, while the violet ones to the residuals (i.e. the background obtained after the subtraction of resolved sources) for the different configurations.}
    \label{fig:residual}
\end{figure}

\paragraph*{BNS residual mergers: }
The background, as it is defined,  contains only non-resolved gravitational waves. In that sense, the residual background (background obtained after subtraction of all resolved sources\footnote{We assumed an SNR threshold of 12 to define a source as resolved. The resolved sources are then subtracted to obtain the residual background.}) will reveal the BNSs contributions to the background. Figure \ref{fig:residual} shows the residual backgrounds for each type of binary and some of the configurations. As for BNSs, they are more difficult to be resolved due to their low masses, and their gravitational waves will mostly remain in the background and dominate it \cite{Perigois:2020ymr,Perigois:2021ovr}.

\subsection{Impact of correlated magnetic, seismic and Newtonian noise}\label{sect:corrnoise}


Since searches for a stochastic gravitational-wave background (SGWB)~\cite{Christensen:2018iqi} typically rely on cross-correlating data from two or more detectors, they are also susceptible to correlated noise. In this section we will discuss the impact of different noise sources on the ET and how they affect the different configurations, that is triangular versus 2L configuration. In Section~\ref{sec:SeismicAndNN} we will discuss the effect of correlations in the seismic and accompanying Newtonian noise (\acrshort{nn}), whereas Section~\ref{sec:Magnetic} focuses on correlations in magnetic field fluctuations.

In the following sections, we have used the ET-D design sensitivity in a triangular design with arm lengths of 10~km. However, the discussion depends minimally on this assumption and is a good representation for the comparison of the triangular versus 2L configurations regardless of the exact arm length and sensitivity.

At the end of the section, in Table \ref{tab:CorrelatedNoiseOverview} the frequency regions of searches for an SGWB affected by correlated noise are summarized.
We want to point out that here we only discuss the effect of fundamental noise sources, which implies that infrastructural noise sources will have to be reduced to a similar or lower level than the fundamental noise sources. This is, to some extent, discussed for magnetic noise in 
Section~\ref{sec:Magnetic}; for all noise sources, i.e magnetic, seismic and Newtonian noise, this should be investigated further.
Furthermore, we stress that most of these correlated noise investigations focus on the effects of the search for an isotropic SGWB. However, as shown in a recent study on the effect of magnetic glitches due to lightning strikes \cite{Janssens:2022tdj}, some correlated transient events might also affect other searches, which should be further investigated for seismic and Newtonian noise.

\subsubsection{Seismic and Newtonian Noise}
\label{sec:SeismicAndNN}

During the last decade, much research has focused on understanding the effect of seismic and Newtonian noise on the sensitivity of  ET \cite{Badaracco:2019vjq,NN_Sardinia2020,10.1785/0220200186,Bader:2022tdz,Koley:2022wpe}. In a recent study the possibility of correlated seismic and Newtonian noise is investigated for the ET triangular configuration~\cite{Janssens:2022xmo}. 
Searches for a SGWB using data from the second-generation gravitational-wave interferometric detectors, LIGO and Virgo, are considered to have no correlations in the seismic ambient field due to their large separation and the damping of seismic waves. However, in the triangular design of the ET  the input and output mirrors of two interferometers are proposed to be separated by distances of only several hundred of  meters. As illustrated in Fig. 11 of~\cite{Janssens:2022xmo}, there are five coupling locations between two ET interferometers, of which two are between aligned mirrors and the remaining three between mirrors rotated by $60^{\circ}$. Based on the ET design report~\cite{ETdesignRep} the horizontal separation between those mirrors is about 300--500~m for the aligned  mirrors and 330--560~m for the other pairs~\cite{Janssens:2022xmo}.

To investigate correlations between seismic measurements over a scale of several hundreds of meters,  ref.~\cite{Janssens:2022xmo} analyzed data at the Earth's surface from geophones deployed near Terziet (NL)~\cite{3T_network} and underground data from seismometers in the former Homestake mine (South Dakota, USA)~\cite{10.1785/0220170228}. 
The study concludes that at the surface at Terziet significant (above Gaussian noise background) seismic correlations exist between sensors with a separation of several hundreds of meters. The 50th percentile coherence is significant up to about 10~Hz, whereas the 90\% coherence is significant up to 50~Hz, which is the highest recorded frequency. 
For 90\% of the time at Homestake there is significant coherence up to 20~Hz and more than 50\% even up to 40~Hz; that is the highest frequency before the response of the sensors decreases drastically.
Neither the coherence nor the cross-power spectral density (\acrshort{csd}) were  observed to have a dependence on distance (between 200m and 800m).
At both locations night times are observed to be quieter, with lower CSDs. At night there are more periods when there is significant correlated noise. These studies indicate that there is a non-negligible amount of coherence during a significant amount of time (night and day), up to frequencies of 50~Hz. This leads to the conclusion that the effect of these correlated noise sources should be understood, particularly for searches for an SGWB.

Note that in \cite{Janssens:2022xmo} it is emphasized that some correlations are considered to be of anthropogenic origin. Furthermore, some of the power spectral densities (\acrshort{psd}s) observed at the candidate sites are lower than the CSD observed in the Homestake mine, and therefore site-specific studies should be envisioned in the future. However, the results discussed in \cite{Janssens:2022xmo} are expected to give a good indication of the order of magnitude of the possible levels of correlated seismic and Newtonian noise which might be expected in an underground environment. Homestake represents an underground environment where both ambient and anthropogenic noise sources are present. This may also be expected for ET, since its infrastructures could be expected to introduce additional anthropogenic noise, regardless of various attempts to reduce and shield them.

It is possible, based on the measurements of correlations of the seismic noise, to estimate the correlated Newtonian Noise (NN). The NN is a disturbance produced in the gravitational-wave detector by local fluctuations of the gravitational field. Two types of seismic waves produce NN: the first are Rayleigh waves which propagate at the surface of the Earth, and the second are body waves present in the Earth's mantle.
The NN from Rayleigh waves decreases significantly with depth, so their effect  on  ET will be limited if one assumes an underground facility at 300~m. As shown in Fig.~15 of~\cite{Janssens:2022xmo}, the NN from Rayleigh waves significantly impacts the power-law integrated  sensitivity of ET only up to about 5~Hz.\footnote{The  power-law integrated  sensitivity is defined as in \cite{Thrane:2013oya}, see also App.~\ref{app:PLS}. We use an SNR of 1 after one year of data at a 100\% duty cycle.} At higher frequencies, no effect is expected.

The median contribution from correlated body waves, however, is expected to be three to five times higher than the ET sensitivity, in the range of 2 Hz to 10 Hz.
While it is important to investigate the impact of noise on the sensitivity curve, correlated noise can more drastically affect some searches. The search for a SGWB typically relies on the cross-correlation of data from two detectors to decrease the Gaussian detector noise and improve the search sensitivity. 

\begin{figure}
    \centering
    \includegraphics[width=0.9\textwidth]{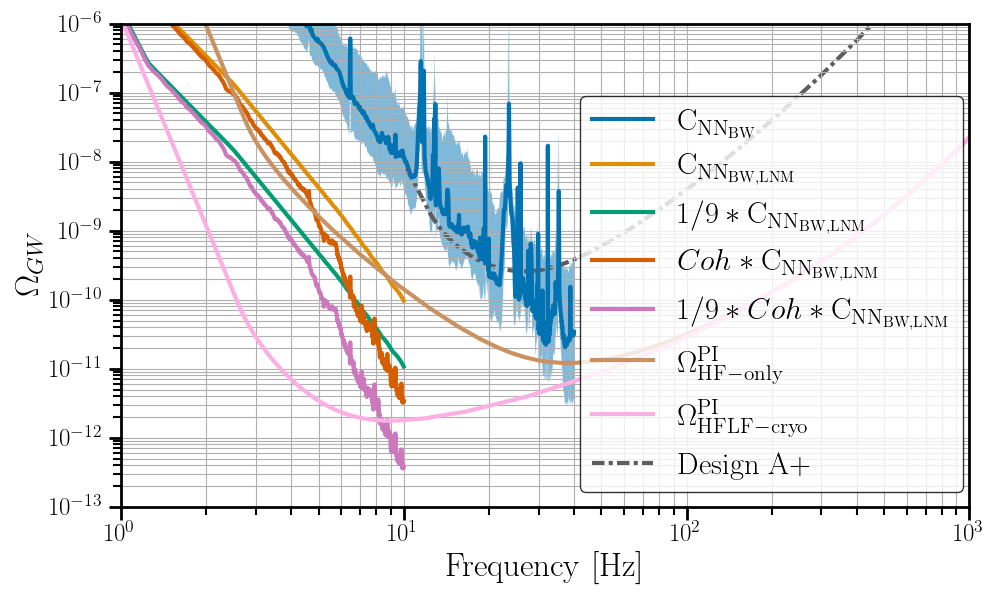}
    \caption{\small The projected impact from correlated NN from body-waves ${\rm C}_{{\rm NN}_{\rm BW}}$, as calculated in Section~III of \cite{Janssens:2022xmo}, is shown in blue. The blue line represents the median value, and the associated surface is delimited by the 10\% and 90\% percentiles. 
    Also curves with  more optimistic assumptions (see the text) are shown.   
    Based on Fig. 16 of \cite{Janssens:2022xmo}.
    \label{fig:StochBudget_NN}}
\end{figure}

Fig.~\ref{fig:StochBudget_NN} presents the expected impact on the search for a SGWB due to NN produced by body waves.\footnote{To produce this figure, for the broadband ($\Omega^{\rm PLS}_{{\rm HF-only}}$, $\Omega^{\rm PLS}_{{\rm HFLF-cryo}}$) sensitivity to an SGWB we assumed one year of observation time (100\% duty cycle). The one-year PLS curve of the A+ design for the LIGO Hanford, LIGO Livingston detectors and AdV+ design for the Virgo detector is represented by the dot-dashed curve. LIGO’s A+ and Virgo’s AdV+ design are the interferometer designs which are aimed to be reached during O5 \cite{KAGRA:2020rdx}. This curve was obtained using the open data provided by the LVK collaborations \cite{O3IsotropicDataset} and was first presented in \cite{KAGRA:2021kbb}.  Note that in this figure we present the 1$\sigma$ PLS, whereas in \cite{KAGRA:2021kbb} the 2$\sigma$ PLS-curve is shown.} It shows that the budget calculated in~\cite{Janssens:2022xmo} affects the SGWB search up to at least 40~Hz (blue band), which is the highest frequency in the analysis. Near 3~Hz, this gives an effect which is larger than the 1~yr ET sensitivity curve for the SGWB (light brown line in Fig.~\ref{fig:StochBudget_NN}) by a factor
$\sim 8\times 10^6$ (90\% percentile), or $\sim 6\times 10^5$ (50\% percentile).

\vspace{1mm}

{\em According to the noise estimate shown in blue in Fig.~\ref{fig:StochBudget_NN}, with the triangular configuration the sensitivity of  ET to an SGWB, at least below ${\cal O}(50-100)$~Hz, would be of the same order of magnitude as LIGO’s A+ and Virgo’s AdV+ design.
Even assuming that an optimistic factor of 10 NN cancellation \cite{Badaracco:2019vjq} could be applied to each interferometer, i.e. a factor of 100 for the SGWB budget, the search for an SGWB would still be impacted up to $\sim$ 30\,Hz.} 
When considering the night-time data with fewer anthropogenic activities, the impact decreases but stays at the same order of magnitude, i.e. a factor $\sim 10^5-10^6$ at 3~Hz~\cite{Janssens:2022xmo}.

In Fig.~\ref{fig:StochBudget_NN} we also provide the result from a more optimistic scenario (pink curve). The construction of this scenario is broken down as follows. We start from the yellow curve, which displays the affect of NN from body waves assuming a seismic spectrum identical to the Peterson low noise model~\cite{Pet1993} and 100\% coherence. 
We then assume a NN cancellation by a factor of 3 (i.e. a factor of 9 on the stochastic budget), which is considered as a realistic target (green curve). The dark orange curve is generated from the assumption that the coherence equals to the median coherence observed in Homestake, as reported in \cite{Janssens:2022xmo}. Finally the pink curve represents the Peterson low noise model multiplied by the coherence observed at Homestake, as well as a factor of 3 NN cancellations for each detector. {\em This pink curve represents the most optimistic scenario that one can achieve. In this case, SGWB searches will still be affected by correlated NN up to $\sim$10Hz.} 

The effect of correlated noise presented above, as well as possible additional effects from infrastructural noise, would only affect the search for an SGWB in the case of the triangular design leading to closely located interferometers.
The problem of co-located correlations (of order a few hundred meters) of the seismic field should represent no problem for very distant L-shaped  interferometers. However, for a triangle one can take advantage of the ``null-channel" to estimate the power spectrum~\cite{Goncharov:2022dgl,Janssens:2022cty}, see the discussion in Section~\ref{sect:nullstream}. In the future, site-specific studies could give more detailed information. Furthermore, it would be important to  estimate the order of magnitude of correlated noise expected by infrastructural noise sources caused by the `typical equipment' needed for an underground, cryogenic gravitational-wave interferometer. 

Finally, the effects of seismic transients such as (micro-)earthquakes should be understood. The latter would not only affect the search for an SGWB but could also be expected to affect searches for other GW signals such as CBCs, bursts or continuous waves if the noise sources are too loud. Further studies on the possible effect of correlated seismic transients would also be advisable. Such a study could be envisioned to investigate the possibility of lock-loss, similar to an earlier study for LIGO~\cite{Coughlin:2016lny}, or the impact on the sensitivity, similar to a study of the effect from earthquakes as part of the O3 environment investigations done at Virgo~\cite{Virgo:2022ypn}. In the studies for ET one should also look into the possibility of correlations, as well as the `transient Newtonian Noise', linked to the transient seismic fields and how these might introduce additional effects on the different signal searches. 

\subsubsection{Magnetic noise}
\label{sec:Magnetic}

One noise source commonly investigated with respect to the search for an SGWB is the Schumann resonances~\cite{Schumann1,Schumann2}. These are standing electromagnetic waves in the cavity formed by the Earth's surface and the ionosphere, and are driven by lightning strikes all across the globe. The fundamental mode has an amplitude between $\sim$ 0.5~pT and $\sim$ 1~pT depending on the location and time of the day and year. The resonances are several Hz wide and have frequencies $\sim$7.8Hz, $\sim$14Hz, $\sim$21Hz, ... 

To explain how magnetic fields couple to the interferometer, it is instructive to understand how the coupling is measured. On a regular basis, strong magnetic fields are generated in the experimental buildings of the current generation of interferometric detectors. By injecting fields many times stronger than the ambient magnetic field one tries to measure their effect on the strain ASD. We define $Y$ and $X$ to be the ASD of the strain and witness channel (magnetometer) respectively. We use the subscript 'inj' and 'bck'  refer to times during the injection, respectively a quiet background time just before/after the injection. Then we can measure the magnetic coupling function $\Kappa$ as follows~\cite{Fiori:2020arj},
\begin{equation}
    \label{eq:MagneticCoupling}
    \Kappa_{\rm measured}(f) = \sqrt{\frac{Y^2_{\rm inj}(f)-Y^2_{\rm bck}(f)}{X^2_{\rm inj}(f)-X^2_{\rm bck}(f)}} ~ .
\end{equation}
The coupling function describes how strong the strain data will be affected given ambient magnetic fields.

In recent work, the effect of the Schumann resonances and other forms of correlated magnetic noise on  ET was investigated~\cite{Janssens:2021cta}. At low frequencies ($<$ 30~Hz) it was shown that the magnetic coupling function has to be improved by two to four orders of magnitude compared to the second-generation detectors Advanced LIGO and Advanced Virgo in order to prevent significant coupling. 
Note that compared to the Advanced Virgo coupling function, a factor of 5 improvement could be expected due to the larger mirror mass of the ET~\cite{Amann:2020jgo,Cirone:2018vdc}. However, this assumes that the same actuator magnets as with Advanced Virgo will be used, which will make the control of the test mass more difficult.
Furthermore, at low frequencies, the Advanced LIGO magnetic coupling is lower compared to the Advanced Virgo magnetic coupling. Therefore the electrostatic actuator design used by Advanced LIGO should be considered for ET. If ET will rather have a magnetic coupling similar to that of Advanced Virgo, the impact of correlated magnetic noise is predicted to be larger and will impact searches at a wider frequency band, up to $f \lesssim$ 50 Hz.
One must also stress the danger that induced eddy currents close-by to sensitive detector parts pose, so these should be minimized as much as possible.

\begin{figure}
    \centering
    \includegraphics[width=0.49\textwidth]{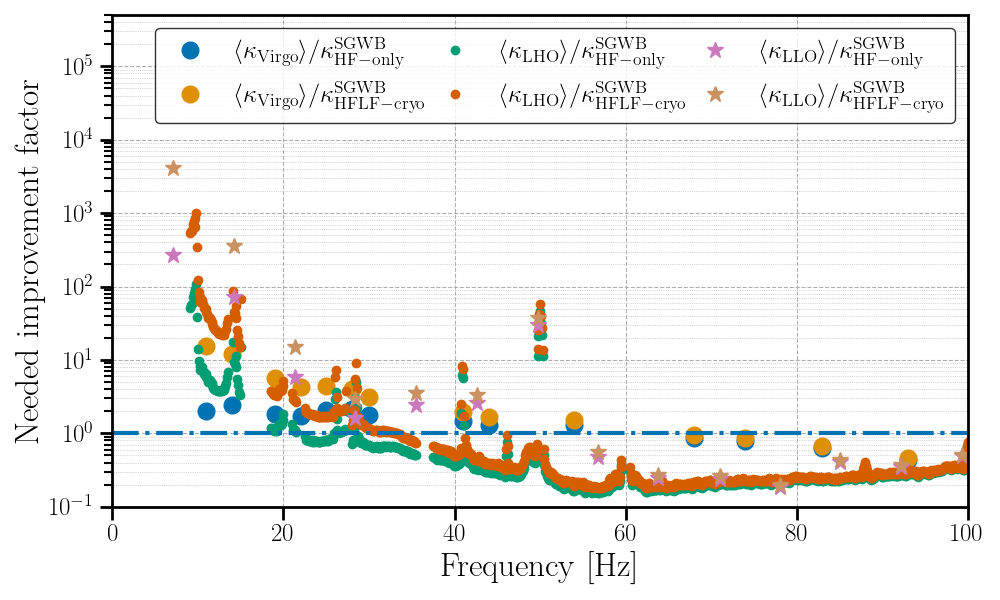}
    \includegraphics[width=0.49\textwidth]{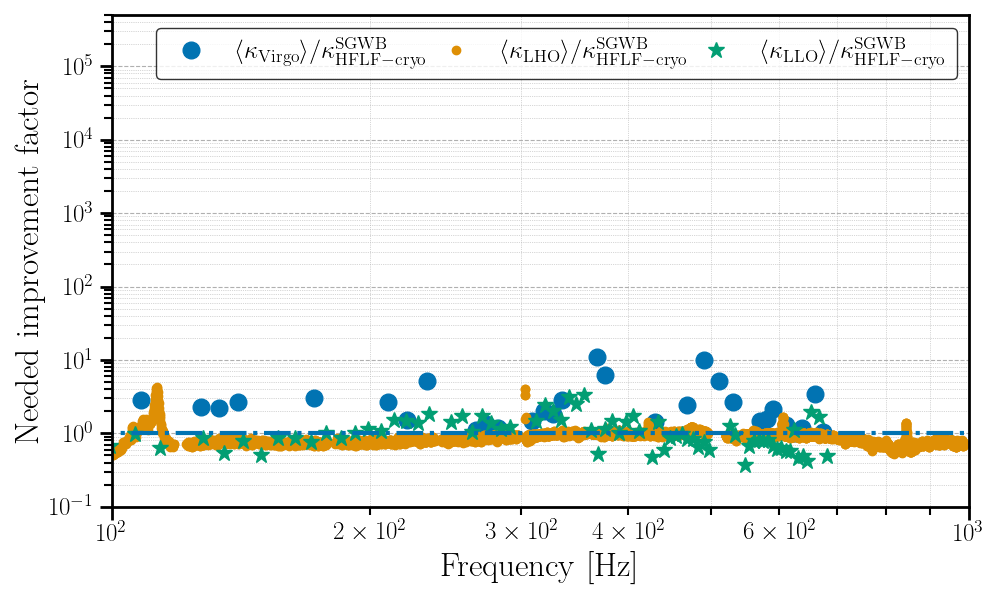}
    \includegraphics[width=0.70\textwidth]{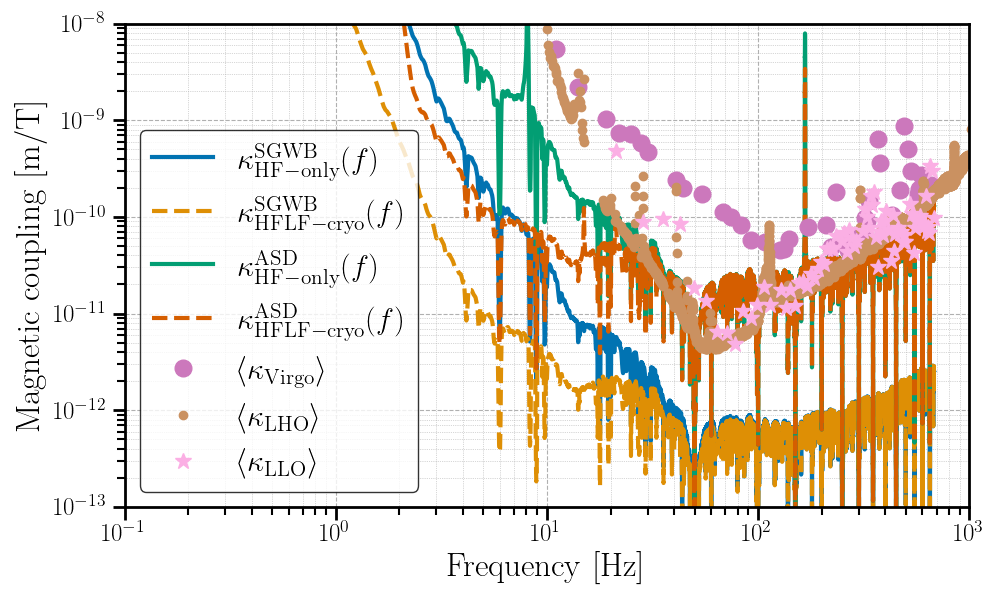}    
    \caption{\small Top panels: needed improvement factor as a function of frequency for the upper limits on the ET magnetic coupling function, such that the isotropic SGWB search is unaffected. LHO and LLO refer respectively to the LIGO Hanford and LIGO Livingston observatories. The low-frequency (left panel) magnetic coupling poses a greater challenge for the operation of ET compared to the high-frequency (right panel) magnetic coupling. In both panels, the dash-dotted blue line indicates the line where no improvement is necessary. This figure is based on the two right panels of Fig. 6 of \cite{Janssens:2021cta}, however, we have used an additional factor of 5 in the reduction due to the larger mirror mass when going from Virgo to ET, in line with earlier work \cite{Cirone:2018vdc,Amann:2020jgo}. This only applies to the left panel, since at high frequencies the coupling is dominated by other mechanisms. We report the upper limits for both the HF-only and HFLF-cryo sensitivity for SGWB searches (in \cite{Janssens:2021cta} they used ET-B and ET-D). Note that, for $f>100~{\rm Hz}$, HF-only and HFLF-cryo have the same sensitivity to a SGWB, so we only present the HFLF results in the top left panel. \\
    Bottom panel: the upper limits on the maximal allowed magnetic coupling to prevent impact on the ASD and/or SGWB searches in the pessimistic scenario where magnetic infrastructural noise, as measured in the central building at Virgo, will be fully correlated between interferometers. The upper limits are provided for both HF-only and HFLF-cryo. The upper limits are compared with the mean observed magnetic coupling at the LIGO and Virgo detectors. Based on Fig. 7 of \cite{Janssens:2021cta}.}
    \label{fig:Magnetic}
\end{figure}

The effect of magnetic noise is shown in Fig.~\ref{fig:Magnetic}, based on Fig. 6 of \cite{Janssens:2021cta}. It presents the ratio of \scalebox{1.5}{$\kappa$}$^{\rm SGWB}_{\rm HF-only}$ and the magnetic coupling measured at LIGO and Virgo. The same reatio is also whon for \scalebox{1.5}{$\kappa$}$^{\rm SGWB}_{\rm HFLF-cryo}$. Here \scalebox{1.5}{$\kappa$}$^{\rm SGWB}_{\rm HF-only}$ and \scalebox{1.5}{$\kappa$}$^{\rm SGWB}_{\rm HFLF-cryo}$ are the magnetic coupling for each ET interferometer such that no effect of correlated magnetic noise is present in the search for an isotropic SGWB using one year of data with 100\% duty cycle at the HF-only, respectively HFLF-cryo design sensitivity. For more details on the computation, we refer the reader to \cite{Janssens:2021cta}. The needed improvement of the magnetic coupling such that magnetic noise is not larger than 1/10 of the ASD \footnote{This is a typical requirement for noise sources which are considered not to be a 'fundamental' noise source, i.e. part of the calculation of the optimal sensitivity curve.}, is roughly a factor 10 less stringent than the SGWB constraint. For more information see Fig.~6 and the accompanying discussion of \cite{Janssens:2021cta}. Note that in \cite{Janssens:2021cta} was not taken into account a factor of 5 of reduction obtained by increasing the mirror mass. In the top left panel of Fig.~\ref{fig:Magnetic} we applied this additional factor-of-5 reduction for Virgo's magnetic coupling below 100 Hz. At higher frequencies, the coupling function is dominated by different mechanisms. We note that in the recent LIGO-Virgo third observing run (O3), for which we use the magnetic coupling function, an additional coupling was observed \cite{Fiori:2020arj} compared to the earlier work describing the inverse proportionality with mirror mass \cite{Cirone:2018vdc}. Therefore it is not guaranteed that this reduction could be achieved if the other coupling mechanisms are not properly understood and mitigated. 

Up to $\sim$ 20 Hz, the observed amplitude of the magnetic CSD is similar, regardless of the separation between the detectors~\cite{Janssens:2022tdj}. At higher frequencies, the magnetic CSD decreases due to stronger attenuation of electromagnetic fields travelling long distances.  Therefore the effect of correlations in magnetic field fluctuations up to $\sim$ 30 Hz on ET is (mainly) independent of the detector configuration. Both the triangular and 2L designs will be impacted at the same order of magnitude.
However, it is important to understand the distance between the detectors and large thunderstorm regions. Regions with large thunderstorm activity are the Americas, sub-Saharan Africa and southeast Asia \cite{boccippio2000regional}, but more research is needed to investigate the `local' activity in Europe and the possible impact on gravitational-wave interferometric detectors. 

{\em To summarize the effect of fundamental magnetic noise on HFLF-cryo ET, we can state that at low frequencies ($<$30 Hz) an effect of up to $\mathcal{O}(10^3)$ will be present on the ASD and up to $\mathcal{O}(10^4)$ on the sensitivity for the SGWB. This is (mainly) independent of the triangular versus 2L configuration, due to the low attenuation of electromagnetic fields at these frequencies. 
Despite the dramatic effects of correlated magnetic noise, one can consider several approaches to reduce the magnetic coupling, as will be further discussed at the end of this section.}

At higher frequencies, minimal effects are expected \cite{Janssens:2021cta}, however, these are more dependent on the configuration. The co-located triangular design will be more susceptible compared to two spatially separated Ls. Even more important is the location of the detector with respect to active thunderstorm regions.

{\em The effect from individual lightning strikes is also predicted to impact the ET sensitivity \cite{Janssens:2022tdj}, in case of similar magnetic coupling as at LIGO and Virgo. A glitch rate of $\mathcal{O}(10^5)$ for correlated magnetic transients per week could be expected, with the potential to affect ET's detector sensitivity $<$50 Hz~\cite{Janssens:2022tdj}. It might also impact background estimation for CBC and burst searches. The 2L configuration is less prone to effects coming from coherent lightning strikes compared to the triangular design.}

Depending on which coupling function is used, the magnetic contamination at low frequencies ($\lesssim $50~Hz) is roughly at the same level (or worse) than the NN from body waves as projected in Fig.~\ref{fig:StochBudget_NN}. However, there is a crucial difference between the two scenarios: the effect of the NN can only be mitigated by noise cancellation methods. For the magnetic noise, noise cancellation is also an option \cite{Coughlin:2018tjc}, however one can also directly attempt to improve the magnetic coupling function. An earlier study investigated the possibility of additional magnetic shielding and considered several configurations of Helmholtz coils or a spherical shell made out of chain-link metal~\cite{Cirone:2019zwq}. This study showed that a factor-of-10 reduction in the magnetic coupling function should be feasible by several different shielding approaches. These results are promising, given the `simple' methods proposed. One can consider doing better by using mu metal or superconducting coils, however, this will come at a significantly larger cost. Furthermore one can seek to reduce the number of magnets (or metallic parts) attached to the lower stages of your suspensions and test masses~\cite{AdvLIGO:2021oxw}. Even better would be to move magnets and metallic structures higher up in the suspension chain, leading to the dampening of the magnetic fields by the respective pendulum stages. Finally, the induction of currents in cables of e.g. the actuator system \cite{AdvLIGO:2021oxw} might be mitigated by using optical fibres where possible, definitely in sensitive locations. More research is needed to obtain reductions in magnetic coupling by different methods. Nevertheless, one could envision substantially decreasing the effect of magnetic fields by using a combination of the proposed methods, which might however come at a non-negligible investment cost. We would also like to point out that the study \cite{Janssens:2021cta} showed that ET's sensitivity curve might be affected up to $\sim$ 15Hz-20Hz (top left panel of Fig. 6 of \cite{Janssens:2021cta}), and some additional shielding and/or redesign of the suspensions is needed to prevent any limitations on ETs sensitivity curve.

The previous discussion assumes all effects from infrastructural noise are overcome. This infrastructural magnetic noise is typically about two orders of magnitude louder than the fundamental magnetic noise; for example, note the effects in second-generation detectors such as Virgo's central building~\cite{Janssens:2021cta}. In the case of a very pessimistic assumption, where the magnetic noise in  ET is the same as in the Virgo central building and it is fully correlated between two ET interferometers, an additional one to two orders of magnitude improvement of the magnetic coupling function will be required. This would only apply to the co-located interferometers, such as the ET triangular configuration. 
Furthermore, this would imply that the entire frequency range over which measurements of the magnetic coupling exist $(\sim 10-675) \, {\rm Hz}$,  would be affected if no improvement in the magnetic coupling is achieved with respect to Advanced LIGO and Advanced Virgo. In that scenario, the magnetic noise has the potential to limit the ET sensitivity, as shown in the bottom panel of Fig.~\ref{fig:Magnetic} \cite{Janssens:2021cta}.

\vspace{1mm}

{\em This would  imply that even greater investments might be needed to sufficiently reduce the magnetic coupling in the case of the triangular configuration, as compared to the 2L configuration.} 
Further research is needed to understand to which level these infrastructural noise sources could be produced and correlated between the ET interferometers.

\begin{table}
    \centering
    \begin{tabular}{|c|c|c|}
    \cline{2-3}
       \multicolumn{1}{c|}{} & \multicolumn{2}{|c|}{SGWB searches}   \\
       \cline{2-3}
       \hline
        Noise type  & Triangle & 2L       \\
       \hline
       \hline
        Seismic ambient &  Yes, $\lesssim$ 4 Hz \tablefootnote{Assuming a horizontal-to-vertical and vertical-to-horizontal coupling $\lesssim 10^{-12}$ above 4Hz. Tilt measurements and coupling was not taken into account. For details see \cite{Janssens:2022xmo}.}  & No   \\ 
        \hline
        \hline
        NN ambient - Rayleigh waves & Yes, $\lesssim$ 5 Hz \tablefootnote{Assuming an underground facility located at a depth of 300m. For details see \cite{Janssens:2022xmo}
        }   & No  \\ 
        \hline
        NN ambient - Body waves &  Yes, $\lesssim$ 40 Hz   & No   \\ 
        \hline
        \hline
        Magnetic ambient & Yes, $\lesssim$ 30Hz, possibly $\gtrsim$ 100Hz \tablefootnote{This assumes a magnetic coupling function similar to LIGO Hanford. For a coupling similar to Virgo magnetic contamination is possible $\lesssim$ 50Hz and $\gtrsim$ 100Hz. Furthermore, additional research is needed with respect to lightning activity in Europe and the effect might be site-dependent. For more details see \cite{Janssens:2021cta} and \cite{Janssens:2022tdj}. }& Yes, $\lesssim$ 30Hz  \\ 
        \hline
    \end{tabular}
    \caption{\small Frequency regions over which searches for an isotropic SGWB would be affected, considering different types of correlated noise and assuming no improvements in the magnetic coupling function with respect to Advanced LIGO and Advanced Virgo. Furthermore, in this Table, we only mention fundamental noise sources  and we neglect the potential effects of local (infrastructural) noise. Whereas more research is needed to address these noise sources in the case of seismic and Newtonian noise, for magnetic noise an initial study \cite{Janssens:2021cta} showed that the correlated noise might affect the entire frequency band for which measurements of the magnetic coupling exist, i.e. between $\sim$ 10 Hz and $\sim$ 675 Hz.  This would only affect the search for an SGWB in the case of the triangular configuration. } 
    \label{tab:CorrelatedNoiseOverview}
\end{table}

\clearpage\newpage

\section{Impacts of  detector designs on specific science cases}\label{sect:ImpactSpecific}

We now discuss how different geometries, or different ASDs, affect specific aspects of the Science Case, examining a broad set of particularly significant questions in fundamental physics, cosmology, and astrophysics.

\subsection{Physics near the BH horizon}


A cornerstone of fundamental physics is testing the nature of compact objects and the gravitational interaction in the relativistic, highly-dynamical regime at the scale of the BH horizon.
In vacuum GR, the Kerr geometry~\citep{Kerr:1963ud} is the unique physically acceptable equilibrium, asymptotically flat BH solution~\citep{Carter:1971zc,Robinson:1975bv,Chrusciel:2012jk}. 
Therefore, deviations from the Kerr metric require either modified gravity or specific matter fields within GR.

In the following, we discuss the impact of detector designs on specific tests of the nature of compact objects and the physics at the BH horizon.

\subsubsection{Testing the GR predictions for space-time dynamics near the horizon}\label{sect:testGRnearhorizon}

The ringdown waveform originates from the perturbed remnant object, and can be approximated as a superposition of damped sinusoids characterized by the complex quasinormal modes~(\acrshort{qnm}s) of the remnant. The frequencies, damping times, amplitudes, and phases of QNMs depend on the binary progenitors and on the underlying theory of gravity.

There are countably infinite QNMs indexed by $(l,m,n)$, where $(l,m)$ denote the angular dependence of the mode and $n=0,1,2$ is the overtone index (with $n=0$ labelling the fundamental tone).
The $(l,m,n)=(2,2,0)$ is the dominant mode in a binary BH ringdown signal and the excitation of the subdominant modes depends on the initial configuration of the progenitor binary~\cite{Kamaretsos:2011um,Kamaretsos:2012bs,Gossan:2011ha,JimenezForteza:2020cve,Ota:2019bzl,Ota:2021ypb,Forteza:2022tgq}.

If the remnant is a Kerr BH and the underlying theory is GR, the entire QNM spectrum is fully characterized by its mass $M_f$ and dimensionless spin $\chi_f$. The {\it BH spectroscopy} program~\citep{Kokkotas:1999bd,Berti:2009kk,Berti:2005ys,Berti:2007zu,Dreyer:2003bv,Gossan:2011ha} aims at detecting more than one QNM,  providing us with multiple independent null-hypothesis tests of GR.
In addition, even if the remnant is described by the Kerr solution but the underlying theory is not GR or the progenitors are not BHs (for
example NSs or more exotic objects), then the QNM amplitudes and phases would be different from the GR prediction~\cite{Forteza:2022tgq}. Thus, measuring the ringdown modes in the post-merger signal of a binary coalescence provides a clean and robust way to test GR and the nature of the remnant.

The fundamental QNM frequency and damping time have been measured by the LVK Collaboration only for a few events, providing an independent measurement of the mass and spin of the remnant which is in agreement with what inferred from the inspiral-merger phase~\citep{LIGOScientific:2021sio}. The first GW event, GW150914, stands out among those with the highest accuracy.

\begin{table}[t]
    \centering
    \begin{tabular}{c||c|c}
         \hline\hline
         $\text{SNR}_{\rm GW150914}$& HFLF-cryo & HF-only  \\ 
         \hline\hline
         $\Delta$-10~km & 141 & 141 \\
         $\Delta$-15~km & 190 & 190 \\
         2L-15~km-$0^{\circ}$ & 196 & 196 \\
         2L-15~km-$45^{\circ}$ & 192 & 192 \\
         2L-20~km-$0^{\circ}$ & 240 & 240 \\
         2L-20~km-$45^{\circ}$ & 235 & 235 \\
         \hline\hline
    \end{tabular}
    \caption{\small Projected ringdown SNR of GW150914 as detected by ET with different detector geometries (left column headers) and PSDs (top row headers). As explained in the main text, we do not observe differences between different PSDs because the signal is dominated by a monochromatic component at a frequency insensitive to the LF part of the PSD.}
    \label{tab:spectro_snr}
\end{table}

If measured with ET, GW150914 would have had a ringdown signal-to-noise ratio between 141 and 240 depending on the detector configuration, see Table~\ref{tab:spectro_snr}. It would therefore constitute a golden event with exquisite precision in the measurability of the QNM spectrum. We can quantify the measurability of a GW150914-like event with ET using a Fisher matrix formalism~\cite{Bhagwat:2021kwv, Berti:2005ys}.
We include the first three subdominant modes $(3,3,0)$, $(2,1,0)$ and $(4,4,0)$, modeling their amplitudes through the numerical fits in~\cite{Gossan:2011ha,Forteza:2022tgq}, and assuming a start time for the ringdown at $t\approx 10M$ after the peak of the waveform.
The relative uncertainties on the dominant frequencies $f_{220}$ and damping times $\tau_{220}$ are prospected to scale with the ringdown SNR as 
\begin{equation}\label{deltafSNRringdown}
    \frac{\Delta f_{220}}{f_{220}}\sim0.2\%\left(\frac{100}{\rm SNR}\right)\,,
    \qquad\frac{\Delta \tau_{220}}{\tau_{220}}\sim2\%\left(\frac{100}{\rm SNR}\right)\,,
\end{equation}
which propagate similar relative errors on the mass and spin of the remnant (assuming the latter to be a Kerr BH) thus giving an estimate of the performance of
consistency tests informed by the inspiral part of the signal~\cite{LIGOScientific:2021sio}.
On the other hand, the subdominant frequencies scale as
\begin{equation}
    \frac{\Delta f_{330}}{f_{330}}\sim2\%\left(\frac{100}{\rm SNR}\right)\,,
    \quad\frac{\Delta f_{210}}{f_{210}}\sim4\%\left(\frac{100}{\rm SNR}\right)\,,
    \quad\frac{\Delta f_{440}}{f_{440}}\sim3\%\left(\frac{100}{\rm SNR}\right)\,. \label{errorsQNMs}
\end{equation}
This shows that, by detecting high-SNR events, and in particular golden events with ${\rm SNR}\geq100$, ET will be able to measure the QNM spectrum at the percent level for individual events. Moreover, by stacking together multiple events (see Table~\ref{tab:spectro_results_1} below), it will be possible to test deviations from the GR predictions at the sub-percent level. 

\begin{figure}[t]
    \centering
    \includegraphics[width=\textwidth]{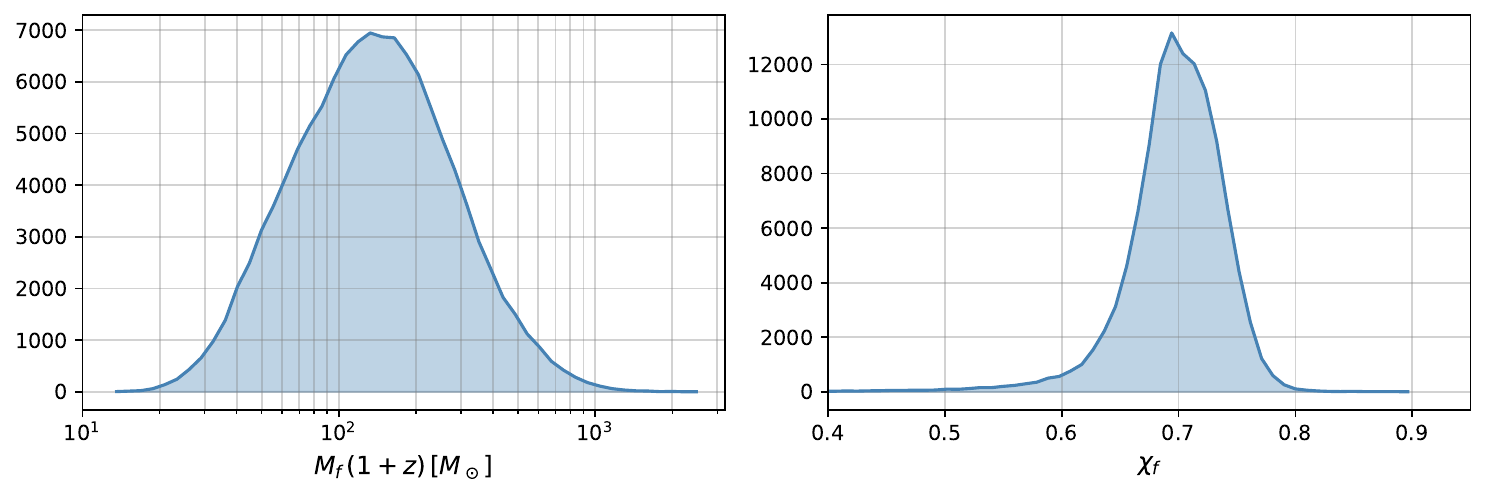}\\
    \includegraphics[width=\textwidth]{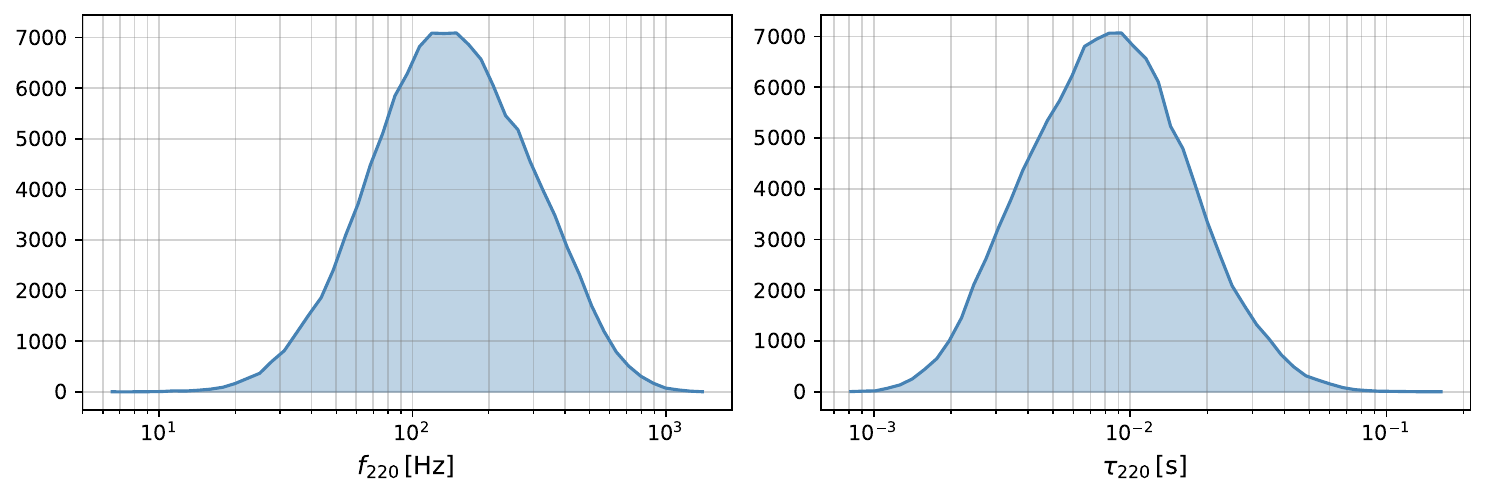}
    \caption{\small Top: Distribution of the redshifted final masses and dimensionless final spins from the simulated BBH events as detailed in Section~\ref{sect:CBC}.
    Bottom: Corresponding distribution of the ringdown frequency and damping time of the dominant $(2,2,0)$ mode. 
    }
    \label{fig:final_properties}
\end{figure}

Next, using the BBH population catalog generated with the {\sc fastcluster} code~\cite{Mapelli:2021syv,Mapelli:2021gyv} as described in Section~\ref{sect:CBC}, we estimate the event rates for BH spectroscopy with different ET designs.
Figure~\ref{fig:final_properties} shows the distributions of source-frame mass $M_f^z=(1+z) M_f$ and $\chi_f$ for the final BH, as predicted from the simulated BBH merger events in the catalog, and computed from the analytical fits in \cite{Barausse:2012qz,Hofmann:2016yih}.\footnote{We also benchmarked them against \texttt{surfinBH} \cite{Varma:2018aht}, which is a machine learning predictor trained on numerical simulations, finding excellent agreement within the common regimes of validity.} 
%
The spin $\chi_f$ clusters around $0.7$ as a consequence of the fact that the majority of the BBH events in the catalog have progenitor mass ratio close to unity (with only a small portion having mass ratio $q\equiv m_1/m_2\gtrsim5$) and moderate progenitor spins.
Furthermore, we see that the distribution of the QNM fundamental frequency peaks at around $100\,{\rm Hz}$, which sets the typical scale of the (almost monochromatic) ringdown signal.


In order to assess the feasibility of BH spectroscopy with different ET configurations, we compute the number of ringdown detections per year. Here a detection corresponds to $\text{SNR}\geq12$ in the ringdown-only portion of the signal. Moreover, we compute the number of ringdown detections per year with ringdonwn signal-to-noise ratio $\text{SNR}\geq50$ and with $\text{SNR}\geq100$, the latter representing golden events for BH spectroscopy. All detector designs are assumed in the HFLF-cryo configuration. {\em We did not find any significant difference when the LF instrument is absent. This is due to the fact that the ringdown signal is almost monochromatic and, as shown in Fig.~\ref{fig:final_properties}, there is a negligible portion of events for frequencies $f_{220}\lesssim 20\,{\rm Hz}$, where the contribution of the LF instrument impacts on the ET sensitivity curve.}
Results are reported in Table~\ref{tab:spectro_results_1} for the particular realization of the catalog considered here, where the rates are subjected to the statistical Poisson counting uncertainty $\sigma(N_{\rm det})\approx\sqrt{N_{\rm \det}}$.
\begin{table}[t]
    \centering
    \begin{tabular}{c||c|c|c|c}
         \hline\hline
          & $N_{\rm det}({\rm SNR}\geq12)$ & $N_{\rm det}({\rm SNR}\geq50)$  & $N_{\rm det}({\rm SNR}\geq100)$ & max(SNR)\\
         \hline\hline
         LVKI-O5 & 15 & 0 & 0 & 38\\
         \hline\hline
         ET \\
         \hline
         $\Delta$-10~km & 4665 & 34 & 3 & 262\\
         $\Delta$-15~km & 11692 & 117 & 13 & 317\\
         2L-15~km-$0^{\circ}$ & 11627 & 110 & 11 & 281\\
         2L-15~km-$45^{\circ}$ & 10175 & 97 & 9 & 330\\
         2L-20~km-$0^{\circ}$ & 18972 & 255 & 21 & 327\\
         2L-20~km-$45^{\circ}$ & 17185 & 228 & 18 & 384\\
         \hline\hline
    \end{tabular}
    \caption{\small Number of ringdown events per year above the detection threshold $({\rm SNR}\geq12)$, with high signal-to-noise ratio $({\rm SNR}\geq50)$, and golden events $({\rm SNR}\geq100)$ for different proposed detector designs (all taken in the HFLF-cryo configuration) and compared to the most optmistic prospect for LVKI in O5. The last column displays the SNR of the loudest event in the catalogue. Here SNR refers to the ringdown part of the signal only.} 
    \label{tab:spectro_results_1}
\end{table}

{\it
In conclusion, for BH spectroscopy, the baseline 10~km triangle configuration  allows for $\sim 34$ high-SNR events and 3 golden events per year. While these numbers are already extremely interesting, both the 15~km triangle design and the 15~km 2L design allow to  increase by a factor $\sim 3-4$  the detection rate w.r.t.~the baseline,
with $90+$ high-SNR events and $9+$ golden events per year and a slight preference for the 15-km triangle. The 20~km 2L configurations show the best performances in terms of detected events, with $\sim240$ high-SNR events\footnote{A back-of-the-envelope estimate, assuming that by stacking $N_{\rm det}$ signals the errors shown in Eq.~\eqref{errorsQNMs} are reduced by a factor $\approx\sqrt{N_{\rm det}}$, suggests that by stacking multiple ringdown detections with ${\rm SNR}>50$, the 2L-15~km-$45^{\circ}$ configuration can reach $\approx 0.1\%$ accuracy for the subdominant QNMs.} and $\sim20$ golden events per year.}
Table~\ref{tab:spectro_results_1} also reports the event rates as measured by the LVKI network in O5. {\em We see that ET will constitute a game changer for probing near-horizon physics.}

\begin{figure}[t]
    \centering
    \includegraphics[width=\textwidth]{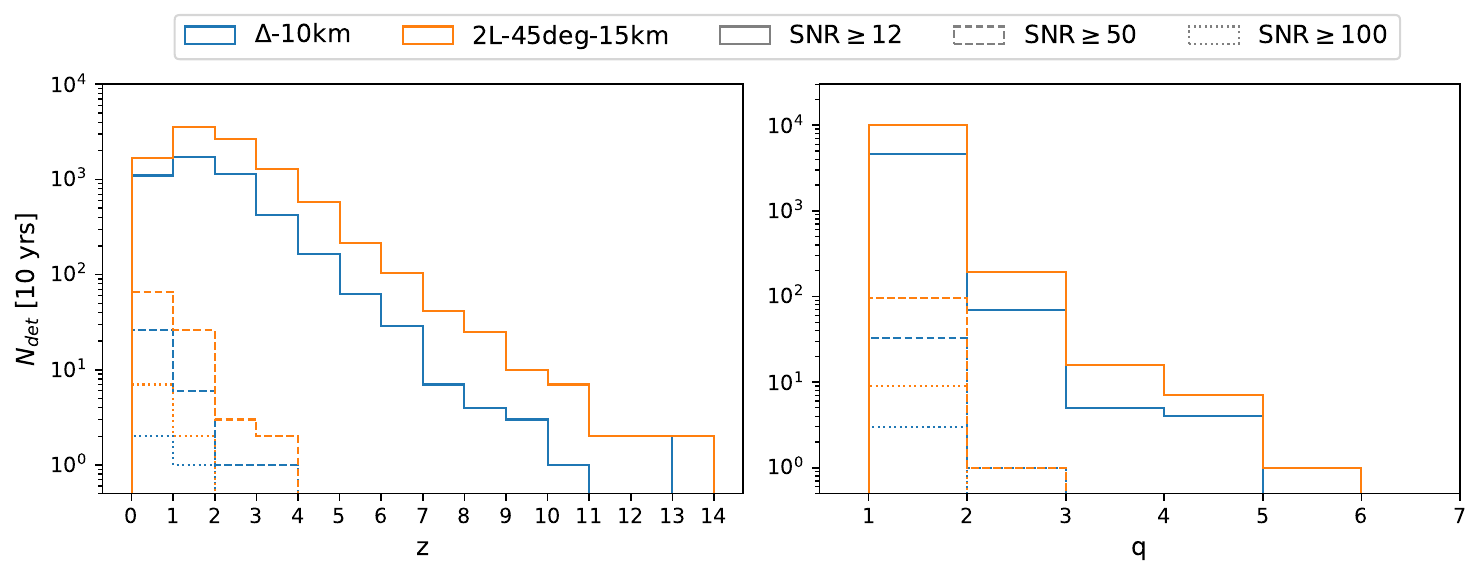}    
    \caption{\small Ringdown event rates as a function of redshift (left) and mass ratio (right) for the baseline 10km triangle and the 15km-$45^{\circ}$-2L configuration. Different colors indicate detectors configurations; continuous, dashed, and dotted contours indicate different ringdown SNR thresholds.}
    \label{fig:spectro_ndet}
\end{figure}

Next, we investigate the redshift reach of BH spectroscopy for the two representative cases of a 10-km triangle and a 15km-$45^{\circ}$-2L configuration. Figure~\ref{fig:spectro_ndet}~(left) shows the event rate per year as a function of the redshift: while ET is capable of detecting ringdown events up to redshift $z\lesssim14$, high-SNR are limited to $z\lesssim4$ and golden events are limited to $z\lesssim2$. Figure~\ref{fig:spectro_ndet}~(right) displays the complementary information of the event rate as a function of the mass ratio: we see that the loud ringdown events are limited to small mass ratios $q\lesssim3$ while high mass-ratio events are detected with moderate or small SNR. Since different mass ratios excite different modes in the ringdown spectrum, high mass-ratio events offer valuable complementary information and it is, therefore, desirable to develop efficient stacking techniques~\cite{Meidam:2014jpa,Yang:2017zxs,Maselli:2019mjd} to take advantage from them.

Finally, in Table~\ref{tab:spectro_results_2} we also computed the event rates when ET operates in synergy with a single 40~km CE detector and with a network of one 40~km CE and one 20~km CE. 
The results show that joint detections by ET in its baseline configuration and CE produce ringdown rates comparable with ET alone in its 2L-20~km configurations,
and  that a 15~km or 20~km ET in synergy with CE will constitute a major improvement for BH spectroscopy as it will result in between 30 to 50 golden events per year. We also observe that the differences among the results, induced by different choices of the ET geometry, remain quite significant even when ET is in a network with 1CE or  with 2CE.

\begin{table}[t]
    \centering
    \begin{tabular}{c||c|c|c|c}
         \hline\hline
          ET (+1CE) & $N_{\rm det}({\rm SNR}\geq12)$ & $N_{\rm det}({\rm SNR}\geq50)$ & $N_{\rm det}({\rm SNR}\geq100)$ & max(SNR)\\
         \hline\hline
         $\Delta$-10~km & 17268 & 188 & 18 & 298\\
         $\Delta$-15~km & 23634 & 311 & 29 & 350\\
         2L-15~km-$0^{\circ}$ & 23342 & 311 & 29 & 317\\
         2L-15~km-$45^{\circ}$ & 22262 & 290 & 27 & 362\\
         2L-20~km-$0^{\circ}$ & 29307 & 504 & 41 & 359\\
         2L-20~km-$45^{\circ}$ & 28126 & 439 & 38 & 412\\
         \hline\hline
         ET (+2CE) &  &  & \\
         \hline
         $\Delta$-10~km & 20990 & 268 & 25 & 308\\
         $\Delta$-15~km & 27065 & 383 & 38 & 359\\
         2L-15~km-$0^{\circ}$ & 26691 & 384 & 37 & 327\\
         2L-15~km-$45^{\circ}$ & 25749 & 357 & 37 & 370\\
         2L-20~km-$0^{\circ}$ & 32266 & 597 & 49 & 368\\
         2L-20~km-$45^{\circ}$ & 31197 & 531 & 45 & 420\\
         \hline\hline
    \end{tabular}
    \caption{\small Same as Table~\ref{tab:spectro_results_1} but with a single 40~km CE detector (top) or with one 40~km CE detector and one 20~km CE detector (bottom) added to the network.}
    \label{tab:spectro_results_2}
\end{table}
%



\subsubsection{Searching for echoes and near-horizon structures}\label{sect:echoes}



Exotic compact objects (\acrshort{eco}s) are horizonless objects predicted in certain quantum-gravity extensions of GR and in the presence of exotic matter fields~\cite{Giudice:2016zpa,Cardoso:2019rvt,Abedi:2020ujo}. %
Some broad motivations for ECOs are (see~\cite{Cardoso:2019rvt} for a review): i) resolving the singularities inevitably present inside BHs within GR; ii) evading deep conceptual conundrums associated with the presence of a horizon (most notably, the information loss paradox); or iii) simply providing effective models for new species of compact objects that might co-exist in the universe along with BHs and NSs.

Several models of ECOs have been conceived, including BH ``fuzzball" microstates in string theories~\cite{Mathur:2005zp,Bena:2022ldq} and boson stars as self-gravitating objects made of massive bosonic fields minimally coupled to GR~\cite{Jetzer:1991jr,Liebling:2012fv}.
One can devise a model-independent framework~\cite{Maggio:2021ans} in which the ECO properties are parametrized in terms of their effective radius $r_0 = r_+ (1 + \epsilon)$ (where $r_+$ is the would-be location of the Kerr BH horizon and $\epsilon$ is a dimensionless closeness parameter), and the reflectivity ${\cal R}$ at the effective radius with a generic phase $\phi(\mathcal{R})$. The BH case is described by a null reflectivity at the horizon for any frequency, whereas an ECO reflects part of the incoming radiation, i.e. ${\cal R}\neq0$ (the reflectivity is generically complex and frequency dependent).
Because of this reflectivity, ECOs emit a different GW signal relative to the BH case. One of their smoking guns is the emission of a modulated train of GW echoes in the late-time ringdown stage of a compact binary coalescence, associated to signals reflected off the object interior~\cite{Cardoso:2016rao,Cardoso:2017cqb}.
When an ECO remnant with $\epsilon \ll 1$ forms, a ringdown analogous to the BH one is emitted at early times, followed by a new signal in the form of late-time echoes. Several searches for GW echoes in LVK data have been performed, claiming no evidence for their observation~\cite{Westerweck:2017hus,Lo:2018sep,Uchikata:2019frs,Tsang:2019zra,LIGOScientific:2021sio} (although some results are controversial, see~\cite{Abedi:2020ujo}). 

\begin{figure}
    \centering
    \includegraphics[width=0.5\textwidth]{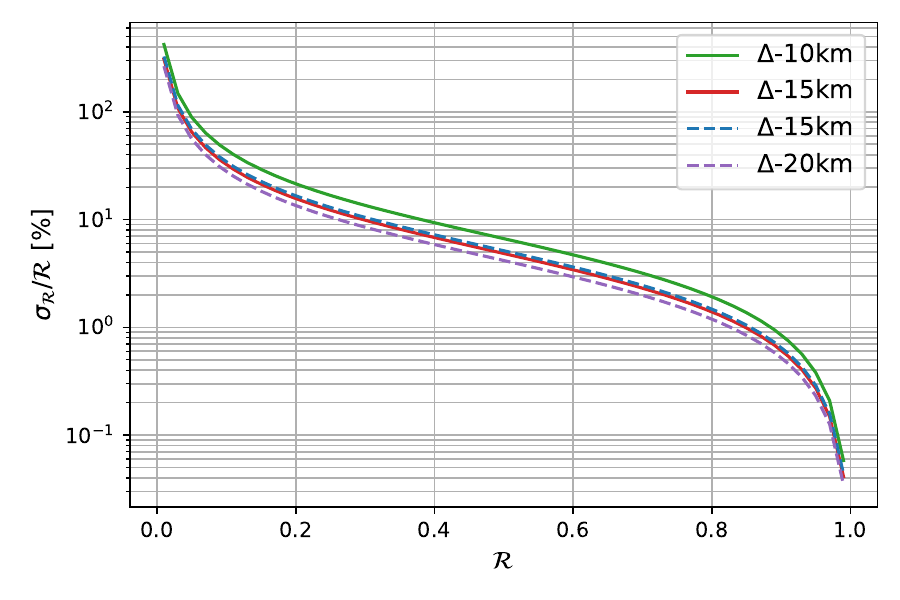}
    \caption{\small Fractional percentage errors on the ECO reflectivity ${\cal R}$ for a GW150914-like event placed at $d_L=1$ Gpc with final mass $M_f=70 M_\odot$ and final spin $\chi_f =0.68$. The detector is taken in the cryogenic HFLF configuration and errors are averaged over the sky position and the inclination angle. There are no significant deviations in the figure for the HF-only  configuration. The Fisher-matrix parameters are $\{ M_f, \chi_f, \mathcal{R}, \phi({\cal R}), \epsilon \}$, where $\mathcal{R}$ is generically a complex number, and the amplitude and phases $\{{\cal A}_{lmn},\phi_{lmn}\}$ of the excited modes.}
    \label{fig:echo}
\end{figure}

Here, we assess the detectability of GW echoes with different ET configurations.
We focus on a GW150914-like system (final mass $M_f=70 M_\odot$ and final spin $\chi_f =0.68$) placed at a fiducial distance $d_L=1$ Gpc. 
%
We use the frequency-domain echo template developed in Ref.~\cite{Maggio:2019zyv} for a representative choice $\epsilon=10^{-5}$ where we vary the reflectivity in the range ${\cal R}\in[0.01,0.99]$ and set $\phi(\mathcal{R})=0$.
Figure \ref{fig:echo} shows the fractional percentage errors (obtained with a Fisher matrix analysis) on the reflectivity of an ECO for different ET configurations. The fractional error is a proxy for the detectability of ECOs, since the BH case corresponds to ${\cal R}=0$ and therefore any putative measurement of ${\cal R}$ should exclude ${\cal R}=0$ with the highest confidence level.
We report the values of $\sigma_{\cal R}/{\cal R}$ after averaging over the sky position and the inclination angle. We take the detectors in the cryogenic HFLF configuration but, {\em as in the ringdown analysis reported in Section~\ref{sect:testGRnearhorizon}, we do not find significant deviations when the LF instrument is not operating.}

\textit{The main result is that the detector configuration  affects the accuracy on the reflectivity of compact objects by a factor of $\sim2$ between the 10 km designs and the 20 km designs, with the 15 km designs representing the best compromise.}

This is further quantified in Table~\ref{tab:echo} for three values of the reflectivity (small, medium and large, respectively). We can ascribe these differences to the larger SNR of longer arm length configurations. The distribution of the SNR follows the same statistics of the ringdown SNR as reported in Table~\ref{tab:spectro_results_1} and Fig.~\ref{fig:spectro_ndet}  
and therefore the same considerations apply about the advantage of a 15-km design relative to the baseline 10-km design.

\begin{table}[hb]
    \centering
    \begin{tabular}{||c||c|c|c||}
         \hline\hline
          & \multicolumn{3}{c||}{$\sigma_{\cal R}/{\cal R}~[\%]$}\\
           & ${\cal R}=0.01$ & ${\cal R}=0.5$ & ${\cal R}=0.99$\\
         \hline
         $\Delta$-10~km & 422 & 7 & 0.06\\
         $\Delta$-15~km & 308 & 5 & 0.04\\
         2L-15~km & 326 & 5 & 0.04\\
         2L-20~km & 265 & 4 & 0.03\\
         \hline\hline
    \end{tabular}
    \caption{\small Fractional percentage errors on the measurability of the ECO reflectivity ${\cal R}$ for small (${\cal R}=0.01$), medium (${\cal R}=0.5$) and large (${\cal R}=0.99$) reflectivity for the same system and detector configuration of Fig.~\ref{fig:echo}.}
    \label{tab:echo}
\end{table}

\subsubsection{Constraining tidal effects and multipolar structure}\label{sec:tidaleffects}

The multipolar structure of a Kerr BH can be elegantly written as~\cite{Hansen:1974zz},
\begin{equation}
\label{eq:nohair}
M_{\ell}^{\rm BH}+ i S_\ell^{\rm BH}=M^{\ell+1}\left(i \chi\right)^\ell\,,
\end{equation}
where ${\chi:=J/M^2}$ is the dimensionless spin, $M=M_{0 0}$, $J=S_{1 0}$, $M_\ell=M_{\ell 0}$, $S_\ell=S_{\ell 0} $,
with $M_{lm}$ and $S_{lm}$ being the mass and current multipole-moment tensors, respectively.
For a Kerr BH, the only nonvanishing multipole moments have $m=0$ and the mass (current) multipole moments vanish when $\ell$ is odd (even). These properties are a consequence of the  \emph{axial} and \emph{equatorial symmetry} of the Kerr solution.

In general, objects other than BHs violate these symmetries and the unique relation~\eqref{eq:nohair} between all multipole moments and the BH mass and spin. Their multipolar structure can be schematically written as
\begin{equation}
\label{eq:momentsECOs}
M_{\ell m}=M^{\rm BH}_{\ell} + \delta M_{\ell m}\quad\,,\quad
S_{\ell m}=S^{\rm BH}_{\ell} + \delta S_{\ell m}\,,
\end{equation}
where $\delta M_{\ell m}$ and $\delta S_{\ell m}$ are some model-dependent corrections to the mass and current multipole moments. 

\begin{figure*}[ht]
\centering
\includegraphics[width=16cm]{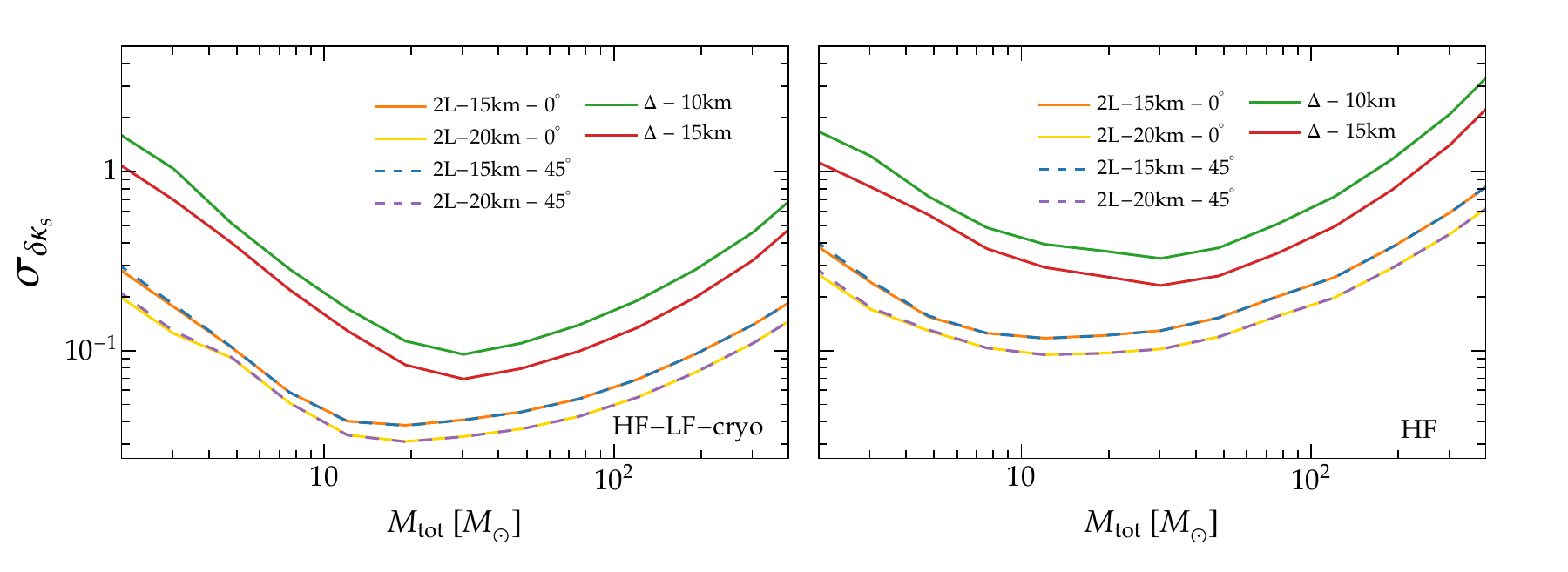}
	\caption{\small 
	1-$\sigma$ uncertainty on the parameter  controlling deviations on the symmetric combination of the spin-induced 
	ECO quadrupole moment, $\kappa_s=1+\delta\kappa_s$, with 
	$\kappa_s=1$ corresponding to binaries with Kerr BHs. We assume the Kerr case ($\delta\kappa_s=0$) and optimally-oriented sources with mass ratio $q=2$, non-precessing spins $\chi_1=0.9$, $\chi_2=0.7$ at a luminosity distance of $d_L=100\,{\rm Mpc}$. The optimal orientation depends on the detector location and configuration. The Fisher-matrix parameters are $ \{{\cal M}_c, \eta, d_L, \theta, \phi, \iota, \psi, t_c, \Phi_c, \chi_{1,z}, 
	\chi_{2,z},\delta\kappa_s$.\}}
\label{fig:err_quadrupole}
\end{figure*}

Generically, the mass quadrupole moment $M_{2m}$ is the dominant multipolar contribution, entering the inspiral GW signal at second PN order~\cite{Krishnendu:2017shb}. When $m\neq0$, this term can induce binary precession even in the absence of progenitor spins~\cite{Loutrel:2022ant}.
Deviations from standard BH predictions can be parameterized by defining $M_{20}=-\kappa M^3 \chi^2$, where $\kappa=1$ corresponds to the Kerr case, while $\kappa(\chi)$ in all other cases. At the 2PN order, 
contributions from the quadrupole moments of the binary 
components enter the waveform through symmetric and 
antisymmetric combinations given by 
$\kappa_s=(\kappa_1+\kappa_2)/2$ and  
$\kappa_a=(\kappa_1-\kappa_2)/2$, respectively. The latter 
can be recast in terms of deviations from the Kerr 
baseline as $\kappa_s=1+\delta\kappa_s$, and 
$\kappa_a=\delta\kappa_a$. 
Figure~\ref{fig:err_quadrupole} shows the absolute errors on 
the parameter $\delta\kappa_s$ as a function of the binary total 
mass, for representative systems with mass ratio $q=2$ and spins $(\chi_1,\chi_2)=(0.9,0.7)$ aligned to the orbital angular momentum (and therefore nonprecessing). We use a TaylorF2 waveform model, truncating 
the inspiral phase at the \acrshort{isco} of the Kerr remnant, also including
corrections induced by the self-force and the spin of 
the less massive object~\cite{Favata:2010ic}.

We find that the differences on the uncertainties obtained assuming the triangle and the L-shaped 
configurations are typically of a factor two  in favor of the triangle, across the range of masses we considered.


Another smoking-gun deviation from the standard Kerr BH in GR is given by the tidal deformability, considered in Section~\ref{sect:PEBNS} for BNSs. 
A remarkable result in GR is that the tidal deformability of BHs is precisely zero~\cite{Binnington:2009bb,Damour:2009vw,Gurlebeck:2015xpa}, 
at variance with any other object in GR or with BHs in modified gravity~\cite{Pani:2015tga,Uchikata:2016qku,Porto:2016zng,Cardoso:2017cfl,Giddings:2019ujs,Sennett:2017etc}.

\begin{figure*}[t]
	\centering
	\includegraphics[width=1\textwidth]{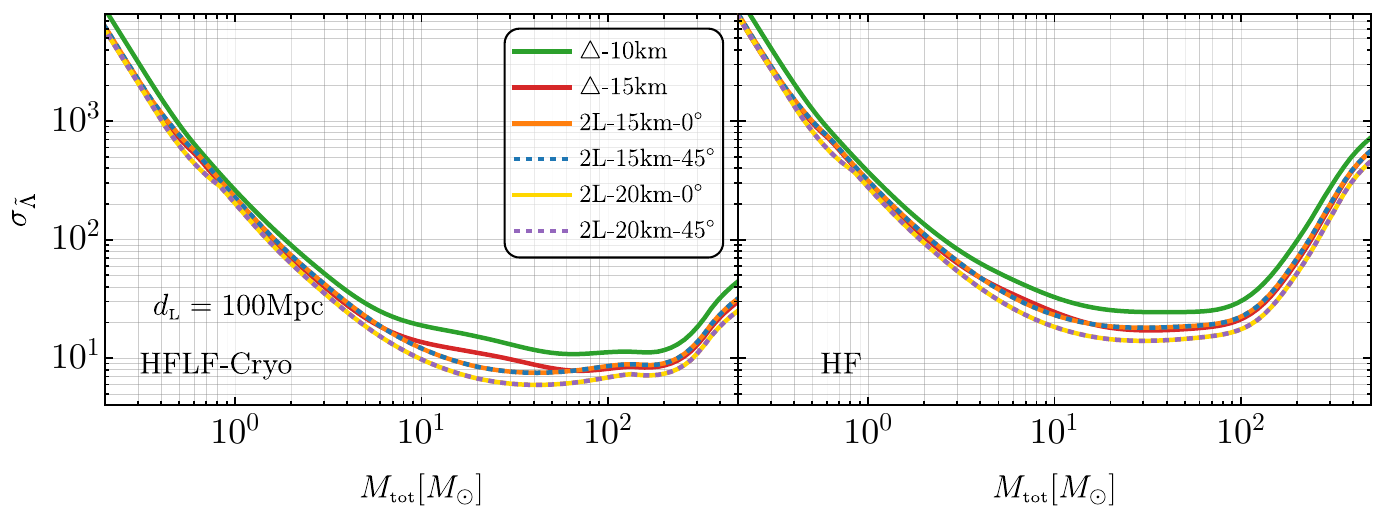}
	\caption{\small 
	Absolute (1-$\sigma$) uncertainty on the tidal deformability $\tilde \Lambda$
	as a function of total mass $M_\text{\tiny tot}$ in the source frame for the various configurations.
	We assume negligible deformability of the source, nonspinning equal-mass binaries, and an optimally oriented source at a distance $d_L=100\,{\rm Mpc}$. The optimal orientation depends on the detector location and configuration. The Fisher-matrix parameters are $ \{{\cal M}_c, \eta, d_L, \theta, \phi, \iota, \psi, t_c, \Phi_c, \chi_{1,z}, \chi_{2,z}, \tilde{\Lambda}, \delta \tilde{\Lambda}\}$.
	These results complement what presented in Section~\ref{sect:PEBNS} for BNSs to the case of ECOs or subsolar BHs (i.e. PBHs) potentially observable with ET, see more details in Section~\ref{sec:otherPBHs}. }
\label{fig:err_tidal_deformability}
\end{figure*}

Figure~\ref{fig:err_tidal_deformability} shows the absolute error on the binary tidal parameter $\tilde \Lambda$ as 
a function of the source frame total mass, for non-spinning, equal-mass, optimally oriented binaries, at luminosity distance $d_L=100\,{\rm Mpc}$. We assume the TaylorF2 waveform,\footnote{We checked that, at the level of the comparison between different configurations, assuming the IMRPhenomD\_NRTidalv2 waveform approximant gives analogous results. 
However, in the range of masses of its validity, the latter provides a smaller uncertainty on $\tilde \Lambda$ due to the fact that the waveform (calibrated on BNS coalescences) is valid up to higher frequencies, as well as to a dependence of the amplitude on the deformability (which is only included in the phase of TaylorF2).} cutting the inspiral phase at the ISCO frequency of a remnant Schwarzschild BH with mass $\approx M_{\rm tot}$.
Extending the Fisher analysis to higher frequencies would significantly improve measurement errors but would require specific ECO models.

{\it We conclude that different ET configurations do not impact significantly on the accuracy in the measurement of the tidal deformability for $M_{\rm tot}\lesssim {\cal O}(M_\odot)$, while differences are slightly larger, within typically a factor of order 2,  for heavier binaries (with the 2L configurations performing again better than the triangle ones). }

\subsection{Nuclear physics}

One of the longstanding open challenges in nuclear physics is to determine the properties of matter at supranuclear densities.  Given the extremely high central densities in neutron stars and the possibility to form merger remnants that not only probe the highest densities in our Universe before black hole formation but also matter at nonzero temperature, the Einstein Telescope will provide important nuclear-physics insights, see e.g., Refs.~\cite{Maggiore:2019uih,Weih:2019xvw,Bauswein:2018bma,Carson:2019xxz,Pacilio:2021jmq,Smith:2021bqc,Gupta:2022qgg,Breschi:2022ens,Puecher:2022oiz,Wijngaarden:2022sah}. In particular, it will be possible to probe the cold EoS from the inspiral phase with unprecedented accuracy and furthermore, in contrast to existing gravitational-wave detectors, there is a high chance of detecting gravitational-wave emission from the merger and its remnant which not only probes the cold EoS but would also allow us to place constraints at the EoS at temperatures up to $\sim 50 \rm MeV$; cf.~e.g.\ \cite{Baiotti:2016qnr} and references therein. 
For these reasons, it is essential to understand to which extent the different detector configurations impact possible nuclear-physics constraints. 

\subsubsection{Radius estimation from Fisher-matrix computation}
\label{sec:Fisher_nuclear}


We start by assessing the accuracy of constraining the EoS by providing estimates for the uncertainty in the inferred radius based on the Fisher matrix approach. The results used here differ from those in Section~\ref{sect:PEBNS} due to the assumption of a common EoS for the NSs when producing the events catalog. In particular, we adopt the DD-LZ1 model\footnote{We point out that the EoSs employed here do not contain phase transitions, however, we expect that the presence of a strong phase transition will likely influence the accuracy to which we can measure the neutron star radius.} \citep{Wei:2020kfb, Xia:2022pja, Xia:2022dvw} and compute the adimensional tidal deformability parameters of the objects from their source-frame masses $m$, using the $\Lambda(m)$ relation predicted by this EoS. The choice of DD-LZ1 is dictated by its high allowed maximum mass (of $2.56\,{\rm M}_{\odot}$); this makes it consistent  with the  BNS population model  used throughout this work which, as mentioned in Section~\ref{sect:CBC}, assumes a source-frame mass distribution of the component objects  uniform  between $1.1\,{\rm M}_{\odot}$ and $2.5\,{\rm M}_{\odot}$.  {Moreover, the behavior of DD-LZ1 complies with all the present constraints from low energy nuclear physics \citep{Margueron:2017eqc}}. As in Section~\ref{sect:PEBNS},  the results are produced using the \texttt{GWFAST} package \cite{Iacovelli:2022bbs,Iacovelli:2022mbg} and adopting the IMRPhenomD\_NRTidalv2 waveform model. {A detailed study of the effect of the injection EoS and waveform model is ongoing.}

\begin{figure}[t]
\hspace{-1.0cm}
\begin{tabular}{l@{\hskip -.01cm}l@{\hskip +.01cm}l}
     \includegraphics[width=5.7cm]{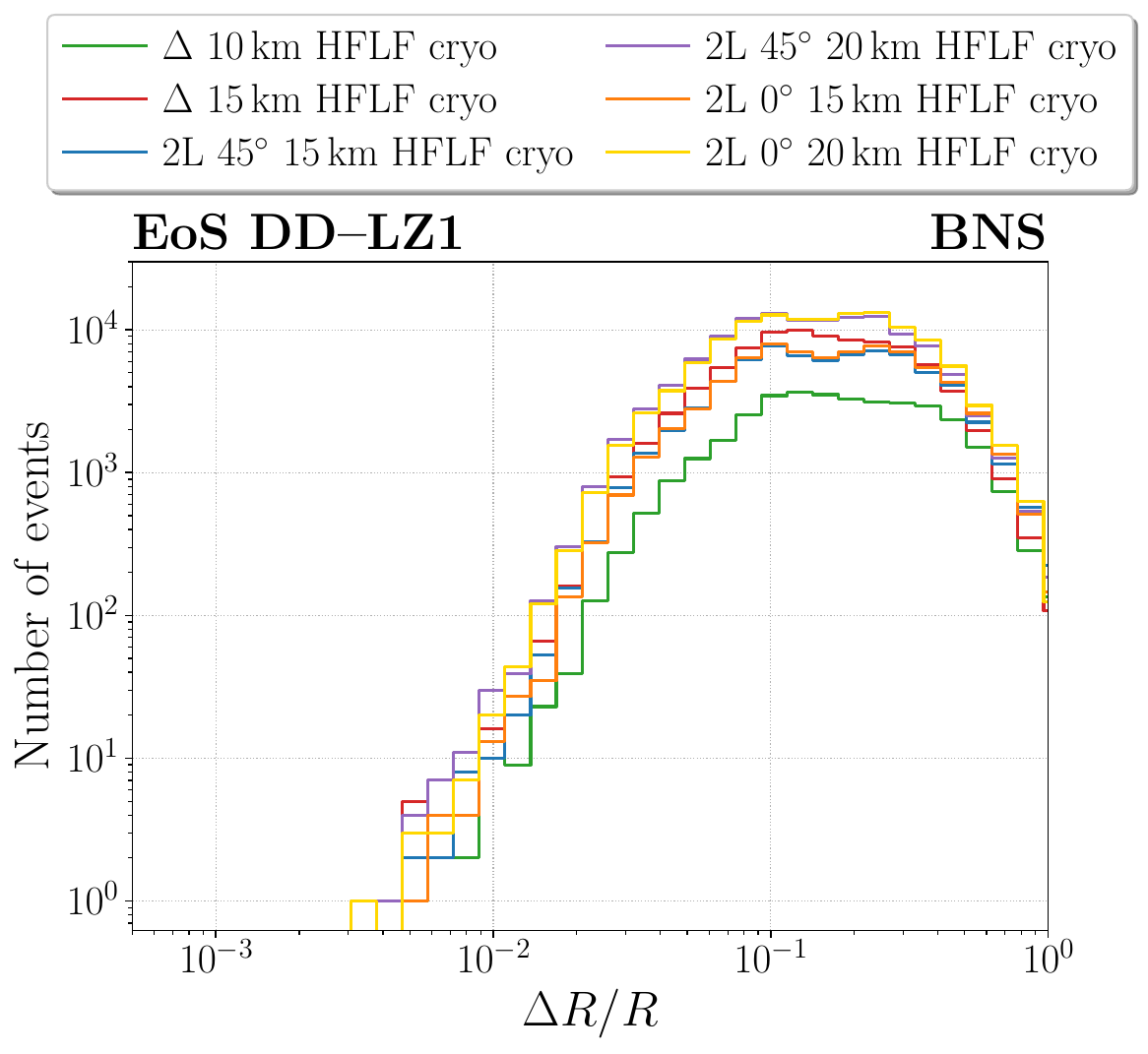} & 
     \includegraphics[width=5.4cm]{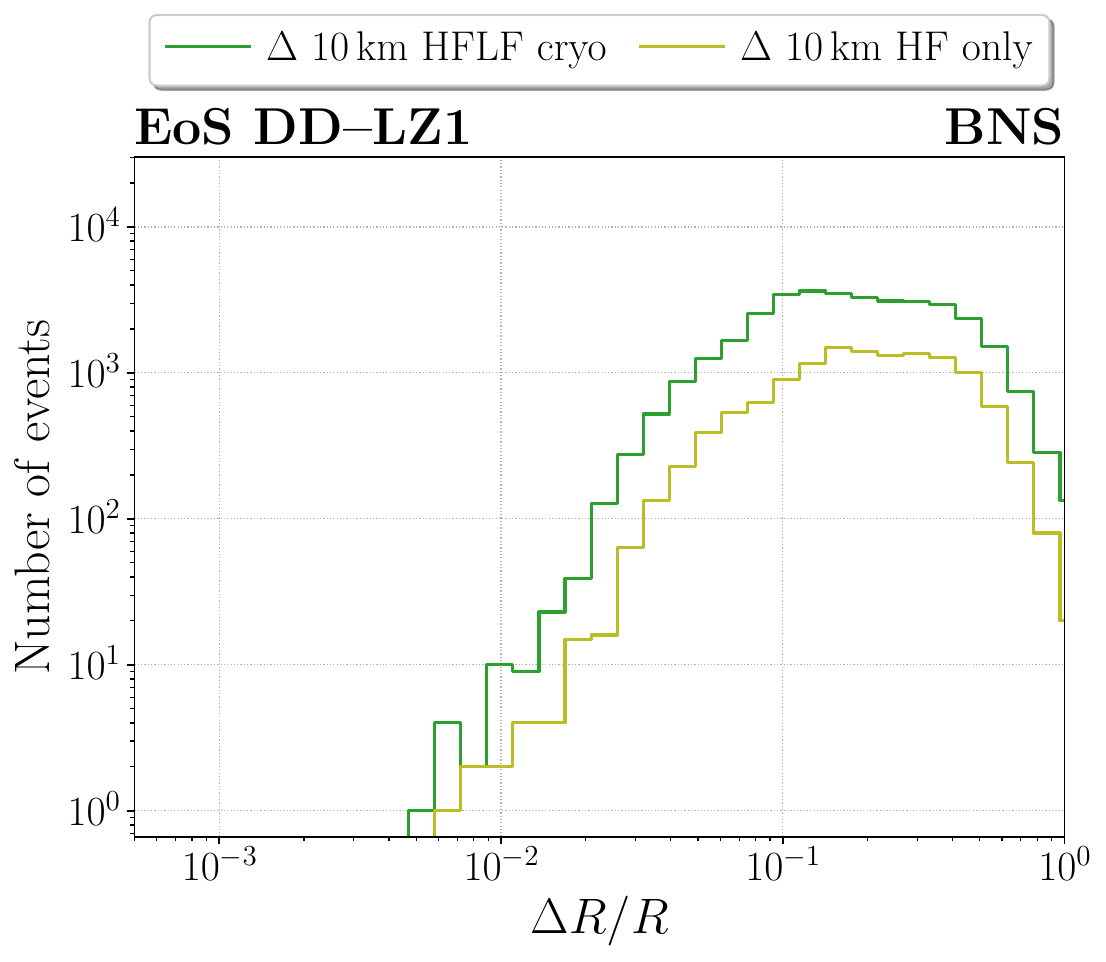} & \includegraphics[width=5.6cm]{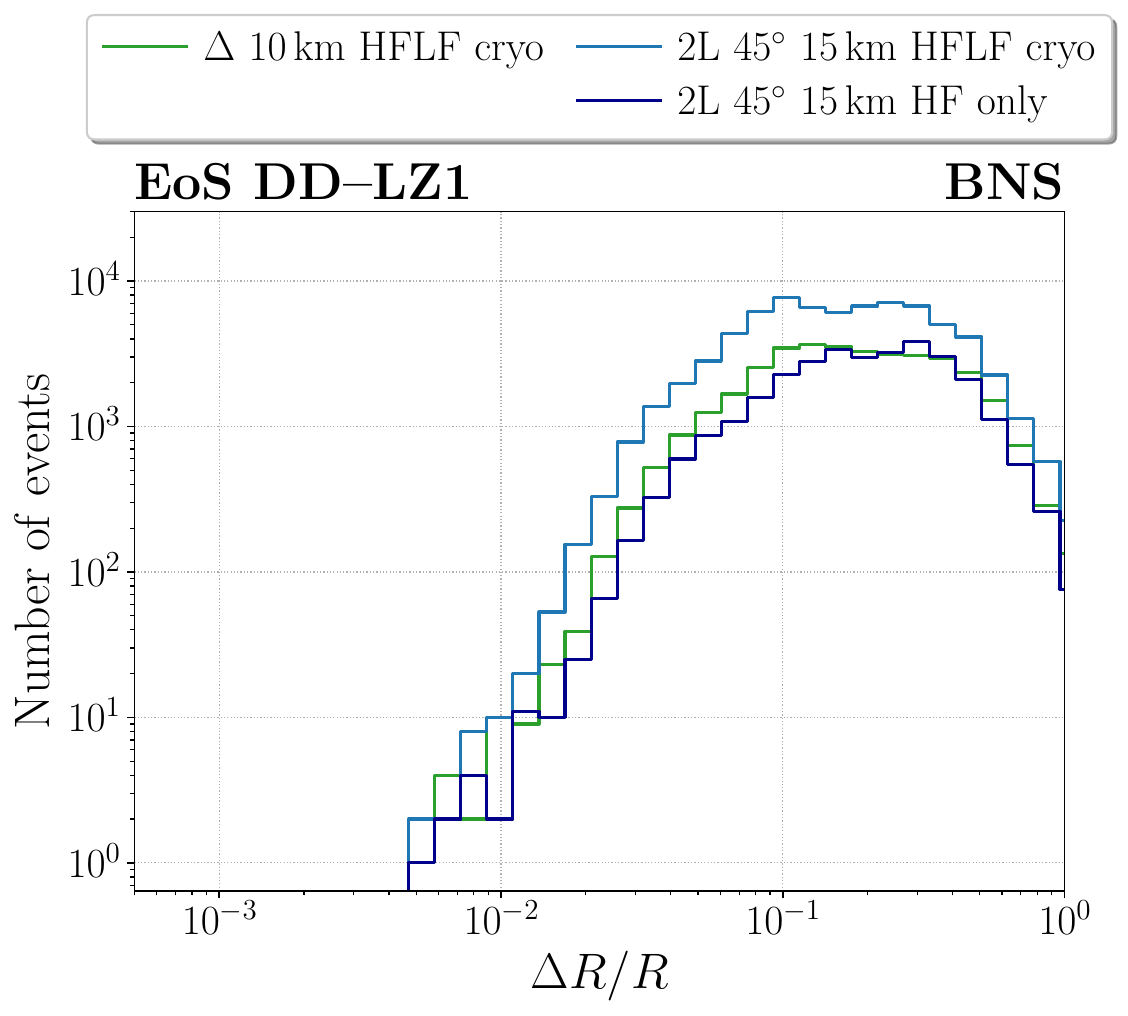} \\
\end{tabular}
    \caption{\small Distribution of the relative errors on the NS radius attainable from the adimensional tidal deformability combination $\tilde{\Lambda}$, as measured by different detector configurations and sensitivity curves. In particular, the left panel shows the results for the six geometries considered, all with their best sensitivity, the central panel for the 10~km triangle with the two different ASDs, and the right panel for the 2L with 15~km arms at $45^{\circ}$ and the two ASDs considered.}
    \label{fig:ET_allgeom_ASD_BNS_Hist_DelRovR}
\end{figure}

We use two ways to transform the error of the tidal deformability $\tilde{\Lambda}$ [defined in \eq{deftildeLambda}] into the error on the measurement of the NS radius $R$. 
The first approach assumes that $\tilde{\Lambda} \propto R^6$, which leads to
\begin{equation}\label{eq:DelR_DelLam_Rel}
    \frac{\Delta R}{R} = \frac{1}{6} \frac{\Delta \tilde{\Lambda}}{\tilde{\Lambda}},
\end{equation}
and arises from relations derived in Ref.~\cite{De:2018uhw}. This allows us directly to compute the individual radius uncertainty $\Delta R/R$ for all detections. We summarize these results in Fig.~\ref{fig:ET_allgeom_ASD_BNS_Hist_DelRovR}. {\em Comparing different configurations (left panel), we find that the main limiting factor is the arm-length of the gravitational-wave detectors, so the 10~km triangle performs less well than a 15~km 2L.}  Consistently with a number of other results in Section~\ref{sect:CBC}, the right panel of Fig.~\ref{fig:ET_allgeom_ASD_BNS_Hist_DelRovR} shows
that the 2L with 15~km arms at $45^{\circ}$ in the HF-only configuration performs at a level comparable to the 10~km triangle in the full HFLF~cryo configuration. 

An important consequence of the large number of BNS detections with ET is that we can noticeably improve the EoS constraints by stacking information from multiple detections at different masses, which enables constraining the EoS at different densities. 
To quantify this, we  perform a simple analysis in which we assume a common radius of 12km radius for all neutron stars.  While this assumption is a priori unrealistic, it nevertheless captures the feature that  
different neutron star measurements will probe the same underlying hyperparameters characterizing the EoS.
To obtain a combined ansatz, we compute the average $\overline{(\Delta \tilde{\Lambda}/\tilde{\Lambda})}$ and use 
\begin{equation}\label{eq:DelR_All_DelLam_Def}
  \Delta  R_{\rm all} = \frac{1}{6\sqrt{N_{\rm det}}} \overline{\left(\frac{\Delta \tilde{\Lambda}}{\tilde{\Lambda}}\right)}\ 12 {\rm km}.
\end{equation}
We present results for 
$R_{\rm all}$ in Table~\ref{tab:div6:Fisher_radius} for all ET configurations, from which we see that {\em loosing the LF instrument, the constraints gets worse by about a factor of 2, irrespectively of the geometry.}

In Table~\ref{tab:div6:Fisher_radius_CE},  complementing the results in Section~\ref{sect:3Gnetwork},  we also report the results for some ET geometries, all with their full HFLF cryogenic ASD, working in a network  with 1 or 2 CE detectors. To improve the reliability of the Fisher matrix approach, we restrict the analysis to events with a relative error on the adimensional tidal deformability of $\Delta\tilde{\Lambda}/\tilde{\Lambda} \leq 50\%$. This threshold is chosen to have a large number of events in each configuration and with each ASD, and also reflects the fact that the loudest events -- with smaller errors -- will be the ones providing most of the information. We emphasize that this error represents only the statistical uncertainty; however, systematic uncertainties, e.g., due to modeling errors in the gravitational waveform will also be important~\cite{Kunert:2021hgm}. Moreover, we stress that the results shown in Fig.~\ref{fig:ET_allgeom_ASD_BNS_Hist_DelRovR} and in Tabs.~\ref{tab:div6:Fisher_radius} and \ref{tab:div6:Fisher_radius_CE} depend on the choice of the EoS. From the dependence of the tidal deformability on the star's compactness, it is evident that, with increasing mass, the tidal deformability decreases. Thus EoS models predicting large maximum masses in general also predict very low values of $\Lambda$ close to the maximum mass, which is difficult to constrain. Indeed, DD-LZ1 predicts tidal deformability as small as $\Lambda\sim 5$ for $m\simeq2.56\,{\rm M}_{\odot}$. Assuming a different EoS [which might require a lowering of the maximum mass in the mass distribution to accommodate the maximum mass allowed by the EoS] could result in tighter or broader constraints. {This will be discussed in further details in section \ref{sect:FullPENuclPhys}.} Furthermore, the uniform distribution adopted in Section~\ref{sect:PEBNS} predicts smaller values for the error on the radius, but using an EoS enables including the physical effect of the tidal deformability parameter decreasing with increasing mass. The results presented here should be taken only as indicative, however, they still provide a relevant benchmark for comparing different detector configurations.

\begin{table}[tb]
\centering
\begin{tabular}{|l|cc|cc|}
\hline
\multirow{2}{*}{\diagbox[height=2.27\line]{Geometry}{ASD}} & \multicolumn{2}{c|}{HFLF cryo}  & \multicolumn{2}{c|}{HF only} \\ 
 &  
\multicolumn{1}{c}{$N_{\rm det}^{\Delta\tilde{\Lambda}/\tilde{\Lambda} \leq 0.5}$} & \multicolumn{1}{c|}{$\Delta R_{\rm all}$ [m]} &
\multicolumn{1}{c}{$N_{\rm det}^{\Delta\tilde{\Lambda}/\tilde{\Lambda} \leq 0.5}$} & \multicolumn{1}{c|}{$\Delta R_{\rm all}$ [m]} \\
\hline
$\Delta$ 10~km & 4878 & 10.0 & 1412 & 19.2\\
$\Delta$ 15~km & 15285 & 5.7 & 4047 & 11.3\\
2L 15~km $45^{\circ}$ & 12013 & 6.4  & 3194 & 12.4\\ 
2L 20~km $45^{\circ}$ & 25489 & 4.4 & 7387 & 8.2\\
2L 15~km $0^{\circ}$ & 11884 & 6.5 & 3357 & 12.1 \\
2L 20~km $0^{\circ}$ &  23988 & 4.5 & 7081 & 8.3 \\
1L 20~km & 5464 & 9.4 & 1493 & 19.1 \\ \hline
\end{tabular}
\caption{\small Number of detections, $N_{\rm det}$ with $\Delta\tilde{\Lambda}/\tilde{\Lambda} \leq 50\%$, and statistical error on the NS radius obtained combining all of them according to \eqref{eq:DelR_All_DelLam_Def}, $\Delta R_{\rm all}$ (in meters), for various geometries and ASDs.
\label{tab:div6:Fisher_radius}}

\end{table}

\begin{table}[t]
\centering
\begin{tabular}{|l|cc|cc|}
\hline
\multirow{2}{*}{\diagbox[height=2.27\line]{Geometry}{ASD}} & \multicolumn{2}{c|}{HFLF cryo + 1CE} & \multicolumn{2}{c|}{HFLF cryo + 2CE}  \\ 
 &  
\multicolumn{1}{c}{$N_{\rm det}^{\Delta\tilde{\Lambda}/\tilde{\Lambda} \leq 0.5}$} & \multicolumn{1}{c|}{$\Delta R_{\rm all}$ [m]} & \multicolumn{1}{c}{$N_{\rm det}^{\Delta\tilde{\Lambda}/\tilde{\Lambda} \leq 0.5}$} & \multicolumn{1}{c|}{$\Delta R_{\rm all}$ [m]}\\
\hline
$\Delta$ 10~km & 24129 & 4.6 & 35694 & 3.8 \\
2L 15~km $45^{\circ}$ & 35172 & 3.8 & 47791 & 3.2  \\ 
2L 15~km $0^{\circ}$ & 35217 & 3.8 & 48581 & 3.2 \\
\hline
\end{tabular}
\caption{\small Same as in Table \ref{tab:div6:Fisher_radius} with some selected geometries, all with the full HFLF cryogenic ASD, in combination with 1 CE detector with a length of 40\,km and 2 CE detectors, one with a length of 40\,km and the other of 20\,km. 
\label{tab:div6:Fisher_radius_CE}}

\end{table}

Our second approach is based on first generating a prior distribution of EoS by Monte-Carlo sampling of a large parameter set of 10 independent, uniformly distributed empirical parameters. These parameters characterize the density dependence of the energy in the symmetric matter (i.e.\ equal number of protons and neutrons) and of the symmetry energy (i.e.\ the variation of binding energy as a function of the neutron-to-proton ratio). 
{They are given by the successive derivatives with respect to the density of the symmetric matter and symmetry energy functionals, calculated at the equilibrium density of symmetric matter. } Their prior distribution is consistent
with the present empirical knowledge for a large set of nuclear data~\cite{Margueron:2017eqc}. The use of the same functional to describe the core and the inhomogeneous crust~\cite{Carreau:2019zdy} guarantees a consistent estimation of the crust-core transition inside the neutron star and thus consistent predictions for its radius. This approach enables incorporating priors from nuclear physics on the EoS and including the uncertainties at high densities. The only limiting assumption is that matter is composed of charged leptons, nucleons and nuclei only, and in particular that no first-order phase transition occurs.
There might also be some dependence on the different approximations applied to treat inhomogeneous matter, e.g. assuming vanishing temperature, which could slightly modify the crust properties (see e.g. \cite{Barba-Gonzalez:2022pkn}). However, this should not affect the comparison of different configurations. The assumed nuclear prior complies with the chiral EFT energy per particle band for symmetric and pure neutron 
matter as given by \cite{Drischler:2015eba} for baryon number densities $ 0.02 \le n_B \le 0.18  \mathrm{fm}^{-3}$, see \cite{Thi:2021jhz} for details of the current implementation.

\begin{figure}[t]{}
    \centering
{\includegraphics[width = .5 \columnwidth]{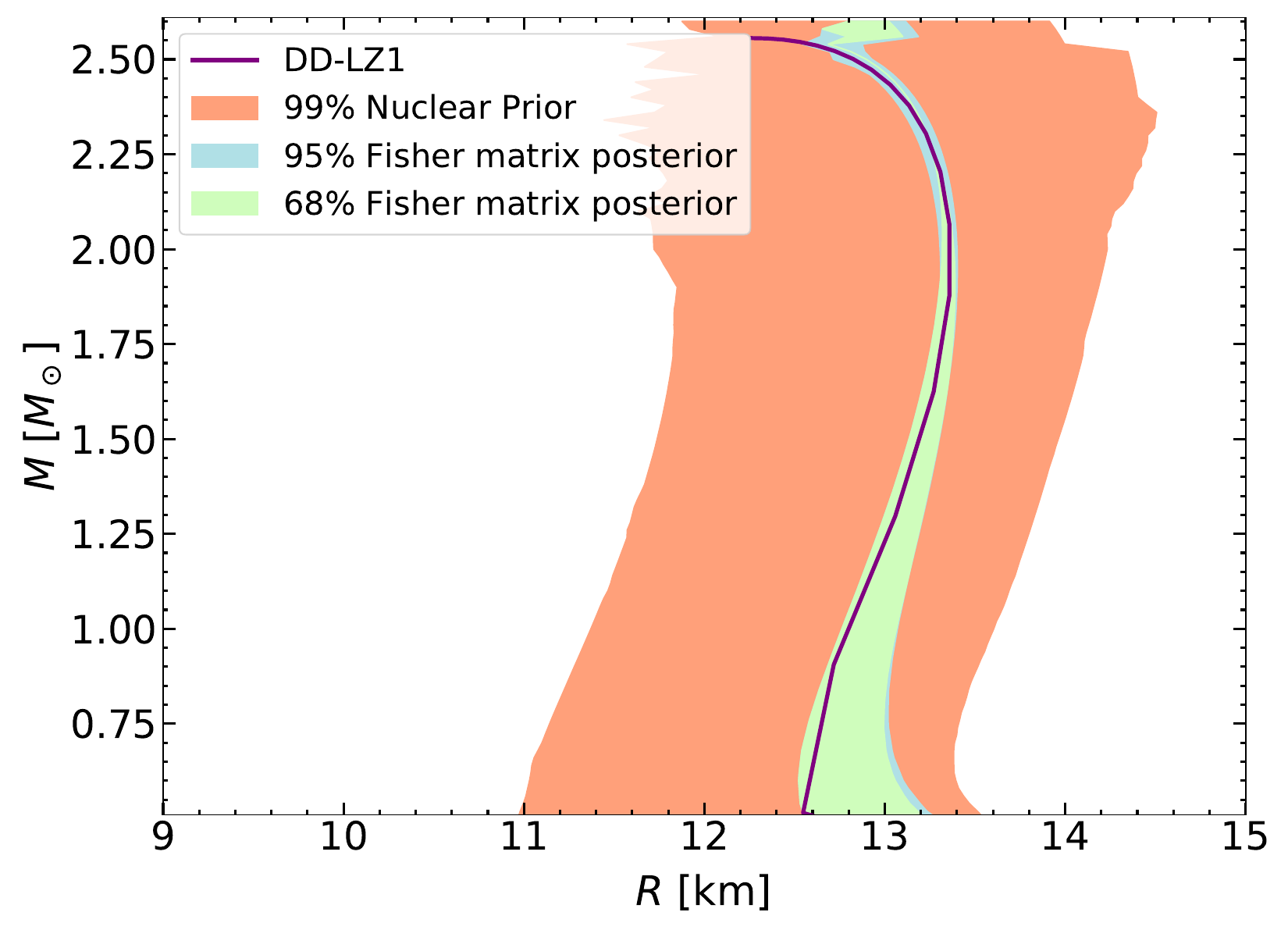}}
        \caption{\small \label{MRFischer}
Posterior distribution for NS mass-radius relation obtained assuming 5970 detections with geometry 2L -20
km with aligned arms. For comparison, the results obtained from the nuclear prior as well
the underlying DD-LZ1 EoS model used in the Fisher matrix calculations are
displayed.
        }
\end{figure}

In Figure \ref{MRFischer} we show the mass-radius distribution for the nuclear prior described above along with the posteriors at 68\% and 95\% confidence intervals assuming 5970 simulated ``observations" obtained from a Fisher
matrix calculation. We also indicated the  DD-LZ1 EoS used as input EoS in the
calculation. {The full compatibility of the DD-LZ1 model with nuclear physics constraints is shown by the fact that the injected mass-radius relation sits in the middle of the nuclear informed prior.} This particular simulation was performed
with geometry 2L -20 km with aligned arms (cryogenic), but we have checked
that 
different configurations do not make any
significant difference in the outcome, {provided that the number of detections exceeds  $N_{det}\sim 6000$}.

To understand the impact of different number of detections using the same
geometry as above, we summarize the $1\sigma$ uncertainty on radii for 1.4 and 2.0 solar
mass NSs with 54, 592 and 5970 detections in Table \ref{err_RFischer}.
\begin{table}[t]
\centering
\begin{tabular}{ccc}
\hline
        $N_{det}$     & ${R_{1.4M_{\odot}}}^{+\Delta {R_+}}_{-\Delta {R_-}}$ [km]  &
        ${R_{2.0M_{\odot}}}^{+\Delta {R_+}}_{-\Delta {R_-}}$ [km] \\
\hline
        Prior & $12.983^{+0.420}_{-0.420}$ & $13.156^{+0.447}_{-0.454}$\\
\hline
        54 & $13.163^{+0.221}_{-0.227}$ & $13.358^{+0.234}_{-0.242}$ \\
\hline
        592 & $13.146^{+0.122}_{-0.136}$ & $13.355^{+0.099}_{-0.083}$ \\
\hline
        5970 & $13.107^{+0.148}_{-0.037}$ & $13.332^{+0.050}_{-0.013}$ \\
        \hline
\end{tabular}
\caption{\small \label{err_RFischer}
        Number of detections, $N_{det}$ along with  $1\sigma$  uncertainty on radii for
        1.4 and 2.0 solar mass NSs obtained for geometry 2L-20km-$0^{\circ}$ using nuclear prior.
        }
\end{table}
We can observe that the uncertainty on the radii only roughly
follows the Poissonian law as described in eq.~(\ref{eq:DelR_All_DelLam_Def}). The reason is probably that the radius
is not a direct observable but, rather, is obtained in a convoluted way from
the \acrshort{tov} equation and the underlying hypothesis on the EoS formalism. {As a general statement, we can see that the Poisson hypothesis tends to underestimate the radius uncertainty from the Fisher matrix analysis, unless the number of detections is extremely large (more than $N_{det}\sim 6000$). Moreover, we should also stress that the uncertainty estimations of Table~\ref{err_RFischer} might be underestimated themselves, because of the restrictive hypothesis of a continuous equation of state at any density}. Nevertheless,
these results give us an indicative impact of ET measurements on the radii uncertainties
using a fairly general EoS, within the hypothesis of purely nucleonic degrees of freedom.

\subsubsection{Full parameter estimation results}\label{sect:FullPENuclPhys}

In addition to our estimates based on the Fisher matrix computations, we also perform full parameter estimation studies. 
\begin{figure}
    \centering
    \vspace{-.5cm}\includegraphics[width=0.51\textwidth]{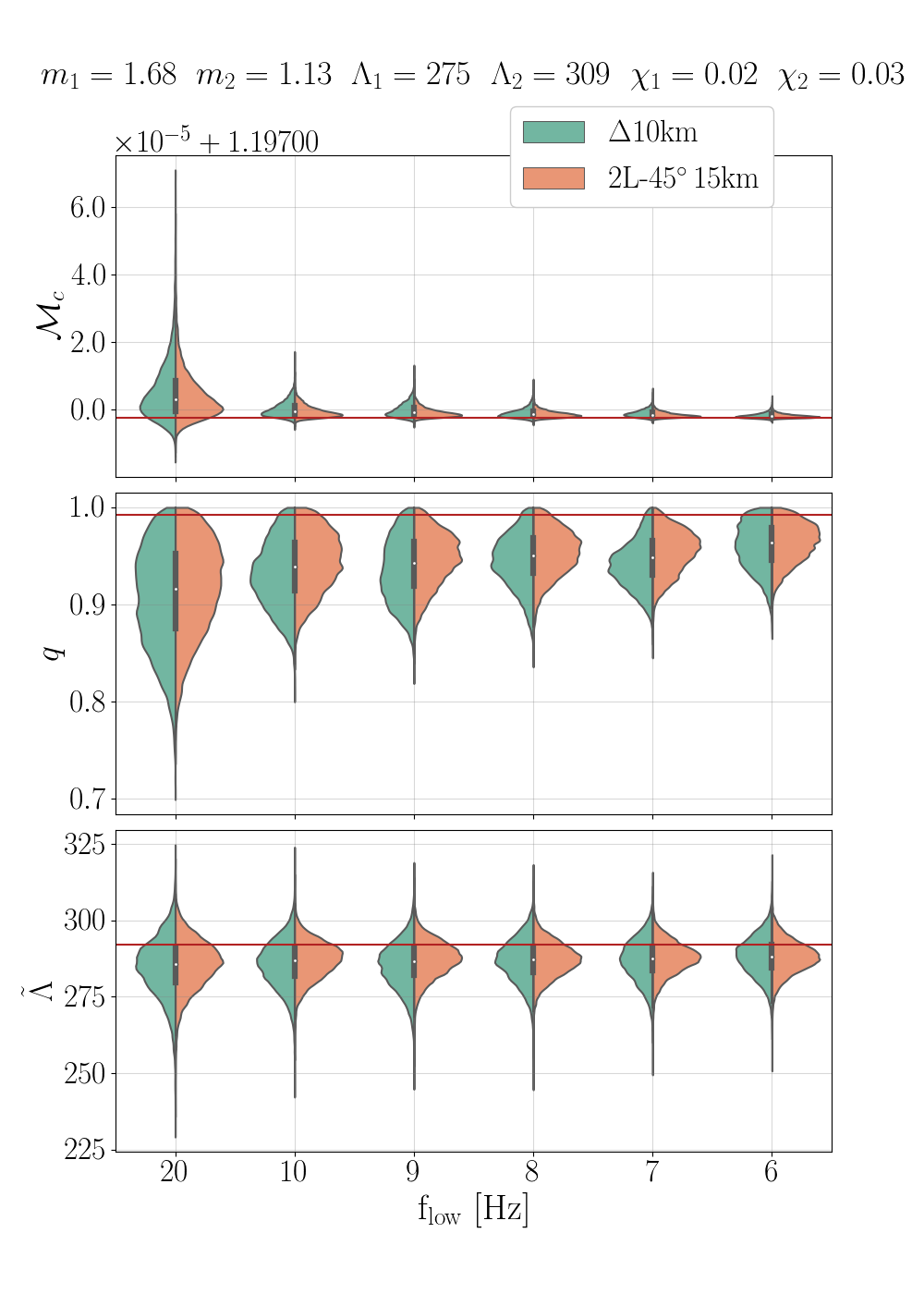}
        \includegraphics[width=0.48\textwidth]{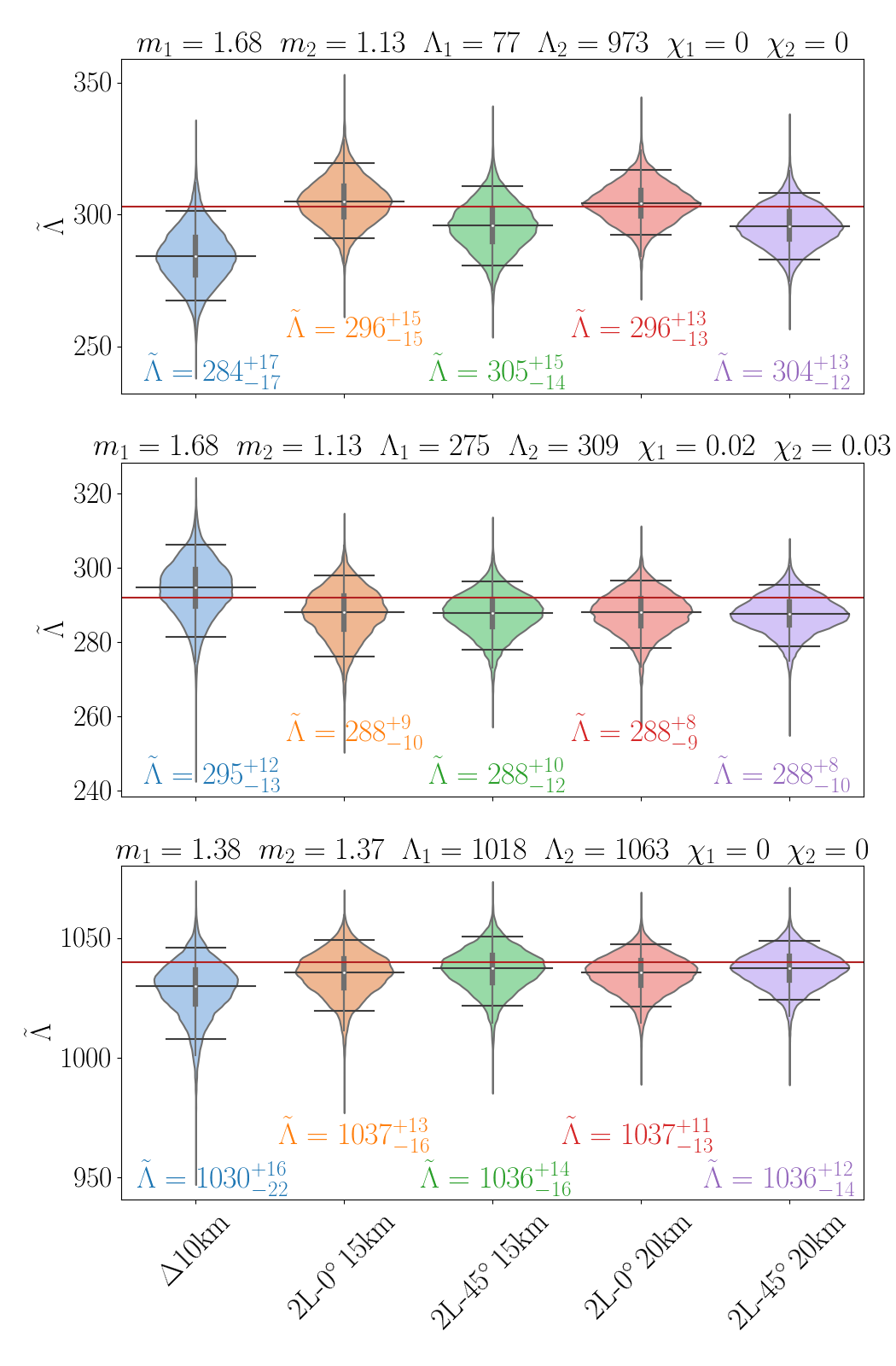}
    \caption{\small Left: Constraints on the chirp mass $\mathcal{M}_c$, mass ratio $q$, and tidal deformability $\tilde{\Lambda}$ for a binary neutron star system with 
             parameters listed at the top of the panel. A decreasing minimum frequency noticeably increases the accuracy of measuring the chirp mass and also 
             has an important effect on determining the mass ratio but has a smaller effect on the measurement of the tidal deformability. Right: Constraints on the tidal deformability for three different physical systems and different detector designs using an initial frequency of $6\rm Hz$.  
             }
    \label{fig:div6:PE}
\end{figure}

{For this analysis, we consider three different physical systems (A,B,C)
corresponding to different input equations of state (APR4 and H4), mass
ratios (symmetric or asymmetric), and spins (zero or non-zero). The values
of those injection parameters can be seen on the different panels of
Fig.~\ref{fig:div6:PE}, and on Tab.~\ref{tab:div6_PE_sources}.
The choice of the particular systems follows the injection study
of~\cite{LIGOScientific:2018hze} done by the LVK.}
\begin{table}[tbh]  
  \setlength\extrarowheight{4pt}  
  \begin{tabular}{ |c|c|c|c|c|c|c| }      
\hline            
Name   & Injection model & $m_1$, $m_2$ [$\rm M_{\odot}$] & $\mathcal{M}_c\,[\rm M_{\odot}]$ & $\Lambda_1$, $\Lambda_2$ &
$\tilde{\Lambda}$ & $\chi_1$, $\chi_2$  \\ [0.5ex]      
\hline      
Source A & APR4 & 1.68, 1.13 & 1.19479 & 77, 973  & 303 & 0,0\\      
Source B & APR4 & 1.38, 1.37 &  1.19700 & 275, 309 & 292 & 0.02, 0.03 \\      
Source C & H4 & 1.38, 1.37 & 1.19700  & 1018, 1063  & 1040 & 0, 0 \\            
\hline   \end{tabular}  
  \caption{\small Source properties used for the full PE injections in
Sec.~\ref{sect:FullPENuclPhys}.}

   \label{tab:div6_PE_sources}
\end{table}

For these studies, we use the relative binning technique~\cite{Zackay:2018qdy, Dai:2018dca, Leslie:2021ssu} 
to reduce the computational cost of our analysis.
Figure~\ref{fig:div6:PE} shows some of our main findings, and we refer to \cite{Puecher_in_prep} for more details. We analyze signals injected in zero noise, to avoid the impact of fluctuations from noise when analyzing different starting frequencies.
The left panel shows the impact of varying the initial frequency on the recovery of the chirp mass, the mass ratio, and the tidal deformability. 
Overall the width of the 90\% confidence interval for the recovery of the chirp mass reduces by more than an order of magnitude when going from 20 to 6Hz, 
and also the size of the 90\% confidence interval for the mass ratio shrinks by a factor of 2. 
However, the constraint on the tidal deformability reduces only by about 25\%.  
{\em Comparing the $\Delta \rm 10 km$ and the 2L $15\rm km$ setup, we find that the 2L configuration is overall better for measuring tidal deformability by about 25\%, 
where this difference is more or less independent of the initial frequency. 
As shown in ~\cite{Puecher_in_prep}, this is due mainly to the different arm lengths.
}
In the right panel of Fig.~\ref{fig:div6:PE}, we compare the $\tilde{\Lambda}$ posteriors for three different physical systems, and five different detector configurations. In this case, we use an initial frequency ${\rm f_{low}}=10~{\rm Hz}$ for all the runs, and a noise simulated with the HFLF cryogenic sensitivity. We find that in general, the aligned or misaligned 2L detector configurations lead to similar results. The 2L configurations result in tighter bounds on $\tilde{\Lambda}$ than the $\Delta \rm 10 km$ one; as above, these improvements arise mainly from the different arm lengths. It is also worth pointing out that the error on $\tilde{\Lambda}$ {for the three selected sources}, as reported in the right panel of Fig.~\ref{fig:div6:PE}, is consistent with that found by Fisher matrix calculations {within a few percent, for the various detector configurations considered}.

\subsubsection{Connected uncertainty of nuclear-physics parameters}
\begin{table}[t]
\begin{tabular}{|c|c|c|c|c|c|c|c|}
\hline
	source & Geometry & $n_{sat}$ & $E_{sat}$ & $K_{sat}$ & $E_{sym}$ & 
	$L_{sym}$ & $K_{sym}$  \\
\hline
	source-A & 2L 20km $0^{\circ}$ & $0.164^{+0.005}_{-0.004}$ & $-16.18^{+0.41}_{-0.40}$ & 
	$228^{+34}_{-24}$ &  $31.65^{+2.85}_{-2.15}$ & $40.9^{+9.1}_{-13.9}$ & $-262^{+61}_{-70}$\\
 & $\Delta$ 10km & $0.164^{+005}_{-007}$ & $-16.14^{+0.44}_{-0.42}$ & $229^{+27}_{-25}$ & 
		 $31.59^{+3.41}_{-2.99}$ & $40.1^{+14.8}_{-15.1}$ & $-269^{+83}_{-70}$\\
\hline 
	source-B &  2L 20km $0^{\circ}$ & $0.162^{+0.007}_{-0.007}$ & $-16.06^{+0.54}_{-0.47}$ & 
	$227^{+29}_{-27}$ & $30.91^{+2.69}_{-2.81}$ & $42.0^{+21.9}_{-19.0}$ & 
	$-197^{+155}_{-141}$  \\
	& $\Delta$ 10km & $0.163^{+0.006}_{-0.007}$ & $-16.05^{+0.53}_{-0.49}$ & 
	$227^{+27}_{-26}$ &	 $30.86^{+3.34}_{-2.86}$ & $40.9^{+17.1}_{-18.9}$ & $-208^{+165}_{-137}$\\
\hline 
	source-C &  2L 20km $0^{\circ}$ & $0.155^{+0.005}_{-0.004}$ & $-15.94^{+0.44}_{-0.56}$ & 
	$236^{+21}_{-17}$ & $30.96^{+1.54}_{-2.06}$ & $70.3^{+11.7}_{-10.3}$ & 
	$74^{+78}_{-87}$\\
  &  $\Delta$ 10km  & $0.156^{+0.004}_{-0.005}$ & $-15.92^{+0.47}_{-0.49}$ & 
	$236^{+23}_{-19}$ & $30.92^{+1.58}_{-2.02}$ & $70.8^{+11.2}_{-11.8}$ & 
	$74^{+78}_{-87}$\\

 \hline
\end{tabular}
\caption{\small \label{tab:NMP}
	Values for the nuclear-matter parameters up to second order in units of 
	MeV except for $n_{sat}$, which is in units of 
	fm$^{-3}$. For ET, the two configurations leading to the smallest and largest uncertainties have been chosen; all other configurations give uncertainties in-between these.  }
\end{table}

\begin{table}
\centering
\begin{tabular}{|c|c|c|}
\hline
	source & Geometry & $R_{1.4M_{\odot}}~[{\rm km}]$    \\
\hline
	source-A &  2L 20km $0^{\circ}$ & $11.44^{+0.13}_{-0.13}$    \\

  & $\Delta$ 10km & $11.34^{+0.15}_{-0.15}$   \\
\hline
	source-B &  2L 20km $0^{\circ}$ & $11.30^{+0.19}_{-0.21}$   \\
 	& $\Delta$ 10km & $11.32^{+0.20}_{-0.20}$   \\
\hline
	source-C &  2L 20km $0^{\circ}$ & $13.69^{+0.07}_{-0.06}$   
	\\
  & $\Delta$ 10 km& $13.68^{+0.06}_{-0.07}$  \\ 
  \hline
\end{tabular}
\caption{\small \label{tab:PE_NSprop}
	Radii (km) for 1.4$M_{\odot}$ 
 neutron stars as obtained for different geometries and different sources from the full parameter estimation runs under the nucleonic hypothesis. Again, as in table~\ref{tab:NMP}, only the geometries leading to the smallest and largest uncertainties, respectively, are shown here. 	}
\end{table}

\begin{figure}[t]
\includegraphics[width = .5\columnwidth]{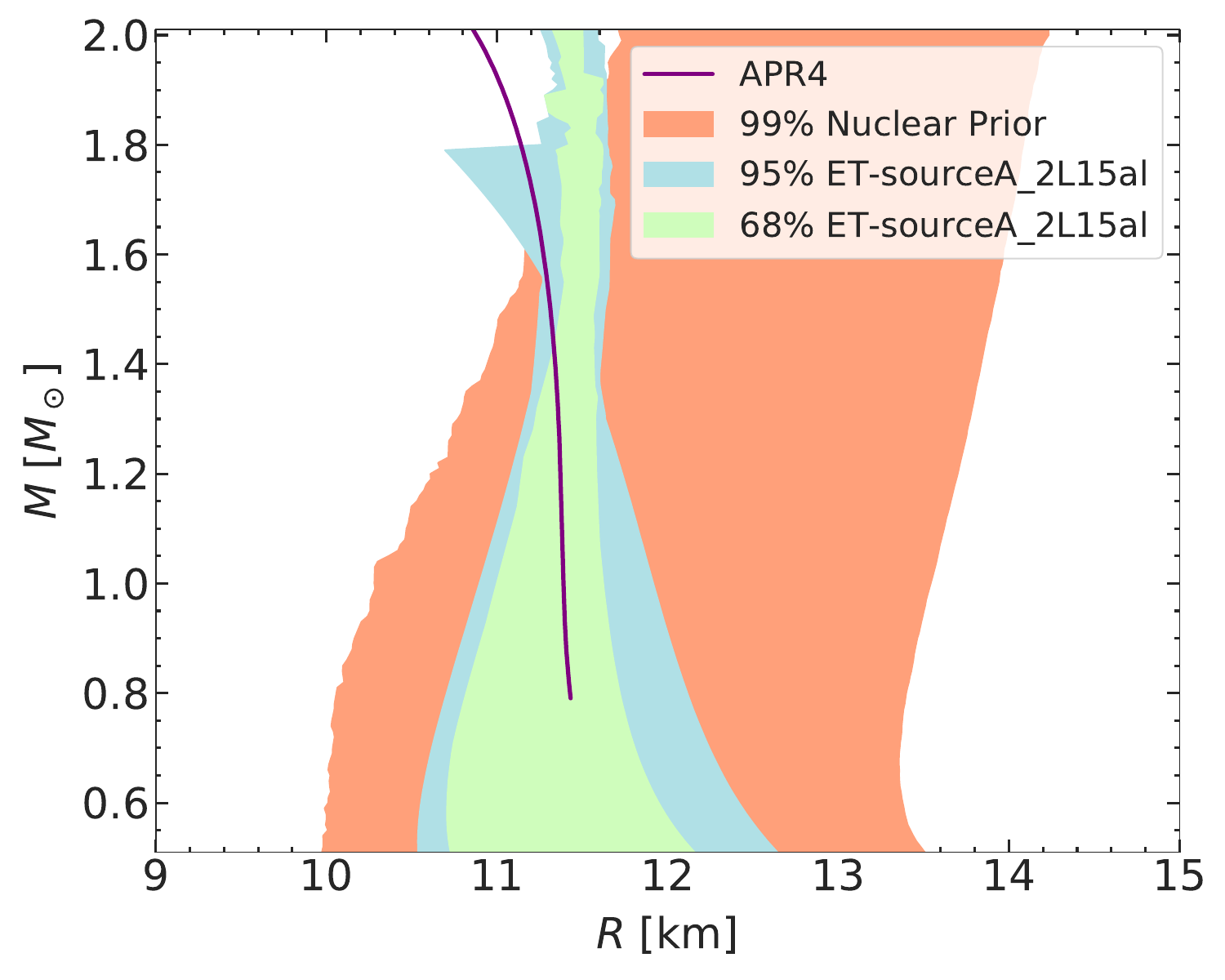}
\includegraphics[height = 2.68in,width=.5\columnwidth]{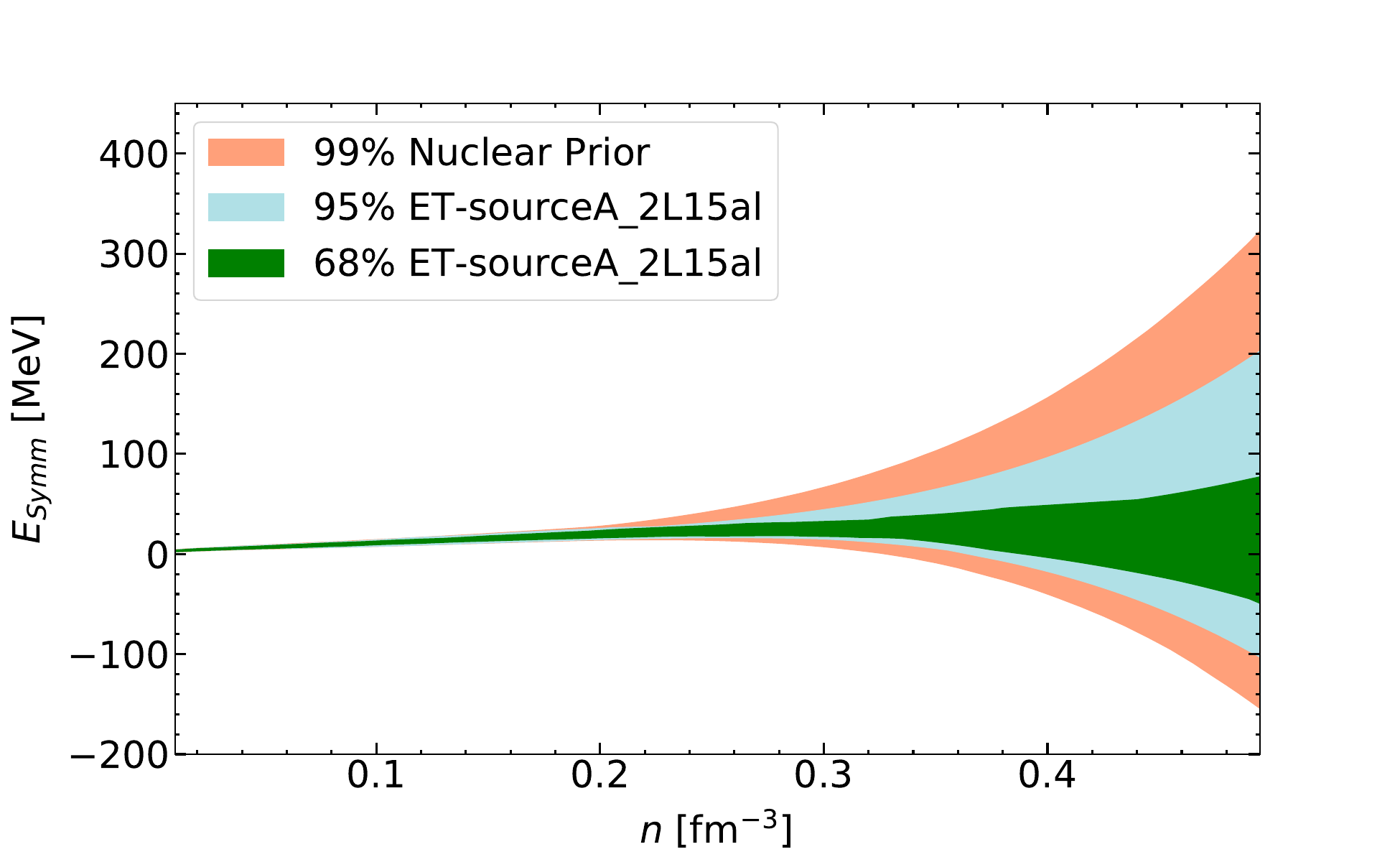}
\caption{\small \label{fig:MR_Esym_sourceA}Posterior distribution for NS mass-radius relation (left) and the nuclear symmetry energy as function of baryon number density (right) for source A with geometry 2L -15 km with aligned arms. For comparison, the results obtained from the nuclear prior are also shown.
}
\end{figure}

An immediate question is to what extent the different proposed configurations of the instrument can improve on our knowledge of nuclear matter and NS properties. In Section~\ref{sec:Fisher_nuclear} we have already discussed estimates for the uncertainties on the NS radius resulting from the different detector configurations. Here we will continue the discussion by analysing the full parameter estimation results for the three sources discussed above. As above, we employ the prior on the EoS from present-day nuclear physics knowledge~\cite{Thi:2021jhz} as a baseline. 
  {Because of the nuclear physics constraints, a number of popular equations of state might be only marginally compatible with our nuclear physics informed prior. This is particularly the case of the APR4 model, that was never confronted to nuclear physics data. This can be seen from the NS mass-radius relation shown in Fig.~\ref{fig:MR_Esym_sourceA} (left panel) for source A and the 2L geometry with 15 km arm length. In this figure, we can see that the APR model predicts too small radii for the heaviest neutron stars, with respect to our nuclear prior \footnote{In this figure, in order to allow for slightly smaller radii, we have relaxed the 2 solar mass constraint for the nuclear prior which is applied in Fig.\ref{MRFischer} above.}. This explains why the injection value is not precisely recovered in the posterior distribution. For this reason, the radius uncertainty estimation will not be reliable for the higher masses and is not analyzed in the following. }
In table~\ref{tab:NMP} we summarise the posterior results for the different nuclear-matter parameters up to second order at saturation density with the three sources for the two configurations of the instrument leading to the smallest and largest uncertainty, respectively.
 {These parameters represent the equilibrium density ($n_{sat}$) and energy ($E_{sat}$), the symmetry energy at saturation ($E_{sym}$) and its slope ($L_{sym}$), and the nuclear matter compressibility in the isoscalar ($K_{sat}$) and isovector sector ($K_{sym})$.} 
For the symmetry energy, the posterior as a function of density is shown in Fig.~\ref{fig:MR_Esym_sourceA} (right panel) for source A and the 2L geometry with 15 km arm length. The results indicate first of all that the injected properties are {reasonably} well recovered, {even if the input model is only marginally compatible with our nuclear physics informed prior.} On the scale of this figure, the differences between the proposed configurations of the instrument are insignificant and the 2L geometry is only slightly favored concerning the constraints on the nuclear-matter parameters. 

In Table~\ref{tab:PE_NSprop} we summarise the posterior results for the NS radius at the fiducial mass 
1.4 $M_\odot$. 
with the three sources for the two configurations of the instrument leading to the smallest and largest uncertainty, respectively. 
{We can see that the uncertainties depend stronger on the injection model than on the configuration of the instrument.}   The absolute values are different from those obtained with the Fisher matrix approach, and reported in Tab.~\ref{tab:div6:Fisher_radius} and \ref{err_RFischer}. This is essentially due to the fact that the results in Tab.~\ref{tab:PE_NSprop} are obtained on single events with very high SNR, which thus result in tight constraints on the source parameters, while the ones in Sec.~\ref{sec:Fisher_nuclear} are produced from runs on a full realistic population of sources, the vast majority of which have much higher statistical uncertainties, resulting in looser constraints on the EoS parameters. Also, we again stress that the results presented in Tab.~\ref{tab:div6:Fisher_radius} are obtained through the extremely simplified metric in Eq.~\ref{eq:DelR_All_DelLam_Def}, and should thus only be considered as indicative for the comparison of the various detector configurations.

{To conclude, even if the quantitative uncertainties depend on many not fully controlled factors, from the NS mass-radius posterior shown} 
in Fig.~\ref{fig:MR_Esym_sourceA} (left panel), 
{\em the improvement with respect to current knowledge by Einstein Telescope is clearly seen.} In addition, the present analysis assumes at each time one detected source, i.e., an inspiral with two given masses, whereas the Einstein Telescope will detect tens of thousands of different BNS merger events with a distribution of NS mass. This staggering will considerably reduce the uncertainty on the density dependence of the dense matter EoS and hence the nuclear-matter parameters and the information on the NS radius; see also the discussion in Section~\ref{sec:Fisher_nuclear}. {\em The impact of the instrument geometry on these conclusions is only marginal;
as above, the results indicate that the 2L geometry is favored, but the difference with the other geometries is not significant.}

\subsubsection{Postmerger detectability}

A BNS system typically merges at a frequency of $\sim$ 1 kHz. After the merger, there are a number of physical processes, such as thermal effects, turbulences, phase transitions, magnetohydrodynamical instabilities, dissipative processes etc.~\cite{Bauswein:2018bma, Most:2018eaw, Siegel:2013nrw, Alford:2017rxf, Radice:2017zta, Shibata:2017xht, DePietri:2018tpx}, which drive the postmerger physics leading to different remnants, depending on the masses of the neutron stars and EOS of the system. The merger remnant can be either a black hole, a hypermassive NS (\acrshort{hmns}), a supramassive NS (\acrshort{smns}) or a massive NS. If the mass of the merger remnant is above a threshold mass it will collapse to a BH~\cite{Bauswein:2013jpa, Koppel:2019pys, Agathos:2019sah}. 
Otherwise, there is a post-merger gravitational-wave signal that can last over a time scale of about 10ms up to several 100ms~\cite{Bauswein:2011tp} emitted at frequencies of $\sim 2-4~{\rm kHz}$.
While the postmerger signal from a BNS merger is not expected to be detectable for 2G detectors due to their insufficient sensitivity at high frequencies, it will be accessible to 3G detectors. 
Due to the complicated physical effects involved, postmerger signals are most accurately represented by numerical relativity (\acrshort{nr}) simulations. For this study we will use NR waveforms~\cite{Takami:2014tva,Clark:2015zxa,Tsang:2019esi,Wijngaarden:2022sah,Breschi:2022xnc}.  

To determine the detectability of the postmerger signal with different detector configurations, we will compute and compare the SNR from the postmerger signal for a set of NR BNS waveforms as seen by the various configurations considered in this study. The postmerger SNR is defined as the SNR of the signal from the merger frequency~\cite{Dietrich:2018uni} of the BNS system, up to 8192 Hz. We use a set of 6 BNS-NR waveforms from the SACRA catalogue~\cite{Kiuchi:2019kzt}, the masses and tidal deformabilities of which are given in Table~\ref{tab:tab_bns_waveforms}. 
For each detector geometry and PSD configuration, and each NR waveform, we use the same random set of 60 (ra, dec, $\psi$) values and 
quote the averaged postmerger SNR ($\rho_{\rm pm}^{\mathrm{Avg}}$) for each of the NR waveforms.  

\begin{table}
    \centering
    \begin{tabular}{|c|c|c|c|c|}
        \hline
         Configuration name & mass1 & mass2 & $\Lambda_1$ & $\Lambda_2$ \\
         \hline
         \hline
         15H\_135\_135\_00155 & 1.35 & 1.35 & 1211 & 1211  \\
         \hline
         125H\_107\_146\_0015 & 1.07 & 1.46 & 3196 & 535  \\
         \hline
         H\_117\_156\_00155 & 1.17 & 1.56 & 1415 & 238  \\
         \hline
         H\_135\_135\_00155 & 1.35 & 1.35 & 607 & 607  \\
         \hline
         125H\_121\_151\_00155 & 1.21 & 1.51 & 1621 & 435  \\
         \hline
         H\_118\_155\_00155 & 1.18 & 1.55 & 1354 & 249  \\
         \hline
    \end{tabular}
    \caption{\small Table of the BNS-NR waveforms and their physical parameters used to compute the postmerger SNR.}
    \label{tab:tab_bns_waveforms}
\end{table}

\begin{figure}
    \centering
    \includegraphics[width=0.99\textwidth]{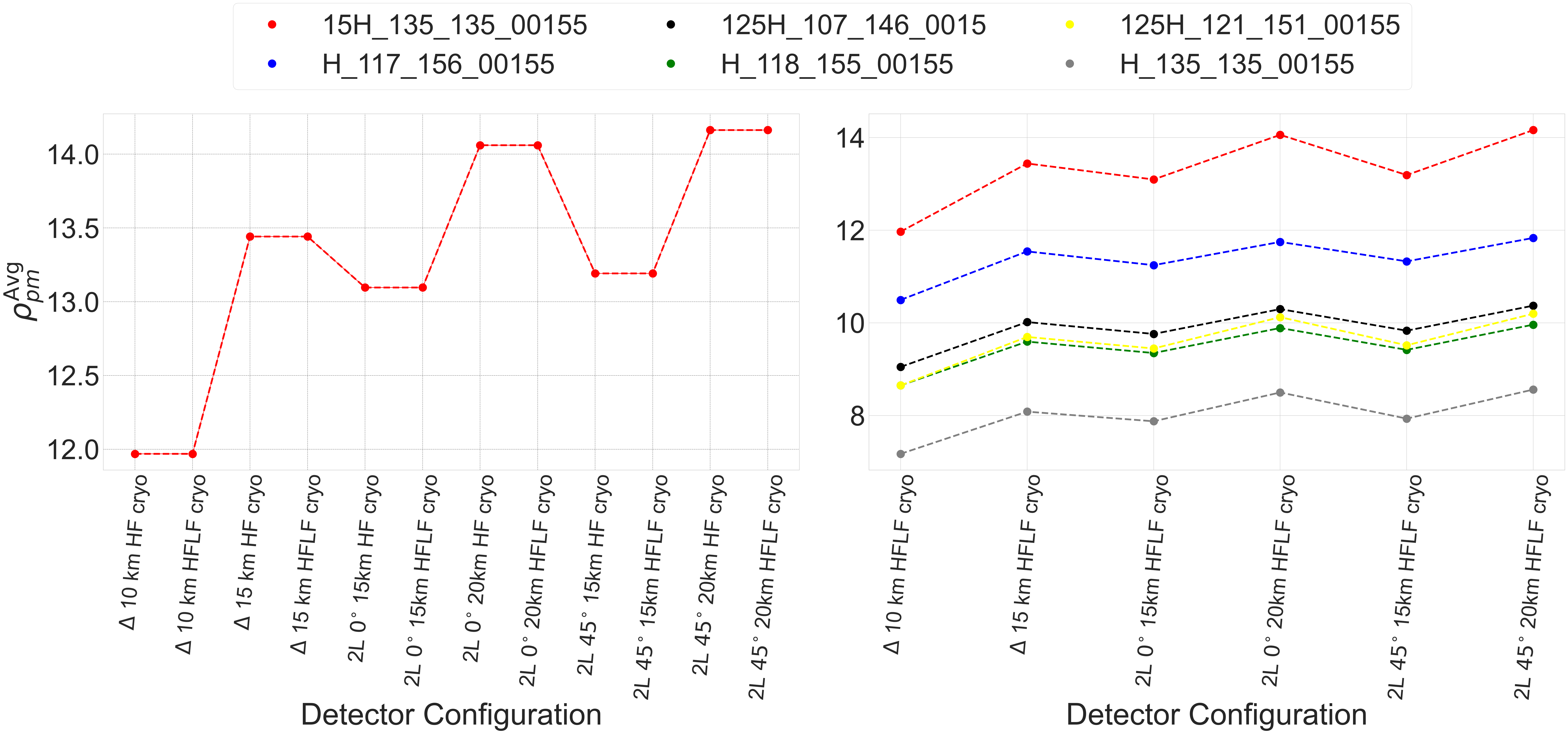} 
    \caption{\small Left: Averaged postmerger SNR for the 15H\_135\_135\_00155 system for different detector configuration and PSD combinations.
    Right: Averaged postmerger SNR for all the BNS-NR systems as given in Table~\ref{tab:tab_bns_waveforms} for different detector configuration and HLFL cryo PSD. Each BNS-NR system is plotted with a different color with the detector configuration given on the $x$-axis. }
    \label{fig:post_merger_fig1}
    \label{fig:post_merger_fig2}
\end{figure}

The left panel of Fig.~\ref{fig:post_merger_fig1} shows $\rho_{\rm pm}^{\mathrm{Avg}}$ for all the detector geometry -- PSD combinations. for the system  15H\_135\_135\_00155. As the high-frequency sensitivity of the HFLF cryo and HF-only PSDs are quite similar, the postmerger SNRs for these two 
PSDs are almost equal. We see the majority of variation occurring with varying arm-length for a given detector shape and changing detector shapes (triangular vs. 2L configurations). The maximum $\rho_{\rm pm}^{\mathrm{Avg}}$ for this system is seen by the 2L $45^{\circ}$ 20km HLFL  configuration which is $\sim$ 20\% larger than the minimum value, which corresponds the $\Delta$ 10km HF configuration.  

The right panel of Fig.~\ref{fig:post_merger_fig2} shows $\rho_{\rm pm}^{\mathrm{Avg}}$ for all the BNS-NR systems for the different detector configurations and only the HFLF cryo PSD. For all the BNS-NR systems, 2L 20km $45^{\circ}$ gives the highest postmerger SNR. For $\rho_{\rm pm}^{\mathrm{Avg}}$, there is not much difference between the SNR for 2L $0^{\circ}$ and $45^{\circ}$ ($\sim$ 0.7 \%) irrespective of the arm length. Going from 15km 2L to 20km 2L gives a $\sim$ 6\% increase in $\rho_{\rm pm}^{\mathrm{Avg}}$. Between $\Delta$ 10 km and $\Delta$ 15 km, we see $\sim$ 11\% increase. 

{\em These results show that the performances of the different geometries are relatively similar within a $10-20\%$ level, with 2L $45^{\circ}$ 20km detector configuration being the  best for detecting postmerger signals.} 

\subsubsection{Conclusions:  nuclear physics with ET}
The Einstein Telescope will significantly advance our ability to constrain fundamental nuclear-physics properties, going well beyond the capabilities with any existing observatory or experiment, and providing complementary constraints to explore large domains of the QCD phase diagram. For example, the upgrade of the FAIR at the GSI Helmholtz Centre for Heavy Ion Research will probe matter which is only slightly neutron-rich and at lower baryon densities and higher temperatures than what is expected for ET, which can reach the most extreme densities. Furthermore, the X-ray observatory NICER and proposed future instruments, such as eXTP, Strobe-X or ATHENA will be able to improve on current knowledge of cold dense matter by observing more pulsars and increasing the length of the observational data period and thus obtain NS radii with a precision of a few percent, similar to the anticipated improvements with the existing advanced LIGO and advanced Virgo detectors (even in combination with KAGRA and LIGO India) or proposed upgrades such as LIGO A$\#$ and Virgo\_next. ET, in addition to being able to determine NS radii with a statistical precision of sub-percent order due to the immense statistics with $10^5$ mergers expected
per year, will also be the only instrument able to add to the zero temperature $\beta$-equilibrated NS matter the information on the hot ultra-dense matter from the postmerger phase. {\em With regards to the payoffs for nuclear physics,  there is no significant difference between the different detector configurations, with longer arm-lengths leading to slightly better results.}

\subsection{Population studies}
\subsubsection{Merger rate reconstruction}


\begin{figure}
    \centering
    \includegraphics[keepaspectratio, width=12cm]{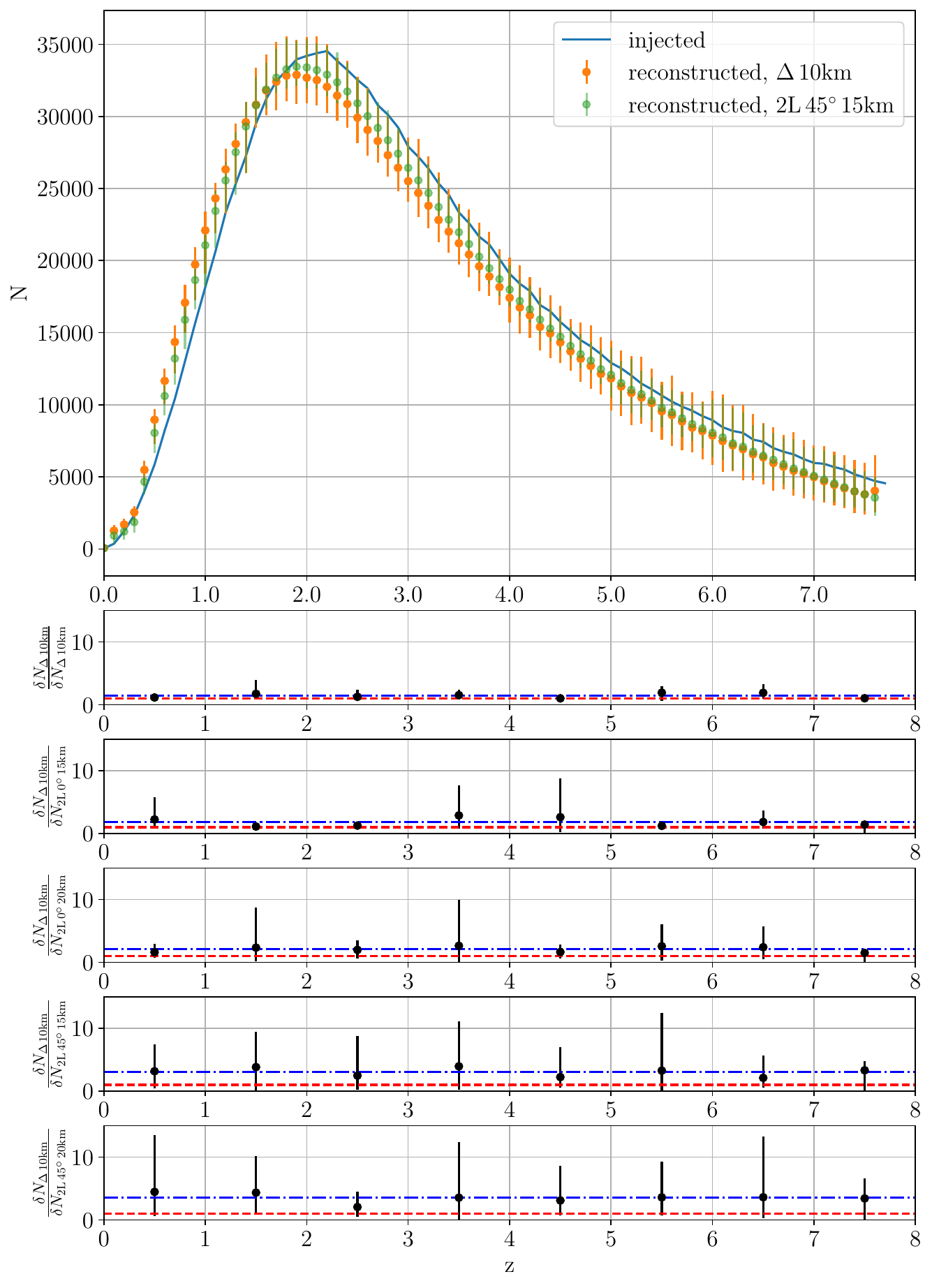}
    \caption{\small \emph{Top:} Representation of the reconstructed merger rate with the $90\%$ confidence interval for 10 years of observation of ET in a triangular shape with 10km arms (orange) and  ET with 2 L-shaped detectors with 15km arms at $45^{\circ}$ (green). The blue line represents the true merger rate distribution. The two configurations enable one to reconstruct the merger rate but the accuracy is better for the 2 L-shaped detectors. \emph{Bottom:} Ratio of the width of the $90\%$ confidence interval for the triangular 10km configuration with the other possible configuration. The red line corresponds to the case where the two detectors are equivalent, while the blue line is the average ratio. The hierarchy to be seen in this figure is that the 2-L aligned detectors are better than the triangular ones, and the 2L with arms at $45^{\circ}$ are more accurate than the 2 L's with arms at $0^{\circ}$. For a given configuration, the increase in arm length gives a small improvement, mainly due to the increase in SNR.}
    \label{fig:MergerRate2LvsTriangle}
\end{figure}

Since ET will have a long distance range, it will see binary coalescences out to cosmological distances. Therefore, it will probe the merger rate up to high redshifts. Knowing how the BBHs are distributed in the Universe would help constrain their various formation channels, leading to a better understanding of how such binaries form~\cite{Oguri:2018muv, Artale2019, Santoliquido:2021, Mapelli2021, Mapelli:2021gyv}. Different models exist, often assuming that the merger rate follows the star formation rate with some additional contribution coming from primordial black holes. Models differ in the details but most of them agree on the general shape of the distribution:  the merger rate increases up to a redshift between 1 and 2 before decaying for higher redshifts (see~\cite{Oguri:2018muv, Mapelli2021} for examples). The merger rate  as a function of the redshift is written  as~\cite{Regimbau:2009rk}
\begin{equation}\label{eq:MergerRateFctRedshift}
    \frac{dR}{dz}(z) = \rho(z) \frac{dV}{dz} \, ,
\end{equation}
where $V(z)$ is the comoving volume up to redshift $z$ and
\begin{equation}\label{eq:FormationRate}
    \rho(z) \propto \int_{t^{min}_d}^{+\infty} \frac{\rho_{*}(z_f(z, t_d))}{1+z_f(z, t_d)} P(t_d) dt_d \, 
\end{equation}
is the binary formation rate, where $\rho_{*}$ is the star formation rate, $z$ is the redshift of interest, $z_f$ is the redshift at which the binary is formed, $t_d$ is the time delay between the formation of the binary and its merger, and $P(t_d)$ is the probability to have a given time delay. 

To evaluate how well different configurations for ET would be able to identify the merger rate density, we follow a procedure similar to~\cite{VanDenBroeck:2013uza}, with a theoretical population made of 10 years of BBHs as given by~\cite{Mapelli2021}, and discussed in Section~\ref{sect:CBC}. Then, we start by assigning a random sky location to the binaries. We compute their (network) SNRs and keep only the events with an SNR higher than 9 for a given configuration. The others are considered non-detectable. These events are then used to estimate the detector's efficiency as a function of the redshift. For each event, we perform a Fisher matrix analysis to obtain a multivariate Gaussian likelihood and sample from it to have an approximation for the measurements~\cite{Gupta:2022qgg}. For each detectable BBH, we make a Gaussian centered on the mean value of the recovered posterior (which corresponds roughly to the injected value since the result of a Fisher matrix is centered on the true value), with a width equal to the standard deviation of the luminosity distance posterior obtained from sampling the Fisher likelihood. We then draw a \emph{measured} luminosity distance. This is done for all the detectable BBHs. Then, we convert the luminosity distances to redshifts and count the number of events that we have in the redshift bins. This gives an observed merger rate, which is then corrected by the efficiency to account for the number of mergers we theoretically expect to miss for each redshift bin. We repeat the drawing of measurements 250 times to get error bars on the observed merger rate. 

The top panel of Figure~\ref{fig:MergerRate2LvsTriangle} shows the reconstructed merger rate for 10 years of detections with  ET seen as a single triangle with 10 km arms versus ET as 2 L's with 15km arms and a $45^\circ$ rotation between the two detectors. The other panels show the ratio in accuracy between the single triangle with 10 km arms and the other configurations considered in this work. {\em Although the two configurations lead to a correct reconstruction of the merger rate, one sees that the configuration made of 2 {\rm L}-shaped detectors leads to a more accurate reconstruction} as it has smaller error bars and closer central values. The increase in the ratio is such that a hierarchy can be established between the different configurations. 2 L's are better than a single triangle and L's rotated by $45^\circ$ are better than aligned ones. Table~\ref{tab:MeanRatios} shows the mean ratio found between the detectors. The improvement from one detector configuration to the other is driven by a reduction in error on the luminosity distance. On the other hand, for a given configuration, 5 km longer arms are slightly better, which can be understood by the increased SNR of events. {\em Nevertheless, it is still possible to have a good grasp of the merger rate density for all the configurations within 10 years of observation.} This should be quite helpful in the quest to reveal the formation channels for BBH mergers.

\begin{table}
    \centering
    \begin{tabular}{l r}
    \hline
    \hline
    Other configuration & Ratio width 90\% confidence interval\\
    \hline
    \hline
         &  \\
     $\Delta$ 15 km & 1.2 \\
     2 L $0^\circ$ 15 km & 1.8 \\
     2 L $0^\circ$ 20km & 2.0 \\
     2 L $45^\circ$ 15 km & 2.9 \\
     2 L $45^\circ$ 20 km & 3.3 \\ 
     & \\
     
     \hline 
     \hline
    \end{tabular}
    \caption{\small Mean values for the ratios of the 90\% confidence intervals in the merger rate reconstruction between ET triangle with 10 km arm-lengths and the other possible configuration envisaged in this work. 2 L's are more accurate than one triangle, with L's rotated by $45^\circ$ the one with respect to the other being better. This happens because of the reduced error on the sky location. For a given configuration, increasing the arm length by 5km does not strongly increase the accuracy since the only source of improvement is the increased SNR for the observed events.}
    \label{tab:MeanRatios}
\end{table}


\subsubsection{Constraints on PBHs from high-redshift mergers}\label{sect:PBHhighz}

Primordial black holes (\acrshort{pbh}s) may form in the early universe and represent a dark matter (\acrshort{dm}) candidate (see e.g. Ref.~\cite{Green:2020jor} for a review).  
Many constraints were set on their abundance in a wide range of masses \cite{Carr:2020gox} and recent GW detections suggest that, within standard scenarios, PBHs can only account for a subdominant DM component in the stellar mass range \cite{Bird:2016dcv,Clesse:2016vqa,Sasaki:2016jop,Wang:2016ana, Ali-Haimoud:2017rtz,Chen:2018czv,Raidal:2018bbj,Vaskonen:2019jpv,DeLuca:2020qqa,Bhagwat:2020bzh,Hall:2020daa,DeLuca:2020jug,Wong:2020yig,Hutsi:2020sol,Kritos:2020wcl,Deng:2021ezy,Kimura:2021sqz,DeLuca:2021wjr,Bavera:2021wmw,Chen:2021nxo,Franciolini:2021tla,Mukherjee:2021ags,
Wang:2022nml,Franciolini:2022ewd,Zheng:2022wqo}. Nonetheless, a discovery of a sub-population of PBHs would have fundamental consequences for our understanding of cosmology and particle physics (see e.g. \cite{Byrnes:2012yx,Boldrini:2019isx,Bertone:2019vsk,Volonteri:2010wz,Franciolini:2022tfm}).

One ``smoking-gun" signature of the primordial scenario is the production of high redshift mergers, which are unique targets of 3G detectors. 
The PBH model predicts a merger rate density that grows monotonically with redshift~\cite{Nakamura:2016hna,Ali-Haimoud:2017rtz,Raidal:2018bbj,Chen:2019irf,DeLuca:2020qqa}, 
with a power-law scaling of the form $R_\text{\tiny PBH} (z) \propto[t(z)/t(z=0)]^{-34/37}$,
where $t(z)$ indicates the age of the universe at redshift $z$.
This rate evolution extends up to redshifts $z\gtrsim{\cal O}(10^3)$. 
On the contrary, even though large uncertainties are still present,
astrophysical-origin mergers are generically not expected to occur at $z\gtrsim 30$ in standard cosmologies \cite{Koushiappas:2017kqm} (see also \cite{Schneider:1999us,Schneider:2001bu,Schneider:2003em,Bromm:2005ep,Tornatore:2007ds,Trenti:2009cj,deSouza:2011ea,Mocz:2019uyd,Kinugawa:2014zha,Kinugawa:2015nla,Hartwig:2016nde,Belczynski:2016ieo,Inayoshi:2017mrs,Liu:2020lmi,Liu:2020ufc,Kinugawa:2020ego,Tanikawa:2020cca,Singh:2021zah}).

Recently, it was shown that a PBH population which may be compatible with a subset of the recent LVK observations would also lead to a large number of high-redshift detections at 3G experiments \cite{DeLuca:2021wjr}. 
Observations of such distant events may be characterised by large measurement uncertainties on the inferred luminosity distance due to the low SNR \cite{Ng:2021sqn,Franciolini:2021xbq,Martinelli:2022elq,Ng:2022vbz}. However, focusing on constraining the merger rate evolution at redshift larger than ${\cal O}(10)$ may allow probing PBH populations up to abundances as low as $f_{\rm PBH} \approx 10^{-5}$ \cite{Ng:2022agi} in the solar mass range (even including the contamination of a population of Pop-III binaries), which would improve current capabilities by orders of magnitude.  Therefore, the search for high-redshift primordial mergers is one of the scientific targets of the ET experiment. 

In order to test the performance of different configurations to search and constrain a PBH population, we consider the benchmark PBH model resulting from the analysis in Ref.~\cite{Franciolini:2021tla}. 
The PBH model hyper-parameters are: 
$[M_c, \sigma]$ determining the PBH mass distribution, modeled as a lognormal function with mean $M_c$ and width $\sigma$; 
the abundance $f_\PBH$ normalised with respect to the dark matter; 
the redshift $z_\text{\tiny cut-off}$ characterising PBH accretion \cite{DeLuca:2020bjf}.
These are assumed to be given by the population inference result of Ref.~\cite{Franciolini:2021tla}, constraining the potential contribution from a PBH subpopulation to the \acrshort{gwtc}-2 catalog and representing an upper bound on the PBH abundance in this mass range. 
In particular, we adopt $ M_c = 34.54 M_\odot$, $\sigma = 0.41$, $f_\PBH = 10^{-3.64}$ and $z_\text{\tiny cut-off} = 23.90$, such that the PBH channel may be adequate in explaining around $(1-21)\%$ of the detections in the O1/O2/O3a run of LVC, and given the associated astrophysical models considered, the mass gap event GW190521. 

We compute the number of detectable PBH events, $N_\text{\tiny det}$, by estimating the detector network selection bias based on an injection of $N=10^6$ events at redshift  $z \geq 10$.
We randomise event directions and orientations, and we assume negligible PBH spins (compatible with the prediction of the standard scenario at high redshift \cite{DeLuca:2019buf,DeLuca:2020bjf,DeLuca:2020qqa}). The computation is performed using the IMRPhenomHM waveform model. We set a detection threshold ${\rm SNR}\geq  9$.
The minimum testable abundance is conservatively estimated by requiring at least one detectable event per year with $z \geq 30$ \cite{DeLuca:2021wjr,DeLuca:2021hde}. 
This simplified approach provides an estimate of the ET capabilities which is consistent with actual population analyses, as performed in Ref.~\cite{Ng:2022agi}. 

We report the result of the analysis for the various configurations in Tables~\ref{tabresPBH1}, \ref{tabresPBH2} and \ref{tabresPBH3}, 
detailing how the detection prospects change as a function of the configuration geometry, the inclusion of a single 40~km CE detector in the network, and removal of the low-frequency instrument, respectively.
In each table, we indicate the number of detectable events above redshift $z>10$ and $z>30$, alongside the estimated minimum testable abundance. 
{\em The comparison of Tables~\ref{tabresPBH1} and \ref{tabresPBH3} shows the important of the LF instrument for the observation of PBH at large redshifts. In particular, without the LF instrument, there are no detection at $z\geq 30$, for all the the geometries considered.}

\begin{table*}
\centering
\footnotesize
\renewcommand{\arraystretch}{1.2}
\begin{tabular}{|l|c|c|c|}
\hline
\hline
Configuration  & 
 $N_\text{\tiny det}(z>10)\,[{\rm 1/yr}]$  & 
 $N_\text{\tiny det}(z>30)\,[{\rm 1/yr}]$   &
 $f_\text{\tiny PBH} ^\text{\tiny constrained}  \, [\times10^{-5}] $  
\\
\hline
\hline
{\bf $\Delta$-10km} &   {\bf 1140} & {\bf 77} & {\bf 2.61}
    \\    
\hline 
$\Delta$-15km &         1764 & 261 & 1.42
    \\   
     \hline
2L-15km-$0^\circ$ &                    1597 & 238 & 1.48
    \\   
 \hline 
2L-15km-45$^\circ$ &                   1651 & 221 & 1.54
    \\   
\hline 
2L-20km-$0^\circ$ &                   1984 & 434 & 1.10
    \\   
\hline 
2L-20km-45$^\circ$ &                  2080 & 416 & 1.12
    \\   
  \hline
  \hline 
\end{tabular}
\vspace{0.2cm}
\caption{\small Capabilities for constraining high-redshift PBH mergers of the various ET geometries (all taken with their HFLFcryo ASD). From left to right, columns report the number of detectable events per year at redshift larger than $10$ or $30$ assuming the PBH population from Ref.~\cite{Franciolini:2021tla}. The rightmost column reports the minimum testable PBH abundance, see more details in the main text.  In bold we indicate the reference configuration.
}\label{tabresPBH1}
\end{table*}

\begin{table*}
\centering
\footnotesize
\renewcommand{\arraystretch}{1.2}
\begin{tabular}{|l|c|c|c|}
\hline
\hline
Configuration  & 
 $N_\text{\tiny det}(z>10)\,[{\rm 1/yr}]$  & 
 $N_\text{\tiny det}(z>30)\,[{\rm 1/yr}]$   &
 $f_\text{\tiny PBH} ^\text{\tiny constrained}  \, [\times10^{-5}] $  
\\
 \hline 
  \hline 
CE40km &                        			1373 & 47 & 3.34
    \\   
 \hline 
$\Delta$-10km + CE40km &                1940 & 180 & 1.71
    \\   
  \hline 
$\Delta$-15km  + CE40km &               2276 & 372 & 1.19
    \\   
 \hline 
2L-15km-45$^\circ$ + CE40km &          2210 & 333 & 1.26
    \\   
 \hline 
2L-20km-45$^\circ$ + CE40km & 		2476 & 522 & 1.00
\\
\hline
\hline
\end{tabular}
\vspace{0.2cm}
\caption{\small Same as Table~\ref{tabresPBH1} for four ET geometries in a network configurations including one 40km Cosmic Explorer. ET is only taken here in its HFLFcryo configuration.} \label{tabresPBH2}
\end{table*}

\begin{table*}
\centering
\footnotesize
\renewcommand{\arraystretch}{1.2}
\begin{tabular}{|l|c|c|c|}
\hline
\hline
Configuration  & 
 $N_\text{\tiny det}(z>10)\,[{\rm 1/yr}]$  & 
 $N_\text{\tiny det}(z>30)\,[{\rm 1/yr}]$   &
 $f_\text{\tiny PBH} ^\text{\tiny constrained}  \, [\times10^{-5}] $  
\\
\hline
\hline 
$\Delta$-10km-HF &          		15 & 0.00 & -                 
 \\ 
  \hline 
$\Delta$-15km-HF &                           85 & 0.00 & -
 \\ 
  \hline
2L-15km-0$^\circ$-HF &                       75 & 0.00 & -
 \\ 
 \hline 
2L-15km-45$^\circ$-HF &                      69 & 0.00 & -
 \\ 
   \hline 
2L-20km-0$^\circ$-HF &                      178 & 0.00 & -
 \\ 
  \hline 
2L-20km-45$^\circ$-HF  &                     170 & 0.00 & -
 \\ 
\hline
\hline
\end{tabular}
\vspace{0.2cm}
\caption{\small Same as Table~\ref{tabresPBH1} for  configurations with the HF-only ASD. }\label{tabresPBH3}
\end{table*}

\begin{figure*}[ht]
	\centering
	\includegraphics[width=0.85\textwidth]{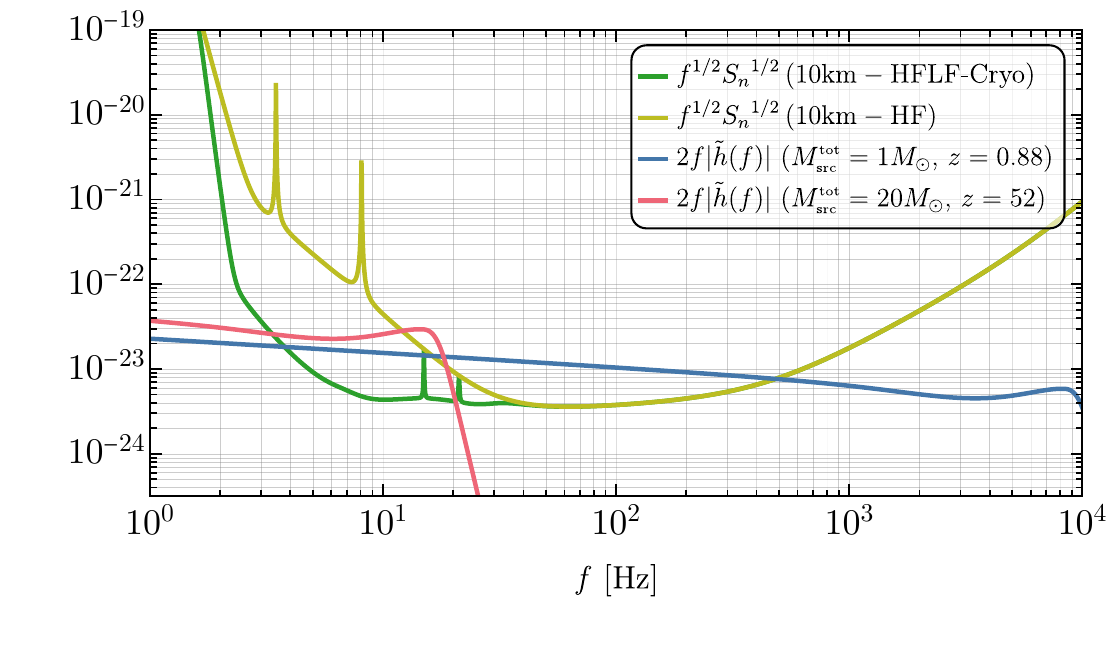}
	\caption{\small Comparison between the amplitude of a PBH binary signal 
 $2f|\tilde h(f)|$
 and the ET detector sensitivity curves $(f S_n)^{1/2}$
 shown in green (10km-HFLF-Cryo) 
	or yellow (10km-HF). 
	We consider a distant merger with source frame mass $M_\text{\tiny src}^\text{\tiny tot} =20 M_\odot$ at $z = 52$. Also, we show a subsolar merger with $M_\text{\tiny src}^\text{\tiny tot} =1 M_\odot$ at $z = 0.88$. 
	In both cases, we assume $q=1$, optimal orientation of the binary and the redshift is fixed in order to have SNR=9 in the $\Delta$-10km-HFLF-Cryo configuration.}
\label{fig:LFvsPBHs_signal}
\end{figure*}

The importance of the LF instrument when searching high redshift mergers can be intuitively understood by looking at Fig.~\ref{fig:LFvsPBHs_signal}, where we compare the signal amplitude for a merger with source frame mass $M_\text{\tiny src}^\text{\tiny tot} =20 M_\odot$ located close to the horizon (i.e. having SNR $=9$ in the $\Delta$-10km-HFLF-Cryo configuration). Due to the large source redshift, the detector frame mass is redshifted by a large factor, and the majority of SNR can only accumulate in the LF portion of the sensitivity curve.

\begin{figure*}[ht]
	\centering
	\includegraphics[width=1\textwidth]{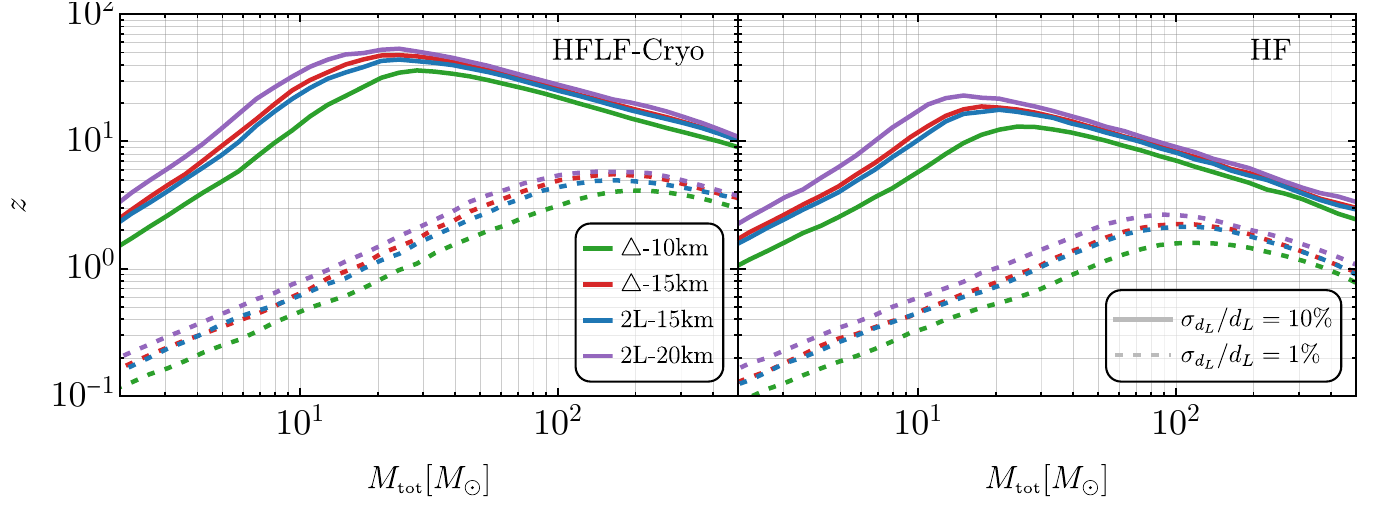}
	\caption{\small Contour lines of constant relative error on $d_L$ as a function of source frame total mass $M_\text{\tiny tot}$ and source redshift $z$. Different colors indicate different configurations while each panel (from left to right) assume HFLF-Cryo and HF, respectively. The solid (dashed) lines corresponds to $\sigma_{d_L}/{d_L} = 10\%$ (1\%). We assume equal masses $m_1 = m_2$ and optimal orientation of the binary. As a consequence of setting $\iota = 0$, aligned and misaligned 2L configurations result in the same performance for this kind of estimation.}
\label{fig:distance_error_highz}
\end{figure*}

In order to fully exploit the high redshift smoking-gun signature of a PBH population of mergers, sufficient precision on the source distance is also necessary, to distinguish PBHs from potential Pop-III contaminants.
The performance of ET on single-event-identification was studied in details Ref.~\cite{Ng:2021sqn,Ng:2022vbz} adopting a Bayesian parameter estimation. A limited precision on the source redshift 
may also impact the ability to reconstruct the merger rate evolution 
at high redshift within a population analysis, as shown in Refs.~\cite{Ng:2022agi,Martinelli:2022elq}. 
Based on a Fisher Matrix analysis, 
we compare the performance of different configurations in Fig.~\ref{fig:distance_error_highz}, which reports the maximum source redshift above which a relative precision on $d_L$ better than $10\%$ and $1\%$ cannot be achieved. 
The relative error $\sigma_{d_L}/{d_L}$ is highly impacted by the performance of the detector configurations at low frequencies
and follows the general trend observed in the behaviour of the horizon (see Fig.~\ref{fig:Detector_Horizons_BBH_AllConf}) and number of detectable high redshift PBH events (i.e. Tables~\ref{tabresPBH1}, \ref{tabresPBH2} and \ref{tabresPBH3}). We refer to \cite{Franciolini_in_prep} for further details on the analysis.

As far as the comparison of the performance of the various configurations is concerned, the results can be summarised as follows. 
\emph{
i)
The 10~km triangle is the one that performs less well. In its full HFLF-cryo configuration, for the assumed PBH population, the 2L-15km-$45^{\circ}$ configuration would detect about three times more events with $z>30$ than the 10~km triangle.
ii) 
A dramatic reduction in the number of high-redshift event   is observed in the absence of the LF instrument, to the extent that no events are detected at redshifts $z>30$, where a detection is identified as a PBH. This is also reflected in the behaviour of the uncertainties on the luminosity distance.} This is due to the reduced capabilities at small frequencies, where most of the high-$z$ events are expected, as they are characterised by a large detector frame total mass.

\subsubsection{Other PBH signatures} \label{sec:otherPBHs}

In this section, we analyse some other observables which may be crucial to discover/constrain the population of PBHs 
(see Ref.~\cite{Franciolini:2021xbq} for a systematic overview and \cite{Franciolini_in_prep} for further details on the present analysis).

$\circ$ {\bf Subsolar mass range.}
The distribution of PBH masses $m_\PBH$ is determined by the characteristic size and statistical properties of the density perturbations, corresponding to curvature perturbations generated during the inflationary epoch. Within this standard formation scenario, $m_\PBH$ is related to the mass contained in the cosmological horizon at the time of the collapse, 
and a wider range of masses is accessible compared to astrophysical BHs~\cite{Sasaki:2018dmp}.
PBHs, in particular, can have subsolar masses, which are not expected from standard stellar evolution, while they can also populate the astrophysical mass gaps~\cite{Clesse:2020ghq,DeLuca:2020sae,Franciolini:2022tfm}. 
For masses $m_i \gtrsim 3 M_\odot$ PBHs should be distinguished from stellar-origin BHs by mass-spin measurements \cite{DeLuca:2020bjf,Franciolini:2021xbq,Franciolini:2022iaa}.
The range $M_\odot \lesssim m_i \lesssim 3M_\odot  $ can be distinguished from NS thanks to tidal deformability measurements (see Section~\ref{sec:tidaleffects}). Finally, 
the subsolar range represents a smoking-gun signature of the non-standard, i.e. potentially primordial,  origin of a binary.\footnote{See, however, Ref.~\cite{Shandera:2018xkn} for models in which subsolar BHs are born out of dark sector interactions.
Also, in this case, the detection of subsolar mergers would 
indicate new physics.}$^{,}$\footnote{
While in the main text we focus on nearly equal mass mergers with $q \approx 1$, other potentially interesting systems for the detection of subsolar objects are intermediate mass ratio inspirals, given the high precision in measuring the mass of the light object at subpercent level~\cite{Barsanti:2021ydd}. 
Due to the relevance of the low-frequency range for these observations, the implementation of the LF interferometer could in principle enhance the capability of detecting these systems.}

\begin{figure*}[t]
	\centering
	\includegraphics[width=1\textwidth]{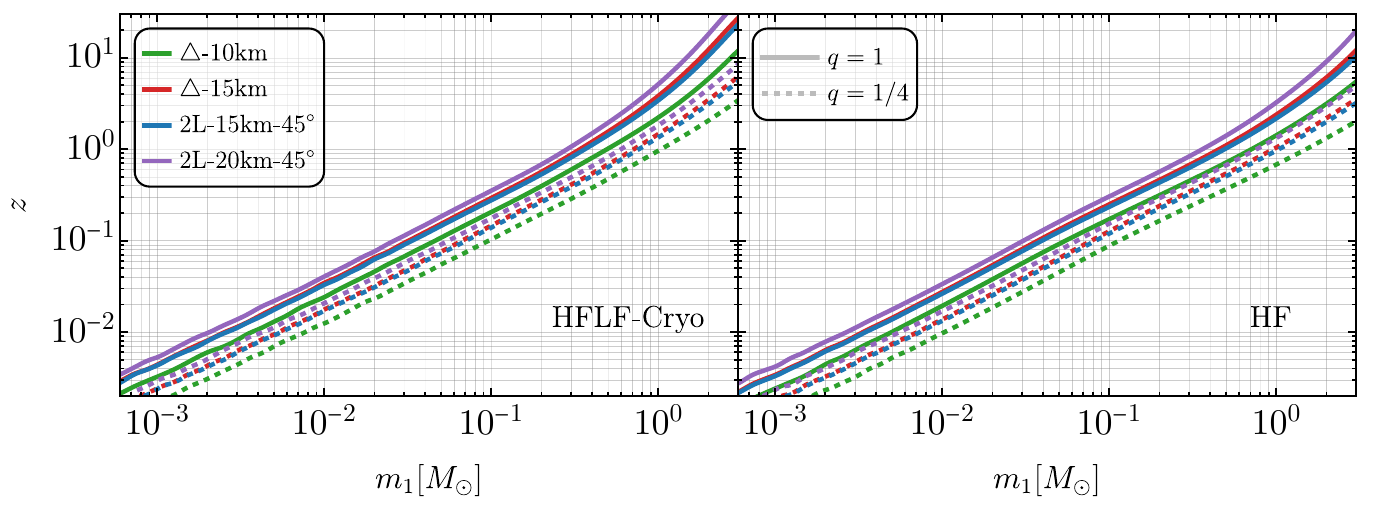}
	\caption{\small 
	Detection horizons for non-spinning PBH binaries assuming a mass ratio $q=1$ or $q=1/4$, for the various detector configurations considered. This plot extends Fig.~\ref{fig:Detector_Horizons_BBH_AllConf} in the sub-solar mass range, of interest for PBHs.  }
\label{fig:subsolarhorizon}
\end{figure*}

The Einstein Telescope will dramatically extend the reach of current technology when searching for mergers with at least one component below the Chandrasekhar limit~\cite{DeLuca:2021hde,Pujolas:2021yaw}. In Fig.~\ref{fig:subsolarhorizon} we show the detection horizon 
in the subsolar mass range, 
 extending Fig.~\ref{fig:Detector_Horizons_BBH_AllConf} in the range of masses where only PBHs, 
  white dwarfs, brown dwarfs, or exotic compact objects~\cite{Cardoso:2019rvt} (e.g. boson stars~\cite{Guo:2019sns}) can appear.
{\em As a large portion of the SNR is built up in the HF range of the sensitivity curve, 
the horizons are only slightly dependent on the LF instrument}.
This feature is shown in Fig.~\ref{fig:LFvsPBHs_signal}, where we compare the signal amplitude for a subsolar merger with source frame mass $M_\text{\tiny src}^\text{\tiny tot} =1 M_\odot$ located close to the horizon (i.e. having SNR $=9$ in the $\Delta$-10km-HFLF-Cryo configuration).

\begin{figure*}[ht]
	\centering
	\includegraphics[width=1\textwidth]{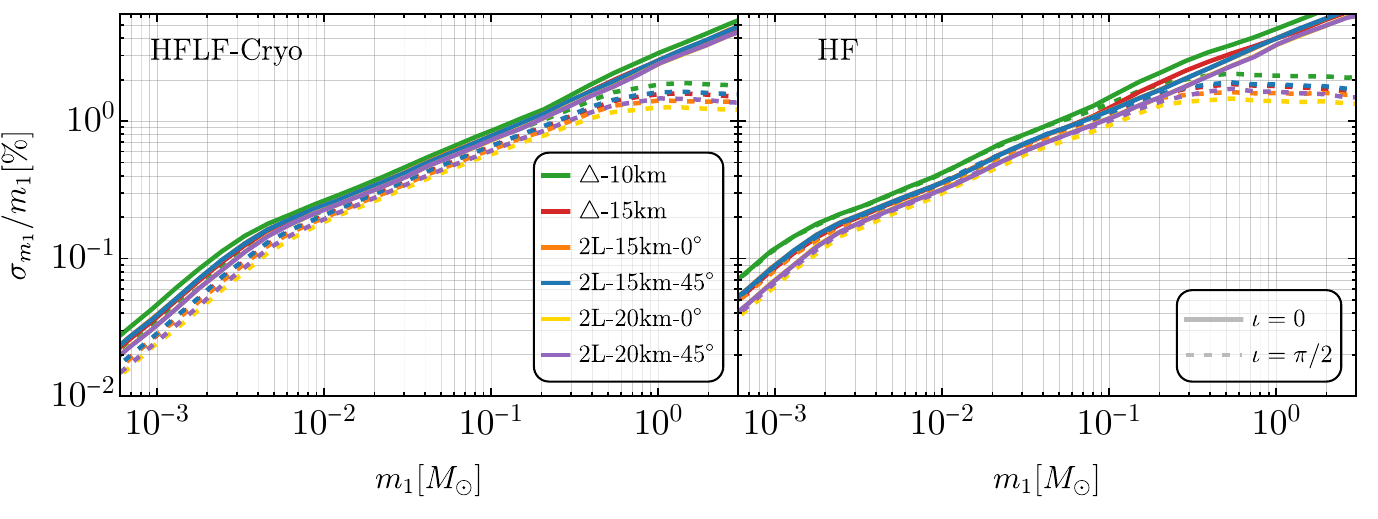}
	\includegraphics[width=1\textwidth]{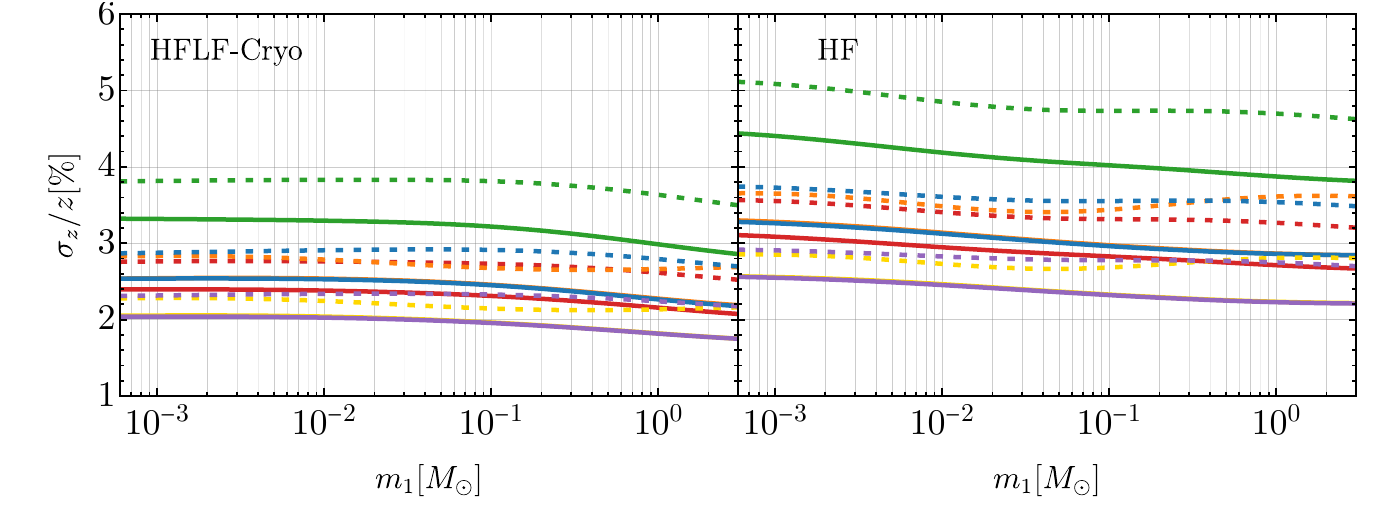}
	\caption{\small  
	Relative error on the primary source frame mass (top) and source redshift (bottom) as a function of $m_1$ in the subsolar range. 
	We consider a source located at a distance such that the SNR = 30 in the $\Delta\text{-10km}$ HFLF-Cryo configuration.}
\label{fig:subsolar}
\end{figure*}

\begin{figure*}[tp]
	\centering
	\includegraphics[width=1\textwidth]{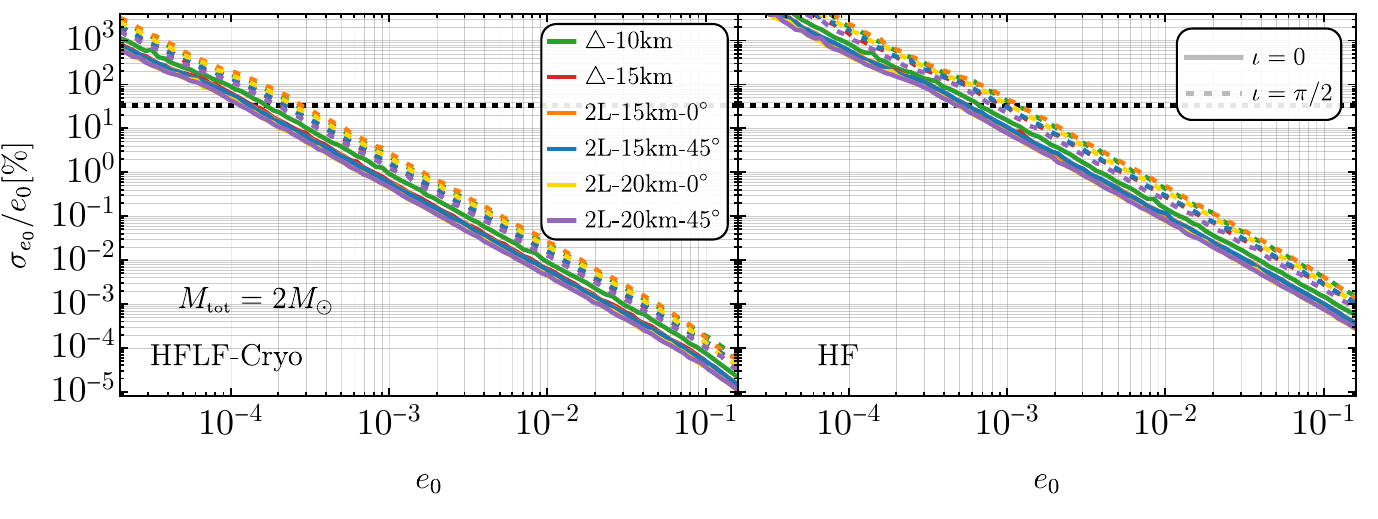}
	\includegraphics[width=1\textwidth]{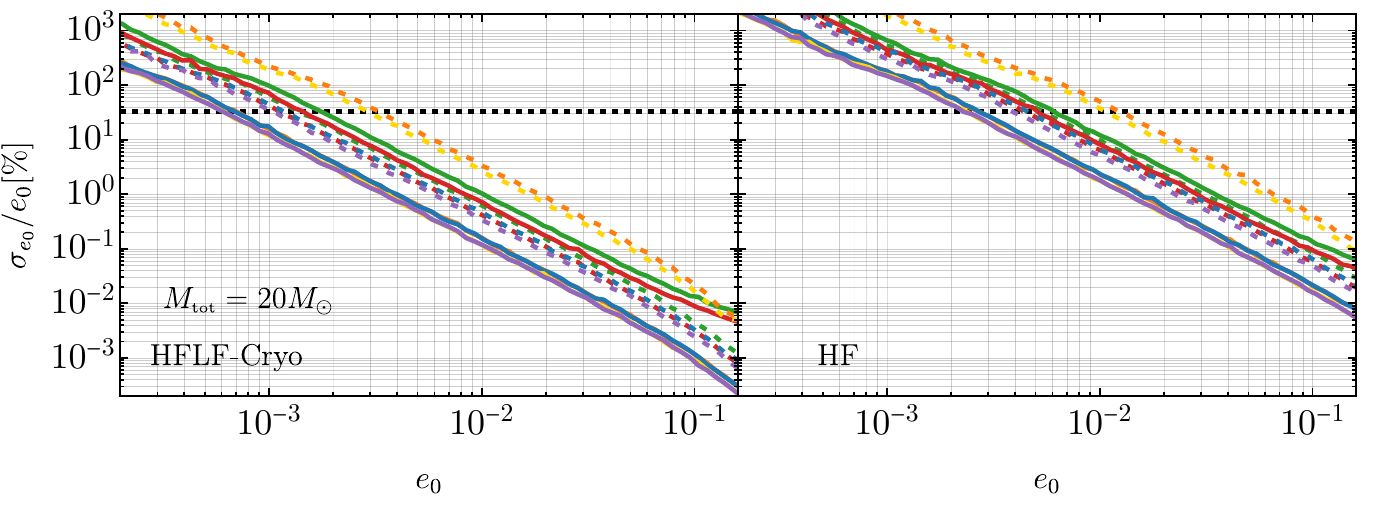}
	\caption{\small  
	Relative error on the eccentricity $e_0$ at $f_0 = 10 {\rm Hz}$ as a function of its injected value for 
	a binary with a total source frame mass of 
	$M_\text{\tiny tot} = 2 M_\odot$ and $d_L=100$~Mpc in the upper panel, 
	while $M_\text{\tiny tot} = 20 M_\odot$ and $d_L=500$~Mpc
	in the lower panel. 
 For the range of eccentricities in the horizontal axis (that are too large for PBH binaries), one is able to exclude  (at $3\sigma$ level) a PBH origin for events that fall below the horizontal black dashed line.}
\label{fig:eccentricity}
\end{figure*}

In Fig.~\ref{fig:subsolar} we compare the performance of different ET configurations when measuring the source frame mass and distance of subsolar PBH mergers.
In order to perform a proper comparison between different designs, 
for each mass $m_1$, we arbitrarily assume a source located at a distance fixed in order to have SNR$=30$ in the $\Delta\text{-10km}$ 
HFLF-cryo configuration. 
We adopt the IMRPhenomHM waveform model and test both face-on ($\iota = 0$) and edge-on ($\iota = \pi/2$) binaries. 
{\em In this case we conclude that the loss of the LF sensitivity results only in a minor reduction of the uncertainty on $m_1$ and $z$.
The length of the arms, which leads to a larger SNR, is the main factor responsible for the different performance of the various designs.}

$\circ$ {\bf Eccentricity measurements:}
Another key prediction of the PBH model involves the eccentricity $e$. 
While formed with large eccentricity at high redshift \cite{Nakamura:1997sm,Ioka:1998nz}, PBH binaries then have enough time to circularize before the GW signal can enter the observation band of current and future detectors \cite{Franciolini:2021xbq}.
Therefore, PBH binary candidates must have small eccentricities. 
Let us mention here that, on a more general scope, 
the measurement of eccentricity may also inform us about the possible astrophysical formation pathway of a binary. 
While isolated formation channels predict small values of $e$ in the observable range of frequencies of 3G detectors, dynamical channels 
 predicts a fraction ${\cal O}(10\%)$ with $e>0.1$ \cite{Mandel:2018hfr,Benacquista:2011kv,Bae:2013fna,Rodriguez:2017pec,Zevin:2018kzq,Mapelli:2021gyv,Kritos:2022ggc}.
This can be used to distinguish the origin of compact objects (see e.g. \cite{Nishizawa:2016jji,Nishizawa:2016eza,Zevin:2021rtf,Favata:2021vhw}).

To test the sensitivity of the various configurations to the measurement of a possible orbital eccentricity, we use the TaylorF2 inspiral--only waveform model \cite{Buonanno:2009zt, Ajith:2011ec, Mishra:2016whh} with the extension presented in \cite{Moore:2016qxz} to account for a small eccentricity in the orbit. Using as reference values for the total source--frame mass $M_{\rm tot} = 2 M_\odot$ and $20  M_\odot$, for each detector configuration, we compute the relative error that can be attained on equal mass, non spinning systems, as a function of the eccentricity
$e_0$ defined at $f_{0,\rm ecc} = 10$Hz. The distance to the source is fixed to $d_L=100$~Mpc for the case 
$M_{\rm tot} = {2}{\msun}$ and to $d_L={500}$~Mpc for the other.  The results for the relative errors attainable on $e_0$ are reported in Fig.~\ref{fig:eccentricity}. 
{\em We notice, in particular, the great relevance of the LF instrument to perform eccentricity measurements.} This can be traced to the fact that eccentricity gives larger effects during the inspiral (i.e. at low frequencies) since when going closer to a merger, a binary system tends to circularize, with $e_0\propto f_{\rm GW}^{-19/18}$.
The range of values of $e_0$ shown on the horizontal axis of Fig.~\ref{fig:eccentricity} corresponds to eccentricities that are too large for a PBH binary, that are expected to have $e_0\sim 10^{-6}$ when they reach  $f=10$~Hz \cite{Franciolini:2021xbq}. Therefore, in this range, if at the same time the relative error on $e_0$ for a given detection is sufficiently small, e.g. below the horizontal dashed line in the figure  (which corresponds to $\Delta e_0/e_0 = 0.33$), one would be able to exclude that this event has  a primordial origin.

The information 
in Figs.~\ref{fig:subsolarhorizon},~\ref{fig:subsolar} and \ref{fig:eccentricity} can be summarised as follows: \emph{i) among the considered geometries, the triangle is the less performing giving e.g., uncertainties on the estimates of redshift about 1.4 times larger as compared to the 2L configurations, with both the considered ASDs. ii) the LF instrument turns out to be of great relevance for the reconstruction of the eccentricity (with a gain in relative error that can be larger than one order of magnitude) and to a lesser extent also in the reconstruction of the masses and distances of sub-solar objects that, thanks to the LF sensitivity, can stay in the band 
for more cycles.}

\subsection{Cosmology} 

\subsubsection{Hubble parameter and dark energy from joint GW/EM detections}

In the context of cosmological studies, a key property  of compact binary coalescences is that, from their GW signal, one can reconstruct the luminosity distance  to the source~\cite{Schutz:1986gp}. Compact binary coalescences are then typically called `standard sirens', by analogy with the `standard candles' that provide absolute cosmological distance measurements from electromagnetic observations. Since the GW signal does not provide  a direct measurement of the redshift, the ideal situation to constrain cosmological parameters is to measure the redshift of the host galaxy of the gravitational-wave source. Due to the poor localization capability of the gravitational-wave detectors, the host galaxy is typically identified only if we  detect an electromagnetic counterpart (with some possible exceptions, see Section~\ref{sect:highmassevents}). There are, however, a variety of statistical methods, based either on spatial correlations with galaxies or large-scale structures, or on features in the mass distribution of the sources, that can be used to extract cosmological information even in the absence of an electromagnetic counterpart~\cite{Holz:2005df,Dalal:2006qt,MacLeod:2007jd,Nissanke:2009kt,Cutler:2009qv,Messenger:2011gi,Taylor:2011fs,Taylor:2012db,DelPozzo:2015bna,Chen:2017rfc,Feeney:2018mkj,Gray:2019ksv,Farr:2019twy,Hannuksela:2020xor,Wang:2020xwn,Mukherjee:2020hyn,Finke:2021aom,Ye:2021klk,Chatterjee:2021xrm,Finke:2021znb,Finke:2021eio,LIGOScientific:2021aug,Palmese:2021mjm,Mancarella:2021ecn,Jin:2022qnj,Mukherjee:2022afz,Ghosh:2022muc,Shiralilou:2022urk,Mancarella:2022cnu,Gair:2022zsa}. 
Standard sirens, either with counterpart or combined to statistical methods,  allow us to measure the Hubble parameter $H_0$ at small redshifts 
while, at the larger redshifts that become accessible to a 3G detector, we can also explore the dark energy (\acrshort{de}) sector, through the study of the DE equation of state  
and of modified GW propagation.

In this section we explore the relative potential of different ET configurations for  cosmology, limiting ourselves to the study of GW sources with an electromagnetic counterpart, while in Sections~\ref{sect:H0BNStidal} and
\ref{sect:highmassevents} we will discuss  alternative possibilities.
For this purpose,  among the instruments discussed in Section~\ref{sect:MMOGRB}, the most useful are those which enable to obtain precise (spectroscopic) measurement of the host galaxy redshift.
In particular, following the analysis in~\cite{Belgacem:2019tbw}, we consider joint GW/EM detections between ET and THESEUS. We consider this instrument, for the detection of the associated short GRB, since the localisation capabilities ($\sim$ arcmin/arcsec) of both the XGIS and SXI instruments of THESEUS facilitate the follow-up by ground-based telescopes, guaranteeing a high chance of obtaining the redshift determination (see the discussion in Section~\ref{sect:MMOGRB}). We consider only the prompt emission (see Tables~\ref{tab_jointprompt_cryo} and \ref{tab_jointprompt_HF} for the number of joint detection per year). We further consider the joint detections between ET and VRO, which is able to observe the kilonova emission produced by BNS mergers in the optical band and give a precise localization for accurate redshift measurements. As discussed in Section~\ref{sect:MMO}, this allows us to study multi-messenger prospects for ET both at the high redhsifts probed by GRB observations and at $z\, \lsim \,  (0.3-0.4)$, where it will be possible to detect kilonova emission using wide field telescopes.

We stress  that the absolute estimates for the accuracy that we will obtain below for $H_0$, as well as for the parameters describing the DE equation of state and modified GW propagation introduced in the following, should not be understood as  forecasts for the overall capability of ET for measuring these cosmological parameters. 
We are focusing here only on standard sirens  
observed from two specific instruments (in this case, THESEUS and VRO). Apart from the fact that the results also rely on assumptions on their performances and specific observational strategies, 
other EM observatories are expected to operate together with ET, increasing the sample of detections and/or adding further complementary information.
On top of this, `dark sirens', i.e. coalescing binaries without an observed electromagnetic counterpart, can provide very significant information through various statistical techniques, as mentioned above. In particular, the correlation with galaxy catalogs already provides valuable cosmological information at the level of current LVC data: the statistical method
has been applied to extract $H_0$ from the recent LIGO/Virgo detections in a number of papers~\cite{LIGOScientific:2018gmd,Soares-Santos:2019irc,LIGOScientific:2019zcs,DES:2020nay,Finke:2021aom, LIGOScientific:2021aug,Palmese:2021mjm}, and to studies of the DE sector in~\cite{Finke:2021aom,Mancarella:2021ecn,Ezquiaga:2021ayr,Leyde:2022orh}. 
These studies in the ET era will greatly benefit from Euclid and the Large Synoptic Survey Telescope (\acrshort{lsst}) surveys by the VRO which will map billion of galaxies giving more precise photometric redshifts, and spectroscopic surveys by multi-object survey spectrographs such as WAVE or 4MOST.
Furthermore, several other ideas have been put forward to extract cosmological information from ET, see e.g. \cite{Messenger:2011gi,Taylor:2011fs,Taylor:2012db,Farr:2019twy,Hannuksela:2020xor,Finke:2021znb,Finke:2021eio,Ezquiaga:2021ayr,Balaudo:2022znx}.
Therefore, an assessment of the full capabilities of ET for estimating cosmological parameters must take into account all these opportunities. Here, we focus on two examples, the joint GW/EM detections by  ET+THESEUS and by ET+VRO, with the  purpose of evaluating  the relative performances of different configurations of ET in these specific situations, rather than  the absolute  precision to which cosmological parameters can be reconstructed at ET.

For the ET+THESEUS analysis, the population of BNSs and the associated EM signal at high energies are generated following the approach in ref.~\cite{Ronchini:2022gwk}. Following the same approach adopted in Section~\ref{sect:MMO}, we fix the  EM signal for each BNS of the population, for the entire set of simulations corresponding to the different ET configurations and geometries. As a reference, we consider 5 years of observations, and 
we select from our simulation a sample of events, such that among them there are 75 joint detections with the configuration 2L 20 km HFLF cryo (see Table~\ref{tab_jointprompt_cryo}). Considering the same BNS sample, we then compute the number of joint detections for all the other configurations. With this method, we are correctly evaluating the capabilities of the ET configurations considering all of them to have the same total observational time, the same BNS sample and the same EM signal. In this derivation, the duty cycle of each ET configuration is taken into account. 

We have performed parameter estimation using both \texttt{GWFISH} \cite{Dupletsa:2022wke} and  \texttt{GWFAST}~\cite{Iacovelli:2022bbs,Iacovelli:2022mbg},
finding full agreement. 
A delicate point \label{foot:2suSNR} of the parameter estimation procedure is that the events that give a detectable GRB are close to face-on, where the degeneracy between the luminosity distance $d_L$ and the inclination angle $\iota$ results in large errors on $d_L$, and often unreliable inversion of the Fisher matrices. This can be cured by imposing a prior on $\iota$, which reflects the information that, if a jet was observed from cosmological distances, the binary was close to face-on. A full analysis of the corresponding selection effects is quite non-trivial, and is in progress. For this study, we observe that all the GRBs of our catalog have  $|\iota| < (5-10)\, {\rm deg}$.\footnote{More precisely, to take into account the periodicity of $\iota$, they have 
$ \min\(\iota, 180^{\circ}-\iota\)<(5-10) \, {\rm deg}$.} We also keep fixed the position of the source in the sky, having assumed that the host galaxy has been identified.
We then use the  simple (and common) prescription of setting $\Delta d_L/d_L=2/{\rm SNR}$. This expression,
with a factor of two included to be conservative, corresponds to the large SNR limit of the Fisher-matrix analysis when all other parameters are fixed. Since $\iota$ and the position in the sky are the parameters that are most degenerate with $d_L$, once these are kept fixed or strongly constrained by priors, this limiting value can become a reasonable  approximation. A more detailed study on this  is in progress.
We then add in quadrature an error 
$\Delta d_L/d_L=0.05 z$ due to lensing, as in ~\cite{Sathyaprakash:2009xt,Zhao:2010sz} (see also \cite{Kalomenopoulos:2020klp} for more recent justification). 
From the value of $z$ of a joint GW+EM event of our catalog, we then compute the nominal value of its luminosity distance using a fiducial $\Lambda$CDM model (we use the fiducial values $H_0 = 67.66\, {\rm km}\, {\rm s}^{-1}\, {\rm Mpc}^{-1}$
and $\oma = 0.31$, from {\em Planck-2018} \cite{Planck:2018vyg}), and the error $\Delta d_L$ from the above procedure and then, in order to simulate the realistic setting of an experiment, we scatter the `observed' value of $d_L$ using a Gaussian distribution with the given $\Delta d_L$.

In a similar way, we study the capabilities of the synergy between ET and VRO. The population of sources and their respective KN emission are generated through the procedure described in Section~\ref{sect:MMO}. In this case, the number of events considered corresponds to one year of observations, rather than 5, due to the larger number of potential joint detections (see Section~\ref{sect:MMO} and Table~\ref{tab:jointKNFull} for details). We consider the observational strategy consisting on following-up all the events localized with $\Delta\Omega_{90\%}<40~{\rm deg}^2$, two filter ({\em g} and {\em i}) observations repeated the first and second night after the merger and an exposure time for each pointing of 600 s.  
In particular, for each configuration we consider a number of sources matching the ones reported in column 3 of Table~\ref{tab:jointKNFull}. Note that, thanks to its isotropic nature, the KN emission can be observed irrespectively of the binary inclination, resulting in a sample of events with better conditioning for the Fisher matrix analysis, thanks to a milder degeneracy between $\iota$ and $d_L$. 

We now present the results obtained for $(H_0,\Omega_{m,0})$, for the DE equation of state, and for modified GW propagation.

\paragraph{Hubble constant.}\label{sect:H0}
In a flat $\Lambda$CDM model, the  relation between the luminosity distance $d_L$ and the redshift is 
\be\label{dLemLCDM}
d_L(z)=\frac{c}{H_0}\, (1+z) \, \int_0^z\, 
\frac{d\tilde{z}}{\sqrt{\oma (1+\tilde{z})^3 +\ola }}\, ,
\ee
where $\oma$  is the present matter  fraction and  $\ola$ is the energy density fraction associated with the cosmological constant (we neglect the contribution from radiation, which  is completely negligible at the redshifts relevant for standard sirens. In this approximation $\ola =1-\oma$ for a flat $\Lambda$CDM cosmology). At low redshift this reduces to Hubble's law $d_{L}(z)\simeq (c/H_0)z$, so standard sirens at low redshift can allow us to measure  $H_0$.

The first measurement of $H_0$ from a standard siren with counterpart has been obtained from  
GW170817~\cite{Abbott:2017xzu}. However, the error from this single detection is still too large to discriminate between the value of $H_0$ obtained from  late-Universe probes~\cite{Riess:2019cxk,Wong:2019kwg}, and that  inferred from early-Universe probes assuming $\Lambda$CDM~\cite{Aghanim:2018eyx,Abbott:2018xao}, which are currently  in disagreement at  $5.3\sigma$ level.

\begin{figure}[tp]
	\centering
	\includegraphics[width=10cm]{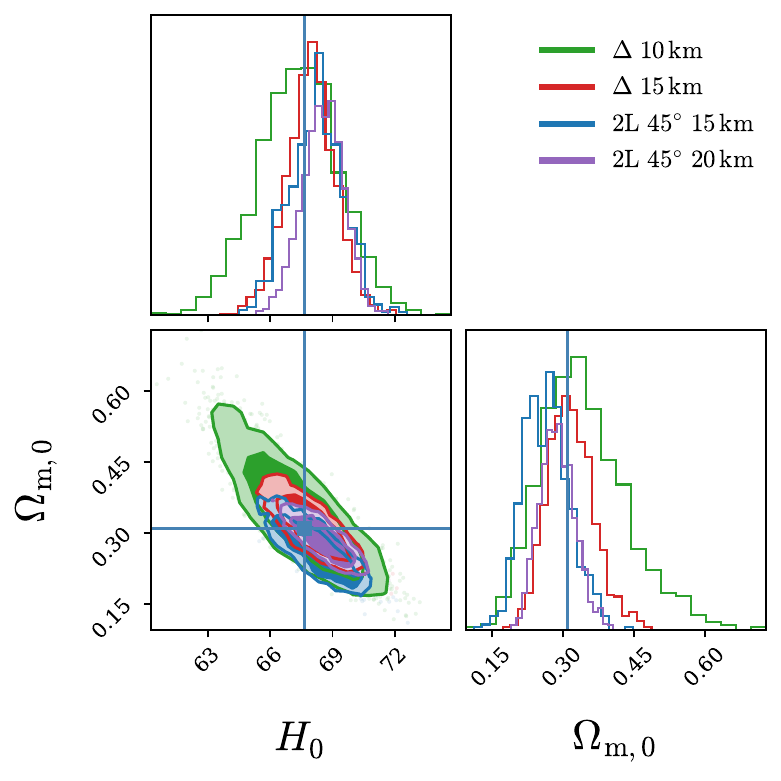}
	\caption{\small Reconstruction of the parameters $H_0$ and $\Omega_M$ in $\Lambda$CDM, from the joint GW+EM events obtained with ET+THESEUS in 5 yr of observations, for the different geometries of ET shown, all with their HFLF-cryo sensitivity.}
\label{fig:corner_all_H0Om}
\end{figure}

\begin{figure}[tp]
	\hspace{-1.6cm}
	\includegraphics[width=9.2cm]{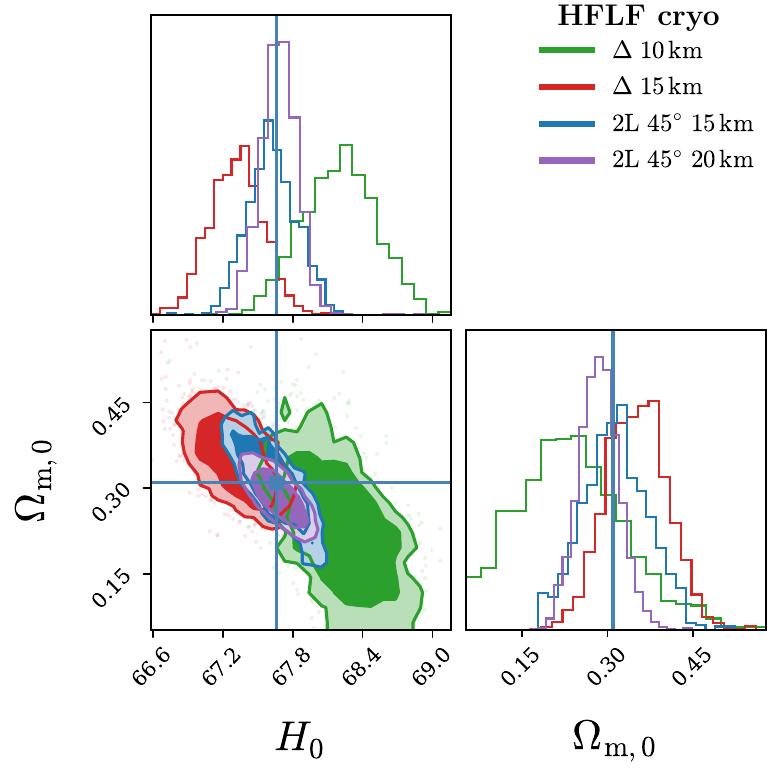}
	\hfill
	\includegraphics[width=9.2cm]{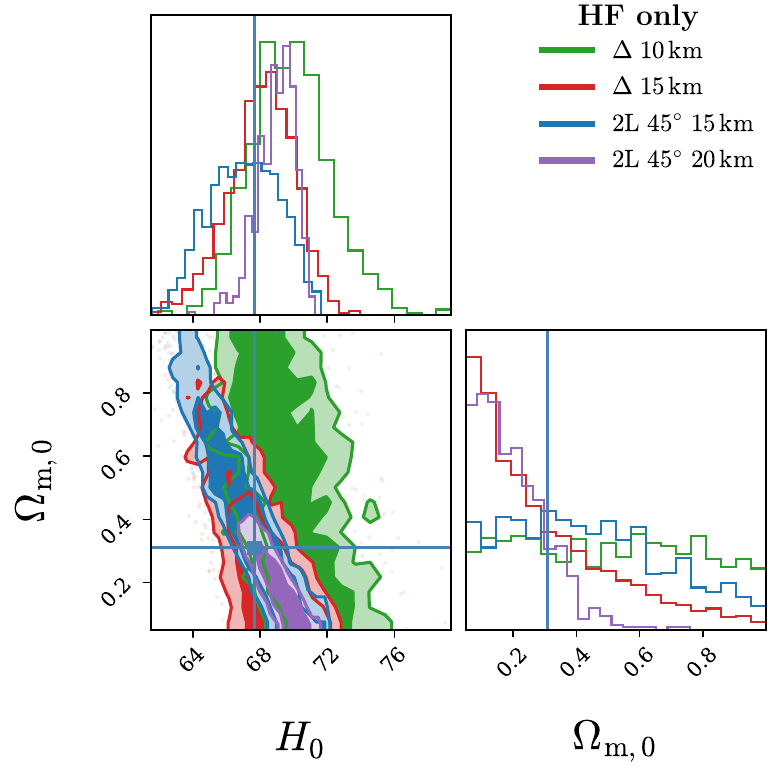}
	\caption{\small Reconstruction of the parameters $H_0$ and $\Omega_M$ in $\Lambda$CDM, from the joint GW+EM events obtained with ET+VRO in 1 yr of observations, for the different geometries of ET shown, in the left panel with their HFLF-cryo sensitivity and in the right panel with the HF instrument only.}
\label{fig:corner_all_H0Om_VRO}
\end{figure}

\begin{table}
\centering
\begin{tabular}{|l|c|c|}
\hline
\hline
Configuration       &  $\Delta H_0/H_0$ &  $\Delta \oma/\oma$  \\  \hline \hline
$\Delta$-10km       &  $0.057$          & $0.546$              \\  \hline 
$\Delta$-15km       &  $0.035$          & $0.290$              \\  \hline
2L-15km-45$^\circ$  &  $0.040$          & $0.370$              \\  \hline  
2L-20km-45$^\circ$  &  $0.029$          & $0.276$              \\  \hline \hline 
\end{tabular}
\vspace{0.2cm}
\caption{\small Relative errors on $H_0$ and $\oma$ in $\Lambda$CDM (median and symmetric 68\% CI), from the joint GW+EM events obtained with ET+THESEUS, for the different geometries of ET shown, all with their HFLF-cryo sensitivity. We stress that no prior from electromagnetic observations, such as  CMB+\acrshort{bao}+\acrshort{sn}e, is used here; with such priors, the accuracy on $H_0$ becomes sub-percent.}
\label{tab:H0Oma}
\end{table}

\begin{table}
\centering
\begin{tabular}{|l|c|c|}

\hline
\hline
\multicolumn{3}{|c|}{\textbf{HFLF cryogenic}}\\
\hline
Configuration       &  $\Delta H_0/H_0$ &  $\Delta \oma/\oma$  \\  \hline \hline
$\Delta$-10km       &  $0.009$          & $0.832$              \\  \hline 
$\Delta$-15km       &  $0.007$          & $0.303$              \\  \hline
2L-15km-45$^\circ$  &  $0.006$          & $0.370$              \\  \hline  
2L-20km-45$^\circ$  &  $0.004$          & $0.243$              \\  \hline \hline 
\end{tabular}
\hspace{.7cm}
\begin{tabular}{|l|c|c|}

\hline
\hline
\multicolumn{3}{|c|}{\textbf{HF only}}\\
\hline
Configuration       &  $\Delta H_0/H_0$ &  $\Delta \oma/\oma$  \\  \hline \hline
$\Delta$-10km       &  $0.065$          & $1.23$              \\  \hline 
$\Delta$-15km       &  $0.057$          & $1.86$              \\  \hline
2L-15km-45$^\circ$  &  $0.066$          & $1.31$              \\  \hline  
2L-20km-45$^\circ$  &  $0.031$          & $1.22$              \\  \hline \hline 
\end{tabular}
\vspace{0.2cm}
\caption{\small Relative errors on $H_0$ and $\oma$ in $\Lambda$CDM (median and symmetric 68\% CI), from the joint GW+EM events obtained with ET+VRO in 1~yr of observations, for the different geometries of ET shown, in the left table with their HFLF-cryo sensitivity and in the right table with the HF instrument only.}
\label{tab:H0Oma_KN}
\end{table}

Our results for $H_0$ and $\oma $ are shown in Fig.~\ref{fig:corner_all_H0Om}  and Table~\ref{tab:H0Oma} for ET+THESEUS (that shows the result of a typical realization of the catalog of joint detections, among several that we have generated), 
and in Fig.~\ref{fig:corner_all_H0Om_VRO} and Table~\ref{tab:H0Oma_KN} for ET+VRO.
We should stress that these are the results obtained from standard sirens only, without any prior from cosmological electromagnetic probes. Of course, CMB, BAO, SNe, structure formation, and other electromagnetic  probes provide much stronger constraints on $\oma$; assuming $\Lambda$CDM, {\em Planck-2018} fixes $\oma =0.315\pm 0.007$~\cite{Planck:2018vyg}, i.e.
$\Delta\oma/\oma\simeq 0.022$, much smaller than what can be obtained from standard sirens alone. In principle, one could then combine cosmological and standard sirens data, as in \cite{Belgacem:2019tbw}, reducing significantly the error on $H_0$.\footnote{The limits on $H_0$ reported here, obtained  from GWs+GRB only, are consistent  with the results in \cite{Belgacem:2019tbw}, after taking into account 
that we are considering 5~years of data (which is more consistent with updated estimates for the expected duration of the THESEUS mission), while  the data in \cite{Belgacem:2019tbw} were relative to a 10~yr period, and that we use a model for short GRBs normalized to the observations (which takes into account the fact that not all BNS merger produce a GRB), resulting in a smaller number of joint GW+GRB detections. As shown in \cite{Belgacem:2019tbw}, if one further combines the constraints from joint GW+EM detection with purely electromagnetic observations from CMB, BAO, SNe, etc, the accuracy on $H_0$ reaches the  subpercent level.
} 
In any case, for our present purpose of comparing the different geometries,  Figs.~\ref{fig:corner_all_H0Om} and \ref{fig:corner_all_H0Om_VRO} and Tables~\ref{tab:H0Oma} and \ref{tab:H0Oma_KN} contain the information that we need, on the relative performances of the different ET configurations.

We see that the best results for $H_0$  are obtained by ET+VRO, whose detections are potentially more numerous and at small redshift, where the impact of $H_0$ is bigger. However, we notice that the results obtained for THESEUS are very solid and conservative: they are based on the current observations of short GRBs and consider only the prompt emission (including the afterglow detections by XGIS and SXI will increase the detections per year by a factor of about 3). In contrast, the VRO estimates strongly depend on the BNS merger rate normalization (which, on the basis of the GW observations by the LIGO and Virgo detectors, can be one order of magnitude smaller or larger).

Concerning the comparison of configurations  we find that {\em the  triangle-15km and  the {\rm 2L}-15km-$45^\circ$ have quite similar accuracy on $H_0$ for this test, and improve by a factor  $\sim (1.3-1.6)$ the result of the 10~km triangle, for  {\rm ET+THESEUS} and {\rm ET+VRO} (with {\rm ET} taken in both cases in the {\rm HFLF-cryo} configuration). From the {\rm ET+VRO} results, we also see the huge impact of the {\rm LF} instrument, which allows us to observe a larger number of sources with good localization and results in constraints more than a factor of 7 tighter on $H_0$ and more than a factor of 2 tighter on $\Omega_{m,0}$.}

Observe that the reduction of joint ET+VRO detections is dramatic if we lose the low-frequency: only 6\% of the events detected by the triangle-10~km HFLF-cryo,   and only 11\% of those detected by the 2L-15km HFLF-cryo,  are observed by the corresponding  HF-only configurations, see columns 4 of Table~\ref{tab:jointKNHF} and Table~\ref{tab:jointKNFull}. In contrast, for the joint ET+THESEUS detections the reduction is much smaller: 60\% of the events detected by the triangle-10~km HFLF-cryo and 70\% of those detected by the 2L-15km HFLF-cryo are observed by the corresponding HF-only configurations, see columns 4 of Table~\ref{tab:jointKNHF} and Table~\ref{tab:jointKNFull}. Taking into account that  the 
detections are distributed in the same redshift range (see for example Fig.~\ref{fig:HERMESETHFjoint}), the observations of THESEUS together with ET-HF become comparable to those obtained by THESEUS together with  ET at full HFLF-cryo sensitivity, in a observation time longer by  a factor $\sim 1.5$ (which could also be partly compensated if the duty cycle of the HF-only instrument will be larger than that of the full HFLF-cryo configuration).

\paragraph{Dark energy equation of state}\label{sect:DEEoS}

An accurate measurement of the Hubble parameter with GWs would be of course very interesting a priori. However, it is in principle possible (although by no means guaranteed) that, by the time that 3G detectors will be operative, the discrepancy on $H_0$ between late-Universe and early-Universe probes might have already found a resolution. More specific to 3G detectors is the possibility of testing the cosmological model with GWs up to moderate and large redshifts, where signatures of a dynamical dark energy could be found. 

In general, on cosmological scales, one  performs a separation between the
homogeneous Friedmann-Robertson-Walker (\acrshort{frw}) background, and  scalar, vector and tensor perturbations  over it; tensor perturbations, when they are well inside the cosmological horizon, are just GWs propagating on the FRW background. Whenever gravity is modified on cosmological scales, both the background evolution and the dynamics in the sectors of scalar and tensor perturbations are modified (vector perturbations usually decay and are irrelevant).
Here we discuss the effect of dark energy on the background evolution, which, as we will recall, can be described by an effective dark energy (DE) equation of state, while below we will discuss the modification of the tensor sector related to modified GW propagation.

\begin{figure}[t]
	\centering
	\includegraphics[width=10cm]{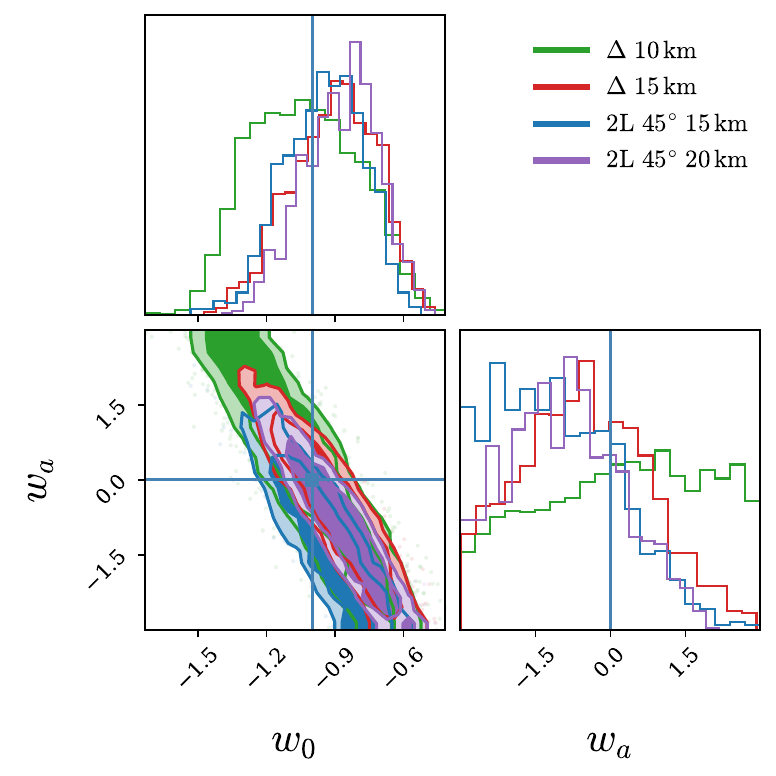}
	\caption{\small Reconstruction of the DE equation of state parameters $w_0$ and $w_a$, from the joint GW+EM events obtained with ET+THESEUS in 5 yr of observations, for the different geometries of ET shown, all with their HFLF-cryo sensitivity.}
\label{fig:corner_all_w0wa}
\end{figure}

\begin{table*}
\centering
\begin{tabular}{|l|c|c|}
\hline
\hline
Configuration       &  $\Delta w_0$ & $\Delta w_a$ \\  \hline \hline
$\Delta$-10km       &  $0.49$          & $3.81$              \\  \hline 
$\Delta$-15km       &  $0.40$          & $2.65$              \\  \hline
2L-15km-45$^\circ$  &  $0.35$          & $2.55$              \\  \hline  
2L-20km-45$^\circ$  &  $0.34$          & $2.40$              \\  \hline \hline 
\end{tabular}
\vspace{0.2cm}
\caption{\small Absolute errors on the DE equation of state parameters $w_0$ and $w_a$ (symmetric 68\% CI), from the joint GW+EM events obtained with ET+THESEUS, for the different geometries of ET shown, all with their HFLF-cryo sensitivity.}
\label{tab:w0wa}
\end{table*}

At the background level, the effect of 
a dynamical dark energy component is 
described by the DE density $\rde(z)$ and by its pressure 
$p_{\rm DE}(z)$. Equivalently, one can introduce the DE 
equation of state $\wde(z)$, defined by $p_{\rm DE}(z)=\wde(z)\rde(z)$. The standard $\Lambda$CDM model is recovered for $\wde(z)=-1$. Using the conservation of the DE energy-momentum tensor, one finds that
the DE density is given as a function of redshift by 
\be\label{4rdewdeproofs}
\rde(z)  =\rho_0 \Omega_{\rm DE}\, \exp\left\{ 3\int_{0}^z\, \frac{d\tilde{z}}{1+\tilde{z}}\, [1+\wde(\tilde{z})]\right\}\, ,
\ee
where $\Omega_{\rm DE}=\rde(0)/\rho_0$ is the DE density fraction and $\rho_0=3H_0^2/(8\pi G)$ is the critical density.
At the background level, the properties of DE are therefore encoded just in one function $\wde(z)$.
The corresponding expression for the luminosity distance (neglecting again radiation)  is
\be\label{dLemmod}
d_L(z)=\frac{c}{H_0}\, (1+z) \,\int_0^z\, 
\frac{d\tilde{z}}{\sqrt{\oma (1+\tilde{z})^3+ \rde(\tilde{z})/\rho_0}}\, ,
\ee
and reduces to \eq{dLemLCDM} when $\rde(\tilde{z})$ is a constant, i.e. when $\wde(z)=-1$.
Any redshift dependence of $\rde$, i.e. any deviation  of the DE equation of state from the $\Lambda$CDM value $\wde=-1$, would  provide evidence for a dynamical dark energy.

\begin{figure}[tp]
	\hspace{-1.65cm}
	\includegraphics[width=9.2cm]{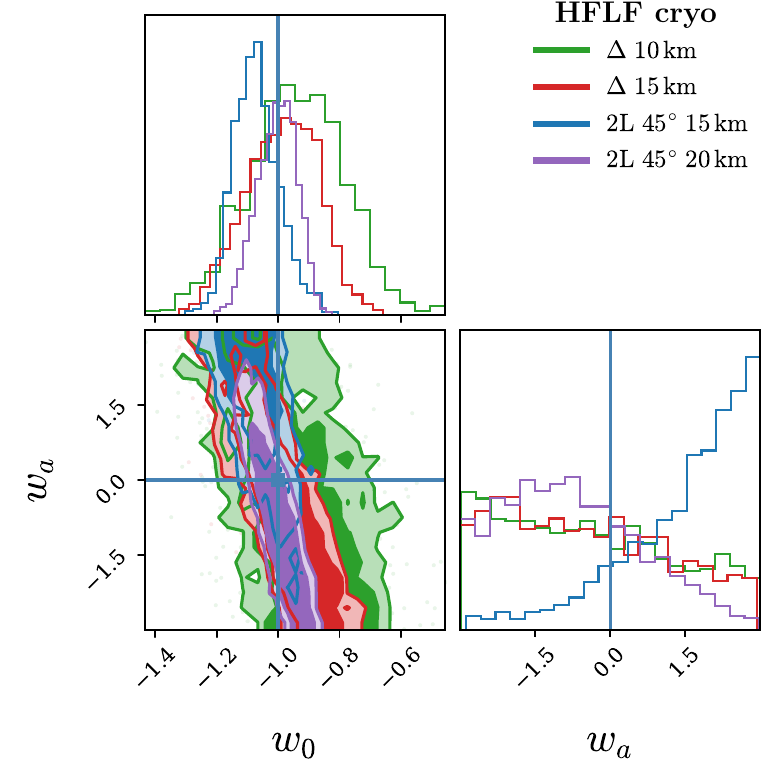}
	\hfill
	\includegraphics[width=9.2cm]{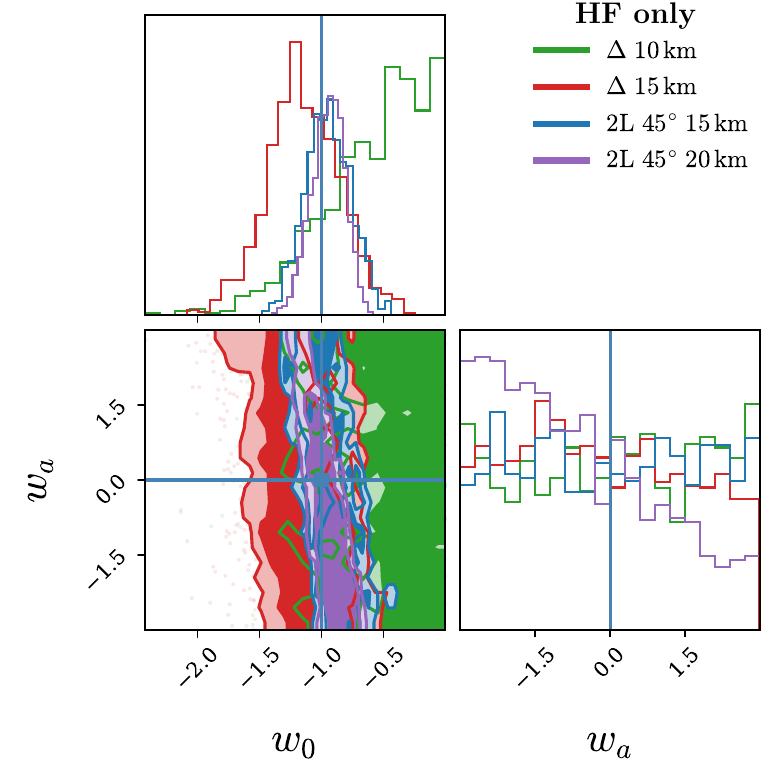}
	\caption{\small Reconstruction of the DE equation of state parameters $w_0$ and $w_a$, from the joint GW+EM events obtained with ET+VRO in 1 yr of observations, for the different geometries of ET shown, in the left panel with their HFLF-cryo sensitivity and in the right panel with the HF instrument only.}
\label{fig:corner_all_w0wa_KN}
\end{figure}

\begin{table}
\centering
\begin{tabular}{|l|c|c|}

\hline
\hline
\multicolumn{3}{|c|}{\textbf{HFLF cryogenic}}\\
\hline
Configuration       &  $\Delta w_0$ &  $\Delta w_a$  \\  \hline \hline
$\Delta$-10km       &  $0.31$          & $3.92$              \\  \hline 
$\Delta$-15km       &  $0.23$          & $3.69$              \\  \hline
2L-15km-45$^\circ$  &  $0.14$          & $2.49$              \\  \hline  
2L-20km-45$^\circ$  &  $0.13$          & $2.89$              \\  \hline \hline 
\end{tabular}
\hspace{.7cm}
\begin{tabular}{|l|c|c|}
\hline
\hline
\multicolumn{3}{|c|}{\textbf{HF only}}\\
\hline
Configuration       &  $\Delta w_0$ &  $\Delta w_a$  \\  \hline \hline
$\Delta$-10km       &  $0.85$          & $4.26$              \\  \hline 
$\Delta$-15km       &  $0.56$          & $3.84$              \\  \hline
2L-15km-45$^\circ$  &  $0.37$          & $4.13$              \\  \hline  
2L-20km-45$^\circ$  &  $0.27$          & $3.52$              \\  \hline \hline 
\end{tabular}
\vspace{0.2cm}
\caption{\small Absolute errors on the DE equation of state parameters $w_0$ and $w_a$ (symmetric 68\% CI), from the joint GW+EM events obtained with ET+VRO, for the different geometries of ET shown, in the left table with their HFLF-cryo sensitivity and in the right table with the HF instrument only.}
\label{tab:w0wa_KN}
\end{table}

In general, it is difficult to extract from the data a full function of redshift, such as $\wde(z)$, and a parametrization in terms of a small number of parameters is useful. For the DE equation of state a  standard choice is the $(w_0,w_a)$ parametrization \cite{Chevallier:2000qy,Linder:2002et},
\be\label{w0wa}
w_{\rm DE}(z)= w_0+\frac{z}{1+z} w_a\, .
\ee 
Combining \eq{dLemmod} with \eqs{4rdewdeproofs}{w0wa}, from the measurements of the luminosity distance and redshift of an ensemble of sources we can perform the inference on 
the cosmological parameters $H_0,\oma,w_0,w_a$. The constraining power of GWs alone is too weak to provide useful information on all these parameters (as we already saw just for $(H_0,\oma)$ in Table~\ref{tab:H0Oma}), so one is obliged to combine standard sirens with other cosmological probes. 
In this context, it is then common to assume that $H_0,\oma$ are simply fixed, or strongly constrained by priors obtained from electromagnetic probes, and keep only  $(w_0,w_a)$ in the inference.
Several forecasts of $(w_0,w_a)$ for ET  already exist in the literature~\cite{Sathyaprakash:2009xt,Zhao:2010sz,Belgacem:2018lbp,Belgacem:2019tbw}. Here we present updated results using our catalog of joint ET+THESEUS detections, for the various ET geometries considered, as well as ET+VRO for the various geometries and the two ASDs (as we saw above, in the case of ET+THESEUS, the  loss of the LF instrument is much less dramatic, and basically can be compensated by increasing by a factor $\sim 1.5$ the observation time, which does not affect the relative performances of the geometries). The results are shown in Fig.~\ref{fig:corner_all_w0wa} and Table~\ref{tab:w0wa} for ET+THESEUS and in Fig.~\ref{fig:corner_all_w0wa_KN} and Table~\ref{tab:w0wa_KN} for ET+VRO.\footnote{The results for  ET+THESEUS, obtained with 
\texttt{GWFAST}, have been double-checked using 
\texttt{TiDoFM}, finding broad agreement.}

{\em For {\rm ET+THESEUS}, we see that the differences in performances are not large between the various geometries: the {\rm 2L}~15km~$45^{\circ}$ and the
{\rm 2L}~20km~$45^{\circ}$ are quite comparable and slightly better than the 15-km triangle, while the 10~km triangle performs less well (e.g. by a factor 1.4 on $w_0$, with respect to the {\rm 2L}~15km~$45^{\circ}$).  For {\rm ET+VRO}
the situation is similar, with slightly larger differences (the 10~km triangle performs less well, with respect to the {\rm 2L}~15km~$45^{\circ}$, by a factor 2.2 on $w_0$).
We also  notice again the improvement brought by the LF instrument for the VRO+ET case, with results tighter on $w_0$ by a factor  $\sim (2.4-2.7)$, depending on the geometry.}

\paragraph{Modified GW propagation}\label{sect:Xi0}

As we have discussed, the DE equation of state characterizes the impact of dark energy on the background evolution and, as such, can also be studied with purely electromagnetic probes.  
Current {\em Planck-2018} analysis from CMB, combined with SNe and BAO, including both $w_0$ and $w_a$ in the inference, find $w_0=-0.961\pm 0.077$ and $w_a=-0.28^{+0.31}_{-0.27}$ ($68\%$ c.l.)~\cite{Planck:2018vyg}. 
While GW observation can provide an independent probe of these quantities, it is not evident that even 3G detectors will be able to improve quantitatively on these accuracies. As we see from Tables~\ref{tab:w0wa} and \ref{tab:w0wa_KN},
this is certainly not the case for the specific example of joint ET+THESEUS or ET+VRO observations that we are considering, where we are well above the $\sim 7\%$ accuracy on $w_0$ that we already have from CMB+BAO+SNe. For instance (recalling that in $\Lambda$CDM
$w_0=-1$, so the absolute errors on $w_0$ reported in Tables~\ref {tab:w0wa_KN} and \ref {tab:w0wa} are the same as the relative errors), for ET+VRO the configuration 2L-15km-$45^{\circ}$ reaches an accuracy on $w_0$ of $14\%$, while the 10~km triangle of $31\%$.

As emphasized in
\cite{Belgacem:2017ihm,Belgacem:2018lbp}, in the tensor sector
the situation is different, and potentially much more interesting,  for two reasons: (1) Modified GW propagation can only be tested from GW observations at large redshift, and therefore current limits (see below) are much broader. (2) There exist phenomenologically viable and theoretically motivated  modified gravity models~\cite{Maggiore:2013mea} that comply with all observational bounds at the background level and in the scalar perturbation sector (where, therefore, they differ from $\Lambda$CDM by at most a few percent level) and, nevertheless, in the tensor sector predict deviations from GR and $\Lambda$CDM that could be as large as $80\%$~\cite{Belgacem:2019lwx,Belgacem:2020pdz}.

The effect is due to a modification of the equation that governs the  propagation of tensor perturbations over FRW. In GR, this is given by 
\be\label{4eqtensorsect}
\tilde{h}''_A+2{\cal H}\tilde{h}'_A+c^2k^2\tilde{h}_A=0\, ,
\ee
where $\tilde{h}_A(\eta, \vk)$ is the Fourier-transformed GW amplitude,  $A=+,\times$ labels the two GW polarizations, the prime denotes the derivative with respect to cosmic time $\eta$ [which is  defined by $d\eta=dt/a(t)$], $a(\eta)$ is the FRW scale factor, and 
${\cal H}=a'/a$. 
Whenever gravity is modified on cosmological scales, unavoidably this propagation equation is also modified, as has been seen on many explicit 
examples~\cite{Saltas:2014dha,Lombriser:2015sxa,Amendola:2017ovw,Nishizawa:2017nef,Arai:2017hxj,Nishizawa:2019rra,Belgacem:2019lwx,Belgacem:2020pdz,LISACosmologyWorkingGroup:2019mwx}.
A change in  the coefficient of the $k^2\tilde{h}_A$ term induces a speed of GWs, $c_{\rm gw}$, different from that of light. After the observation of GW170817, this is  now excluded  at a level  $|c_{\rm gw}-c|/c< {\cal O}(10^{-15})$ \cite{Monitor:2017mdv}. However, the modified gravity models that pass this constraint still, in general, induce a change in the `friction term', so the propagation equation for tensor modes becomes, 
\be\label{prophmodgrav}
\tilde{h}''_A  +2 {\cal H}[1-\delta(\eta)] \tilde{h}'_A+c^2k^2\tilde{h}_A=0\, ,
\ee
for some function  $\delta(\eta)$ that encodes the modifications from GR.
In GR, using \eq{4eqtensorsect}, one finds that the GW amplitude decreases over cosmological distances as the inverse of the FRW scale factor.  From this one can show that, after propagation from the source to the  observer,   the amplitude of the GW from a compact binary coalescence is proportional to the inverse of the luminosity distance to the source (see, e.g.  Section 4.1.4 of \cite{Maggiore:1900zz}). This is  at the origin of the fact that compact binaries are standard sirens, i.e. that the luminosity distance of the source can be extracted from their gravitational signal. However, when the GW propagation is rather governed by \eq{prophmodgrav}, this result is affected. It can then be shown that the quantity extracted from GW observations is no longer the standard luminosity distance $d_L(z)$ of the source [that, in this context, we will denote by $\dem(z)$, since this is the quantity that would be measured, for instance, using the  electromagnetic signal from a counterpart]. Rather, the quantity extracted from GW observation is a  `GW luminosity distance'  $\dgw(z)$
\cite{Belgacem:2017ihm}, related to $\dem(z)$ by~\cite{Belgacem:2017ihm,Belgacem:2018lbp}
\be\label{dLgwdLem}
\dgw(z)=\dem(z)\exp\left\{-\int_0^z \,\frac{dz'}{1+z'}\,\delta(z')\right\}\, ,
\ee
where the function $\delta$ that appears in \eq{prophmodgrav} has now been written as a function of redshift. 

\begin{figure}[t]
	\centering
	\includegraphics[width=10cm]{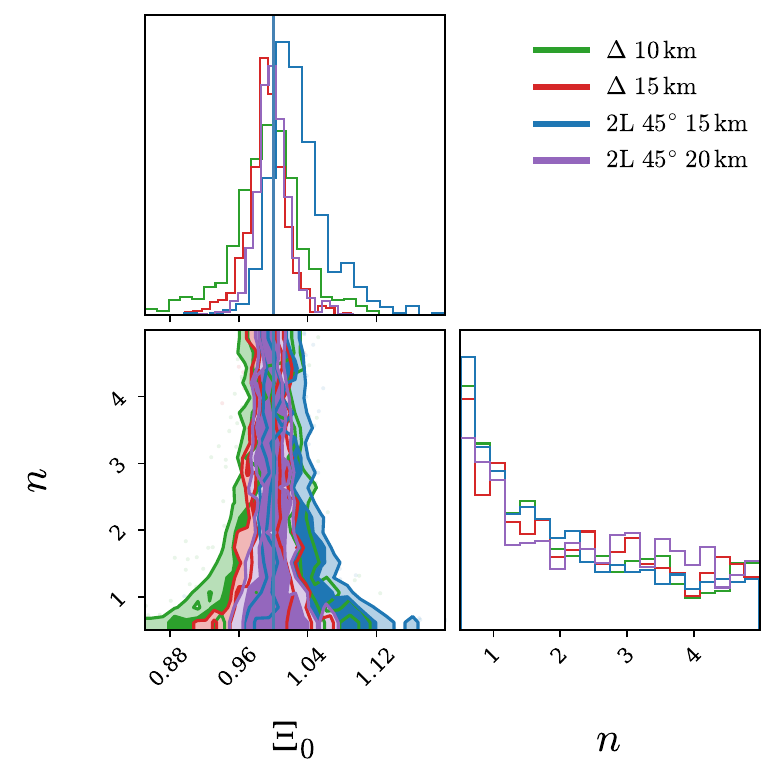}
	\caption{\small Reconstruction of the parameters $\Xi_0$ and $n$ in modified gravity, from the joint GW+EM events obtained with ET+THESEUS in 5 yr of observations, for the different geometries of ET shown, all with their HFLF-cryo sensitivity.}
\label{fig:corner_all_Xi0n}
\end{figure}

\begin{figure}[tp]
	\hspace{-1.65cm}
	\includegraphics[width=9.2cm]{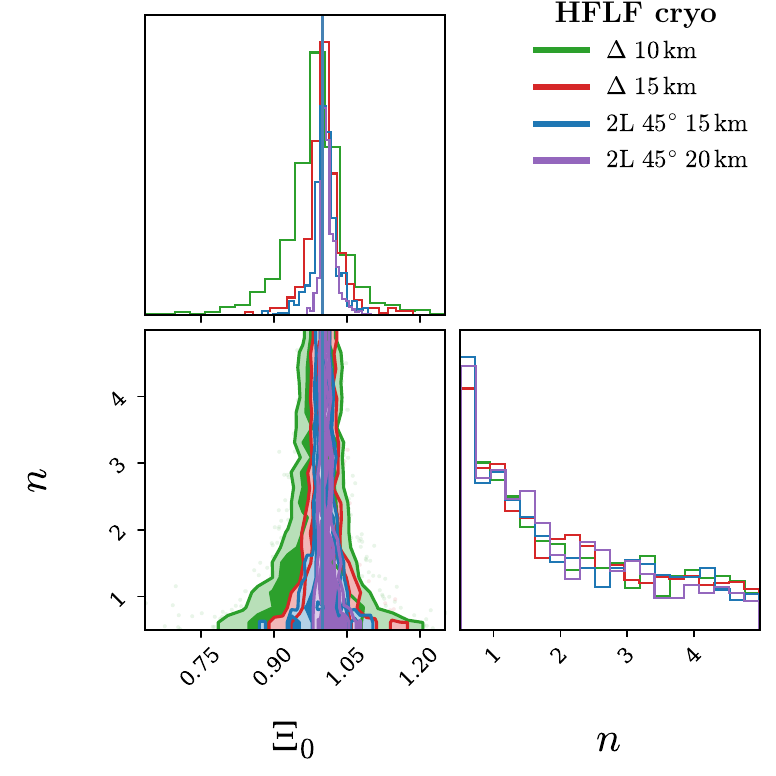}
	\hfill
	\includegraphics[width=9.2cm]{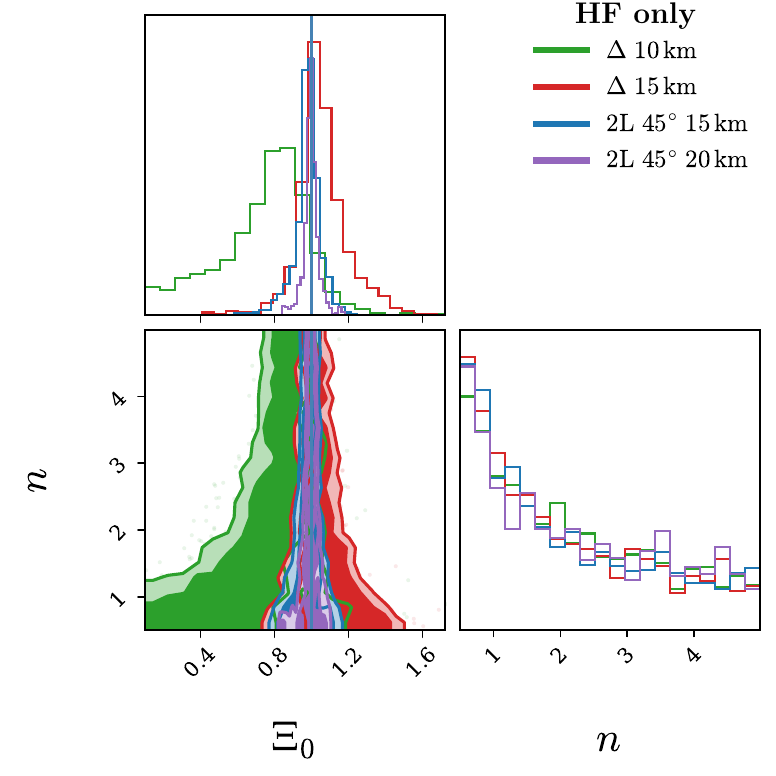}
	\caption{\small Reconstruction of the parameters $\Xi_0$ and $n$ in modified gravity, from the joint GW+EM events obtained with ET+VRO in 1 yr of observations, for the different geometries of ET shown, in the left panel with their HFLF-cryo sensitivity and in the right panel with the HF instrument only.}
\label{fig:corner_all_Xi0n_VRO}
\end{figure}

\begin{table}[ht]
\centering
\begin{tabular}{|l|c|c|}
\hline
\hline
Configuration       &  $\Delta \Xi_0/\Xi_0$   &  $\Delta n$                  \\  \hline \hline
$\Delta$-10km       &  $0.071$          & $2.87$             \\  \hline 
$\Delta$-15km       &  $0.040$          & $3.01$              \\  \hline
2L-15km-45$^\circ$  &  $0.049$          & $2.91$              \\  \hline  
2L-20km-45$^\circ$  &  $0.030$          & $3.12$              \\  \hline \hline 
\end{tabular}
\vspace{0.2cm}
\caption{\small Errors on the parameters $\Xi_0$ and $n$ that describe modified GW propagation (median and symmetric 68\% CI), from the joint GW+EM events obtained with ET+THESEUS, for the different geometries of ET shown, all with their HFLF-cryo sensitivity.}
\label{tab:Xi0n}
\end{table}

\begin{table}
\centering
\begin{tabular}{|l|c|c|}

\hline
\hline
\multicolumn{3}{|c|}{\textbf{HFLF cryogenic}}\\
\hline
Configuration       &  $\Delta \Xi_0/\Xi_0$ &  $\Delta n$  \\  \hline \hline
$\Delta$-10km       &  $0.097$          & $3.04$              \\  \hline 
$\Delta$-15km       &  $0.051$          & $2.95$              \\  \hline
2L-15km-45$^\circ$  &  $0.038$          & $2.95$              \\  \hline  
2L-20km-45$^\circ$  &  $0.027$          & $2.59$              \\  \hline \hline 
\end{tabular}
\hspace{.7cm}
\begin{tabular}{|l|c|c|}

\hline
\hline
\multicolumn{3}{|c|}{\textbf{HF only}}\\
\hline
Configuration       &  $\Delta \Xi_0/\Xi_0$ &  $\Delta n$  \\  \hline \hline
$\Delta$-10km       &  $0.617$          & $2.80$              \\  \hline 
$\Delta$-15km       &  $0.207$          & $2.87$              \\  \hline
2L-15km-45$^\circ$  &  $0.108$          & $2.88$              \\  \hline  
2L-20km-45$^\circ$  &  $0.063$          & $3.00$              \\  \hline \hline 
\end{tabular}
\vspace{0.2cm}
\caption{\small Errors on the parameters $\Xi_0$ and $n$ that describe modified GW propagation (median and symmetric 68\% CI), from the joint GW+EM events obtained with ET+VRO, for the different geometries of ET shown, in the left table with their HFLF-cryo sensitivity and in the right table with the HF instrument only.}
\label{tab:Xi0n_KN}
\end{table}

Similarly to what we have discussed for the DE equation of state, inference on a full function of redshift such as $\d(z)$ is difficult (although it can be performed, to some extent, with the technique of Gaussian process reconstruction~\cite{Belgacem:2019zzu}), and it is convenient to introduce a parametrization. 
A very convenient choice, in terms of two parameters $(\Xi_0,n)$, has been proposed in \cite{Belgacem:2018lbp}. Rather than parametrizing $\delta(z)$, it is simpler to parametrize directly the ratio $\dgw(z)/\dem(z)$ (which is also the directly observed quantity), in the form
\be\label{eq:fit}
\frac{d_L^{\,\rm gw}(z)}{d_L^{\,\rm em}(z)}=\Xi_0 +\frac{1-\Xi_0}{(1+z)^n}\, .
\ee
This parametrization reproduces  the fact that, at $z= 0$, $d_L^{\,\rm gw}/d_L^{\,\rm em}=1$ since,  as the distance to the source goes to zero, there can be no effect from modified propagation. 
In the opposite limit of large redshifts, in contrast, \eq{eq:fit} predicts that $d_L^{\,\rm gw}/d_L^{\,\rm em}$ approaches a constant value $\Xi_0$. This is motivated by the fact that,
in  typical  modified gravity  models, the deviations from GR only appear in the recent cosmological epoch, so $\delta(z)$ goes to zero at large redshift, and  the integral in \eq{dLgwdLem}
saturates to a constant. The parametrization (\ref{eq:fit}), 
interpolates between these two limiting behaviors, with a power-law determined by $n$, and GR corresponds to $\Xi_0=1$ (for any $n$).
This simple parametrization has been shown to work remarkably well for practically all best-studied modified gravity models~\cite{LISACosmologyWorkingGroup:2019mwx}. A particularly interesting example is given by the model proposed in \cite{Maggiore:2013mea} (see \cite{Maggiore:2016gpx,Belgacem:2020pdz} for reviews) that, while complying with all current cosmological constraints,  predicts modified GW propagation with a value of $\Xi_0$ that 
(depending on a single parameter related to the initial conditions of the cosmological evolution) 
can differ from the GR value $\Xi_0=1$ by an amount  between a few percent up to a value
$\Xi_0\simeq 1.8$~\cite{Belgacem:2019lwx,Belgacem:2020pdz},  that would correspond to a $80\%$ deviation from GR.
Limits on $\Xi_0$ have already been obtained from the current set of LVC detections, either from GW170817 as a standard siren with counterpart~\cite{Belgacem:2018lbp}, or using  BBH dark sirens from the O1, O2 and O3a LIGO/Virgo run, and performing a correlation  with  galaxy catalogs. This results in the value $\Xi_0=2.1^{+3.2}_{-1.2}$~\cite{Finke:2021aom}.\footnote{An even more stringent result is obtained if one accepts the  tentative  identification of the flare  ZTF19abanrhr as the electromagnetic counterpart of the BBH coalescence GW190521. Then, the analysis in \cite{Finke:2021aom} gives  
$\Xi_0=1.8^{+0.9}_{-0.6}$. 
($68\%$ c.l.). This is consistent with similar results obtained in \cite{Mastrogiovanni:2020mvm}.} Using instead a joint cosmology-population  analysis  that exploits the mass scales in the BBH mass function~\cite{Farr:2019twy,Ezquiaga:2021ayr} and the GWTC-3 catalog of detections \cite{LIGOScientific:2021djp} one finds
the  constraints $\Xi_0 = 1.2\pm 0.7$~\cite{Mancarella:2021ecn}.

Here we show the constraint on $\Xi_0$ that can be obtained from the joint set of GW and electromagnetic observations obtained from ET+THESEUS and ET+VRO. As we have stressed at the beginning of this section, the absolute estimates for the accuracy that we will present should not be understood as the forecast for the overall capability of ET for measuring these cosmological parameters, since we are 
studying the information that could be obtained 
from two examples, namely the joint detections of ET+THESEUS and ET+VRO,
and, as discussed at the beginning of this section, there are 
other ways of obtaining information on $\Xi_0$ from the ET data. Again,  here we are mainly interested  in the relative performances of the different ET configurations on some concrete examples.

Fig.~\ref{fig:corner_all_Xi0n} and Table~\ref{tab:Xi0n} show the results for the parameters $\Xi_0$ and $n$ that characterize modified GW propagation from the ET+THESEUS joint detections, while Fig.~\ref{fig:corner_all_Xi0n_VRO}, and Table~\ref{tab:Xi0n_KN} show the results for ET+VRO. {\em The hierarchy of performances is  similar to that for $H_0$, with the 2L~20~km providing the best results, followed by the 15~km triangle and the 2L~15~km with similar performances (with 2L-15~km better than triangle-15km for ET+VRO), and the 10~km triangle providing the less good performances. For ET+VRO, the huge improvement as a consequence of the inclusion of the LF instrument is also apparent.}

Observe that, in view of the discussion above, the accuracy that can be obtained from $\Xi_0$ even from these two specific examples, which ranges between $3\%$ and $10\%$ depending on the geometry with the full sensitivity curve, is already very interesting in absolute terms, since current $1\sigma$ limits are at the $60\%-100\%$ level, and furthermore there are models that predicts a potentially much higher signal.

\subsubsection{Hubble parameter and dark energy from BNS tidal deformability}\label{sect:H0BNStidal}

 BNS have an intrinsic mass scale and can only exist in a narrow range of masses. This mass scale is imprinted in the tidal interaction between the component NSs. Therefore, if the nuclear EoS is known, one can determine the source-frame masses by a measurement of the tidal deformability. This, in turn, would allow the measurement of the redshift directly from a GW observation because it is the redshifted mass that is inferred from the point-particle approximation of the waveform. Such a method was first proposed in~\cite{Messenger:2011gi} and further explored in~\cite{Messenger:2013fya,Li:2013via}. A measurement of the Hubble constant using a known relationship between the tidal parameter and source-frame mass was explored in~\cite{DelPozzo:2015bna,Chatterjee:2021xrm,Shiralilou:2022urk} while~\cite{Ghosh:2022muc} showed that one can simultaneously estimate both the nuclear EoS and the Hubble constant using future observatories. A measurement of the dark energy EoS was explored in~\cite{Wang:2020xwn,Jin:2022qnj}. 

In this section, we explore the potential of different ET configurations to constrain the expansion history of the Universe assuming that the nuclear EoS is known. It is found in \cite{Chatterjee:2021xrm} that up to a 15\% uncertainty in the knowledge of the EoS does not affect the measurement of the Hubble constant in a meaningful manner. We use the TaylorF2 waveform model augmented with the 5PN and 6PN tidal terms in the phase, terminating the signal at the ISCO frequency corresponding to the total mass of the binary. In Section.~\ref{sec:tidaleffects}, it was found that the results using this model versus an inspiral-merger-ringdown (\acrshort{imr}) waveform gives similar results. Since the analysis in this section is along similar lines, we expect the same to hold here too.
Additionally, we assume the DD-LZ1 EoS for the NS for reasons elucidated in Section~\ref{sec:Fisher_nuclear}. We fit the logarithm (base 10) of the tidal deformability as a function of the mass of the NS using a fourth-order polynomial given by
\begin{equation}
    \log_{10}\Lambda(m) = -0.3550 \, m^4 + 2.162 \, m^3 - 4.652 \, m^2 + 2.514 \, m + 3.892\, ,
\end{equation}
where $m$ is in units of $\msun$.
We verify that the fit reproduces the slope of the curve accurately with maximum errors at a few percent. This is crucial because it is the slope of the curve that contributes to the Fisher errors on the redshift. 

The Fisher errors from the $d_L$--$z$ space are then propagated to the space of cosmological parameters, $\Vec{\phi}$, via another Fisher matrix given by~\cite{Heavens:2014xba}
\begin{equation}
\label{eq:cosmo_fisher}
    \mathcal{G}_{ij} = \sum_{k=1}^N \frac{1}{\sigma_{d_L,k}^2} \frac{\partial d_L^k(z)}{\partial \phi^i} \frac{\partial d_L^k(z)}{\partial \phi^j} \,,
\end{equation}
where $N$ is the total number of observations in the catalog and $\sigma_{d_L,k}^2$ is the total luminosity distance error for the k-th event given by
\begin{equation}
    (\sigma_{d_L})^2 = (\sigma_{d_L}^{h})^2 + (\sigma_{d_L}^{z})^2.
\end{equation}
Here, $\sigma_{d_L}^{h}$ is the contribution to the luminosity distance error due to the error in the GW amplitude while $\sigma_{d_L}^{z}$ is that due to the error in the redshift measurement, given by
\begin{equation}
\label{eq:errz_dl}
    \sigma_{D_L}^{z} = \left|\frac{\partial D_L}{\partial z}\right| \sigma_{z}.
\end{equation}
In writing \eq{eq:cosmo_fisher}, we have neglected the correlations in the $d_L$--$z$ space for simplicity.

\paragraph{Hubble constant.}

\begin{figure*}
    \includegraphics[width=8cm]{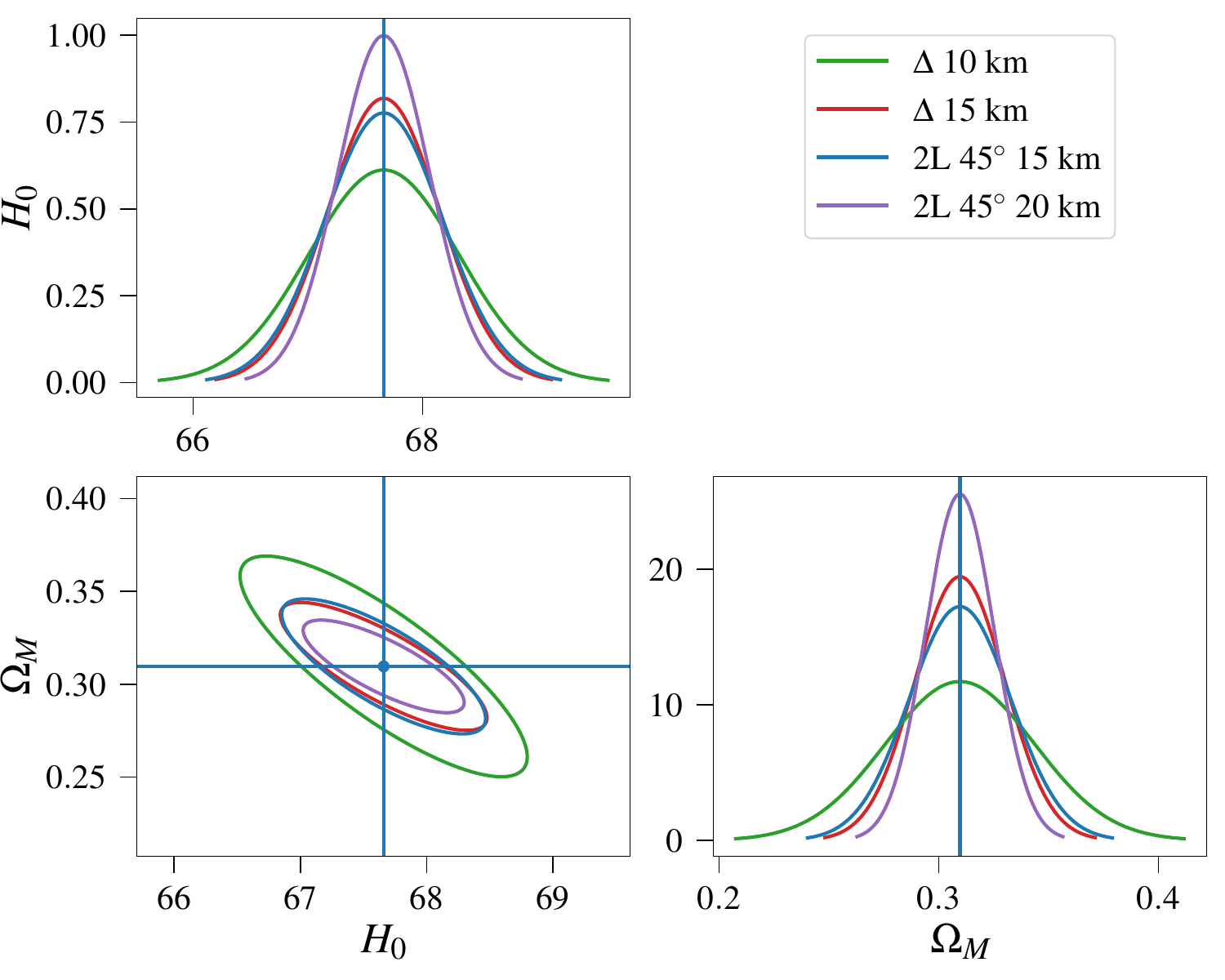}
    \includegraphics[width=8cm]{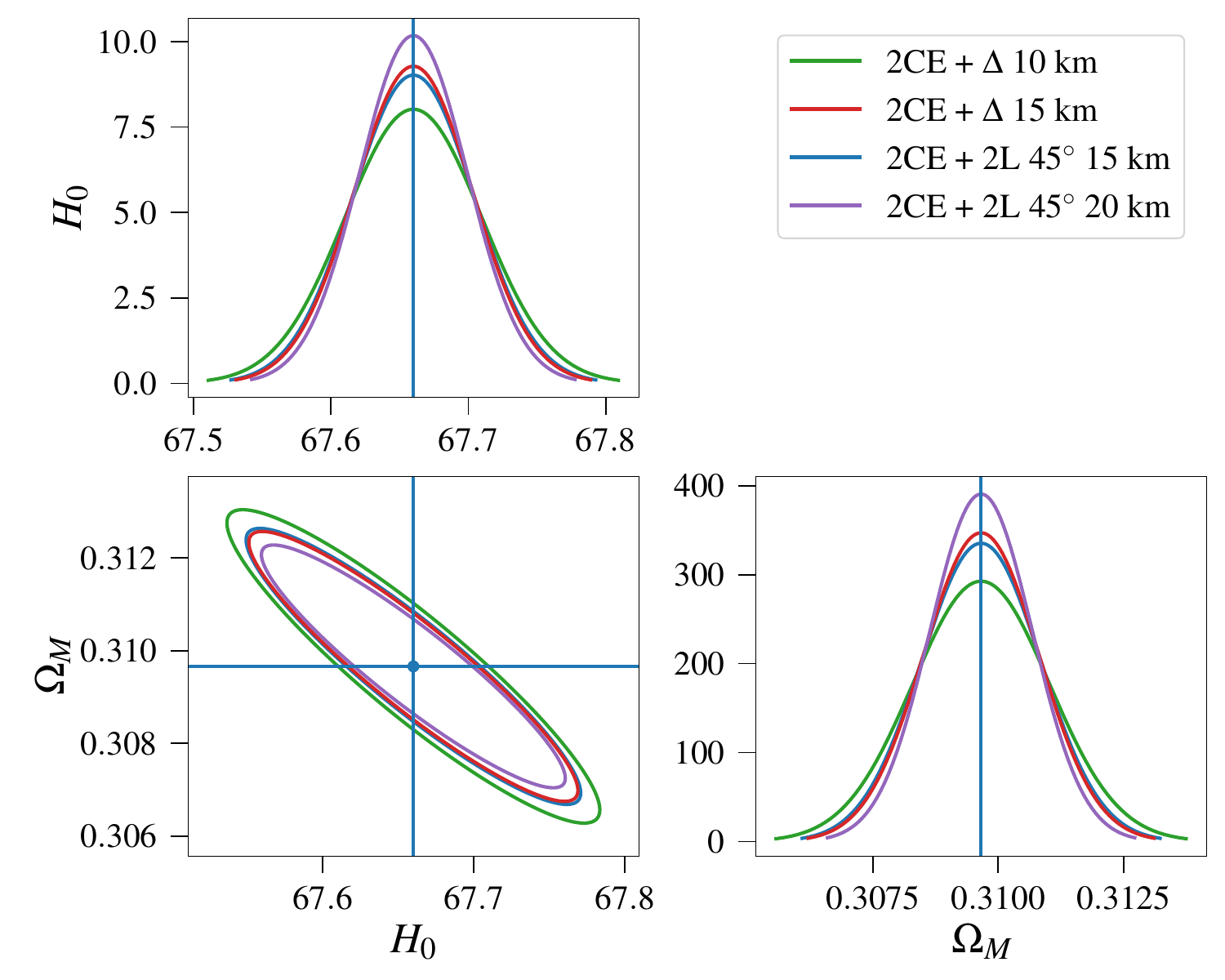}
    \caption{\small Constraints on the parameters $H_0$ and $\Omega_{M}$ in $\Lambda$CDM model using one year GW observations from BNS alone for different ET geometries. The covariance for a standalone ET is shown in the left panel while ET in a network with  2CE is shown in the right panel. $H_0$ is measured in ${\rm km}\, {\rm s}^{-1}\, {\rm Mpc}^{-1}$.}
    \label{fig:lcdm}
\end{figure*}

The results for $H_0$ and $\Omega_M$ are shown in Fig.~\ref{fig:lcdm} and Tab.~\ref{tab:lcdm}. A description of $\Lambda$CDM cosmology is given in Section~\ref{sect:H0}. We see that a single year of observing run can achieve sub-percent errors on $H_0$ with ET alone, albeit the errors improve by an order of magnitude when ET operates in a network with 2CE. The ET configuration with two 20km L-shaped detectors at $45^{\circ}$ gives the best constraints. Similar trends follow for the dark matter energy density parameter, $\Omega_M$, which can be constrained at percent level precision, except for the case of a standalone 10km triangular ET which can only achieve an accuracy of $\sim10\%$.

\begin{table*}[t]
\begin{tabular}{|l|c|c|}
\hline
\hline
Configuration       &  $\Delta H_0/H_0$   &  $\Delta \Omega_M/\Omega_M$                  \\  \hline \hline
$\Delta$-10km       & $9.63\times10^{-3}$ & $1.10\times10^{-1}$ \\  \hline 
$\Delta$-15km       & $7.20\times10^{-3}$ & $6.62\times10^{-2}$  \\  \hline
2L-15km-45$^\circ$  & $7.59\times10^{-3}$ & $7.47\times10^{-2}$  \\  \hline    
2L-20km-45$^\circ$  & $5.90\times10^{-3}$ & $5.04\times10^{-2}$  \\  \hline
\end{tabular}
\quad
\begin{tabular}{|l|c|c|}
\hline
\hline
Configuration       &  $\Delta H_0/H_0$   &  $\Delta \Omega_M/\Omega_M$                  \\  \hline \hline
$\Delta$-10km  +2CE     & $7.35\times10^{-4}$ & $4.40\times10^{-3}$ \\  \hline 
$\Delta$-15km  +2CE      & $6.35\times10^{-4}$ & $3.71\times10^{-3}$  \\  \hline
2L-15km-45$^\circ$ +2CE  & $6.54\times10^{-4}$ & $3.84\times10^{-3}$  \\  \hline   
2L-20km-45$^\circ$ +2CE  & $5.79\times10^{-4}$ & $3.30\times10^{-3}$  \\  \hline \hline
\end{tabular}
\vspace{0.2cm}
\caption{\small Standard deviation on the parameters $H_0$ and $\Omega_M$ in $\Lambda$CDM using one year of GW observations from BNS alone for different geometries of ET alone (left) and ET in a network with 2CE (right). ET is always taken with the full HFLF-cryo sensitivity.}
\label{tab:lcdm}
\end{table*}

\paragraph{Dark energy equation of state.}

\begin{figure*}
    \includegraphics[width=8cm]{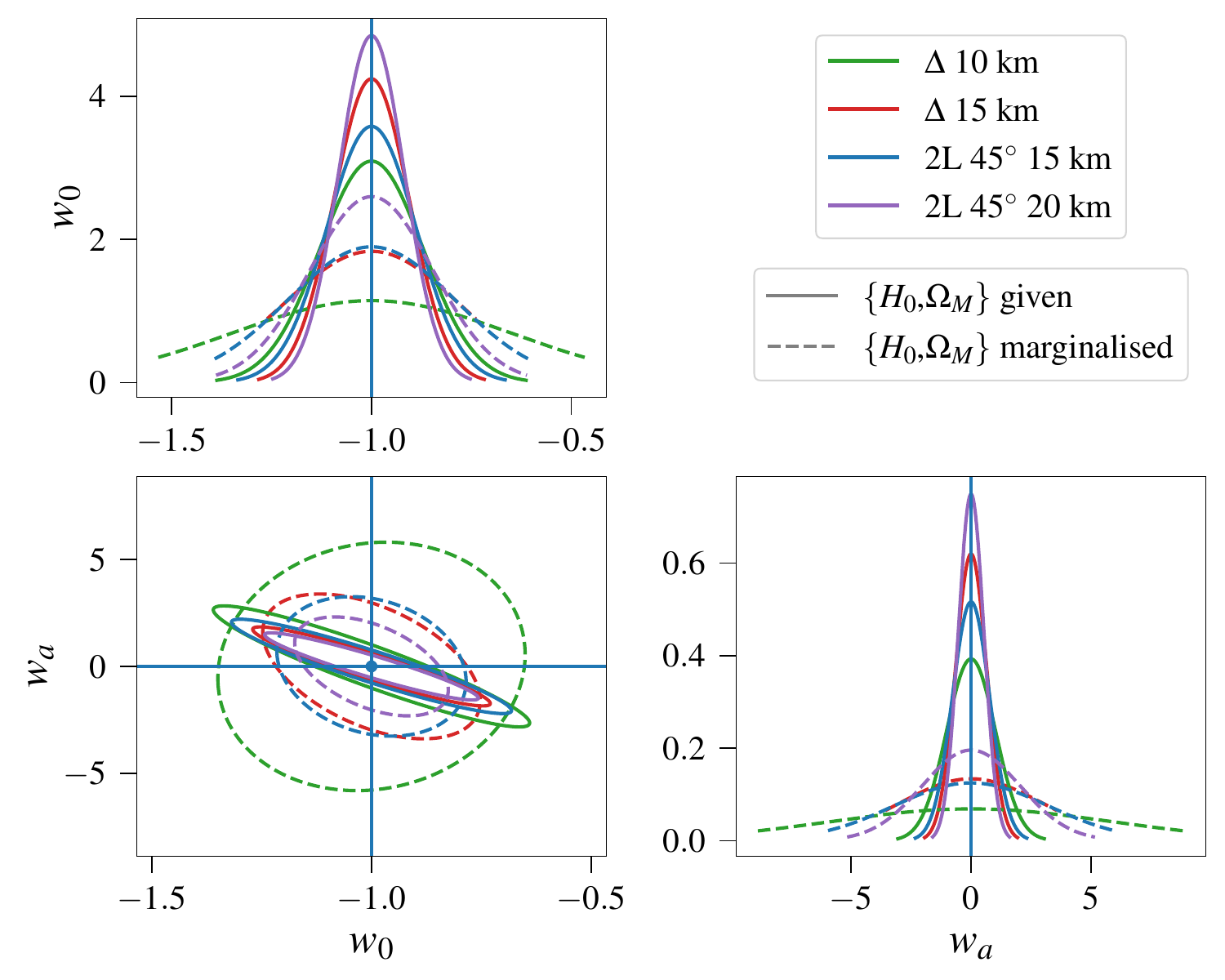}
    \includegraphics[width=8cm]{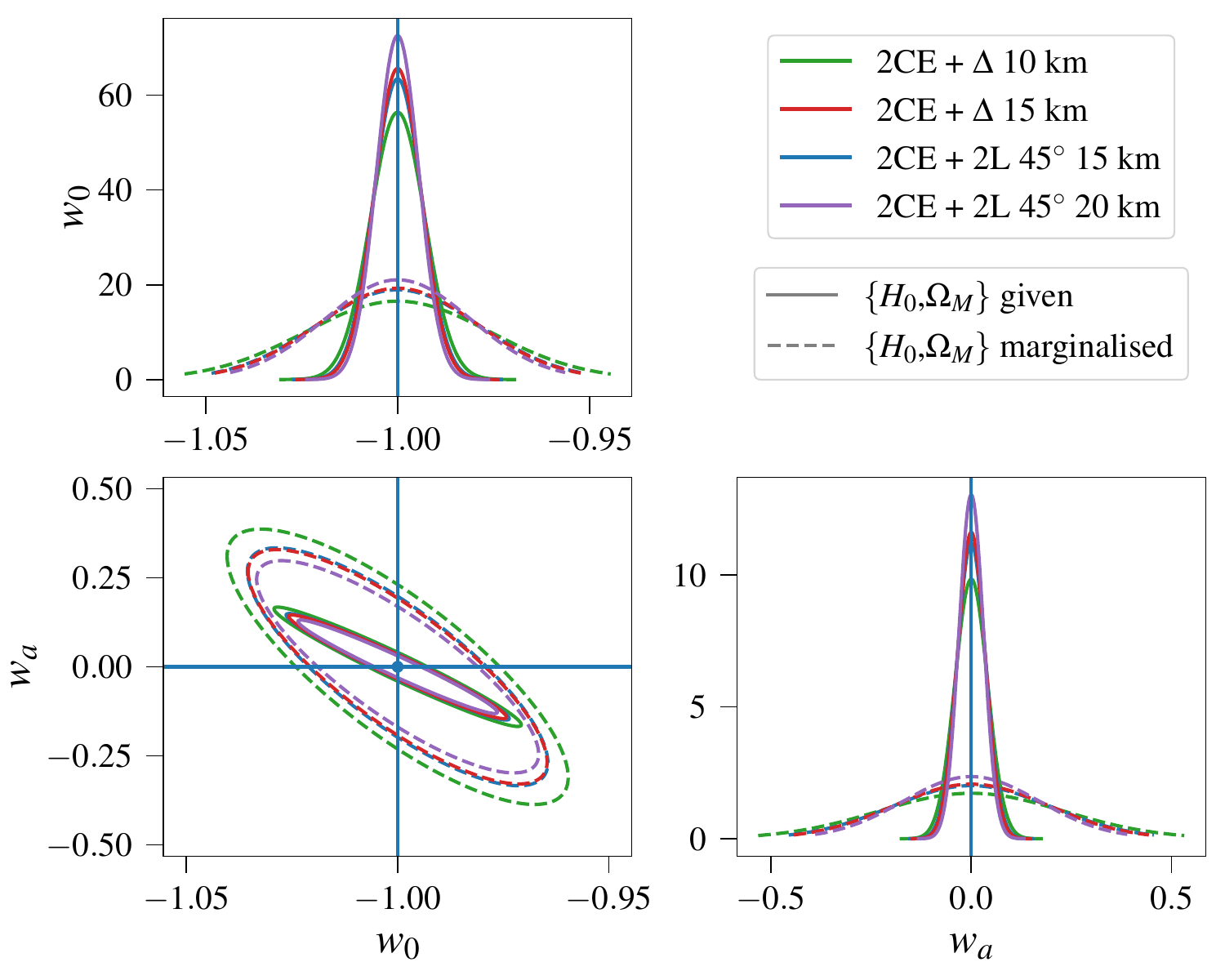}
    \caption{\small Constraints on the $w_0$--$w_a$ plane in a dynamical dark energy model using one year GW observations from BNS alone for different ET geometries. The covariance for a standalone ET is shown in the left panel while ET in a network with 2CE is shown in the right panel.}
    \label{fig:w0wacdm}
\end{figure*}

The results for the dark energy equation of state parameters are shown in Fig.~\ref{fig:w0wacdm} and the $1\sigma$ errors tabulated in Tab.~\ref{tab:w0wacdm}. A description of the dark energy model considered here is in Section~\ref{sect:DEEoS}. We see that the 2L 20km at $45^{\circ}$  obtains the best constraints on the parameters, albeit the EoS parameter $w_a$ is not constrained by a standalone ET if one marginalises over the $\Lambda$CDM parameters. In such a scenario, it is common to fix the $\Lambda$CDM parameters to that measured with EM probes and measure only the dark energy parameters using GWs. In this case, except for the single 10km triangular ET configuration, the EoS parameters can be constrained.

\begin{table}[t]
\begin{tabular}{|l|c|c|}
\hline
\hline
Configuration       &  $\Delta w_0$   &  $\Delta w_a$                  \\  \hline \hline
$\Delta$-10km       & $0.348$ & $5.79$ \\  \hline 
$\Delta$-15km       & $0.217$ & $2.98$ \\  \hline
2L-15km-45$^\circ$  & $0.210$ & $3.19$ \\  \hline  
2L-20km-45$^\circ$  & $0.153$ & $2.04$ \\  \hline \hline 
\end{tabular}
\quad\quad
\begin{tabular}{|l|c|c|}
\hline
\hline
Configuration         &  $\Delta w_0$   &  $\Delta w_a$             \\  \hline \hline
$\Delta$-10km  +2CE      & $0.0241$ & $0.231$ \\  \hline 
$\Delta$-15km  +2CE      & $0.0207$ & $0.193$ \\  \hline
2L-15km-45$^\circ$ +2CE  & $0.0211$ & $0.198$ \\  \hline  
2L-20km-45$^\circ$ +2CE  & $0.0189$ & $0.169$ \\  \hline
\hline 
\end{tabular}

\begin{tabular}{|l|c|c|}
\hline
\hline
Configuration       &  $\Delta w_0$   &  $\Delta w_a$                  \\  \hline \hline
$\Delta$-10km       & $0.129$ & $1.02$ \\  \hline 
$\Delta$-15km       & $0.0940$ & $0.644$ \\  \hline
2L-15km-45$^\circ$  & $0.111$ & $0.773$ \\  \hline  
2L-20km-45$^\circ$  & $0.0823$ & $0.532$ \\  \hline
\hline 
\end{tabular}
\quad
\begin{tabular}{|l|c|c|}
\hline
\hline
Configuration       &  $\Delta w_0$   &  $\Delta w_a$                  \\  \hline \hline
$\Delta$-10km  +2CE      & $7.07\times10^{-3}$ & $0.0405$ \\  \hline 
$\Delta$-15km  +2CE      & $6.08\times10^{-3}$ & $0.0343$ \\  \hline
2L-15km-45$^\circ$ +2CE  & $6.29\times10^{-3}$ & $0.0356$ \\  \hline  
2L-20km-45$^\circ$ +2CE  & $5.50\times10^{-3}$ & $0.0306$ \\  \hline \hline 
\end{tabular}
\vspace{0.2cm}
\caption{\small Errors on the parameters $w_0$ and $w_a$ ($68\%$ c.l.) in a dynamical dark energy model using one year of GW observations from BNS alone for different geometries of ET alone (left) and ET in a network with 2CE (right), all with their HFLF-cryo sensitivity. The top panel shows the constraints marginalised over the parameters $H_0$ and $\Omega_M$ whereas, in the bottom panel, the $\Lambda$CDM parameters are assumed to be known a priori.}
\label{tab:w0wacdm}
\end{table}

\paragraph{Modified GW propagation.}

\begin{figure*}[t]
    \includegraphics[width=8cm]{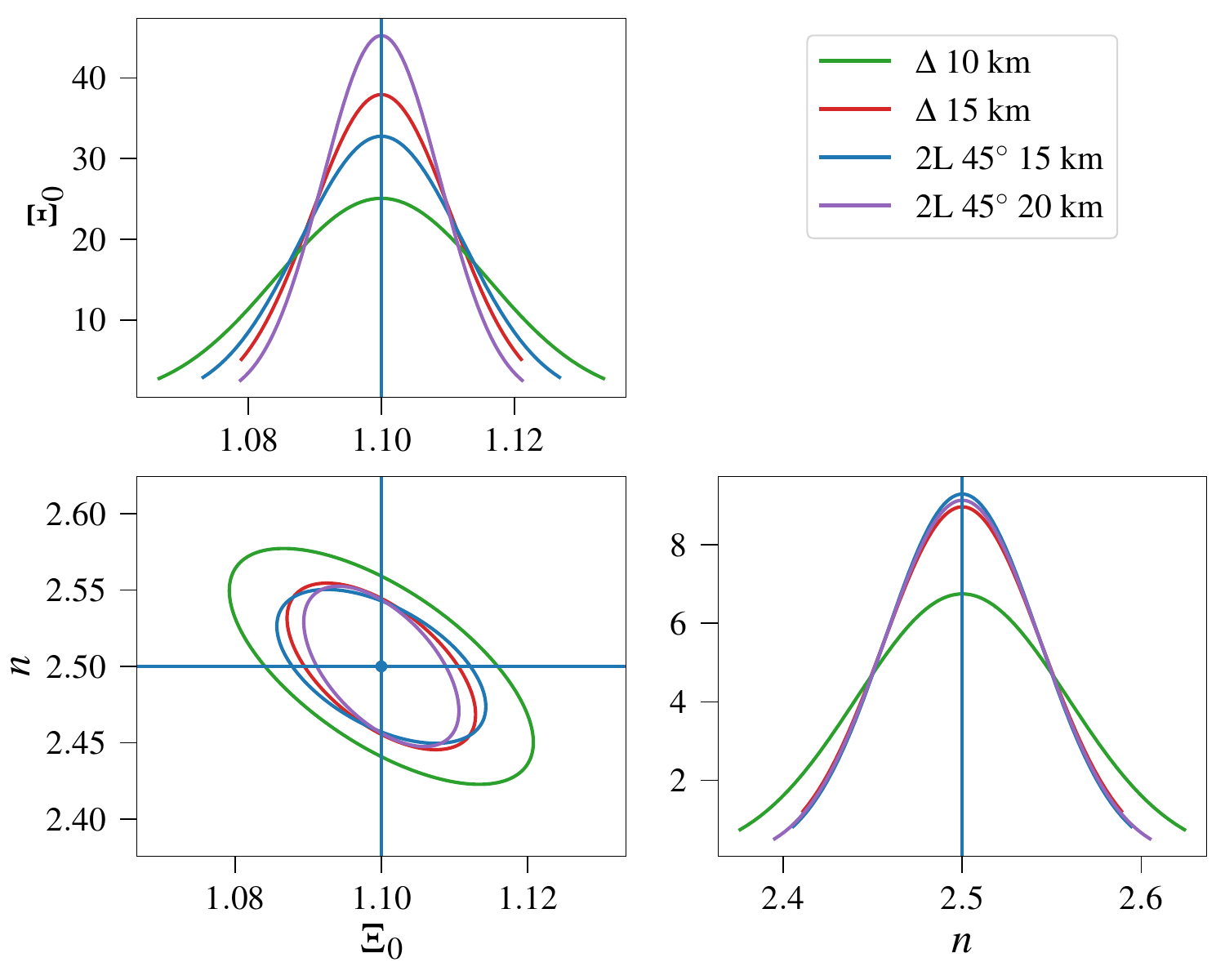}
    \includegraphics[width=8cm]{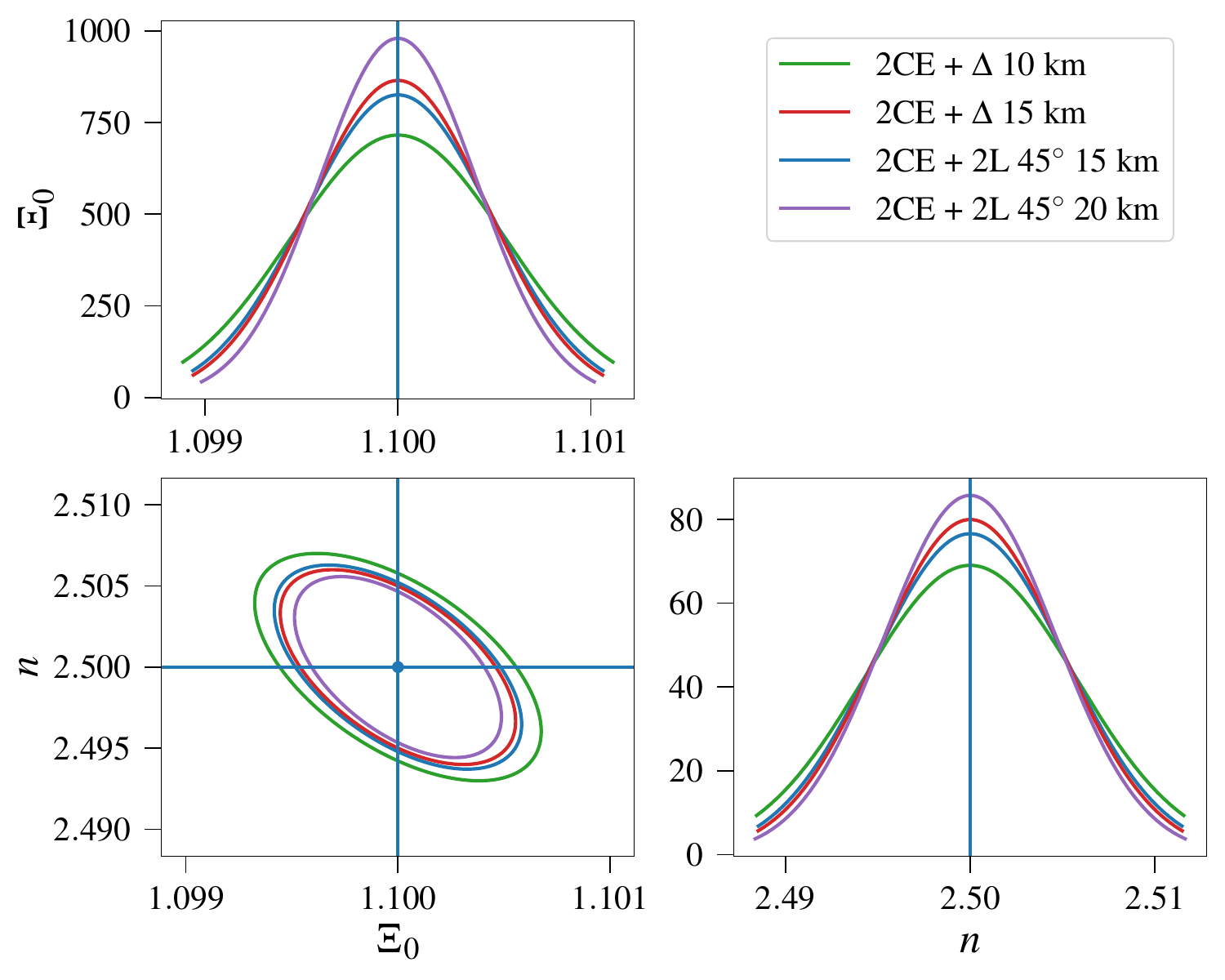}
    \caption{\small Constraints on the $\Xi_0$--$n$ plane in a modified gravity model with modified tensor perturbations using one year GW observations from BNS alone for different ET geometries. The covariance for a standalone ET is shown in the left panel while ET in a network with  2CE is shown in the right panel. The $\Lambda$CDM parameters are assumed to be known a priori.}
    \label{fig:xi0ncdm}
\end{figure*}

We next consider modified gravitational wave propagation, using the parametrization (\ref{eq:fit}).
The bounds on the corresponding parameters $\Xi_0$ and $n$ are shown in Fig.~\ref{fig:xi0ncdm} and the corresponding error (at $68\%$ c.l.) are tabulated in Tab.~\ref{tab:xi0ncdm}. A  difference from the analysis of the previous section using EM counterparts is that we assume fiducial values of $\Xi_0=1.1$ and $n=2.5$. As described in Section~\ref{sect:Xi0}, these values are consistent with all current observations. We do this because the form of the parameterization (\ref{eq:fit}) renders the Fisher matrix (\ref{eq:cosmo_fisher}) singular for GR value of $\Xi_0=1$ and for $n=0$. Any value of $n$ away from 
$n=0$ is sufficient to make the Fisher matrix well-defined. We choose $n=2.5$ for reasons outlined in~\cite{Belgacem:2019tbw}. We  see that the hierarchy among the configurations
is analogous to that found in Table~\ref{tab:Xi0n}, with the
2L~20km with arms at $45^{\circ}$  providing the best results, followed by 
2L~15km with arms at $45^{\circ}$ and triangle~15km, that gives quite similar results, 
while the triangular 10km ET gives the largest errors. 

\begin{table}[t]
\begin{tabular}{|l|c|c|}
\hline
\hline
Configuration       &  $\Delta \Xi_0/\Xi_0$   &  $\Delta n$                  \\  \hline \hline
$\Delta$-10km       & $1.06\times10^{-2}$ & $5.91\times10^{-2}$ \\  \hline 
$\Delta$-15km       & $7.01\times10^{-3}$ & $4.45\times10^{-2}$ \\  \hline
2L-15km-45$^\circ$  & $8.11\times10^{-3}$ & $4.29\times10^{-2}$ \\  \hline  
2L-20km-45$^\circ$  & $5.88\times10^{-3}$ & $4.37\times10^{-2}$ \\  \hline \hline 
\end{tabular}
\quad
\begin{tabular}{|l|c|c|}
\hline
\hline
Configuration       &  $\Delta \Xi_0/\Xi_0$   &  $\Delta n$                  \\  \hline \hline
$\Delta$-10km  +2CE      & $3.71\times10^{-4}$ & $5.78\times10^{-3}$ \\  \hline 
$\Delta$-15km   +2CE     & $3.07\times10^{-4}$ & $4.99\times10^{-3}$ \\  \hline
2L-15km-45$^\circ$  +2CE & $3.22\times10^{-4}$ & $5.21\times10^{-3}$ \\  \hline  
2L-20km-45$^\circ$  +2CE & $2.71\times10^{-4}$ & $4.99\times10^{-3}$ \\  \hline \hline 
\end{tabular}
\vspace{0.2cm}
\caption{\small Errors (at $68\%$ c.l.) on the parameters $\Xi_0$ and $n$  that describe modified GW propagation, using one year of GW observations from BNS alone for different geometries of ET alone (left) and ET in a network with  2CE (right), all with their HFLF-cryo sensitivity. The $\Lambda$CDM parameters are assumed to be known a priori.}
\label{tab:xi0ncdm}
\end{table}

\subsubsection{Hubble parameter from high-mass ratio events}\label{sect:highmassevents}

A main limitation for the reconstruction of the luminosity distance $d_L$ is
the fact that it is highly correlated with the angle $\iota$ between the binary's orbital angular momentum and the observer's  line-of-sight.\footnote{For precessing binaries $\iota$ would be a function of time since the orbital angular momentum precesses about the total angular momentum, which includes the companions' spin angular momenta. For such systems, $\iota$ should be defined at some fixed, but arbitrary, instant of time.}  Purely based on the signal-to-noise ratio of an event, for large SNR we should expect that the luminosity distance $d_L$ of a source is measured to an accuracy of $\Delta d_L/d_L = 1/{\rm SNR}.$ However, 
this limit is never reached since this correlation (as well as, to lesser extent, the correlation with sky location) degrades the accuracy, typically leading to error in the distance measurement an order of magnitude larger \cite{Ajith:2009fz} (unless the  angle $\iota$ and the sky location of the source can be fixed by the observation of an associated GRB, see the discussion  on page~\pageref{foot:2suSNR}). For instance, when the inclination angle changes from 0 to $\pi/6$ radians, the relative amplitudes of the $h_+$ and $h_\times$ polarizations of the dominant quadrupole mode change by no more than 1\%,  while the overall amplitude reduces by 12\%. This change in amplitude can be entirely compensated by moving the binary 12\% closer to the observer without an appreciable change in the polarization content of the observed signal.

This scenario changes dramatically for high-mass ratio events from either neutron star-black hole or double black hole binaries. Significant energy in the higher-order modes can break the distance-inclination angle degeneracy since the polarization modes have a different dependence on the inclination angle.  In a recent paper  \cite{Borhanian:2020vyr} it has been shown that high-mass ratio events with large SNR would not only break the $d_L$-$\iota$ degeneracy, but a small population of such events would also localize the source to a region small enough to contain, on average, just one $L_*$ galaxy. This helps in obtaining the redshift of the source from the galaxy, although the source itself may not have an EM afterglow. As already mentioned, these events are called \textit{dark sirens} and can measure the Hubble constant without the need for an extra distance calibrator. Some of these events, albeit very rare, could measure the Hubble constant to an accuracy of better than 2\% with a single event with no known systematics and are called \emph{golden $H_0$ events.} This method could be one of the most \emph{precise} and \emph{accurate} way of measuring the Hubble constant. 

\begin{figure}
    \centering
    \includegraphics[width=\textwidth]{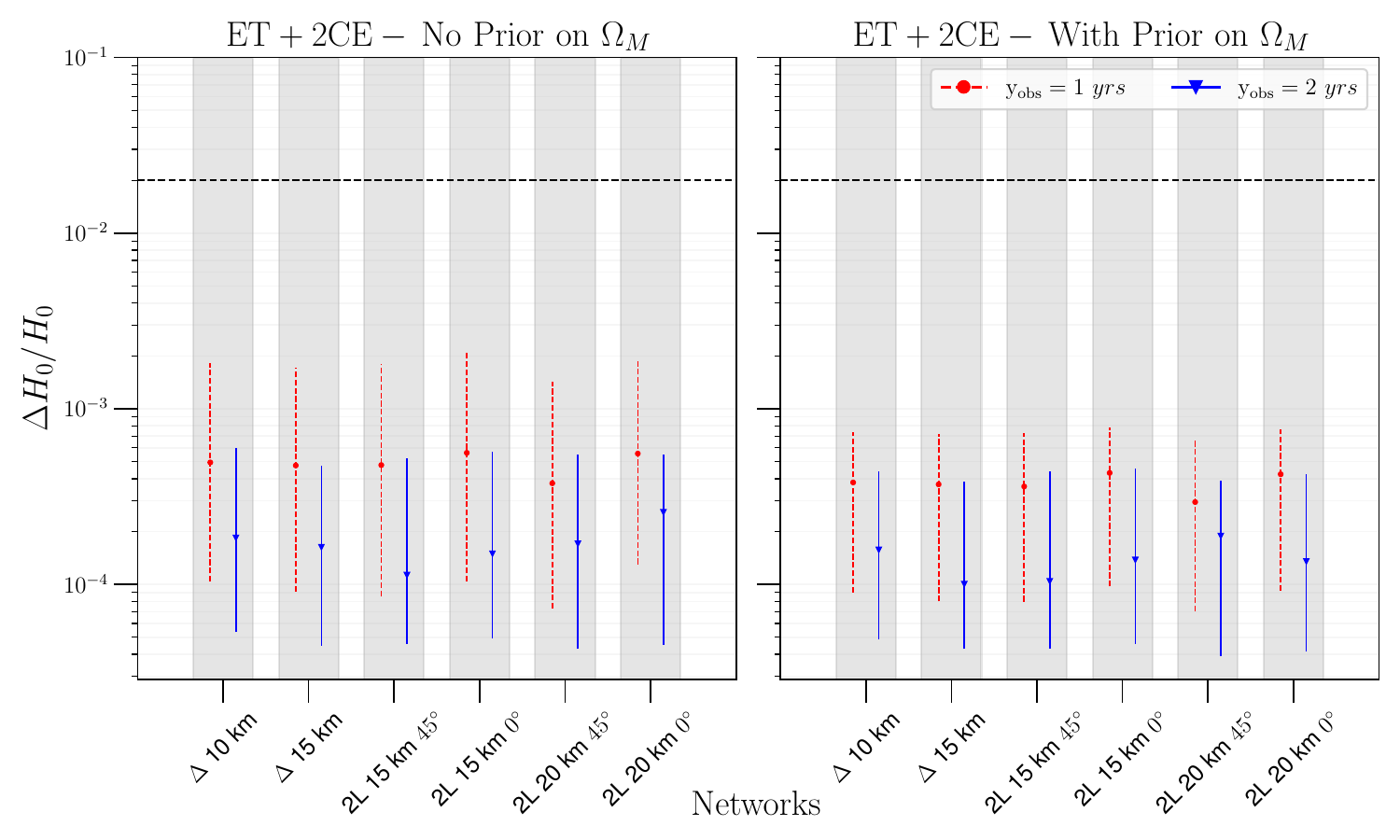}
    \caption{\small The accuracy with which the Hubble constant can be measured by different detector networks by high-mass ratio binary black holes, 
    with events corresponding to 1 and 2 yrs of observation time picked randomly from the 10-year catalog of BBH events located within $z = 0.1$. In the left plot, $H_0$ is measured with no prior imposed on the dark matter energy density $\Omega_M$, while for the right plot we assume  a gaussian prior with a width of 0.017. The markers show the median value of the fractional error in $H_0$ and the error bars denote the $68\%$ confidence region.}
    \label{fig:H0 dark sirens}
\end{figure}

Since BBHs are not likely to be associated with an EM counterpart, the only way to identify the true host is to localize the source to a single galaxy. For NSBH events, however, we expect EM afterglows if the mass ratio is not too small (say, greater than about 1:4). While asymmetric masses in a NSBH binary could help break the distance-inclination degeneracy, the host galaxy could be identified from the kilonova. Additionally, for NSBH binaries with large mass asymmetry, it is also possible to identify the host within the sky localization from gravitational wave observations. In this sense, NSBH binaries could be used as both bright and dark sirens or \emph{gray} sirens.

For BBH systems, we restrict our population to events within the redshift  $z=0.1$, or $476$ Mpc. This leaves us with $86$ BBH mergers in a span of $10$ years. Among these, we choose those events for which the sky area is constrained to better than $0.04$ $\mbox{deg}^2$, as we expect only one $L_{*}$ galaxy to lie in this sky patch up to a redshift of 0.1 \cite{Singer:2016eax}. For the events that qualify, we obtain the measurement errors in luminosity distance by calculating the Fisher information matrix. Assuming the $\Lambda$CDM cosmology, the errors in $d_L$ can be converted to measurement uncertainty in $H_0$ after accounting for the measurement errors in the matter density, $\Omega_M$. We present the bounds on $H_0$ for two cases: without taking a prior on $\Omega_M$, and with a gaussian prior on $\Omega_M$. The width of the gaussian prior is fixed to the sum of the squares of the bounds placed on $\Omega_M$ by Planck \cite{Planck:2018vyg} and SH0ES \cite{Riess:2021jrx}, which is equal to $0.017$. Using binary black holes as dark sirens, the accuracy with which $H_0$ would be measured by different detector configurations in a network of ET and two CE detectors, one of which is a 40 km arm length detector and the other 20 km, is shown in figure \ref{fig:H0 dark sirens}.

\begin{figure}
    \centering
    \includegraphics[width=\textwidth]{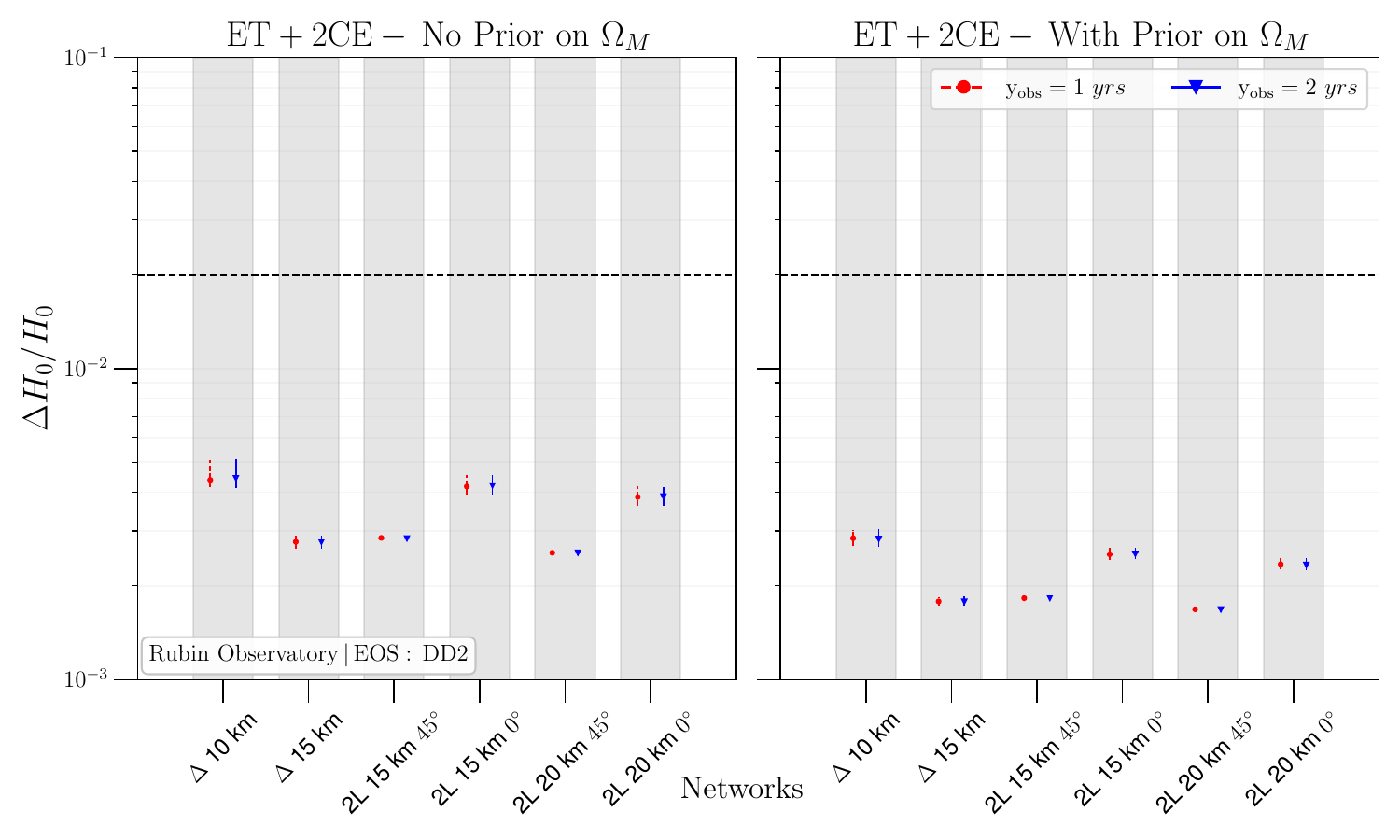}
    \caption{\small Same as Fig.\,\ref{fig:H0 dark sirens} except the results are obtained for the NSBH population instead of the BBH population. Moreover, the results were obtained by combining measurement accuracies from systems that produce kilonovae in addition to those that identify a unique host within the localization error box.}
    \label{fig:H0 gray sirens}
\end{figure}

For NSBH systems, the Hubble constant can be measured with gravitational waves both without (dark sirens) and with (bright sirens) electromagnetic counterpart. From the $10$ year population, we choose events that qualify as dark or bright sirens (and make sure to not double-count events that qualify as both). The NSBH events that qualify as dark sirens are chosen in the same way as was done for BBH systems. For bright sirens, we only consider events that lie within a redshift of $z=0.5$, as we do not expect to detect the kilonovae from events that lie farther. In addition, only those events are selected for which the sky position is measured better than $9.6\,\mbox{deg}^2$, i.e. the field of view of the Rubin observatory. For these events, we generate the kilonova light curves with DD2 as the neutron star equation of state using the numerical recipes from \cite{Kruger:2020gig,Raaijmakers:2021slr}. We claim that a particular kilonova is detected with the Rubin observatory by using the same criteria that were used in section \ref{sect:MMOkilonova}, i.e. the luminosity of the kilonova should be brighter than the limiting magnitude (corresponding to 600s of exposure) of the g and the i filters on two consecutive nights. We also assume a $50\%$ duty cycle for the Rubin observatory. The events for which the corresponding kilonova is detected qualify as bright sirens and are considered for the measurement of the Hubble constant. The luminosity distance errors for both dark and bright siren events are combined and the measurement errors for $\Omega_M$ are accounted for in the same way as was done for BBH dark siren events. The accuracy in measurement of $H_0$ using this gray siren approach with NSBH systems is shown in figure \ref{fig:H0 gray sirens}.

\begin{table}[t] 
  \centering
    \begin{tabular}{|c|cc|cc|}
    \hline \hline
    \multirow{2}{*}{Configuration} & \multicolumn{2}{c|}{BBH} & \multicolumn{2}{c|}{NSBH} \\
    & With 2CE & Without 2CE & With 2CE & Without 2CE \\
    \hline \hline
    $\Delta$ 10 km & 68 & 5  & 21  & 0  \\
    $\Delta$ 15 km & 74 &  7 & 30  & 0   \\
    2L 15 km $45^{\circ}$ & 73 & 4  & 25  & 1   \\
    2L 15 km $0^{\circ}$ & 69 & 0  & 21  & 0   \\
    2L 20 km $45^{\circ}$ & 77 & 8  & 37  & 1   \\
    2L 20 km $0^{\circ}$ & 70 & 0 & 22  & 0   \\
    \hline \hline
    \end{tabular}
    \caption{\small \label{tab:h0_golden_events}The number of dark siren BBH golden events and gray siren NSBH golden events for different ET detector configurations in an observation time of $10$ years.} 
\end{table}

Among the events that contribute to the measurement of the Hubble constant, there are certain \textit{golden} events for which the accuracy in the measurement of the luminosity distance is better than $2\%$. Such events  can individually measure $H_0$ with a precision better than $2\%$ and resolve the Hubble tension. The number of such golden binaries is given in table \ref{tab:h0_golden_events},  both for the dark siren case with BBHs, and for the gray siren case with NSBH systems, and an observation span of $10$ years. {\em We see that, when  in a network with 2CE detectors, the performances of all the ET geometries considered are comparable in the measurement of the Hubble constant using the dark and gray siren approaches. In contrast, for ET alone, the L-shaped configurations with parallel arms perform significantly worse compared to the other configurations, as they are unable to detect a single golden event in an observing time of 10 years.}
     \label{fig: LSS Xis}

\subsection{Cosmological stochastic backgrounds}\label{sect:cosmoback}

In this section we consider the performances of the various ET geometries for the detection 
of specific examples of stochastic backgrounds of cosmological origin. In Section~\ref{sect:cosmicstrings} we compare the ET sensitivities with the predictions from models of cosmic strings, while in Section~\ref{sect:phasetransitions} we consider the stochastic background produced by a phase transition in the early Universe. Finally, in Section~\ref{sect:sourceseparation}, we study the prospects for separating these cosmological backgrounds from the astrophysical stochastic background.

\subsubsection{Cosmic Strings}\label{sect:cosmicstrings}

Phase transitions followed by spontaneously broken symmetries may lead to the production of topological defects as relics of the previous more symmetric phase of the Universe. Within Grand Unified Theories, it has been shown \cite{Jeannerot:2003qv} that cosmic strings (line-like defects) are generically formed.  At the energy scales relevant to cosmology, the width of a string is negligible relative to its characteristic size. Such strings (Nambu-Goto strings) are parameterised by a single dimensionless quantity, the string tension $G\mu$ related to the energy scale $\eta$ at which cosmic strings are formed through $G\mu \sim 10^{-6}[\eta/(10^{16} {\rm GeV})]^2$ (we set the speed of light $c = 1$).

Cosmic strings emit bursts of beamed gravitational radiation.  The main sources of bursts are kinks, cusps or kink-kink collisions. 
Kinks are discontinuities in the tangent vector of the string that propagate at the speed of light and appear in pairs.
Cusps are points on the string that briefly travel at the speed of light.
The incoherent superposition of these bursts through the history of the Universe produces a strongly non-Gaussian stochastic gravitational wave background~\cite{Damour:2000wa}. 
Occasionally there may also be sharp and high-amplitude bursts of GWs above this stochastic background. The production of gravitational waves by cosmic strings offers a tool to probe particle physics beyond the Standard Model at energy scales much above the ones reached by accelerators.
A non-detection of a stochastic background of gravitational waves imposes bounds on the cosmic string tension and therefore on particle physics models beyond the Standard Model. ET will be able to considerably improve on current 2G bounds \cite{LIGOScientific:2021nrg}.

\begin{figure}
    \centering
    \includegraphics[width = 15cm]{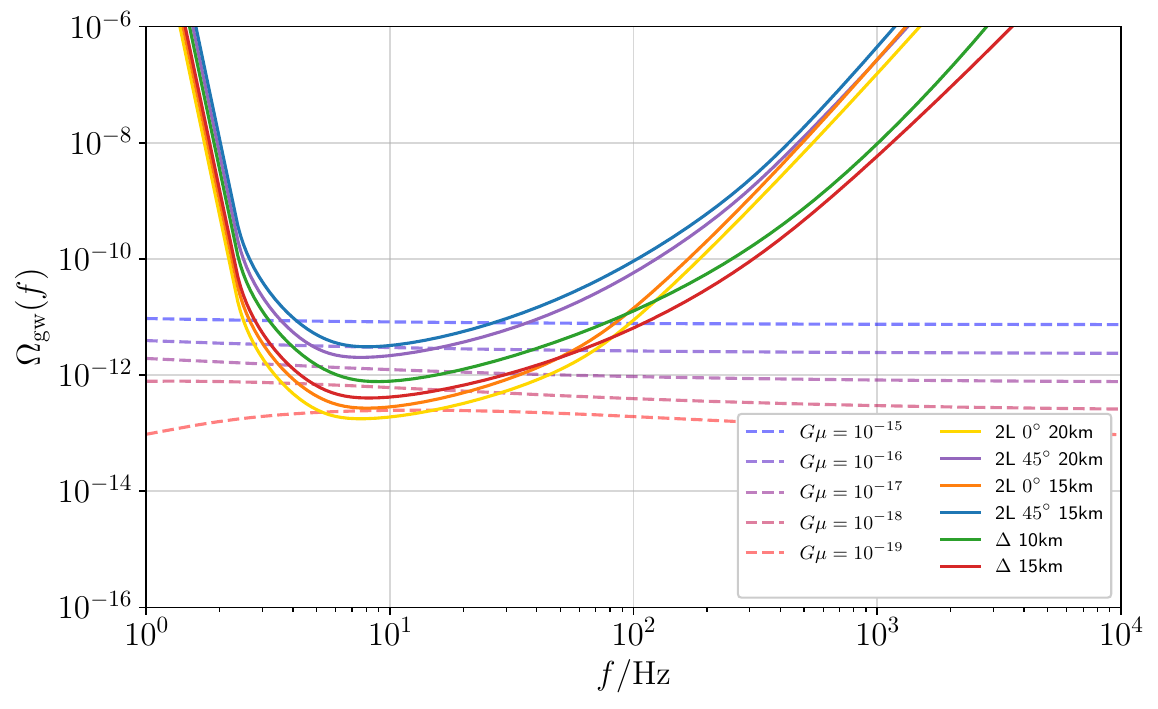}
    \caption{\small Energy density of gravitational waves emitted from cosmic string loops with loop distribution given by model A \cite{Blanco-Pillado:2013qja}. The various curves define the power-law integrated sensitivity for each ET configuration, assuming ${\rm SNR}_{th} = 1$ and $T_{\rm obs}=1$ yr.}
    \label{fig:stringsA}
\end{figure}

\begin{figure}
    \centering
    \includegraphics[width = 15cm]{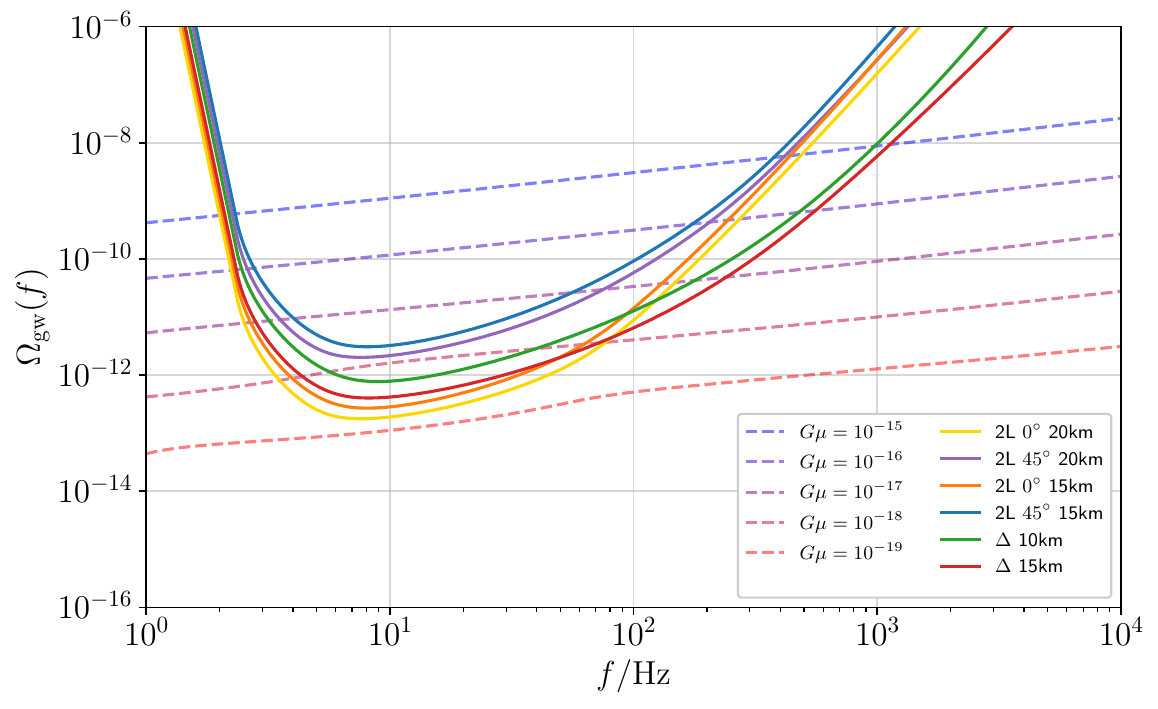}
    \caption{\small Energy density of gravitational waves emitted from cosmic string loops with loop distribution given by model B \cite{Lorenz:2010sm}. The various curves define the power-law integrated sensitivity for each ET configuration, assuming ${\rm SNR}_{th} = 1$ and $T_{\rm obs}=1$ yr.}
    \label{fig:stringsB}
\end{figure}

In Figs.~\ref{fig:stringsA} and  \ref{fig:stringsB} we show the prediction for $\Omega_{\rm GW}(f)$ for two different models of the loop distribution of cosmic strings,  denoted as  
models~A \cite{Blanco-Pillado:2013qja} and model~B \cite{Lorenz:2010sm},  and different values of the string tension $G\mu$. In model A the loop production function can be determined from Nambu-Goto simulations of cosmic string networks. In model B the distribution of non-self-intersecting scaling loops is the extracted quantity from different simulations. Within the latter model, loops are formed at all sizes following a power-law while the scaling loop distribution is cut-off on small scales by the gravitational back-reaction. In model B  are produced many more tiny loops than in the model A. These two models for  the loop distributions have been used by the LVK collaboration to put constraints on cosmic strings from O1, O2 and O3 data \cite{LIGOScientific:2021nrg}.

 The various curves define the power-law integrated sensitivity for each ET configuration, assuming SNR=1 and $T_{\rm obs}=1$ yr. As already remarked in Section~\ref{BackgroundSensitivities}, {\em the triangular configurations perform better at high frequency, because the overlap reduction function  remains constant at high frequencies for co-located detectors, while it falls to zero for widely separated  detectors. The 2L shape configuration does better around a frequency of 10 Hz, but only for co-aligned detectors (this maximises the coherence for the stochastic search). The case of 45 deg 2L configurations is universally the worst.} We conclude that, for the considered loop distributions (models A and B),  the best stochastic constraints on the string tension are $\sim 10^{-19}$, corresponding to an energy scale between $10^8$ GeV and $3\times 10^{10}$ GeV, 6  to 8 orders of magnitude above the electroweak scale. 

\subsubsection{First-order phase transition}\label{sect:phasetransitions}

As the Universe cooled down, it may have undergone a first-order phase transition (\acrshort{fopt}) with bubbles of true vacuum forming, expanding and colliding \cite{Kosowsky:1991ua,Mazumdar:2018dfl,Hindmarsh:2020hop}. This in turn leads to gravitational waves generated from the collision of bubbles, and from the sound waves and turbulence in the surrounding plasma (see e.g. \cite{Caprini:2015zlo,Maggiore:2018sht} for reviews). The frequency of gravitational radiation emitted depends on the temperature. For a temperature of the order $(10^5-10^{10})$ GeV, energy scales inaccessible to particle colliders, we find a SGWB that peaks in the ET frequency range. Here we briefly discuss ET detection prospects of supercooled FOPTs, a special class of FOPTs which is expected to create the ``loudest'' gravitational waves \cite{Ellis:2019oqb,DelleRose:2019pgi}.

\begin{figure}
    \centering
    \includegraphics[width=0.49\linewidth]{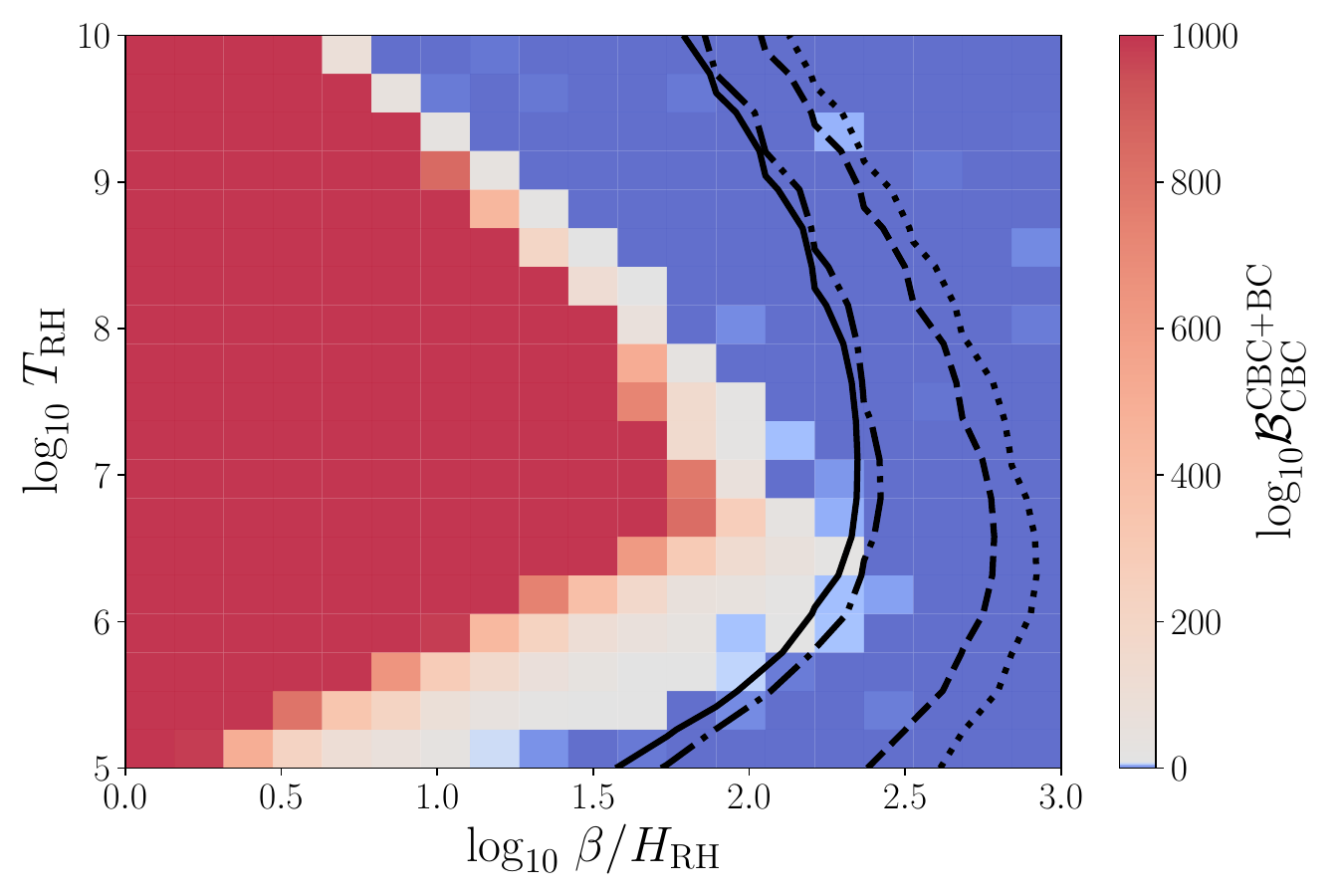}
    \includegraphics[width=0.49\linewidth]{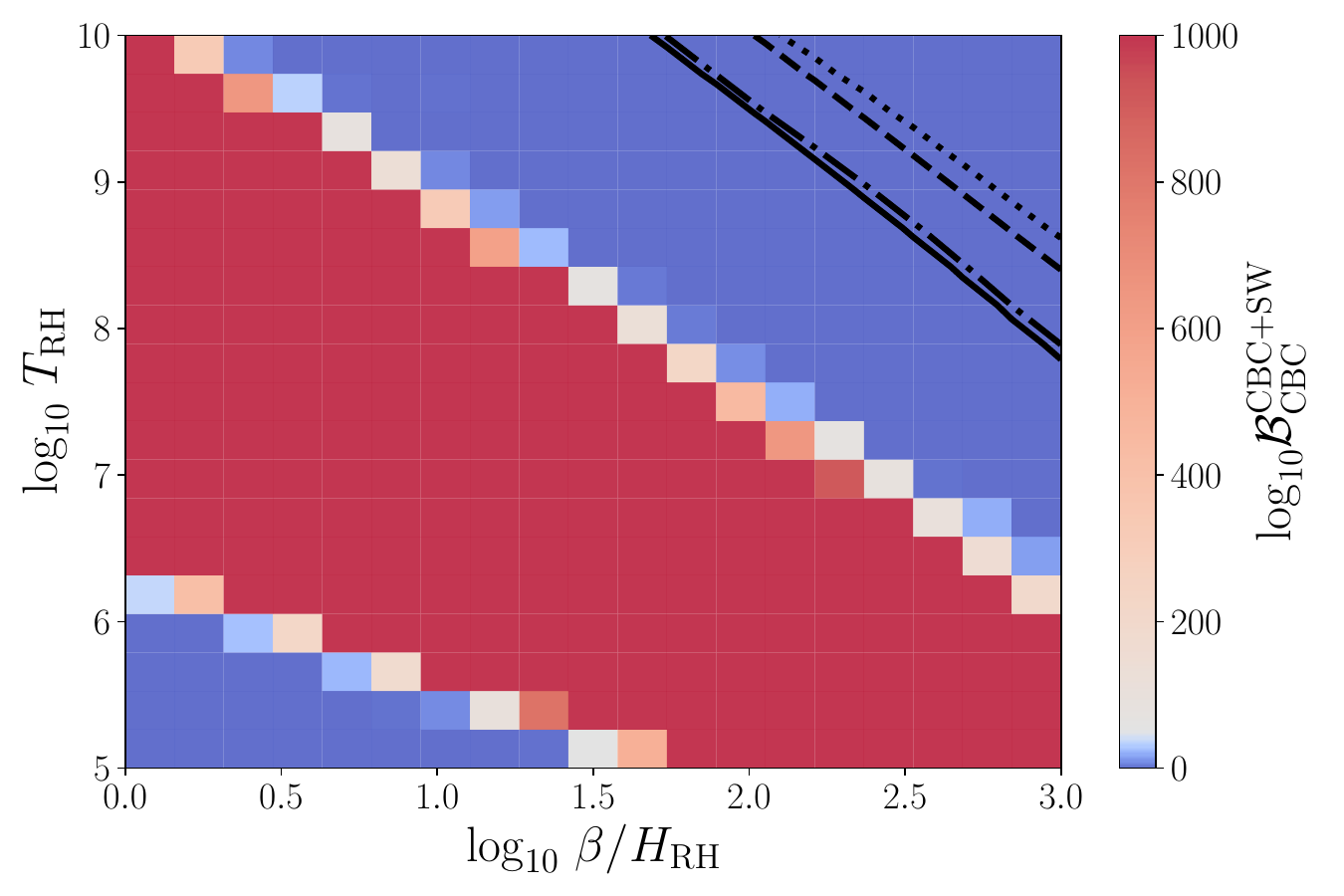}
    \caption{\small Detectability of bubble collisions (left) and sound wave (right) dominated supercooled FOPT with next-generation ET detector. The dashed line is the detection threshold using 10 km triangular ET configuration, the dotted line is 15 km triangular configuration, while the solid and dot-dashed contours are for two L-shaped detectors with 15 km and 20 km arms, respectively.}
    \label{fig:fopt}
\end{figure}

The gravitational wave spectrum from a supercooled FOPT depends only on two parameters, namely the reheating temperature $T_{\rm RH}$ and inverse duration $\beta / H_{\rm RH}$, where we denote $H_{\rm RH} \equiv H(T_{\rm RH})$.\footnote{For simplicity, we assume wall velocity to be equal to the speed of light.} Note that the strength of transition $\alpha$ that affects the SGWB for the more general FOPTs, does not impact the supercooled spectrum. In Fig.~\ref{fig:fopt} we simulate a SGWB containing the residual from unresolved CBCs and an FOPT signal for a range of $T_{\rm RH}$ and $\beta / H_{\rm RH}$ values. The left panel includes a FOPT signal that is dominated by bubble collisions, while the right panel is for a FOPT signal dominated by sound waves. Turbulence is expected to be subdominant to both bubble collisions and sound waves, so we exclude it from our study. It is clear that we will be able to exclude a large part of the parameter space with the next generation of GW detectors.

\subsubsection{Source separation}\label{sect:sourceseparation}


The unresolved astrophysical mergers are expected to be the dominant contribution to the background observed by ET. It is therefore important to subtract the individual sources, as discussed in the previous section, in the hope to reveal the cosmological background. Note that the imperfect subtraction of CBC sources may pose a serious threat to digging out a cosmological background, and extensive research must be done to improve the subtraction methods~\cite{Zhou:2022otw,Zhou:2022nmt}. Here we take as an example two cosmological sources: cosmic strings and an early-Universe first-order phase transition. 

\begin{figure}
    \centering
    \includegraphics[width=0.49\linewidth]{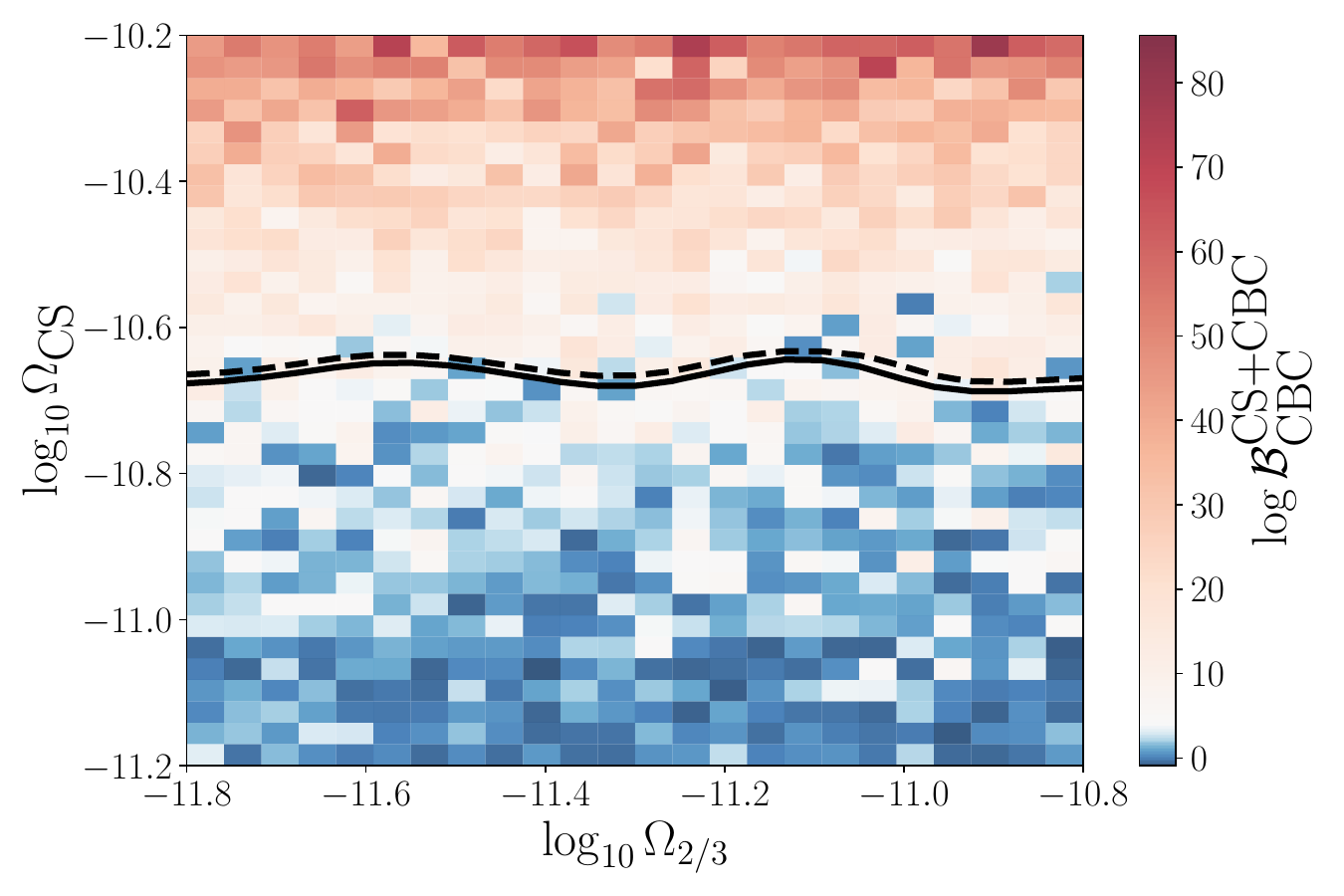}
    \includegraphics[width=0.49\linewidth]{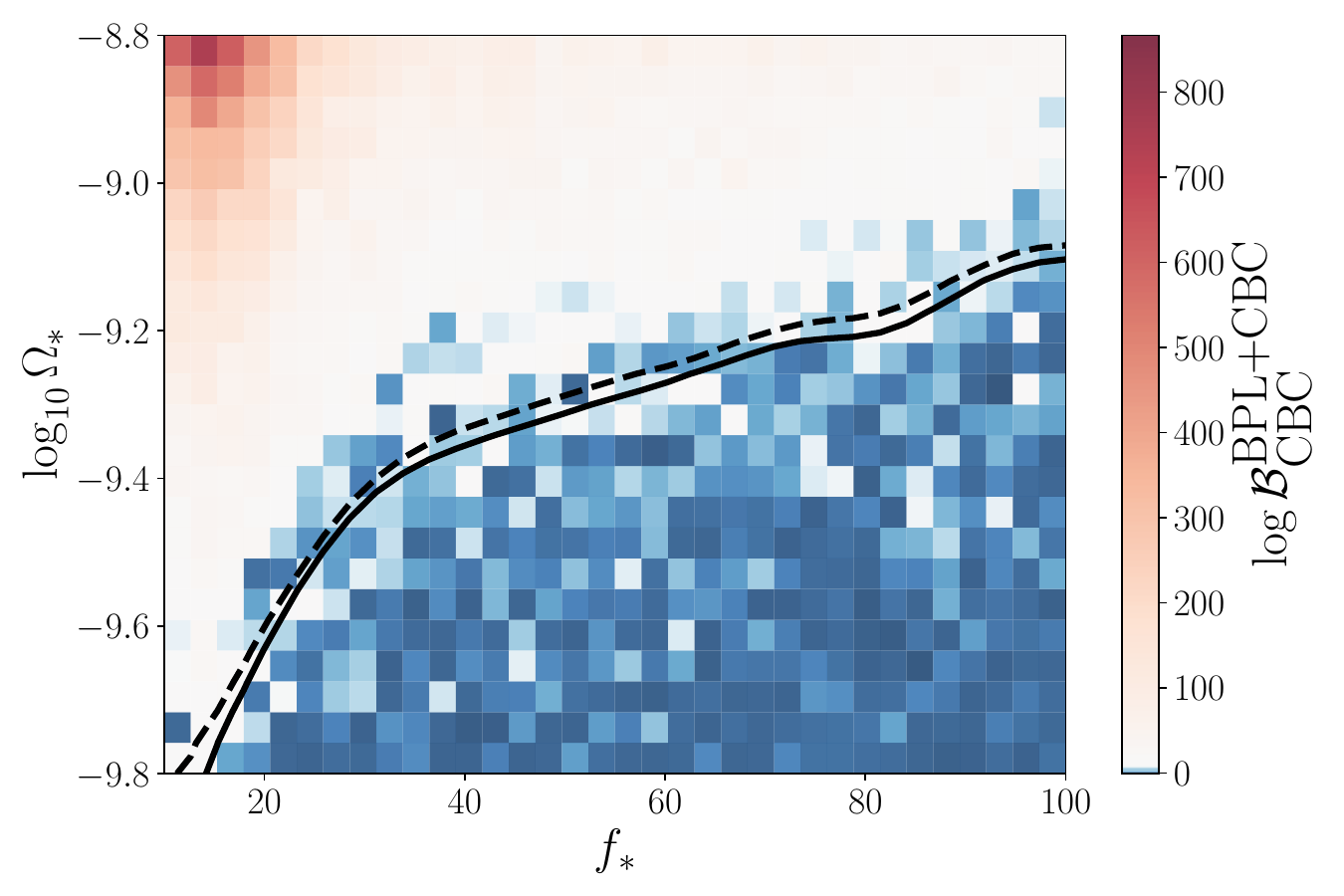}
    \caption{\small Upper limits we can place on the amplitude of the background from cosmic strings (left panel) and a first-order phase transition (right panel) with ET's triangular 10 km configuration (dashed line) and 2L 20 km configuration (solid line). We assume a strong preference for a signal that contains an astrophysical and a cosmological contribution, relative to a signal containing only unresolved astrophysical objects for log Bayes factors greater than 8.}
    \label{fig:source_separation}
\end{figure}

By limiting the study to the frequency range $f \in [10, 100]$ Hz, we can assume the CBC background to follow $f^{2/3}$ and the cosmic strings background to be frequency-independent, i.e. $\Omega_{\rm CS}=\rm{const}$. The first-order phase transition is modeled as a broken power-law (with $\Omega_*$ amplitude and $f_*$ break frequency), with the assumption that the sound-wave component is the dominant contribution to the cosmological SGWB. We simulate astrophysical and cosmological SGWB and perform a model selection study of the signal. For details on the parameter estimation Bayesian tools, see \cite{Martinovic:2020hru}. We compute Bayes factors for a CBC signal to noise, and for a combination of cosmological and CBC signal to noise. Comparing the two Bayes factors demonstrates if one model is preferred over the other, i.e. if we can detect the presence of a cosmological signal. From Fig.~\ref{fig:source_separation} we conclude that a cosmic strings background of amplitude $2.2 \times 10^{-11}$ and a first-order phase transition background of amplitude up to $4.0 \times 10^{-10}$ (at $25\,\rm{Hz}$) will be probed with this 3G detector.

{\em Note that the various ET configuration makes little quantitative difference to the results, but including next-generation Cosmic Explorer can lead to great improvement in SGWB source separation.}

\subsection{Continuous waves}
Broadly speaking, the class of \textit{continuous waves} (\acrshort{cw}) comprises all deterministic\footnote{In contrast to GW backgrounds, which are described by means of stochastic processes.} GW signals that last much longer than typical transient signals, whose durations vary from a fraction of a second to tens or hundreds seconds at the most. No CW has been detected so far, although interesting upper limits have been established, see e.g. \cite{Riles:2022wwz,Piccinni:2022vsd,Tenorio:2021wmz} for recent reviews. The prototypical source of CWs is a spinning neutron star, asymmetric with respect to the rotation axis, that emits a quasi-monochromatic gravitational wave whose frequency changes extremely slowly over time. For such sources, the expected signal duration is much longer than one day, meaning that the frequency modulation due to the Doppler effect, caused by the motion of the Earth, and the phase and amplitude modulations due to the detector radiation pattern functions, play relevant roles. 

Actually, other sources of CW-like signals have been investigated. In particular, \textit{transient CW} signals (\acrshort{tcw}) have been considered, such as those expected to be emitted by a newborn magnetar, in the early stages of its evolution \cite{2001A&A...367..525P,Stella:2005yz,Lander:2019guk,DallOsso:2021xbv}. In this case, the frequency secular variation is caused by a combination of the electromagnetic emission, due to the super-strong magnetic field, and the gravitational-wave emission, due to the strong asymmetry induced by the electromagnetic field itself. As a consequence, the resulting source spin-down is orders of magnitudes faster than for standard neutron stars, producing a signal that stays in the sensitivity band of detectors for hours or days.    

Recently, several theoretical works have shown that GW detectors can also be used to
probe dark matter, see e.g. \cite{Bertone:2019irm} for a comprehensive review. In many of the proposed mechanisms, the resulting signal has a CW-like nature. This is, for instance, the case for depleting scalar or vector ultra-light boson clouds that could form around Kerr black holes through superradiance \cite{Brito:2015oca}, the interaction of dark matter particles with detector components \cite{Pierce:2018xmy},\footnote{In this case the effect of the interaction is not a true GW signal, but is such to induce a relative motion of the optical components which can be searched as if it was a standard gravitational wave.} or the early inspiral of sub-solar mass primordial black holes \cite{Miller:2020kmv}.  

The typical duration of CW signals has some important consequences. First, once detected, a CW source can be monitored for long periods of time, becoming a true laboratory for fundamental physics and astrophysics, and potentially allowing observations of tiny deviations from the assumed models (for instance, a deviation from GR, in the form of non-standard polarizations \cite{Will:2014kxa}). At the same time, signal persistence allows us to reduce the false alarm probability of any candidate virtually to zero, because if a signal is present in a given dataset -- even with low significance -- it will also appear in new datasets, and its significance will increase until a detection can be claimed (of course, this is not necessarily true for tCWs). Moreover, CW source parameters can be measured with very high accuracy, as their uncertainties decrease with the observation time. On the other hand, CW signals are predicted to have small amplitudes, compared to e.g. typical CBC signals, and to search for them can require a huge computational cost. For this reason, the development of increasingly sensitive, robust, and computationally efficient algorithms is a very active field that grows in parallel to current detector upgrades and R\&D for future detectors, such as ET.
In the following, we give more details on different sources of CW-like signals, focusing attention on the impact that different ET configurations have on the chance of detection.
\subsubsection{CWs from spinning neutron stars}
\label{sec:cwns}
Spinning neutron stars (NS), isolated or in a binary system, asymmetric with respect to the rotation axis, emit a persistent gravitational wave signal with frequency in a given ratio to the rotation frequency of the star. 
 
The typical amplitude of CWs signals is difficult to estimate, as it depends on poorly known quantities, especially the star's ellipticity, which is a measure of its degree of asymmetry. 
In the case of a CW emitted by a non-accreting, spinning NS, assuming it is asymmetric with respect to its rotation axis, the signal amplitude is given by
\begin{equation}
  \label{eq:hexpected}
h_0 = \frac{4\pi^2G}{c^4} \frac{\epsilon I_{zz}  f^2}{d}  \approx                  1.06\times10^{-26}\left(\frac{\epsilon}{10^{-6}}\right) 
    \times\left(\frac{I_{\rm zz}}{10^{38}\>\rm{kg\ m}^2}\right)\left(\frac{f}{100\>{\rm  Hz}}\right)^2
\left(\frac{1\>{\rm kpc}}{d}\right), 
\end{equation}
where $d$ is the source distance, $f$ is the GW frequency (equal to two times the star's spin frequency in the prototypical case of a NS rotating around one of its principal axes of inertia), $\epsilon \equiv (I_{\rm xx}-I_{\rm yy})/I_{\rm zz}$ is the ellipticity of the star, and $I_{\rm zz}$ is the moment of inertia of the star with respect to the principal axis aligned with the rotation axis.

The search for CWs from spinning NSs depends on the amount of information we have or we assume to have on the sources. 
For \textit{targeted} searches, we use accurate measurements of the main NS parameters, namely the position, spin frequency and its derivative(s), and Keplerian parameters for NSs in binary systems, obtained from electromagnetic (EM) observations (typically radio, gamma, or X-rays) to perform very sensitive searches based on matched filtering. Data from the detector network can be coherently combined in order to increase the sensitivity; see \cite{LIGOScientific:2021hvc} for the most recent O3 results obtained by the LIGO and Virgo Collaborations for targeted searches. On the other hand, there are \textit{all-sky} searches for NSs without an EM counterpart, in which a portion of the parameter space as large as possible is explored. Due to computational load constraints, such searches are based on hierarchical semi-coherent methods which sacrifice some sensitivity in order to make them computationally feasible. See \cite{KAGRA:2022dwb} for the latest LVK  all-sky search results. Searches covering ``intermediate'' cases are also routinely carried out. For instance, \textit{narrow-band} searches slightly extend the parameter space around the EM-inferred parameters of known pulsars, in order to account for possible small -- fraction of a Hz -- mismatches between the EM and GW emission \cite{LIGOScientific:2021quq}. \textit{Directed searches} assume the source position is known, while the star's rotational parameters are unknown, as in the case of the supernova remnants or the galactic center \cite{LIGOScientific:2021mwx,KAGRA:2022osp}. All-sky and directed searches typically use data from different detectors (or runs of the same detector) in order to make coincidences among outliers, with the aim of suppressing the false alarm probability. Although no detection of CWs has been made so far, interesting upper limits on the star's ellipticity have been placed, see references above. These can be interpreted in terms of constraints on the residual initial star's deformation frozen after crust solidification or as due to an inner magnetic field misaligned with the rotation axis, using a given ratio with the external poloidal one.

 Although the range of CW searches for spinning NSs is very wide, in the following we provide some figures of merit about detection perspectives by ET focusing on two relevant cases, targeted and all-sky searches, that bracket the span of typical searches in terms of dimension of the explored parameters space, sensitivity and computational cost.
 
For targeted searches, we provide some summary results in Tab. \ref{tab:pulsar_eps} where, for different ET configurations, the number of potentially detectable sources (among those contained in the ATNF catalogue\footnote{\url{https://www.atnf.csiro.au/research/pulsar/psrcat/}}) is given, under three different assumptions of the source ellipticity. We consider a total observation time $T_\mathrm{obs}=1$ year, a duty cycle of 85$\%$ and assume to make a fully-coherent search based on matched filtering. In the case of 2L-shaped network, results do not depend on the relative orientation of the two detectors. Columns 2 indicates the number of detectable pulsars under the assumption that they emit at their spin-down limit, i.e. that all their spin-down is due to the emission of GWs. This provides a theoretical upper limit to the emitted strain, given by
\begin{equation}
    h_{0,{\rm sd}}=\frac{1}{d}\left(\frac{5GI_{zz}}{2c^3}\frac{|\dot{f}|}{f}\right)^{1/2} ,
    \label{eq:h0sd}
\end{equation}
where $\dot{f}$ is the signal frequency first time derivative. The corresponding limit on the star ellipticity is
\begin{equation}
\label{eq:eps_sd}
\epsilon_{\rm sd} = \frac{h_{0,{\rm sd}}}{I_{zz}}\left(\frac{c^4 d}{4\pi^2G f^2}\right) \approx 9.46\times 10^{-6}\left(\frac{h_{0,{\rm sd}}}{10^{-25}}\right) \times\left(\frac{10^{38}\,\text{kg}\,\text{m}^2}{I_{zz}}\right)
         \left(\frac{100\,\text{Hz}}{f}\right)^2\left(\frac{d}{1\,\text{kpc}}\right).
\end{equation} 
\begin{table}[t!]
    \caption{\small Expected number of detectable sources, assuming three different conditions for the ellipticity: $\epsilon=\epsilon_{sd}~(n_1)$, $\epsilon=min(\epsilon_{sd},~10^{-6})~(n_2)$, $\epsilon=min(\epsilon_{sd},~10^{-9})~(n_3)$, assuming a total observation time $T_\mathrm{obs}=1$ year and a duty cycle of 85$\%$. For each case, we give in parentheses the minimum and median value of ellipticity for detectable signals.}
    \label{tab:pulsar_eps}
    \resizebox{\columnwidth}{!}{
    \begin{tabular}{|c|c|c|c|}
    \toprule
    Configuration & $n_1$ & $n_2$   & $n_3$ \\ \hline
    $\Delta$ 10km & 866 ($2.5\times 10^{-10},~1.3\times 10^{-4}$) & 180 ($2.5\times 10^{-10},~4.4\times 10^{-9}$) &  19  ($2.5\times 10^{-10},~7.5\times 10^{-10}$)\\ 
    $\Delta$ 10km HF-only & 398 ($2.5\times 10^{-10},~6.2\times 10^{-6}$) & 178 ($2.5\times 10^{-10},~4.4\times 10^{-9}$) & 19 ($2.5\times 10^{-10},~7.5\times 10^{-10}$)\\
    $\Delta$ 15km & 983 ($2.1\times 10^{-10},~1.1\times 10^{-4}$) & 214 ($2.1\times 10^{-10},~4.4\times 10^{-9}$) &  33  ($2.1\times 10^{-10},~7.9\times 10^{-10}$)\\ 
    2L 15km & 959  ($2.0\times 10^{-10},~1.2\times 10^{-4}$) & 206   ($2.0\times 10^{-10},~4.2\times 10^{-9}$) &  29 ($2.0\times 10^{-10},~8.1\times 10^{-10}$) \\
    2L 15km HF-only & 451 ($2.0\times 10^{-10},~5.6\times 10^{-6}$)& 203 ($2.0\times 10^{-10},~4.0\times 10^{-9}$) & 29 ($2.0\times 10^{-10},~8.1\times 10^{-10}$)\\
    2L 20km & 1035  ($1.8\times 10^{-10},~1.1\times 10^{-4}$) & 227   ($1.8\times 10^{-10},~4.3\times 10^{-9}$) &  33 ($1.8\times 10^{-10},~7.3\times 10^{-10}$) \\
    \hline  
    \end{tabular}
    }
\end{table} 
The number of pulsars given in column 3 is computed by assuming that every star's ellipticity is the minimum of the spin-down limited value $\epsilon_{sd}$ and $10^{-6}$, which is of the order of the maximum theoretical ellipticity a NS with a standard EOS could have \cite{Morales:2022wxs}. In column 4, the number of detectable pulsars is computed assuming the minimum ellipticity between  $\epsilon_{sd}$ and $10^{-9}$, which is considered as a plausible value for millisecond pulsars \cite{Woan:2018tey,Soldateschi:2021hfk}. For each case, we also report the minimum and median value of the ellipticity for the potentially detectable sources. 

From this table, we can draw some considerations. {\em The main factor affecting the number of detectable sources is the arm length, rather then the network geometry}. Indeed, by coherently combining individual detector data streams, the triangle configuration would provide a sensitivity gain of $\sqrt{3/2}$, with respect to the 2L network, which is nearly fully balanced by the shape factor, so that at the end what mostly contributes to the sensitivity is the arm length. {\em We also note that, for the more realistic choices on star's ellipticities, corresponding to columns $n_2$ and $n_3$, most of the potentially detectable sources are millisecond or young fast pulsars, spinning at frequencies larger than a few tens of Hertz, for which the low frequency sensitivity of the detectors is not particularly relevant.} The corresponding typical values of the ellipticity are of about $2\times 10^{-10}$, corresponding to ``mountains'' extending from the star's surface not higher than about 1$\mu m$. Lastly, focusing on the most conservative condition, given by column $n_3$, we find that {\em the number of potential detections increases by 50-70$\%$ going from 10km-arm detectors to 15 or 20km-arm detectors.}    

\begin{figure}[t]
    \centering
     \includegraphics[width=0.49\columnwidth]{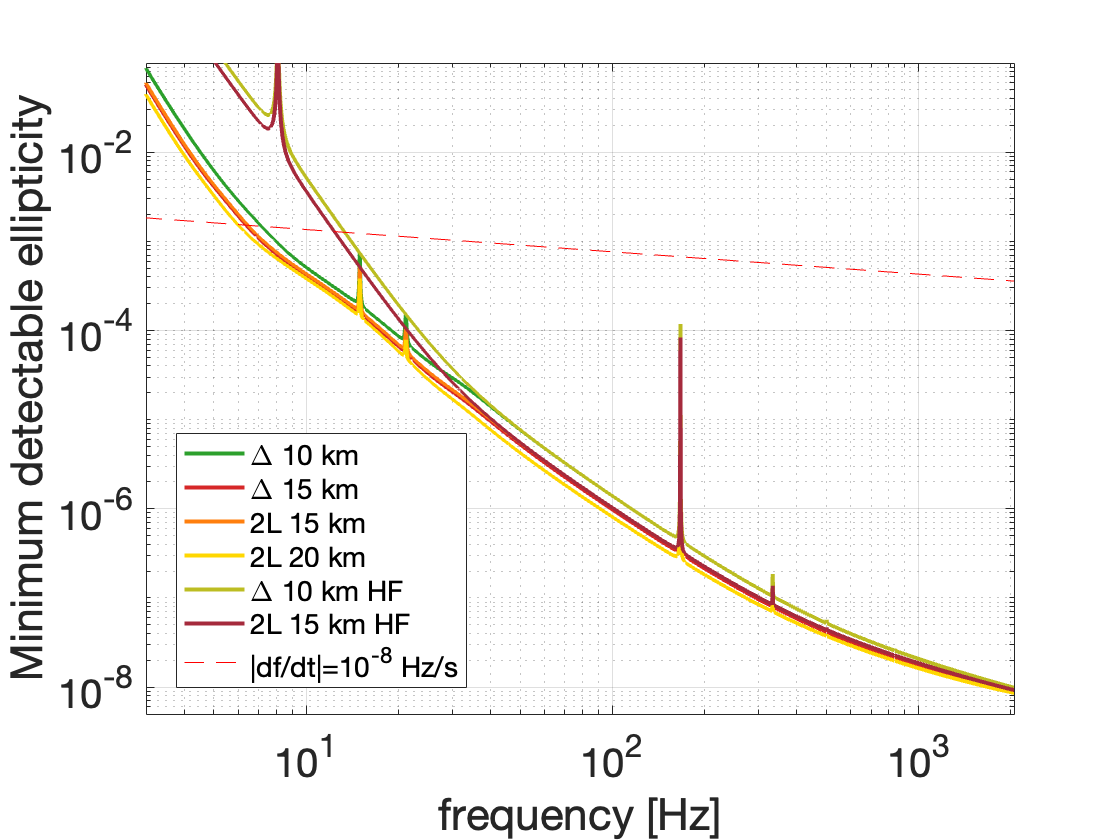}
     \includegraphics[width=0.49\columnwidth]{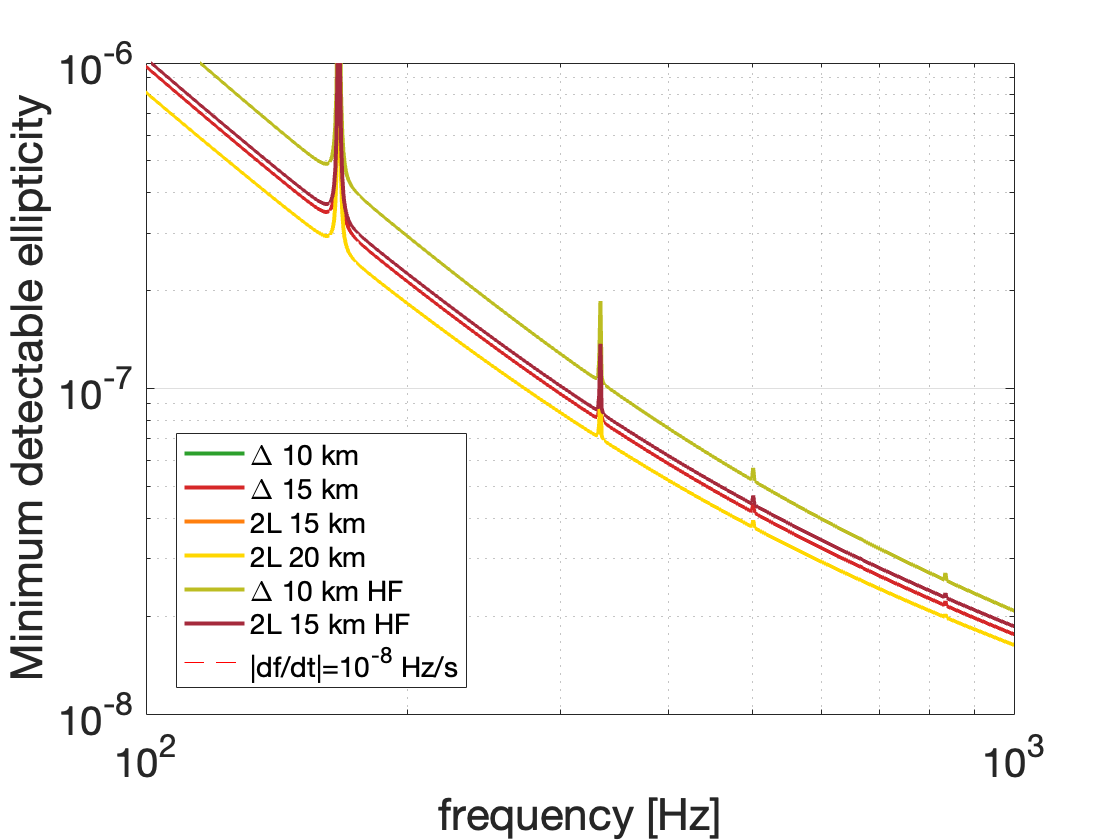}
    \caption{\small ET sensitivity reach to all-sky searches of CWs (95$\%$ C.L.). The various continuous curves are the minimum detectable ellipticity as a function of frequency for different  detector configurations, assuming a source distance of 8 kpc, an observation time $T_{\mathrm{obs}}$=1 yr, a duty cycle D=85$\%$ and data segments duration $T_{\mathrm{FFT}}$=10 days. The dashed curve indicates a constant spin-down of $10^{-8}$ Hz/s. Right plot is a zoom of the left plot in the high frequency region. On the scale of the right plot, the curves corresponding to $\Delta$ 10 km HF and $\Delta$ 10 km coincides, and similarly for those corresponding to the 2L 15 km HF and 2L 15 km configurations.}
    \label{fig:allsky}
\end{figure}

Wide-parameter searches are based on semi-coherent methods in which data segments of given duration are incoherently combined. In principle, they are less sensitive than targeted ones, but can cover a huge parameter space and are intrinsically more robust against non-predicted deviations from the assumed signal model.
Concerning, in particular, all-sky searches, Fig. \ref{fig:allsky} shows, for different detector configurations, the minimum detectable ellipticity as a function of the signal frequency, assuming source rotational evolution is dominated by the emission of GWs and a fixed source distance of 8~kpc. These results have been obtained considering an analysis done by means of the FrequencyHough pipeline \cite{Astone:2014esa}, a semi-coherent procedure routinely used in current CW searches, whose sensitivity can be estimated (under the assumption the noise is Gaussian) as
\begin{equation}
h_{\rm min,95\%}\approx
\frac{4.97}{N^{1/4}}\sqrt{\frac{S_n(f)}{T_{\rm FFT}}}\sqrt{{\rm CR}_{\rm thr}+1.6449},
\label{eq:hul}
\end{equation}
where $N=T_{\rm obs}/T_{\rm FFT}$ is the ratio between the total observation time and the duration of the data segments to be incoherently combined, $S_n(f)$ is the detector average noise power spectrum and ${\rm CR}_{\rm thr}$ is the threshold Critical Ratio to select outliers. In the plot, we used $T_{\rm obs}$=1 year, $T_{\rm FFT}$=10 days and ${\rm CR}_{\rm thr}=3.4$ for triangular configurations and ${\rm CR}_{\rm thr}=4$ for L-shape configurations. 
The presence of more that one detector is exploited in such kind of searches to strongly reduce the false alarm probability. For the standard search parameters considered here, the number of points in the parameter space (sky position, frequency and frequency derivatives) is of the order of $10^{23}$. Assuming that for each detector of the order of $10^{12}$ outliers are 
selected,\footnote{Currently, the typical number of selected outliers is O($10^{10}$), but we devise an increase of two orders of magnitude as reasonable in the ET era.} it is possible to show that double coincidences are enough to reduce the number of candidates to O(10), and that triple coincidences would allow - in principle - to shrink this number to a negligible level, assuming ideal Gaussian noise \cite{Astone:2014esa}. As a matter of fact, in practice we expect a significant number of candidates to survive the coincidence step, due to non-Gaussian instrumental artefacts which, inevitably, pollute the data. Having triple coincidences, as for the triangle configuration, allows to better reduce false candidates with respect to the 2L configuration. This implies that in the former case we can use a slightly lower threshold for outliers selection, with the aim of having a similar number of candidates after the coincidences. A lower threshold, according to \eq{eq:hul}, corresponds to a better sensitivity. Of course the exact value of the threshold would depend on the non-Gaussian, non-stationary features of the actual dataset, which are difficult to predict in advance. As a rule of thumb, we decide to reduce the threshold on the CR from 4 to 3.4 in the case of the triangular configuration. This choice would correspond to a reduction of one order of magnitude in the false alarm probability in presence of Gaussian noise.   

As for targeted search, the relative orientation of the two triangular detector in the 2L-shaped configuration does not impact on the analysis outcome.
{\em A clear feature of these results is that different configurations do not produce very different results, especially in the high frequency range, which corresponds to smaller ellipticities. In general, however, longer-arm detectors perform better than shorter ones, with differences up to a few tens percent. }
For instance, a source located at 8 kpc, and emitting a signal at about 500 Hz, would be detected if its ellipticity is at least $\sim (4-5)\times 10^{-8}$. At a given frequency, the ellipticity scales linearly with the distance. Nearby sources, located within $\sim 1$ kpc, would produce detectable signals with an ellipticity as small as few times per $10^{-9}$, if spinning at around 500 Hz.
Although current searches typically use a shorter data segment duration, values as large as 10 days are reasonable, and likely conservative, for the ET era, given the algorithmic developments and the increase in available computing power. It is interesting to stress, however, that while the sensitivity of semi-coherent searches scales as the fourth root of the segment duration, the dimension of the parameter space scales at least as the duration to the fifth power. This means, for instance, that using segments of 30 days, instead of 10, would provide a theoretical sensitivity gain of about 30$\%$, in front of a computing cost at least $3^5$ times bigger. Moreover, and probably even more importantly, the use of very long segment durations reduces the search robustness with respect to unmodeled signal features, which can translate into a sensitivity loss. As an example, a source may have a significant intrinsic transversal velocity with respect to the line of sight. If not taken into account, it may determine a sub-optimal Doppler correction as a consequence of the change of sky position during the observation time. Another possibility is that a signal may be characterized by some level of frequency wandering due to some unpredicted mechanism, involving the interaction with other objects, or due to intrinsic processes in the source (see e.g. \cite{KAGRA:2022dqk,KAGRA:2021tse} for analysis methods that tackle this issue). It is then important to note that a reasoned balancing among theoretical sensitivity, available computing power and algorithm robustness must be always taken into account when new algorithms for wide-parameter searches are developed.  

\subsubsection{Transient CWs}

Newly-born neutron stars could be extremely deformed and emit very strong GWs immediately after they have formed from a binary neutron star merger or supernova explosion. See \cite{Bochenek:2020xdi,Jordana-Mitjans:2022gxy} and references therein for recent observational evidences.

Such systems are astrophysically very interesting: detecting a remnant of a neutron star merger would help to improve constraints on the equation of state \cite{Margalit:2017dij,Bauswein:2017vtn,Rezzolla:2017aly,Radice:2017lry} compared to only considering the inspiral of the system alone \cite{LIGOScientific:2017ycc,LIGOScientific:2018cki,LIGOScientific:2018hze} in the case of binary neutron star mergers or only electromagnetic observations of the supernova. Furthermore, detecting such a system would be a huge step in multi-messenger astronomy, and would allow us to determine the nature of the remnant, something at which electromagnetic observations can only hint. As we saw from GW170817, astronomers were able to see the remnant in multiple ways --UV, IR, visible -- and each way seemed to imply a different expected remnant. Some studies report the formation of a hypermassive neutron star lasting for $\mathcal{O}$(s) \cite{Kasen:2017sxr,Granot:2017tbr,Granot:2017gwa,Pooley:2017mzo,Matsumoto:2018mra}, while others
saw continued energy emission from a long-lived remnant  \cite{Yu:2017syg,Ai:2018jtv,Geng:2018vaa,Li:2018hzy}. In the latter case, the signal emitted would last longer than the canonical $\mathcal{O}$(s) expected from mergers, but shorter than the durations $\mathcal{O}$(years) expected from much older, asymmetrically rotating neutron stars. This ``intermediate'' regime of signal durations would contain very interesting physics, and yet until recently, not much effort had been made to search for these kinds of sources. 

Such searches for a remnant of GW170817 were performed on O2 data, analyzing data in segments of $\mathcal{O}$(seconds) for burst-like signals and $\mathcal{O}$(hours-days) for tCWs, and set the first-ever upper limits on GW emission from a remnant \cite{LIGOScientific:2017fdd,LIGOScientific:2018urg}. In contrast to ordinary CWs, tCWs last for a much shorter time, have very large spin-downs, and have frequencies that evolve non-linearly over time. Thus, new analysis methods had to be designed to search for these signals \cite{Thrane:2010ri,Mytidis:2015kea,Miller:2018rbg,Sun:2018owi,Oliver:2019ksl,Miller:2019jtp}. 
The sensitivity of the O2 search allowed us to reach distances  $\mathcal{O}$(1) Mpc; however, with ET, we will be able to reach distances of $\mathcal{O}(10-100)$ Mpc, depending on the degree of deformation of the neutron star. To compute this sensitivity, we use an equation similar to \eq{eq:hul}, that has been generalized for power-law signals and gives the maximum distance reach at a given confidence level:
\begin{equation}
d_{\rm max}=4.63\cdot 10^{-9} \text{ m}\left(\frac{I_{\rm zz}}{10^{38} \text{ kg$\cdot$m$^2$}}\right)\left(\frac{\epsilon}{10^{-3}}\right)\frac{1}{N^{1/4}}\frac{T_{\rm FFT}}{\sqrt{T_{\rm obs}}}\left(\sum_i \frac{\mathcal{F}^2_i}{S_n(f_i)}\right)^{1/2} \hspace{-2mm} \left({\rm CR}_{\rm thr}+1.6449\right)^{-1/2}\hspace{-1mm}.
\label{eq:tcwdmax}
\end{equation}
Here, $\mathcal{F}_i=f_i^2$ is short-hand for the $i$th frequency, and the sum goes over all frequencies during $T_{\rm obs}$.

\begin{figure}[t]
    \centering
     \includegraphics[width=0.49\columnwidth]{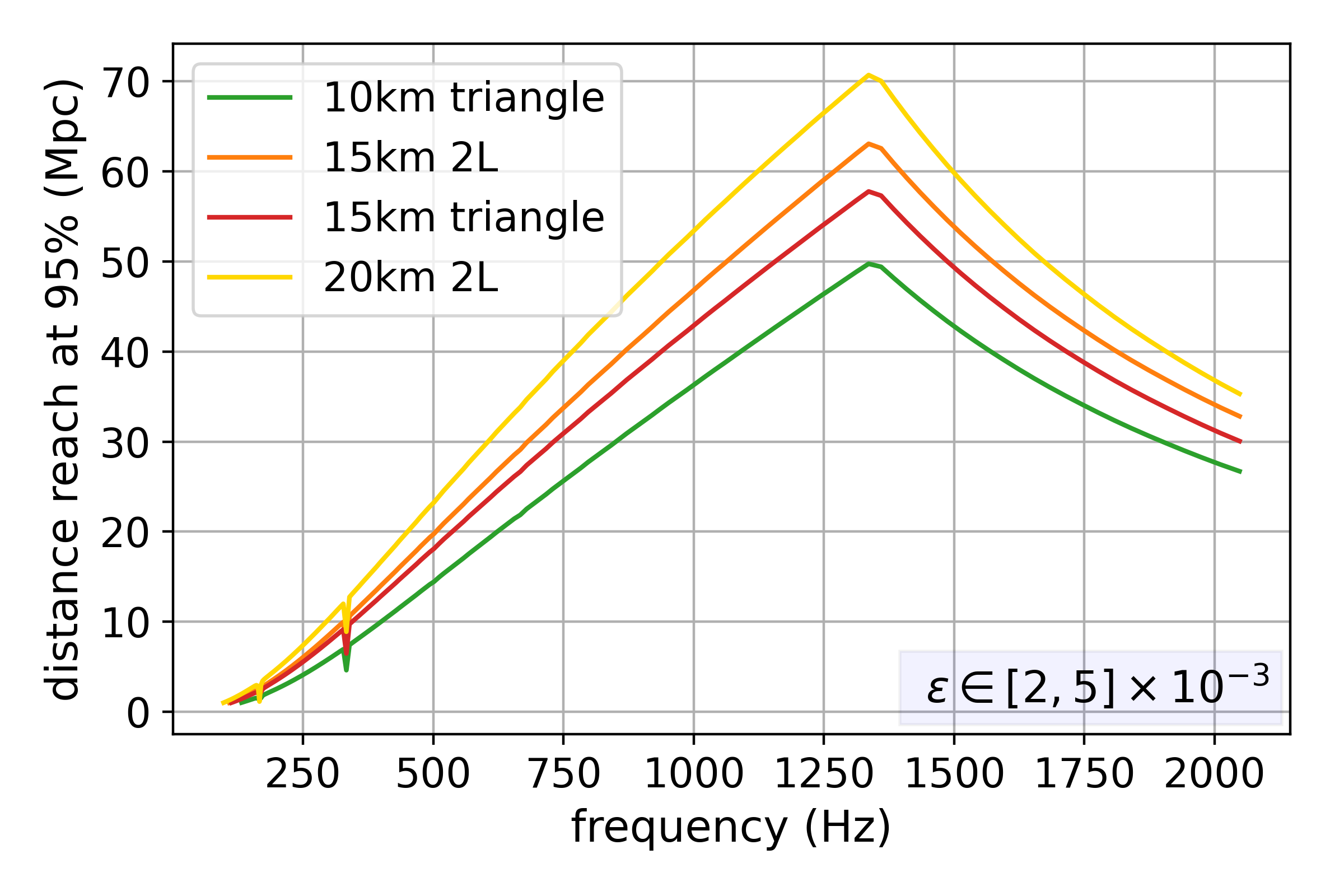}
     \includegraphics[width=0.49\columnwidth]{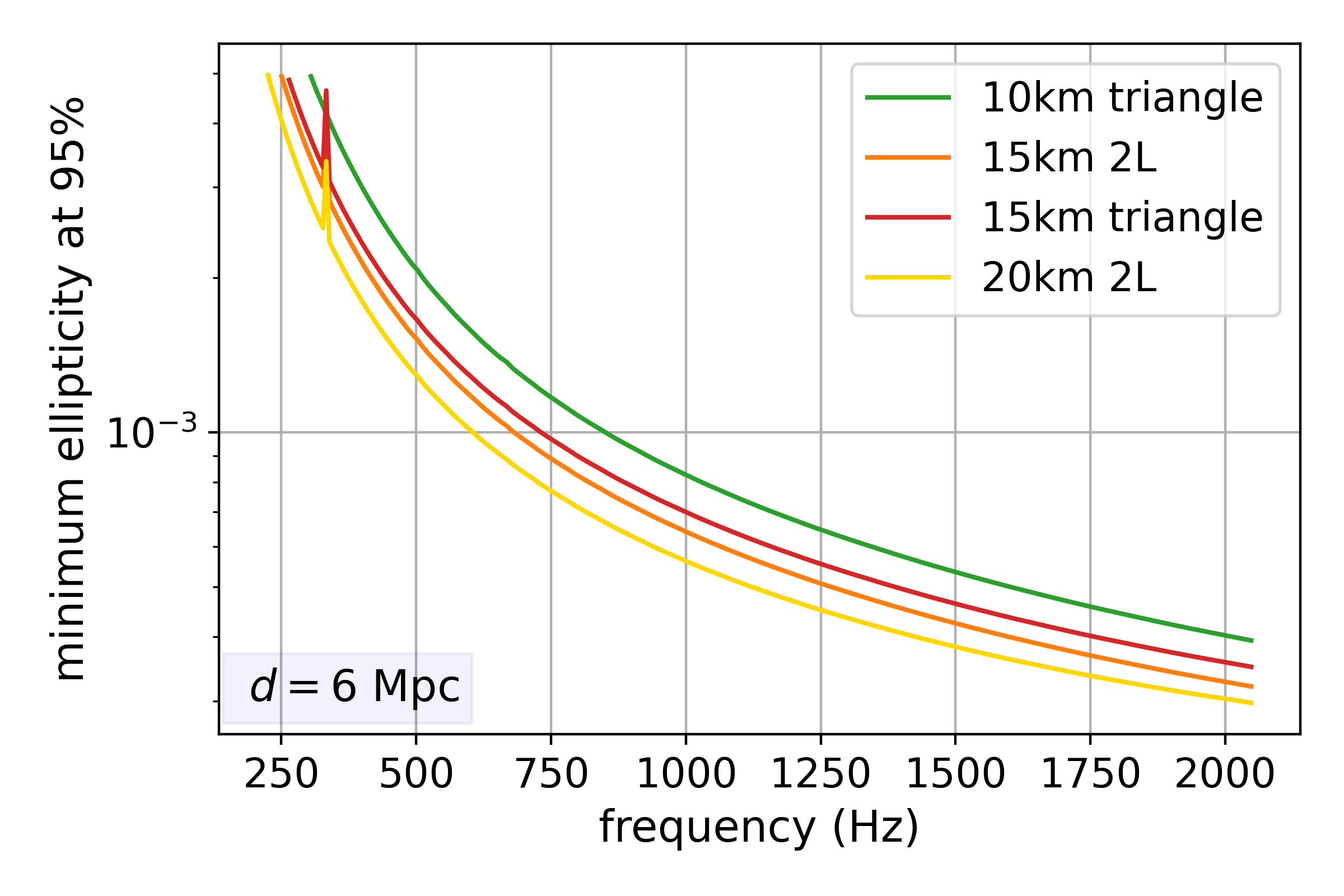}
    \caption{\small Left: ET distance reach to transient CW signals emitted by remnants of binary neutron star mergers or supernovae for various arm lengths and shapes, fixing the configuration to allow both high and low-frequency enhancements. The ellipticites used to construct these curves are the maximum allowed by energy conservation, assuming that all rotational energy is emitted in GWs, and imposing  $\epsilon\leq5\times 10^{-3}$, the largest expected degree of deformation. At frequencies for which the ellipticity given by energy conservation exceeds $5\times 10^{3}$, we impose $\epsilon=5\times 10^{-3}$. Right: ET minimum detectable ellipticity for different configurations for newborn neutron stars 6 Mpc away. Both plots use $\dot{f}_0\leq -1\times 10^{-4}$ Hz/s, $T_{\rm FFT}\sim 70$ s, $T_{\rm obs}=1$ day, and $I_{\rm zz}=10^{38}$ kg$\cdot$ m$^2$. }
    \label{fig:tcwsens}
\end{figure}

In the left-hand panel of Fig.~\ref{fig:tcwsens}, we plot the expected distance reach for different ET configurations, for a braking index $n=5$, i.e. pure GW emission. The ellipticity at each frequency is the maximum allowed by energy conservation, and we restrict to ellipticities less than $5\times 10^{-3}$, the largest expected for newborn neutron stars \cite{DallOsso:2018dos}. However, at frequencies below $\sim 1200$ Hz, the ellipticity given by energy conservation is much larger than that allowed theoretically; hence, we impose $\epsilon=5\times 10^{-3}$ up to 1200 Hz.
In this regime, we see that the distance reach steadily increases with frequency at a fixed $\epsilon=5\times 10^{-3}$, which is consistent with \eq{eq:tcwdmax}. For frequencies greater than $1200$ Hz, the distance reach decreases slightly, since at this point, the ellipticity allowed by energy conservation is in fact less than $5\times 10^{-3}$. 
Magnetars formed in core collapse supernovae look more promising, in terms of event rate, with respect to those following the merger of NS binary systems. The overall core collpase supernova rate is of the order of one event per year within 10~Mpc (4 events/yr within 20 Mpc) \cite{2017APS..APR.C3001G}. The fraction of core collapses leading to the formation of a magnetar is uncertain and, while past estimations pointed to the  $(1-10)\%$ range \cite{Kouveliotou:1998ze}, more recent results, based on observations and a more proper consideration of selection effects, typically provide higher values, even of the order of 40$\%$, see e.g. \cite{Beniamini:2019bga}. This would correspond to a magnetar formation rate of $\sim 2$ per year within 20 Mpc, which is well within the search distance reach for a wide range of initial spin frequencies, for the optimistic deformation rate discussed above.   

In the right-hand panel of Fig.~\ref{fig:tcwsens} we plot the minimum detectable ellipticity at a fixed distance of 6 Mpc, at which we expect to detect magnetars forming at a rate of  one per 5-10 years. At such a distance, the only low-frequency ($\lesssim 20$ Hz) sources that could produce a detectable gravitational-wave signal are those with unphysical ellipticities, i.e. $\epsilon>1$, so we omit those on the plot. We only consider frequencies at which the minimum detectable ellipticity is less than $5\times 10^{-3}$, which is consistent with the left-hand panel. In fact, at such distance, we would be able to detect the tCW from newborn magnetars with ellipticity of the order $10^{-3}-10^{-4}$, much lower than the maximum predicted value.

{\em We see that the differences among configurations are mostly driven by the arm-length, and 
there is little difference between the 2L-15km and 15km triangle}. As in the CW case, we have applied ${\rm CR}_{\rm thr}=4$ for the 2L case, and ${\rm CR}_{\rm thr}=3.4$ in the triangle case to capture the anticipated improvements in both computational power in the 3G era and improved coincidences resulting from a triangle-shaped detector. Furthermore, these plots are created with the full HFLF xylophone configuration. {\em If we were to remove the low-frequency instrument in the left-hand panel, the distance at low frequencies $ \lesssim 100$ Hz (not shown)
decreases from $\mathcal{O}(1-10)$ kpc to $\mathcal{O}(0.1-1)$ kpc. This could be relevant, since the enhancements due to the LF instrument allow us to reach the galactic center at low frequencies.} However, in the right-hand panel, the results do not vary at all, since the ellipticities are already unphysical at frequencies below 30 Hz (not shown).

\subsubsection{Search for dark matter with CWs}
GW detectors could probe the existence of some kinds of dark matter (DM)~\cite{Bertone:2016nfn}. In many cases, DM has been predicted to produce CW-like signals, so that data analysis methods used for the search of ``standard'' CW or tCW have been extended, see e.g. \cite{DAntonio:2018sff}, and applied to the search for DM signatures. 
Ultralight bosons are interesting DM candidates, including dark photons and Quantum ChromoDynamics axions~\cite{Peccei:1977hh,Arvanitaki:2009fg,Arvanitaki:2010sy,Brito:2015oca}.
Potential detectable systems might be boson clouds formed around spinning BHs or vector bosons in form of dark photons. Detectable continuous signals can arise also from compact dark objects or  primordial black holes.
\paragraph{Boson clouds around black holes}

Ultra-light bosons fields present in the nearby regions of a Kerr black hole can clump around it through a superradiant mechanism, forming a BH-boson ``cloud'' system at the expense of the BH angular momentum~\cite{Arvanitaki:2014wva,Arvanitaki:2010sy,Brito:2015oca}. 
This formation channel is maximally efficient when the particles's Compton wavelength is comparable to the size of the BH, 
$\hbar/(m_bc)\simeq GM_{\rm BH}/c^2$,
where $m_b$ is the boson mass, $M_{\rm BH}$ is the black hole mass, $\hbar$ is the reduced Planck's constant, and $c$ is the speed of light.  
Once formed, the cloud starts to dissipate via particle annihilation to gravitons, emitting a quasi-monochromatic and long-duration signal with a frequency dependent mainly on the boson mass, 
\be
f_\text{gw} \approx 483\,{\rm Hz} \left(\frac{m_\text{b}c^2}{10^{-12}~\text{eV}}\right)\, .
\ee
Earth-bound detectors are mostly sensitive to boson masses in the $10^{-14}-10^{-11}~{\rm eV}/c^2$ range~\cite{KAGRA:2021tse}. The BH-boson cloud system is often referred to as a gravitational atom.
The emitted signal is characterized by a spin-up due to the cloud mass decrease, which reduces its binding energy \cite{Baryakhtar:2017ngi,Isi:2018pzk}, as well as to boson self-interaction~\cite{Baryakhtar:2020gao}. 
The GW signal amplitude is 
\be
h(t)=h_0\left(1+t / \tau_{\mathrm{gw}}\right)^{-1}\, ,
\ee
where $h_0$ is mainly determined by the BH and boson masses, the BH initial spin and the distance, through
\begin{equation}
h_{0} \simeq 3 \times 10^{-24}\left(\frac{\alpha}{0.1}\right)^{7}\left(\frac{\chi_{i}-\chi_{c}}{0.5}\right)\left(\frac{M_{\mathrm{BH}}}{10~\mathrm{M}_{\odot}}\right)\left(\frac{1~\mathrm{kpc}}{d}\right)\, ,
\end{equation}
where $\alpha$ is the the fine-structure constant of the gravitational atom, given by
\begin{equation}
    \alpha = \frac{G M_{\rm BH}}{c^3} \frac{m_bc^2}{\hbar}\, . \label{eq:alpha}
\end{equation}
The gravitational-wave timescale $\tau_{\mathrm{gw}}$ is also dependent on the BH mass, the boson mass and the BH initial spin~\cite{Brito:2015oca, KAGRA:2021tse,Isi:2018pzk}. 
The detection of CW signals from boson clouds would allow to establish a fascinating connection among particle physics and black holes. Null results, as those obtained so far, see e.g. \cite{Palomba:2019vxe,Sun:2019mqb,KAGRA:2021tse}, brings anyway valuable constraints on the permitted boson masses.

\paragraph{Dark-matter particles interacting with GW detectors}

Another interesting scientific scenario for ET are ultralight particles 
behaving as a classical field interacting coherently with the atoms of the test masses.
For instance, dark photons coupled to the baryon or baryons minus leptons number $U(1)_B/U(1)_{B-L}$ have been considered in literature. These particles can produce an oscillating force on dark charged objects~\cite{Pierce:2018xmy,Graham:2015ifn,Carney:2019cio}. 
The same type of behavior is expected for tensor particles~\cite{Armaleo:2020efr}. 
Different production mechanisms have been proposed for the dark photon production, such as misalignment mechanism associated with the inflationary epoch, light scalar decay or cosmic strings~\cite{Co:2017mop,Co:2018lka}.
The dark photon DM field oscillations impinge a time-dependent Equivalence Principle-violating force acting on the test masses. This will produce a change in the relative length of the detector's arms, causing a signal strain at the detector output.
The time-dependent force acting on the test masses produces a strain oscillating at the same frequency and phase as the dark photon field~\cite{Pierce:2018xmy,Guo:2019ker,Morisaki:2020gui,Michimura:2020vxn}. It should be noted that a similar signal may be produced in the case of scalar dark matter particles directly interacting with the mirrors. In particular, as described by~\cite{Vermeulen:2021epa}, the interaction of the scalar bosons with the detector beam-splitter will induce oscillations in the thickness of the mirror due to the oscillations of the fundamental constants~\cite{Stadnik:2015xbn,Stadnik:2014tta,Grote:2019uvn}. They could also interact with the particles in the reference cavity in the detector, resulting in a similar measurable oscillation of fundamental constants \cite{Hall:2022zvi}.
The signal frequency is determined by the dark photon mass, $f_{0}=m_{A}c^2/(2\pi\hbar)$, corresponding for Earth-based detectors to dark photon masses in the range $10^{-14}-10^{-11} \mathrm{~eV}/c^2$. In fact, the signal is not exactly monochromatic due to the fact 
dark photon particles follows a Maxwell-Boltzmann velocity distribution.


Taking into account the physics of dark-matter signals, different CW methods based on cross correlation and excess power have been applied to look for these types of signals \cite{Pierce:2018xmy,Guo:2019ker,Miller:2020vsl,Miller:2022wxu}, and have resulted in competitive constraints on $U(1)_B$ coupling using data from the most recent observing run of advanced LIGO/Virgo/KAGRA \cite{Morisaki:2020gui,LIGOScientific:2021ffg}, and on scalar dilaton dark-matter coupling to electrons and photons using GEO600 data \cite{Vermeulen:2021epa}. Furthermore, projected constraints for axions altering the polarization of light shining down the interferometer arms have been produced \cite{Nagano:2019rbw,Nagano:2021kwx}.

When considering the different configurations and geometries, we note that the 2L geometry would experience a larger strain due to dark photons by a factor of $2/\sqrt{3}\sim 1.15$ compared to a triangle-shaped detector, independent of the arm-length \cite{Pierce:2018xmy}. Since any frequency (or mass) for the dark photon or other dark matter particle is possible, both the low- and the high-frequency instruments  of the ET xylophone configuration are important. Furthermore, increased arm length is also the primary way to improve sensitivity to these particles, since they behave just as traditional CWs, though $T_{\rm FFT}$ could not be as long as 10 days because of the frequency modulation induced by the Maxwell-Boltzmann distributed velocities. Finally, the triangle-shaped detector would allow triple coincidences, but since the dark-photon signal would show up looking like ``correlated noise'' between the detectors, having a single triangle-shaped detector in one place may increase the number of correlated noise disturbances, reducing the sensitivity to dark-matter particles.

\paragraph{Inspiraling compact dark objects and PBHs.}

Dark matter can be present in our universe in form of macroscopic objects, we generically refer to as compact dark objects (\acrshort{cdo}s). The actual formation channel and the origin of CDOs is still widely debated. 
CDOs can form pairs and emit an almost monochromatic signal while being far from the coalescence phase. Primordial BHs (PBHs), that we have discussed in detail in Sections~\ref{sect:PBHhighz} and \ref{sec:otherPBHs},  are an example of CDOs. 
In the context of CW signals, binary systems made of sub-solar mass PBHs as well as more generic CDOs, represent potential targets. Indeed, the GW signals emitted by these systems, when the two compact objects are far away from the coalescence and their masses are small enough, say $<10^{-2}~M_\odot$, i.e below the mass range considered in 
Section~\ref{sec:otherPBHs}, can be modeled as CW or tCW signals with a spin-up described by a power law, see e.g.~\cite{Miller:2020kmv,Guo:2022sdd}. 

If the chirp mass is small enough, the frequency can be modeled exactly as the linear frequency Taylor expansion used for standard CW searches. This means that, for instance, a pair of inspiralling PBHs with chirp masses below $10^{-5}~M_{\odot}$ would emit a GW signal indistinguishable from those arising from non-axisymmetric rotating NSs spinning up, and upper limits have already been placed on the fraction of dark matter that PBHs could compose $(f_{\rm PBH})$ using CW results from the O3a and O3 observing runs \cite{Miller:2021knj,KAGRA:2022dwb}. 

These upper limits on $f_{\rm PBH}$ derived from CW searches are of particular importance, since they lie within the ``asteroid-mass'' PBH mass regime. In this mass range, there are almost no existing constraints on $f_{\rm PBH}$ \cite{Green:2020jor}. ET will allow us to probe realistic $f_{\rm PBH}$ in the asteroid-mass regime (currently upper limits on $f_{\rm PBH}>$1 from these searches), and will result in an improvement of 1-2 orders of magnitude, regardless of the detector configuration or geometry \cite{Miller:2020kmv}. The same comparison of different detector designs discussed in Section~\ref{sec:cwns} for all-sky searches would be applicable here, since the constraints on $f_{\rm PBH}$ are derived from all-sky search results.

\subsubsection{Conclusions}

We have considered here the impact of different ET geometries and configurations on the detectability of persistent GW signals. In the case of CWs from slowly spinning down asymmetricaly rotating neutron stars, we find that the number of detectable sources in targeted searches is mostly sensitive to the arm length while, for realistic degrees of deformation to high-frequency enchancements,  the shape of the detector does not have a relevant role. Furthermore, in all-sky searches, we expect to be able to use a  value of $T_{\rm FFT}$ that is $\sim 100 $ times longer than used in analyses today, and thus find that we can constrain the ellipticity of unknown neutron stars to be less than $1\times 10^{-7}$ at frequencies above 300 Hz from the galactic center. This sensitivity is also mostly a function of the arm-length, and does not vary significantly between design choice.

For transient CWs, the arm length is also the most important factor that determines the relativity sensitivity of the different configurations considered, with the low-frequency sensitivity only helping to reach the galactic center at signal frequencies below $\sim 30$ Hz. However, we considered sources with initial spin-downs one to three orders of magnitude smaller than those searched for in \cite{LIGOScientific:2018urg}, determined by the maximum allowed ellipticity predicted theoretically. This limited our distance reach by a couple of orders of magnitude compared to the ellipticities  $10^{-2}-10^{-1}$ considered in \cite{LIGOScientific:2018urg}.

We have hinted at the possibility of constraining exotic physics with CW and tCW searches, and note that the results presented here can be recast in terms of constraints on, e.g. the fraction of dark matter that PBHs could compose, the maximum coupling strength to baryons that dark-matter particle could have, or boson/black hole mass pairs that could be excluded from existing.

All these results must be interpreted carefully, since they assume Gaussian noise and a lack of detector artifacts. In practice, glitches and noise lines will affect the sensitivity to all of these sources. Noise lines will inhibit dark-matter and CW searches the most, since they focus on persistent, long-lived GWs, while glitches are more problematic for the transient CW searches. We have tried to capture the impact of improved coincidences by lowering ${\rm CR}_{\rm thr}$ for the triangle-shaped detector compared to the 2L one; however, these thresholds are subject to change in practice based on available computing power and the characteristics of the noise. We do not consider explicitly how the so-called ``null stream'' would impact the detectability of these sources, though we expect that our methods, already designed to be robust against noise disturbances and glitches, can achieve comparable sensitivities as those obtained in Gaussian noise, based on current GW searches.

\section{The role of the null stream in the triangle-2L comparison}\label{sect:nullstream}

The null stream is a signal-free linear combination of the interferometer strain data, and it plays an important role in the comparison of 2L and triangle ($\Delta$) configurations of ET.
A null stream can be formed for an arbitrary GW detector network with at least three detectors, but generally only for one GW signal at a time~\cite{Guersel:1989th}.
However, the $\Delta$ configuration allows one to form a null stream that cancels out all gravitational-wave signals simultaneously~\cite{Regimbau:2012ir}, because of its closed geometry of component arms and almost negligible light travel times between components.
The null stream in the $\Delta$ configuration (from now on just ``null stream'') has the unique ability to provide access to a signal-free channel, and therefore access to the properties of (incoherent) instrumental noise without contamination from gravitational-wave signals~\cite{Goncharov:2022dgl}.

It should be stressed that inference from data with the null stream noise discarded is identical to coherent inference from the full set of data so that access to the null stream does not directly improve inference~\cite{Wong:2021eun}.
Moreover, a 2L configuration without access to the null stream may still achieve good noise mitigation but requires more complex methods that introduce uncertainty into the process.
Access to the null stream provides a more straightforward way (compared to the 2L configuration) to extract the power-spectral density (PSD) of the detector noise and enables unique data-analysis techniques for improved science extraction.
In the following, we will describe some of these techniques and their impact on science extraction.

\paragraph{Estimation of unbiased noise power spectrum:}
Within the $\Delta$ configuration of ET, the null stream $d_{\rm null}$ can be straightforwardly formed by
\begin{equation}\label{eq:null:et}
	d_\text{null} = d_1 + d_2 + d_3,
\end{equation}
where $d_i$ are the strain data for the individual detectors in the triangle.
If the noise properties are homogeneous and incoherent among these detectors, then the noise PSD of each detector $S^i_n$ can be estimated by
\begin{equation}
	S^i_n = \frac{1}{3}S^{\rm null}_n,
\end{equation}
where $S^{\rm null}_n$ is the PSD of the null stream.
In the absence of a null stream, it is non-trivial to disentangle the detector noise from the unresolvable GW signals, which is also referred to as ``confusion noise''.

Assuming uncorrelated noise, another method to estimate the instrument-noise PSD is by calculating the cross power-spectral density (CSD) of the null stream with the data streams \cite{Goncharov:2022dgl}:
\begin{equation}\label{dnulldi}
	\left< d_{\rm null} d^*_{i} \right> = S^i_n,
\end{equation}
where $S^i_n$ is the PSD of detector $i$. This method does not require the PSD of different interferometers to be the same.

Noise can be produced identically in two interferometers, which would therefore not appear in the null stream and bias PSD estimates that use the null stream. Technically, the bias in PSD estimates is produced by the real part of the complex-valued instrument-noise CSDs between interferometers. The natural magnetic background is a possible source of such noise. Also seismic gravity fluctuations are considered a possible source of identical noise. However, over the $>400\,$m distances between any two test masses of two LF interferometers, this so-called Newtonian noise is most likely weakly correlated \cite{Harms:2019}, which means that one should only expect a mild suppression in the null stream. See Section~\ref{sect:corrnoise} for a more in-depth discussion of these noise sources. Ref.~\cite{Janssens:2022cty} shows that, given knowledge of the correlated/non-identical noise sources (e.g. through witness sensors), unbiased instrument-noise PSD estimates can in principle be obtained even in the presence of correlated noise.

\textit{Impact on science extraction:}
The most direct impact of an unbiased estimate of the noise power spectrum is on the measurement of the stochastic gravitational-wave background.
By subtracting the null stream PSD from the individual detector PSD, one obtains the PSD of the GW signals $S_h$ present in the data
\begin{equation}
\label{eq:nullstreamsubtraction}
	S_h \simeq S^i_n - \frac{1}{3}S^{\rm null}_n,
\end{equation}
where the equality holds when the noise is identical amongst individual detectors (see e.g. Refs.~\cite{Goncharov:2022dgl} and \cite{Janssens:2022cty} for discussion on deviations from this assumption).
$S_h$ can be directly related to the SGWB if the data contains no resolvable signals.
Otherwise, one has to first subtract the loud sources from the data (see e.g. Refs.~\cite{Wu:2022pyg, Sachdev:2020bkk}).

\begin{figure}[t]
	\includegraphics[width=1.0\textwidth]{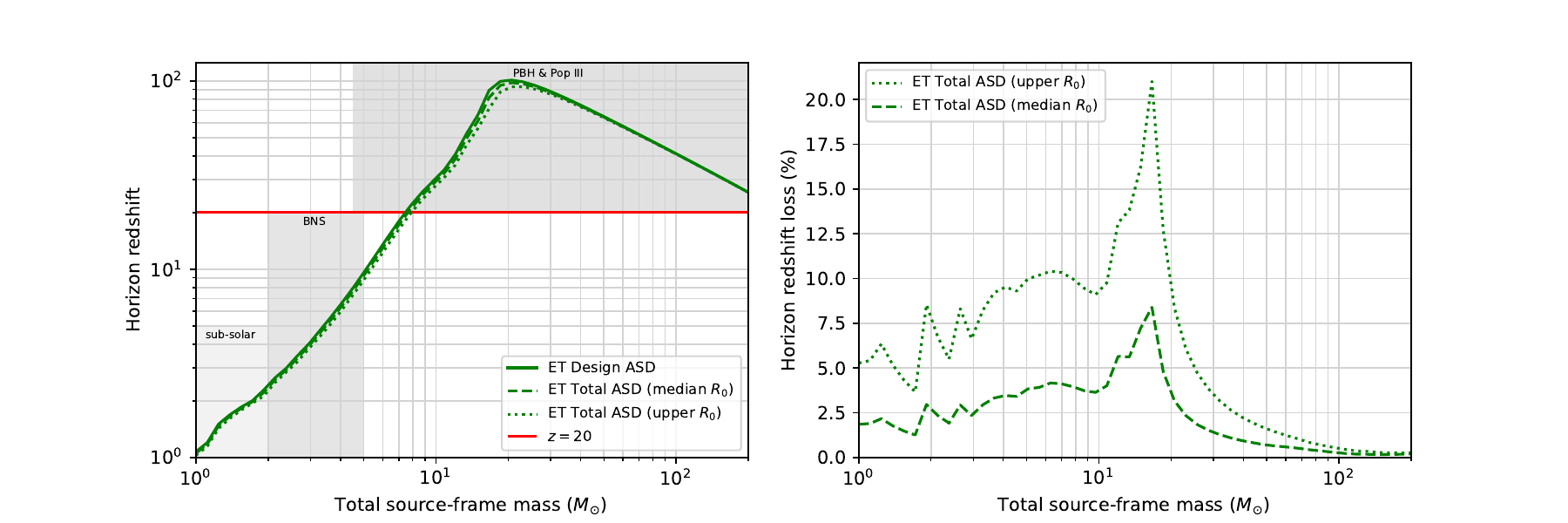}
	\caption{\small The loss in the ET detection horizon, as estimated in~\cite{Wu:2022pyg}, due to the confusion of instrumental noise and compact binary inspirals. Dashed lines show the effect of the median local merger rate and dotted lines represent the upper merger rate. A higher merger rate introduces a stronger confusion noise and hence the higher horizon redshift loss. This effect can be mitigated by the ET null stream available in the $\Delta$ configuration.}
	\label{fig:null:horizon}
\end{figure}

The inability to disentangle detector noise from confusion noise also has the effect of raising the overall perceived noise level.
In a templated GW search, the effect manifests as a loss of matched filtering SNR, and it is demonstrated in Fig.~\ref{fig:null:horizon}.
Fig.~\ref{fig:null:horizon} shows the CBC horizon redshift (left) and its percentage loss (right) as a function of the total source-frame mass and for different merger rates.
The loss in the redshift reach of ET increases with a higher merger rate of compact binaries, and for BBHs it may appear between $\approx 2.5\%$ and $\approx 20\%$, depending also on the mass of the CBC.
For the $\Delta$ configuration, the efficiency for mitigating this effect approaches 100\% as the sensitivity of all ET components approaches the same level.

Furthermore, the confusion noise impacts a calculation of the false alarm rate (\acrshort{far}),  the probability of incorrectly rejecting the noise hypothesis. 
The FAR is calculated by comparing the noise-induced (background) distribution of a detection statistic (e.g. SNR for matched filtering) with the one measured from a signal candidate.
For example, one could perform matched filtering on time shifted data (between the individual detectors) to estimate the background distribution (see e.g. Ref.~\cite{Was:2009vh}).
FAR calculations typically assume that the number of genuine GW detectable signals is low enough not to impact the estimate, which will not be the case for ET anymore.
By estimating the noise background distribution directly from the null stream (e.g. time shifting the null stream with respect to itself), one could mitigate the contamination from confusion noise and thereby improving the detection sensitivity and accuracy.

\paragraph{Mitigation of transient detector glitches:} 
A number of known instrumental noise artefacts such as glitches invalidate the standard assumption that the data is described by a superposition of the Gaussian instrumental noise and GW signals. If unmodeled, they will appear as false positives, whereas modeling glitches introduces additional uncertainties. Unlike GWs, these artefacts will appear as non-Gaussian outliers in the null stream, which provides an opportunity to remove them.
Ref.~\cite{Goncharov:2022dgl} outlines two methods for eliminating glitches with the null stream. First, one can identify glitches by match-filtering them in the null stream. Second, one can identify null stream data segments that are inconsistent with Gaussian noise. The authors show that it is possible to end up with a clean Gaussian background in the $\Delta$ configuration in the limit where ET components approach the same sensitivity.

\textit{Impact on science extraction:} 
The non-Gaussian part of the background causes extended tails in the noise background and therefore limits the FAR.
The exact extent to which this happens will only be known when ET is operational.
Current gravitational wave observatories not only discover more GWs while achieving a better sensitivity but also retract more spurious signals~\cite{Cabero:2020eik}.
This increasing trend might continue down to the design sensitivity of the ET. 
The null stream will reduce the uncertainty associated with modeling glitches, and allow us to mitigate the effects of non-Gaussian noise.
The most prominent benefit of the null stream is expected for high-mass distant BBHs that mimic glitches as well as for signals that are only described by phenomenological models.

\paragraph{Control of known and unknown systematic errors:}
Any errors in the detector calibration can propagate into the null stream to cause incomplete cancellations of gravitational wave amplitudes \cite{Chatterji:2006nh}.
If the signal waveform is a-priori well understood, and if its parameters are well-measured by the network of detectors, then the residual signal in the null stream will be the product (in the frequency domain) of the calibration error and known weighted amounts of signal~\cite{Schutz:2020hyz}.
One can detect this residual by performing matched filtering on the null stream.
The calibration error can be obtained by fitting with a family of specific functions supplemented by the SNR output of the matched filters over a number of detected events.
The ET null stream in the triangle configuration reduces the uncertainty in calibration parameters, which may propagate to GW  signal parameters. However, the level of systematic errors in detector calibration in ET is not known, and even for current detectors it is only known to be less than $2\%$~\cite{Sun:2020wke}. Therefore, it is not clear how many signals in ET will be affected by these systematic errors. Further studies are also necessary to demonstrate the performance of ET null stream calibration compared to other techniques, as well as the impact on the gravitational wave parameter estimation. In the worst case scenario, in 2L detectors calibration uncertainties will have to be estimated simultaneously with gravitational-wave source parameters, which would increase the computational cost of the analyses and introduce additional uncertainty. 
Note also that 
Ref.~\cite{Schutz:2020hyz} estimates that, in a network of at least three L-shaped 3G detectors,  the calibration error can be inferred at the percent level if supplemented with $O(100)$ relatively loud (SNR=20) events.

\section{Summary} \label{sect:summary}

In this work we have performed a  detailed study of the Science Case of ET,  significantly expanding the study in ref.~\cite{Maggiore:2019uih}. We have
considered both  ET in its reference design (a 10~km triangle in a xylophone configuration, with an HF  instrument and a cryogenic LF  instrument), as well as variations on the reference design, both in the geometry (triangle vs 2L, with different arm lengths) and in the ASD (full sensitivity vs HF instrument only).

\vspace{2mm}\noindent
First of all, the significantly more detailed study of the Science Case performed here confirms and provides  more detail and evidence for the  picture summarized in  ref.~\cite{Maggiore:2019uih}: {\em ET, in its reference configuration, is a superb detector, with an extraordinary discovery potential in the domains of astrophysics, cosmology and fundamental physics.} With order $10^4-10^5$ BBH, BNS and NSBH detections per year, it will  address and provide answers 
to an extremely  rich and varied set of scientific questions; at the same time, it will penetrate deeply into unknown territories, where revolutionary discoveries could await for us. 

\vspace{2mm}\noindent
Starting from the extraordinary potential of the reference design, it is, however, still mandatory to study what happens under variations of this baseline. At the very least, this allows us to understand how the different elements of the design, such as its geometry or the relative roles of the  HF and LF instruments, concur to produce the scientific output of ET, and what could be reached in intermediate stages of the commissioning of the detector. Furthermore, this study can suggest directions where changes of the baseline design could results in  improvements of the science output. 
The original structure of the ET design was first laid down 10-15 years ago~\cite{Hild:2008ng,Punturo:2010zz,Hild:2010id},  well before the first detections of GWs. At that time, one was not even sure that there was, out there, a population of BBHs that coalesced within a Hubble time.
Now we have a statistically significant catalog of detections, and building on the understanding that has accumulated thanks to these discoveries and the studies that they have stimulated,  we are now able to present in this paper a study based on astrophysically-motivated populations.\footnote{Last but not least, the setting up and the development of the ET Observational Science Board allowed the formation of a coherent scientific community focused on ET, with the necessary expertise across the rather broad range of subjects needed for such a study.}

\subsection{Comparison of different geometries}

In this paper we have examined several different geometries, comparing two L-shaped detectors on widely separated sites to a single-site triangle, and with different choices of arm-length (and relative orientations for the two L-shaped detectors), to provide a  broad set of options.
We will begin by summarizing the main results from the comparison of the 15~km 2L configuration with the  10~km triangle, in Section~\ref{sect:152L10T}. In Section~\ref{sect:152L15T} we will instead summarize the results of the comparison between the 15~km 2L and the  15~km triangle. In Section~\ref{sect:1L} we also  summarize the results for a single third-generation L-shaped detector. Further elements of the comparison between geometries, including correlated noise, the null stream, etc., are summarized in Section~\ref{sect:further}.

\subsubsection{Comparison between 15~km 2L and 10~km triangle}\label{sect:152L10T}

Focusing first on the  comparison of the 15~km 2L with the  10~km triangle, the main results 
can be summarized as follows.

\begin{itemize}

\item {\bf Binary Black Holes}. The results of Section~\ref{sect:PEBBH} show that, for BBHs, the 
 {\rm 2L} configuration with 15~km arms, oriented at $45^{\circ}$, is superior to a 10~km triangle for the estimation of all parameters, and especially luminosity distance, see  Fig.~\ref{fig:AllGeoms_CumulBBH_NdetScale} in the main text and Tables~\ref{tab:BBHAllConfSNR}--\ref{tab:BNSAllConfSNR} in App.~\ref{app:TablesCBC}.
To put things into perspective, for all parameters (except luminosity distance) the differences are at the level of factors of 2-3, while both configurations outperform LVKI-O5 by orders of magnitudes on all parameters (with less large, but still significant  differences for angular localization), see Figs.~\ref{fig:ETS_T_10km_CumulBBH_NdetScale} and \ref{fig:ETSMR_2L4515_CumulBBH_NdetScale}. For instance, while all ET configurations detect basically the whole  BBH population with ${\rm SNR}\geq 12$, if we restrict to `golden events' with ${\rm SNR}\geq 100$ the 2L-15km-$45^{\circ}$ configuration detects (in our sample realization) 4933 BBH/yr, while the 10~km triangle 2298, and LVKI-O5 only 4. For ${\rm SNR}\geq 200$ these figures becomes 644, 282 and 2, respectively; see Table~\ref{tab:BBHAllConfSNR} in App.~\ref{app:TablesCBC}. For most parameters, as mentioned above, the differences between the 
15~km 2L at $45^{\circ}$ and  the 10~km triangle are at the level of factors 2-3, in favor of the 2L; however, they  are significantly larger for the accuracy on  luminosity distance: in the 2L-15km-$45^{\circ}$ configuration there are $202$  BBH/yr
with error on $d_L$ smaller than $1\%$,  
to be compared with $28$ for the 10~km triangle (and, in our sample realization, just 1  for LVKI-O5). For angular localization, focusing  on the events with $\Delta\Omega_{90\%}\leq 50\, {\rm deg}^2$ 
, we find 10304  BBH/yr for 
2L-15km-$45^{\circ}$, 6064 for the 10~km triangle and 1607 for LVKI-O5, 
while for $\Delta\Omega_{90\%}\leq 10\, {\rm deg}^2$ these numbers become 2124, 914 and 599, respectively;
see  Table~\ref{tab:BBHAllConfDeldLDelOm}.\footnote{Here the difference with LVKI-O5 becomes smaller: the events with very accurate angular localization are in general the closest, so they are accessible also to LVKI-O5, which can have good localization thanks to its  network of five widely separated detectors.} The corresponding difference in the joint accuracies on $d_L$ and angular resolution shows that, also for this metric,  the 2L-15km-$45^{\circ}$ is clearly superior to the 10~km triangle, see
Fig.~\ref{fig:scatter_dLOm_2L4515kmvsT10and15km_BBH}.

In contrast, the 2L configurations with parallel arms are disfavored, because of their worse angular resolution; e.g., for 2L-15km-$0^{\circ}$ there are 3030 BBH/yr with $\Delta\Omega_{90\%}\leq 50\, {\rm deg}^2$), to be compared with 
10304  BBH/yr for  2L-15km-$45^{\circ}$; for $\Delta\Omega_{90\%}\leq 10\, {\rm deg}^2$ the difference is even larger, with 374 BBH/yr for  2L-15km-$0^{\circ}$ (so, almost a factor of two less than LVKI-O5), compared to 2124 for  2L-15km-$45^{\circ}$.

\item {\bf Binary Neutron Stars.} Similar conclusions follow from the study of parameter estimation for BNSs in Section~\ref{sect:PEBNS},  see Fig.~\ref{fig:AllGeoms_CumulBNS_NdetScale} and Tables~\ref{tab:BNSAllConfSNR}--\ref{tab:BNSAllConfDelMcDelLam}. Again the  2L with 15~km arms at $45^{\circ}$ improves on the already remarkable performances of the 10~km triangle, typically by factors of order 2-3, while both outperform LVKI-O5. For instance, 
if we restrict to `golden events' with ${\rm SNR}\geq 50$, the 2L-15km-$45^{\circ}$ configuration detects 1052 events per year, while the 10~km triangle 458 and LVKI-O5 only 3 (for ${\rm SNR}\geq 100$ these figures becomes 134, 57 and 0, respectively).
If we consider, instead, events with error on luminosity distance smaller than $10\%$, there is almost a factor of 10 between 2L-15km-$45^{\circ}$ and the 10~km triangle,
with 479  events/yr for 
2L-15km-$45^{\circ}$ and 52 for the 10~km triangle (and only 1 for LVKI-O5; as always, in particular for small numbers, these are the numbers for our specific sample realization). 
For angular localization, focusing on the events with $\Delta\Omega_{90\%}\leq 100\, {\rm deg}^2$, we find 559  events/yr for 
2L-15km-$45^{\circ}$, 184 for the 10~km triangle and 51 for LVKI-O5, see Table~\ref{tab:BNSAllConfDeldLDelOm} (however, for a handful of very close events, localizable to better than 10~${\rm deg}^2$, the five-detector network LVKI-O5 does better, and we get 25, 8 and 31 events, respectively).    The left panel of 
Fig.~\ref{fig:scatter_dLOm_2L4515kmvsT10and15km_BNS} shows  that, also for BNS,  
the 2L-15km-$45^{\circ}$ is clearly superior to the 10~km triangle for the  joint accuracy on $d_L$ and angular resolution, that determine the localization volume.

The {\rm 2L} configurations with parallel arms has again an  angular localization capability worse than that of the 2L configuration at $45^{\circ}$, but the effect is less strong compared to BBHs (see the  panel for $\Delta\Omega_{90\%}$ in Fig.~\ref{fig:AllGeoms_CumulBNS_NdetScale}).\footnote{This is due to the  long duration of BNS signals, which allow using the rotation of the Earth to localize the signal, also for the parallel setting.} On the other hand, the accuracy on the inclination angle $\iota$  for the parallel configurations is, comparatively, quite poor. Since $\iota$ is degenerate with the luminosity distance $d_L$, also for $d_L$ the parallel 2L configuration has significantly lower  performance compared to the configuration at $45^{\circ}$. For instance, for the 2L-15km-$0^{\circ}$ we get 48 BNS/yr with error on $d_L$ better than $10\%$, a factor of 10 lower than the  479 BNS/yr for
2L-15km-$45^{\circ}$.
As for angular localization, for  2L-15km-$0^{\circ}$ we get 293 BNS/yr localized better than  $100\, {\rm deg}^2$, so slightly better than the 10~km triangle, but about a factor of 2 worse than  the 559 BNS/yr for 2L-15km-$45^{\circ}$; see again Table~\ref{tab:BNSAllConfDeldLDelOm}.

For the parameter $\tilde{\Lambda}$ that encodes the tidal deformability of the neutron stars, we see  from Fig.~\ref{fig:AllGeoms_CumulBNS_NdetScale} and Table~\ref{tab:BNSAllConfDelMcDelLam} that, again, the 2L with 15~km arms  improves by a factor  $\sim 2$ the already remarkable performances of
the 10~km triangle, and both outperform LVKI-O5 by orders of magnitudes. For instance, considering the events with an error on $\tilde{\Lambda}$ smaller than $10\%$, we get 2463 BNS/yr  for 2L-15km-$45^{\circ}$, 1040 for the 10-km triangle, and 2 for LVKI-O5. For events with  an error on $\tilde{\Lambda}$ smaller than $5\%$
these numbers become 200, 96 and 0, respectively.

\item {\bf ET in a network with CE.} When ET is inserted in a network with Cosmic Explorer (either a single CE 40~km detector or two 2 CE detectors, one with 40~km and one with 20~km arms) the performance of the whole ET+CE network is, naturally, somewhat less sensitive to the choice of geometry for ET. However, the differences are still significant, particularly for the ET+1CE network. For instance, for ET+1CE, the number of BBHs with ${\rm SNR} > 200$ is $1.8\times 10^3$ when ET is in the 10~km triangle configuration, and raises to  $2.4\times 10^3$ for 2L-15km-$45^{\circ}$, while the number of BBHs localized better than $10\, {\rm deg}^2$ raises from $3.0\times 10^4$ to $3.6\times 10^4$, and the number of BBHs with luminosity distance measured better than $1\%$ raises from $2.9\times 10^3$ to $4.3\times 10^3$ (or from $5.3\times 10^3$ to $7.0\times 10^3$, for ET+2CE);  
see Fig.~\ref{fig:ETCE_CumulBBH_NdetScale} and Tables~\ref{tab:BBHAllConfSNR}, \ref{tab:BBHAllConfDeldLDelOm}.

For BNSs, at  ET+1CE, the number of detections/yr with ${\rm SNR} > 100$ is 312 when ET is in the  10~km triangle configuration, and raises to   418 for 2L-15km-$45^{\circ}$; the number of BNS/yr localized to better than $10\, {\rm deg}^2$ raises from $2.4\times 10^3$ to $3.8\times 10^3$; the number of BNS/yr with luminosity distance measured better than $1\%$ raises from $4.1\times 10^3$ to $7.9\times 10^3$ (or from $1.4\times 10^4$ to $1.9\times 10^4$, for ET+2CE); and the number of BNS/yr  with tidal deformability measured better than $5\%$ raises from 243 to 400 (or from 337 to 535 for ET+2CE); see Fig.~\ref{fig:ETCE_CumulBNS_NdetScale} and
Tables~\ref{tab:BNSAllConfSNR}--\ref{tab:BNSAllConfDelMcDelLam}.

In general, we see that the choice of geometry still has significant consequences also when ET is in a 3G network with one or two CE.

\item {\bf Multi-messenger astrophysics.}
For multi-messenger astrophysics, the results of Section~\ref{sect:MMO} show that, again, the 2L-15km-$45^{\circ}$ allows us to obtain a further improvement on the already remarkable performances of the 10~km triangle (and is comparable to the 15~km triangle), enabling the  observation of a larger number of well-localized events, up to a larger redshift. The number of short GRBs with an associated GW signal increases by about 30\% making possible a joint detection at larger redshifts, and the number of expected kilonovae counterparts increases by a factor of 2. The better volumetric localization of 2L-15km-$45^{\circ}$ with respect to 10~km triangle, as shown in 
Fig.~\ref{fig:scatter_dLOm_2L4515kmvsT10and15km_BBH}, facilitates the search of the EM counterpart giving the possibility to use galaxy-targeting strategy and enables a more efficient removal of contaminants. For pre-merger alerts with sky-localization smaller than 100 $\rm deg^2$, the 15~km triangle is performing better than the 10~km triangle and the 15 km {\rm 2L} configuration, reaching almost the capability of the {\rm 2L} configuration with 20~km arms, see Table~\ref{table:premergerFull}. 

\item {\bf Stochastic backgrounds.} For stochastic backgrounds, the best configurations are either 2L with parallel arms or the triangles, with the 2L parallel being the best below about 100~Hz and the triangle above 100~Hz. The 2L with arms at $45^{\circ}$ is  worse than the 2L with parallel arms\footnote{See Section~\ref{sect:geometry} for the definition that we use  for the relative angle between the L-shaped detectors, which differs by about $2.51^{\circ}$ from the more standard one using the great circle connecting the two detectors.} below about
$200-300$~Hz, while the two become comparable above $200-300$~Hz (where, however, both are less good than the triangle); see Fig.~\ref{fig:ET_PLS_auto_cross_SNR_1a}. However,
to exploit the better sensitivity (below 200-300 Hz) of 2L-parallel with respect to 2L-$45^{\circ}$ for studies of cosmology, the subtraction of the astrophysical background is necessary, see Section~\ref{sect:sourceseparation}. An imperfect subtraction of CBC sources would pose a serious threat to reaching the nominal sensitivity to a cosmological background (see \cite{Zhou:2022otw,Zhou:2022nmt,Pan:2023naq} for recent discussions), possibly even wiping out any advantage of the 2L-parallel with respect to the 2L-$45^{\circ}$ configuration. 

The ability to resolve the angular distribution of stochastic backgrounds, through their lowest multipoles, is significantly better for the 2L configurations (both aligned and misaligned) compared to the triangle, see Fig.~\ref{fig:Nell}.

\end{itemize}

The analysis in Section~\ref{sect:ImpactSpecific} shows that (as could be expected) the performance of the various geometries,  with respect to the specific science cases studied, broadly follows the indications obtained from the more general metrics studied in Sections~\ref{sect:CBC}--\ref{sect:stochastic}. The 10~km triangle is a game-changer with respect to LVKI-O5, allowing a jump of orders of magnitudes, accessing physics that it is well beyond reach for 2G detectors. 
The 15~km 2L with arms at $45^{\circ}$  provides a further gain with respect to the 10~km triangle, typically by factors of order 2-3, depending on the specific scientific question. Some examples, from the various studies presented in this paper, are as follows. 

\begin{itemize}

\item Concerning tests of GR based on BH quasinormal modes, we see from Table~\ref{tab:spectro_snr} that the ringdown SNR of GW150914 would have had a remarkable value of 141 in the 10~km triangle, which would further raise to  192 in 2L-15km-$45^{\circ}$, inducing a corresponding difference in the reconstruction of ringdown frequencies and damping times, according to \eq{deltafSNRringdown}.  Furthermore, we see from Table~\ref{tab:spectro_results_1} that the 10~km triangle would detect  41 events/yr  with a ringdown SNR higher than 50, that raise to 110 events/yr for
the 2L-15km-$45^{\circ}$  (for a ringdown SNR higher than 100, these numbers become 4 and 10, respectively), resulting in a further difference in performance when stacking different events.

\item For nuclear physics studies, we see from  Table~\ref{tab:div6:Fisher_radius} that the  10~km triangle can reach a remarkable accuracy of 10.0~m on the radius of neutron stars (taken the same for all NS, just as a proxy for the actual hyper-parameters determining the NS equation of state); the 2L-15km-$45^{\circ}$ further brings it down to 6.4~m, thanks to a more accurate reconstruction of tidal deformability for individual events, and the stacking of a larger number of events. For the postmerger SNR, instead, we see from Fig.~\ref{fig:post_merger_fig1} that the differences between different geometries do not exceed the $(10-20)\%$ level, with the hierarchy fixed by the arm length.

\item For the reconstruction of the merger rate, Fig.~\ref{fig:MergerRate2LvsTriangle}  and Table~\ref{tab:MeanRatios} show that both 
the  2L-15km-$45^{\circ}$
configurations and the 10-km triangle lead to a correct reconstruction of the merger rate, although  the 2L-15km-$45^{\circ}$ configuration  leads to a more accurate reconstruction, by an average  factor  $\sim 3$ in the central values.

\item For primordial BHs, assuming the PBH population model mentioned in the text, 
the 10~km triangle would detect 77 events/yr at $z>30$ (a smoking-gun signature for a primordial origin), which is already quite remarkable.
The 2L-15km-$45^{\circ}$ configuration would further rise this by a factor  $\sim 3$, with 238 events/yr,   resulting in a limit on the PBH abundance stronger by a factor about 1.8; see Table~\ref{tabresPBH1}.

\item In cosmology, we see from  Tables~\ref{tab:H0Oma}--\ref{tab:lcdm}
that, for the tests   that we have performed, the results for the accuracy on $H_0$, on the dark energy equation of state and on modified GW propagation  from the various geometries analyzed  are quite similar.\footnote{Here we restricted to the 10 and 15~km triangles and the 15 and 20~km 2L at $45^{\circ}$.} In general,  the 2L-15km-$45^{\circ}$ can  improve the results of the 10~km triangle on the accuracy of $H_0$, $w_0$ or $\Xi_0$   by    factors  ranging between  $\sim (1.3-2.5)$, depending on the test.\footnote{Larger differences,  in favor of the 15km-2L-$45^{\circ}$ with respect to the 10km triangle (and, to some extent, also with respect to the  15km triangle), are expected when extracting cosmological information by
correlating dark sirens with galaxy catalogs, because of the much better volume localization
of the 2L-15km-$45^{\circ}$ configuration, see Fig.~\ref{fig:scatter_dLOm_2L4515kmvsT10and15km_BBH} 
for BBHs, and  
Fig.~\ref{fig:scatter_dLOm_2L4515kmvsT10and15km_BNS} 
for BNSs. A quantitative assessment, however,  also depends on the completeness of the galaxy catalogs that will be available when ET will be in operation.  Work on this is in progress.}

\end{itemize}

\vspace{3mm}
\noindent
To sum up, the first overall conclusion that emerges is that:
 
\vspace{5mm}
\fbox{\parbox{14cm}
{\em 1. All the triangular and 2L geometries that we have investigated can be the baseline for a superb 3G detector, that will allow us to improve by orders of magnitudes compared to 2G detectors, and allow us to penetrate deeply into unknown territories.}
}

\vspace{5mm}\noindent
How far we will penetrate into such territories, however, depends to some extent on the configuration chosen.  From the above discussion, it follows that

\vspace{5mm}

\fbox{\parbox{14cm}
{\em 2a. The 2L-15km-$45^{\circ}$ configuration  in general   offers  better scientific return  with respect to the 10~km triangle,   improving on most figures of merits and scientific cases, by factors typically of order 2-3 on the errors of the relevant parameters.}
}

\subsubsection{Comparison between 15~km 2L and 15~km triangle}\label{sect:152L15T}

We next  summarize   the main results of the comparison  between  the 15~km triangle and the  2L-15km-$45^{\circ}$ configuration, so in this case we change the geometry while keeping the arm-length fixed.  
The main results of this comparison are as follows:

\begin{itemize}

\item For parameter estimation of BBHs, the 2L-15km-$45^{\circ}$ is clearly superior to the 15~km triangle for the estimate of luminosity distance (with 202 events/yr with $d_L$ measured better than $1\%$, against 77 for the 15~km triangle), and quite similar for all other parameters, see
Fig.~\ref{fig:AllGeoms_CumulBBH_NdetScale} and the first row of Fig.~\ref{fig:ET_allgeom_ASD_BBH_distr_vs_z},  and Tables~\ref{tab:BBHAllConfDeldLDelOm} and \ref{tab:BBHAllConfDelMcDelchi} in App.~\ref{app:TablesCBC}. Correspondingly, also the number of sources with good overall volume localization (say, error less than $1\%$ on $d_L$ and angular resolution better than  50 ${\rm deg}^2$),  is  clearly better for the 2L-15km-$45^{\circ}$, see the right panel in 
Fig.~\ref{fig:scatter_dLOm_2L4515kmvsT10and15km_BBH}.

\item The same holds 
for parameter estimation of BNSs,  see Fig.~\ref{fig:AllGeoms_CumulBNS_NdetScale} and the first row of Fig.~\ref{fig:ET_allgeom_ASD_BNS_distr_vs_z}, and 
Tables~\ref{tab:BNSAllConfDeldLDelOm} and \ref{tab:BNSAllConfDelMcDelLam} in App.~\ref{app:TablesCBC}.
In particular, the 2L-15km-$45^{\circ}$ has 479 events with $d_L$ measured better than $10\%$, to be compared with
153 for the 15~km triangle (which become 4328 to be compared with 1756, requiring an error smaller than $30\%$);
see Table~\ref{tab:BNSAllConfDeldLDelOm}. Instead, for angular resolution, the 15~km 2L with arms at $45^{\circ}$ and the 15~km triangle are quite comparable  (e.g., for the BNS localized better than  $100\, {\rm deg}^2$, with a threshold ${\rm SNR}>12$  we find 559  events/yr for 2L-15km-$45^{\circ}$ and 479  for the 15km triangle, see Table~\ref{tab:BNSAllConfDeldLDelOm}, 
which become 644 and 764, respectively, with ${\rm SNR}>8$, see Table~\ref{tab:skyloccryoMM}. At this level, the comparison is of course affected also by sample variance).
The respective joint distribution of events with respect to distance  and angular localization is shown in the right panel of 
Fig.~\ref{fig:scatter_dLOm_2L4515kmvsT10and15km_BNS}, 
which again shows an overall preference for 2L-15km-$45^{\circ}$.
For all other parameters the performances are quite similar.

\item This reflects also on several  aspects of the science case, where we see that the  2L-15km-$45^{\circ}$ and the 15~km triangle have  similar performances; this can be seen, e.g., in  Tables~\ref{tab:skyloccryoMM}, \ref{tab_jointprompt_cryo}
and \ref{tab:jointKNFull} for multi-messenger astrophysics (except for the pre-merger alerts for which the 15~km triangle results are better, as shown in Table~\ref{table:premergerFull}); Tables~\ref{tab:spectro_snr} and \ref{tab:spectro_results_1} for physics near the BH horizon; 
Fig.~\ref{fig:err_tidal_deformability} for tidal deformability of exotic compact objects;
Table~\ref{tab:div6:Fisher_radius} for the measurement of the NS radii; Fig.~\ref{fig:post_merger_fig1} for the post-merger signal of BNS coalescences; Table~\ref{tab:MeanRatios} for the merger  rate reconstruction (where the 
2L-15km-$45^{\circ}$ is a factor of 2-3 better than the 15~km triangle); Table~\ref{tabresPBH1} and Fig.~\ref{fig:distance_error_highz} for primordial  BHs; or Tables~\ref{tab:H0Oma}--\ref{tab:xi0ncdm} for cosmology.

\end{itemize}

The overall conclusion, in this comparison, is that:

\vspace{5mm}

\fbox{\parbox{14cm}
{\em 2b. The 2L-15km-$45^{\circ}$ configuration and the 15~km triangle have  very  similar performances on all parameters both for BBHs and BNSs, except for luminosity distance, where the  2L-15km-$45^{\circ}$ configuration is better by a factor $\sim 3$ in the number events with  accurately measured distance.}
}

\subsubsection{A single L-shaped detector}\label{sect:1L}

Finally, we have considered the performance of a single L-shaped detector with 20~km arms, still with the ET characteristics in terms of ASD. The conclusion is that such a detector, taken as a single detector not inserted into a network with other 3G detectors, is not capable of delivering the science expected from the next generation of detectors. Its capability of angular localization of the source, and of reconstruction of the luminosity distance, would even be very much  inferior to what the LVK network is expected to reach by the end of the O5 run, see the corresponding panels in Figs.~\ref{fig:ETMR_1L_20km_CumulBBH_NdetScale} and \ref{fig:ETMR_1L_20km_CumulBNS_NdetScale} (actually, the number of well-localized sources detected per year would not even really improve on the results already obtained with the current GWTC-3 catalog of detections). This would 
result in a complete loss of all the science case aspects of 3G detectors related to multi-messenger astrophysics and to cosmology. Stochastic backgrounds of GWs, that in the frequency band of ground-based detectors can only be detected by correlating the output of two or more detectors,  would also not be accessible to a single 3G detector, giving up again the corresponding aspects of the science case related to early Universe cosmology and to astrophysics. Furthermore, with a single detector,  unmodeled burst signals and short signals due to massive BBH coalescences would be hard to distinguish from instrumental glitches, rendering problematic any confident detection of transient sources.

Therefore, a single L-shaped detector does not have a valid science case, even with 20~km arms (or longer).
If inserted in an international network with 2CE, such as detector could be a useful addition to the network. However, given the scale of the investment, ET must have a solid science case, independently of the decisions that will be taken by other funding agencies, unrelated to a European project. Therefore:

\vspace{5mm}
\fbox{\parbox{14cm}
{\em 3. A single L-shaped detector is not a viable alternative, independently of arm length. If a single-site solution should be preferred for ET, the detector must necessarily have the triangular geometry.}
}

\vspace{5mm}

\subsubsection{The null stream}\label{sect:nullsummary}

 The triangle has the advantage of having a null stream, where the GW signal cancels. This will certainly be beneficial. However, while the mathematics of the null stream is elegant, when one is confronted to its use  in the concrete setting of the experiment several issues arise, that make it difficult to quantify reliably its impact. 

First of all, the triangle null stream is only operative when the  interferometers in all three arms are on, so it is quite sensitive to the duty cycle. In this work we have assumed an uncorrelated 85\% duty cycle in each L-shaped detector, and in each of the three detectors  composing the triangle, see Section~\ref{sect:CBC}; in this case the null stream will be available for $(0.85)^3=61\%$ of the run time.
Assuming instead  an independent duty cycle of $80\%$ for each arm (the best duty cycle for the O3b LVK run is $79\%$ \cite{aLIGO:2020wna}), implies that the null stream will only be available for $51\%$ of the run time. However, the triangle design is considerably more complex than that of 2G detectors, with two interferometers in each detector (an HF interferometer, and a LF interferometer working at cryogenic temperatures), and it is difficult to reliably estimate what the duty cycle will be. In a (hopefully very pessimistic) scenario where each of the six interferometers has an independent  duty cycle of $80\%$, the null stream would only be available for  $26\%$ of the run time.

An advantage of the null stream is that it provides  a direct  estimate of the instrument-noise PSD, see \eq{dnulldi}. Here, however, enters the assumption that noise in different arms are uncorrelated. Noise  produced identically in two interferometers (as could be the case for the  natural magnetic background or seismic gravity fluctuations) would  not appear in the null stream, and would therefore bias the PSD estimates that use the null stream (unless these noise can be identified and characterized, e.g. with witness sensors). 

Assuming to have obtained an unbiased estimate of the PSD, the most direct impact would be on the measurement of the stochastic gravitational-wave background. An unbiased estimate of the PSD would also be beneficial for compact binary coalescences,  to avoid a reduction of the detection horizon from the confusion noise of the unresolved BNS (that, being long-lasting, can produce a continuous backgrounds, contrary to BBHs that rather produce a `popcorn noise, see Section~\ref{sect:stochastic}). However, here the advantage of the null stream should be weighted against the possibility of making longer arms. Fig.~\ref{fig:null:horizon} shows that, assuming a value of the local rate $R_0$ of BNS mergers equal to the median of the currently allowed range, the reduction in the horizon range for the 2L detectors, due to this confusion noise, would be between $2\%$ and $7.5\%$ for CBCs with  masses below $20\msun$, and negligible above. With extreme assumptions on the rate $R_0$, taken at the upper limit of the currently allowed range, the reduction would be between $5\%$ and $20\%$ for CBCs with  masses below $20\msun$, and again negligible above. The triangle null stream would be able to mitigate this reduction, possibly eliminating it completely, which is certainly very beneficial. On the other hand, we see from Fig.~\ref{fig:Detector_Horizons_Difference}, left panel, that, in the same range of masses, the 2L~15km at $45^{\circ}$ configuration has an horizon distance larger than the 10~km triangle by a factor $\sim (1.4-1.8)$, i.e.  $(140-180)\%$.

The confusion noise also impacts the calculation of the false alarm rate. The null stream  would allow estimating the noise background distribution directly  (e.g. time shifting the null stream with respect to itself). In this way one could obtain more accurate estimates of the false alarm rate.

The null stream can be very effective at subtracting glitches, that would appear as  non-Gaussian outliers in the null stream.
The most prominent benefit in this case  is expected for high-mass distant BBHs, that mimic glitches, as well as for signals that are only described by phenomenological models. In a 2L configuration, one would have to rely only on coincidences between detectors, and environmental sensing.

Another potential virtue of the null stream is that is reduces the uncertainty in the calibration (although it only provides relative calibration errors).  However, the level of systematic errors in detector calibration in ET is not known. As discussed in Section~\ref{sect:nullstream}, further studies are needed to demonstrate the performance of the ET null stream calibration, compared to other calibration techniques. In the worst case scenario, in 2L detectors calibration uncertainties will have to be estimated simultaneously with gravitational-wave source parameters, which would increase the computational cost of the analyses and introduce additional uncertainty. 

{\em In summary, while in the triangle configuration the null stream would certainly be beneficial, its concrete application is subject to several uncertainties that are difficult/impossible to model before building and commissioning the detector, such as the duty cycle, the level of correlated  or identical noise in different arms, the level of non-gaussianity  in the noise or, as for instance in the case of  detector calibration, the state of advancement of other alternative techniques.}

\subsubsection{Further aspects of the triangle-2L comparison}\label{sect:further}

In this section we summarize our discussions of further aspects that are relevant to the comparison between the triangle and the 2L configurations.

\begin{itemize}

\item {\bf Correlated noise.} For the triangle vs. 2L comparison an important aspect,  that  needs further studies, is that the triangle configurations suffer from a potential threat from correlated Newtonian, seismic and magnetic noise, or lightening strikes; while significant uncertainties exist in their estimate, these correlated noise  might 
spoil the sensitivity to stochastic backgrounds below some frequency and, possibly, even to unmodeled bursts. A network of two widely spaced L-shaped detectors is significantly less sensitive to these problems. 

More in detail, an important potential danger, for all triangle configurations, is due to the  fact that some mirrors of  different nested interferometers are  apart by a few hundred meters and, on this scale, correlated seismic and Newtonian noise can be important. This has been studied in detail only recently~\cite{Janssens:2022xmo}. According to that analysis, such correlated noise would swamp the sensitivity to stochastic backgrounds of ET, in its 10~km triangle configuration, up to frequencies of  order $(50-100)$~Hz (blue curve in Fig.~\ref{fig:StochBudget_NN}), to the extent that the actual sensitivity of ET to stochastic backgrounds, up to these frequencies, would not improve on
LIGO’s A+ and Virgo’s AdV+ design. Different assumptions could result in a lower level of noise, also shown in Fig.~\ref{fig:StochBudget_NN}.
With the most  optimistic assumptions, 
the search for a stochastic background would still be impacted up to $\sim 10$~Hz (pink curve in Fig.~\ref{fig:StochBudget_NN}). Such problems arise from correlations on the scale of hundreds of meters, and therefore do  not concern a 2L configuration with widely spaced detectors. 

Magnetic noise will also pose difficulties to the triangle configuration:
in the case of a very pessimistic assumption, where the magnetic noise in  ET is the same as in the Virgo central building and it is fully correlated between two ET interferometers, it will affect  the ASD sensitivity over the entire frequency band, by several orders of magnitudes at low frequencies, $f\,\lsim\, {\cal O}(30)\, {\rm Hz}$, and by a factor 1-10 at higher frequencies, see the lower panel in Fig.~\ref{fig:Magnetic}. Again, this would only apply for the co-located interferometers in the triangular configuration. However, further research is needed to understand to which level these infrastructural noise sources would be correlated.

Compared to the triangular design, the 2L configuration is also less prone to effects coming from coherent lightning strikes,  which could affect not only stochastic background searches, but, potentially, also the search for  unmodeled bursts.

\item {\bf Site optimization for 2L.} Finally, it should  be observed that, in this work, no effort was made to optimize the distance between the two L-shaped detectors, even just within Europe, and we simply used the locations of the two current candidate sites. A longer baseline would improve in particular the localization accuracy, as well as the estimation of the  parameters more strongly correlated with angular localization.\footnote{A third possible candidate site, depending on the outcome of the geological studies, could emerge near Kamenz, in the Lusatia region, Saxony (Germany). The
great circle chord distance between the Sardinia site and the site in the Meuse-Rhine
is 1165.0 km, while that between the Sardinia site  and Kamenz is 1247.3 km, so the two distances are quite comparable, with the Sardinia-Kamenz distance larger by about $7\%$. In contrast, the
great circle chord distance between Kamenz and  the site in Meuse-Rhine is 575.4 km, significantly smaller.}

\end{itemize}

\subsection{The role of the low-frequency sensitivity}

We now draw our conclusions from the comparison between the two different choices of the amplitude spectral density (ASD) that we have studied, namely the full xylophone design, consisting of a high-frequency  instrument together with a low-frequency instrument working at  cryogenic temperature (`HFLF-cryo'),  and the situation in which there is only the high-frequency instrument (`HF-only').
The main messages that emerge from our analysis are the following:

\vspace{5mm}
\fbox{\parbox{14cm}
{\em 4. The low-frequency sensitivity is crucial for exploiting the full scientific potential of ET.
In the HF-only configuration, independently of the geometry chosen, several  crucial scientific targets of the science case would be lost or significantly diminished.}
}

\vspace{5mm}
\noindent
For instance:

\begin{itemize}

\item For all geometries, in the HF-only configuration the number of BNS sources localized to better than $10^2\, {\rm deg}^2$ (which are the sources relevant for multi-messenger studies) degrades catastrophically; with the threshold ${\rm SNR}\geq 8$ that we use for multi-messenger studies,  we see from Tables~\ref{tab:skyloccryoMM} and \ref{tab:skylocHFMM} that, for the 10~km triangle, the number of detections with $\Delta\Omega_{90\%}< 10^2\, {\rm deg}^2$  (and all orbit inclinations) decreases from 280  to
14 BNS/yr, i.e. by a factor of 20 (and from 26 to 7 for events with viewing angle less than $15^{\circ}$), while for the 2L~15km-$45^{\circ}$ it decreases from 644 to 76 (for all  orbit inclinations), or from 68 to 35 (for viewing angle less than $15^{\circ}$). Note that the reduction is more severe for the 10~km triangle than for the 2L~15km-$45^{\circ}$ configuration.
Increasing the SNR threshold the effect is even stronger:
for ${\rm SNR}\geq 12$, the number of events localized to better than $10^2\, {\rm deg}^2$  decreases from 184 to 4 BNS/yr for the 10~km triangle while, for the 2L~15km-$45^{\circ}$, it decreases from 559 to 11; see Table~\ref{tab:BNSAllConfDeldLDelOm}. 

Indeed, the decrease is so severe that the number of BNS/yr localized to better than $10^2\, {\rm deg}^2$ would even become much lower than what   will be obtained already by LVKI during the O5 run, see the panel on $\Delta\Omega_{90\%}$ in Figs.~\ref{fig:ETS_T_10km_CumulBNS_NdetScale} 
and \ref{fig:ETSMR_2L4515_CumulBNS_NdetScale}, and Table~\ref{tab:BNSAllConfDeldLDelOm}:  as we mentioned, the number of BNS/yr with ${\rm SNR}\geq 12$ and $\Delta\Omega_{90\%}< 10^2\, {\rm deg}^2$ would be 11 for 
the 2L~15km-$45^{\circ}$ HF-only, and 4 for the 10~km triangle HF-only; by comparison, for LVKI-O5 we find 51 BNS/yr with ${\rm SNR}\geq 12$ and $\Delta\Omega_{90\%}< 10^2\, {\rm deg}^2$.

The accuracy on the measurement of  the luminosity distance also degrades dramatically. For instance, for the  2L with 15~km arms at $45^{\circ}$, the number of BNS/yr with $d_L$ measured to better than $10\%$ degrades from 479 in the HFLF-cryo configuration, to 
12 in the HF configuration, while for the 10~km triangle it degrades from 52 to 1;
for 2L-15km-$45^{\circ}$, the number of events/yr with $d_L$ measured to better than $5\%$ degrades from ${\cal O}(100)$ in the HFLF-cryo configuration, to about
$1$. The same trend is observed for BNS `golden events', 
see Fig.~\ref{fig:ET_allgeom_ASD_BNS_distr_vs_z}.
The accuracy on many  other parameters would also degrade significantly, see again Figs.~\ref{fig:ETS_T_10km_CumulBNS_NdetScale}
and \ref{fig:ETSMR_2L4515_CumulBNS_NdetScale}. These results  are due to the fact that, with the low-frequency sensitivity ensured by the LF instrument, BNSs stay in the bandwidth for a long time, with a corresponding crucial benefit for parameter estimation and, particularly, localization and the partial disentangling of the polarizations, that allows us to alleviate the degeneracy between $d_L$ and $\cos\iota$. 

\item For the same reason, for BNS, pre-merger alerts for localized events would become impossible without the low-frequency instrument, compare Table~\ref{table:premergerFull} with  Table~\ref{table:premergerHF}. For instance, for the 10~km triangle the total number of BNS/yr detected 30~min before merger would go from 905 (of which 10 localized,  30~min before merger, to better than $10^2\, {\rm deg}^2$ and 85  better than $10^3\, {\rm deg}^2$) to zero; for the 2L~15km-$45^{\circ}$ it would go from 2172 (of which 20 localized,  30~min before merger, to better than $10^2\, {\rm deg}^2$ and 194 better than $10^3\, {\rm deg}^2$) again to zero.
As an example, an event such as GW170817 would enter the ET bandwidth about 1~day before the merger with the full HFLF-cryo sensitivity, but only ${\cal O}(10)$ minutes before the merger with an ASD such that it entered the detector bandwidth only at 10~Hz.

This would have a dramatic impact on the possibility to detect precursors, and to probe the prompt/early counterparts, which provide rich information on the physics acting in the GRB engine and jet launch and kilonova ejecta. Of particular interest are prompt/early detection in the very high-energy gamma rays and ultraviolet. 

\item For the multi-messenger studies, losing the low frequencies reduces the detection of short GRBs with associated gravitational-wave signals; the number of joint detection decreases by about 40\% for the 10~km triangle, by about 30\% for the 15~km triangle and 2L with 15~km arms at $45^{\circ}$, and 20\% for the 2L with 20~km arms at $45^{\circ}$, see 
Tables~\ref{tab_jointprompt_cryo} and \ref{tab_jointprompt_HF}. The reduction is more severe for kilonova counterparts: the triangle HF configurations detect less than 6\% of the kilonovae detected by triangle HFLF-cryo, while the 2L HF-only configurations detect less than 15\% of the KNe detected by the 2L HFLF-cryo configuration, see Tables~\ref{tab:jointKNFull} and \ref{tab:jointKNHF}. Note that, once again, the triangle geometry is more heavily affected than the 2L by the loss of the low frequency instrument.
The reduction of short GRBs with an associated GW signal might impact our understanding of  the physics of the GRB engine and relativistic jets, by requiring more years of joint observations to achieve the same science goals. However for cosmology, nuclear physics, and studies of the heavy-elements nucleosynthesis which rely on kilonova detections, loosing the LF has a major impact on achieving the goals themselves.

It is worth noting that, for short GRBs, the HF-only 2L-15km-$45^{\circ}$ and the HF-only 15km triangle enable the detection of the GW signal for a number of GRBs comparable to the full  HFLF-cryo triangle with 10~km arms.  On the other hand, for the kilonovae counterparts, no HF-only configuration (even  HF-only 2L with 20~km arms at $45^{\circ}$) is comparable to 10~km triangle HFLF-cryo. This is due to a different type of search: while to observe short GRBs, wide FoV satellites in survey mode are used and the temporal coincidence of the GW/GRB signals are used to identify a joint detection, to search for kilonovae the telescopes FoV is typically much smaller  than the GW localization uncertainty and thus requires selecting better-localized events to be followed up. The much-improved localization obtained by accessing low-frequencies then becomes a key parameter for the search.

\item The HF-only configuration has a significantly smaller reach in distance, compared to the HFLF-cryo  configuration, in the whole mass range relevant to 3G detectors,
as can be seen from  the right panels in Figs.~\ref{fig:Detector_Horizons_BBH_AllConf} and \ref{fig:Detector_Horizons_Difference}. 

This has several significant consequences. In the range of masses relevant to BNS, for the 10~km triangle, the redshift to which a   BNS  with optimal orientation and sky location can be detected would decrease from $z\simeq 4$  down to $z\simeq 2$ (or, for the 
2L-15km-$45^{\circ}$ , from $z\simeq 6$ to  $z\simeq 3)$. This is particularly significant since the peak of the star formation rate is around $z\simeq 2-3$, so  studies of demography and population of BNS would be strongly impacted.

The lowest reach in redshift  would also make it impossible to  identify primordial BHs on the basis of the criterion that BBHs at $z\, \gsim\, 30$ cannot be of astrophysical origin.
For instance, for the 10~km triangle, the maximum redshift at which a compact binary coalescence could be seen, for optimal values of the total  mass and optimal orientation and sky location, reduces from $z\simeq 100$ in the HFLF-cryo configuration to $z\simeq 20$ in the HF configuration, see  Fig.~\ref{fig:Detector_Horizons_BBH_AllConf}.  

Intermediate mass BHs, with masses in the range $(10^3-10^4) \msun$, would  also be visible to significantly smaller distances, see again Fig.~\ref{fig:Detector_Horizons_BBH_AllConf}, reducing significantly the chances of discovery. As an example, a BBH with total mass $5\times 10^3\, \msun$ would be visible up to $z\simeq 1.1$ in the HFLF-cryo configurations (almost independently of geometry),  but only up to $z\simeq 0.6$ in the HF only configurations, corresponding to a  reduction by a factor $\sim 5$ in the comoving volume explored and in the corresponding chances of detection.

\item The measurement of the eccentricity of subsolar mass compact objects, an important criterion for assessing or excluding the primordial origin of a BBH, would degrade by more than one order of magnitude, see Fig.~\ref{fig:eccentricity}.

\item  The degradation of the accuracy in luminosity distance and angular localization would basically render impossible all studies of cosmology, such as accurate measurements of $H_0$ and especially studies of the dark energy equation of state and of modified GW propagation, which crucially rely on accuracy in sky localization and luminosity distance for events at large $z$. Similarly,
the exquisite localization accuracy on some high-mass ratio events, needed to extract the Hubble parameter from them, see Section~\ref{sect:highmassevents},  would be impossible without the LF instrument.

\item For a number of other aspects of the science case the loss of the LF instrument would not be as  disruptive,  but would still results in a loss of accuracy on the relevant parameters, by factors of order 2-3. This is the case, for instance, for the possibility of distinguishing BHs from Exotic Compact Objects on the basis of the spin-induced quadrupole moment or of a non-vanishing  tidal deformability (Figs.~\ref{fig:err_quadrupole}  and \ref{fig:err_tidal_deformability}), or for the accuracy on the reconstruction of NS radii, see Table~\ref{tab:div6:Fisher_radius}.
\end{itemize}

\vspace{2mm}\noindent
On the other hand, not all aspects of the Science Case depend on the LF instrument, and 
another important conclusion  that follows from our study is:
\vspace{2mm}

\fbox{\parbox{14cm}
{\em 5. There are some important targets of the Science Case that depend only on the HF sensitivity, and that could be fully reached with an HF-only instrument.}
}

\vspace{3mm}
\noindent
In particular:

\begin{itemize}

\item For multi-messenger astronomy, the number of joint detections of GWs from a BNS merger and 
the associated X-ray afterglow, as could be detected by an instrument such as 
THESEUS in survey mode, is basically independent of the LF instrument; see Table~\ref{tab_jointafterglowsurvey}.

\item Tests of physics near the BH horizon based  on the ringdown signal of the final BH are completely independent of the LF instrument, see Table~\ref{tab:spectro_snr} and the discussion in Section~\ref{sect:testGRnearhorizon}. The same holds for
the search of echoes and near-horizon structures, see Section~\ref{sect:echoes}.

\item The post-merger signal of BNS coalescences, that contains the information of the NS equation of state, is concentrated in the high-frequency region, and 
is completely insensitive to the presence of the LF instrument, see the left panel in Fig.~\ref{fig:post_merger_fig1}.

\item The detectability of sub-solar mass primordial BHs is another example of a metric that is not affected  by the loss of the LF instrument; see Fig.~\ref{fig:subsolar}.

\item Stochastic backgrounds of cosmological origin are in general expected to have a smooth power-like behavior, $\Ogw(f)\propto f^{\alpha}$,  over the bandwidth of  ground based or space-borne detectors (with the exception of backgrounds from cosmological phase transitions, that will be broadly peaked around a characteristic frequency). At the  very low frequencies, say $f\sim 10^{-18}\, {\rm Hz}$, tested by CMB, there are extremely stringent limits $\Ogw\,\lsim\, 10^{-15}$ (see, e.g. Chapter~22 of \cite{Maggiore:2018sht}). Therefore, at the frequencies of ground based detectors, a stochastic background, relic of the early Universe, could only be detectable if it grows with frequency,
$\Ogw(f)\propto f^{\alpha}$ with $\alpha>0$, at least until a frequency where it saturates, followed by a cutoff.
For this reason, the stochastic background predicted by standard single-field slow-roll inflation is not
detectable, since it is basically flat (in fact, slightly red, i.e. $\alpha <0$). However, there are alternative cosmological models that predict spectra with $\alpha >0$, such as the pre-big-bang model~\cite{Brustein:1995ah,Buonanno:1996xc} or axion inflation
(see \cite{Cook:2011hg}, or Fig.~12 of \cite{Maggiore:2019uih}). In these cases, a significant part of the SNR could   accumulate in the high-frequency part of the detector bandwidth, where  the LF instrument does not contribute to the sensitivity to stochastic backgrounds,
see Fig.~\ref{fig:ET_PLS_auto_cross_SNR_1_room}.
A similar behavior takes place, in the ET bandwidth, in the model for   cosmic strings whose predictions are shown in Fig.~\ref{fig:stringsB}.

\end{itemize}

\vspace{2mm}\noindent
A final important conclusion that emerges from our study is that:

\vspace{5mm}
\fbox{\parbox{14cm}
{\em 6. For some important aspects of the Science Case, the 2L with 15~km arms at $45^{\circ}$, already in the HF-only configuration, is  comparable the 10~km triangle in a full HFLF-cryo configuration.}
}

\vspace{5mm}

\noindent
In particular:

\begin{itemize}

\item For parameter estimation of BBHs,  the 2L with 15~km arms at $45^{\circ}$ in the HF-only configuration is  comparable to the 10~km triangle at full HFLF-cryo sensitivity, with better performance of luminosity distance, less good performance on mass reconstruction, and equivalent performances on all other parameters and in SNR distribution, see Fig.~\ref{fig:ETSMR_2L4515_CumulBBH_NdetScale}.
The performance of the 2L-15km-$45^{\circ}$ configuration is also equivalent to that of 
the 10~km triangle for what concerns `golden BBH events', see the lower row of Fig.~\ref{fig:ET_allgeom_ASD_BBH_distr_vs_z}. 

\item From the right panel of Fig.~\ref{fig:ET_allgeom_ASD_BNS_Hist_DelRovR} we see that, for the measurement of the neutron star radii and the  consequences for nuclear physics that we can derive from it, the 2L with 15~km arms at $45^{\circ}$ already in the HF-only configuration has performances very similar   to that of the full 10~km triangle HFLF-cryo.

\item For all the items discussed above, where the LF instrument does not contribute 
(joint GW+X-ray afterglow and, to some extent, GRB detections,  tests of physics near the BH horizon, post-merger signal of BNS coalescences, sub-solar mass BHs, stochastic backgrounds growing as $f^{\alpha}$ with $\alpha>0$), the 2L-15km-$45^{\circ}$ in the HF only configuration will be superior to the 10~km triangle with the full HFLF-cryo sensitivity.

\end{itemize}

\subsection{Conclusions}\label{sect:Conclusions}

The decisions on the ideal design of a detector such as the Einstein Telescope, which is meant to have a leading role in the field of gravitational waves for decades, are part of a complex process. In this study we have considered alternatives to the geometry (a single-site triangle against two widely separated L-shaped interferometers, with different options for the arm-lengths) and to other aspects of the design (considering the baseline design given by a full xylophone configuration with a  high frequency instrument together with a cryogenic low-frequency instrument,  and comparing it to  a high-frequency only instrument).

The first conclusion of this work, which  extends the study presented in~\cite{Maggiore:2019uih},
is  that, whatever geometry is decided between the triangle and 2L configurations considered, ET will clearly be a game-changer. It will  over--perform the network of 2G observatories at their best expected O5 sensitivities by several orders of magnitudes in all relevant metrics, accessing physics that it is well beyond reach for 2G detectors. 
Note, however, that this is only true for the triangle  and for the 2L configurations. A single L-shaped detector, not inserted in a global 3G network,  even with a very long arm length, does not have a viable Science Case, at the level expected from a 3G detector.

The comparison between the 10~km triangle and a 15~km 2L shows that, from the scientific point of view, the 15~km 2L with arms at $45^{\circ}$ is superior on basically all the metrics that we have considered, with the exception of the nominal sensitivity to stochastic backgrounds. However, in this case we have seen that the triangle suffers from a  potentially  very significant treat from correlated noise, particularly at low frequencies.\footnote{Furthermore, more work is needed to asses the effect of an imperfect subtraction of the astrophysical backgrounds, which could limit the sensitivity to cosmological backgrounds to a level fixed in all cases by the accuracy of the subtraction.} For compact binary coalescences, the better performances of the 15~km 2L with arms at $45^{\circ}$ allow us to gain typically further factors of order 2-3 on the number of events which pass given cuts on SNR, or on the  accuracy of reconstruction of various parameters and physical quantities. In fact, we have found that  the 15~km 2L with arms at $45^{\circ}$ is quite comparable to the 15~km triangle: the two have  very similar  performances for the reconstruction of all parameters of compact binary coalescences, except for  luminosity distance, for which  the 15~km 2L with arms at $45^{\circ}$ is significantly better.

Another interesting element that emerged from our study is that the configuration
2L~15km at $45^{\circ}$ gives a better possibility of proceeding in steps, compared to the 10~km triangle. 
Both for the triangle and 2L configurations, the full scientific outcome expected from a 3G detector will only be reached in the full configuration with the HF and the cryogenic LF instrument. However, the 2L~15km at $45^{\circ}$,
already in the HF-only configuration, would be sufficient to obtain some  important  results (which is true only to a lesser extent for the 10~km triangle). It is therefore natural to investigate whether  a staging procedure will be advantageous, where the HF interferometer, which is the simpler to commission, is put into operation first, until the full HFLF-cryo configuration is reached, possibly going through intermediate steps such as, for instance,  an LF instrument working, at first, at room temperatures. These issues, however, are beyond the scope of this study, and should be examined within the context of the ET Instrument Science Board.

Finally, the triangle configuration is  prone to the effect of correlated Newtonian, seismic and magnetic noise on the scale of hundreds of meters. While more work is needed to fully understand them, these effects might pose a risk/challenge to obtaining the target sensitivity of the triangle, particularly in the low-frequency region.


\section*{Acknowledgments}

The research leading to these results has been conceived and developed within the ET Observational Science Board (OSB). We are particularly grateful to Giancarlo Cella, Gianluca Guidi and Frank Ohme for their extremely careful and very useful internal refereeing of the manuscript.
We thank  Marie-Anne Bizouard, Nelson Christensen, Fernando Ferroni,  Harald L\"uck, Mario Martinez, Ed Porter, Michele Punturo, Tania Regimbau, David Shoemaker, Achim Stahl and Patrice Verdier for very useful discussions and feedback. We express special thanks to Jameson G. Rollins for his consultations about the use of PyGWINC. 

The work of  Michele Maggiore, Stefano Foffa, Francesco Iacovelli, Michele Mancarella and Niccol\`o Muttoni  is supported by the  Swiss National Science Foundation, grant 200020$\_$191957, and  by the SwissMap National Center for Competence in Research, and   made use of the Yggdrasil cluster at the University of Geneva.  M.~Branchesi acknowledges financial support from the Italian Ministry of University and Research (MUR) for the PRIN grant METE under contract no. 2020KB33TP.  B.~Banerjee acknowledge financial support from MUR (PRIN 2017 grant 20179ZF5KS). M.~Branchesi and G.~Oganesyan acknowledge financial support from the AHEAD2020 project (grant agreement n. 871158). The Gran Sasso Science Institute group acknowledges Stefano Bagnasco, Federica Legger, Sara Vallero, and the INFN Computing Center of Turin for providing support and computational resources. V. De Luca is supported by funds provided by the Center for Particle Cosmology at the University of Pennsylvania.
Elisa Maggio acknowledges funding from the Deutsche Forschungsgemeinschaft (DFG) - project number: 386119226.
G.Franciolini acknowledges financial support provided under the European
Union's H2020 ERC, Starting Grant agreement no.~DarkGRA--757480 and under the MIUR PRIN programme, and support from the Amaldi Research Center funded by the MIUR program ``Dipartimento di Eccellenza" (CUP:~B81I18001170001).
This work was partly enabled by the UCL Cosmoparticle Initiative.
Katarina Martinovic is supported by King's College London through a Postgraduate International Scholarship.
M. Mapelli, C. P\'{e}rigois and F. Santoliquido acknowledge financial support from the European Research  Council for the ERC Consolidator grant DEMOBLACK, under contract no. 770017. 
M.~Mancarella and C.~Pacilio are supported by European Union's H2020 ERC Starting Grant No.~945155-GWmining and Cariplo Foundation Grant No.~2021-0555. Numerical calculations by C.~Pacilio have been made possible through a CINECA-INFN agreement, providing access to resources on MARCONI at CINECA. 
Anuradha Samajdar thanks the Alexander von Humboldt foundation in Germany for a Humboldt fellowship for postdoctoral researchers. 
Paolo Pani acknowledges financial support provided under the European Union's H2020 ERC, Starting Grant agreement no.~DarkGRA--757480 and under the MIUR PRIN and FARE programmes (GW-NEXT, CUP: B84I20000100001).
BG is supported by the Italian Ministry of Education, University and Research within the PRIN 2017 Research Program Framework, n. 2017SYRTCN.
Mairi Sakellariadou is supported in part by the Science and Technology Facility Council (STFC), United Kingdom, under the research grant ST/P000258/1.
ACJ was supported by the Science and Technology Facilities Council through the UKRI Quantum Technologies for Fundamental Physics Programme [grant number ST/T005904/1]. 
Swetha Bhagwat is supported by the UKRI Stephen Hawking Fellowship with grant ref. no EP/W005727. 
Angelo Ricciardone acknowledges financial support from the Supporting TAlent in ReSearch@University of Padova (STARS@UNIPD) for the project “Constraining Cosmology and Astrophysics with Gravitational Waves, Cosmic Microwave Background and Large-Scale Structure cross-correlations’'.
Kamiel Janssens is supported by FWO-Vlaanderen via grant number 11C5720N.
F.Gulminelli and C.Mondal acknowledge partial support from the In2p3 Master Project NewMAC.
D. Alonso is supported by the Science and Technology Facilities Council through an Ernest Rutherford Fellowship, grant reference ST/P004474.
Ssohrab Borhanian acknowledges support from the Deutsche Forschungsgemeinschaft, DFG, project MEMI number BE 6301/2-1.
Arnab Dhani, Ish Gupta and B.S. Sathyaprakash were supported by NSF grant numbers PHY-2012083, AST-2006384 and PHY-2207638. 
G. Cusin is supported by CNRS and by Swiss National Science Fundation (Ambizione grant --\emph{Gravitational wave propagation in the clustered universe)}.

\appendix

\section{Basic formalism for stochastic backgrounds}\label{app:PLS}

In this appendix we review several basic concepts relevant for the characterization of stochastic GW backgrounds.
At the frequencies of ground-based detectors, to obtain an interesting sensitivity one needs to cross-correlate the output of at least two detectors in a network, see e.g. \cite{Maggiore:1999vm,Romano:2016dpx} for  reviews: as a consequence the observable quantity is not a strain, but rather quantities quadratic in the strain.  One usually introduces a polarisation tensor defined  (in frequency space) as
\be
\tilde{\mathcal{P}}_{ab}=\tilde{h}_a^*\tilde{h}_b\,,
\ee
with $a, b=+, \times$ and we defined the superposition of signals in a given direction $\nv$ and at a given frequency $f$ as
\be\label{tildehnv}
\tilde{h}_{ij}(\nv, f)=\tilde{h}_+(\nv, f)e_{ij}^+(\nv)+\tilde{h}_{\times}(\nv, f)e_{ij}^{\times}(\nv)\,.
\ee
In full generality the polarisation tensor, being a complex $2\times 2$  tensor, can be decomposed in a basis of the identity and Pauli matrices as 
\be\label{tildePnv}
\tilde{\mathcal{P}}_{ab}(\nv, f)=\frac{1}{2}\left[I(\nv, f)1_{a b}+U(\nv, f)\sigma_{ab}^{(1)}  +V(\nv, f)\sigma_{ab}^{(2)}  +Q(\nv, f)\sigma_{ab}^{(3)} \right]\,,
\ee
where the coefficients $I, U, V, Q$ are Stokes parameters describing intensity and polarisation, respectively. Polarisation is expected to be very small if we consider a homogeneous and isotropic AGWB and a CGWB: the amount of polarisation generated via classical diffusion is very small for both background components, see \cite{Cusin:2018avf}. However, there is a non-negligible amount of circular polarization in the AGWB, generated by Poisson fluctuations in the number of unresolved sources, which can be detected by third-generation interferometers with SNR$>1$ (see \cite{ValbusaDallArmi:2023ydl}). Measuring a highly polarised background component would therefore be interesting both for shedding light on some cosmological parity-violating generation mechanisms (see e.g.~\cite{Domcke:2019zls, Martinovic:2021hzy}) or to estimate whether the SGWB comes from a handful of sources or a relatively large population of binaries. Here we focus on the study of intensity and of its spectrum. It is common  in the literature to introduce a dimensionless quantity related to intensity via 
\bees
\Omega_{\rm gw}(\nv, f)&\equiv& 
\frac{4\pi^2f^3}{G\rho_c}I(\nv,f)\nn\\
&=&\frac{1}{\rho_c}\frac{d\rho_{\rm gw}(\nv, f)}{d\ln f\, d^2\nv}\,,
\label{eq:Abkg:def:App}
\ees
where the second equality clarifies the physical meaning of this observable: it corresponds to the background energy density, $\rho_{\rm gw}(\nv, f)$, per unit of logarithmic frequency and unit solid angle (normalised to the critical density of the Universe today, $\rho_c$), that reaches the observed from a direction $\nv$.

To characterize the detector sensitivity to stochastic backgrounds, we use the Power-Law integrated  Sensitivity curve (PLS), introduced for the first time in \cite{Thrane:2013oya} for a network of L-shaped detectors, and we extend it to the case of triangular detectors, following \cite{Alonso:2020rar}. When dealing with triangular detectors (three nested detectors), one has to keep in mind that the noise between two detectors with an arm in common is not uncorrelated, hence the correlation has to be accounted for when computing the PLS. 
Here we assume that the noise in ET
is 20\% correlated between detectors with an arm in common. As we discuss in 
Section~\ref{sect:corrnoise}, see in particular Fig.~\ref{fig:StochBudget_NN}, this is a value intermediate between  the extremes of modeled noise correlations;  we chose here a frequency-independent correlation for simplicity.

Let us assume to have a network of $N$ detectors. 
The normalized overlap reduction function (see~\cite{Flanagan:1993ix, Maggiore:2007ulw, Thrane:2013oya}) for a detector pair $\{A,B\}$ in the network is defined as
\begin{equation}\label{eq: norm ORF}
\gamma_{AB}(f)=\frac{5}{8\pi}\int \mathrm{d}^2\nv\left[F^+_A\left(\nv\right)F^+_B\left(\nv\right)+F^\times_A\left(\nv\right)F^\times_B\left(\nv\right)\right]\,e^{i2\pi \frac{f}{c}\nv\cdot{\bf \Delta x}_{AB}}\,,
\end{equation}
where ${\bf \Delta x}_{AB}={\bf x}_B-{\bf x}_A$ is the separation between the two detectors $A$ and $B$.

\noindent The normalization factor $5/(8\pi)$ in Eq. (\ref{eq: norm ORF}) is conventionally introduced so that $\gamma_{AB}(f)=1$ for two co-located and perfectly aligned detectors with orthogonal arms. The quantities $F^+_A\left(\nv\right)$ and $F^\times_A\left(\nv\right)$ are the detector $A$ pattern functions (and similarly for $B$). They are determined by the detector $A$ response tensor ${\bf{D}}_A$ and the polarisation tensors  ${\bf{e}}^{+,\times}$ as $F^+_A\left(\nv\right)=\text{Tr}\left[{\bf{D}}_A~{\bf{e}}^{+}\left(\nv\right)\right]$ and $F^\times_A\left(\nv\right)=\text{Tr}\left[{\bf{D}}_A~{\bf{e}}^{\times}\left(\nv\right)\right]$.

\noindent For a Michelson interferometer with arms pointing along the directions  ${\bf u}_A$ and ${\bf v}_A$, the detector $A$ response tensor has components 
$D^{ij}_A$ given by (see e.g.~\cite{Flanagan:1993ix})
\be
D^{ij}_A=(u_A^i u_A^j-v_A^i v_A^j)/2\,.
\ee

We start with the effective noise power spectral density. In full generality, this is given by
\begin{equation}\label{Sf}
S_{\text{eff}}(f)=\left[\mathlarger{\mathlarger{\sum}}_{A,B,C>B,D>A} (N_{f}^{-1})^{AB}~\gamma_{BC}(f)~(N_{f}^{-1})^{CD}~\gamma_{DA}(f)\right]^{-1/2}\,,
\end{equation}
where $N_{f}^{-1}$ is the inverse of the noise covariance matrix of the network at each frequency $f$. The restrictions on the sum in Eq.~(\ref{Sf}) ensure that only a cross-correlation search at unordered pairs of distinct detectors is considered, thus excluding auto-correlations (and without double counting).

\noindent Notice that in the case of uncorrelated detector noises, the noise matrix is diagonal
\begin{equation}
(N^{-1}_{f})^{AB}=\frac{\delta^{AB}}{N_{f}^{AA}}\,,
\end{equation}
where $N_{f}^{AA}$ is the noise power spectral density (PSD) of the detector $A$. In this case Eq.~(\ref{Sf}) can be simplified to:
\begin{equation}
S_{\text{eff}}(f)=\left[\sum_{A=1}^N \sum_{C>A}^N \frac{\gamma_{AC}^2(f)}{N_{f}^{AA} N_{f}^{CC}}\right]^{-1/2}\,.
\end{equation}
However, for triangular detectors, we need to know both the noise variance (diagonal terms) and the cross-detector noise covariance (off-diagonal).

We define an effective dimensionless noise energy spectrum for $S_{\text{eff}}(f)$ in Eq.~(\ref{Sf}) as
\begin{equation}\label{eq: Omega eff noise}
\Omega_{\text{eff}}(f)=\frac{10\pi^2}{3 H_0^2}f^3 S_{\text{eff}}(f)\,
\end{equation}
and consider an isotropic and unpolarized stochastic GW background with fractional energy density contribution $\Omega_{\text{gw}}(f)$ with respect to the critical energy density $\rho_c$ of the Universe today.
With the definition~(\ref{eq: Omega eff noise}), the integrated signal-to-noise ratio $\rho$ for a cross-correlation search at the network in a frequency range from $f_{\rm min}$ to $f_{\rm max}$, for a coincident observation time $T$, is~\cite{Maggiore:2007ulw}
\begin{equation}\label{rho stoch}
\rho=\left[2 T\int_{f_{\text{min}}}^{f_{\text{max}}} \mathrm{d}f~\frac{\Omega_{\text{gw}}^2(f)}{\Omega_{\text{eff}}^2(f)} \right]^{1/2}\,.
\end{equation}
For a set of power-law indices $\beta$, we write
\be\label{Omegabetafref}
\Omega_{\text{gw}}(f)=\Omega_{\beta}(f/f_{\text{ref}})^{\beta}\, 
\ee
where $f_{\text{ref}}$ is a reference frequency such that, at $f=f_{\text{ref}}$, $\Omega_{\text{gw}}(f)$ has the value $\Omega_{\beta}$, and we
compute the value of the amplitude $\Omega_{\beta}$ such that the integrated signal-to-noise ratio $\rho$ has some fixed value.  Using Eq.~(\ref{rho stoch}), this is given by
\begin{equation}\label{Omegabetarho}
\Omega_{\beta}=\frac{\rho}{\sqrt{2 T}}\left[\int_{f_{\text{min}}}^{f_{\text{max}}}\mathrm{d}f~\frac{(f/f_{\text{ref}})^{2\beta}}{\Omega_{\text{eff}}^2(f)}\right]^{-1/2}\,.
\end{equation}
In the following we will set  $\rho=1$.
For each pair of values $(\beta, \Omega_{\beta})$, we compute $\Omega_{\text{GW}}=\Omega_{\beta}(f/f_{\text{ref}})^{\beta}$. The envelope is the power-law integrated sensitivity curve. Formally it is given by
\begin{equation}
\Omega_{\text{PLS}}(f)=\text{max}_{\beta}\left[\Omega_{\beta}\left(\frac{f}{f_{\text{ref}}}\right)^{\beta}\right]\,.
\label{eq:Omega_PI}
\end{equation}
Any line (on a log-log plot) that is tangent to the power-law integrated sensitivity curve corresponds to a gravitational-wave background power-law
spectrum with an integrated signal-to-noise ratio $\rho=1$.
This implies that, if the curve for a predicted background lies everywhere below the sensitivity curve, then $\rho<1$ for
such a background. Note that, from the definition (\ref{Omegabetafref}) of $\Omega_{\beta}$ and 
$f_{\text{ref}}$, a change in the choice of $f_{\text{ref}}$ is compensated by a corresponding change of $\Omega_{\beta}$, such that $\Omega_{\beta}f_{\text{ref}}^{-\beta}$ remains constant. Therefore, the value of $\Omega_{\beta}$ obtained from \eq{Omegabetarho} depends on the arbitrary choice of 
$f_{\text{ref}}$, but $\Omega_{\text{PLS}}(f)$, given by \eq{eq:Omega_PI}, does not.

\section{Sensitivity to stochastic backgrounds of misaligned 2L configurations} 
\label{app:StocasticMisalignement}

Focusing on the stochastic gravitational-wave background, we highlight below the effect of the orientation of the two L-shaped 
interferometers on the peak sensitivity. As discussed in Section~\ref{sect:geometry}, in this paper we have fixed the arms of the Sardinia interferometer so that they point locally East and North, and rotate anti-clockwise the arms of the Netherlands interferometer by an angle $\alpha$ with respect to the local East-North directions. The angle $\alpha=0$ differs from the usual definition of the alignment of two interferometers with respect to the great circle~\cite{Flanagan:1993ix,Christensen:1996da} by 2.51 degrees and is clearly not the maximum possible alignment, due to the curvature of the Earth between the two sites.

\begin{figure}[t]
    \centering
    \includegraphics[width=1.0\textwidth]{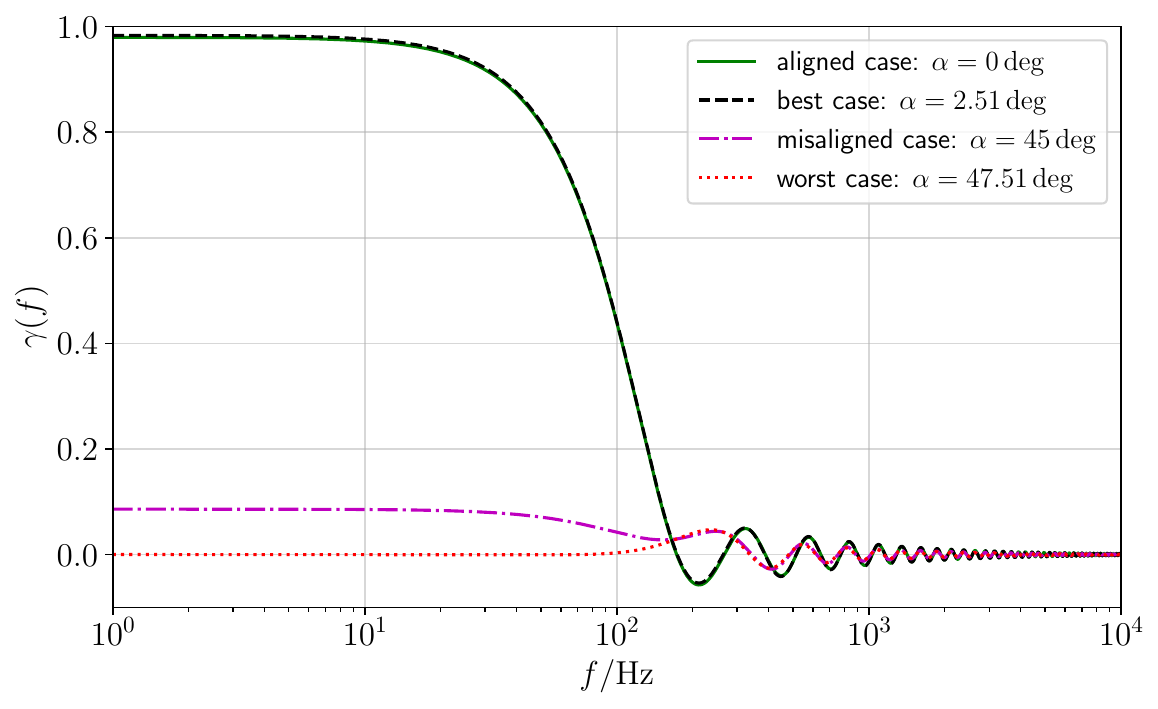}
    \caption{\small Normalized overlap reduction function (ORF) $\gamma(f)$ for different values of the angle $\alpha$.
    }
    \label{fig:fig1}
\end{figure}

\begin{figure}[t]
    \centering
    \includegraphics[width=1.0\textwidth]{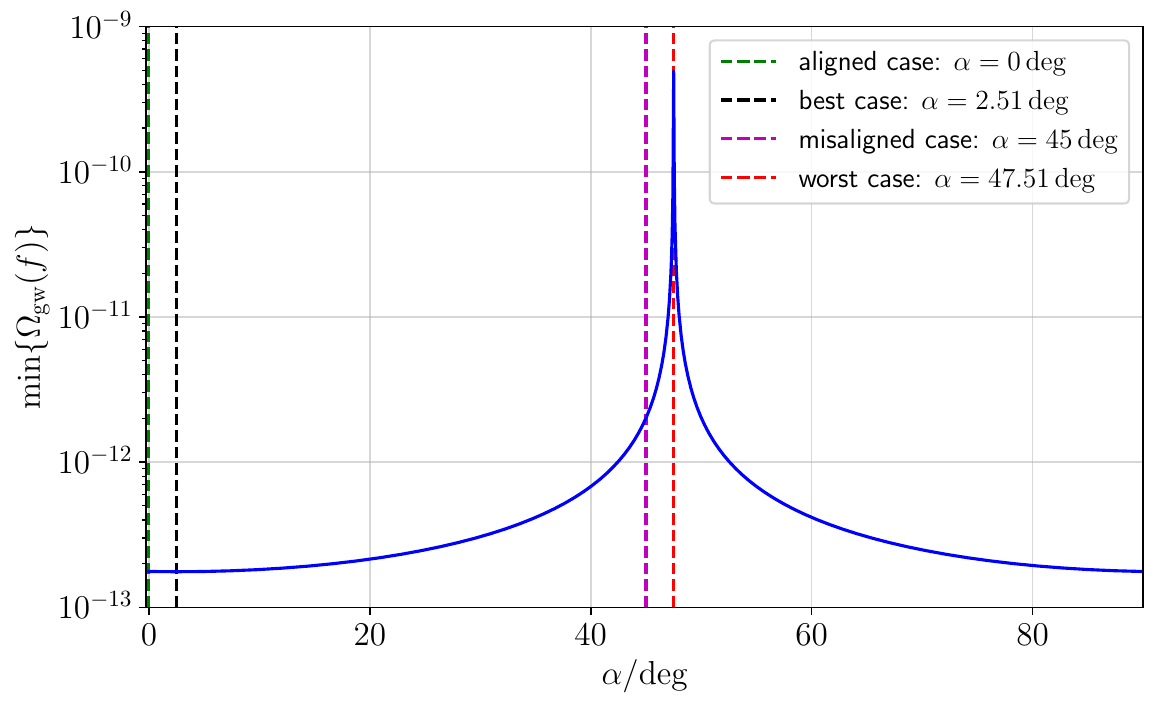}
    \caption{\small Peak sensitivity to the stochastic gravitational-wave background as a function  of the angle $\alpha$ in the $2{\rm L}$-20km-HFLF-Cryo configuration.
    The SNR threshold used is $\rho=1$ and the observation time is $T=1~{\rm yr}$. 
    }
    \label{fig:fig2}
\end{figure}

In Fig.~\ref{fig:fig1} we show the normalized overlap reduction function (\acrshort{orf}) defined in Eq.~(\ref{eq: norm ORF}) as a function of frequency, for the  $\alpha=0$ deg and $\alpha=45$ deg cases, which are the examples considered in the main text. However, the best and worst cases for a stochastic search are the values $\alpha=2.51$ deg and $\alpha=47.51$ deg, respectively. Clearly, while varying the angle $\alpha$ near the 0 deg case has very little impact,  the consequences of varying $\alpha$ just a few degrees around the 45 deg are very important. These results reflects into a significant dependence on  $\alpha$ of the peak sensitivity to the stochastic gravitational-wave background.

Figure~\ref{fig:fig2} 
illustrates the peak sensitivity to the stochastic gravitational-wave background (minimum value of the power-law integrated curve) as a function of the angle $\alpha$. The choices $\alpha = 0$ deg and 45 deg, considered in the main text of this document, are not respectively the best and worst cases for stochastic searches. Varying the angle $\alpha$ near the 0 deg case has very little impact, but varying $\alpha$ just a few degrees around the 45 deg case varies the peak sensitivity to the stochastic gravitational-wave background by orders of magnitude. As one can clearly see, the best and worst cases for stochastic searches, are  when $\alpha = 2.51$ deg and 47.51 deg, respectively. 

To be strictly optimal the alignment of two interferometers would
have the two detectors such that each had an arm along the great circle connecting them for 0 deg (which is how the two LIGO detectors are aligned), and rotating one by 45 deg for the minimal value~\cite{Christensen:1996da}. For such a minimal alignment the overlap reduction function would be exactly zero in the limit   $f\Delta x\ra 0$, where $\Delta x$ is the separation between the two detectors,  and very small basically for all frequencies, as we see from Fig.~\ref{fig:fig1}. This would also give the best polarization separation for the CBC searches.

\section{Tables of figures of merit for BBHs and BNSs}\label{app:TablesCBC}

To make the comparison among different geometries and ASDs easier, we  provide here tables with the number of sources detected by each configuration with various cuts on the match-filtered SNR, or on measurement errors on some of the most relevant quantities, corresponding to the results shown in Figs.~\ref{fig:AllGeoms_CumulBBH_NdetScale}, \ref{fig:ETS_T_10km_CumulBBH_NdetScale}, \ref{fig:ETSMR_2L4515_CumulBBH_NdetScale}
and \ref{fig:ET_allgeom_ASD_BBH_distr_vs_z} for BBHs, and to Figs.~\ref{fig:AllGeoms_CumulBNS_NdetScale}, \ref{fig:ETS_T_10km_CumulBNS_NdetScale}, \ref{fig:ETSMR_2L4515_CumulBNS_NdetScale} 
and \ref{fig:ET_allgeom_ASD_BNS_distr_vs_z} for BNS.
We further show the same quantities for the networks ET+1CE and ET+2CE, 
 corresponding to the results shown in Figs.~\ref{fig:ETCE_CumulBBH_NdetScale}  and \ref{fig:ETCE_CumulBNS_NdetScale}. 

In particular, in Table~\ref{tab:BBHAllConfSNR} we show the number of detected BBH sources with different cuts in SNR, ranging from 8 to 200; in Table~\ref{tab:BBHAllConfDeldLDelOm} with cuts on the relative error on the luminosity distance and the 90\% sky localisation [notice that the parameter estimation  analysis has been performed only for the events detected with $\rm SNR\geq12$]; and in Table~\ref{tab:BBHAllConfDelMcDelchi} with cuts on the relative error on the detector-frame chirp mass and the error on the spin magnitude of the primary object. 

In Table~\ref{tab:BNSAllConfSNR} we report the number of BNS sources detected with different cuts in SNR, in Table~\ref{tab:BNSAllConfDeldLDelOm} with cuts on the luminosity distance relative error and 90\% sky localisation area,\footnote{Some differences with the results shown in Tables~\ref{tab:skyloccryoMM} and \ref{tab:skylocHFMM} are  due to the different choices for the cut in SNR for the parameter estimation analysis; in Section~\ref{sect:CBC} (to which the result of this appendix refer) was used $\rm SNR\geq12$, while in Section~\ref{sect:MMO} the threshold was lowered to 
$\rm SNR\geq8$ because of the increased statistical significance provided by to the coincidence with an electromagnetic counterpart. As can be seen from Table~\ref{tab:BNSAllConfSNR}, passing from one cut to the other, the number of detections changes by more than factor of 2, in particular for the HF-only configurations. Part of the discrepancy is also due to the different inversion procedure adopted in the two sections, as already outlined in footnote~\ref{footnote:Fisher_inversion_issues}, with some events being discarded in the results presented here and in Section~\ref{sect:CBC}, while their Fisher matrices are regularized in the ones reported in Section~\ref{sect:MMO}, resulting in higher numbers.} and in Table~\ref{tab:BNSAllConfDelMcDelLam} with cuts on the relative errors on the chirp mass and adimensional tidal deformability parameter $\tilde{\Lambda}$ [defined in Eq.~\ref{deftildeLambda}]. 

\begin{table}[htbp]
    \hspace{-1.cm}
    \begin{tabular}{||l||*{5}{c|}|}
\hline\hline
Configuration & ${\rm SNR} \geq 8$ & ${\rm SNR} \geq 12$ & ${\rm SNR} \geq 50$ & ${\rm SNR} \geq 100$ & ${\rm SNR} \geq 200$ \\
\hline\hline
$\Delta$-10km-HFLF-Cryo & 103\,528 & 87\,568 & 13\,674 & 2298 & 282 \\
$\Delta$-15km-HFLF-Cryo & 111\,231 & 101\,308 & 26\,092 & 5730 & 759 \\
2L-15km-$45^{\circ}$-HFLF-Cryo & 107\,661 & 97\,205 & 23\,491 & 4933 & 644 \\
2L-20km-$45^{\circ}$-HFLF-Cryo & 110\,698 & 103\,773 & 34\,009 & 8828 & 1267 \\
2L-15km-$0^{\circ}$-HFLF-Cryo & 104\,935 & 94\,015 & 24\,088 & 5143 & 642 \\
2L-20km-$0^{\circ}$-HFLF-Cryo & 106\,417 & 98\,274 & 32\,915 & 8551 & 1246 \\
\hline\hline
$\Delta$-10km-HF & 87\,125 & 65\,092 & 5595 & 773 & 98 \\
$\Delta$-15km-HF & 102\,149 & 85\,698 & 13\,697 & 2360 & 292 \\
2L-15km-$45^{\circ}$-HF & 97\,881 & 81\,210 & 12\,089 & 1987 & 248 \\
2L-20km-$45^{\circ}$-HF & 105\,032 & 93\,050 & 20\,551 & 4144 & 515 \\
2L-15km-$0^{\circ}$-HF & 89\,707 & 73\,696 & 10\,688 & 1732 & 201 \\
2L-20km-$0^{\circ}$-HF & 104\,558 & 92\,308 & 21\,970 & 4540 & 569 \\
\hline\hline
$\Delta$-10km-HFLF-Cryo+CE-40km & 115\,179 & 110\,118 & 44\,676 & 12\,590 & 1805 \\
2L-15km-$45^{\circ}$-HFLF-Cryo+CE-40km & 116\,328 & 112\,661 & 50\,947 & 15\,545 & 2355 \\
2L-15km-$0^{\circ}$-HFLF-Cryo+CE-40km & 114\,816 & 110\,265 & 49\,034 & 14\,820 & 2243 \\
$\Delta$-10km-HFLF-Cryo+2CE & 117\,045 & 113\,910 & 52\,092 & 16\,109 & 2505 \\
2L-15km-$45^{\circ}$-HFLF-Cryo+2CE & 117\,436 & 115\,166 & 57\,678 & 19\,028 & 3126 \\
2L-15km-$0^{\circ}$-HFLF-Cryo+2CE & 116\,639 & 113\,597 & 55\,218 & 17\,849 & 2917 \\
\hline\hline
LVKI-O5 & 8603 & 2861 & 47 & 4 & 2 \\
\hline\hline
\end{tabular}
    \caption{\small Summary table of the number of BBH detections with various cuts in SNR for the considered ET geometries and ASDs, also in combination with one 40~km CE detector or two CE detectors and for a LVKI network during O5.}
    \label{tab:BBHAllConfSNR}
\end{table}

\begin{table}[htbp]
    \vspace{-.3cm}
    \hspace{-1.8cm}
    \begin{tabular}{||l||*{4}{c|}|}
\hline\hline
Configuration & $\Delta d_L/d_L \leq 0.1$ & $\Delta d_L/d_L \leq 0.01$ & $\Delta \Omega_{90\%} \leq 50\,{\rm deg}^2$ & $\Delta \Omega_{90\%} \leq 10\,{\rm deg}^2$ \\
\hline\hline
$\Delta$-10km-HFLF-Cryo & 10\,969 & 28 & 6064 & 914 \\
$\Delta$-15km-HFLF-Cryo & 17\,321 & 77 & 10\,470 & 2273 \\
2L-15km-$45^{\circ}$-HFLF-Cryo & 22\,237 & 202 & 10\,304 & 2124 \\
2L-20km-$45^{\circ}$-HFLF-Cryo & 28\,801 & 365 & 14\,920 & 3648 \\
2L-15km-$0^{\circ}$-HFLF-Cryo & 13\,865 & 79 & 3030 & 374 \\
2L-20km-$0^{\circ}$-HFLF-Cryo & 17\,008 & 144 & 4706 & 608 \\
\hline\hline
$\Delta$-10km-HF & 3919 & 6 & 2409 & 281 \\
$\Delta$-15km-HF & 8083 & 26 & 5156 & 817 \\
2L-15km-$45^{\circ}$-HF & 11\,193 & 56 & 5263 & 835 \\
2L-20km-$45^{\circ}$-HF & 16\,155 & 113 & 8448 & 1566 \\
2L-15km-$0^{\circ}$-HF & 4111 & 17 & 1054 & 120 \\
2L-20km-$0^{\circ}$-HF & 9693 & 57 & 2936 & 362 \\
\hline\hline
$\Delta$-10km-HFLF-Cryo+CE-40km & 80\,676 & 2901 & 69\,268 & 29\,924 \\
2L-15km-$45^{\circ}$-HFLF-Cryo+CE-40km & 82\,358 & 4301 & 73\,164 & 36\,457 \\
2L-15km-$0^{\circ}$-HFLF-Cryo+CE-40km & 64\,471 & 2995 & 57\,497 & 25\,782 \\
$\Delta$-10km-HFLF-Cryo+2CE & 101\,912 & 5250 & 90\,889 & 46\,744 \\
2L-15km-$45^{\circ}$-HFLF-Cryo+2CE & 104\,289 & 7006 & 95\,387 & 54\,640 \\
2L-15km-$0^{\circ}$-HFLF-Cryo+2CE & 93\,832 & 5824 & 85\,267 & 45\,429 \\
\hline\hline
LVKI-O5 & 767 & 1 & 1607 & 599 \\
\hline\hline
\end{tabular}
    \caption{\small Same as Table~\ref{tab:BBHAllConfSNR},  for BBH sources, with cuts on the relative error on the luminosity distance and on the 90\% sky localisation area.}
    \label{tab:BBHAllConfDeldLDelOm}
    \vspace{-.2cm}
\end{table}

\begin{table}[htbp]
    \hspace{-1.45cm}
 \begin{tabular}{||l||*{4}{c|}|}
\hline\hline
Configuration & $\Delta {\cal M}_c/{\cal M}_c \leq 10^{-3}$ & $\Delta {\cal M}_c/{\cal M}_c \leq 10^{-4}$ & $\Delta \chi_1 \leq 0.05$ & $\Delta \chi_1 \leq 0.01$ \\
\hline\hline
$\Delta$-10km-HFLF-Cryo & 48\,922 & 4549 & 27\,877 & 2811 \\
$\Delta$-15km-HFLF-Cryo & 64\,469 & 7703 & 41\,612 & 4856 \\
2L-15km-$45^{\circ}$-HFLF-Cryo & 58\,371 & 6456 & 35\,943 & 3958 \\
2L-20km-$45^{\circ}$-HFLF-Cryo & 67\,999 & 9073 & 45\,666 & 5706 \\
2L-15km-$0^{\circ}$-HFLF-Cryo & 57\,330 & 6472 & 33\,236 & 3653 \\
2L-20km-$0^{\circ}$-HFLF-Cryo & 63\,154 & 8279 & 40\,068 & 4935 \\
\hline\hline
$\Delta$-10km-HF & 21\,146 & 1580 & 11\,715 & 1438 \\
$\Delta$-15km-HF & 32\,643 & 2818 & 19\,956 & 2564 \\
2L-15km-$45^{\circ}$-HF & 28\,442 & 2405 & 16\,382 & 2025 \\
2L-20km-$45^{\circ}$-HF & 36\,969 & 3547 & 23\,205 & 2940 \\
2L-15km-$0^{\circ}$-HF & 25\,863 & 2146 & 13\,669 & 1652 \\
2L-20km-$0^{\circ}$-HF & 38\,537 & 3740 & 23\,065 & 2935 \\
\hline\hline
$\Delta$-10km-HFLF-Cryo+CE-40km & 73\,785 & 8901 & 54\,789 & 6649 \\
2L-15km-$45^{\circ}$-HFLF-Cryo+CE-40km & 79\,425 & 11\,187 & 60\,709 & 7988 \\
2L-15km-$0^{\circ}$-HFLF-Cryo+CE-40km & 77\,772 & 10\,885 & 57\,471 & 7519 \\
$\Delta$-10km-HFLF-Cryo+2CE & 79\,895 & 10\,523 & 62\,826 & 8226 \\
2L-15km-$45^{\circ}$-HFLF-Cryo+2CE & 84\,507 & 12\,703 & 68\,382 & 9557 \\
2L-15km-$0^{\circ}$-HFLF-Cryo+2CE & 82\,473 & 12\,307 & 65\,287 & 9012 \\
\hline\hline
LVKI-O5 & 78 & 1 & 155 & 20 \\
\hline\hline
\end{tabular}
    \caption{\small Same as Table~\ref{tab:BBHAllConfSNR},  for BBH sources, with cuts on the relative error on the detector-frame chirp mass and on the error on the spin magnitude of the primary object.}
    \label{tab:BBHAllConfDelMcDelchi}
    \vspace{-.2cm}
\end{table}

\begin{table}[t!]
    \hspace{-1.cm}
    \begin{tabular}{||l||*{5}{c|}|}
\hline\hline
Configuration & ${\rm SNR} \geq 8$ & ${\rm SNR} \geq 12$ & ${\rm SNR} \geq 50$ & ${\rm SNR} \geq 100$ & ${\rm SNR} \geq 150$ \\
\hline\hline
$\Delta$-10km-HFLF-Cryo & 107\,902 & 36\,985 & 458 & 57 & 19 \\
$\Delta$-15km-HFLF-Cryo & 213\,583 & 89\,910 & 1206 & 159 & 38 \\
2L-15km-$45^{\circ}$-HFLF-Cryo & 190\,528 & 77\,458 & 1052 & 134 & 33 \\
2L-20km-$45^{\circ}$-HFLF-Cryo & 275\,595 & 129\,821 & 2018 & 243 & 64 \\
2L-15km-$0^{\circ}$-HFLF-Cryo & 192\,030 & 78\,675 & 1040 & 136 & 33 \\
2L-20km-$0^{\circ}$-HFLF-Cryo & 274\,395 & 132\,486 & 2048 & 250 & 65 \\
\hline\hline
$\Delta$-10km-HF & 44\,713 & 13\,410 & 166 & 18 & 9 \\
$\Delta$-15km-HF & 116\,349 & 41\,181 & 516 & 55 & 17 \\
2L-15km-$45^{\circ}$-HF & 101\,550 & 34\,956 & 447 & 52 & 15 \\
2L-20km-$45^{\circ}$-HF & 176\,396 & 70\,441 & 961 & 115 & 32 \\
2L-15km-$0^{\circ}$-HF & 103\,539 & 35\,817 & 443 & 57 & 17 \\
2L-20km-$0^{\circ}$-HF & 184\,799 & 74\,805 & 989 & 124 & 37 \\
\hline\hline
$\Delta$-10km-HFLF-Cryo+CE-40km & 348\,434 & 177\,925 & 2836 & 312 & 87 \\
2L-15km-$45^{\circ}$-HFLF-Cryo+CE-40km & 392\,680 & 212\,260 & 3677 & 418 & 116 \\
2L-15km-$0^{\circ}$-HFLF-Cryo+CE-40km & 402\,234 & 220\,023 & 3770 & 414 & 119 \\
$\Delta$-10km-HFLF-Cryo+2CE & 406\,630 & 220\,725 & 3961 & 436 & 120 \\
2L-15km-$45^{\circ}$-HFLF-Cryo+2CE & 442\,526 & 252\,136 & 4900 & 559 & 152 \\
2L-15km-$0^{\circ}$-HFLF-Cryo+2CE & 448\,798 & 258\,615 & 4974 & 531 & 162 \\
\hline\hline
LVKI-O5 & 250 & 71 & 3 & 0 & 0 \\
\hline\hline
\end{tabular}
    \caption{\small Same as Table~\ref{tab:BBHAllConfSNR} for BNS sources.}
    \label{tab:BNSAllConfSNR}
\end{table}

\begin{table}[htbp]
    \hspace{-1.8cm}
    \begin{tabular}{||l||*{4}{c|}|}
\hline\hline
Configuration & $\Delta d_L/d_L \leq 0.3$ & $\Delta d_L/d_L \leq 0.1$ & $\Delta \Omega_{90\%} \leq 100\,{\rm deg}^2$ & $\Delta \Omega_{90\%} \leq 10\,{\rm deg}^2$ \\
\hline\hline
$\Delta$-10km-HFLF-Cryo & 748 & 52 & 184 & 8 \\
$\Delta$-15km-HFLF-Cryo & 1756 & 153 & 479 & 23 \\
2L-15km-$45^{\circ}$-HFLF-Cryo & 4328 & 479 & 559 & 25 \\
2L-20km-$45^{\circ}$-HFLF-Cryo & 7821 & 919 & 1028 & 43 \\
2L-15km-$0^{\circ}$-HFLF-Cryo & 774 & 48 & 293 & 12 \\
2L-20km-$0^{\circ}$-HFLF-Cryo & 1499 & 104 & 565 & 23 \\
\hline\hline
$\Delta$-10km-HF & 4 & 1 & 4 & 0 \\
$\Delta$-15km-HF & 7 & 1 & 11 & 1 \\
2L-15km-$45^{\circ}$-HF & 126 & 12 & 11 & 0 \\
2L-20km-$45^{\circ}$-HF & 262 & 22 & 24 & 1 \\
2L-15km-$0^{\circ}$-HF & 20 & 1 & 11 & 1 \\
2L-20km-$0^{\circ}$-HF & 28 & 2 & 24 & 1 \\
\hline\hline
$\Delta$-10km-HFLF-Cryo+CE-40km & 32\,053 & 4100 & 54\,994 & 2427 \\
2L-15km-$45^{\circ}$-HFLF-Cryo+CE-40km & 45\,252 & 7949 & 75\,828 & 3838 \\
2L-15km-$0^{\circ}$-HFLF-Cryo+CE-40km & 16\,999 & 2079 & 29\,821 & 1515 \\
$\Delta$-10km-HFLF-Cryo+2CE & 72\,335 & 13\,630 & 112\,705 & 6570 \\
2L-15km-$45^{\circ}$-HFLF-Cryo+2CE & 89\,877 & 19\,129 & 145\,272 & 9841 \\
2L-15km-$0^{\circ}$-HFLF-Cryo+2CE & 78\,798 & 14\,909 & 125\,640 & 7592 \\
\hline\hline
LVKI-O5 & 12 & 1 & 51 & 31 \\
\hline\hline
\end{tabular}
    \caption{\small Same as Table~\ref{tab:BBHAllConfDeldLDelOm} for BNS sources.}
    \label{tab:BNSAllConfDeldLDelOm}
\end{table}

\begin{table}[ht]
    \hspace{-1.7cm}
    \begin{tabular}{||l||*{4}{c|}|}
\hline\hline
Configuration & $\Delta {\cal M}_c/{\cal M}_c \leq 10^{-3}$ & $\Delta {\cal M}_c/{\cal M}_c \leq 10^{-4}$ & $\Delta \tilde{\Lambda}/\tilde{\Lambda} \leq 0.1$ & $\Delta \tilde{\Lambda}/\tilde{\Lambda} \leq 0.05$ \\
\hline\hline
$\Delta$-10km-HFLF-Cryo & 35\,127 & 18\,401 & 1040 & 96 \\
$\Delta$-15km-HFLF-Cryo & 84\,835 & 39\,607 & 2783 & 227 \\
2L-15km-$45^{\circ}$-HFLF-Cryo & 68\,391 & 36\,645 & 2463 & 200 \\
2L-20km-$45^{\circ}$-HFLF-Cryo & 115\,695 & 59\,051 & 5189 & 386 \\
2L-15km-$0^{\circ}$-HFLF-Cryo & 67\,023 & 29\,813 & 2225 & 179 \\
2L-20km-$0^{\circ}$-HFLF-Cryo & 113\,726 & 49\,445 & 4703 & 374 \\
\hline\hline
$\Delta$-10km-HF & 10\,831 & 687 & 248 & 21 \\
$\Delta$-15km-HF & 31\,663 & 1548 & 667 & 45 \\
2L-15km-$45^{\circ}$-HF & 25\,058 & 2033 & 634 & 47 \\
2L-20km-$45^{\circ}$-HF & 50\,464 & 3644 & 1494 & 90 \\
2L-15km-$0^{\circ}$-HF & 22\,839 & 1463 & 445 & 36 \\
2L-20km-$0^{\circ}$-HF & 42\,731 & 2292 & 780 & 72 \\
\hline\hline
$\Delta$-10km-HFLF-Cryo+CE-40km & 171\,997 & 46\,008 & 2669 & 243 \\
2L-15km-$45^{\circ}$-HFLF-Cryo+CE-40km & 197\,631 & 66\,214 & 4753 & 400 \\
2L-15km-$0^{\circ}$-HFLF-Cryo+CE-40km & 198\,171 & 63\,219 & 4488 & 371 \\
$\Delta$-10km-HFLF-Cryo+2CE & 216\,875 & 58\,522 & 3847 & 337 \\
2L-15km-$45^{\circ}$-HFLF-Cryo+2CE & 244\,198 & 80\,406 & 6319 & 535 \\
2L-15km-$0^{\circ}$-HFLF-Cryo+2CE & 250\,033 & 79\,746 & 6106 & 505 \\
\hline\hline
LVKI-O5 & 54 & 2 & 2 & 0 \\
\hline\hline
\end{tabular}
    \caption{\small Same as Table~\ref{tab:BBHAllConfDelMcDelchi}, for BNS sources, with cuts on the relative error on the detector-frame chirp mass and on the relative error on the tidal deformability parameter $\tilde{\Lambda}$.}
    \label{tab:BNSAllConfDelMcDelLam}
\end{table}

\clearpage

\section{Correlation between parameters for typical events}\label{app:corrBBHs}

In this appendix we discuss the correlation between parameters,  in the BBH parameter estimation. This is instructive to obtain a physical understanding of some of the results presented in the main text. In particular, we wish to understand why the 15~km triangle and the 15~km 2L at $45^{\circ}$ have performances very similar   for the reconstruction  of all parameters, except for luminosity distance, where the 15~km 2L at $45^{\circ}$ is clearly superior, see  Fig.~\ref{fig:AllGeoms_CumulBBH_NdetScale} and  Tables~\ref{tab:BBHAllConfDeldLDelOm}, \ref{tab:BBHAllConfDelMcDelchi}. To this purpose, we consider in detail some selected events (the correlation patterns that we will discuss are in fact generic). We select some BBH event with very high SNR, so that the Fisher matrix analysis is expected to be more reliable. As a first example we  consider, in our ensemble of detected events, a `light' BBH event  (that we denote as ``event 1'') with source-frame masses $m_1\simeq 7.9\msun$, $m_2=7.6\msun$, at a distance $d_L\simeq 0.30$~Gpc ($z\simeq 0.065$). Due to its close distance, its SNR is very high, and is 513 in the 15~km triangle and 478 in the 15~km 2L at $45^{\circ}$. 

\begin{figure}[t]
\centering
\includegraphics[width=1.0\textwidth]{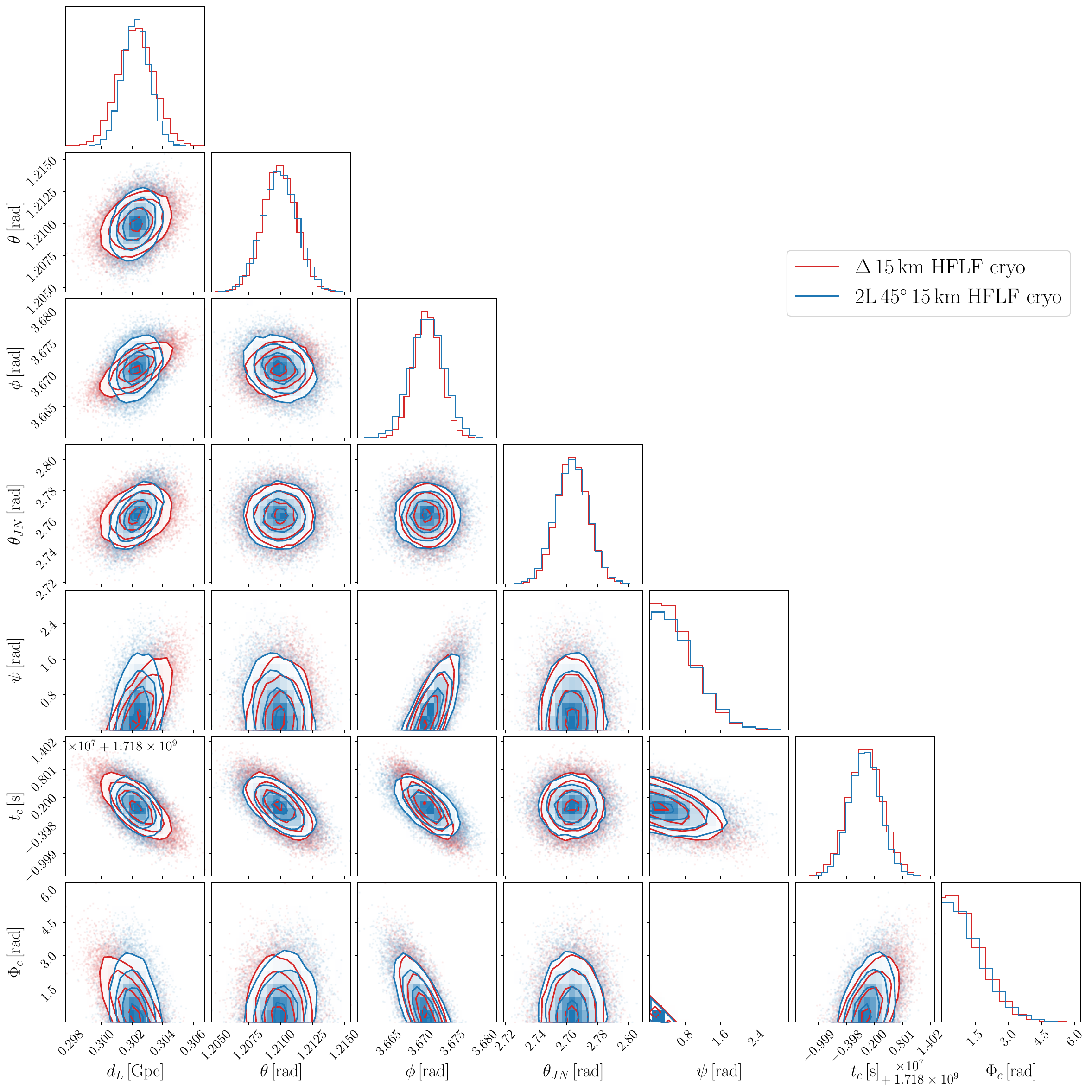}
\caption{\small A corner plot showing the correlations between various parameters for a BBH event with light masses and very close distance (``event 1", see the text).}
\label{fig:Tvs2Lcontours_ev1}
\end{figure}

In Fig.~\ref{fig:Tvs2Lcontours_ev1} we show the correlation between some parameters. For readability,   within the 15-dimensional parameter space that we use for BBHs,  we restrict   to the most interesting correlations, and we compare the results for the 15~km 2L at $45^{\circ}$ with the 15~km triangle, both in their full HFLF-cryo configuration. The interesting point is that, in the correlation of various parameters with the luminosity distance, the contours for the 15~km triangle (red) are  in general more tilted than the blue contour referring to the 15~km 2L at $45^{\circ}$. In Fig.~\ref{fig:Tvs2Lcontours_ev1} we see this in the correlation with all the parameters shown. This implies that, when one marginalizes over these parameters, the same  error on these parameters induces a larger marginalized error on $d_L$. We see indeed that, for all parameters except $d_L$, for this event the marginalized errors are the same for the 15~km triangle and the 15~km 2L at $45^{\circ}$, while the marginalized distribution for $d_L$ is narrower for the 15~km 2L at $45^{\circ}$. Observe that this happens despite the fact that, for this event, the SNR in the triangle was higher than in the 15~km 2L at $45^{\circ}$. 

\begin{figure}[t]
\centering
\includegraphics[width=1.0\textwidth]{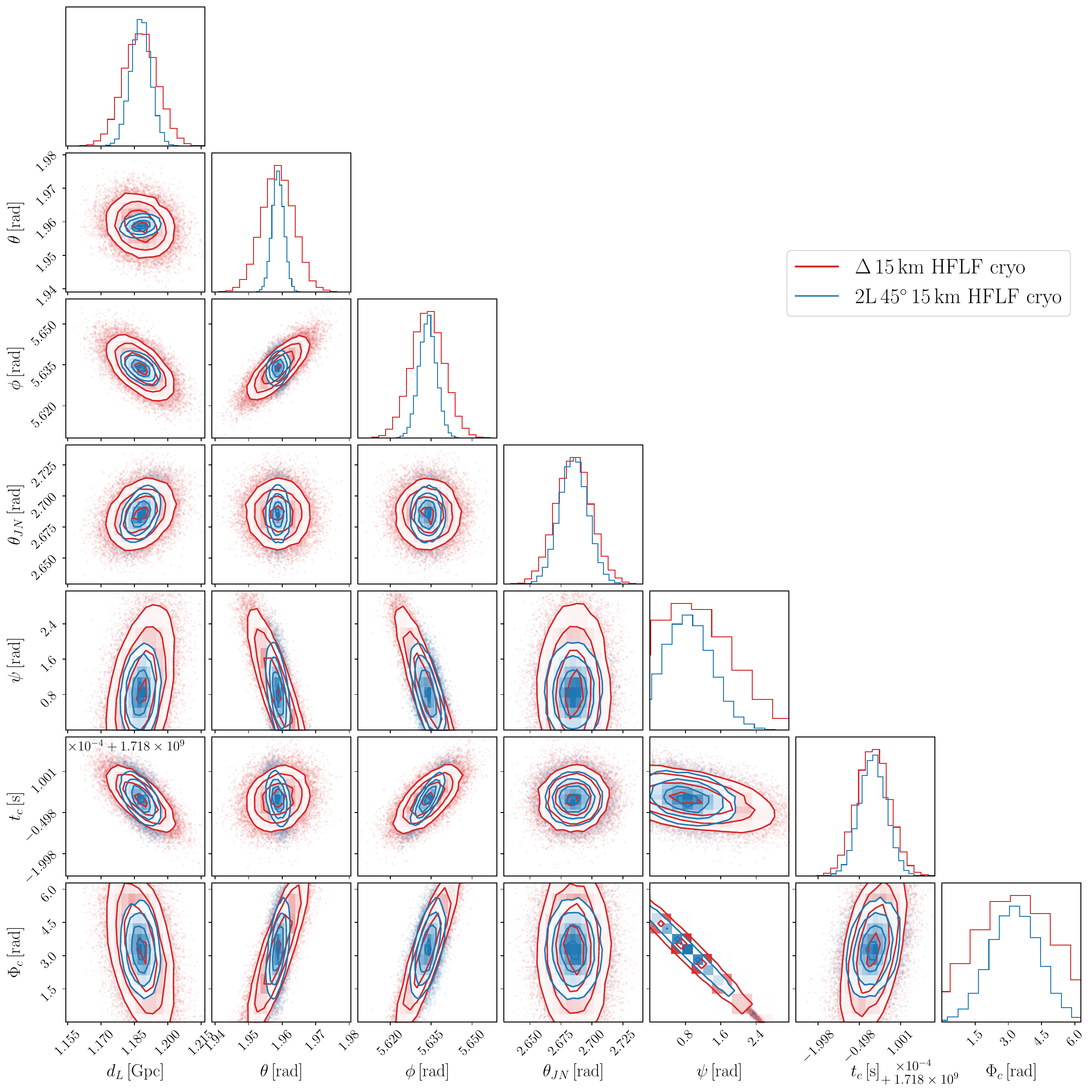}
\caption{\small As in Fig.~\ref{fig:Tvs2Lcontours_ev1} for a BBH event with heavier  masses and greater distance (``event 2", see the text).}
\label{fig:Tvs2Lcontours_ev2}
\end{figure}

Fig.~\ref{fig:Tvs2Lcontours_ev2} shows the analogous results for an event  (``event 2'') with source-frame masses $m_1\simeq 29.0\msun$, $m_2=24.6\msun$, at a distance $d_L\simeq 1.19$~Gpc ($z\simeq 0.23$). Its SNR is again very high, 424 in the 15~km triangle and 318 in the 15~km 2L at $45^{\circ}$.  As we reminded in Section~\ref{sect:geometry}, the differences in the SNR between configurations, alone, is not a good indicator of the relative performances of parameter estimation (the most obvious example being that, on an ensemble of events, a 2L parallel  configuration has higher SNR but worse angular localization than a 2L at $45^{\circ}$). This is also clearly visible from this plot: in this case, despite the larger SNR in the 15~km triangle, most parameters are significantly better estimated by the  15~km 2L at $45^{\circ}$. We also see that, again, the correlation contours of $d_L$ with various parameters are more tilted in the triangle case, contributing to enlarging the marginalized posterior of $d_L$. This effect is also particularly evident, for this event, in the correlation between $\theta$ and $\phi$.

\clearpage

\printglossary[type=\acronymtype]

\clearpage

\bibliographystyle{utphys}
\bibliography{CoBAScience.bib} 
\end{document}